\documentclass[11pt,letterpaper]{article}
\usepackage[margin=0.82in]{geometry}
\usepackage[T1]{fontenc}
\usepackage[utf8]{inputenc}
\usepackage{lmodern,microtype}
\usepackage{amsmath,amssymb,amsthm,mathtools}
\usepackage{booktabs,tabularx,array,longtable}
\usepackage[dvipsnames]{xcolor}
\usepackage{enumitem,seqsplit}
\usepackage{graphicx,fancyvrb}
\usepackage{fancyhdr}
\usepackage[colorlinks=true,allcolors=blue,hypertexnames=false]{hyperref}
\usepackage{url,xurl}
\usepackage{appendix}

\definecolor{navy}{HTML}{17365D}
\definecolor{green}{HTML}{176B45}
\definecolor{amber}{HTML}{9A6700}
\definecolor{redx}{HTML}{A12622}
\definecolor{softamber}{HTML}{FFF5D6}

\newcommand{\HH}{H_2}
\newcommand{\Rphi}{R_{\phi}}
\newcommand{\Target}{\mathcal T}
\newcommand{\Gap}{G}
\newcommand{\artifact}[1]{\path{#1}}
\newcommand{\hashcode}[1]{{\footnotesize\ttfamily\seqsplit{#1}}}
\newcommand{\level}[1]{\textsc{#1}}
\newcommand{\F}{\overline F}
\newcommand{\Ph}{\overline\Phi}
\newcommand{\et}{\overline\eta}
\newcommand{\kap}{\overline\kappa}
\newcommand{\RR}{R_\phi}

\newcommand{\atanh}{\operatorname{atanh}}
\newcommand{\asinh}{\operatorname{asinh}}

\newcommand{\code}[1]{\path{#1}}

\newcommand{\Hd}{H_2}
\newcommand{\GDB}{G_{\mathrm{dbl}}}
\newcommand{\GHF}{G_{\mathrm{HF}}}

\newtheorem{theorem}{Theorem}[section]
\newtheorem{lemma}[theorem]{Lemma}
\newtheorem{proposition}[theorem]{Proposition}
\newtheorem{corollary}[theorem]{Corollary}
\theoremstyle{definition}

\theoremstyle{remark}
\newtheorem{remark}[theorem]{Remark}

\newcommand{\E}{\mathbb E}

\providecommand{\tightlist}{\setlength{\itemsep}{0pt}\setlength{\parskip}{0pt}}

\newcommand{\sourcefile}[1]{\par\smallskip\noindent\textit{Certificate source:} \path{#1}.\par\smallskip}
\ifdefined\DeclareUnicodeCharacter
\DeclareUnicodeCharacter{2212}{-}
\DeclareUnicodeCharacter{2013}{--}
\DeclareUnicodeCharacter{2014}{---}
\DeclareUnicodeCharacter{2018}{`}
\DeclareUnicodeCharacter{2019}{'}
\DeclareUnicodeCharacter{201C}{``}
\DeclareUnicodeCharacter{201D}{''}
\DeclareUnicodeCharacter{2264}{\ensuremath{\le}}
\DeclareUnicodeCharacter{2265}{\ensuremath{\ge}}
\DeclareUnicodeCharacter{2192}{\ensuremath{\to}}
\DeclareUnicodeCharacter{21D2}{\ensuremath{\Rightarrow}}
\DeclareUnicodeCharacter{2208}{\ensuremath{\in}}
\DeclareUnicodeCharacter{221E}{\ensuremath{\infty}}
\DeclareUnicodeCharacter{03B6}{\ensuremath{\zeta}}
\DeclareUnicodeCharacter{03C6}{\ensuremath{\phi}}
\DeclareUnicodeCharacter{03D5}{\ensuremath{\phi}}
\DeclareUnicodeCharacter{03C8}{\ensuremath{\psi}}
\DeclareUnicodeCharacter{03B7}{\ensuremath{\eta}}
\DeclareUnicodeCharacter{03B4}{\ensuremath{\delta}}
\DeclareUnicodeCharacter{0394}{\ensuremath{\Delta}}
\DeclareUnicodeCharacter{03C1}{\ensuremath{\rho}}
\DeclareUnicodeCharacter{03C4}{\ensuremath{\tau}}
\DeclareUnicodeCharacter{03BA}{\ensuremath{\kappa}}
\DeclareUnicodeCharacter{03BC}{\ensuremath{\mu}}
\DeclareUnicodeCharacter{03BB}{\ensuremath{\lambda}}
\DeclareUnicodeCharacter{03B5}{\ensuremath{\epsilon}}
\DeclareUnicodeCharacter{03B3}{\ensuremath{\gamma}}
\DeclareUnicodeCharacter{03B2}{\ensuremath{\beta}}
\DeclareUnicodeCharacter{03B1}{\ensuremath{\alpha}}
\DeclareUnicodeCharacter{03BD}{\ensuremath{\nu}}
\DeclareUnicodeCharacter{03A6}{\ensuremath{\Phi}}
\DeclareUnicodeCharacter{00B2}{\ensuremath{{}^2}}
\DeclareUnicodeCharacter{00D7}{\ensuremath{\times}}
\DeclareUnicodeCharacter{2248}{\ensuremath{\approx}}
\fi

\makeatletter
\renewcommand{\@pnumwidth}{3.0em}
\renewcommand{\@tocrmarg}{4.0em}
\renewcommand*{\l@section}{\@dottedtocline{1}{1.5em}{3.6em}}
\renewcommand*{\l@subsection}{\@dottedtocline{2}{5.1em}{4.2em}}
\let\orig@l@part\l@part
\renewcommand*{\l@part}[2]{\orig@l@part{#1}{#2}\nobreak\vskip 4pt\relax}
\makeatother
\newcommand{\startmanuscriptpart}[3]{%
  \clearpage
  \fancyhead[L]{\small\color{navy}#2}%
  \fancyhead[R]{\small\color{navy}#3}%
  \part{#1}}
\newcommand{\startsupplement}[1]{%
  \clearpage
  \setcounter{section}{0}\setcounter{equation}{0}%
  \setcounter{table}{0}\setcounter{figure}{0}%
  \renewcommand{\thesection}{S.\arabic{section}}%
  \renewcommand{\theequation}{S.\arabic{equation}}%
  \renewcommand{\thetable}{S.\arabic{table}}%
  \renewcommand{\thefigure}{S.\arabic{figure}}%
  \phantomsection\addcontentsline{toc}{part}{#1}%
  \part*{#1}}
\hypersetup{pdfauthor={Zijie Chen, Amin Gohari, Adel Javanmard, Honghao Lin,
Vahab Mirrokni, Chandra Nair, David P. Woodruff}}
\hypersetup{pdftitle={A Proof of the Most Informative Boolean Function Conjecture}}
\begin{document}
\title{A Proof of the Most Informative Boolean Function Conjecture}

\author{%
  Zijie Chen\textsuperscript{1}\qquad
  Amin Gohari\textsuperscript{1}\qquad
  Adel Javanmard\textsuperscript{2,3}\qquad
  Honghao Lin\textsuperscript{2}\\[6pt]
  Vahab Mirrokni\textsuperscript{2}\qquad
  Chandra Nair\textsuperscript{1}\qquad
  David P.~Woodruff\textsuperscript{2,4}\\[14pt]
  {\small\textsuperscript{1}Department of Information Engineering,
   The Chinese University of Hong Kong}\\[2pt]
  {\small\textsuperscript{2}Google\qquad
   \textsuperscript{3}University of Southern California\qquad
      \textsuperscript{4}Carnegie Mellon University}\\[10pt]
  {\footnotesize\texttt{\{zijie,agohari,chandra\}@ie.cuhk.edu.hk}}\\[1pt]
  {\footnotesize\texttt{\{adeljavanmard,honghaol,mirrokni,woodruffd\}@google.com}}}

\date{}
\maketitle

\begin{abstract}
Let $X$ be uniform on $\{-1,1\}^n$, let $Y$ be obtained by passing its
coordinates independently through a binary symmetric channel with
crossover probability $p$, and let $g:\{-1,1\}^n\to\{0,1\}$ be a Boolean function. We give a
computer-assisted proof of the Courtade--Kumar conjecture
$I(g(X);Y)\le1-H_2(p)$, where $H_2$ is binary entropy, with equality attained by dictator functions.
The present work builds on the differential-equation method, itself a limiting form of the
auxiliary-receiver approach in network information theory using a continuum
of degraded receivers.
The proof proceeds from a local inequality to a dimension-independent
bound on entropy production. Differentiation along the Boolean noise
semigroup expresses entropy production as an average of edge costs.
The key estimate is therefore an unrestricted Bellman inequality with
two mean constraints and two entropy constraints, allowing arbitrary
couplings of the edge variables. 

This paper and its supplement provide the proofs and computational verification records. The document is lengthy because it is designed to be entirely self-contained, deriving all proofs from first principles and reproducing the proofs of cited results. We also provide a self-contained \href{http://github.com/dpwoodru/general-courtade-kumar-lean/blob/main/short-summary.pdf}{expository note}
explaining the reduction to a low-dimensional inequality
and the ideas behind the key lower bounds.
\end{abstract}
\begin{center}
  \emph{The entire proof, including all numerical certificates,
  has been formally verified in Lean end-to-end. 
  The lean formalization is available
  \href{https://github.com/dpwoodru/general-courtade-kumar-lean}{online}.}
\end{center}
\clearpage\tableofcontents\clearpage
\section{Introduction}\label{sec:introduction}
Let $X$ be uniform on $\{-1,1\}^n$, and let $Y$ be the output of a binary
symmetric channel with crossover probability $p$ whose input is $X$. How
much information about $Y$ can a single bit $g(X)$ carry? A dictator
function $g(x)=(1+x_i)/2$ gives $I(g(X);Y)=1-H_2(p)$, where $H_2$ is the
binary entropy. Courtade and Kumar conjectured that no Boolean function
does better, in any dimension
\cite{KumarCourtade2013,CourtadeKumar2014}. This is the most informative
Boolean function conjecture, or Courtade--Kumar (CK) conjecture. We prove
it.

\begin{theorem}[Courtade--Kumar inequality]\label{unbal:thm:ck}
Let $n\ge1$, let $X$ be uniform on $\{-1,1\}^{n}$, and let $Y$ be obtained
by flipping each coordinate of $X$ independently with probability
$p\in[0,1]$. Every Boolean function $g:\{-1,1\}^{n}\to\{0,1\}$ satisfies
\begin{equation}\label{eq:CK-conjecture}
 I(g(X);Y)\le1-H_2(p),\qquad
 H_2(p)=-p\log_2p-(1-p)\log_2(1-p).
\end{equation}
Equality is attained by dictator functions $g(x)=(1+x_i)/2$ and their complements.
\end{theorem}

\paragraph{Previous work.}
Courtade and Kumar verified the conjecture numerically for Boolean
functions of dimension at most seven
\cite{KumarCourtade2013,CourtadeKumar2014}. It has since been studied with
tools from information theory, discrete probability, Fourier analysis, and
isoperimetry on the Boolean cube.

Several variants and equivalent formulations are known. Anantharam, Gohari,
Kamath, and Nair considered a weaker two-function variant, which asks
whether, for $(X,Y)$ as in Theorem~\ref{unbal:thm:ck} and arbitrary Boolean
functions $f$ and $g$, the mutual information $I(f(X);g(Y))$ is maximized by
dictator functions. They related this variant to a finite-dimensional
conjecture involving chordal slopes of hypercontractivity ribbons and
certain information-theoretic functionals
\cite{AnantharamGohariKamathNair2013}. The two-function variant was later
proved by Pichler, Piantanida, and Matz by a different method
\cite{pichler2018dictator}; the chordal-slope conjecture would give a
separate proof. Li and M\'edard studied a related \(q\)-stability problem
and conjectured dictator optimality for balanced Boolean functions when
\(1\le q\le2\) \cite{li2021boolean}. Barnes and \"Ozg\"ur showed that a
symmetrized form of the Li--M\'edard conjecture is equivalent to the
balanced Courtade--Kumar conjecture \cite{barnes2020courtade}.

Other work gave bounds, or proved the conjecture for part of the parameter
range. Ordentlich, Shayevitz, and Weinstein obtained an improved universal
upper bound for balanced Boolean functions \cite{ordentlich2016improved}.
Samorodnitsky proved the conjecture when the absolute channel correlation
\(|1-2p|\) is smaller than a universal positive constant, and developed
further estimates for the entropy of noisy Boolean functions
\cite{samorodnitsky2016entropy}. Yu introduced a general framework for
\(\Phi\)-stability and related Boolean-function conjectures, and obtained
additional ranges of parameters for which dictator functions are optimal
\cite{yu2023phi}. In later work, Yu established local optimality of
dictator functions for a broad class of stability functionals and, by
combining this with computer-assisted estimates, extended the range of
channel parameters for which the balanced conjecture is verified
\cite{yu2024localoptimalitydictatorfunctions}. Javanmard and Woodruff
proved a sharp bound on the sum of coordinate-wise mutual informations,
valid without a balance assumption, and obtained an optimal second-order
error term in the high-noise entropy expansion
\cite{javanmard2026progresscourtadekumarconjectureoptimal}.

Stronger conjectures have also been studied. Anantharam, Bogdanov,
Chakrabarti, Jayram, and Nair proposed a Hellinger-type inequality that
implies the Courtade--Kumar bound and is related to sharp isoperimetric and
sensitivity inequalities on the cube \cite{anantharam2017hellinger}. Chen
and Nair introduced a family of entropy inequalities that interpolates
between the mutual-information and Hellinger formulations, and established
a monotone ordering within that family \cite{chen2024optimality}.

\paragraph{Approach and contributions.}
We follow the differential-equation method of Chen, Gohari, and Nair
\cite{ChenGohariNair2025}, a limiting form of the auxiliary-receiver
approach of Gohari and Nair \cite{gon21}.\footnote{The original
auxiliary-receiver approach in \cite{gon21} suggests considering
\(I(g(X);Y)-I(g(X);Z)\) for an auxiliary receiver \(Z\), and using
Gallager's past/future receiver symbols to single-letterize the resulting
expressions. In \cite{ChenGohariNair2025}, the authors assume that \(Z\) is
very close to \(Y\), so that \(I(g(X);Y)-I(g(X);Z)\) corresponds to the
derivative of the mutual-information term. They then replace the standard
past/future single-letterization argument with an induction argument
applied directly to this derivative.}
The method follows the conditional entropy of \(g(X)\) given the channel
output as the noise increases. It reduces a lower bound on this entropy, in
every dimension, to a single inequality for pairs of \([0,1]\)-valued
random variables with prescribed means and prescribed expected entropies.
We call it the \emph{Bellman inequality} for a function \(A\) of a mean and
an entropy, and we call such a function a \emph{candidate}; the definitions
are in Section~\ref{sec:intro-local}. In \cite{ChenGohariNair2025} a
candidate \(\phi\) was proposed, and its Bellman inequality was shown to
imply the conjecture for balanced functions. The inequality itself was left
there as a conjecture
\cite[Conjecture~3]{ChenGohariNair2025}.

This paper makes two contributions. First, we prove the Bellman inequality
for \(\phi\) (Theorem~\ref{bal:thm:global-unrestricted-target}), which
settles the balanced case. Second, we introduce the candidate
\(B=\max\{\phi,\psi\}\) defined in \eqref{eq:intro-candidates}, in which
\(\psi\) accounts for the entropy deficit \(1-H_2(\mathbb Eg(X))\) of a
biased function, and we prove its Bellman inequality
(Theorem~\ref{thm:intro-local}); this gives Theorem~\ref{unbal:thm:ck}.
A pointwise maximum of candidates that satisfy the Bellman inequality
satisfies it as well \cite{ChenGohariNair2025}, but \(\psi\) alone does not
satisfy it,
so Theorem~\ref{thm:intro-local} does not follow from the result for
\(\phi\). Both proofs combine analytic estimates with interval arithmetic
on compact parameter boxes. Section~\ref{sec:intro-local} states the local
inequality and deduces Theorem~\ref{unbal:thm:ck} from it,
Section~\ref{sec:intro-overview} outlines the proof of the local
inequality, and Section~\ref{sec:intro-organization} describes the
computer-assisted part and the organization of the paper.

For a concise account of the reduction to a low-dimensional
inequality and the ideas behind the key lower bounds, see our
\href{https://github.com/dpwoodru/general-courtade-kumar-lean/blob/main/short-summary.pdf}{short, self-contained expository note}.

\paragraph{Lean formalization}
The proof, including the analytic reductions and all numerical certificates,
is verified end-to-end in Lean. The formalization and instructions for
reproducing the verification are available in the
\href{https://github.com/dpwoodru/general-courtade-kumar-lean}{GitHub repository}.
Supplementary materials supporting the computer-assisted parts of the proof
are also
\href{https://drive.google.com/file/d/1wcudPRFLoLAfuFFUW4fRxVt-VKqAJ5O0/view?usp=sharing}{available online}.

\paragraph{Independent work.}
While preparing this manuscript, we learned through personal communication of
independent work by Vu Khac Ky and Tuan Tran~\cite{KyTran2026}, who have
obtained a proof of the conjecture.
Their proof is substantially different from ours: it develops a new
entropy-production and spectral framework, whereas our proof builds on a
differential-equation approach. We thank Vu Khac Ky and Tuan Tran for their
collegiality in coordinating the simultaneous posting of the two manuscripts
on arXiv.

Shortly after we posted our paper on arXiv, two other groups contacted us. Matthew Ho and a collaborator shared, in private communication, an outline of a Bellman induction bound without its proof, an AI-generated document they were using to study the inductive hypothesis, and a link to a folder of Lean files. They indicated that a human-readable write-up was still in preparation and expected in approximately two weeks.
Separately, Mahdavifar and Beirami \cite{mahdavifar2026} shared a manuscript proving the Courtade-Kumar conjecture. Their disclosure states that generative AI systems produced its detailed mathematical derivations, with the authors providing high-level direction, feedback, and final review. Mahdavifar and Beirami also studied the multibit extension of the Courtade--Kumar conjecture. Chandar and Tchamkerten had previously shown that the \(k\)-bit extension of the Courtade--Kumar conjecture has counterexamples for \(k \geq 11\)~\cite{chandar2014}. Mahdavifar and Beirami construct counterexamples for \(k \geq 8\) and conjecture that the extension holds for \(2 \leq k \leq 7\).

\subsection{The local inequality}\label{sec:intro-local}

This subsection states the local inequality and shows how
Theorem~\ref{unbal:thm:ck} follows from it. The framework, up to and
including the candidate \(\phi\), is that of \cite{ChenGohariNair2025}; the
candidate \(B\) and Theorem~\ref{thm:intro-local} are new. There are four
steps. (i)~By \eqref{eq:intro-mi-identity}, it suffices to bound the
conditional entropy of \(g(X)\) given \(Y\) from below. (ii)~Along the noise
semigroup, this entropy grows at a rate equal to the edge energy
\(\mathcal E_n\) of the posterior probabilities; see
\eqref{eq:intro-entropy-production}. (iii)~Splitting the cube along one
coordinate expresses \(\mathcal E_n\) through the energies of the two halves
and a two-point cost \eqref{eq:intro-energy-split}. If a function \(A\) of
the mean and the average entropy satisfies the Bellman inequality
\(\zeta\ge R_A\), then induction on \(n\) gives \(\mathcal E_n\ge A\), and
hence the differential inequality \eqref{eq:intro-ode}. (iv)~For \(A=B\),
this inequality is compared with the entropy of a single bit sent through
the same channel, which is the dictator case. The proof of
Theorem~\ref{unbal:thm:ck} therefore rests on
Theorem~\ref{thm:intro-local}.

Throughout, \(J(t)=H_2'(t)=\log_2\frac{1-t}{t}\), and
\(H_2^{-1}:[0,1]\to[0,1/2]\) is the inverse of \(H_2\) on the lower half
of the unit interval.

\paragraph{Entropy along the noise semigroup.}
For \(\rho\in[0,1]\), let \(T_\rho\) be the noise operator: for a function
\(v\) on \(\{-1,1\}^n\),
\((T_\rho v)(y)=\mathbb E[v(X)\mid Y=y]\), where \(Y\) is obtained from the
uniform input \(X\) by flipping each coordinate independently with
probability \((1-\rho)/2\). If \(g\) is Boolean and \(0\le p\le1/2\), then
\((T_{1-2p}g)(y)\) is the posterior probability that \(g(X)=1\) given
\(Y=y\), and therefore
\begin{equation}\label{eq:intro-mi-identity}
 I(g(X);Y)=H_2(\mathbb Eg(X))-\mathbb EH_2\bigl((T_{1-2p}g)(Y)\bigr).
\end{equation}
The case \(p>1/2\) reduces to the case \(1-p\) by complementing the channel
output.
The first term of \eqref{eq:intro-mi-identity} depends only on the mean of
\(g\). Theorem~\ref{unbal:thm:ck} is thus a lower bound on the conditional entropy
\(\mathbb EH_2((T_{1-2p}g)(Y))\), and, as in \cite{ChenGohariNair2025}, we
obtain it by following this quantity along the semigroup
\(T_{e^{-2t}}\), \(t\ge0\).

For an array \(v:\{-1,1\}^n\to(0,1)\), a coordinate \(i\), and
\(y_{-i}\in\{-1,1\}^{n-1}\), let \(v_i^{\pm}(y_{-i})\) be the value of \(v\)
at the point whose \(i\)th coordinate is \(\pm1\) and whose other
coordinates are \(y_{-i}\). The \emph{edge cost} and the \emph{edge energy}
of \(v\) are
\begin{equation}\label{eq:intro-edge-energy}
\begin{gathered}
 j(u,w)=\frac12(u-w)\bigl(J(w)-J(u)\bigr)
       =\frac12\bigl(D_2(u\Vert w)+D_2(w\Vert u)\bigr),\\
 \mathcal E_n(v)=\sum_{i=1}^n2^{-(n-1)}\sum_{y_{-i}}
   j\bigl(v_i^-(y_{-i}),v_i^+(y_{-i})\bigr),
\end{gathered}
\end{equation}
where \(D_2\) is the binary relative entropy in bits; in particular
\(j\ge0\). Let \(v_0:\{-1,1\}^n\to(0,1)\), put \(v_t=T_{e^{-2t}}v_0\), and let
\(\gamma(t)=2^{-n}\sum_yH_2(v_t(y))\) be the average entropy of \(v_t\).
The average of \(v_t\) does not depend on \(t\), and differentiation gives
\begin{equation}\label{eq:intro-entropy-production}
 \gamma'(t)=\mathcal E_n(v_t);
\end{equation}
see the proof of
Corollary~\ref{bal:cor:unrestricted-differential-induction}. The problem is
therefore to bound the edge energy of an arbitrary array from below by a
function of its average and its average entropy.

\paragraph{The two-point problem.}
Split the cube along its last coordinate. Let \(v^-\) and \(v^+\) be the
restrictions of \(v\) to the two sections, let \(Y'\) be uniform on
\(\{-1,1\}^{n-1}\), and put \(U=v^-(Y')\) and \(W=v^+(Y')\). Then
\begin{equation}\label{eq:intro-energy-split}
 \mathcal E_n(v)=\frac12\mathcal E_{n-1}(v^-)+\frac12\mathcal E_{n-1}(v^+)
 +\mathbb Ej(U,W).
\end{equation}
The induction on \(n\) below retains four statistics of the pair
\((U,W)\), the \emph{four-moment tuple}
\[
 M=(a,b,e,f)=\bigl(\mathbb EU,\ \mathbb EW,\ \mathbb EH_2(U),\
 \mathbb EH_2(W)\bigr).
\]
We define
\begin{equation}\label{eq:intro-zeta}
 \zeta(M)=\inf\bigl\{\mathbb Ej(U,W):\ (U,W)\text{ is a pair of }
 [0,1]\text{-valued random variables with four-moment tuple }M\bigr\}.
\end{equation}
The infimum is over all joint laws. The two sections of an arbitrary array
are not ordered, so \(U\le W\) is not assumed. On the boundary of the unit
square, \(j(0,0)=j(1,1)=0\) and \(j=+\infty\) elsewhere. A tuple \(M\) is
\emph{feasible}, that is, realized by some law, exactly when
\(0\le a,b\le1\), \(0\le e\le H_2(a)\), and \(0\le f\le H_2(b)\)
(Lemma~\ref{bal:lem:unrestricted-feasible}).

\paragraph{The Bellman inequality.}
Let \(A(m,e)\) be a function of a mean \(m\in(0,1)\) and an entropy
\(0<e\le H_2(m)\). For a feasible tuple with \(0<a,b<1\) and \(e,f>0\), put
\begin{equation}\label{eq:intro-increment}
 R_A(M)=A\Bigl(\frac{a+b}2,\frac{e+f}2\Bigr)
        -\frac{A(a,e)+A(b,f)}2.
\end{equation}
We say that \(A\) satisfies the \emph{Bellman inequality} if
\(\zeta(M)\ge R_A(M)\) for every such \(M\); the terminology is that of the
Bellman function method, in which an inequality in all dimensions is
obtained by induction from a two-point inequality. The functions \(A\)
used below are called \emph{candidates}.

Suppose that \(A\) satisfies the Bellman inequality and that
\(A(m,H_2(m))=0\). Then, for every \(n\ge0\) and every array
\(v:\{-1,1\}^n\to(0,1)\),
\[
 \mathcal E_n(v)\ge A\Bigl(2^{-n}\sum_yv(y),\ 2^{-n}\sum_yH_2(v(y))\Bigr).
\]
For \(n=0\) the energy is zero, and the right side is \(A(m,H_2(m))=0\).
For \(n\ge1\), let \(m_\pm\) and \(e_\pm\) be the average and the average
entropy of \(v^\pm\). The tuple \(M=(m_-,m_+,e_-,e_+)\) is feasible, with
means in \((0,1)\) and positive entropies, because \(v\) takes values in
\((0,1)\). By \eqref{eq:intro-energy-split}, the induction hypothesis for
\(v^-\) and \(v^+\), and \(\mathbb Ej(U,W)\ge\zeta(M)\ge R_A(M)\),
\[
 \mathcal E_n(v)\ge\frac12A(m_-,e_-)+\frac12A(m_+,e_+)+R_A(M)
 =A\Bigl(\frac{m_-+m_+}2,\frac{e_-+e_+}2\Bigr),
\]
and the two arguments on the right are the average and the average entropy
of \(v\). This is carried out in
Lemmas~\ref{bal:lem:unrestricted-static-induction}
and~\ref{unbal:lem:static} for the functions \(\phi\) and \(B\) defined
below. Together with
\eqref{eq:intro-entropy-production} it yields the differential inequality
\begin{equation}\label{eq:intro-ode}
 \gamma'(t)\ge A\Bigl(2^{-n}\sum_yv_0(y),\ \gamma(t)\Bigr).
\end{equation}

\paragraph{The candidates.}
For \(0<e\le1\), let
\[
 \eta(e)=\bigl(1-2H_2^{-1}(e)\bigr)J\bigl(H_2^{-1}(e)\bigr),
\]
so that \(\eta(1)=0\). This is the entropy production of a single bit: if
\(r(t)\in(0,1/2)\) is the crossover probability of a bit at time \(t\), so
that \(r'=1-2r\), then
\(\frac{d}{dt}H_2(r(t))=\eta\bigl(H_2(r(t))\bigr)\). For \(s>0\) and
\(e>0\), let \(\tau\in(0,1/2)\) be the unique solution of
\(sH_2(\tau)=e(1-2\tau)\) and put \(F(s,e)=sJ(\tau)\); set \(F(0,e)=0\). The two
candidates and their maximum are
\begin{equation}\label{eq:intro-candidates}
 \phi(m,e)=\eta(e)-F(|1-2m|,e),\qquad
 \psi(m,e)=\eta\bigl(e+1-H_2(m)\bigr),\qquad
 B=\max\{\phi,\psi\},
\end{equation}
for \(0<m<1\) and \(0<e\le H_2(m)\). All three vanish at \(e=H_2(m)\), and
all three equal \(\eta(e)\) at \(m=1/2\). The candidate \(\phi\) is the one
proposed in \cite{ChenGohariNair2025}; the choice of \(B\) was found in the
course of this work. Our main technical result is the following.

\begin{theorem}[Local inequality]\label{thm:intro-local}
For every \(0<a,b<1\), \(0<e\le H_2(a)\), and \(0<f\le H_2(b)\),
\[
 \zeta(a,b,e,f)\ge R_B(a,b,e,f).
\]
\end{theorem}

This is Theorem~\ref{unbal:thm:global}, where it is called the hybrid
Bellman inequality because \(B\) combines two candidates. The corresponding
statement for
\(\phi\) alone, \(\zeta\ge R_\phi\), is
Theorem~\ref{bal:thm:global-unrestricted-target}; it is proved first and is
used in the proof of Theorem~\ref{thm:intro-local}.

\paragraph{From the local inequality to Theorem~\ref{unbal:thm:ck}.}
Let \(g\) be Boolean, fix \(0<\varepsilon<1/2\), and apply
\eqref{eq:intro-ode} with \(A=B\) and
\(v_0=h_\varepsilon=\varepsilon+(1-2\varepsilon)g\). Write \(m_\varepsilon\) for
the average of \(h_\varepsilon\) and \(\gamma_\varepsilon(t)\) for the
average entropy of \(T_{e^{-2t}}h_\varepsilon\). Since \(B\ge\psi\), the
function
\(\delta_\varepsilon(t)=\gamma_\varepsilon(t)+1-H_2(m_\varepsilon)\)
satisfies
\[
 \delta_\varepsilon'(t)\ge\eta\bigl(\delta_\varepsilon(t)\bigr),
 \qquad \delta_\varepsilon(0)\ge H_2(\varepsilon).
\]
The single-bit entropy \(H_2(r_\varepsilon(t))\) with
\(r_\varepsilon(t)=(1-e^{-2t}(1-2\varepsilon))/2\) satisfies the corresponding
equation and starts at \(H_2(\varepsilon)\), and a comparison gives
\(\delta_\varepsilon(t)\ge H_2(r_\varepsilon(t))\).
On the other hand, \(v_{\varepsilon,t}=T_{e^{-2t}}h_\varepsilon\) is the
posterior probability that \(g(X)\oplus N_\varepsilon=1\), where
\(N_\varepsilon\) is an independent Bernoulli\((\varepsilon)\) bit, so, as in
\eqref{eq:intro-mi-identity}, the mutual information between this bit and
the channel output equals \(1-\delta_\varepsilon(t)\). Letting
\(\varepsilon\downarrow0\) and choosing \(t\) with \((1-e^{-2t})/2=p\)
proves \eqref{eq:CK-conjecture} for \(0<p<1/2\); the other values of \(p\)
are elementary. The details are in Section~\ref{unbal:sec:main}.

The two candidates play different roles. The comparison above uses only
\(B\ge\psi\): the shift \(1-H_2(m)\), the output entropy deficit at mean
\(m\), is what makes the differential inequality for \(\delta_\varepsilon\)
independent of the mean of \(g\). Since \(\psi\) alone does not satisfy the
Bellman inequality, \(\phi\) is needed as well. It enters through the
proof of Theorem~\ref{thm:intro-local}, which uses
Theorem~\ref{bal:thm:global-unrestricted-target} wherever
\(\phi\ge\psi\) at the averaged moments; see
Section~\ref{sec:intro-overview}. When \(g\) is balanced,
\(m_\varepsilon=1/2\) and \(\phi\) alone suffices;
this gives Theorem~\ref{bal:thm:CK}, the balanced case of
Theorem~\ref{unbal:thm:ck}.

\subsection{Outline of the proof of the local inequality}
\label{sec:intro-overview}
Part~\ref{part:balanced} proves the Bellman inequality for \(\phi\), and
Part~\ref{part:unbalanced} derives Theorem~\ref{thm:intro-local} from it.

\paragraph{Lower bounds for \(\zeta\).}
Since the objective and the four moment constraints are linear in the
law of $(U,W)$, a standard Carath\'eodory--extreme-point argument for
moment problems shows that the infimum defining $\zeta(M)$ is attained
by a probability measure supported on at most five atoms
\[
    (u_1,w_1),\ldots,(u_5,w_5)\in[0,1]^2.
\]
Indeed, the feasible measures are specified by four moment constraints,
in addition to the normalisation of the total mass. Nevertheless, this
finite-dimensional reduction is not by itself tractable: five atoms
contribute ten location variables, and their five probabilities
contribute four independent weight variables, resulting in an
optimization problem of dimension as large as fourteen. Rather than
attempting to analyse this fourteen-dimensional problem directly, we
develop explicit analytic lower bounds on $\zeta(M)$ that depend only on
the four-moment tuple $M$.
Three kinds are used. The first is the four-moment
lower bound \(L_4\le\zeta\) of
Theorem~\ref{bal:thm:four-moment-lower-bound}. The function \(L_4(M)\),
defined in \eqref{bal:eq:L4}, is the sum of
\(F\bigl(|a-b|,(e+f)/2\bigr)\) and an explicit function of \((e,f)\) that
is nonnegative, jointly convex, and zero on the diagonal. For the proof one
subtracts \(F\bigl(|u-w|,(H_2(u)+H_2(w))/2\bigr)\) from \(j(u,w)\) and
bounds the remainder from below by its value at the atom obtained by
reflecting \(u\) and \(w\) into \([0,1/2]\); Jensen's inequality, the bound
\(\mathbb E|U-W|\ge|\mathbb EU-\mathbb EW|\), and the monotonicity of \(F\)
in its first argument then give a bound in terms of the four moments. The
second kind consists of supporting planes. For a law
carried by the corners \((0,0)\), \((1,1)\) and one interior atom, there
is, under explicit conditions on the atom, an affine function of
\(\bigl(u,w,H_2(u),H_2(w)\bigr)\) that lies below
\(j\) on the unit square and touches it at these three points. Integrating
it gives a lower bound for \(\zeta\) at every tuple with the same entropies
and the same difference of means, together with a description of the
tuples at which this bound is attained
(Proposition~\ref{bal:prop:fixed-difference-minimum} and
Theorem~\ref{bal:thm:general-endpoint-region}). The third is a consequence
of the log-sum inequality that is affine in the differences
\(H_2(a)-e\) and \(H_2(b)-f\) (Theorem~\ref{unbal:ss:logsum}).

\paragraph{The Bellman inequality for \(\phi\).}
Consider the distance of each of the two means to the nearer of the
points \(0\) and \(1\). When the sum of these two distances is less than
\(S=10^{-4}\), the inequality \(\zeta\ge R_\phi\) is proved directly
(Theorem~\ref{bal:thm:small-boundary}). Elsewhere we prove the stronger
statement
\[
 G=L_4-R_\phi\ge0
\]
(Theorem~\ref{bal:thm:pure-gap}). Both terms of \(G\) are explicit, and
\(G\) is unchanged when either mean is reflected about \(1/2\)
(Lemma~\ref{bal:lem:pure-gap-reflection}), so the two means may be placed
in \([0,1/2]\). The proof is by contradiction, starting from a negative
minimum of \(G\). Two preliminary
reductions (Lemmas~\ref{bal:lem:global-entropy-rearrangement}
and~\ref{bal:lem:maximal-continuity}) order the two entropies and place
them in \((0,1)\). With the entropies fixed, the admissible pairs of means
form a compact set, on which a negative minimum of \(G\) would be attained.
An interior
critical point is impossible (Theorem~\ref{bal:thm:e8-main}). The boundary
consists of the interface with the first region
(Section~\ref{bal:sec:seam}), the pairs with equal means or with a mean
equal to \(1/2\), and the faces \(e=H_2(a)\) and \(f=H_2(b)\). On such a
face one of the two random variables is constant, and the face is called
a \emph{deterministic entropy cap}. These cases are treated in
Sections~\ref{bal:sec:inputs} and~\ref{bal:sec:retained}, and
Theorem~\ref{bal:thm:global-unrestricted-target} combines the two regions.

\paragraph{From \(\phi\) to \(B\).}
Fix a tuple \(M\) and write \((m,E)=((a+b)/2,(e+f)/2)\). If
\(B(m,E)=\phi(m,E)\), then \(R_B(M)\le R_\phi(M)\), because \(B\ge\phi\) at
\((a,e)\) and at \((b,f)\), and Part~\ref{part:balanced} applies. If
\(B(m,E)=\psi(m,E)\), then \(R_B(M)\le R_\psi(M)\), and convexity of
\(P(x)=\eta(1-x)\) bounds \(R_\psi(M)\) by a quantity that depends on
\(e\) and \(f\) only through \(E\); see \eqref{unbal:eq:split-target}. A
sharper bound of the same kind is used when \(e\) is close to \(H_2(a)\)
or \(f\) is close to \(H_2(b)\). In most regions it then suffices to show
that one of the lower bounds above
dominates this quantity; in the others, \(\phi(m,E)>\psi(m,E)\) is proved,
or \(\zeta\) is compared with \(R_B\) directly. The argument is divided
according to the position of the means: on the same side of \(1/2\)
(Theorem~\ref{unbal:ss:main}) or on opposite sides
(Theorem~\ref{unbal:thm:opposite}). In the second case the pairs with both
means in \([1/10,9/10]\) are covered by Theorem~\ref{unbal:thm:central},
and the remaining pairs are treated in
Section~\ref{unbal:sec:opposite-completion}. In each case analytic
estimates cover the singular parts of the domain, where a mean tends to
\(0\) or \(1\), an entropy tends to \(0\), or a ratio of parameters is
unbounded. What remains is a compact parameter box.
Section~\ref{unbal:sec:assembly} combines the cases.

\subsection{Computation and organization of the paper}
\label{sec:intro-organization}
The computer-assisted steps concern explicit inequalities in a small
number of real parameters on compact boxes, each stated together with its
domain. Such a box is called a \emph{root}. A partition of the root into
subboxes with rational endpoints establishes coverage. On every subbox,
outward-rounded interval arithmetic proves one of a stated list of
sufficient inequalities, or shows that the subbox does not meet the
feasible domain. The record of the partition and of these evaluations is
called a \emph{certificate}, and an inequality between explicit constants
that is verified in the same arithmetic is called a \emph{directed
comparison}. Part~\ref{part:computation} describes this protocol and
lists, for each result that depends on a computation, the domain, the
verified inequality, and the location of the record.

Part~\ref{part:balanced} consists of
Sections~\ref{bal:sec:definitions}--\ref{bal:sec:framework}.
Section~\ref{bal:sec:definitions} fixes the notation.
Section~\ref{bal:sec:inputs} contains the lower bounds for \(\zeta\) and
the boundary estimates, with the proofs that are not deferred to the
supplement. Section~\ref{bal:sec:orientation} describes the symmetries,
the reductions of the entropy coordinates, and the division of the
parameter domain.
Sections~\ref{bal:sec:seam}--\ref{bal:sec:global-target} prove
\(\zeta\ge R_\phi\), and Section~\ref{bal:sec:framework} deduces the
balanced case of Theorem~\ref{unbal:thm:ck}. In
Part~\ref{part:unbalanced}, Section~\ref{unbal:sec:main} states
Theorem~\ref{unbal:thm:global} and deduces Theorem~\ref{unbal:thm:ck} from
it, Section~\ref{unbal:sec:inputs} lists the results of
Part~\ref{part:balanced} that are used,
Sections~\ref{unbal:ss:section}--\ref{unbal:sec:opposite-completion} prove
the estimates for \(B\) and treat the regions described above, and
Section~\ref{unbal:sec:assembly}
completes the proof. The supplement, which follows the references and
whose sections are numbered S.1, S.2, and so on, contains the longer
analytic arguments, the proofs for the individual regions, and the
specifications and records of the computations.

Unless a calculation states otherwise, all entropies, divergences, and
mutual informations are measured in bits, and \(\ln\) denotes the natural
logarithm. Where natural units are used, \(h=(\ln2)H_2\) and
\(j_e=(\ln2)j\), so that entropy coordinates and costs are converted
together.

\startmanuscriptpart{Balanced Case}{Balanced Bellman inequality}{Courtade--Kumar}
\label{part:balanced}
\section{Notation and unrestricted Bellman formulation}
\label{bal:sec:definitions}

This section fixes the notation of Part~\ref{part:balanced} and gives the
precise form of the definitions in Section~\ref{sec:intro-local}. The binary
entropy \(H_2\) is written \(\HH\), its inverse on \([0,1/2]\) is written
\(\iota\), and a four-moment tuple is written
\(M=(\mu_u,\mu_w,e_u,e_w)\); in the introduction and in
Part~\ref{part:unbalanced} the same tuple is written \((a,b,e,f)\). Unless
natural units are explicitly introduced, entropies and costs are in bits:
\(\log=\log_2\), while \(\ln\) always denotes the natural logarithm.  Put
\[
 \HH(t)=-t\log t-(1-t)\log(1-t),\qquad
 J(t)=\HH'(t)=\log\frac{1-t}{t},
 \qquad 0<t<1.
\]
Let
\[
 \iota(e)=\HH^{-1}(e)\in[0,\tfrac12],\qquad 0\le e\le1,
\]
where the lower inverse is intended.  Define the symmetrized edge cost
\begin{equation}
 j(u,w)=\frac12(u-w)\bigl(J(w)-J(u)\bigr)
       =\frac12\bigl(D_2(u\|w)+D_2(w\|u)\bigr)\ge0.
 \label{bal:eq:edge-cost}
\end{equation}
At the diagonal corners use \(j(0,0)=j(1,1)=0\); a cross-corner atom has
infinite cost in the lower-semicontinuous extension.

Define
\[
 \mathcal L(t)=\frac{2\HH(t)}{1-2t},\qquad 0\le t<\frac12,
\]
and
\begin{equation}
 \eta(e)=(1-2\iota(e))J(\iota(e)),\qquad
 F(s,e)=sJ\!\left(\mathcal L^{-1}\!\left(\frac{2e}{s}\right)\right)
 \quad(s>0,e>0).
 \label{bal:eq:eta-F}
\end{equation}
We use the lower-semicontinuous perspective extension
\[
 F(0,e)=0\quad(e\ge0),\qquad F(s,0)=+\infty\quad(s>0).
\]
In particular, \(F(0,0)=0\). For \(e>0\), set
\[
 \Phi(s,e)=\eta(e)-F(s,e),\qquad s\ge0.
\]
Equivalently, with
\[
 \Gamma(z)=zJ\!\left(\mathcal L^{-1}(1/z)\right),\qquad
 \Theta=\Gamma',\qquad Q=\Theta^{-1},
\]
one has the homogeneous form
\[
 F(s,e)=2e\,\Gamma\!\left(\frac{s}{2e}\right).
\]
The functions \(\Theta\) and \(Q\) are taken on the branches fixed in
Supplementary Theorem~\ref{bal:imp:e8:thm:onevar}.
An even extension in the first coordinate is used only where it is
explicitly declared.  The balanced candidate is
\begin{equation}
 \phi(m,e)=\Phi(|1-2m|,e)
          =\eta(e)-F(|1-2m|,e),
 \qquad 0<e<\HH(m),
 \label{bal:eq:balanced-candidate}
\end{equation}
with \(\phi(m,\HH(m))=0\).  On the zero-entropy boundary set
$\phi(0,0)=\phi(1,0)=0$ and $\phi(m,0)=+\infty$ for $0<m<1$.
These are boundary values; they are not obtained termwise from the formula
for $\Phi$.  At a tuple with a zero entropy coordinate the Bellman
inequality is read in the addition form
\eqref{bal:eq:zero-bellman-addition}, which avoids the difference
$\infty-\infty$.  Such tuples are treated in
Theorem~\ref{bal:thm:zero-target}.  The only other place where a zero
entropy occurs is the boundary of the region studied in
Section~\ref{bal:sec:seam}; it is treated in Lemma~\ref{bal:lem:seam-zero}.

For arbitrary labeled random variables \(U,W\in[0,1]\), define
\[
 M=(\mu_u,\mu_w,e_u,e_w)
   =(\mathbb EU,\mathbb EW,\mathbb E\HH(U),\mathbb E\HH(W))
\]
and
\begin{align}
 \zeta(M)
 &=\inf\{\mathbb E j(U,W):\text{the four moments of }(U,W)\text{ are }M\},
 \label{bal:eq:unrestricted-zeta}\\
 \Rphi(M)
 &=\phi\!\left(\frac{\mu_u+\mu_w}{2},
               \frac{e_u+e_w}{2}\right)
   -\frac12\phi(\mu_u,e_u)-\frac12\phi(\mu_w,e_w).
 \label{bal:eq:Rphi}
\end{align}
Both quantities are invariant under the exchange
\((\mu_u,e_u)\leftrightarrow(\mu_w,e_w)\).  Throughout,
\emph{unrestricted} means that the infimum in
\eqref{bal:eq:unrestricted-zeta} is over all joint laws of \((U,W)\).  The
information-theoretic conclusion of Part~\ref{part:balanced} concerns
balanced Boolean functions.

\begin{lemma}[Unrestricted feasibility]
\label{bal:lem:unrestricted-feasible}
A tuple \(M=(\mu_u,\mu_w,e_u,e_w)\) is feasible for \(\zeta\) if and only if
\[
 0\le\mu_u,\mu_w\le1,
 \qquad 0\le e_u\le\HH(\mu_u),
 \qquad 0\le e_w\le\HH(\mu_w).
\]
\end{lemma}
\begin{proof}
Necessity follows from nonnegativity and concavity of binary entropy.  For
the converse, fix \(0<\mu<1\) and \(0\le e\le\HH(\mu)\), and put
\(\lambda=e/\HH(\mu)\).  Let \(V=\mu\) with probability \(\lambda\),
and with the remaining probability let \(V\) be a Bernoulli random variable
of mean \(\mu\).  Then
\(\mathbb EV=\mu\) and
\(\mathbb E\HH(V)=\lambda\HH(\mu)=e\).  The endpoint means are immediate.
Construct the two marginals in this way and couple them independently.
\end{proof}

Put
\begin{equation}
 \kappa(u,w)=j(u,w)-F\!\left(|u-w|,
                    \frac{\HH(u)+\HH(w)}2\right),
 \qquad
 g(e_1,e_2)=\kappa(\iota(e_1),\iota(e_2)).
 \label{bal:eq:kappa-g}
\end{equation}
Its natural closure satisfies
\[
 g(0,0)=0,\qquad
 g(0,e)=g(e,0)=+\infty\ (e>0),\qquad
 g(e,e)=0.
\]
The four-moment lower bound is
\begin{equation}
 L_4(M)=F\!\left(|\mu_w-\mu_u|,\frac{e_u+e_w}{2}\right)
        +g(e_u,e_w).
 \label{bal:eq:L4}
\end{equation}
For positive entropy coordinates, set
\begin{equation}
 \Target(M)=\zeta(M)-\Rphi(M),\qquad
 \Gap(M)=L_4(M)-\Rphi(M).
 \label{bal:eq:two-gaps}
\end{equation}
At each singular endpoint used below, the value of the whole expression is
given by the estimate cited there, so that no difference of infinite
quantities occurs.  Where all terms are finite,
\begin{equation}
 \Target=(\zeta-L_4)+\Gap.
 \label{bal:eq:gap-decomposition}
\end{equation}

Let \(X\) be uniform on \(\{-1,+1\}^n\), and let \(Y\) be obtained from
\(X\) by a BSC with crossover \(p_{\mathrm{BSC}}\).  The balanced
Courtade--Kumar assertion is
\begin{equation}
 I(f(X);Y)\le1-\HH(p_{\mathrm{BSC}})
 \label{bal:eq:bck}
\end{equation}
for every Boolean \(f\) satisfying \(\mathbb Ef(X)=1/2\).

\begin{lemma}[Comparison of Bellman and pure-gap inequalities]
If a Bellman lower bound \(B\le\zeta\) satisfies \(B-\Rphi\ge0\) on a set,
then \(\Target\ge0\) on that set.  The inequality
\(\Target\ge0\) alone does not imply \(\Gap=L_4-\Rphi\ge0\).
\end{lemma}
\begin{proof}
The first statement is immediate from
\(\Target=(\zeta-B)+(B-\Rphi)\).  For the second, the nonnegative term
\(\zeta-L_4\) in \eqref{bal:eq:gap-decomposition} can compensate for a negative
value of \(\Gap\); hence \(\Target\ge0\) does not imply \(\Gap\ge0\).
\end{proof}

Because \(\Gap\ge0\) is a stronger statement than \(\Target\ge0\), we record
which of the two a given result proves. In the tables below,
\(\level{target}\) marks a proof of \(\Target\ge0\) at positive entropy (or
of \eqref{bal:eq:zero-bellman-addition} at zero entropy),
\(\level{pure-g}\) marks a proof of \(\Gap\ge0\), and
\(\level{reduction}\) marks a result that excludes a possible location of a
negative minimum of \(\Gap\).

\section{Lower bounds and boundary estimates}
\label{bal:sec:inputs}

This section states the results on which the proof of \(\zeta\ge\Rphi\)
rests. Those that are not proved here are proved in the supplementary
sections cited with them, and the protocol for the computer-assisted ones
is described in Part~\ref{part:computation}.

The results appear in the order in which they are used. The four-moment
lower bound reduces \(\Target\ge0\) to the explicit inequality
\(\Gap\ge0\). The endpoint results give supporting planes and characterize
equality. The mean reduction and the boundary estimates then locate, and
exclude, a possible negative minimum of \(\Gap\). These results are
combined in Section~\ref{bal:sec:retained}, after the interface between
the two regions of the proof has been treated in
Section~\ref{bal:sec:seam}.

\begin{theorem}[Global four-moment lower bound]
\label{bal:thm:four-moment-lower-bound}
For every feasible four-moment tuple, \(\zeta\ge L_4\), with the extended-real
boundary conventions used in \eqref{bal:eq:L4}.
\end{theorem}
\begin{proof}
A representing law with infinite expected cost satisfies the required
lower bound automatically. Consider a law with finite expected cost.
Almost surely, its atoms lie in \((0,1)^2\) or at the diagonal corners
\((0,0),(1,1)\), since every other boundary atom has infinite cost.
For an interior atom \((u,w)\), write
\[
 j(u,w)
 =F\!\left(|u-w|,\frac{\HH(u)+\HH(w)}2\right)
  +\kappa(u,w).
\]
The reflection theorem proved in Supplementary Section~\ref{bal:imp:reflection}, after
exchanging the two atom labels when necessary, compares the actual atom to
its lower-half entropy representatives
and gives
\[
 \kappa(u,w)\ge g(\HH(u),\HH(w)).
\]
This is generally an inequality for cross-half atoms, not an exact
decomposition through \(g\). At each diagonal corner, the pointwise bound
\(j(u,w)\ge F(|u-w|,(\HH(u)+\HH(w))/2)+g(\HH(u),\HH(w))\)
holds with both sides zero. The closed perspective extension of \(F\) is
convex, and Supplementary Theorem~\ref{bal:imp:joint:thm:main} establishes
convexity of the extended-real closure of \(g\) on the closed entropy square.
Both functions are nonnegative. Jensen's inequality for these closed convex
functions therefore applies directly to the finite-cost law and gives
\[
\begin{aligned}
 \mathbb Ej(U,W)
 &\ge F\!\left(\mathbb E|U-W|,
       \frac{\mathbb E\HH(U)+\mathbb E\HH(W)}2\right)\\
 &\quad +g(\mathbb E\HH(U),\mathbb E\HH(W))\\
 &\ge F\!\left(|\mu_w-\mu_u|,\frac{e_u+e_w}{2}\right)+g(e_u,e_w)
 =L_4(M).
\end{aligned}
\]
The second inequality uses
\(\mathbb E|U-W|\ge|\mathbb EU-\mathbb EW|\) and monotonicity in the
first perspective coordinate.  Indeed, from
\(F(s,e)=2e\,\Gamma(s/(2e))\) one has
\(F_s(s,e)=\Theta(s/(2e))\ge0\) on the certified branch.
The closed convex extensions already include the maximal-entropy sides.
For clarity, the zero-entropy cases can also be checked directly. If
\(e_u=0\), then \(U\in\{0,1\}\) almost surely, and finite expected cost
forces \(W=U\) almost surely. Thus a tuple with a zero entropy coordinate
can have finite cost only if \(e_u=e_w=0\) and \(\mu_u=\mu_w\).
In that case the diagonal Bernoulli law gives \(\zeta=L_4=0\).
All other feasible tuples with a zero entropy coordinate have
\(\zeta=L_4=+\infty\) by the stated boundary values. Exchanging the
variables handles \(e_w=0\). Taking the infimum proves the theorem.
\end{proof}

\begin{lemma}[Negative-sublevel compactness]
\label{bal:lem:negsubcompact}
Let \(K\subset\mathbb R^k\) be compact, let \(\Omega\subset K\) be relatively open, and let
\(\Gap\) be continuous on \(\Omega\).  Assume every sequence
\(x_n\in\Omega\) with \(x_n\to K\setminus\Omega\) satisfies
\(\liminf\Gap(x_n)\ge0\).  If \(\Gap\) is negative somewhere, it attains a
negative minimum in \(\Omega\).  If the minimizer is an interior point of
\(\Omega\) in \(\mathbb R^k\) and \(\Gap\) is \(C^1\) near it,
then its gradient vanishes by Fermat's rule.
\end{lemma}
\begin{proof}
Choose \(\Gap(x_0)<\alpha<0\).  The sublevel
\[
 A_\alpha=\{x\in\Omega:\Gap(x)\le\alpha\}
\]
is nonempty.  Every sequence in \(A_\alpha\) has a subsequence converging in
\(K\); the liminf assumption excludes a limit in \(K\setminus\Omega\), and
continuity closes \(A_\alpha\) inside \(\Omega\).  Thus \(A_\alpha\) is
compact.  Its minimum is below \(\alpha\), while \(\Gap>\alpha\) outside
\(A_\alpha\).  Under the additional interior hypothesis, a Euclidean
ball about the minimizer lies in \(\Omega\), so the unconstrained
Fermat rule gives \(\nabla\Gap=0\).
\end{proof}

Set $S=10^{-4}$.  After the reductions of
Section~\ref{bal:sec:orientation}, the two means are represented by their
distances $a\le c$ to the nearer of the points $0$ and $1$, the
\emph{lower-half means}.  The case $a+c<S$ is treated directly by
Theorem~\ref{bal:thm:small-boundary} below.  The region $a+c\ge S$ is
called the \emph{retained region}, and the interface $a+c=S$ is called the
\emph{seam}.  A face $e=\HH(a)$ or $f=\HH(c)$, on which one of the two
marginals is deterministic, is called a \emph{deterministic entropy cap};
on such a cap the means are relabeled by the capped marginal, as explained
before Theorem~\ref{bal:thm:boundary-estimates}.

The minimum argument in the retained region takes no limit in the entropy
coordinates.  After fixing \(e,f\in(0,1)\), it minimizes only over
\[
 K_{e,f}=\left\{(a,c):0\le a\le c\le\frac12, a+c\ge S,
                  e\le\HH(a), f\le\HH(c)\right\}.
\]
This set is compact, and \(\Gap(a,c,e,f)\) is continuous on it and \(C^1\)
on its relative interior.  Hence, if \(\Gap\) takes a negative value at
these fixed entropies, a negative minimum over the means is attained.  The
boundary of \(K_{e,f}\) consists of the seam, the pairs with equal means or
with a mean equal to \(1/2\), and the deterministic entropy caps.  On the
seam itself, compactness follows from the entropy constraints in
\eqref{bal:eq:seam-cell}, and the zero-entropy limits are treated in
Lemma~\ref{bal:lem:seam-zero}.

\begin{theorem}[Small-boundary Bellman inequality]
\label{bal:thm:small-boundary}
The target \(\Target\ge0\) holds on the unconditional same-side branch
\[
 0<\mu_u\le\mu_w\le10^{-2}
\]
and on the unconditional opposite-side branch
\[
 0<\mu_u<\tfrac12<\mu_w<1,
 \qquad \mu_u+(1-\mu_w)\le10^{-4},
\]
for every feasible positive-entropy four-moment tuple, including deterministic
entropy caps.  Here feasibility is unrestricted: the representing law need not
obey any pointwise order.  Full label exchange and complement give the
corresponding upper orientations; the zero-entropy limits are included by the
endpoint estimates below.
\end{theorem}

\begin{proposition}[Exact endpoint minimum and four-moment equality]
\label{bal:prop:fixed-difference-minimum}
Use natural logarithms and put
\[
 h(t)=-t\ln t-(1-t)\ln(1-t),\qquad
 J_e(t)=\ln\frac{1-t}{t},\qquad
 j_e(x,y)=\frac12(x-y)\{J_e(y)-J_e(x)\}.
\]
At the boundary use the lower-semicontinuous extended Jeffreys cost:
\(j_e(0,0)=j_e(1,1)=0\), and \(j_e=+\infty\) at every other boundary point.
Fix \(e_u,e_w>0\) and \(D\in(0,1)\), and define
\[
 \mathfrak Z_e(D;e_u,e_w)
 :=\inf\left\{\mathbb E j_e(U,W):
 \begin{array}{c}
 U,W\in[0,1],\quad \mathbb E(W-U)=D,\\
 \mathbb Eh(U)=e_u,\quad \mathbb Eh(W)=e_w
 \end{array}\right\}.
\]
Suppose there are \(p\in(0,1]\) and \(u,v\in(0,1/2)\) such that
\begin{equation}
 D=p(1-u-v),\qquad e_u=ph(u),\qquad e_w=ph(v).
 \label{bal:eq:fixed-difference-envelope-system}
\end{equation}
Then this triple is unique and
\begin{equation}
 \boxed{\mathfrak Z_e(D;e_u,e_w)=p\,j_e(u,1-v).}
 \label{bal:eq:fixed-difference-exact-value}
\end{equation}
The infimum is attained.  More precisely, if \(q=1-p\), then for every
\(\alpha\in[0,1]\) the law
\begin{equation}
 \begin{array}{c|ccc}
 (U,W)&(0,0)&(u,1-v)&(1,1)\\ \hline
 \text{mass}&q(1-\alpha)&p&q\alpha
 \end{array}
 \label{bal:eq:fixed-difference-extremizer}
\end{equation}
has the prescribed entropy moments and mean difference and attains
\eqref{bal:eq:fixed-difference-exact-value}.  Equivalently,
\[
 \mathfrak Z_e(D;e_u,e_w)
 =\inf_{m_w-m_u=D}\zeta_e(m_u,m_w,e_u,e_w).
\]
Here \(\zeta_e\) denotes the same unrestricted four-moment Bellman value as
\(\zeta\), with the entropy and edge cost written in natural-log units.

More strongly, for every feasible \(M=(\mu_u,\mu_w,e_u,e_w)\) with
\(\mu_w-\mu_u=D\),
\[
 \zeta_e(M)\ge p\,j_e(u,1-v),
\]
and equality holds if and only if
\begin{equation}
 \boxed{pu\le\mu_u\le pu+1-p.}
 \label{bal:eq:fixed-moment-equality-window}
\end{equation}
Equivalently, \(pu\le\mu_u\) and \(pv\le1-\mu_w\).  In that case the
unique optimizing probability law has masses
\begin{equation}
 \boxed{\quad
 p_{00}=1-\mu_w-pv,\qquad p_{\mathrm{int}}=p,\qquad
 p_{11}=\mu_u-pu
 \quad}
 \label{bal:eq:fixed-moment-unique-masses}
\end{equation}
at \((0,0),(u,1-v),(1,1)\), respectively.  When \(p<1\),
\[
 \alpha=\frac{\mu_u-pu}{1-p}
        =1-\frac{1-\mu_w-pv}{1-p}.
\]
When \(p=1\), the condition reduces to
\((\mu_u,\mu_w)=(u,1-v)\); both endpoint masses vanish and \(\alpha\)
has no effect on the law.  If the equality window fails, the lower bound
is strict, with \(+\infty\) allowed as the Bellman value.

For an explicit existence test, assume \(e_u,e_w\in(0,\ln2)\), put
\(p_{\min}=\max\{e_u,e_w\}/\ln2\), and, for
\(p\in[p_{\min},1]\), define
\[
 u(p)=h^{-1}(e_u/p),\qquad v(p)=h^{-1}(e_w/p),\qquad
 \mathfrak f(p)=p\{1-u(p)-v(p)\},
\]
where \(h^{-1}:[0,\ln2]\to[0,1/2]\) is the lower inverse.
The hypothesis holds exactly when
\begin{equation}
 \boxed{\mathfrak f(p_{\min})<D\le\mathfrak f(1).}
 \label{bal:eq:fixed-difference-range}
\end{equation}
The lower endpoint is excluded because at least one contact coordinate
then equals \(1/2\).  On \((p_{\min},1]\), the preimage is unique because
\begin{equation}
 \mathfrak f'(p)=1-u-v+\frac{h(u)}{J_e(u)}
                         +\frac{h(v)}{J_e(v)}>0.
 \label{bal:eq:fixed-difference-monotonicity}
\end{equation}
\end{proposition}

\begin{proof}
\emph{Proof outline.}
We construct a plane touching the edge cost at the prescribed interior
atom and at the two deterministic corners. The argument then has four
steps: prove uniqueness of its stationary contact, use quotient
minimization to establish a global minorant, integrate to identify the
optimal value, and analyze the zero set to characterize equality for
the prescribed four moments.

Equation \eqref{bal:eq:fixed-difference-monotonicity}, obtained by implicit
differentiation, proves uniqueness.  Let \(d=1-u-v>0\), and define
\(C_u,C_w\) by
\begin{equation}
 \begin{pmatrix}-\ln u&-\ln(1-v)\\-\ln(1-u)&-\ln v\end{pmatrix}
 \binom{C_u}{C_w}
 =\binom{d^2/[2u(1-v)]}{d^2/[2(1-u)v]}.
 \label{bal:eq:fixed-difference-C-system}
\end{equation}
Both coefficients are positive.  Indeed, the determinant is positive and
\[
 \frac{-\ln(1-v)}{-\ln v}
 <\frac{(1-u)v}{u(1-v)}
 <\frac{-\ln u}{-\ln(1-u)},
\]
because \(-z\ln z>-(1-z)\ln(1-z)\) for \(0<z<1/2\); Cramer's rule gives
the signs.

Choose
\[
 c=-\partial_xj_e(u,1-v)-C_uJ_e(u)
\]
and, with the sign convention \(D=\mathbb E(W-U)>0\), put
\begin{equation}
 Q_c(x,y)=j_e(x,y)-c(y-x)+C_uh(x)+C_wh(y).
 \label{bal:eq:fixed-difference-plane}
\end{equation}
Thus the coefficient of \(y-x\) is \(-c\).  Writing instead
\(j_e-c(x-y)+C_uh(x)+C_wh(y)\) uses the opposite sign convention for
\(c\).
We prove
\begin{equation}
 Q_c(x,y)\ge0\qquad((x,y)\in[0,1]^2).
 \label{bal:eq:fixed-difference-plane-global}
\end{equation}
\smallskip\noindent\emph{Step 1: uniqueness of a stationary contact.}
We prove a scalar uniqueness statement for arbitrary nonnegative entropy
coefficients. It will identify every possible interior minimum of the
supporting-plane quotient.

For \(A_0,B_0\ge0\), define
\begin{align*}
 I_1(x,y)&=A_0\ln x+B_0\ln y+\frac{(x-y)^2}{2xy},\\
 I_0(x,y)&=A_0\ln(1-x)+B_0\ln(1-y)
              +\frac{(x-y)^2}{2(1-x)(1-y)}.
\end{align*}
If \(A_0+B_0>0\), the system \(I_1=K_1\), \(I_0=K_0\), with
\(K_0,K_1\ge0\), has at most one solution \(0<x<y<1\).  To see this, set
\[
 r=x/y,\qquad s=(1-y)/(1-x),\qquad G_0(t)=\frac{(1-t)^2}{2t}.
\]
The equations become
\begin{align*}
 F_1(r,s)&=G_0(r)+A_0\ln r+(A_0+B_0)
              \ln\frac{1-s}{1-rs}=K_1,\\
 F_0(r,s)&=G_0(s)+B_0\ln s+(A_0+B_0)
              \ln\frac{1-r}{1-rs}=K_0.
\end{align*}
Since
\[
 (F_1)_s=-\frac{(A_0+B_0)(1-r)}{(1-s)(1-rs)}<0,
\]
the first equation determines a single level curve \(s=s(r)\).  Its
physical domain is one interval because, for
\(f_{A_0}(r)=G_0(r)+A_0\ln r\),
\[
 f_{A_0}'(r)=\frac{r^2+2A_0r-1}{2r^2}.
\]
Along that curve,
\[
 \frac{d}{dr}F_0(r,s(r))
 =\frac{(1-s)(1-rs)}{4(A_0+B_0)r^2s^2(1-r)}\,\mathcal R,
\]
where
\[
 \mathcal R=(1-r^2)(1-s^2)
 -2(1+rs)(A_0r+B_0s)+4A_0B_0rs.
\]
If \(F_1,F_0\ge0\), the inequalities
\(\ln t<2(t-1)/(t+1)\) and \(\ln t<t-1\) imply \(L_1,L_2\ge0\), where
\begin{align*}
 L_1&=(1-r^2)(1-rs)-2A_0r(2+s-rs)-2B_0rs(1+r),\\
 L_2&=(1-s^2)(1-rs)-2A_0rs(1+s)-2B_0s(2+r-rs).
\end{align*}
Indeed,
\[
 F_1\le\frac{1-r}{2r(1+r)(1-rs)}L_1,
 \qquad
 F_0\le\frac{1-s}{2s(1+s)(1-rs)}L_2.
\]
On a hypothetical fold \(\mathcal R=0\), put
\[
 P=1+rs-2A_0r,\qquad Q=1+rs-2B_0s.
\]
Then \(PQ=(r+s)^2\).  The identity
\[
 L_1=(2+s-rs)P+r(1+r)Q
 -(1+s)(2r^2+rs+r+1-r^2s)
\]
excludes \(P,Q<0\), so \(P=t(r+s)\), \(Q=(r+s)/t\) for some \(t>0\).
Since \(P,Q\le1+rs\),
\[
 t+t^{-1}\le\frac{1+rs}{r+s}+\frac{r+s}{1+rs}.
\]
Consequently
\[
 (1-s)L_1+(1-r)L_2
 \le-\frac{(1-r)(1-s)(1-rs)^2(r+s)}{1+rs}<0,
\]
contradicting \(L_1,L_2\ge0\).  Thus the curve has no fold where
\(F_0\ge0\).  At its initial endpoint, writing
\[
 y_0=\frac{1-s}{1-rs}
 =\exp\!\left(\frac{K_1-G_0(r)-A_0\ln r}{A_0+B_0}\right),
 \qquad x_0=ry_0,
\]
then, as \(r\downarrow0\), \(y_0\) decays faster than every power of \(r\),
and
\[
 F_0=-A_0ry_0-B_0y_0+O(y_0^2+r^2y_0^2)<0.
\]
If \(B_0=0\), then \(y_0/r\to0\) makes the first term dominant.  At the
terminal endpoint, \(y_0\uparrow1\), \(s\downarrow0\), and
\(G_0(s)\sim(2s)^{-1}\) dominates the logarithmic terms, so
\(F_0\to+\infty\).  In the sole limiting case \(A_0=K_1=0\), one has
\(r_*=1\), \(1-y_0\asymp(1-r)^2\), and \(s\asymp1-r\), giving the same
terminal limit.  Since the derivative cannot vanish where \(F_0\ge0\),
the nonnegative set has no bounded component and its terminal component is
strictly increasing.  Hence every level \(K_0\ge0\) is met at most once.

\smallskip\noindent\emph{Step 2: global validity of the plane.}
Apply this uniqueness result with \(A_0=C_u\), \(B_0=C_w\), and
\(K_0=K_1=0\).  Equation \eqref{bal:eq:fixed-difference-C-system} says that
\((u,1-v)\) is its strict solution.  Minimize
\[
 \mathcal Q(x,y)=
 \frac{j_e(x,y)+C_uh(x)+C_wh(y)}{y-x},\qquad 0<x<y<1.
\]
The quotient tends to \(+\infty\) at every boundary approach:
when \(y-x\to0\) away from the two deterministic corners, its positive
entropy numerator stays bounded away from zero; when \(x,y\to0\), use
\(\mathcal Q\ge C_wh(y)/y\to+\infty\); when \(x,y\to1\), use
\(\mathcal Q\ge C_uh(x)/(1-x)\to+\infty\).
At the remaining boundary approaches use
\(j_e(x,y)/(y-x)=\{J_e(x)-J_e(y)\}/2\to+\infty\).
Thus its sublevel sets are compact inside \(0<x<y<1\).
At a critical point, with \(c_*=\mathcal Q(x,y)\), direct differentiation and
\[
 h(t)=-\ln(1-t)+tJ_e(t)=-\ln t-(1-t)J_e(t)
\]
give \(Q_{c_*}=0\), \(\nabla Q_{c_*}=0\).  More explicitly, at every
interior point and for every \(c_*\),
\[
 Q_{c_*}-x(Q_{c_*})_x-y(Q_{c_*})_y=-I_0,
 \qquad
 Q_{c_*}+(1-x)(Q_{c_*})_x+(1-y)(Q_{c_*})_y=-I_1.
\]
Subtracting these identities gives
\((Q_{c_*})_x+(Q_{c_*})_y=I_0-I_1\); at a stationary point they give
\(Q_{c_*}=-I_0=-I_1\).  Thus critical points of the quotient solve
\(I_0=I_1=0\).
At \((u,1-v)\), the chosen \(c\) gives \(Q_c=0\) and \(\nabla Q_c=0\).
If \(Q_c\) were negative in \(0<x<y<1\), then \(\mathcal Q\) would have
a second interior critical point below its contact value, contradicting the
uniqueness just proved.  Hence \(Q_c\ge0\) on that triangle.  The contact
identity is
\[
 j_e(u,1-v)=cd-C_uh(u)-C_wh(v),
\]
so
\[
 c=\frac{j_e(u,1-v)+C_uh(u)+C_wh(v)}d>0.
\]
On \(y\le x\), all terms in
\(j_e(x,y)+c(x-y)+C_uh(x)+C_wh(y)\) are nonnegative.  The boundary values
at \((0,0)\) and \((1,1)\) are zero; at every other boundary point
\(j_e=+\infty\).  This proves \eqref{bal:eq:fixed-difference-plane-global}.

\smallskip\noindent\emph{Step 3: the optimal value and attainment.}
Integration against any admissible law gives
\[
 \mathbb E j_e(U,W)
 \ge cD-C_ue_u-C_we_w
 =p\{cd-C_uh(u)-C_wh(v)\}=p j_e(u,1-v).
\]
Conversely, the law \eqref{bal:eq:fixed-difference-extremizer} has the prescribed
moments and mean difference, and its only positive-cost atom is
\((u,1-v)\).  Its cost is \(p j_e(u,1-v)\), proving equality and attainment.

\smallskip\noindent\emph{Step 4: equality for the prescribed four moments.}
To determine equality for prescribed four moments, first note that the zero
set of \(Q_c\) is exactly
\[
 \{(0,0),(u,1-v),(1,1)\}.
\]
Indeed, any zero in \(0<x<y<1\) is a minimum of the smooth nonnegative
function \(Q_c\), so its gradient vanishes and \(I_0=I_1=0\);
the uniqueness argument above forces the stated contact.
On \(y\le x\) in the open square, the positive entropy coefficients exclude
zeros, and the boundary was already checked.

The set of probability laws on \([0,1]^2\) with the prescribed four moments
is weakly compact, since the four moment functions are continuous.
The integral of the nonnegative lower-semicontinuous cost \(j_e\) is
lower semicontinuous.  Thus whenever the Bellman value is finite, a minimizing
law exists.  Equality in the integrated plane bound forces this law to be
supported on the three zeros.  The difference constraint then fixes its
interior mass as \(D/(1-u-v)=p\), and the two means fix the endpoint masses
as in \eqref{bal:eq:fixed-moment-unique-masses}.  Their nonnegativity is exactly
\eqref{bal:eq:fixed-moment-equality-window}; their sum is \(1-p\).
Conversely these nonnegative masses give the required law and attain the
bound.  This proves the necessity, sufficiency, uniqueness, and strictness
claims.  Finally \(\mathfrak f\) is continuous at \(p_{\min}\), so its strict
monotonicity proves \eqref{bal:eq:fixed-difference-range}.
\end{proof}

\begin{proposition}[Complete moment criterion for the strict cross-half endpoint representation]
\label{bal:prop:endpoint-moment-region}
Write \(\iota_{\mathrm n}=h^{-1}:[0,\ln2]\to[0,1/2]\).  Given
\[
 0<\mu_u<\mu_w<1,\qquad e_u,e_w\in(0,\ln2),
\]
put \(D=\mu_w-\mu_u\), \(E=\max(e_u,e_w)\), \(e=\min(e_u,e_w)\), and
\[
 p_0=\frac E{\ln2},\qquad
 D_{\mathrm{cross}}=p_0\left\{\frac12-\iota_{\mathrm n}\!\left(\frac e{p_0}\right)\right\},
 \qquad
 D_1=1-\iota_{\mathrm n}(e_u)-\iota_{\mathrm n}(e_w).
\]
Equations \eqref{bal:eq:fixed-difference-envelope-system} and
\eqref{bal:eq:fixed-difference-extremizer} represent exactly these four moments
with \(0<u,v<1/2\), \(0<p\le1\), \(0\le\alpha\le1\) if and only if:
\begin{enumerate}
\item \(D_{\mathrm{cross}}<D\le D_1\).
\item For the unique root \(p\in(p_0,1]\) of
\[
 f(p):=p\{1-\iota_{\mathrm n}(e_u/p)-\iota_{\mathrm n}(e_w/p)\}=D,
\]
one has
\[
 \boxed{\quad p\iota_{\mathrm n}(e_u/p)\le\mu_u,\qquad
               p\iota_{\mathrm n}(e_w/p)\le1-\mu_w.\quad}
\]
\end{enumerate}
These conditions imply ordinary four-moment feasibility; it need not be
assumed separately.  The contact is \(u=\iota_{\mathrm n}(e_u/p)\), \(v=\iota_{\mathrm n}(e_w/p)\),
and the endpoint masses are given by
\eqref{bal:eq:fixed-moment-unique-masses}.  The exact value is
\[
 \zeta_e(M)
 =\frac D2\{J_e(u)+J_e(v)\}.
\]
At fixed \(D,e_u,e_w\) in the range above, the equality region is the closed
line segment
\[
 (\mu_u,\mu_w)=(pu+r,\ p(1-v)+r),\qquad 0\le r\le1-p.
\]
Equivalently its midpoint-coordinate description is
\[
 \left|\mu_u+\mu_w-1-p(u-v)\right|\le1-p.
\]
\end{proposition}
\begin{proof}
At \(p=p_0\), at least one of \(\iota_{\mathrm n}(e_u/p)\) and \(\iota_{\mathrm n}(e_w/p)\) equals
\(1/2\); hence \(f(p_0)=D_{\mathrm{cross}}\), while \(f(1)=D_1\).
Proposition~\ref{bal:prop:fixed-difference-minimum} proves the first condition
is necessary and sufficient for the strict contact triple.  For that triple,
the endpoint masses satisfy
\[
 (\mu_u-pu)+(1-\mu_w-pv)
 =1-D-p(u+v)=1-p.
\]
They are nonnegative exactly under the second condition, and then the
three-atom law has precisely the prescribed means and entropies.
The exact value follows from
\(j_e(u,1-v)=(1-u-v)\{J_e(u)+J_e(v)\}/2\).
The line-segment and midpoint descriptions follow by setting
\(r=\mu_u-pu\).  When \(p<1\), \(r=(1-p)\alpha\); for \(p=1\) the segment
is a single point.
\end{proof}

\begin{remark}[An alternative test using only inverse functions and endpoint evaluations]
\label{bal:rem:endpoint-threshold-test}
For a condition that avoids solving the difference equation, define
\[
 r_h(t)=\frac{h(t)}t,\qquad 0<t\le\frac12.
\]
This decreases continuously from \(+\infty\) to \(2\ln2\), because
\(r_h'(t)=\ln(1-t)/t^2<0\).  For \(e>0,a>0\), set
\[
 P(e,a)=
 \begin{cases}
 e/\ln2,&e\le2a\ln2,\\[1mm]
 a/r_h^{-1}(e/a),&e>2a\ln2.
 \end{cases}
 \qquad
 R=\max\{P(e_u,\mu_u),P(e_w,1-\mu_w)\}.
\]
Under the assumptions of Proposition~\ref{bal:prop:endpoint-moment-region},
its two conditions are equivalently
\[
 \boxed{\quad R\le1,\qquad D_{\mathrm{cross}}<D\le D_1,\qquad f(R)\le D.\quad}
\]
Here \(R\ge p_0\), so \(f(R)\) is defined whenever \(R\le1\).
To verify the criterion, write \(A_e(p)=p\iota_{\mathrm n}(e/p)\) for \(p\ge e/\ln2\).
At the left endpoint \(A_e=e/(2\ln2)\), and for \(p>e/\ln2\),
\[
 A_e'(p)=\iota_{\mathrm n}(e/p)-\frac{h(\iota_{\mathrm n}(e/p))}{J_e(\iota_{\mathrm n}(e/p))}
 =\frac{\ln(1-\iota_{\mathrm n}(e/p))}{J_e(\iota_{\mathrm n}(e/p))}<0.
\]
Also \(A_e(p)\to0\) as \(p\to\infty\).  If \(a\ge e/(2\ln2)\),
then \(A_e(p)\le a\) throughout its domain.  Otherwise its unique equality
point is \(p=a/r_h^{-1}(e/a)\).  Thus
\[
 A_e(p)\le a
 \quad\Longleftrightarrow\quad p\ge P(e,a)
 \qquad(p\ge e/\ln2).
\]
The two endpoint-mass conditions are therefore exactly \(p\ge R\),
equivalent to \(D=f(p)\ge f(R)\) by strict monotonicity.
\end{remark}

\begin{remark}[Including a contact at \(1/2\)]
\label{bal:rem:endpoint-half-contact}
The same exact-value and optimizer conclusions hold for
\(0<u,v\le1/2\), provided \(u+v<1\).  Accordingly, allowing
\(e_u,e_w\in(0,\ln2]\), the complete moment criterion becomes
\[
 0<D,\qquad D_{\mathrm{cross}}\le D\le D_1,\qquad
 p\iota_{\mathrm n}(e_u/p)\le\mu_u,\qquad p\iota_{\mathrm n}(e_w/p)\le1-\mu_w,
\]
where \(p\) is the unique root in the closed interval \([p_0,1]\).
If \(p_0=1\), this interval is a singleton and the condition forces \(p=1\).
The alternative threshold test becomes
\[
 R\le1,\qquad 0<D,\qquad f(R)\le D\le f(1).
\]
To justify the extension, note that at \(u=1/2,v<1/2\) the determinant
in \eqref{bal:eq:fixed-difference-C-system} is
\(\ln2\,\ln((1-v)/v)>0\), and both coefficients remain positive by Cramer's
rule.  The ratio inequalities in that proof remain strict because
\(v<1/2\); the case \(v=1/2,u<1/2\) is symmetric.
The contact is still strictly inside \(0<x<y<1\), so the full-square plane
proof and its zero-set proof apply without change.  Continuity and strict
monotonicity on \(p>p_0\) prove the closed range condition, without using
the derivative formula at \(p_0\).
The case \(u=v=1/2\) has \(D=0\) and is excluded here.
\end{remark}

\begin{corollary}[Symmetric means and equal entropies]
\label{bal:cor:endpoint-symmetric}
For \(0<m<1/2\) and \(0<e\le h(m)\), let \(t\in(0,1/2)\) be the unique
solution of
\[
 \frac{h(t)}{1-2t}=\frac e{1-2m}.
\]
Then, with \(p=(1-2m)/(1-2t)\le1\),
\[
 \boxed{\zeta_e(m,1-m,e,e)=(1-2m)J_e(t).}
\]
The optimizing law has masses \((1-p)/2,p,(1-p)/2\) at
\((0,0),(t,1-t),(1,1)\).  Thus \(\alpha=1/2\) when \(p<1\).
\end{corollary}
\begin{proof}
Here \(D_{\mathrm{cross}}=0\), and \(D_1=1-2\iota_{\mathrm n}(e)\ge1-2m\), so the unique strict contact
exists with \(u=v=t\).  Its difference equation gives
\(pt=(p-(1-2m))/2\).  Consequently
\[
 m-pt=(1-p)/2=(1-(1-m))-pt,
\]
which verifies both endpoint-mass conditions, including the cap \(p=1\).
The displayed scalar equation and value follow immediately.
\end{proof}

This is the natural-log form of \cite[Theorem~4, arXiv v2]{ChenGohariNair2025},
whose statement uses bits.  The present four-moment result allows unequal
entropies and means that need not be complementary.  We use the closed-square
extended-cost convention, so the endpoint law is an attained optimizer.
In the manuscript's bit normalization, the same characterization holds after
replacing \(h,\iota_{\mathrm n},J_e,\zeta_e\) by \(H_2,\iota,J,\zeta\) and \(\ln2\) by \(1\).

Proposition~\ref{bal:prop:endpoint-moment-region} characterizes the \emph{strict}
endpoint representation, and Remark~\ref{bal:rem:endpoint-half-contact}
extends it to a contact at \(1/2\) with nonzero difference.
For \(\mu_w<\mu_u\), exchange the two
complete mean--entropy labels.  When the difference triple exists but its
four-moment endpoint masses have the wrong sign, Proposition~\ref{bal:prop:fixed-difference-minimum}
still gives a strict lower bound.  The opposite-side hypotheses of
Theorem~\ref{bal:thm:small-boundary} guarantee that the triple exists, as proved in
Supplementary Section~\ref{supp:small-boundary-proof}.

The cross-half calculation gives both a global lower bound and a test for
equality. For a general contact \(0<x<y<1\) we treat the two questions
separately: first the existence of the contact, and then the sign
conditions on the plane coefficients and on the endpoint masses that decide
optimality and equality.

\subsection{General endpoint optimizers}

\begin{proposition}[All oriented endpoint contacts]
\label{bal:prop:all-oriented-endpoint-contacts}
Let $h(t)=-t\ln t-(1-t)\ln(1-t)$, let
$\iota_{\mathrm n}=h^{-1}:[0,\ln2]\to[0,1/2]$ be the lower inverse, and write
$J_e(t)=\ln((1-t)/t)$ and
$j_e(x,y)=(y-x)(J_e(x)-J_e(y))/2$.  Fix $0<e_u,e_w<\ln2$
and $D>0$.  Put
\[
 p_0=\frac{\max(e_u,e_w)}{\ln2},\qquad
 D_-=|\iota_{\mathrm n}(e_u)-\iota_{\mathrm n}(e_w)|,\qquad
 D_+=1-\iota_{\mathrm n}(e_u)-\iota_{\mathrm n}(e_w),
\]
\[
 D_{\mathrm{cross}}=p_0\left\{\frac12-\iota_{\mathrm n}\left(\frac{\min(e_u,e_w)}{p_0}\right)\right\}.
\]
There is a triple $0<p\le1$, $0<x<y<1$ satisfying
\begin{equation}
 e_u=ph(x),\qquad e_w=ph(y),\qquad D=p(y-x)
 \label{bal:eq:all-oriented-contact-system}
\end{equation}
if and only if $D_-\le D\le D_+$.  This triple is unique.
It is obtained from one of the following monotone scalar equations, with
$p\in[p_0,1]$:
\[
\begin{array}{c|c|c}
\text{range}&(x,y)&D\text{ as a function of }p\\ \hline
D\ge D_{\mathrm{cross}}&(\iota_{\mathrm n}(e_u/p),\,1-\iota_{\mathrm n}(e_w/p))
 &p\{1-\iota_{\mathrm n}(e_u/p)-\iota_{\mathrm n}(e_w/p)\}\\
D<D_{\mathrm{cross}},\ e_u<e_w&(\iota_{\mathrm n}(e_u/p),\,\iota_{\mathrm n}(e_w/p))
 &p\{\iota_{\mathrm n}(e_w/p)-\iota_{\mathrm n}(e_u/p)\}\\
D<D_{\mathrm{cross}},\ e_u>e_w&(1-\iota_{\mathrm n}(e_u/p),\,1-\iota_{\mathrm n}(e_w/p))
 &p\{\iota_{\mathrm n}(e_u/p)-\iota_{\mathrm n}(e_w/p)\}.
\end{array}
\]
At $D=D_{\mathrm{cross}}>0$, the applicable descriptions meet at $p=p_0$ and one
contact coordinate is $1/2$.  If $e_u=e_w$, then $D_-=D_{\mathrm{cross}}=0$ and only
the first row is relevant for $D>0$.
\end{proposition}
\begin{proof}
The entropy equations imply $p\ge p_0$ and force each coordinate onto
one of its two inverse-entropy branches.  The ordering $x<y$ excludes
the upper--lower branch, leaving exactly the three rows displayed.
The cross-half difference is strictly increasing on $(p_0,1]$,
by the derivative already computed in
Proposition~\ref{bal:prop:fixed-difference-minimum}; its endpoint values are
$D_{\mathrm{cross}},D_+$.  For a lower-branch coordinate $z=\iota_{\mathrm n}(e/p)$, put
\[
 k_{\mathrm{end}}(z)=z-\frac{h(z)}{J_e(z)}.
\]
Implicit differentiation gives
\[
 \frac{d}{dp}\{p\iota_{\mathrm n}(e/p)\}=k_{\mathrm{end}}(z),\qquad
 k_{\mathrm{end}}'(z)=-\frac{h(z)}{z(1-z)J_e(z)^2}<0
 \quad(0<z<1/2).
\]
Consequently the same-side difference
$p|\iota_{\mathrm n}(e_w/p)-\iota_{\mathrm n}(e_u/p)|$ is strictly decreasing when the entropies differ.
Its endpoint values are $D_{\mathrm{cross}},D_-$.  Continuity handles $p=p_0$ without
requiring differentiability there.  The two ranges meet only at their
common half-contact, proving existence and uniqueness.
\end{proof}

\begin{theorem}[Exact general endpoint region and genuine three-atom regime]
\label{bal:thm:general-endpoint-region}
Let $M=(\mu_u,\mu_w,e_u,e_w)$, where $0<\mu_u<\mu_w<1$,
$0<e_u,e_w<\ln2$, and $D=\mu_w-\mu_u$.  If the contact triple
$(p,x,y)$ in Proposition~\ref{bal:prop:all-oriented-endpoint-contacts}
exists, define $A,B$ by
\begin{equation}
 \begin{pmatrix}
 -\ln x&-\ln y\\
 -\ln(1-x)&-\ln(1-y)
 \end{pmatrix}
 \binom A B
 =\binom{(y-x)^2/(2xy)}{(y-x)^2/(2(1-x)(1-y))}.
 \label{bal:eq:general-endpoint-coefficients}
\end{equation}
The matrix has positive determinant.  The condition $A,B\ge0$ is
equivalent to the explicit ratio test
\begin{equation}
 \boxed{
 \frac{-\ln y}{-\ln(1-y)}
 \ \le\ \frac{(1-x)(1-y)}{xy}
 \ \le\ \frac{-\ln x}{-\ln(1-x)}.}
 \label{bal:eq:general-endpoint-ratio-test}
\end{equation}
If this test holds, then for every feasible $M$ with these entropy
moments and difference,
\[
 \zeta_e(M)\ge p\,j_e(x,y).
\]
Equality holds if and only if
\begin{equation}
 q_{00}:=1-\mu_w-p(1-y)\ge0,\qquad
 q_{11}:=\mu_u-px\ge0.
 \label{bal:eq:general-endpoint-masses}
\end{equation}
When equality holds, the unique optimizing law places masses
$q_{00},p,q_{11}$ at $(0,0),(x,y),(1,1)$, respectively, and
\[
 \boxed{\zeta_e(M)=p\,j_e(x,y)
       =\frac D2\{J_e(x)-J_e(y)\}.}
\]
In particular, a \emph{genuine three-atom} optimizer of this form,
with both endpoint masses positive, exists if and only if the unique
contact triple exists, the ratio test holds, and both inequalities in
\eqref{bal:eq:general-endpoint-masses} are strict.  Zero endpoint masses
give overlaps with the two-atom regime; $p=1$ gives a one-atom law.
Necessity of the ratio test is asserted only when both corner masses are
positive.  A degenerate one- or two-atom endpoint law can be optimal even
when that test fails.
\end{theorem}
\begin{proof}
\emph{Proof outline.}
We first extend the supporting-plane argument to arbitrary oriented
contacts with nonnegative entropy coefficients, including the cases in
which one coefficient vanishes. Its zero set gives the equality and mass
conditions. To prove the converse for a genuine three-atom optimizer, we
obtain a supporting plane from a Bellman subgradient and use its behavior
near a deterministic corner to force the coefficient signs.

The function $(-\ln t)/(-\ln(1-t))$ is strictly decreasing on $(0,1)$,
so the determinant in \eqref{bal:eq:general-endpoint-coefficients} is positive.
Cramer's rule gives \eqref{bal:eq:general-endpoint-ratio-test}.  Whenever
$A,B\ge0$, the positive right-hand side also gives $A+B>0$.

The stationary-contact uniqueness argument in the proof of
Proposition~\ref{bal:prop:fixed-difference-minimum} was proved for arbitrary
nonnegative entropy coefficients, not just cross-half contacts.  It says
that the two equations $I_0=I_1=0$ have at most one solution $0<x<y<1$.
The present coefficient system makes the prescribed $(x,y)$ such a
solution.  This yields the same global supporting plane
\[
 Q(s,t)=j_e(s,t)-c(t-s)+Ah(s)+Bh(t)\ge0,
 \qquad
 c=\frac{j_e(x,y)+Ah(x)+Bh(y)}{y-x}>0.
\]
For completeness, allowing $A=0$ or $B=0$ causes no loss of the
compactness used in that argument.  The quotient
\[
 \frac{j_e(s,t)+Ah(s)+Bh(t)}{t-s},\qquad 0<s<t<1,
\]
still tends to infinity at every boundary approach.  Near $(0,0)$,
if $s/t\to0$, its Jeffreys term tends to infinity; otherwise $s/t$ is
bounded below along a subsequence and the entropy contribution is
bounded below by a positive multiple of $\ln(1/t)$.  The argument near
$(1,1)$ uses $(1-t)/(1-s)$ in the same way.  Along the interior diagonal,
$(A+B)h(s)>0$, and all remaining boundary approaches have divergent
Jeffreys quotient.  Thus a quotient minimum exists in the open triangle;
stationary-contact uniqueness makes $(x,y)$ its unique minimum.  On
$t\le s$, all terms of $Q=j_e+c(s-t)+Ah(s)+Bh(t)$ are nonnegative and
there are no interior zeros.  The full zero set is therefore exactly
\[
 \{(0,0),(x,y),(1,1)\}.
\]
Integrating gives the lower bound.  The contact entropy equations fix
the interior mass at $p$, while the means force precisely
\eqref{bal:eq:general-endpoint-masses}; these masses sum to $1-p$.
Nonnegativity is thus sufficient and necessary for equality.  Necessity
uses existence of a minimizing law at finite Bellman value, by weak
compactness and lower semicontinuity, exactly as in
Proposition~\ref{bal:prop:fixed-difference-minimum}.  The zero set also proves
uniqueness and strictness when the mass condition fails.

\smallskip\noindent\emph{Necessity of the coefficient condition.}
It remains to prove necessity of $A,B\ge0$ for a genuine three-atom
optimizer; sufficiency was just proved.  Such a law satisfies
$0<e_u<h(\mu_u)$ and $0<e_w<h(\mu_w)$ by strict concavity of entropy,
so $M$ is in the open interior of the ordinary four-moment feasible set.
The Bellman value is convex and finite on this interior.  Finiteness
follows by representing each marginal mean and strictly intermediate
entropy using finitely many points in $(0,1)$ and taking the product
coupling.  Hence $\zeta_e$ has a subgradient at $M$, furnishing an affine
minorant of $j_e$ in the four moment functions.  The subgradient inequality
applies also at boundary moment tuples: mix such a tuple with $M$, apply
the interior inequality, and use convexity.  In particular it applies to
every deterministic pair law, so the minorant is global on the square.  Positive mass at both
deterministic corners forces this affine function to have zero constant
term and opposite mean coefficients.  It is therefore
$c(t-s)-Ah(s)-Bh(t)$ and touches $j_e$ at the interior atom.  The contact
and stationarity identities force exactly
\eqref{bal:eq:general-endpoint-coefficients}.

To obtain the signs, set $(s,t)=(a\varepsilon,b\varepsilon)$ with any
fixed $a,b>0$ and let $\varepsilon\downarrow0$.  The supporting-plane
gap has the expansion
\[
 Q(a\varepsilon,b\varepsilon)
  =\varepsilon\ln(1/\varepsilon)(Aa+Bb)+O(\varepsilon).
\]
Its nonnegativity for every $a,b>0$ forces $A,B\ge0$.
Finally, any genuine optimizer of the stated form must satisfy the
entropy and difference equations, so uniqueness of the contact triple
already proved excludes other endpoint contacts.
\end{proof}

\begin{remark}[A three-atom optimizer with both contact coordinates below \(1/2\)]
The earlier lower-inverse criterion only covers $x\le1/2\le y$.
For example, take $(x,y)=(1/5,2/5)$, $p=1/2$, and both endpoint masses
equal to $1/4$.  Then
\[
 M=\left(\frac7{20},\frac9{20},
          \frac12h(1/5),\frac12h(2/5)\right),\qquad D=\frac1{10}.
\]
Here the central ratio in \eqref{bal:eq:general-endpoint-ratio-test} is $6$,
while
\[
 \frac{\ln(5/2)}{\ln(5/3)}<2<6
 <\frac{\ln5}{\ln(5/4)}.
\]
The first strict inequality follows from $5/2<(5/3)^2$, and the last
from $(5/4)^6<5$.  Thus this is a unique genuine three-atom optimizer
whose interior contact is strictly on one side of $1/2$.  The uniqueness
of the oriented contact triple excludes a second, cross-half
representation of these same entropy moments and difference.
In this example $J_e(1/5)-J_e(2/5)=\ln(8/3)$, so
\[
 \zeta_e(M)=\frac1{20}\ln\frac83.
\]
\end{remark}

\begin{remark}[Orientation and the zero-difference diagonal case]
Exchange the two complete mean--entropy labels when $\mu_w<\mu_u$.
The preceding classification concerns $D>0$ and positive entropy moments;
it does not assert uniqueness on the diagonal $D=0$.  If
$\mu_u=\mu_w=m$ and $e_u=e_w=e\in[0,h(m)]$, then $\zeta_e(M)=0$.
For $0<m<1$, this value is achieved on at most two diagonal points:
choose $t\in[m,1]$ with $m h(t)/t=e$, and put mass $m/t$ at $(t,t)$
and the remaining mass at $(0,0)$.  Such $t$ exists because
$h(t)/t$ decreases continuously, with derivative $\ln(1-t)/t^2$,
from $h(m)/m$ to zero.  If the entropy moments differ, no law on the
diagonal can realize them, and no endpoint law with one interior atom
can have $D=0$.
\end{remark}

The full proof of Theorem~\ref{bal:thm:small-boundary}, including its
scalar interval estimates and endpoint bounds, is given in
Supplementary Section~\ref{supp:small-boundary-proof}.

\begin{theorem}[Zero-entropy boundary inequality]
\label{bal:thm:zero-target}
If a feasible four-moment tuple has $e_u=0$ or $e_w=0$, its Bellman
inequality holds in the well-defined addition form
\begin{equation}\label{bal:eq:zero-bellman-addition}
 \zeta(M)+\tfrac12\phi(\mu_u,e_u)+\tfrac12\phi(\mu_w,e_w)
 \ge\phi\!\left(\frac{\mu_u+\mu_w}{2},\frac{e_u+e_w}{2}\right).
\end{equation}
Moreover, $\zeta(M)=0$ if $e_u=e_w=0$ and $\mu_u=\mu_w$;
at every other such tuple, $\zeta(M)=+\infty$.
\end{theorem}
\begin{proof}
Suppose \(e_u=\mathbb E\HH(U)=0\).  Since binary entropy is nonnegative and
vanishes only at \(0,1\), one has \(U\in\{0,1\}\) almost surely.  For an
endpoint \(u\in\{0,1\}\), the lower-semicontinuous edge cost
\(j(u,w)\) is finite only when \(w=u\).  Hence every finite-cost
representing law has \(W=U\in\{0,1\}\) pointwise, without any ordering
assumption.  Consequently both entropy moments vanish and the means
agree.  Conversely, for every common mean $m$, the law $U=W$ with
$U\sim\operatorname{Bernoulli}(m)$ has these moments and zero cost, so
$\zeta=0$.  In this case the two sides of
\eqref{bal:eq:zero-bellman-addition} are both $\phi(m,0)$, either zero or
$+\infty$, and the extended-real order is well-defined.  At every other
tuple with $e_u=0$, all representing laws have infinite cost, so the left
side is $+\infty$ and the addition-form inequality again holds.  All
candidate values being added are nonnegative.  The case $e_w=0$ is
symmetric.  The proof does not use $\Rphi$ or $\Target$ at the boundary,
where they need not be defined as differences of extended-real values.

\end{proof}

\begin{theorem}[Mean-interior reduction]\label{bal:thm:e8-main}
At fixed positive strictly submaximal entropies, a smooth mean-interior
critical point of \(\Gap\) cannot occur.  The boundary of that fixed-entropy
mean problem consists of
\[
 \mu_u=\mu_w,\qquad \mu_u=\frac12,\qquad \mu_w=\frac12,\qquad
 \HH(\mu_u)=e_u,\qquad \HH(\mu_w)=e_w.
\]
Thus the reduction returns equal means, a half mean, or a deterministic
entropy cap.  It makes no assertion about zero or maximal entropy, or the outer
interface \(a+c=S\).
\end{theorem}
\begin{proof}
In either open mean-order chamber, a critical point would satisfy the
stationarity equations \eqref{bal:imp:e8:eq:stationarity} and the
ratio identity \eqref{bal:imp:e8:eq:ratio-identity}. The substitutions in
Supplementary Section~\ref{bal:imp:e8} then force equality in
\eqref{bal:imp:e8:eq:eq8}, contrary to the strict inequality proved in
Theorem~\ref{bal:imp:e8:thm:eq8}. The detailed calculation is given in
Corollary~\ref{bal:imp:e8:cor:reduction}. Continuity on the fixed-entropy
feasible mean rectangle therefore leaves only its boundary and the
equal-mean interface, which are the families listed above.
\end{proof}

\begin{lemma}[Fixed-entropy mean partition]
\label{bal:lem:outer-minimum-partition}
Fix \(e,f\in(0,1)\).  A negative value on \(K_{e,f}\) has a negative
mean-minimum which lies on \(a+c=S\), at equal or half means, on a
deterministic entropy cap, or in the smooth fixed-entropy mean-interior where
Theorem~\ref{bal:thm:e8-main} applies.
\end{lemma}
\begin{proof}
Compactness and continuity on \(K_{e,f}\) give attainment.  Its defining
closed inequalities show that the mean boundary consists exactly of the
interface \(a+c=S\), mean-order/half-mean faces, and the two marginal entropy
caps.  At every remaining point Fermat's rule applies, and Theorem~\ref{bal:thm:e8-main} is the exact
fixed-entropy interior reduction.
\end{proof}

\paragraph{Coordinates on a deterministic cap.}
\label{bal:par:cap-coordinates}
After exchanging the complete marginal labels if necessary, let $a$ be
the capped mean and $c$ the other mean, and put $b=\iota(e_w)$.  Thus
\[
 M_{\rm cap}(a,b,c)=(a,c,\HH(a),\HH(b)),\qquad
 h=\frac{\HH(a)+\HH(b)}2,
 \quad 0<a,c\le\tfrac12,\quad0<b\le c.
\]
Here $(a,c)$ need not be ordered: the label $a$ identifies the capped
marginal.  In particular, the radial ordering $b<c<a$ is compatible with
the ordered coordinates used before this relabeling.  The retained
condition is $a+c\ge S$.  Direct substitution gives
\begin{equation}\label{bal:eq:cap-gap-definition}
\begin{aligned}
 \Gap_{\rm cap}(a,b,c)&:=\Gap(M_{\rm cap}(a,b,c))\\
 &=j(a,b)-F(|a-b|,h)+F(|a-c|,h)\\
 &\quad-\Phi(1-a-c,h)+\tfrac12\phi(c,\HH(b)).
\end{aligned}
\end{equation}
We used $\phi(a,\HH(a))=0$.  A stationary $c$-fiber below means
that $a,b$ are fixed and $\partial_c\Gap_{\rm cap}=0$ at an interior
point of the feasible retained interval for $c$.  The three strict
orderings of $a,b,c$ have names, recorded in \eqref{bal:eq:three-caps}:
Case~E is $a<b<c$, the radial ordering is $b<c<a$, and the middle ordering
is $b<a<c$.

\begin{theorem}[Boundary and stationary-point estimates]\label{bal:thm:boundary-estimates}
The zero-entropy assertion uses the addition-form Bellman inequality.
All other value assertions concern the pure gap $\Gap=L_4-\Rphi$
at positive feasible entropy coordinates. The double-cap and
cap--half-mean estimates follow from Theorem~\ref{bal:hyp:SCplus}.
\begin{enumerate}[label=\textup{(\alph*)}]
\item \textbf{Equal-mean lemma:} \(\Gap\ge0\) at equal means.
\item \textbf{Double-cap theorem:} clause (iv) of Theorem~\ref{bal:hyp:SCplus} proves
\(\Gap\ge0\) on every strict retained lower-half deterministic double cap;
its boundary is treated analytically.
\item \textbf{Cap--half-mean theorem:} clause (v) of Theorem~\ref{bal:hyp:SCplus} proves the direct
pure-gap value at the Case-E cap--half-mean intersection.
\item \textbf{Zero-entropy theorem:} an original feasible tuple with a zero entropy moment
satisfies the addition-form Bellman inequality
\eqref{bal:eq:zero-bellman-addition}.
\item \textbf{Zero-entropy seam lemma:} the zero-entropy closure of the fixed-sum seam pure
gap has no finite negative value.
\item \textbf{Maximal-entropy continuity lemma:} after pointwise entropy rearrangement, a negative
positive maximal-entropy value has a negative marginal-feasible continuation
with both entropies strictly submaximal.
\item \textbf{Half-mean estimate:} a smooth adverse-entropy half-mean point with both
positive entropy constraints strict and the other mean fixed-entropy
interior cannot be a local negative minimum; its
transverse scalar core is
\[
 D_0(s)=Q'(s)Q(2s)-2Q'(0)\bigl(Q(2s)-Q(s)\bigr)>0\quad(s>0),
\]
where \(Q=\Theta^{-1}\).
\item \textbf{Stationary cap reduction:} on the retained Case-E cap \(0<a<b<c<1/2\), every
admissible interior stationary solution of the one-dimensional \(c\)-fiber
has \(\Gap\ge0\). The chart, certificate inequalities, and exact
coverage argument are given in Supplementary
Section~\ref{bal:sec:casee-stationary}.
\item \textbf{Radial stationary-point estimate:} on the retained radial cap
\(0<b<c<a<1/2\), \(a+c\ge S\), every admissible interior stationary
solution of the one-dimensional \(c\)-fiber has \(\Gap\ge0\).
The reduction, terminal acceptance inequalities, and boundary attachments
are given in Supplementary Section~\ref{bal:sec:radial-stationary}.
\item \textbf{Middle-cap inequality:} on the middle cap \(0<b<a<c<1/2\), analytic
subadditivity of \(\Theta\) gives \(\Gap\ge0\).
\end{enumerate}
\end{theorem}

\begin{lemma}[Global equal-entropy inequality]
\label{bal:lem:global-equal-entropy}
If \(e_u=e_w=e>0\), then \(\Gap\ge0\).
\end{lemma}
\begin{proof}
Put
\[
 x=|\mu_w-\mu_u|,\qquad r=|1-\mu_u-\mu_w|.
\]
The two individual radial coordinates satisfy the multiset identity
\[
 \{|1-2\mu_u|,|1-2\mu_w|\}=\{r+x,|r-x|\}.
\]
Since \(g(e,e)=0\) and \(\Phi=\eta-F\), direct expansion of \(\Rphi\) gives
\[
 \Gap
 =F(x,e)+F(r,e)
  -\frac12F(r+x,e)-\frac12F(|r-x|,e).
\]
Supplementary Lemma~\ref{bal:aux:fourpoint} proves this four-point perspective inequality
directly from the decreasing second derivative of the perspective, whose
profile signs are established in Supplementary Section~\ref{bal:imp:e8}.
The zero-entropy endpoint is not part of this lemma; it is covered by
Theorem~\ref{bal:thm:zero-target}.
\end{proof}

\begin{lemma}[Global equal-mean inequality]
\label{bal:lem:global-equal-mean}
For a feasible tuple with $e_u,e_w>0$ and $\mu_u=\mu_w=m$,
one has $\Gap\ge0$.
\end{lemma}
\begin{proof}
Let \(h=(e_u+e_w)/2\) and \(s=|1-2m|\).  Since \(F(0,h)=0\),
direct expansion gives
\[
 \Gap(m,m,e_u,e_w)
 =g(e_u,e_w)+\frac12\Phi(s,e_u)+\frac12\Phi(s,e_w)-\Phi(s,h).
\]
The first term is nonnegative, and certified convexity of
\(r\mapsto\Phi(s,r)\) makes the remaining Jensen difference nonnegative.
Here \(g\ge0\) follows from joint convexity, symmetry, and its zero
diagonal in Supplementary Section~\ref{bal:imp:jointconvexity}; the full entropy-convexity
argument for \(\Phi\) is given in Supplementary Theorem~\ref{bal:aux:entropy}.
\end{proof}

\begin{proposition}[Analytic double-cap entropy bound]
\label{bal:prop:double-cap-entropy}
Let $0<a\le b\le1/2$, let
$M=(a,b,\HH(a),\HH(b))$, and put
\[
 \Delta_H(a,b)=\HH\!\left(\frac{a+b}{2}\right)
                       -\frac{\HH(a)+\HH(b)}2.
\]
Then
\[
 \boxed{L_4(M)=j(a,b)\ge4\Delta_H(a,b).}
\]
The inequality is strict when $a\ne b$, and the constant $4$ is sharp.
In particular, if $b\le1/100$, the established same-side bound gives
\[
 \boxed{L_4(M)\ge4\Delta_H(a,b)\ge R_\phi(M).}
\]
This proves the pure-gap inequality on the entire small-mean double cap,
including $a+b=S$ and the remaining portion $a+b>S$ with $b\le1/100$.
\end{proposition}
\begin{proof}
Since $\iota(\HH(a))=a$ and $\iota(\HH(b))=b$, the definitions give,
with $h=(\HH(a)+\HH(b))/2$,
\[
 L_4(M)=F(b-a,h)+\kappa(a,b)=j(a,b).
\]
The comparison $j\ge4\Delta_H$ is already proved in
\eqref{bal:eq:sb-edge-jensen}, independently of the small-mean restriction.
Here is another short proof that also identifies its equality cases.
Let $P=(a,1-a)$, $Q=(b,1-b)$, and, for each coordinate, set
$s_i=P_i+Q_i$ and $r_i=(P_i-Q_i)/s_i$.  Direct expansion yields
\[
 (\ln2)\{j(a,b)-4\Delta_H(a,b)\}
 =\sum_{i=1}^2s_i\ell(r_i),\qquad
 \ell(r)=-\ln(1-r^2)-r\operatorname{artanh}r.
\]
Since
\[
 \ell(0)=\ell'(0)=0,\qquad
 \ell''(r)=\frac{2r^2}{(1-r^2)^2}\ge0,
\]
every summand is nonnegative and is positive when $r_i\ne0$.
Thus equality holds exactly at $a=b$.  For any fixed interior $a$,
Taylor expansion as $b\to a$ gives $j(a,b)/\Delta_H(a,b)\to4$,
proving sharpness.  This argument proves the scalar comparison on the
whole probability interval, with boundary cases interpreted by limits.
Simultaneous complementation gives the same $L_4=j$ conclusion when both
means are in the upper half.

For $a,b\le1/100$, the same-side proof of
Theorem~\ref{bal:thm:small-boundary} establishes
$R_\phi(M)\le4\Delta_H(a,b)$, including both entropy caps by continuity.
Combining the inequalities proves the final assertion directly at the
level of $L_4-R_\phi$.
\end{proof}

\begin{theorem}[Generic double-cap inequality]
\label{bal:thm:double-cap}
By Theorem~\ref{bal:hyp:SCplus}, for
\[
 0<a<b<\frac12,\qquad a+b>S,
\]
the deterministic tuple
\[
 (\mu_u,\mu_w,e_u,e_w)=(a,b,\HH(a),\HH(b))
\]
has \(\Gap\ge0\).
\end{theorem}
\begin{proof}
Because \(a,b\) are the lower entropy representatives,
\[
 L_4=F\!\left(b-a,\frac{\HH(a)+\HH(b)}2\right)
      +\kappa(a,b)=j(a,b).
\]
If $b\le1/100$, Proposition~\ref{bal:prop:double-cap-entropy} already proves
the result without Theorem~\ref{bal:hyp:SCplus}.  For the remaining $b>1/100$, clause (iv) of Theorem~\ref{bal:hyp:SCplus}
is exactly
\[
 \Gap=j(a,b)-\Phi\!\left(1-a-b,
              \frac{\HH(a)+\HH(b)}2\right)\ge0.
\]
The boundary cases are treated elsewhere: \(a=b\) by
Lemma~\ref{bal:lem:global-equal-mean}, and \(a+b=S\) by the argument for the
double-cap edge in Lemma~\ref{bal:lem:deterministic-seam-facets}.  At
\(b=1/2\) the entropy \(\HH(b)\) is maximal, and
Theorem~\ref{bal:thm:pure-gap} handles this case through the continuity
reduction of Lemma~\ref{bal:lem:maximal-continuity}.  The face \(a=0\) has
a zero entropy moment and is covered by Theorem~\ref{bal:thm:zero-target}.
\end{proof}

\begin{theorem}[Cap--half-mean inequality]
\label{bal:thm:cap-half-mean}
By Theorem~\ref{bal:hyp:SCplus}, for \(0<a<b<1/2\), put
\[
 h=\frac{\HH(a)+\HH(b)}2,
 \qquad x=\frac12-a.
\]
Then the Case-E endpoint with deterministic first marginal and second mean
\(1/2\) satisfies
\[
 \Gap_{\rm HF}(a,b)
 =j(a,b)-F(b-a,h)+2F(x,h)-\eta(h)
   +\frac12\eta(\HH(b))\ge0.
\]
\end{theorem}
\begin{proof}
The displayed identity is direct substitution in \(L_4-\Rphi\); its
nonnegativity is exactly clause (v) of Theorem~\ref{bal:hyp:SCplus}.
\end{proof}

The following theorem collects the inequalities on the seam and at the
endpoints of the caps. Their proofs are given in
Supplementary Sections~\ref{bal:sec:seam-normalization}--\ref{bal:sec:cap-sharp},
and the records of the computations in Supplementary
Section~\ref{bal:sec:verification}. The stationary Case-E and radial
estimates in Theorem~\ref{bal:thm:boundary-estimates}(h) and (i) are proved
in the consecutive Supplementary Sections~\ref{bal:sec:casee-stationary}
and~\ref{bal:sec:radial-stationary}.

\begin{theorem}[Seam and endpoint inequalities]
\label{bal:hyp:SCplus}
\label{bal:thm:SCplus}
Let \(S=10^{-4}\),
\[
 a=\frac{S-y}{2},\qquad c=\frac{S+y}{2}.
\]
The following five conclusions hold:
\begin{enumerate}[label=\textup{(\roman*)}]
\item on
\[
 0<y<S,\quad 0<e_u<e_w,\quad
 e_u<\HH(a),\quad e_w<\HH(c),
\]
with \(e_u=h+d,e_w=h-d\),
and with the frozen-coordinate conventions
\[
 \Gap_y=\left.\partial_y\Gap\right|_{h,d},\qquad
 \Gap_d=\left.\partial_d\Gap\right|_{y,h},\qquad
 \Gap_{e_u}=\left.\partial_{e_u}\Gap\right|_{y,e_w},
\]
\[
 \Gap<0,\qquad \Gap_y=0\Longrightarrow
 \bigl(\Gap_d<0\bigr)\ \text{or}\ \bigl(\Gap_{e_u}<0\bigr);
\]
\item on \(U^\circ\), namely
\[
 0<y<S,\quad e_u=\HH(a),\quad
 \HH(a)<e_w<\HH(c),
\]
one has \(\Gap\ge0\);
\item on \(W^\circ\), namely
\[
 0<y<S,\quad 0<e_u<\HH(a),\quad e_w=\HH(c),
\]
one has \(\Gap\ge0\).
\item on the generic retained double-deterministic domain
\[
 0<a<b<\frac12,\qquad a+b>S,\qquad
 h=\frac{\HH(a)+\HH(b)}2,
\]
one has the direct value inequality
\[
 \Gap_{\rm dbl}(a,b)
 :=j(a,b)-\Phi(1-a-b,h)\ge0.
\]
The portion $b\le1/100$ is already proved analytically by
Proposition~\ref{bal:prop:double-cap-entropy}; the complementary region is proved
by the R1 reduction and interval certificate in
Supplementary Section~\ref{bal:sec:doublecomplete}.
\item on the Case-E cap--half-mean intersection
\[
 0<a<b<\frac12,\qquad
 h=\frac{\HH(a)+\HH(b)}2,\qquad x=\frac12-a,
\]
one has the direct value inequality
\[
 \Gap_{\rm HF}(a,b)
 :=j(a,b)-F(b-a,h)+2F(x,h)-\eta(h)
   +\frac12\eta(\HH(b))\ge0.
\]
\end{enumerate}
The last four clauses concern the pure gap \(L_4-\Rphi\).
\end{theorem}

For a clause that depends on a computation, the certificate specifies an
exact partition of the domain and outward-rounded enclosures on every cell.
Supplementary Section~\ref{bal:sec:verification} identifies the recorded
executions, structural audits, and arithmetic replays, with their respective
coverage. Repetition at a higher precision is distinguished from verification
by an independently implemented arithmetic kernel.
Proposition~\ref{bal:prop:cap-clause-redundancy} shows that a single
inequality on the deterministic caps would imply clauses (ii)--(v).  This
implication is not used in the proofs below.

\begin{proof}
The clauses serve two purposes in the minimum arguments of
Sections~\ref{bal:sec:seam} and~\ref{bal:sec:retained}. Clause~(i)
excludes a negative stationary point in the interior of the seam; the
other clauses give values of \(\Gap\) on the deterministic facets of the
seam and at the endpoints of the caps.

For clause~(i), the normalization in Supplementary
Section~\ref{bal:sec:seam-normalization} preserves the signs of
the gap and all three frozen partial derivatives. The analytic
low-entropy and equality collars, together with the exact two-chart
cover in Supplementary
Sections~\ref{bal:sec:low}--\ref{bal:sec:union}, establish
the stated minimum-exclusion implication. In particular, the
premise $\Gap<0$ is retained when a region is covered by a
nonnegative value bound.

Clauses~(ii) and~(iii) are proved in Supplementary
Sections~\ref{bal:sec:ii} and~\ref{bal:sec:iii},
respectively. Clause~(iv) follows from the small-mean argument
and the complete R1 reduction and certificate in Supplementary
Section~\ref{bal:sec:doublecomplete}. Clause~(v) follows from
clause~(iv) and Supplementary Theorem~\ref{bal:thm:m1m2},
which combines the M1 and M2 estimates with the analytic
large-logit region. The exact computational domains, acceptance
conditions, and verification records are specified in
Supplementary Section~\ref{bal:sec:verification}.
\end{proof}

\begin{proposition}[A full cap theorem discharges the seam and endpoint clauses (ii)--(v)]
\label{bal:prop:cap-clause-redundancy}
Let $S=10^{-4}$.  Suppose the following pointwise cap theorem is available:
\begin{equation}
\begin{gathered}
0<\mu_u,\mu_w\le\tfrac12,\qquad \mu_u+\mu_w\ge S,\\
e_u=\HH(\mu_u),\qquad 0<e_w\le\HH(\mu_w)
\quad\Longrightarrow\quad \Gap(M)\ge0.
\end{gathered}
\tag{CAP}\label{bal:eq:full-cap-input}
\end{equation}
Then clauses (ii)--(v) of Theorem~\ref{bal:hyp:SCplus} follow, so only
clause (i) requires an additional mathematical input.  In particular,
no separate numerical certification of the four value clauses is needed.
The theorem must include both orders of the means; the cap may belong to
the smaller or larger mean.
\end{proposition}
\begin{proof}
The four substitutions are exact:
\begin{center}
\begin{tabular}{lll}
\toprule
Clause & Tuple to which CAP is applied & Sum of means\\
\midrule
(ii) & $(a,c,\HH(a),e_w)$ & $a+c=S$\\
(iii) & $(c,a,\HH(c),e_u)$, after label exchange & $c+a=S$\\
(iv) & $(a,b,\HH(a),\HH(b))$ & $a+b>S$\\
(v) & $(a,\tfrac12,\HH(a),\HH(b))$ & $a+\tfrac12>S$\\
\bottomrule
\end{tabular}
\end{center}
Full label exchange preserves $L_4$, $R_\phi$, and feasibility.  In
clause (v), $0<b<1/2$ implies $0<\HH(b)<\HH(1/2)=1$.
Thus every row satisfies CAP and gives precisely the value inequality
in that clause.  The equal-sum boundary is explicitly included in CAP.
If CAP is stated only at strict positive-entropy endpoints, its finite
continuous extension to the displayed second-cap and half-mean endpoints
suffices; this extension must be part of its proof.
\end{proof}

Within clause (i), the alternatives provide a strict descent direction
at a hypothetical negative stationary point: either the entropy-difference
derivative $\Gap_d$ or the entropy derivative $\Gap_{e_u}$ is negative, with the other independent
coordinates held fixed as specified in the clause. The normalized derivative
conventions and the exact covering argument are given in Supplementary
Sections~\ref{bal:sec:seam-normalization} and~\ref{bal:sec:union}.
Each derivative test is used only on the region where its sign is certified.

\section{Symmetries and decomposition of the parameter domain}
\label{bal:sec:orientation}

The unrestricted Bellman target has the exact label-exchange and simultaneous
complement symmetries
\begin{align}
 \mathcal E(\mu_u,\mu_w,e_u,e_w)
   &=(\mu_w,\mu_u,e_w,e_u),\label{bal:eq:label-exchange}\\
 \mathcal K(\mu_u,\mu_w,e_u,e_w)
   &=(1-\mu_u,1-\mu_w,e_u,e_w).\label{bal:eq:simultaneous-complement}
\end{align}
They are induced respectively by
\((U,W)\mapsto(W,U)\) and
\((U,W)\mapsto(1-U,1-W)\), and preserve
\(\zeta,\Rphi,L_4\), and \(\Gap\).  We use \(\mathcal E\) first whenever
necessary to arrange \(\mu_u\le\mu_w\).  The derived map
\begin{equation}
 \mathcal C(\mu_u,\mu_w,e_u,e_w)
 =(1-\mu_w,1-\mu_u,e_w,e_u).
 \label{bal:eq:full-complement}
\end{equation}
is \(\mathcal K\circ\mathcal E\); it preserves the chosen mean order as
well as \(\Target\) and \(\Gap\) in \eqref{bal:eq:two-gaps}.

For completeness, push an arbitrary representing law forward by
\((U,W)\mapsto(W,U)\) for \(\mathcal E\), and by
\((U,W)\mapsto(1-U,1-W)\) for \(\mathcal K\).  Each map is its own inverse.
The identities
\[
 j(w,u)=j(u,w),\qquad j(1-u,1-w)=j(u,w),
 \qquad \HH(1-t)=\HH(t),\qquad \phi(1-m,e)=\phi(m,e)
\]
prove the asserted invariances of \(\zeta\) and \(\Rphi\); the formulas for
\(L_4\) and \(\Gap\) give their invariance immediately.

The explicit function \(\Gap\) has one more symmetry, the reflection of a
single mean about \(1/2\).  We do not claim that \(\zeta\) or \(\Target\) is
invariant under this reflection, and it is applied to \(\Gap\) only.

\begin{lemma}[Individual mean reflection for the pure gap]
\label{bal:lem:pure-gap-reflection}
Reflecting either mean about \(1/2\), while retaining its attached entropy,
preserves \(\Gap\), after full relabeling if needed.
\end{lemma}
\begin{proof}
Use the global aggregate coordinates
\[
 x=\mu_u+\mu_w-1,\qquad y=\mu_w-\mu_u,\qquad
 h=\frac{e_u+e_w}{2},\qquad d=\frac{e_u-e_w}{2}.
\]
In these coordinates,
\[
\begin{aligned}
 \Gap(x,y,h,d)={}&F(|y|,h)+g(h+d,h-d)-\Phi(|x|,h)\\
 &+\frac12\Phi(|y-x|,h+d)
  +\frac12\Phi(|x+y|,h-d).
\end{aligned}
\]
Reflecting \(\mu_u\) sends \((x,y)\) to \((y,x)\), whereas reflecting
\(\mu_w\) sends \((x,y)\) to \((-y,-x)\).  In either case the two outer
radial arguments are preserved with their attached entropy labels.  Moreover
\[
 F(|y|,h)-\Phi(|x|,h)
 =F(|y|,h)+F(|x|,h)-\eta(h)
\]
is symmetric in \(|x|,|y|\).  Thus the complete gap is unchanged.  Full
relabeling swaps both means and both entropy coordinates and also preserves
the expression.
\end{proof}

Accordingly, in the case decomposition for \(\Gap\), define
\[
 r_u=\min\{\mu_u,1-\mu_u\},\qquad
 r_w=\min\{\mu_w,1-\mu_w\},
\]
reflect the means algebraically using Lemma~\ref{bal:lem:pure-gap-reflection},
and sort the two complete mean--entropy labels to obtain
\(0\le a\le c\le1/2\).  This is the canonical same-side lower-half chart
used by Theorem~\ref{bal:thm:e8-main} and Theorem~\ref{bal:thm:pure-gap}.

For the pointwise pure-gap reduction, enlarge to the marginal-physical box
\[
 \mathcal R_S=\left\{(a,c,e,f):
 0\le a\le c\le\frac12, a+c\ge S,
 0<e\le\HH(a), 0<f\le\HH(c)\right\}.
\]
This larger box is used only in the following rearrangement lemma, which
is a pointwise comparison.

\begin{lemma}[Global entropy rearrangement]
\label{bal:lem:global-entropy-rearrangement}
For every \((a,c,e,f)\in\mathcal R_S\) with \(e>f\),
\[
 \Gap(a,c,e,f)\ge\Gap(a,c,f,e).
\]
Consequently every negative positive-entropy value has a negative
representative in the chamber \(e\le f\).
\end{lemma}
\begin{proof}
Suppose \(e>f\), and put
\[
 s_+=1-2a,\qquad s_-=1-2c.
\]
Then \(s_+\ge s_-\).  The swapped entropy pair \((f,e)\) remains
marginal-physical because
\(f<e\le\HH(a)\le\HH(c)\).  The terms in the global gap other than its two
outer \(\Phi\)-terms are invariant under the swap.  For \(s,r>0\), the
identity \(F(s,r)=2r\Gamma(s/(2r))\), with \(z=s/(2r)\), gives
\[
 F_{sr}(s,r)=-\frac{z}{r}\Theta'(z)<0,
 \qquad \Phi_{sr}(s,r)>0.
\]
At \(s=0\), the certified continuous one-sided extension satisfies
\(\Phi_{sr}(0,r)=0\).  Hence \(\Phi_{sr}\ge0\) on the entire closed
integration rectangle, including the possible endpoint \(c=1/2\).
Consequently
\[
 \Gap(a,c,e,f)-\Gap(a,c,f,e)
 =\frac12\int_{s_-}^{s_+}\int_f^e
       \Phi_{sr}(s,r)\,dr\,ds\ge0.
\]
Thus every value in the \(e>f\) chamber is bounded below by a value in the
\(e\le f\) chamber.
\end{proof}

\begin{lemma}[Maximal-entropy continuity reduction]
\label{bal:lem:maximal-continuity}
Let \((a,c,e,f)\in\mathcal R_S\) satisfy \(e\le f\).  If \(\Gap<0\), then
there is a negative marginal-physical point with the same means and with both
entropy coordinates in \((0,1)\).
\end{lemma}
\begin{proof}
Only \(f=1\) requires attention.  Marginal feasibility then forces
\(c=1/2\).  If also \(e=1\), feasibility and \(a\le c\) force
\(a=c=1/2\), where Lemma~\ref{bal:lem:global-equal-mean} gives \(\Gap\ge0\).  Otherwise \(0<e<1\).  Choose
\(e<f_n<1\) with \(f_n\uparrow1\), leaving \((a,c,e)\) fixed.  Every
\((a,c,e,f_n)\) remains marginal-physical.  The elementary endpoint
continuity of \(\iota,\eta,F,\Phi\), together with
\[
 g(e,f_n)=\kappa\bigl(\iota(e),\iota(f_n)\bigr)
       \longrightarrow g(e,1),
\]
gives \(\Gap(a,c,e,f_n)\to\Gap(a,c,e,1)<0\).  Hence a sufficiently large
\(n\) supplies the asserted strictly submaximal negative point.

The endpoint extension is included in Supplementary Section~\ref{bal:imp:jointconvexity}.
\end{proof}

In the case decomposition for \(\Target\), the reflection of a single mean
is not available.  First apply the exact label exchange \(\mathcal E\), if needed, so that
\(\mu_u\le\mu_w\).  The resulting mean-canonical pair is assigned to the
same lower distances as follows.
\begin{enumerate}
\item If \(\mu_w\le1/2\), take the same-side lift
\((\mu_u,\mu_w)=(a,c)\).
\item If \(\mu_u\ge1/2\), first apply \(\mathcal C\), then use the preceding
row.
\item If \(\mu_u\le1/2\le\mu_w\), put \(r=\mu_u\) and \(s=1-\mu_w\).
If \(r\le s\), take \(a=r,c=s\) and the opposite-side lift
\((\mu_u,\mu_w)=(a,1-c)\).  If \(s<r\), first apply \(\mathcal C\); the new
opposite-side lift is again \((a,1-c)\), now with \(a=s,c=r\) and the two
entropy coordinates exchanged.
\end{enumerate}
These cases are disjoint after giving equality to the earlier row.  Thus the
scalar \(a+c=r_u+r_w\) is defined for every feasible tuple after exact label
canonicalization.  When this scalar is used for \(\Gap\), the means are
reflected by Lemma~\ref{bal:lem:pure-gap-reflection}; when it is used for
\(\Target\), the same-side or opposite-side lift is kept.

For fixed entropy coordinates \(e,f\), the complete target orbits are
\[
\begin{aligned}
 \mathcal O_{\rm same}(a,c;e,f)=\{&
 (a,c,e,f),(c,a,f,e),\\[-2pt]
 &(1-a,1-c,e,f),(1-c,1-a,f,e)\},\\
 \mathcal O_{\rm opp}(a,c;e,f)=\{&
 (a,1-c,e,f),(1-c,a,f,e),\\[-2pt]
 &(1-a,c,e,f),(c,1-a,f,e)\}.
\end{aligned}
\]
They are generated by \(\mathcal E\) and \(\mathcal K\) alone, which shows
that every labeled orientation is recovered from its canonical form by
these two symmetries of \(\zeta\).

On a pure-gap deterministic chart, the complete exchange is
\begin{equation}
 (y,h,d;a,c;e,f;r_u,r_w)
 \longmapsto(-y,h,-d;c,a;f,e;r_w,r_u),
 \qquad r_u=\iota(e),\quad r_w=\iota(f).
 \label{bal:eq:full-exchange}
\end{equation}
It swaps the two means, both entropy coordinates, and all derived inverse
coordinates.  The exchange reverses the chosen order of the means, so each
use is followed by the canonicalization above.

The primary global domain decomposition is the following precedence list.
\begin{center}
\small
\begin{tabularx}{\textwidth}{>{\raggedright\arraybackslash}p{0.09\textwidth}
 >{\raggedright\arraybackslash}p{0.31\textwidth}
 >{\raggedright\arraybackslash}p{0.17\textwidth}X}
\toprule
Order & Canonical domain & Required level & Primary estimate \\
\midrule
1 & an original entropy coordinate is zero
& \level{target} & Theorem~\ref{bal:thm:zero-target}, before any pure-gap canonicalization \\
2 & positive entropies and \(a+c<S\), \(S=10^{-4}\)
& \level{target} & Theorem~\ref{bal:thm:small-boundary} under the same-/opposite-side lift \\
3 & positive entropies and \(a+c=S\)
& \level{pure-g} & Theorem~\ref{bal:thm:seam-closure};
Theorem~\ref{bal:thm:small-boundary} also proves \(\Target\ge0\) there \\
4 & positive maximal entropy
& \level{reduction} & Lemma~\ref{bal:lem:maximal-continuity} reduces a putative negative value to strictly
submaximal positive entropies; the double-maximal point is covered by Lemma~\ref{bal:lem:global-equal-mean} \\
5a & \(a+c>S\), strictly submaximal positive equal entropies
& \level{pure-g} & Lemma~\ref{bal:lem:global-equal-entropy} \\
5b & \(a+c>S\), strictly submaximal positive entropies, equal means
& \level{pure-g} & Lemma~\ref{bal:lem:global-equal-mean} \\
5c & \(a+c>S\), strictly submaximal positive entropies, a deterministic cap
& \level{reduction} then endpoint estimates & stationary cap reduction, the radial stationary-point estimate, or the middle-cap inequality plus
Lemma~\ref{bal:lem:global-equal-entropy}, the cap-fiber lemma, and clauses (ii)--(v) of Theorem~\ref{bal:hyp:SCplus} \\
5d & \(a+c>S\), a smooth half mean away from a deterministic cap
& \level{reduction} & the half-mean estimate excludes a local negative minimizer; the
cap--half-mean intersection remains in row 5c \\
6 & \(a+c>S\), smooth mean and strictly submaximal entropy interior
& \level{reduction} then \level{pure-g} & Theorem~\ref{bal:thm:e8-main} reduces to one of the preceding
boundary estimates \\
\bottomrule
\end{tabularx}
\end{center}

In the cap coordinates of \eqref{bal:eq:cap-gap-definition}, the three
strict deterministic orderings are exactly
\begin{equation}
 a<b<c\quad\text{(Case E)},\qquad
 b<c<a\quad\text{(radial)},\qquad
 b<a<c\quad\text{(analytic middle)}.
 \label{bal:eq:three-caps}
\end{equation}
Ties belong to equal-entropy, equal-mean, half-mean, or double-cap endpoint
estimates.  Thus
\eqref{bal:eq:three-caps} and the tie strata exhaust every deterministic cap.

\begin{proposition}[Exhaustion of the parameter domain]
\label{bal:prop:owner-exhaustion}
The precedence table above assigns every feasible unrestricted tuple to one
primary domain row, and every retained-region boundary alternative produced
by Theorem~\ref{bal:thm:e8-main} to one applicable pure-gap estimate or minimum-exclusion estimate.
\end{proposition}
\begin{proof}
First remove original zero-entropy tuples by Theorem~\ref{bal:thm:zero-target} at target level, and apply
\(\mathcal E\), if needed, to arrange \(\mu_u\le\mu_w\).  For a remaining
positive-entropy tuple, the two lower distances \(r_u,r_w\) exist uniquely and their
sorted values \(a\le c\) are unique, including ties.  The four target lifts
listed above, together with exact label exchange, prove that this
lower-distance record has a complete inverse to the original mean/side
orientation.

Exactly one of
\[
 a+c<S,\qquad a+c=S,\qquad a+c>S
\]
holds. The first is the row for Theorem~\ref{bal:thm:small-boundary}. The second is the seam row. In the
third, pointwise entropy rearrangement fixes the entropy orientation, and
Lemma~\ref{bal:lem:maximal-continuity} reduces any remaining negative maximal-entropy value to strictly
submaximal positive entropies.  Lemma~\ref{bal:lem:outer-minimum-partition} then partitions the compact
fixed-entropy mean set.  Theorem~\ref{bal:thm:e8-main} excludes its smooth interior or returns an equal
mean, a half mean, or a deterministic cap.  Endpoint precedence removes all ties.
On a strict deterministic cap, the relative order of the three distinct
lower-half means is exactly one of \eqref{bal:eq:three-caps}. The stationary cap reduction, the radial stationary-point estimate,
and the middle-cap inequality, together with the cap-fiber endpoint estimates, exhaust that
alternative.

Disjointness follows from the strict inequalities in the displayed
trichotomy and \eqref{bal:eq:three-caps}; all equality cases were assigned before
the strict rows.  The closed boxes of an interval computation may overlap
across these cases; this does not affect the decomposition.
\end{proof}

\section{The pure gap on the seam}
\label{bal:sec:seam}
This section proves \(\Gap\ge0\) on the seam \(a+c=S\), the boundary value
that the minimum argument of Section~\ref{bal:sec:retained} requires. We
first prove \(\Gap\ge0\) on every facet of the seam, including its
zero-entropy limits. A negative minimum would then lie in the interior of
the seam, where clause~(i) of Theorem~\ref{bal:hyp:SCplus} gives a
contradiction.

Put \(S=10^{-4}\), \(A=1-S\), and
\[
 a=\frac{S-y}{2},\qquad c=\frac{S+y}{2},\qquad
 e=e_u=h+d,\qquad f=e_w=h-d.
\]
The full marginal-physical entropy box initially has no ordering between
\(e\) and \(f\). The pure gap on this seam is
\begin{equation}
\begin{aligned}
 \Gap(y,h,d)={}&F(y,h)+g(h+d,h-d)-\Phi(A,h)\\
 &+\frac12\Phi(A+y,h+d)+\frac12\Phi(A-y,h-d),
\end{aligned}
 \label{bal:eq:seam-gap}
\end{equation}
where \(\Phi(s,r)=\eta(r)-F(s,r)\).

\begin{lemma}[Entropy rearrangement on the seam]
\label{bal:lem:entropy-rearrangement}
For \(y>0\) and positive entropies, every value in the \(e>f\) chamber is
bounded below by its swapped value in the \(e\le f\) chamber.  The zero
boundary is treated separately.
\end{lemma}
\begin{proof}
This is Lemma~\ref{bal:lem:global-entropy-rearrangement} with
\(a=(S-y)/2\), \(c=(S+y)/2\), so
\(s_+=A+y\) and \(s_-=A-y\).
\end{proof}

Consequently the closed seam cell is
\begin{equation}
 \mathcal S_{\rm seam}=\{(y,e,f):0\le y\le S,\ 0\le e\le f,\
                  e\le\HH(a),\ f\le\HH(c)\}.
 \label{bal:eq:seam-cell}
\end{equation}
The strict orientation \(e<f\) has \(d<0\), and
\(\Gap_d=\Gap_{e_u}-\Gap_{e_w}\).

There are five irreducible facets:
\[
 M:y=0,\qquad Z:e=0,\qquad T:e=f,\qquad
 U:e=\HH(a),\qquad W:f=\HH(c).
\]
The set \(y=S\) is not a sixth facet: there \(a=0\), hence \(e=0\), and the
section collapses to an edge.

\begin{lemma}[Equal-mean facet]
On \(M\), \(\Gap\ge0\).
\end{lemma}
\begin{proof}
At \(y=0\), with \(h=(e+f)/2\), \eqref{bal:eq:seam-gap} becomes
\[
 \Gap(0,e,f)=g(e,f)+\frac12\Phi(A,e)+\frac12\Phi(A,f)-\Phi(A,h).
\]
The certified joint-convexity theorem gives \(g(e,f)\ge0\), while the
certified entropy-curvature theorem gives \(\Phi_{rr}(A,r)>0\) in the
interior of its domain, by Supplementary Section~\ref{bal:imp:jointconvexity} and
Supplementary Theorem~\ref{bal:aux:entropy}.
Jensen's inequality proves the claim, and the endpoint follows by
the declared lower-semicontinuous extension.
\end{proof}

\begin{lemma}[Zero-entropy facet]
\label{bal:lem:seam-zero}
On \(Z\) with \(f>0\), the lower-semicontinuous value is \(\Gap=+\infty\).
On the double-zero edge the value is zero at \(y=0\) and \(+\infty\) for
\(y>0\).
\end{lemma}
\begin{proof}
The global joint-convexity closure gives
\[
 g(0,f)=+\infty\quad(f>0),\qquad g(0,0)=0.
\]
The intrinsic same-side estimate on this seam is
\[
 \Rphi\le4\Delta_H(a,c),
 \qquad
 \Delta_H(a,c)=\HH(S/2)-\frac{\HH(a)+\HH(c)}2.
\]
This estimate is proved in Supplementary Section~\ref{supp:small-boundary-proof}
for every feasible same-side pair with $0\le a\le c\le1/100$.
The present seam lies in that domain because $c\le S<1/100$.
Since \(L_4=F(y,h)+g(e,f)\), it follows for interior approximants that
\begin{equation}
 \Gap\ge F(y,h)+g(e,f)-4\Delta_H(a,c).
 \label{bal:eq:seam-zero-lower}
\end{equation}
Thus \(e\downarrow0<f\) forces \(\Gap\to+\infty\).

If \((y_n,e_n,f_n)\to(y_0,0,0)\) with \(y_0>0\), set
\(v_n=\mathcal L^{-1}(2h_n/y_n)\).  Then \(v_n\downarrow0\), so
\(F(y_n,h_n)=y_nJ(v_n)\to+\infty\); 
\eqref{bal:eq:seam-zero-lower} proves the double-zero claim.  If \(y_0=0\),
then \(F\ge0\), \(g\ge0\), and \(\Delta_H(a_n,c_n)\to0\) give
\(\liminf\Gap\ge0\).  Along \(y=0,e=f\downarrow0\), the gap is identically
zero, so the lower-semicontinuous value at the origin is exactly zero.  The
same alternatives settle \(y=S\), where feasibility forces \(e=0\).
\end{proof}

\begin{lemma}[Equal-entropy facet]
On \(T\) with positive common entropy, \(\Gap\ge0\), with the same conclusion
under lower-semicontinuous endpoint extension.
\end{lemma}
\begin{proof}
This is the specialization of
Lemma~\ref{bal:lem:global-equal-entropy} to the seam.  Equivalently, here
\(x=y\) and \(r=A\), so its four-point expression is exactly the expansion
of \eqref{bal:eq:seam-gap}.
\end{proof}

\begin{lemma}[Deterministic seam facets]
\label{bal:lem:deterministic-seam-facets}
By Theorem~\ref{bal:hyp:SCplus}, the facets \(U^\circ\) and
\(W^\circ\) have \(\Gap\ge0\).  Their common double-cap edge also has
\(\Gap\ge0\).
\end{lemma}
\begin{proof}
On \(U^\circ\), set \(b=\iota(f)\).  Since
\(\HH(a)=e<f<\HH(c)\), one has \(0<a<b<c<1/2\), the Case-E ordering.
On \(W^\circ\), set \(r_u=\iota(e)<a<c\).  After the complete map
\eqref{bal:eq:full-exchange} and naming the deterministic marginal first,
\[
 (a^\sharp,b^\sharp,c^\sharp)=(c,r_u,a),
 \qquad0<b^\sharp<c^\sharp<a^\sharp<1/2,
\]
the radial ordering.  Clauses (ii) and (iii) of
Theorem~\ref{bal:hyp:SCplus} directly prove the required pure-gap inequalities
on these two facets.

On the common edge \(e=\HH(a),f=\HH(c)\), put
\(h=(\HH(a)+\HH(c))/2\).  The lower-half identities give
\[
 L_4=F(c-a,h)+\kappa(a,c)=j(a,c).
\]
Since $0<a\le c\le S<1/100$, Proposition~\ref{bal:prop:double-cap-entropy}
gives the direct chain
\[
 L_4=j(a,c)\ge4\Delta_H(a,c)\ge R_\phi.
\]
Thus $\Gap\ge0$ on the positive-entropy double-cap edge.  The collapsed
zero endpoint is covered by Lemma~\ref{bal:lem:seam-zero}.  The argument for
this edge does not use clause (iv) of Theorem~\ref{bal:hyp:SCplus}.
\end{proof}

The preceding lemmas cover every boundary stratum of
\(\mathcal S_{\rm seam}\), including
all pairwise and higher intersections.  We now use clause (i) of
Theorem~\ref{bal:hyp:SCplus} for its strict interior.

\begin{theorem}[Same-gap seam closure]\label{bal:thm:seam-closure}
By Theorem~\ref{bal:hyp:SCplus}, one has \(\Gap\ge0\) on all of
\(\mathcal S_{\rm seam}\).
\end{theorem}
\begin{proof}
The facet estimates give the boundary-liminf hypothesis of
Lemma~\ref{bal:lem:negsubcompact}; the strict gap is continuous and \(C^1\).
Thus any negative value produces a negative minimum in the strict cell.
This minimizer lies in the open three-coordinate strict seam, so
Fermat's rule gives
\[
 \Gap_y=\Gap_{e_u}=\Gap_{e_w}=0.
\]
Because \(e_u=h+d\) and \(e_w=h-d\),
\[
 \Gap_d=\Gap_{e_u}-\Gap_{e_w}=0.
\]
Clause (i) of Theorem~\ref{bal:hyp:SCplus} instead gives either
\(\Gap_d<0\) or
\(\Gap_{e_u}<0\), a contradiction.
\end{proof}

\begin{remark}[A direct proof of \(\Target\ge0\) on the seam]
Theorem~\ref{bal:thm:small-boundary} also proves \(\Target\ge0\) on the entire seam \(a+c=S\): for
the same-side lift \(a,c\le S<10^{-2}\), and for the opposite-side lift
\(\mu_u+(1-\mu_w)=a+c=S\).  This does not replace
Theorem~\ref{bal:thm:seam-closure}: the minimum argument of
Section~\ref{bal:sec:retained} needs the stronger statement \(\Gap\ge0\) on
the seam.
\end{remark}

\section{The pure-gap inequality}
\label{bal:sec:retained}

After Theorem~\ref{bal:thm:e8-main} and
Lemma~\ref{bal:lem:global-equal-mean}, the deterministic caps and the half
means remain. We treat them first and then prove the pure-gap inequality.

\begin{lemma}[Middle deterministic ordering]
On \(0<b<a<c<1/2\), \(\Gap\ge0\).
\end{lemma}
\begin{proof}
Put
\[
 E=\HH(a)+\HH(b),\quad
 U_0=\frac{c-a}{E},\quad
 V_0=\frac{1-a-c}{E},\quad
 W_0=\frac{1-2c}{2\HH(b)}.
\]
Then
\[
 U_0+W_0-V_0
 =\frac{(1-2c)(\HH(a)-\HH(b))}{2\HH(b)E}\ge0.
\]
The certified function \(\Theta\) is increasing, concave, satisfies
\(\Theta(0)=0\), and is therefore subadditive.  Hence
\[
 \Theta(V_0)\le\Theta(U_0+W_0)
 \le\Theta(U_0)+\Theta(W_0).
\]
For the deterministic-cap gap,
\[
 \partial_c\Gap_{\rm cap}
 =\Theta(U_0)-\Theta(V_0)+\Theta(W_0)\ge0.
\]
Thus
\(\Gap_{\rm cap}(a,b,c)\ge\Gap_{\rm cap}(a,b,a)\ge0\), where \(c=a\)
is the equal-mean estimate.
\end{proof}

\begin{lemma}[Acyclic Case-E and radial cap-minimizer bridge]
\label{bal:lem:cap-minimizer-bridge}
The stationary estimates in Theorem~\ref{bal:thm:boundary-estimates}(h)--(i),
together with Theorem~\ref{bal:hyp:SCplus} and the analytic endpoint
estimates, exclude a negative fixed-entropy mean-minimum on a Case-E or
radial deterministic cap.
\end{lemma}
\begin{proof}
For fixed \((a,b)\), the Case-E gap is a continuous one-variable function on
the restricted fiber
\[
 \max\{b,S-a\}\le c\le\frac12.
\]
Put \(E=\HH(a)+\HH(b)\).  A fiber minimizer is either at an endpoint or
satisfies the certified stationary equation
\[
 \Theta\!\left(\frac{1-a-c}{E}\right)
 =\Theta\!\left(\frac{c-a}{E}\right)
  +\Theta\!\left(\frac{1-2c}{2\HH(b)}\right).
\]
The stationary cap reduction proves nonnegativity at every interior stationary solution.  Whenever
\(c=b\) belongs to the restricted fiber, \(a+b\ge S\).  If \(a+b>S\), Theorem~\ref{bal:thm:double-cap}
covers the double cap; if \(a+b=S\), it is the double-cap edge of the seam,
covered by Lemma~\ref{bal:lem:deterministic-seam-facets}.  At \(c=1/2\), the present
deterministic-cap intersection is covered directly by Theorem~\ref{bal:thm:cap-half-mean} (clause (v) of Theorem~\ref{bal:hyp:SCplus}).
Indeed, with \(h=(\HH(a)+\HH(b))/2\) and
\(x=1/2-a\), direct substitution gives exactly \(\Gap_{\rm HF}(a,b)\).
If the cutoff \(a+c\ge S\) is active, \(c=S-a>b\) is exactly the
\(U^\circ\) domain in clause (ii) of Theorem~\ref{bal:hyp:SCplus}.

For the radial ordering, the same one-dimensional minimum argument applies on
\(\max\{b,S-a\}\le c\le a\).  At an interior stationary point the radial
stationary-point estimate gives \(\Gap\ge0\).  At \(c=b\), the same dichotomy applies: Theorem~\ref{bal:thm:double-cap}
covers \(a+b>S\), and Lemma~\ref{bal:lem:deterministic-seam-facets} covers \(a+b=S\).
The endpoint $c=a$ is covered by Lemma~\ref{bal:lem:global-equal-mean}.  When \(c=S-a>b\) is active, it is exactly the
\(W^\circ\) domain in clause (iii) of Theorem~\ref{bal:hyp:SCplus}.  Thus every alternative
contradicts the existence of a negative fixed-entropy mean-minimum.  The
stationary-point estimates are used only at interior stationary points, so
there is no circularity with the endpoint results.
\end{proof}

\begin{lemma}[Half-mean gate]
The smooth half-mean alternatives produced by Theorem~\ref{bal:thm:e8-main} with the other mean
fixed-entropy interior cannot be local negative minimizers.  The deterministic
cap--half-mean intersection is covered by clause (v) of Theorem~\ref{bal:hyp:SCplus}.
\end{lemma}
\begin{proof}
The half-mean estimate proves \(D_0(s)>0\) for every \(s>0\), using an analytic estimate near zero, a compact
outward-rounded interval certificate with an independent verification run, and an analytic
tail.  With \(s=\Theta(x)\) and
\[
 y=Q(2s)=\Theta^{-1}(2\Theta(x)),
\]
the inverse-function identities \(Q'(s)=1/\Theta'(x)\) and
\(Q'(0)=1/\Theta'(0)\) show that \(D_0(s)>0\) is equivalent to
\[
 2\Theta'(x)\left(1-\frac{x}{y}\right)<\Theta'(0).
\]
At a nonendpoint stationary half-mean point, stationarity in the other mean
first gives the relation \(2\Theta(x)=\Theta(y)\); its necessary transverse
second-order condition is the opposite of the displayed strict inequality.
Thus no such smooth half-mean point can be a negative minimum.
The fixed-entropy endpoint is treated by the deterministic-cap estimates.
\end{proof}

\begin{theorem}[Pure-gap inequality]
\label{bal:thm:pure-gap}
The analytic reductions, the boundary and stationary-point estimates
of Theorem~\ref{bal:thm:boundary-estimates}, and
Theorem~\ref{bal:hyp:SCplus} imply that every canonical feasible tuple
with $a+c\ge S$ and positive entropy coordinates satisfies $\Gap\ge0$.
\end{theorem}
\begin{proof}
Suppose a positive-entropy value is negative.  Apply
Lemma~\ref{bal:lem:global-entropy-rearrangement} pointwise to obtain a negative
marginal-physical point in the chamber \(e\le f\).  If a maximal entropy is
present, Lemma~\ref{bal:lem:maximal-continuity} supplies a negative point at the same means with both entropies
strictly submaximal.  Thus in all cases there is a negative point with
\(e,f\in(0,1)\).  If \(e=f\), Lemma~\ref{bal:lem:global-equal-entropy} already contradicts negativity.  Hence
assume \(e<f\), freeze this pair \((e,f)\), and minimize
only over the compact mean set \(K_{e,f}\).  By
Lemma~\ref{bal:lem:outer-minimum-partition}, the attained negative mean-minimum is
on the interface \(a+c=S\), at equal or half means, on a deterministic cap,
or in the smooth fixed-entropy mean-interior.

The interface $a+c=S$ is covered by Theorem~\ref{bal:thm:seam-closure}, and equal means by Lemma~\ref{bal:lem:global-equal-mean}.  At a smooth
half mean with the other mean fixed-entropy interior, the half-mean estimate gives the
transverse contradiction.  A half mean simultaneously lying on a
deterministic cap is covered by Theorem~\ref{bal:thm:cap-half-mean}.  Theorem~\ref{bal:thm:e8-main} excludes the remaining smooth mean-interior.

A strict deterministic cap has exactly one ordering in
\eqref{bal:eq:three-caps}.  The Case-E and radial orderings are excluded by
Lemma~\ref{bal:lem:cap-minimizer-bridge}, using the stationary cap and radial estimates only for interior
fiber stationarity and Theorem~\ref{bal:hyp:SCplus} for the endpoints on the seam and at a half mean.  The
middle ordering is covered by the lemma on the middle deterministic ordering.  Tie strata are covered by
Lemma~\ref{bal:lem:global-equal-entropy}, Theorem~\ref{bal:thm:double-cap}, Lemma~\ref{bal:lem:global-equal-mean}, or Theorem~\ref{bal:thm:cap-half-mean} according as \(a=b\), \(b=c\), \(a=c\), or the
half-mean cap is active.  Every possible location contradicts the choice of
the fixed-entropy mean-minimum.
\end{proof}

\section{The Bellman inequality for the balanced candidate}
\label{bal:sec:global-target}

\begin{theorem}[Unrestricted balanced Bellman inequality]
\label{bal:thm:global-unrestricted-target}
The analytic reductions, the boundary and stationary-point estimates of
Theorem~\ref{bal:thm:boundary-estimates}, and Theorem~\ref{bal:hyp:SCplus}
imply that every feasible four-moment tuple with \(e_u,e_w>0\) satisfies
\[
 \zeta(M)\ge\Rphi(M).
\]
\end{theorem}
\begin{proof}
Apply the exact canonicalization and domain exhaustion of
Section~\ref{bal:sec:orientation}, in particular
Proposition~\ref{bal:prop:owner-exhaustion}.
Both entropy coordinates are positive by hypothesis.  Use only the exact
target symmetries \(\mathcal E\) and
\(\mathcal K\) to select the appropriate same- or opposite-side lift.
Collapsed-mean intersections are assigned to Lemma~\ref{bal:lem:global-equal-mean} and the explicit endpoint
values.  If \(a+c<S\), Theorem~\ref{bal:thm:small-boundary} proves the target directly on that
lift, and the exact symmetries return the conclusion to the original tuple.

If \(a+c\ge S\), Theorem~\ref{bal:thm:pure-gap} gives
\(\Gap=L_4-\Rphi\ge0\) on the canonical same-side lower-half record.
Lemma~\ref{bal:lem:pure-gap-reflection} and full relabeling transfer this
identical gap value back to the original tuple; the reflections are applied
to \(\Gap\) only.
Theorem~\ref{bal:thm:four-moment-lower-bound} on that original tuple gives \(\zeta-L_4\ge0\).
Therefore \eqref{bal:eq:gap-decomposition} yields
\(\Target\ge0\).  The two branches are exhaustive, and equality belongs to
the \(a+c\ge S\) pure-gap branch.
\end{proof}

\section{From the unrestricted Bellman theorem to balanced CK}
\label{bal:sec:framework}
This section converts the local four-moment inequality into the
information-theoretic statement in three steps. Static induction bounds
the total edge energy, differentiation identifies that energy with entropy
production, and a scalar comparison gives the desired conditional entropy
bound. Regularization makes the differentiation legitimate for Boolean
outputs and is removed at the end.

The induction is the reason why \(\zeta\) is defined as an infimum over all
joint laws: the two sections of an arbitrary array are not pointwise
ordered.

\begin{lemma}[Static induction for the balanced candidate \cite{ChenGohariNair2025}]
\label{bal:lem:unrestricted-static-induction}
Assume
\begin{equation}
 \zeta(M)\ge \Rphi(M)
 \label{bal:eq:unrestricted-bellman-assumption}
\end{equation}
for every feasible four-moment tuple \(M\) with positive entropy coordinates,
and assume
\(\phi(m,\HH(m))=0\).  For an arbitrary array
\(v:\{-1,+1\}^n\to(0,1)\), define
\[
 \mathcal E_n(v)
 =\sum_{i=1}^n2^{-(n-1)}
   \sum_{y_{-i}\in\{-1,+1\}^{n-1}}
   j\bigl(v_i^-(y_{-i}),v_i^+(y_{-i})\bigr),
\]
where
\[
 v_i^\pm(y_{-i})
 =v(y_1,\ldots,y_{i-1},\pm1,y_{i+1},\ldots,y_n).
\]
Then
\begin{equation}
 \mathcal E_n(v)
 \ge
 \phi\!\left(2^{-n}\sum_yv(y),
              2^{-n}\sum_y\HH(v(y))\right).
 \label{bal:eq:unrestricted-static-conclusion}
\end{equation}
\end{lemma}

\begin{proof}
Induct on \(n\).  For \(n=0\), the energy is zero and the assertion is
\(0\ge\phi(v,\HH(v))=0\).  For the induction step split \(v\) according
to its last coordinate, writing \(v^-\) and \(v^+\) for the two sections.
With \(Y'\) uniform on \(\{-1,+1\}^{n-1}\), put
\[
 m_\pm=\mathbb Ev^\pm(Y'),
 \qquad e_\pm=\mathbb E\HH(v^\pm(Y')).
\]
The cube-edge energy decomposes exactly as
\begin{equation}
 \mathcal E_n(v)
 =\frac12\mathcal E_{n-1}(v^-)
  +\frac12\mathcal E_{n-1}(v^+)
  +\mathbb E j\bigl(v^-(Y'),v^+(Y')\bigr).
 \label{bal:eq:energy-section-decomposition}
\end{equation}
The induction hypothesis applies separately to both sections and gives
\(\mathcal E_{n-1}(v^\pm)\ge\phi(m_\pm,e_\pm)\).

Now set \(U=v^-(Y')\) and \(W=v^+(Y')\).  No relation between \(U\) and
\(W\) is required.  Because the array takes values in \((0,1)\), both
section entropy means are strictly positive.  Their joint law witnesses feasibility of
\(M=(m_-,m_+,e_-,e_+)\), so
\[
 \mathbb E j(U,W)\ge\zeta(M)\ge\Rphi(M).
\]
Expanding
\[
 \Rphi(M)=
 \phi\!\left(\frac{m_-+m_+}{2},\frac{e_-+e_+}{2}\right)
 -\frac12\phi(m_-,e_-)-\frac12\phi(m_+,e_+)
\]
inside \eqref{bal:eq:energy-section-decomposition} cancels the two section
terms and proves \eqref{bal:eq:unrestricted-static-conclusion}.
\end{proof}

\begin{corollary}[Unrestricted differential induction \cite{ChenGohariNair2025}]
\label{bal:cor:unrestricted-differential-induction}
Let \(h:\{-1,+1\}^n\to(0,1)\) be arbitrary, let
\(\rho_t=e^{-2t}\), and put
\[
 v_t=T_{\rho_t}h,
 \qquad
 \gamma(t)=2^{-n}\sum_y\HH(v_t(y)).
\]
Under the hypothesis of Lemma~\ref{bal:lem:unrestricted-static-induction},
\begin{equation}
 \gamma'(t)=\mathcal E_n(v_t)
 \ge\phi\!\left(\mathbb Eh,\gamma(t)\right).
 \label{bal:eq:unrestricted-differential-induction}
\end{equation}
\end{corollary}

\begin{proof}
The product noise semigroup with \(\rho_t=e^{-2t}\) satisfies
\[
 \partial_t v_t(y)
 =\sum_{i=1}^n\bigl(v_t(y^{(i)})-v_t(y)\bigr),
\]
where \(y^{(i)}\) is obtained by flipping coordinate \(i\).  Since
\(\HH'=J\), differentiating \(\gamma\) and pairing the two orientations
of every \(i\)-edge with endpoint values \(u,w\) gives
\[
 2^{-n}\{J(u)(w-u)+J(w)(u-w)\}
 =2^{-(n-1)}j(u,w).
\]
Summing proves \(\gamma'(t)=\mathcal E_n(v_t)\).  The noise operator
preserves the uniform mean, so
\(2^{-n}\sum_yv_t(y)=\mathbb Eh\).  Apply
Lemma~\ref{bal:lem:unrestricted-static-induction} directly to the possibly
nonmonotone array \(v_t\).
\end{proof}

\begin{theorem}[Unrestricted Bellman theorem implies balanced CK \cite{ChenGohariNair2025}]
\label{bal:thm:unrestricted-framework}
Assume
\[
 \zeta(M)\ge\Rphi(M)
\]
for every feasible four-moment tuple \(M\) with positive entropy coordinates,
and assume
\[
 \phi(m,\HH(m))=0,
 \qquad \phi\!\left(\frac12,e\right)=\eta(e).
\]
Then, for every \(n\), every balanced Boolean function
\(f:\{-1,+1\}^n\to\{0,1\}\), and every BSC crossover probability
\(p_{\mathrm{BSC}}\in[0,1]\),
\[
 I(f(X);Y)\le1-\HH(p_{\mathrm{BSC}}).
\]
\end{theorem}

\begin{proof}
\emph{Proof outline.}
We regularize the Boolean output so that the entropy derivative is
finite, and apply differential induction along the noise semigroup.
At mean one half the candidate reduces to the scalar profile, whose
comparison equation can be integrated explicitly. We then remove the
regularization and treat the channel endpoints by continuity or symmetry.

First suppose \(0<p_{\mathrm{BSC}}<1/2\).  For \(t\ge0\), put
\[
 \rho_t=e^{-2t},\qquad p_t=\frac{1-\rho_t}{2},
\]
and let \(Y_t\) be the product-BSC output with crossover \(p_t\).  Choose
\(t\) so that \(p_t=p_{\mathrm{BSC}}\).

Fix \(0<\epsilon<1/2\), let
\(N_\epsilon\sim\operatorname{Bernoulli}(\epsilon)\) be independent of
\(X\) and the channel, and define
\[
 F_\epsilon=f(X)\mathbin{\oplus}N_\epsilon,
 \qquad
 h_\epsilon(x)=\epsilon+(1-2\epsilon)f(x).
\]
The array \(h_\epsilon\) takes values in
\([\epsilon,1-\epsilon]\), need not be monotone, and satisfies
\(\mathbb Eh_\epsilon=1/2\) because \(f\) is balanced.  Its posterior
array is
\[
 v_{\epsilon,t}=T_{\rho_t}h_\epsilon,
\]
and hence
\[
 \gamma_\epsilon(t):=H(F_\epsilon\mid Y_t)
 =2^{-n}\sum_y\HH(v_{\epsilon,t}(y)).
\]
Corollary~\ref{bal:cor:unrestricted-differential-induction} gives
\begin{equation}
 \gamma_\epsilon'(t)
 \ge\phi\!\left(\frac12,\gamma_\epsilon(t)\right)
 =\eta(\gamma_\epsilon(t)),
 \qquad
 \gamma_\epsilon(0)=\HH(\epsilon).
 \label{bal:eq:balanced-ode-inequality}
\end{equation}

Define
\[
 r_\epsilon(t)=\frac{1-e^{-2t}(1-2\epsilon)}2.
\]
Then \(r_\epsilon(0)=\epsilon\),
\(r_\epsilon'=1-2r_\epsilon\), and
\[
 \frac{d}{dt}\HH(r_\epsilon(t))
 =(1-2r_\epsilon)J(r_\epsilon)
 =\eta(\HH(r_\epsilon(t))).
\]
If \(\gamma_\epsilon\) reaches \(1\), the comparison below is immediate.
Otherwise \(\eta(\gamma_\epsilon)>0\) on the interval under consideration,
so division by \(\eta\) is legitimate.  For completeness, separation of
variables gives
\[
 \int_{\HH(\epsilon)}^{\HH(r)}\frac{du}{\eta(u)}
 =\int_\epsilon^r\frac{dq}{1-2q}
 =\frac12\ln\frac{1-2\epsilon}{1-2r}.
\]
Thus the scalar comparison for \eqref{bal:eq:balanced-ode-inequality} yields
\begin{equation}
 \gamma_\epsilon(t)\ge
 \HH\!\left(\frac{1-e^{-2t}(1-2\epsilon)}2\right).
 \label{bal:eq:regularized-entropy-comparison}
\end{equation}
The integral also shows that no factor of \(\ln2\) is missing: the common
base-two derivative \(J\) cancels before the natural logarithm appears.

Letting \(\epsilon\downarrow0\), finite-alphabet continuity of conditional
entropy gives
\[
 H(f(X)\mid Y_t)\ge\HH(p_t).
\]
Because \(f\) is balanced, \(H(f(X))=1\), and therefore
\[
 I(f(X);Y_t)=1-H(f(X)\mid Y_t)\le1-\HH(p_t).
\]
This proves the claim for \(0<p_{\mathrm{BSC}}<1/2\).  At
\(p_{\mathrm{BSC}}=0\), use \(I(f(X);X)\le H(f(X))=1\); at
\(p_{\mathrm{BSC}}=1/2\), the channel output is independent of \(X\).
If \(p_{\mathrm{BSC}}>1/2\), then \(-Y\) is the output of a BSC with
crossover \(1-p_{\mathrm{BSC}}<1/2\); the bijection \(Y\mapsto-Y\)
preserves mutual information and
\(\HH(1-p_{\mathrm{BSC}})=\HH(p_{\mathrm{BSC}})\).
\end{proof}

\begin{theorem}[Balanced Courtade--Kumar theorem]
\label{bal:thm:CK}
For every \(n\), every balanced Boolean function
\(f:\{-1,+1\}^n\to\{0,1\}\), and every BSC crossover probability
\(p_{\mathrm{BSC}}\in[0,1]\),
\[
 I(f(X);Y)\le1-\HH(p_{\mathrm{BSC}}).
\]
\end{theorem}
\begin{proof}
By Theorem~\ref{bal:thm:global-unrestricted-target}, \(\zeta\ge\Rphi\) at
every feasible tuple with positive entropy coordinates, and
\(\phi(m,\HH(m))=0\) and \(\phi(1/2,e)=\eta(e)\) by the definition of
\(\phi\).  Theorem~\ref{bal:thm:unrestricted-framework} therefore applies to
every balanced Boolean function.
\end{proof}

\startmanuscriptpart{Unbalanced Case}{Unbalanced Bellman inequality}{Courtade--Kumar}
\label{part:unbalanced}
This part proves the local inequality for \(B=\max\{\phi,\psi\}\),
Theorem~\ref{unbal:thm:global}, and deduces Theorem~\ref{unbal:thm:ck} from
it. It uses Theorem~\ref{bal:thm:global-unrestricted-target} and several
lower bounds for \(\zeta\) from Part~\ref{part:balanced};
Section~\ref{unbal:sec:inputs} lists them.

\section{The theorem and its information-theoretic deduction}
\label{unbal:sec:main}
We state the Bellman inequality for \(B\) and show how it implies
Theorem~\ref{unbal:thm:ck}. The inequality itself is proved in
Sections~\ref{unbal:ss:section}--\ref{unbal:sec:assembly}.

We first recall the notation. In this part the binary entropy \(\HH=H_2\)
is written \(H\), its lower inverse \(\iota\) is written \(H^{-1}\), and the
four-moment tuple is written \(M=(a,b,e,f)\). The functions \(J\), \(\eta\),
\(F\), \(\phi\), \(j\), and \(\zeta\) are those of
Section~\ref{bal:sec:definitions}; the definition of \(F\) below is
equivalent to \eqref{bal:eq:eta-F}, because
\(\mathcal L(v)=2H(v)/(1-2v)\). The function \(C\), the candidate \(\psi\),
and the maximum \(B\) are new.

Let $X$ be uniform on $\{-1,1\}^n$. Let $Y$ be obtained from $X$ by
independently flipping each coordinate with probability $p$.
Binary entropy is measured in bits:
\[
 H(x)=-x\log_2x-(1-x)\log_2(1-x),\qquad
 J(x)=H'(x)=\log_2\frac{1-x}{x},\qquad L=\ln2.
\]
The entropy inverse $H^{-1}$ always denotes its lower branch on $[0,1/2]$.
For $0<h\le1$, define
\[
 \eta(h)=(1-2H^{-1}(h))J(H^{-1}(h)),\qquad \eta(1)=0,
 \qquad C(r)=1-H\left(\frac{1-r}{2}\right).
\]
For $z>0$ and $h>0$, let $v\in(0,1/2)$ be the unique solution of
$zH(v)=h(1-2v)$ and put $F(z,h)=zJ(v)$; set $F(0,h)=0$.
The two candidates and their maximum are
\[
 \begin{gathered}
 \phi(m,h)=\eta(h)-F(|1-2m|,h),\qquad
 \psi(m,h)=\eta(h+1-H(m)),\\
 B(m,h)=\max\{\phi(m,h),\psi(m,h)\}.
 \end{gathered}
\]
They are evaluated on $0<m<1$ and $0<h\le H(m)$.

For a four-moment tuple $M=(a,b,e,f)$, write
\[
 m=\frac{a+b}{2},\quad E=\frac{e+f}{2},\quad
 d=|a-b|,\quad q=|1-a-b|,
\]
\[
 C_0=\frac{H(a)+H(b)}2,\quad
 \Delta=H(m)-C_0,\quad s=C_0-E,\quad I=\Delta+s,
 \qquad P(t)=\eta(1-t).
\]
We call \((m,E)\) the \emph{parent} of the tuple \(M\) and \((a,e)\),
\((b,f)\) its \emph{children}; \(\phi\) or \(\psi\) is \emph{active} at a point
if it attains the maximum \(B\) there. For a candidate $A$, the increment
\eqref{eq:intro-increment} is
\[
 R_A(M)=A(m,E)-\frac{A(a,e)+A(b,f)}2.
\]
The cost is
\[
 j(u,w)=\frac12(w-u)[J(u)-J(w)]
       =\frac12\{D_2(u\Vert w)+D_2(w\Vert u)\}.
\]
Its boundary values are the lower-semicontinuous extension: zero at
$(0,0),(1,1)$ and infinite at the other boundary points.
The unrestricted value $\zeta(M)$ is the infimum of $\E j(U,W)$ over all
joint laws satisfying
\[
 \E U=a,\qquad \E W=b,\qquad \E H(U)=e,\qquad \E H(W)=f.
\]
No pointwise ordering of $U,W$ is imposed.

\begin{theorem}[Global hybrid Bellman inequality]
\label{unbal:thm:global}
For every
\[
 0<a,b<1,\qquad 0<e\le H(a),\qquad 0<f\le H(b),
\]
one has $\zeta(a,b,e,f)\ge R_B(a,b,e,f)$.
\end{theorem}

The proof is completed in Section~\ref{unbal:sec:assembly}. The
computations on which it depends are listed in
Part~\ref{part:computation}, and their records are in Supplementary
Section~\ref{supp:verification}.

We now deduce Theorem~\ref{unbal:thm:ck} from
Theorem~\ref{unbal:thm:global}, including the deterministic and zero-noise
limits.

\begin{lemma}[Static induction for the hybrid candidate]
\label{unbal:lem:static}
For every array $v:\{-1,1\}^n\to(0,1)$, define
\[
 \mathcal E_n(v)=\sum_{i=1}^n2^{-(n-1)}
 \sum_{y_{-i}}j\bigl(v_i^-(y_{-i}),v_i^+(y_{-i})\bigr).
\]
Theorem~\ref{unbal:thm:global} implies
$\mathcal E_n(v)\ge B(\E v,\E H(v))$.
\end{lemma}
\begin{proof}
Both candidates vanish at the entropy cap:
$\phi(m,H(m))=0$ and $\psi(m,H(m))=\eta(1)=0$.
For the first identity, if $m\le1/2$ the defining $F$-contact is $v=m$;
the other half follows by complement, with the continuous midpoint value.
The assertion is therefore true for $n=0$.
Split the last coordinate and let $a,b$ be the two section means and $e,f$
their mean entropies. The energy is the average of the two section energies
plus the expected cross-edge cost. The section bounds from induction give
$[B(a,e)+B(b,f)]/2$. The unrestricted law of the section pair gives cross-edge
cost at least $\zeta(M)\ge R_B(M)$. Cancellation proves the claim.
Every section has interior mean and positive entropy because the original
array takes values in $(0,1)$. Ordering of its two sections is unnecessary.
\end{proof}

\begin{proof}[Proof of Theorem~\ref{unbal:thm:ck}]
\emph{Proof outline.}
Regularization allows the hybrid energy bound to be applied along
the noise semigroup. Adding the constant output-entropy deficit turns
the conditional entropy into a scalar quantity satisfying the same
comparison inequality as in the balanced case. After integration, its
complement is the mutual information; removing regularization and
handling the channel endpoints completes the argument.

Fix $0<\varepsilon<1/2$ and regularize
\[
 h_\varepsilon(x)=\varepsilon+(1-2\varepsilon)g(x),\quad
 m_\varepsilon=\E h_\varepsilon,
 \quad v_{\varepsilon,t}=T_{e^{-2t}}h_\varepsilon,
 \quad\gamma_\varepsilon(t)=\E H(v_{\varepsilon,t}).
\]
Here $T_\rho$ is the noise operator whose coordinate crossover is
$(1-\rho)/2$. Every entry of $v_{\varepsilon,t}$ lies in
$[\varepsilon,1-\varepsilon]$. The generator at correlation $e^{-2t}$ is
$\sum_i(v(x^{(i)})-v(x))$. Pairing the two orientations of each cube edge
therefore gives the exact identity
\[
 \gamma_\varepsilon'(t)=\mathcal E_n(v_{\varepsilon,t})
 \ge B(m_\varepsilon,\gamma_\varepsilon(t))
 \ge\eta\bigl(\gamma_\varepsilon(t)+1-H(m_\varepsilon)\bigr).
\]
The factor $2^{-(n-1)}$ and the factor $1/2$ in $j$ agree with this
generator normalization.
Let $\delta_\varepsilon(t)=\gamma_\varepsilon(t)+1-H(m_\varepsilon)$.
The mean is preserved, so
\[
 \delta_\varepsilon'(t)\ge\eta(\delta_\varepsilon(t)),\qquad
 H(\varepsilon)\le\delta_\varepsilon(0)
 =H(\varepsilon)+1-H(m_\varepsilon)\le1.
\]
The last bound uses $m_\varepsilon\in[\varepsilon,1-\varepsilon]$;
concavity of $H$ gives $\delta_\varepsilon(t)\le1$ at every time.
If it reaches 1, the mutual information is zero. Otherwise, separation
of variables, using $\eta>0$ on $(0,1)$, gives
\[
 \delta_\varepsilon(t)\ge
 H\left(\frac{1-e^{-2t}(1-2\varepsilon)}2\right).
\]
For completeness, the primitive is
\[
 \int_{H(\varepsilon)}^z\frac{dh}{\eta(h)}
 =\frac12\ln\frac{1-2\varepsilon}{1-2H^{-1}(z)}.
\]
It confirms both the exponent $e^{-2t}$ and the bit normalization.
Let $G_\varepsilon=g(X)\mathbin{\mathrm{XOR}}N_\varepsilon$, where
$N_\varepsilon$ is an independent Bernoulli$(\varepsilon)$ bit. Uniformity
of $X$ and symmetry of the noise kernel identify $v_{\varepsilon,t}(Y_t)$
with the posterior probability of $G_\varepsilon=1$. Consequently
\[
 I(G_\varepsilon;Y_t)=H(m_\varepsilon)-\gamma_\varepsilon(t)
 =1-\delta_\varepsilon(t).
\]
Letting $\varepsilon\downarrow0$ and using continuity of finite-alphabet
entropy proves the result for $0<p<1/2$. At $p=0$, use
$I(g(X);X)=H(g(X))\le1$; at $p=1/2$, use independence. Complementing
every output coordinate reduces $p>1/2$ to $1-p$ without changing mutual
information or $H(p)$.
\end{proof}

\section{Results from the balanced case}
\label{unbal:sec:inputs}

The proofs in this part use the following results of
Part~\ref{part:balanced} and of its supplementary sections.

\begin{enumerate}
\item Theorem~\ref{bal:thm:global-unrestricted-target} is the Bellman
inequality for \(\phi\),
\begin{equation}\label{unbal:eq:oldbellman}
 \zeta(M)\ge R_\phi(M)
 \quad\text{for every feasible }M\text{ with }e,f>0.
\end{equation}
The means are arbitrary.
\item Theorem~\ref{bal:thm:four-moment-lower-bound} and Supplementary Section~\ref{bal:imp:jointconvexity} give
$\zeta\ge L_4=F(d,E)+g(e,f)$, with $g\ge0$, hence
\begin{equation}\label{unbal:eq:radiallower}
 \zeta(M)\ge F(d,E).
\end{equation}
The nonnegativity of $g$ follows from its convexity, its symmetry, and
its vanishing on the diagonal.
\item Proposition~\ref{bal:prop:fixed-difference-minimum} gives the
supporting-plane lower bound for \(\zeta\) whenever its contact triple
exists. The conditions on the endpoint masses in that proposition
characterize equality; the lower bound holds without them.
Theorem~\ref{bal:thm:general-endpoint-region} gives the corresponding bound
for a general contact with nonnegative entropy coefficients. Both results
are stated in natural units; dividing entropies and costs by $L=\ln2$
converts them to bits.
\item Supplementary Theorem~\ref{bal:imp:e8:thm:onevar} and
Supplementary Section~\ref{bal:aux:sec:fourpoint} give the monotonicity
of $F$ and the concavity of $z\mapsto F(\sqrt z,h)$. Supplementary
Theorem~\ref{bal:aux:thm:entropyconvexity} proves that
$\Phi(r,h)=\eta(h)-F(r,h)$ is convex in $h$ on its feasible interval.
Where an argument below needs curvature information for entropies above
an individual entropy cap, a separate estimate is proved there.
\end{enumerate}

The remaining estimates of this part are proved below or in the
supplement: the log-sum lower bounds, the inequalities for the profile
$P$, the bound $rF_{rr}\le13/6$, the endpoint estimate with its
entropy-split gain, the central mean regions, and the two outer covers.
The table in Part~\ref{part:computation} gives the location of each proof
and of each computation.

\subsection{The maximum candidate and its symmetries}

At each tuple, consider the candidate that is active at the parent. If it
is $\phi$, then $R_B\le R_\phi$, since $B\ge\phi$ at both children, and
Theorem~\ref{bal:thm:global-unrestricted-target} gives $\zeta\ge R_B$. If
it is $\psi$, then $R_B\le R_\psi$, and it suffices to prove
$\zeta\ge R_\psi$; alternatively one may keep $B$ or $\phi$ at the children
and compare $\zeta$ with $R_B$ directly. In particular, $\zeta\ge R_\psi$
is needed only where $\psi$ is active at the parent. The inequality
$\phi(m,E)\ge\psi(m,E)$ is called \emph{parent dominance}; where it holds,
the first case applies.

Only the symmetries
\[
 (a,b,e,f)\mapsto(b,a,f,e),\qquad
 (a,b,e,f)\mapsto(1-a,1-b,e,f)
\]
are used. Both preserve the set of feasible laws, the cost, and $R_B$.
They place every tuple in the canonical orientation
\begin{equation}\label{unbal:eq:canonical}
 0<a\le b<1,\qquad a+b\le1.
\end{equation}
The reflection of a single mean about $1/2$ is not used here.

\subsection{Eliminating the entropy split}

The entropy deficits of the children are $I_a=H(a)-e$ and $I_b=H(b)-f$,
so $I_a+I_b=2s$. Convexity of $P$, proved in Lemma~\ref{unbal:ss:profile}, gives
\begin{equation}\label{unbal:eq:split-target}
 R_\psi=P(\Delta+s)-\frac{P(I_a)+P(I_b)}2
 \le P(\Delta+s)-P(s).
\end{equation}
Most of the arguments below bound \(\zeta\) from below by this last
expression, uniformly over all feasible splits of the entropy between the
two children. Two arguments proceed differently. Near an entropy cap of a
single child, the opposite-side argument keeps track of the feasible
imbalance of $I_a$ and $I_b$. The argument for \(d/E\ge8\)
(Theorem~\ref{unbal:ar:eight}) keeps \(\phi\) at the children and uses the
logarithmic gain of the endpoint bound under an unequal split. Each of
these gives a sufficient condition for \(\zeta\ge R_B\).

The protocol for the computations is described in
Part~\ref{part:computation}. The limits and the interfaces of the compact
domains treated by computation are covered by analytic estimates and by
closed, overlapping domains.

\section{Completion for all same-side mean pairs}\label{unbal:ss:section}

This section proves \(\zeta\ge R_B\) for every pair of interior means on the
same side of \(1/2\).  The argument uses the inequality
\eqref{unbal:eq:oldbellman} for \(\phi\), the lower bound
\eqref{unbal:eq:radiallower}, and Theorem~\ref{unbal:thm:central} for the
central square, which is proved in Section~\ref{unbal:sec:central}. It also
uses the small-mean theorem, Theorem~\ref{unbal:thm:smallmean}, whose proof
in Section~\ref{unbal:ar:section} rests on the log-sum inequality
established below.

\begin{theorem}[All same-side means]\label{unbal:ss:main}\label{unbal:thm:sameside}
Let \(0<a,b<1\) and suppose that
\[
 (a-\tfrac12)(b-\tfrac12)\geq 0,\qquad
 0<e\leq H(a),\qquad 0<f\leq H(b).
\]
Then
\begin{equation}\label{unbal:ss:conclusion}
 \zeta(a,b,e,f)\geq R_B(a,b,e,f).
\end{equation}
There is no lower bound on the distance between the means, on either
positive entropy, or on the ratio of the entropies.
\end{theorem}

The proof of Theorem~\ref{unbal:ss:main} appears at the end of the section.
All logarithms without an indicated base are natural, and \(L=\ln 2\).
In addition to the notation of Section~\ref{unbal:sec:main}, write
\[
 \begin{gathered}
 \nu_z=z(1-z),\qquad
 I_a=H(a)-e,\qquad I_b=H(b)-f,\\
 \overline H=\frac{H(a)+H(b)}2,\qquad I=\Delta+s=H(m)-E.
 \end{gathered}
\]
Thus \(s=(I_a+I_b)/2=\overline H-E\).  Positive feasibility implies
\(0\leq I_a,I_b,s,I<1\).

\subsection{A global lower bound from the log-sum inequality}
\label{unbal:ss:logsum-section}

In this subsection the means may be anywhere in \((0,1)\).  Define
\begin{equation}\label{unbal:ss:kappa}
 \kappa(z)=
 \min\left\{
 \frac{z}{-(1-z)\ln(1-z)},\
 \frac{1-z}{-z\ln z}
 \right\},
 \qquad
 V=\max\{a,b\}\bigl(1-\min\{a,b\}\bigr).
\end{equation}

\begin{theorem}[Global log-sum lower bound]\label{unbal:ss:logsum}
For interior means and feasible entropy moments,
\begin{equation}\label{unbal:ss:chi-bound}
 \zeta(a,b,e,f)\geq
 j(a,b)+\frac{(a-b)^2}{4V}
 \bigl(\kappa(a)I_a+\kappa(b)I_b\bigr).
\end{equation}
Moreover \(\kappa(z)\geq1\).  In particular, after arranging \(a\leq b\),
\begin{equation}\label{unbal:ss:beta-bound}
 \zeta(a,b,e,f)\geq j(a,b)+\beta d^2s,
 \qquad
 \beta=\frac{\min\{\kappa(a),\kappa(b)\}}{2b(1-a)}.
\end{equation}
The weaker choice \(\kappa\equiv1\) is also valid.
\end{theorem}

\begin{proof}
\emph{Proof outline.}
Subtract the tangent plane at the prescribed means and express the
remainder as a sum of divergences between positive masses. The log-sum
inequality and an elementary logarithmic bound turn this remainder into
a quadratic fluctuation bound. We then compare entropy deficits with
variances, take expectations, and minimize over all representing laws.

Fix interior anchors \(a,b\), and subtract the tangent plane to \(j\):
\[
 Q_{a,b}(u,w)=j(u,w)-j(a,b)
       -j_u(a,b)(u-a)-j_w(a,b)(w-b).
\]
For positive, not necessarily normalized masses, put
\[
 {\cal D}(x\Vert y)=\frac{x\ln(x/y)-x+y}{L}.
\]
Set
\[
 r_1=\frac ua,\quad r_0=\frac{1-u}{1-a},\qquad
 t_1=\frac wb,\quad t_0=\frac{1-w}{1-b}.
\]
Direct subtraction of the tangent plane gives the exact identity
\begin{align}
 2Q_{a,b}(u,w)
 ={}&a{\cal D}(r_1\Vert t_1)
       +(1-a){\cal D}(r_0\Vert t_0) \notag\\
    &+b{\cal D}(t_1\Vert r_1)
       +(1-b){\cal D}(t_0\Vert r_0).
 \label{unbal:ss:bregman}
\end{align}
For example, decompose
\[
 j(u,w)=j_0(u,w)+j_0(1-u,1-w),\qquad
 j_0(x,y)=\frac{(x-y)\ln(x/y)}{2L}.
\]
The homogeneity of \(j_0\) gives the first and third terms of
\eqref{unbal:ss:bregman}; applying the same calculation to the complementary
masses gives the other two terms.

The log-sum inequality applied separately to the two rows of
\eqref{unbal:ss:bregman} yields
\begin{equation}\label{unbal:ss:two-masses}
 Q_{a,b}(u,w)\geq
 \frac12\bigl({\cal D}(1\Vert T)+{\cal D}(1\Vert S)\bigr),
\end{equation}
where
\[
 T=1+\frac{(a-b)(w-b)}{\nu_b},\qquad
 S=1-\frac{(a-b)(u-a)}{\nu_a}.
\]
Here the log-sum inequality is the joint convexity inequality for
\({\cal D}\), obtained from the convexity of \(x\ln x\) by taking its
perspective.

For \(z>-1\), integration of \(t/(1+t)\) from \(0\) to \(z\), with
the two signs of \(z\) treated separately, gives
\begin{equation}\label{unbal:ss:log-quadratic}
 z-\ln(1+z)\geq
 \frac{z^2}{2\max\{1,1+z\}}.
\end{equation}
The constraints \(0\leq u,w\leq1\) imply
\[
 \max\{1,T\}\leq \frac{V}{\nu_b},\qquad
 \max\{1,S\}\leq \frac{V}{\nu_a}.
\]
For \(a\leq b\), for instance, the relevant maxima are
\((1-a)/(1-b)\) and \(b/a\), respectively.  Consequently
\begin{equation}\label{unbal:ss:pointwise-variance}
 Q_{a,b}(u,w)\geq
 \frac{(a-b)^2}{4LV}
 \left(\frac{(u-a)^2}{\nu_a}+
       \frac{(w-b)^2}{\nu_b}\right).
\end{equation}

We now convert variance to entropy deficit.  Write \(D_2\) for binary
relative entropy in bits.  Taylor's integral formula gives the
continuous extension
\begin{equation}\label{unbal:ss:kl-ratio}
 \frac{D_2(x\Vert a)}{(x-a)^2}
 =\frac1L\int_0^1
 \frac{1-t}{[a+t(x-a)][1-a-t(x-a)]}\,dt.
\end{equation}
The integrand is convex as a function of \(x\), since
\(1/[y(1-y)]=1/y+1/(1-y)\) is convex.  Its integral therefore attains
its maximum on \([0,1]\) at an endpoint.  Thus
\[
 D_2(x\Vert a)\leq \Gamma_a(x-a)^2,\qquad
 \Gamma_a=
 \max\left\{\frac{-\ln(1-a)}{La^2},
             \frac{-\ln a}{L(1-a)^2}\right\}.
\]
The identity
\[
 \frac1{L\nu_a\Gamma_a}=\kappa(a)
\]
is exactly \eqref{unbal:ss:kappa}.  Also, \(\ln y\leq y-1\) gives
\[
 D_2(x\Vert a)\leq\frac{(x-a)^2}{L\nu_a},
\]
so \(\Gamma_a\leq1/(L\nu_a)\) and \(\kappa(a)\geq1\).

Let \((U,W)\) represent the four prescribed moments.  The tangent terms
have zero expectation, and
\[
 \mathbf E D_2(U\Vert a)=H(a)-\mathbf E H(U)=I_a.
\]
Hence
\(\operatorname{Var}(U)\geq I_a/\Gamma_a
=L\nu_a\kappa(a)I_a\), and similarly for \(W\).  Taking expectations
in \eqref{unbal:ss:pointwise-variance} proves \eqref{unbal:ss:chi-bound}; then take
the infimum over representing laws.  The estimate extends to the
finite-cost boundary atoms \((0,0)\) and \((1,1)\) by continuity.
At the other boundary points the cost is infinite, so the inequality
is immediate.  Infinite-cost laws are likewise harmless.
Finally \(I_a,I_b\geq0\) and \(I_a+I_b=2s\), which gives
\eqref{unbal:ss:beta-bound}.
\end{proof}

\subsection{Same-side estimates and finite certification}
The entropy-profile and contact derivatives, the mean-ratio tail, and the
finite-box inequalities are proved in Supplementary Section~\ref{supp:same-side}.
In particular, Lemma~\ref{unbal:ss:tail} treats the unbounded ratio limit,
Proposition~\ref{unbal:ss:box-tests} gives the sufficient inequalities on
the compact box that remains, and
Proposition~\ref{unbal:ss:certificate} verifies them on a complete
rational partition of that box.

\begin{proof}[Proof of Theorem~\ref{unbal:ss:main}]
By full mean--entropy label exchange and simultaneous complementation,
it is enough to treat \(0<a\leq b\leq1/2\).
The exact diagonal is treated in Supplementary Section~\ref{supp:same-side}.
If \(a/b\leq2^{-32}\), use Lemma~\ref{unbal:ss:tail}.
If \(m\leq1/32\), use Theorem~\ref{unbal:thm:smallmean}.
Every remaining tuple has \(a/b\geq2^{-32}\), \(b\geq1/32\), and
the parametrization \eqref{unbal:ss:coordinates} in \eqref{unbal:ss:root}.
Proposition~\ref{unbal:ss:certificate}, interpreted by
Proposition~\ref{unbal:ss:box-tests}, proves \eqref{unbal:ss:conclusion}
there.  The interfaces overlap at equality.

The reduction \eqref{unbal:ss:target-upper} and all the comparisons
hold for every feasible split with the prescribed \(E\), so the ratio of
the two entropies is unrestricted.
The face \(t=1\) is part of the closed finite cover, while the
parent comparisons handle its boxes approaching \(t=0\) on their
positive-entropy intersections.  The unbounded geometric limit
\(a/b\to0\) is covered by Lemma~\ref{unbal:ss:tail}, and the limit
\(m\to0\) by Theorem~\ref{unbal:thm:smallmean}.  This exhausts all
positive feasible same-side tuples and proves the theorem.
\end{proof}

\section{Uniform estimates for the global cover}
The global cover uses the mixed derivative bound of
Lemma~\ref{unbal:lem:global-mixed}, the endpoint entropy-imbalance gain,
the low-information and normalized zero-entropy estimates, the opposite
deterministic-corner estimate, and the mean-contact supporting planes.
Their statements, domains, and proofs appear in
Supplementary Section~\ref{supp:uniform-estimates}. The endpoint bound
holds for all means: the conditions on the endpoint masses characterize
equality and are not needed for the supporting plane.

\section{Analytic reductions for the unbalanced candidate}
\label{unbal:ar:section}
This section proves \(\zeta\ge R_B\) on the parts of the parameter domain
where a parameter, or a ratio of parameters, is unbounded or degenerates.
We begin with small means, then compare \(\phi\) and \(\psi\) at the parent
and control arbitrary entropy splits. These estimates cover the
low-entropy regions and a strip of extreme means, and leave compact
domains for Sections~\ref{unbal:sec:central}
and~\ref{unbal:sec:opposite-completion}.

All means are interior, and all entropy coordinates are positive and
feasible. The inequalities are uniform in the ratios of the parameters; in
particular, neither entropy is assumed to be bounded away from zero. We
use
\(L=\ln 2\),
\[
 \Phi(r,h)=\eta(h)-F(r,h),\qquad
 \overline H=\frac{H(a)+H(b)}2,\qquad
 s=\overline H-E,\qquad \Delta=H(m)-\overline H.
\]
The function \(C_n\) below uses natural logarithms, whereas \(H\), \(C\),
\(j\), and \(\eta\) use bits:
\begin{equation}
 C_n(r)=r\operatorname{atanh}r+\tfrac12\ln(1-r^2)
 =\sum_{n\ge1}\frac{r^{2n}}{2n(2n-1)},\qquad C_n(r)=L C(r).
 \label{unbal:ar:cn}
\end{equation}
Its continuous values are \(C_n(0)=0\) and \(C_n(1)=L\).
As explained in Section~\ref{unbal:sec:inputs}, parent dominance gives
\(\zeta\ge R_B\) by Theorem~\ref{bal:thm:global-unrestricted-target}. Where
\(\psi\) is active at the parent, it is enough to prove either
\(\zeta\ge R_\psi\) or the sharper inequality obtained by retaining suitable
lower bounds for \(B\) at the children.

\subsection{A uniform small-mean region}

\begin{theorem}[Small-mean region]\label{unbal:thm:smallmean}
For every \(0<a\le b<1\) with \(a+b\le1/16\), and every positive feasible
entropy pair,
\[
 \zeta(a,b,e,f)\ge R_\psi(a,b,e,f).
\]
Consequently \(\zeta\ge R_B\) there, using the unrestricted
\(R_\phi\) inequality of Theorem~\ref{bal:thm:global-unrestricted-target}. Complete label exchange and simultaneous complement
also give the corresponding region \(a+b\ge31/16\).
\end{theorem}

\begin{proof}
\emph{Proof outline.}
The log-sum bound and entropy-split convexity reduce the problem to
a concave function of the total entropy deficit. Its value at one
endpoint is already nonnegative, so only the other endpoint remains.
We estimate that endpoint in normalized mean coordinates, prove the
small-separation range analytically, and certify a scalar inequality on
the remaining compact interval.

The log-sum estimate of Theorem~\ref{unbal:ss:logsum}, together with the convexity
of \(P\), gives
\begin{equation}
 \zeta\ge j(a,b)+\alpha s,\qquad
 R_\psi\le D_\Delta(s):=P(s+\Delta)-P(s),\qquad
 \alpha=\frac{(b-a)^2}{2b(1-a)}.
 \label{unbal:ar:smallmean_start}
\end{equation}
We also use the scalar facts \(P'(0)=4\), \(P''>0\), increasing \(P''\),
and \(j(a,b)\ge P(\Delta)\), which are proved in Supplementary
Section~\ref{unbal:app:profile_logsum}.
If \(a=b\), then \(\Delta=0\), so the result is immediate. Suppose
\(a<b\). The function
\[
 g(s)=j(a,b)+\alpha s-D_\Delta(s),\qquad 0\le s\le\overline H,
\]
is concave, since
\(D_\Delta''(s)=P''(s+\Delta)-P''(s)\ge0\).
Moreover \(g(0)\ge0\). It therefore suffices to prove
\(g(\overline H)\ge0\). Evaluating this scalar function at
\(s=\overline H\) does not require evaluating \(\phi\) at zero entropy.

Put
\[
 r=\frac{b-a}{a+b},\qquad k=\frac m{1-m},\qquad K=\frac1{31}.
\]
Here \(0<r<1\), \(0<k\le K\). Exact entropy identities yield
\begin{align}
 L\Delta&=mC_n(r)+(1-m)C_n(kr),\label{unbal:ar:mean_delta}\\
 Lj(a,b)&=2mr\operatorname{atanh}r
       +2(1-m)kr\operatorname{atanh}(kr).\label{unbal:ar:mean_j}
\end{align}
The nonnegative series coefficients in \eqref{unbal:ar:cn} show that
\(C_n(kr)\le k^2C_n(r)\) and \(C_n(r)\le Lr^2\). Thus
\begin{equation}
 \Delta\le\frac{k C_n(r)}L\le kr^2\le Kr^2.
 \label{unbal:ar:delta_upper}
\end{equation}
We claim the stronger mean-cost estimate
\begin{equation}
 \frac{j(a,b)}\Delta\ge
 4+\frac23(1-k+k^2)r^2.
 \label{unbal:ar:j_bonus}
\end{equation}
Indeed, coefficient comparison gives
\[
 2x\operatorname{atanh}x-4C_n(x)\ge\frac23x^2C_n(x).
\]
At degree \(x^{2n}\), \(n\ge2\), the required comparison reduces to
\[
 6(n-1)^2(2n-3)-n(2n-1)
   =(n-2)(12n^2-20n+9)\ge0.
\]
Write \(z=C_n(kr)/C_n(r)\le k^2\). The weighted factor multiplying
\((2/3)r^2\) after applying this comparison to
\eqref{unbal:ar:mean_j} is
\[
 \frac{1+kz}{1+z/k}\ge\frac{1+k^3}{1+k}=1-k+k^2,
\]
because this ratio decreases with \(z\) for \(k\le1\). This proves
\eqref{unbal:ar:j_bonus}. In particular its quadratic coefficient is at least
\[
 A=\frac23(1-K+K^2)=\frac{1862}{2883}.
\]
Also, using
\(b(1-a)=m(1-m)(1+r)(1+kr)\) and
\eqref{unbal:ar:delta_upper},
\begin{equation}
 \frac\alpha\Delta\ge
 \gamma(r):=\frac{2Lr^2}{(1+r)(1+Kr)C_n(r)}.
 \label{unbal:ar:gamma}
\end{equation}
The function \(\gamma\) decreases on \([0,1]\): each of the factors in
its denominator after dividing by \(r^2\) increases. Its continuous
value at zero is \(4L<14/5\).

The proof uses the following directed comparisons:
\begin{equation}
 H(1/32)<h_*:=\frac{201}{1000},\qquad
 \frac{P'(H(1/32))-4}{H(1/32)}<c_*:=\frac{23}{10},\qquad
 \gamma(1/6)>c_*.
 \label{unbal:ar:small_fixed}
\end{equation}
Since \(P''\) increases, the secant
\((P'(h)-P'(0))/h\) increases. For \(h=H(m)\),
\eqref{unbal:ar:small_fixed} therefore gives
\(P'(h)\le4+c_*h\). Now
\(\overline H=h-\Delta\) and
\(D_\Delta(\overline H)\le\Delta P'(h)\), so
\begin{equation}
 \frac{g(\overline H)}\Delta
 \ge (A-K\gamma(r))r^2+(\gamma(r)-c_*)h.
 \label{unbal:ar:small_g}
\end{equation}
If \(\gamma(r)\ge c_*\), this is nonnegative because
\(A-K\gamma(r)\ge A-14/155>0\). This includes \(0<r\le1/6\).
If \(\gamma(r)<c_*\), use \(h\le h_*\) in the negative-coefficient
term. The sufficient remaining comparison is
\begin{equation}
 W(r):=Ar^2-c_*h_*+\gamma(r)(h_*-Kr^2)>0,
 \qquad \frac16\le r\le1.
 \label{unbal:ar:small_W}
\end{equation}
Notice \(h_*-Kr^2>0\) throughout this interval.

On each rational interval $[r_-,r_+]$, the interval evaluation of
\eqref{unbal:ar:small_W} uses the monotonicity of $\gamma$ and the continuous
value $C_n(1)=L$. Directed enclosures prove the required strict sign on a
complete partition of $[1/6,1]$, together with the directed comparisons
\eqref{unbal:ar:small_fixed}; the records are listed in
Supplementary Section~\ref{supp:verification}.
Thus \(g(\overline H)\ge0\). Concavity of \(g\), then
\eqref{unbal:ar:smallmean_start}, proves the assertion for every feasible split.
\end{proof}

\subsection{Scalar estimates for parent comparisons}
\label{unbal:ar:parent_scalars}

\begin{lemma}[Integrated entropy derivative]\label{unbal:ar:eta_integral}
For \(0<h\le1\),
\[
 -\eta'(h)\ge 2+\frac1{Lh}.
\]
Consequently, if \(u\ge0\) and \(h+u\le1\),
\begin{equation}
 \eta(h)-\eta(h+u)\ge2u+\frac1L\ln(1+u/h).
 \label{unbal:ar:eta_log}
\end{equation}
Moreover
\begin{equation}
 \frac{q^2}{2L}\le C(q)\le\frac{q^2}{2L(1-q^2)}
 \qquad(0\le q<1).
 \label{unbal:ar:C_bounds}
\end{equation}
\end{lemma}

\begin{proof}
Set \(r=1-2H^{-1}(h)\), \(A_r=\operatorname{atanh}r\),
\(\nu=1-r^2\), and \(h_n=Lh\). The function
\[
 T(r)=2rh_n-\nu A_r
\]
vanishes at zero and tends to zero as \(r\to1\), while
\(T'(r)=2h_n-1\) strictly decreases from \(2L-1>0\) to \(-1\).
Hence \(T\ge0\), or \(h_n\ge\nu A_r/(2r)\).
Direct differentiation gives
\[
 -\eta'(h)=2+\frac{2r}{\nu A_r}
 \ge2+\frac1{Lh}.
\]
The zero-radius value follows by continuity. Integration proves
\eqref{unbal:ar:eta_log}. Finally,
\(C(0)=C'(0)=0\) and \(C''(q)=1/[L(1-q^2)]\);
two integrations prove \eqref{unbal:ar:C_bounds}.
\end{proof}

\begin{theorem}[Factor-sixteen parent comparison]\label{unbal:ar:parent16}
Every feasible parent satisfying
\[
 0<q\le\frac12,\qquad x:=q/E\ge16
\]
obeys \(\phi(m,E)>\psi(m,E)\). Thus the hybrid inequality holds for
every feasible child split in this parent region.
\end{theorem}

\begin{proof}
\emph{Proof outline.}
The factor-eight parent theorem covers the smaller parent radius.
For the remaining radii, a derivative bound on the contact logit proves
parent dominance throughout an unbounded ratio tail. Monotonicity then
provides a sufficient interval inequality on the compact rectangle left
between that tail and the claimed threshold.

We use the factor-eight parent theorem in
Supplementary Section~\ref{unbal:app:parent_eight} for
\(0<q\le2/5\), \(q/E\ge8\). It remains to treat larger \(q\).
Define the contact \(v=v(x)\in(0,1/2)\) by
\(xH(v)=1-2v\), and write
\[
 r=1-2v,\quad \ell=\ln\frac{1-v}{v},\quad
 \nu=4v(1-v),\quad h_n=LH(v),\quad
 K_v=h_n+r\ell/2=\ell/2-\ln(1-v).
\]
Implicit differentiation gives
\begin{equation}
 x\ell'(x)=\frac{2rh_n}{\nu K_v}.
 \label{unbal:ar:ell_derivative}
\end{equation}
Indeed, \(x=Lr/h_n\),
\(d\ell/dv=-4/\nu\), and
\(d\ln x/dv=-2K_v/(rh_n)\).
Since \(-\ln(1-v)\le v/(1-v)\),
\[
 h_n\le v\bigl(\ell+(1-v)^{-1}\bigr),\qquad K_v\ge\ell/2.
\]
Equation~\eqref{unbal:ar:ell_derivative} yields
\begin{equation}
 x\ell'(x)\le\frac1{1-v}+\frac1{(1-v)^2\ell}.
 \label{unbal:ar:ell_upper}
\end{equation}
The directed comparisons
\[
 8H(1/50)>1-2/50,\qquad \ln49>19/5
\]
imply \(v<1/50\), \(\ell>19/5\) for \(x\ge8\), since
\((1-2v)/H(v)\) strictly decreases. Hence
\begin{equation}
 x\ell'(x)\le\frac{50}{49}+\frac{2500}{2401}\frac5{19}<\frac43
 \qquad(x\ge8).
 \label{unbal:ar:ell_fourthirds}
\end{equation}
The last comparison is rational and exact.

Lemma~\ref{unbal:ar:eta_integral} gives
\[
 \eta(E)-\eta(E+C(q))\ge\frac1L\ln\left(1+\frac{qx}{2L}\right).
\]
Also \(F(q,E)=q\ell(x)/L\). For \(k>0\),
\(\ln(1+kq)/q\) decreases in \(q>0\); consequently
\begin{equation}
 \frac Lq[\phi(m,E)-\psi(m,E)]
 \ge G(x):=2\ln\left(1+\frac{x}{4L}\right)-\ell(x).
 \label{unbal:ar:parent_G}
\end{equation}
By \eqref{unbal:ar:ell_fourthirds},
\[
 G'(x)\ge\frac2{x+4L}-\frac4{3x}
 =\frac{2x-16L}{3x(x+4L)}>0 \qquad(x>8L).
\]
The directed comparison \(G(128)>36/100\) therefore proves the entire
unbounded tail \(x\ge128\), with no entropy cutoff.

The remaining root rectangle is
\[
 (q,x)\in[2/5,1/2]\times[16,128].
\]
On a rational box \([q_-,q_+]\times[x_-,x_+]\), set
\(E_*=q_+/x_-\). Convexity of \(\eta\) makes
\(\eta(E)-\eta(E+c)\) decreasing in \(E\) and increasing in \(c\).
The actual parent difference is therefore bounded below by
\begin{equation}
 \eta(E_*)-\eta(E_*+C(q_-))
       -q_+\frac{F(x_+,1)}{x_+}.
 \label{unbal:ar:parent16_box}
\end{equation}
The last term is a valid upper bound because \(\ell\) increases.
Every entropy argument in this comparison is strictly in \((0,1)\)
on the root. Directed interval evaluation proves positivity on a complete
rational partition; inverse and contact roots are enclosed by verified
signs. The certificate is listed in Supplementary Section~\ref{supp:verification}.
Together with the factor-eight range and the analytic tail, this proves
the theorem.
\end{proof}

\subsection{A factor-eight theorem without a common entropy cap}
\label{unbal:ar:eight_section}

The next theorem allows \(E\) to be larger than the entropy cap of one of
the children, as happens in the outer opposite-side region.
The scalar-profile bound of Lemma~\ref{unbal:lem:global-mixed} is
\[
 0\le rF_{rr}(r,h)\le\kappa_0:=\frac{13}{6},
\]
We also use the decreasing second derivative of
\(r\mapsto F(r,h)\), one of the results listed in
Section~\ref{unbal:sec:inputs}, and the endpoint estimate of
Lemma~\ref{unbal:lem:endpoint-loggain}:
\begin{equation}
 \zeta\ge B_{\rm end}\ge
 F(d,E)+\frac d{2L}\mathcal J(\tau),\qquad
 \mathcal J(\tau)=-\ln(1-\tau^2),\qquad
 \tau=\frac{e-f}{2E},
 \label{unbal:ar:endpoint_gain}
\end{equation}
when \(d/E\ge5\) and the opposite-side contact is feasible. The proof of
\eqref{unbal:ar:endpoint_gain} in
Supplementary Lemma~\ref{unbal:lem:endpoint-loggain} requires this ratio
and contact feasibility; it does not require \(E\le\min\{H(a),H(b)\}\).

\begin{theorem}[Factor-eight estimate for arbitrary feasible splits]
\label{unbal:ar:eight}
For arbitrary interior means and positive feasible entropy coordinates,
\[
 0<E\le\frac{11}{200},\qquad d\ge8E,\qquad q\le8E
 \quad\Longrightarrow\quad \zeta\ge R_B.
\]
If \(\psi\) is active at the parent and \(q>0\), then more precisely
\begin{equation}
 B_{\rm end}-R_B\ge\frac{21}{1000}\frac{q^2}{E}.
 \label{unbal:ar:eight_margin}
\end{equation}
No common child entropy cap is assumed.
\end{theorem}

\begin{proof}
\emph{Proof outline.}
We first prove an entropy-curvature bound on the extended domain needed
for unequal child entropies. It yields a signed split correction, which
the endpoint gain absorbs up to an explicit loss. Radial and parent
comparisons then bound the remaining terms on the same scale; their
constants leave the stated positive margin. The closing branch argument
transfers this comparison to the hybrid candidate.

\smallskip\noindent\emph{Entropy curvature on the extended domain.}
Let \(v=H^{-1}(h)\), \(\ell=\ln((1-v)/v)\), \(r=1-2v\), and
\(h_n=Lh\). Exact differentiation gives
\[
 h^2\eta''(h)=
 \frac{h_n^2[(1+r^2)\ell-2r]}
 {2Lv^2(1-v)^2\ell^3}.
\]
The identity \(h_n=v\ell-\ln(1-v)\) implies
\(h_n\ge v(\ell+1)\). Put \(t=1/\ell\). Using
\[
 \frac{1+r^2}{2(1-v)^2}=1+\frac{v^2}{(1-v)^2},\qquad
 \frac r{(1-v)^2}=1-\frac{v^2}{(1-v)^2},
\]
we obtain
\[
 h^2\eta''(h)\ge\frac1L(1+t)^2(1-t)\ge\frac1L
 \quad\text{if }0<t\le\frac12.
\]
The last inequality is \(t(1-t-t^2)\ge0\). For \(h\le11/100\),
the entropy chord bound gives
\(H(1/16)\ge1/8>11/100\), hence \(v<1/16\) and
\(\ell>\ln15>2\). Thus
\begin{equation}
 \eta''(h)\ge\frac1{Lh^2}\qquad(0<h\le11/100).
 \label{unbal:ar:sharp_eta_curvature}
\end{equation}
Homogeneity gives
\(F_{hh}(r,h)=(r^2/h^2)F_{rr}(r,h)\le\kappa_0r/h^2\).
It follows that
\begin{equation}
 \Phi_{hh}(r,h)\ge\frac{1/L-\kappa_0r}{h^2},
 \qquad 0<h\le11/100.
 \label{unbal:ar:extended_curvature}
\end{equation}
Here \(\Phi(r,h)=\eta(h)-F(r,h)\) is well defined for all
\(r\ge0\), \(0<h<1\), including values above the feasible entropy
cap belonging to \(r\). On the interval $0<h\le11/100$, which contains
all entropy arguments used here, Equation~\eqref{unbal:ar:extended_curvature}
proves convexity of
\(\Phi(r,h)+(1/L-\kappa_0r)\ln h\).
The coefficient can be negative; the displayed second-derivative
inequality still proves convexity.

\smallskip\noindent\emph{Exact signed split correction.}
The hypotheses imply \(q\le d\), so after complete label exchange and
simultaneous complement we may take
\(a\le1/2\le b\) and \(q=1-a-b\ge0\).
The radial coordinates belonging to \(e,f\) are respectively
\(r_1=d+q\) and \(r_2=d-q\). Both \(e=E(1+\tau)\) and
\(f=E(1-\tau)\) lie in \((0,2E]\subset(0,11/100]\).
Convexity in \eqref{unbal:ar:extended_curvature} gives, for
\(\gamma_i=1/L-\kappa_0r_i\),
\[
 \Phi(r_i,h)\ge\Phi(r_i,E)+\Phi_h(r_i,E)(h-E)
  +\gamma_i\left(\frac{h-E}E-\ln\frac hE\right).
\]
Write \(A_e=\Phi_h(r_1,E)-\Phi_h(r_2,E)\).
Since \(\Phi_{rh}=rF_{rr}/h\),
\begin{equation}
 0\le A_e\le\frac{2\kappa_0q}{E}.
 \label{unbal:ar:A_e}
\end{equation}
Averaging the preceding supporting inequalities, with
\(\gamma_0=1/L-\kappa_0d\), gives the exact correction
\begin{align}
 \frac{\Phi(r_1,e)+\Phi(r_2,f)}2
 \ge{}&\frac{\Phi(r_1,E)+\Phi(r_2,E)}2
       +\frac{E\tau A_e}{2}
       +\frac{\gamma_0}{2}\mathcal J(\tau)\notag\\
 &+\kappa_0q(\operatorname{atanh}\tau-\tau).
 \label{unbal:ar:signed_split}
\end{align}
The endpoint contact needed for \eqref{unbal:ar:endpoint_gain} exists:
\(d>E\) exceeds its lower threshold, and opposite-side feasibility gives
\(d\le1-H^{-1}(e)-H^{-1}(f)\). Its supporting plane remains a global
lower bound even when the associated three-atom law would have inadmissible
deterministic masses; equality of that law is not used.

Combining \eqref{unbal:ar:signed_split} and \eqref{unbal:ar:endpoint_gain}, all
split-dependent terms are bounded below by
\[
 c\mathcal J(\tau)+\frac{E\tau A_e}{2}
   +\kappa_0q(\operatorname{atanh}\tau-\tau),\qquad
 c=\frac{1+d}{2L}-\frac{\kappa_0d}{2}.
\]
Here \(c>0\). For \(\tau\ge0\), the expression is nonnegative.
For \(\tau=-t<0\), inequality~\eqref{unbal:ar:A_e} cancels the linear
term with the required sign, leaving the lower bound
\begin{equation}
 c\mathcal J(t)-\kappa_0q\operatorname{atanh}t.
 \label{unbal:ar:negative_split}
\end{equation}
For \(z=\kappa_0q/(2c)<1\), the exact infimum of
\eqref{unbal:ar:negative_split} over \(0\le t<1\) is
\begin{equation}
 -2c C_n(z).
 \label{unbal:ar:split_inf}
\end{equation}
To check this, substitute \(t=\tanh u\). The expression becomes
\(2c\ln\cosh u-\kappa_0qu\); its unique minimum satisfies
\(\tanh u=z\), and substitution gives \eqref{unbal:ar:split_inf}.

The mean constraint \(d+q\le1\), the inequality \(q\le11/25\),
and the negative coefficient of \(d\) in \(c\) imply
\begin{equation}
 c\ge c_0(q):=\frac1L-\frac{\kappa_0}2+
      \left(\frac{\kappa_0}2-\frac1{2L}\right)q.
 \label{unbal:ar:c0}
\end{equation}
Both coefficients in this affine function are positive.
The ratio \(z_0(q)=\kappa_0q/(2c_0(q))\) increases, and a directed
comparison gives
\(z_0(11/25)<0.919079<1\). Thus \eqref{unbal:ar:split_inf} applies on the
full domain. For fixed \(q\),
\[
 \frac{d}{dc}\left[2cC_n\left(\frac{\kappa_0q}{2c}\right)\right]
 =\ln(1-z^2)\le0,
\]
with equality precisely when \(q=0\). Replacing \(c\) by \(c_0\) therefore gives a valid upper bound for
the loss. The positive series in \eqref{unbal:ar:cn} implies that
\(C_n(z)/z^2\) increases and has limit \(1/2\) at zero. On a
rational interval \([q_-,q_+]\), a uniform upper bound is accordingly
\[
 \frac E{q^2}\,2c_0(q)C_n(z_0(q))
 \le \frac{11}{200}\frac{\kappa_0^2}{2c_0(q_-)}
         \frac{C_n(z_0(q_+))}{z_0(q_+)^2}.
\]
The directed interval certificate proves
\begin{equation}
 \frac E{q^2}\,2c_0(q)C_n(z_0(q))<\frac15
 \label{unbal:ar:split_loss}
\end{equation}
on \(0<E\le11/200\), \(0\le q\le11/25\); see Supplementary
Section~\ref{supp:verification}. All expressions at \(q=0\) use the stated
continuous extension.

\smallskip\noindent\emph{Radial and parent comparisons.}
Set \(f_0(x)=F(x,1)\), so \(F(r,h)=h f_0(r/h)\).
The decreasing second derivative of \(F\) implies
\(F_r(r,h)\ge rF_{rr}(r,h)\). Thus
\(z\mapsto F(\sqrt z,h)\) is increasing and concave, and
\(f_0'(x)/(2x)\) decreases. Jensen's inequality followed by the
tangent inequality gives
\begin{align}
 \frac{F(d+q,E)+F(d-q,E)}2-F(d,E)
 &\le F(\sqrt{d^2+q^2},E)-F(d,E)\notag\\
 &\le\frac{f_0'(d/E)}{2(d/E)}\frac{q^2}{E}
 \le\frac{479}{1000}\frac{q^2}{E}.
 \label{unbal:ar:radial_eight}
\end{align}
The last step uses \(d/E\ge8\) and the directed comparison
\(f_0'(8)/16<479/1000\). For reproducibility, if
\(8H(v)=1-2v\) and \(\ell=\ln((1-v)/v)\), its checked formula is
\[
 f_0'(8)=\frac\ell L+
 \frac{1-2v}{v(1-v)(8\ell+2L)}.
\]

The remaining parent estimate is
\begin{equation}
 \eta(E)-\eta(E+C(q))\ge\frac7{10}\frac{q^2}{E},
 \qquad 0<E\le11/200,\quad0<q\le8E.
 \label{unbal:ar:parent_seven_tenths}
\end{equation}
We prove it analytically for small \(E\) and by a computation on the
remaining compact range.
If \(E\le10^{-4}\), then by \eqref{unbal:ar:C_bounds},
\[
 \frac{C(q)}E\le\rho_*:=
 \frac{64\cdot10^{-4}}{2L(1-64\cdot10^{-8})}.
\]
Use \(\ln(1+u)\ge u/(1+u)\) in \eqref{unbal:ar:eta_log} and drop
its positive \(2C(q)\) term to obtain
\[
 \frac E{q^2}[\eta(E)-\eta(E+C(q))]
 \ge\frac1{2L^2(1+\rho_*)}>1.0359>\frac7{10}.
\]
A directed comparison verifies this strict inequality between constants.

For the compact part put \(q=yE\) and use the exact root
\([10^{-4},11/200]\times[0,8]\) in \((E,y)\).
Two sufficient directed lower bounds are used. First, the identity
\[
 -\eta'(h)=2+\frac{1-2v}{v(1-v)\ln((1-v)/v)},\qquad v=H^{-1}(h),
\]
and \(-\ln(1-v)\ge v\) imply
\[
 -\eta'(h)\ge2+\frac1h\frac{1-2v}{L(1-v)}
                       \left(1+\frac1{\ln((1-v)/v)}\right).
\]
For a box with entropy lower endpoint \(E_-\), entropy upper endpoint
\(E_+\), and radial enclosure \([q_-,q_+]\), enclose
\(u=C(q)/E\) above by \(u_+\) and set
\[
 v_-=H^{-1}(E_-),\qquad
 v_+=H^{-1}(E_++C(q_+)),\qquad
 \beta_-=
 \frac{1-2v_+}{L(1-v_+)}
       \left(1+\frac1{\ln((1-v_-)/v_-)}\right).
\]
Integration and monotonicity of \(\ln(1+u)/u\) show that a lower
bound for the normalized left side in \eqref{unbal:ar:parent_seven_tenths} is
\begin{equation}
 \frac{1/2+q_-^2/12+q_-^4/30+q_-^6/56}{L}
 \left(2E_-+\beta_-\frac{\ln(1+u_+)}{u_+}\right).
 \label{unbal:ar:parent_log_box}
\end{equation}
The ratio at \(u_+=0\) is one. The polynomial factor comes from the
positive series for \(C(q)/q^2\), so this rule includes the \(q=0\)
face without a singular division.
Second, for \(q_->0\), convexity of \(\eta\) provides the alternative
lower bound
\begin{equation}
 \frac{E_-}{q_+^2}
 [\eta(E_+)-\eta(E_++C(q_-))].
 \label{unbal:ar:parent_direct_box}
\end{equation}
On every cell of a complete rational partition, interval evaluation proves
that at least one of \eqref{unbal:ar:parent_log_box} and
\eqref{unbal:ar:parent_direct_box} is at least $7/10$.
Directed comparisons also verify the constants used in the analytic tail
and in the radial estimate; see Supplementary Section~\ref{supp:verification}. No condition
$q\le2/5$ is imposed in this check.

Finally suppose \(\psi\) is active at the parent. Each child value of
\(B\) dominates its corresponding \(\Phi\) value, so
\eqref{unbal:ar:signed_split}--\eqref{unbal:ar:parent_seven_tenths} give
\[
 B_{\rm end}-R_B\ge
 \left(\frac7{10}-\frac{479}{1000}-\frac15\right)\frac{q^2}{E}
 =\frac{21}{1000}\frac{q^2}{E}.
\]
When \(\phi\) is active, the \(R_\phi\) inequality of
Theorem~\ref{bal:thm:global-unrestricted-target} applies.
At \(q=0\), the parent candidates agree exactly. The boundary \(q=d\),
where one child mean equals \(1/2\), is included in the same radial and
curvature argument, with \(F(0,h)=0\). This completes the proof.
\end{proof}

\subsection{The outer opposite-side region at small positive entropy}

\begin{corollary}\label{unbal:ar:outer_lowentropy}
Suppose the means lie on opposite sides of \(1/2\), with at least
one outside \([1/10,9/10]\). Then every positive feasible entropy pair
with \(E\le1/25\) satisfies \(\zeta\ge R_B\).
\end{corollary}

\begin{proof}
Complete label exchange and simultaneous complement show that an outer
opposite-side pair has \(d\ge2/5\) and \(q\le1/2\).
Thus \(d\ge2/5>8E\).
If \(q\le2/5\) and \(q\ge8E\), apply the factor-eight parent
theorem in Supplementary Section~\ref{unbal:app:parent_eight}.
If \(q\le8E\), apply Theorem~\ref{unbal:ar:eight}; it requires
no shared entropy cap. These alternatives overlap on their boundary.

For \(2/5\le q\le1/2\) and \(E\le1/40\),
Theorem~\ref{unbal:ar:parent16} applies because \(q/E\ge16\).
The remaining rectangle is
\[
 (q,E)\in[2/5,1/2]\times[1/40,1/25].
\]
For a rational box \([q_-,q_+]\times[E_-,E_+]\), a lower bound for
\(\phi(m,E)-\psi(m,E)\) is
\begin{equation}
 \eta(E_+)-\eta(E_++C(q_-))-F(q_+,E_-).
 \label{unbal:ar:highq_box}
\end{equation}
This uses convexity and monotonicity of \(\eta\), increasing \(C\),
and the fact that \(F\) increases in its first argument and decreases
in its second. Interval evaluation proves a strict positive lower bound
on a complete rational partition; see Supplementary Section~\ref{supp:verification}.
Parent dominance then proves the hybrid inequality, completing all cases.
\end{proof}

\subsection{Removing the full zero-entropy boundary}

\begin{theorem}[Global low-entropy region]\label{unbal:ar:global_lowentropy}
Every positive feasible four-moment tuple with \(E\le10^{-6}\)
satisfies \(\zeta\ge R_B\), without restrictions on either mean or
on the entropy ratio.
\end{theorem}

\begin{proof}
\emph{Proof outline.}
We remove small means and use the available parent or normalized
low-entropy comparisons when the parent radius is small. In the remaining
radius range, an explicit lower bound on the balanced parent is compared
with an upper bound on the entropy-deficit parent. Monotonicity reduces
that last comparison to a finite interval partition.

Use complete label exchange and simultaneous complement to arrange
\(0<a\le b<1\), \(a+b\le1\). If \(a+b\le1/16\), apply
Theorem~\ref{unbal:thm:smallmean}. Otherwise \(q=1-a-b<15/16\).
The \(q=0\) case is exact parent equality. For \(0<q\le2/5\)
and \(q\ge8E\), the factor-eight parent theorem in
Supplementary Section~\ref{unbal:app:parent_eight} applies.
If \(q\le8E\), then \(q\le32E\), so
Supplementary Theorem~\ref{unbal:thm:normalized-low-entropy} applies.
That theorem covers
\(E\le10^{-6}\), \(q\le32E\), including arbitrary entropy ratios
and the opposite corner.

It remains to treat \(2/5\le q\le15/16\). We first prove the
all-\(q\) parent lower bound used in the finite check. Put
\(v=H^{-1}(E)\), \(r=1-2v\). If \(E<1-q\), the entropy chord
bound gives \(v\le E/2\), hence \(r\ge1-E>q\).
At the defining contact for \(F(q,E)\), its radial coordinate is at
most \(r\), because \(z/H((1-z)/2)\) increases. Consequently
\(F(q,E)\le qJ(v)\), and
\begin{equation}
 \phi(m,E)\ge(r-q)J(v)
 \ge A_q(E):=(1-q-E)\log_2\frac{2-E}{E}.
 \label{unbal:ar:allq_parent}
\end{equation}
All factors used here are nonnegative. Also
\(\psi(m,E)\le K(q):=\eta(C(q))\).
Our domain satisfies \(E\le10^{-6}<1/16\le1-q\).
The function \(A_q(E)\) decreases separately in \(q,E\), and
\(K(q)\) decreases in \(q\). Therefore a sufficient comparison on
\([q_-,q_+]\subset[2/5,15/16]\) is
\begin{equation}
 A_{q_+}(10^{-6})-K(q_-)>0.
 \label{unbal:ar:global_lowentropy_box}
\end{equation}
The directed interval certificate proves \eqref{unbal:ar:global_lowentropy_box}
on a complete rational partition; see Supplementary Section~\ref{supp:verification}.
Thus \(\phi\) dominates at the parent in the remaining range, proving
the theorem.
\end{proof}

\subsection{A uniform extreme-mean opposite-side strip}

\begin{theorem}[Extreme opposite-side strip]\label{unbal:ar:extreme_strip}
For every positive feasible entropy pair,
\[
 0<a\le2^{-28},\qquad \frac12\le b<1
 \quad\Longrightarrow\quad \zeta(a,b,e,f)\ge R_B(a,b,e,f).
\]
Complete label exchange and simultaneous complement give the corresponding
symmetric strips.
\end{theorem}

\begin{proof}
\emph{Proof outline.}
First remove the opposite deterministic corner and the branch where
the balanced candidate is active at the parent. On the remaining branch,
nonnegative child values give an entropy-independent upper bound for the
target. Monotonicity lowers the mean cost to the outer edge of the strip,
so a one-variable interval comparison covers every remaining entropy split.

Besides the \(R_\phi\) inequality in
Theorem~\ref{bal:thm:global-unrestricted-target}, use
Supplementary Theorem~\ref{unbal:thm:opposite-corner}, which covers
\(a+1-b\le2^{-13}\) for every positive feasible entropy pair, and
the factor-eight parent theorem in
Supplementary Section~\ref{unbal:app:parent_eight}.
Joint convexity of relative entropy gives the deterministic mean bound
\[
 \zeta\ge j(a,b)=\tfrac12[D(a\Vert b)+D(b\Vert a)].
\]
If \(\phi\) is active at the parent, the target follows at once.
Suppose \(\psi\) is active. Feasibility implies
\(\phi\ge0\) at each child: if \(v=H^{-1}(h)\), the feasible radial
coordinate satisfies \(|1-2m|\le1-2v\), while
\(F(1-2v,h)=\eta(h)\). Thus \(B\ge0\) at the children and
\begin{equation}
 R_B\le\eta(E+C(q))\le\eta(C(q))\quad(q>0).
 \label{unbal:ar:strip_parent}
\end{equation}
When \(q\le2/5\), a \(\psi\)-active parent must additionally have
\(E>q/8\), by factor-eight parent dominance. In that case
\begin{equation}
 R_B\le\eta(q/8+C(q)).
 \label{unbal:ar:strip_parent8}
\end{equation}

Put \(A_*=2^{-28}\), \(K_*=2^{-13}\), and \(x=1-b\).
After removing the corner of Supplementary Theorem~\ref{unbal:thm:opposite-corner}, \(x>K_*-a\ge K_*-A_*\),
\(x\le1/2\), and
\(q=1-a-b=x-a>K_*-2A_*>0\).
For \(a<b\),
\[
 \partial_a j(a,b)
 =-\tfrac12[J(a)-J(b)]+\tfrac12(b-a)J'(a)<0.
\]
It follows that
\[
 j(a,1-x)\ge j(A_*,1-x)
 =\tfrac12(1-x-A_*)[J(A_*)+J(x)].
\]
Both factors on the right are positive and decrease with \(x\) on
\([K_*-A_*,1/2]\). For a rational interval \([x_-,x_+]\), therefore,
\begin{equation}
 j(a,1-x)\ge\tfrac12(1-x_+-A_*)[J(A_*)+J(x_+)].
 \label{unbal:ar:strip_cost}
\end{equation}
Since \(q\ge x_--A_*>0\), the target is at most
\(\eta(C(x_--A_*))\) by \eqref{unbal:ar:strip_parent}.
If \(x_+\le2/5\), then \(q\le2/5\), and the sharper upper bound
from \eqref{unbal:ar:strip_parent8} is
\[
 \eta((x_--A_*)/8+C(x_--A_*)).
\]
Thus a single strict comparison with \eqref{unbal:ar:strip_cost} proves
all means and entropy splits represented by the interval.

The compact interval is the union of the exactly adjacent intervals
\[
 [K_*-A_*,2K_*],\quad
 [2^{-k},2^{-k+1}]\ (k=12,11,\ldots,3),\quad
 [1/4,2/5],\quad[2/5,1/2].
\]
The first twelve use the sharper target upper bound, and the last uses
\eqref{unbal:ar:strip_parent}. The interval certificate verifies these
bounds on a complete subdivision, including all shared endpoints.
Contact and entropy inverse enclosures are verified by
directed signs. Thus \(j>R_B\) throughout the \(\psi\)-parent branch
outside this corner. The \(\phi\)-parent branch and the corner
complete the proof.
\end{proof}

Theorems~\ref{unbal:thm:smallmean}, \ref{unbal:ar:global_lowentropy}, and
\ref{unbal:ar:extreme_strip} remove unbounded mean and entropy limits using
uniform inequalities. Theorems~\ref{unbal:ar:parent16}, \ref{unbal:ar:eight}, and
Corollary~\ref{unbal:ar:outer_lowentropy} give further sufficient
inequalities for the outer opposite-side computation of
Section~\ref{unbal:sec:opposite-completion}. The records of the
computations used in this section are in
Supplementary Section~\ref{supp:verification}.

\section{The central mean square}
\label{unbal:sec:central}

\begin{theorem}[Central square, all positive feasible entropies]
\label{unbal:thm:central}
For $a,b\in[1/10,9/10]$ and positive feasible entropy coordinates,
$\zeta(a,b,e,f)\ge R_B(a,b,e,f)$.
\end{theorem}

The proofs of the results quoted in the table below are given in
Supplementary Section~\ref{unbal:app:modules}.

\begin{proof}
\emph{Proof outline.}
The central same-side theorem handles one half of the square. For
opposite-side means, we divide by mean separation, total entropy, and
then separation-to-entropy ratio. The table records the result used on
each region; the subsequent checks explain why the regions include all
interfaces and all feasible entropy splits.

The central same-side half-square theorem in
Supplementary Section~\ref{unbal:app:no_separation} treats all positive
feasible entropy pairs for means on the same side of $1/2$.
For opposite-side means, exchange the complete child
labels so that $a\le1/2\le b$. Then $q\le d$ and $q+d\le4/5$, hence
$q\le2/5$. The following closed regions exhaust the remaining tuples.

\begin{center}\small
\begin{tabular}{p{0.36\textwidth}>{\raggedright\arraybackslash}p{0.56\textwidth}}
\toprule
Region & Result used \\
\midrule
$d\le1/50$ & Central diagonal band, Supplementary Theorem~\ref{unbal:thm:no_separation:A} \\
$d\ge1/50$, $E\ge11/200$ & Moderate-entropy cover, Supplementary Section~\ref{unbal:app:central_opposite_moderate} \\
$d\ge1/50$, $E\le11/200$, $d\le4E$ & Central small-ratio theorem, including its analytic tail, Supplementary Theorem~\ref{unbal:thm:small_ratio:4} \\
$d\ge1/50$, $E\le11/200$, $4E\le d\le8E$ & Low-entropy transition cover, Supplementary Section~\ref{unbal:app:central_transition} \\
$E\le11/200$, $d\ge8E$, $q\ge8E$ & Factor-eight parent theorem, Supplementary Section~\ref{unbal:app:parent_eight}, since $q\le2/5$ \\
$E\le11/200$, $d\ge8E$, $q\le8E$ & Eight-ratio theorem, Supplementary Theorem~\ref{unbal:thm:central_eight:1} \\
\bottomrule
\end{tabular}
\end{center}

The diagonal band includes the exact diagonal. The small-ratio theorem
includes $E\downarrow0$; its finite cover is joined to its analytic tail.
In the transition row, $d\ge1/50$ and $d/E\le8$ imply $E\ge1/400$, which
is the lower endpoint of the finite chart and discards no tuple. In the last
row one may use either the central eight-ratio theorem in
Supplementary Section~\ref{unbal:app:central_eight} or the
stronger cap-free Theorem~\ref{unbal:ar:eight}.

All interfaces overlap at equality. If $q=d$, one mean is $1/2$ and the
same-side half-square theorem also applies; $q=0$ has exact parent equality.
The moderate-entropy cover uses the full feasible range
$E=11/200+t(C_0-11/200)$, $0\le t\le1$, using the cap estimates on their
stated subdomains. Entropy-split convexity and the supporting-plane comparisons
of Supplementary Section~\ref{unbal:app:modules} make every row uniform in the split of $2E$.
Every row is therefore a proved region, and their union establishes
the theorem.
\end{proof}

\section{Completion of every opposite-side mean pair}
\label{unbal:sec:opposite-completion}

We now prove \(\zeta\ge R_B\) for every pair of interior means on opposite
sides of \(1/2\), using the results listed in
Section~\ref{unbal:sec:inputs}. The entropy caps of the individual children
and arbitrarily unequal positive entropies are included.

Throughout this section, complete child-label exchange and simultaneous
complementation put the means in the orientation
\begin{equation}\label{unbal:eq:opp-orientation}
 0<a\le\frac12\le b<1,\qquad a\le1-b.
\end{equation}
Thus $q=1-a-b\ge0$ and $d=b-a$. We retain
\[
 C_0=\frac{H(a)+H(b)}2,\qquad E=\frac{e+f}2,\qquad
 s=C_0-E,\qquad \Delta=H((a+b)/2)-C_0,\qquad I=\Delta+s.
\]
Set
\[
 A_*:=2^{-28},\qquad K_*:=2^{-13},\qquad \epsilon_*:=10^{-6}.
\]
These constants delimit the regions treated analytically and the compact
box treated by computation; Theorem~\ref{unbal:thm:opposite} has no
hypothesis involving them.

\subsection{The compact theorem and its exact domain}
The argument on the compact domain has three parts: coordinates in which
every remaining feasible tuple lies in one box, sufficient inequalities
that can be verified on a subbox, and a complete partition of the box into
subboxes on which one of them holds. We give them in this order.

\begin{theorem}[Outer opposite-side compact domain]
\label{unbal:thm:outeropposite}
Assume
\[
 A_*\le a\le\frac1{10},\qquad
 \frac12\le b\le1-a,\qquad
 0<e\le H(a),\quad 0<f\le H(b).
\]
If $C_0>\epsilon_*$ and $\epsilon_*\le E\le C_0$, then
\[
 \zeta(a,b,e,f)\ge R_B(a,b,e,f).
\]
\end{theorem}

The proof of this theorem is a complete directed-interval cover, whose
acceptance inequalities are established below and in
Supplementary Section~\ref{unbal:app:modules}. Its exact rational root is
\begin{equation}\label{unbal:eq:opp-root}
 (u,v,t)\in[3,28]\times[1,28]\times[0,1],
\end{equation}
with the map
\begin{equation}\label{unbal:eq:opp-map}
 a=2^{-u},\qquad b=1-2^{-v},\qquad
 E=\epsilon_*+t(C_0-\epsilon_*).
\end{equation}
Only its constrained intersection $a\le1/10$, $a\le1-b$ is needed.
Indeed, a tuple in Theorem~\ref{unbal:thm:outeropposite} has
\[
 3< -\log_2a\le28,\qquad
 1\le-\log_2(1-b)\le28,\qquad
 t=\frac{E-\epsilon_*}{C_0-\epsilon_*}\in[0,1].
\]
Consequently, no feasible entropy interval is omitted by this
parametrization. In particular, $t=1$ includes the full entropy-cap face.

For a rational subbox of \eqref{unbal:eq:opp-root}, directed exponentiation
produces enclosing intervals for $a$ and $x=1-b$. Write these as
$[a_-,a_+]$ and $[x_-,x_+]$. The constrained intersection is empty if
$a_->1/10$ or $x_+<a_-$. Otherwise the verification program retains the enclosure
\[
 a\in[a_-,\min(a_+,1/10)],\qquad
 b\in[1-x_+,\min(1-x_-,1-a_-)].
\]
An empty resulting interval is also discarded. These are necessary
constraints, so this operation removes no relevant tuple. It may retain
extra points, which only weakens subsequent interval bounds. Every
relevant pair satisfies $d\ge2/5$.

\subsection{Analytic regions in the compact cover}

A subbox may be accepted because it lies in a region where
\(\zeta\ge R_B\) has already been proved. The regions used in this way are
as follows.

The opposite deterministic corner
\[
 a+1-b\le K_*
\]
is covered by Theorem~\ref{unbal:thm:opposite-corner}, for every positive feasible
entropy pair. The region $I\le1/100$ is covered by
Theorem~\ref{unbal:thm:low-information}, with no mean restriction.
Parent dominance also suffices, because $\phi(m,E)\ge\psi(m,E)$ implies
$R_B\le R_\phi$. The parent criteria used here are
\begin{align}
 0<q\le\frac25,\quad q\ge8E
   &\quad\Longrightarrow\quad \phi(m,E)>\psi(m,E),
       \label{unbal:eq:opp-parent8}\\
 0<q\le\frac12,\quad q\ge16E
   &\quad\Longrightarrow\quad \phi(m,E)>\psi(m,E).
       \label{unbal:eq:opp-parent16}
\end{align}
The first is the factor-eight comparison in
Supplementary Section~\ref{unbal:app:parent_eight}; the second is
Theorem~\ref{unbal:ar:parent16}. At $q=0$ the parent candidates agree exactly.

A useful union removes the interface $q=8E$. If
\begin{equation}\label{unbal:eq:opp-eight-union}
 E\le\frac{11}{200},\qquad d\ge8E,\qquad q\le\frac25,
\end{equation}
then either \eqref{unbal:eq:opp-parent8} applies or $q\le8E$ and the
cap-free eight-ratio theorem, Theorem~\ref{unbal:ar:eight}, applies.
The latter theorem has no assumption that $E$ lies below both child
entropy caps. Thus \eqref{unbal:eq:opp-eight-union} treats every feasible split.

In fact the complete low-entropy union is stronger:
\begin{equation}\label{unbal:eq:opp-lowentropy-owner}
 \text{outer opposite-side means},\quad E\le\frac1{25}
 \quad\Longrightarrow\quad \zeta\ge R_B.
\end{equation}
This is Corollary~\ref{unbal:ar:outer_lowentropy}. To recall the case
division in its proof, outer opposite-side means have $d\ge2/5$ and $q\le1/2$. For
$E\le1/25$, therefore, $d\ge8E$. The range $q\le2/5$ is covered by
\eqref{unbal:eq:opp-eight-union}. When $q\ge2/5$ and $E\le1/40$,
\eqref{unbal:eq:opp-parent16} applies. The remaining rectangle
$q\in[2/5,1/2]$, $E\in[1/40,1/25]$ is covered by the computation in the proof of that corollary.

\paragraph{Restricting the additional comparisons to high entropy.}
Every additional comparison below is combined with
\eqref{unbal:eq:opp-lowentropy-owner}. If an entire box lies in $E\le1/25$,
that corollary gives the required bound. Otherwise the lower endpoint
of the entropy enclosure is raised to an outward lower bound for $1/25$,
and only the remaining $E\ge1/25$ intersection is checked. The raw enclosure
$s=(C_0-\epsilon_*)(1-t)$ remains a conservative bound on that
intersection. The rectangle inequalities use the original, unmodified mean--$t$ rectangle.

The conclusion on a low-entropy box is $\zeta\ge R_B$.
A separate $R_\psi$ comparison is required only on the complementary
high-entropy intersection. This is precisely the assertion needed by the
hybrid theorem.

\subsection{Direct parent comparisons}

Two simple comparisons are useful beyond the named regions. Since
$\eta$ is decreasing and convex, the function
\[
 D(E,c):=\eta(E)-\eta(E+c)
\]
decreases in $E$ and increases in $c$. Also $C(q)$ increases for
$q\ge0$, while $F(q,E)$ increases in $q$ and decreases in $E$.
Consequently a box with enclosures $E\in[E_-,E_+]$ and
$q\in[q_-,q_+]$ is covered by parent dominance whenever
\begin{equation}\label{unbal:eq:opp-parent-direct}
 \eta(E_+)-\eta(E_++C(q_-))-F(q_+,E_-)\ge0.
\end{equation}
The verification program uses this expression only when its arguments lie in their
stated entropy domains.

Alternatively, suppose $\psi$ is active at the parent. Feasible children
have $B\ge0$, so
\[
 R_B\le\eta(E+C(q))\le\eta(E_-+C(q_-)).
\]
Joint convexity of the Jeffreys cost gives $\zeta\ge j(a,b)$.
For ordered means, $j$ decreases in its first argument and increases
in its second. Thus
\begin{equation}\label{unbal:eq:opp-parent-envelope}
 j(a_+,b_-)\ge\eta(E_-+C(q_-))
\end{equation}
is another sufficient comparison. A $\phi$-active parent is already
covered by Theorem~\ref{bal:thm:global-unrestricted-target}. Neither test requires a
bound on the entropy ratio.

\subsection{The entropy-cap slope comparison}

Let $A,D$ be the entropy coefficients of the supporting plane at
$(a,b)$ from Lemma~\ref{unbal:lem:global-cap-plane}. The verification program determines
them from \eqref{unbal:eq:ret-cap-matrix} and verifies their nonnegativity
before use. Set $k=\min(A,D)$. That plane and entropy convexity give
\[
 \zeta-R_\psi\ge G(s):=j(a,b)+2ks-P(\Delta+s)+P(s).
\]
The exact cap baseline is $G(0)=j(a,b)-P(\Delta)\ge0$.
If $I_+$ is an upper bound on the actual parent deficit and
\begin{equation}\label{unbal:eq:opp-cap-slope}
 4+2k_-\ge P'(I_+),
\end{equation}
then, for $0\le z\le s$,
\[
 G'(z)=2k-P'(\Delta+z)+P'(z)
       \ge2k_- -P'(I_+)+4\ge0.
\]
Thus $G(s)\ge0$. This includes exact cap equality and avoids dividing
by a vanishing cap gap. The rescaled system is solved for $A_0=A/d^2$ and
$\widehat D=D/d^2$, and uses
$k_- = d_-^2\min((A_0)_-,(\widehat D)_-)$ in
\eqref{unbal:eq:opp-cap-slope}; the factor $d^2$ is kept in the comparison.

\subsection{Keeping the feasible entropy imbalance}

The unrestricted entropy-split bound $R_\psi\le P(I)-P(s)$ can lose
information when one child mean is extreme. Write
\[
 I_a=H(a)-e,\qquad I_b=H(b)-f,
 \qquad I_a+I_b=2s,
\]
so $0\le I_a\le H(a)$ and $0\le I_b\le H(b)$. Put
$c_*=\min(H(a),H(b))$. For fixed $s$, the smallest possible average
of the child profile values is
\begin{equation}\label{unbal:eq:opp-feasible-child-min}
 \frac{P(I_a)+P(I_b)}2
 \ge\frac{P(s-r)+P(s+r)}2,
 \qquad r=\max(0,s-c_*).
\end{equation}
Indeed, a convex function makes the symmetric average increase with
the absolute imbalance. The closest feasible split to equal deficits
has that imbalance $r$; feasibility of the total deficit ensures
that the other cap is respected.

The right side of \eqref{unbal:eq:opp-feasible-child-min} increases in $s$
and in $r\ge0$. Hence a box gives the uniform lower bound
\[
 L_{\rm child}:=\frac{P(s_- -r_-)+P(s_-+r_-)}2,
 \qquad r_-=\max(0,s_- -c_+),\quad
 c_+=\min(H(a)_+,H(b)_+).
\]
Whenever the arguments are in $[0,1)$, this yields
\begin{equation}\label{unbal:eq:opp-feasible-target}
 R_\psi\le P(I_+)-L_{\rm child}.
\end{equation}
The verification program compares this upper bound with $j(a_+,b_-)$ and with
$F(d_-,E_+)$.

There is a further lower-bound gain if $E_->c_+$ and
$d_-\ge5E_+$. With $\tau=(e-f)/(2E)$, at least one child entropy is
at most $c_*$, so
\[
 |\tau|\ge1-\frac{c_*}{E}\ge1-\frac{c_+}{E_-}.
\]
Lemma~\ref{unbal:lem:endpoint-loggain} therefore gives
\begin{equation}\label{unbal:eq:opp-forced-gain}
 \zeta\ge F(d_-,E_+)
   +\frac{d_-}{2L}
      \left[-\ln\left(1-\left(1-\frac{c_+}{E_-}\right)^2\right)\right].
\end{equation}
Opposite-side feasibility and $d>E$ guarantee the contact needed by
that lemma. Equality of the endpoint lower bound with the four-moment
infimum is not required. Comparing \eqref{unbal:eq:opp-forced-gain} with
\eqref{unbal:eq:opp-feasible-target} treats the full permitted entropy split,
including cases where $E$ exceeds a child's cap.

\subsection{Rectangle inequalities and entropy endpoints}

The remaining sufficient inequalities, called \emph{modules} in the
supplement, use the global scalar, contact-plane, and log-sum bounds proved
earlier. Their exact interval formulas and
mean-derivative calculations are reproduced in
Supplementary Section~\ref{unbal:app:modules}. We describe their common mathematical role
here to specify what the certificate asserts.

A symmetric contact $z\in(0,1/2)$ gives the global plane
\[
 \zeta\ge c_zd-2A_zE,
\qquad
 A_z=\frac{(1-2z)^2}{4z(1-z)K_z},\quad
 K_z=-\frac12\ln(z(1-z)),\quad
 c_z=J(z)+\frac{2A_zH(z)}{1-2z}.
\]
For fixed means and fixed contact, the function
\[
 G_z(E)=c_zd-2A_zE-P(H((a+b)/2)-E)+P(C_0-E)
\]
is concave in $E$, because $P''$ increases and $\Delta\ge0$.
A nonnegative lower bound at both entropy endpoints therefore proves
the whole entropy interval. A bound for the mean derivatives extends
these endpoint comparisons from the center of a mean rectangle to
its entire area; this is the endpoint-plane Taylor module.

The shifted log-sum module uses a fixed global affine minorant
\[
 \zeta\ge k+u_0a+v_0b-\alpha e-\delta f.
\]
The pointwise remainder proof supplies this plane at its specified
rational anchors. The inequality $P''\ge8L/3$ then gives
\begin{align*}
 \zeta-R_\psi\ge{}&k+u_0a+v_0b
       +\frac{\delta-\alpha}{2}\,[H(a)-H(b)]
       -(\alpha+\delta)E\\
 &{}-P(I)+P(s)-\frac{3(\alpha-\delta)^2}{16L}.
\end{align*}
For fixed means and fixed plane, this expression is again concave in
$E$. Its two entropy endpoints and rigorous mean-derivative bounds
therefore give another whole-rectangle proof. Both endpoints concern
one fixed lower-bound function; no interpolation between unrelated
candidate branches is made.

The direct and normalized $F$ and log-sum comparisons use global
scalar monotonicity and interval bounds. On these outer mean rectangles,
the mean separation is bounded below by $2/5$; no estimate restricted to
the central square is applied outside its domain.

\subsection{Finite partition and proof of the compact theorem}

The verification program constructs an exact binary subdivision of
\eqref{unbal:eq:opp-root}. Each accepted partition cell has one of the sufficient
proofs just described, or has an empty constrained intersection.
Directed exponentiation encloses the mean map; its exact dyadic
interval endpoints are passed to the rectangle inequalities.
Every entropy inverse or scalar contact used in a verification test
has a verified enclosing bracket. Floating-point searches only
propose contacts or anchors.

The resulting finite partition certifies every constrained point of
the root. The partition and arithmetic records are described in
Supplementary Section~\ref{supp:verification}.

\begin{proof}[Proof of Theorem~\ref{unbal:thm:outeropposite}]
Every tuple in the theorem maps into \eqref{unbal:eq:opp-root} by
\eqref{unbal:eq:opp-map}, and the entire feasible intersection is included.
The complete partition supplies a valid inequality for that tuple.
Direct $R_\psi$ comparisons transfer in the $\psi$-active parent
branch; a $\phi$-active parent is covered by the unrestricted
$R_\phi$ theorem. The low-entropy union and the direct hybrid
comparisons already prove $R_B$ on their stated intersections.
Thus every tuple in the theorem satisfies $\zeta\ge R_B$.
\end{proof}

\subsection{Exhaustion of all opposite-side tuples}

\begin{theorem}[All opposite-side means]\label{unbal:thm:opposite}
Every tuple with
\[
 0<a,b<1,\qquad
 (a-\tfrac12)(b-\tfrac12)\le0,\qquad
 0<e\le H(a),\quad 0<f\le H(b)
\]
satisfies $\zeta(a,b,e,f)\ge R_B(a,b,e,f)$.
\end{theorem}

\begin{proof}
Use label exchange and simultaneous complementation to impose
\eqref{unbal:eq:opp-orientation}.
If $a\ge1/10$, then $b\le1-a\le9/10$, so both means are central
and Theorem~\ref{unbal:thm:central} applies.

Suppose $a\le1/10$. If $a\le A_*$, the uniform opposite-side
strip, Theorem~\ref{unbal:ar:extreme_strip}, applies for every positive
feasible entropy pair. Its proof includes the opposite corner of
Supplementary Theorem~\ref{unbal:thm:opposite-corner}
and the certified one-dimensional comparison for the remaining strip.

It remains that $A_*\le a\le1/10$. If $E\le1/25$, apply
\eqref{unbal:eq:opp-lowentropy-owner}. Otherwise
$C_0\ge E>1/25>\epsilon_*$, and
Theorem~\ref{unbal:thm:outeropposite} applies. These cases are exhaustive;
all interfaces overlap at equality. In particular, neither a mean
boundary limit nor an entropy-cap face leaves an additional case.
\end{proof}

\section{Assembly of the global inequality}
\label{unbal:sec:assembly}

We now prove Theorem~\ref{unbal:thm:global}.

\begin{proof}[Proof of Theorem~\ref{unbal:thm:global}]
\emph{Proof outline.}
We assemble the result separately for same-side and opposite-side
means after canonical orientation. In each case, analytic boundary
regions are removed first, and the remaining tuples are placed in the
appropriate finite chart. Checking the coordinate ranges and the full
feasible entropy interval ensures that the certified covers leave no
additional case before symmetry returns the result to the original tuple.

Use the two symmetries to impose \eqref{unbal:eq:canonical}. First suppose
$b\le1/2$. If $a+b\le1/16$, apply Theorem~\ref{unbal:thm:smallmean}. If
$a/b\le2^{-32}$, apply Lemma~\ref{unbal:ss:tail}. Every other tuple
belongs to the exact finite chart
\[
 x=-\log_2(a/b)\in[0,32],\qquad b\in[1/32,1/2],\qquad
 t=E/C_0\in(0,1],\quad a=b2^{-x}.
\]
Indeed $b\ge(a+b)/2>1/32$. Proposition~\ref{unbal:ss:certificate} certifies the entire closed
root $[0,32]\times[1/32,1/2]\times[0,1]$, with the positive-entropy
interpretation at $t=0$. This proves
the hybrid inequality for every same-side tuple, including all cap interfaces and
arbitrarily small mean differences.

Suppose instead that $b\ge1/2$. If $a\ge1/10$, canonicalization gives
$b\le1-a\le9/10$, and Theorem~\ref{unbal:thm:central} applies. We may therefore
assume $a\le1/10$. Theorem~\ref{unbal:ar:extreme_strip} treats
$a\le2^{-28}$, Supplementary Theorem~\ref{unbal:thm:opposite-corner}
treats $a+1-b\le2^{-13}$, and Theorem~\ref{unbal:ar:global_lowentropy}
treats $E\le10^{-6}$.
For every tuple outside these regions, let
\[
 u=-\log_2a,\qquad v=-\log_2(1-b),\qquad
 \epsilon=10^{-6},\qquad t=\frac{E-\epsilon}{C_0-\epsilon}.
\]
The exact bounding root is
\[
 (u,v,t)\in[3,28]\times[1,28]\times[0,1],\qquad
 a=2^{-u},\quad b=1-2^{-v},\quad
 E=\epsilon+t(C_0-\epsilon).
\]
Here $a>2^{-28}$ and $a\le1/10$ imply $3<u<28$;
$a\le1-b\le1/2$ implies $1\le v<28$. Outside the corner,
at least one of $a,1-b$ exceeds $2^{-14}$, so
\[
 C_0\ge\tfrac12 H(2^{-14})\ge2^{-14}>\epsilon.
\]
Thus the last coordinate is well-defined and lies in $[0,1]$; no entropy
interval is lost. The complete outer opposite partition covers the feasible
intersection of this root, as proved in
Theorem~\ref{unbal:thm:outeropposite}.

These cases cover every canonically oriented tuple. In each case the
comparison at the parent described in Section~\ref{unbal:sec:inputs} gives
the hybrid inequality, and the two symmetries return the result to the
original tuple. This proves the global inequality.
\end{proof}

Theorem~\ref{unbal:thm:global} concerns tuples with positive entropies, and
this is all that the proof of Theorem~\ref{unbal:thm:ck} requires. That
proof regularizes the Boolean array, so that every tuple to which the
Bellman inequality is applied has positive entropy coordinates and
interior means, even when the original array has deterministic sections;
continuity then removes the regularization.

Together with the static induction and semigroup argument in
Section~\ref{unbal:sec:main}, this proves Theorem~\ref{unbal:thm:ck}.

\startmanuscriptpart{Computer-assisted verification}{}{}
\label{part:computation}\label{sec:computation}

\section{The role of the computational lemmas}
\label{sec:numerical-certificate-sufficiency}

The computational lemmas supply explicit finite-dimensional
inequalities used in the analytic proof. Their role is to establish
the global hybrid Bellman inequality
\[
  \zeta(M)\ge R_B(M)
\]
for every tuple $M=(a,b,e,f)$ with
$0<a,b<1$, $0<e\le H(a)$, and $0<f\le H(b)$.
The required inputs include the boundary and stationary-point
estimates for the $\phi$ candidate, the auxiliary scalar estimates
and remainder bounds, and the same-side and opposite-side
certificates. The final two covers therefore form part of the
computational argument, together with the computational lemmas
used in their supporting analytic reductions.

Theorem~\ref{bal:thm:global-unrestricted-target} supplies
$\zeta\ge R_\phi$ on the full positive-entropy feasible domain.
At a parent where $\phi(m,E)\ge\psi(m,E)$, the inequalities
$B(a,e)\ge\phi(a,e)$ and $B(b,f)\ge\phi(b,f)$ imply
\[
  R_B(M)\le R_\phi(M)\le\zeta(M).
\]
At a $\psi$-active parent, the regional arguments use either
a sufficient $R_\psi$ comparison or a direct hybrid comparison
retaining the stated child terms. Thus the required conclusion
throughout is $\zeta\ge R_B$; a global inequality
$\zeta\ge R_\psi$ is not an additional premise.

For each terminal box in the final covers, the verification
establishes that its relevant feasible intersection is empty,
that an already proved regional theorem applies, or that one
of the sufficient comparison inequalities holds.
The implications of these tests are proved in
Proposition~\ref{unbal:ss:box-tests} and
Section~\ref{unbal:sec:opposite-completion}.
Each conclusion holds for every feasible entropy split
$(e,f)$ represented by the box, including the individual
entropy-cap cases. A box assigned to an established regional
theorem uses that theorem together with verified domain
inclusion. Boxes touching zero entropy are interpreted on
their positive-entropy feasible intersections.

The exact partitions cover the remaining compact domains.
Together with the boundary estimates and the two symmetries,
they exhaust all positive-entropy feasible tuples, as shown in
Section~\ref{unbal:sec:assembly}, and hence establish
Theorem~\ref{unbal:thm:global}.
The static cube induction and entropy-flow argument in
Section~\ref{unbal:sec:main} then give
Theorem~\ref{unbal:thm:ck}.
Regularization keeps every application of the Bellman
inequality within its positive-entropy domain, and continuity
removes the regularization. This final passage requires
no further numerical certificate.

\section{Validated numerical protocol}
The computer-assisted propositions concern explicitly specified compact
parameter domains after analytic estimates have removed their singular
limits. A certificate partitions a rational root box by exact rational
bisection. For each terminal box, outward-rounded interval arithmetic
proves one of the sufficient inequalities stated in the corresponding
proposition, or proves that the box has empty intersection with the
feasible domain. Each inequality is verified on the entire feasible
intersection of that box, never at sample points.

Coverage and arithmetic are verified separately. Reconstructing the
binary subdivision tree verifies coverage: every internal node has the
two children of one common split, terminal boxes have no descendants,
and the root and interfaces are included. Validated evaluation then
checks the analytic condition attached to every terminal box. When a
coordinate map is transcendental, its outward enclosure is used before
testing the transformed domain. The limits that a compact box excludes are
covered by the analytic boundary estimates, on closed domains that overlap
the box.

Entropy inverses and contact roots are enclosed by directed sign brackets.
Floating-point approximations may propose brackets or subdivision choices;
they never certify a sign. Series evaluations include rigorous remainder
bounds. The balanced implementations use Arb ball arithmetic
\cite{Johansson2017}, direct MPFR interval calculations, and
\texttt{mpmath.iv}, as identified separately in
Supplementary Section~\ref{bal:imp:trust}.  The extension also uses
\texttt{mpmath.iv} and exact rational geometry. The detailed precision,
implementation, and checking scope of each component are specified in
the supplement.
The inference relies on the correctness of those interval enclosures and
the certifying programs.

\section{Results that depend on computation}
The table lists each result that depends on a computation, with its
domain, the quantity that is verified, and the location of the detailed
record. For the same-side and opposite-side boxes, the inequalities are
asserted only on the intersection of the box with the feasible domain
defined in the corresponding proof.
\begingroup\small
\setlength{\tabcolsep}{4pt}
\begin{longtable}{@{}>{\raggedright\arraybackslash}p{0.22\textwidth}>{\raggedright\arraybackslash}p{0.28\textwidth}>{\raggedright\arraybackslash}p{0.24\textwidth}>{\raggedright\arraybackslash}p{0.19\textwidth}@{}}
\toprule
Result & Domain & Certified quantity & Detailed record\\
\midrule\endhead
Small-boundary theorem, \ref{bal:thm:small-boundary}
& The two domains in the theorem
& Scalar estimates implying $\zeta\ge R_\phi$
& \ref{supp:small-boundary-proof}\\[3pt]
Seam and endpoint theorem, \ref{bal:hyp:SCplus}
& Fixed-sum seam and deterministic-cap domains in its five clauses
& Minimum exclusion and $L_4-R_\phi\ge0$
& \ref{bal:sec:verification}\\[3pt]
Stationary Case E, Theorem~\ref{bal:thm:boundary-estimates}(h)
& $0<a<b<c<1/2$, $a+c\ge10^{-4}$, $\partial_cG_{\rm cap}=0$
& Pure gap $L_4-R_\phi\ge0$
& \ref{bal:sec:casee-stationary}\\[3pt]
Radial stationary estimate, Theorem~\ref{bal:thm:boundary-estimates}(i)
& $0<b<c<a<1/2$, $a+c\ge10^{-4}$, $\partial_cG_{\rm cap}=0$
& Pure gap $L_4-R_\phi\ge0$
& \ref{bal:sec:radial-stationary}\\[3pt]
Mean-interior reduction, \ref{bal:thm:e8-main}
& Positive strictly submaximal entropies
& Scalar derivative inequality and analytic endpoint bounds
& \ref{bal:imp:e8}, \ref{bal:sec:verification}\\[3pt]
Same-side certificate, \ref{unbal:ss:certificate}
& $[0,32]\times[1/32,1/2]\times[0,1]$ in $(x,b,t)$
& Sufficient inequalities in \ref{unbal:ss:box-tests}
& \ref{supp:same-side}\\[3pt]
Outer opposite-side theorem, \ref{unbal:thm:outeropposite}
& $[3,28]\times[1,28]\times[0,1]$ in $(u,v,t)$
& Hybrid or parent comparison on each feasible cell
& \ref{supp:verification}\\[3pt]
Mixed-derivative bound, \ref{unbal:lem:global-mixed}
& $v\in[1/22,1/3]$, with analytic tails
& $zF_{zz}\le13/6$
& \ref{supp:uniform-estimates}\\[3pt]
Scalar reductions, Section~\ref{unbal:ar:section}
& The compact roots and tails stated in that section
& Small-mean, parent, split, and extreme-mean inequalities
& \ref{supp:verification}, \ref{supp:release-notes}\\[3pt]
Central square, \ref{unbal:thm:central}
& $a,b\in[1/10,9/10]$, all positive feasible entropies
& Same-side and opposite-side component inequalities
& \ref{unbal:app:modules}\\
\bottomrule
\end{longtable}
\endgroup
The identifiers of the certificates, the partition sizes, the file names
with their digests, and the commands are given in the supplementary
sections on verification and on the released files.

\section{Reproducibility and supplementary materials}
The supplement, \emph{Supplementary Proofs and Verification Records},
which follows the references, provides the technical proofs and certificate specifications, together
with the archive inventory and reproduction commands in
Supplementary Sections~\ref{bal:sec:records}, \ref{supp:verification}, and
\ref{unbal:app:inventory}. Those records distinguish the exact checks of
the partitions, the arithmetic verification runs, and the integrity checks
of the source files, and they state the scope of the computations on which
each result depends. Supplementary Section~\ref{bal:sec:verification}
reports, for each computation, the original run, the check of the
partition structure, and the extent of the independent arithmetic rerun.
\startmanuscriptpart{Human--AI collaboration}{}{}

This work was developed through an extensive collaboration among the authors and AI systems. It involved a group of researchers from Google who led the AI interaction, and a research team from the Chinese University of Hong Kong (CUHK) who provided human insights. Both human and AI contributions were critical to this work.

The CUHK team had previously developed a framework~\cite{ChenGohariNair2025} that reduces the problem to a finite-dimensional optimization problem, contingent upon the correct formulation of a Bellman candidate function.

On the Google side, researchers were already investigating the problem~\cite{javanmard2026progresscourtadekumarconjectureoptimal}, which was also highlighted in an AI acceleration paper~\cite{acceleration-paper} as an open problem the team was attempting to solve using AI. As part of this effort, the Google team used Gemini within the Stellar Colosseum framework~\cite{stellar-colloseum-paper}, where it identified counterexamples near the boundary of the four-dimensional parameter domain, challenging the generalized statement formulated as Conjecture 5 in Appendix B of~\cite{ChenGohariNair2025}.

The Google team reached out to the CUHK team with these counterexamples. To address them, the CUHK team developed new lower bounds, suggesting the potential feasibility of the approach. Notably, these counterexamples did not refute the principal conjecture (Conjecture 3 of~\cite{ChenGohariNair2025}); instead, they revealed a limitation in the proposed generalization and demonstrated the necessity of developing stronger lower bounds. This interaction sparked ongoing exchanges and a formal collaboration between the two teams.

The CUHK team supplied the max-phi Bellman function for the full unbalanced case, reducing the conjecture to a well-defined 14-variable optimization problem, thereby providing an overall proof strategy. The AI then led the effort to partition the space into different regions, developing novel, non-trivial theoretical ideas and numerical interval arithmetic calculations to solve the problem.

The overwhelming majority of the novel ideas and results in this paper were generated by AI. Throughout the process, the CUHK team provided crucial insights and feedback to the AI model, guiding its reasoning. In particular, one principal lower bound was directly provided by the CUHK team.

In certain phases of the project, effective AI-assisted exploration required human direction, such as specifying relevant parameterizations or identifying the estimates needed to close remaining regions. In other phases---such as transitioning from the balanced to the unbalanced case (Part~\ref{part:unbalanced})---the model operated entirely without human input.

A central part of this collaboration took place within the Stellar Colosseum framework~\cite{stellar-colloseum-paper}, which provided a multi-agent environment for sustained mathematical exploration. The AI proved particularly effective at exploring a vast space of possible inequalities and parameter regimes, identifying difficult boundary configurations, and producing or improving candidate lower bounds proposed by human collaborators. Additional ideas, verification, and critiques were contributed by human collaborators working with Anthropic and OpenAI models.

Stellar Colosseum is available externally as the Long Proof pattern in Antigravity's Teamwork framework~\cite{AGY26teamwork}. While some collaborative features currently remain internal, the Google team plans to integrate more of these capabilities into Antigravity, with the goal of establishing it as a robust collaboration platform for external mathematicians.

\vspace{-5pt}
\paragraph{Lean formalization.}
The formalization was developed using a workflow
involving a variety of AI models. The computational
tasks were run on Google's distributed computing infrastructure.
These systems assisted with reasoning, Lean formalization,
generator development, orchestration, debugging, and verification
audits. The Lean formalization is available
\href{https://github.com/dpwoodru/general-courtade-kumar-lean}{online}.
The complete formal proof was checked end-to-end by Lean's kernel.

\phantomsection

\startsupplement{Supplementary Proofs and Verification Records}
\section*{Scope and reading guide}
This supplement contains the full technical proofs relocated from the
main text, the component verification specifications, and the repository
release documentation. Its sections are numbered S.1, S.2, and so on.
Ordinary equations are numbered (S.1), (S.2), and so on; named conditions
use their supplementary section number followed by an identifier, such as
(S.11.R1). A number without the S prefix refers to the main text. Background references are collected in the bibliography,
which precedes the supplement. Definitions are those of the main text
unless a section explicitly introduces local notation or natural-logarithm
units.

Section~\ref{bal:sec:verification} distinguishes the original numerical
executions, structural audits, and full or partial arithmetic replays.
The final source-locator section identifies the associated computational
records. The conditional support-classification discussion is separate
from the proof of the main theorem.
For a first reading, the following dependency guide complements the
table of contents at the beginning of the paper.
\begin{itemize}
\item The direct small-boundary argument is in
Section~\ref{supp:small-boundary-proof}. The same-side and uniform estimates
used by the unbalanced extension follow in
Sections~\ref{supp:same-side} and~\ref{supp:uniform-estimates}.
\item The seam and deterministic-cap arguments run from
Section~\ref{bal:sec:seam-normalization} through
Section~\ref{bal:sec:cap-sharp}. Their flow is boundary control, exclusion
of strict seam minima, and completion of the two cap clauses.
\item The reflection, entropy-convexity, and inverse-profile inputs are
proved in Sections~\ref{bal:imp:components} and~\ref{bal:aux:sec:components}.
They are followed by the stationary Case-E and radial cap proofs in
Sections~\ref{bal:sec:casee-stationary} and~\ref{bal:sec:radial-stationary},
which establish Theorem~\ref{bal:thm:boundary-estimates}(h) and (i).
The computational records follow these proofs in
Section~\ref{bal:sec:verification}.
\item Section~\ref{unbal:app:modules} contains the detailed regional
arguments for the unbalanced extension. Its component results are assembled
in Section~\ref{unbal:app:central_assembly}; the main text supplies the
final global assembly.
\end{itemize}
The certificate and release sections document the evidence for the finite
steps. The conditional support classification in
Section~\ref{supp:structural} can be read independently of these proof chains.

\section{Proof of the small-boundary Bellman inequality}
\label{supp:small-boundary-proof}

\begin{proof}[Proof of Theorem~\ref{bal:thm:small-boundary}]
\emph{Proof outline.}
We work in natural units and treat the two mean orientations separately.
For same-side means, the mean edge cost dominates four times the entropy
Jensen gap; convexity of a corrected Bellman candidate gives the matching
upper bound. For opposite-side means, we construct an endpoint contact,
write its excess over the target in one displacement variable, and control
that excess using its value, slope, and curvature. In both branches the
scalar certificates are stated before they are used, and the final
continuity and symmetry arguments supply the boundary orientations.

We include the proof because this theorem is a target-level owner and cannot
be replaced by a citation to a pure-gap statement.  In this proof only, all
entropies and costs are written with natural logarithms.  Thus
\[
 h(t)=(\ln2)\HH(t),\qquad J_e(t)=(\ln2)J(t),\qquad
 j_e(x,y)=\frac12(x-y)\{J_e(y)-J_e(x)\}.
\]
At endpoints, \(j_e\) denotes the lower-semicontinuous extended Jeffreys
divergence
\(\tfrac12\{D_e(x\Vert y)+D_e(y\Vert x)\}\), where
\[
 D_e(x\Vert y)
 =x\ln\frac{x}{y}+(1-x)\ln\frac{1-x}{1-y}
\]
is binary relative entropy with natural logarithms and its standard
lower-semicontinuous extended-value conventions.
Put
\[
 L_e(t)=\frac{2h(t)}{1-2t},\quad
 \eta_e(e)=\{1-2h^{-1}(e)\}J_e(h^{-1}(e)),\quad
 F_e(s,e)=sJ_e\!\left(L_e^{-1}\!\left(\frac{2e}{s}\right)\right)
 \quad(s>0),
\]
We use its even continuous extension in the first coordinate,
\(F_e(-s,e)=F_e(s,e)\), with \(F_e(0,e)=0\).  Both inverses are the lower
branches.  Set
\[
 \phi_e(m,e)=\eta_e(e)-F_e(|1-2m|,e).
\]
If \(M_e=(\mu_u,\mu_w,(\ln2)e_u,(\ln2)e_w)\), then, letting
\(\zeta_e\) denote the same four-moment infimum as \(\zeta\), with
\(j,\HH\) replaced by \(j_e,h\), one has
\[
 \zeta_e(M_e)=(\ln2)\zeta(M),\qquad
 R_{\phi,e}(M_e)=(\ln2)\Rphi(M).
\]
After this scaling we relabel \(M_e\) as \(M\), so explicitly
\[
 R_{\phi,e}(M)=
 \phi_e\!\left(\frac{\mu_u+\mu_w}{2},\frac{e_u+e_w}{2}\right)
 -\frac12\phi_e(\mu_u,e_u)-\frac12\phi_e(\mu_w,e_w).
\]
Multiplication by the positive constant \(1/\ln2\) returns the normalization
used in the rest of the manuscript and does not change any inequality.

\smallskip
\noindent\emph{I.  The same-side branch.}
Write
\[
 h(t)=-t\ln t-(1-t)\ln(1-t),\qquad
 J_e(t)=\ln\frac{1-t}{t}.
\]
Joint convexity of relative entropy gives, for every feasible representing law,
\begin{equation}
 \zeta_e(M)\ge j_e(\mu_u,\mu_w).
 \label{bal:eq:sb-divergence}
\end{equation}
Indeed, apply Jensen separately to
\(D(U\Vert W)\) and \(D(W\Vert U)\), then average.

\smallskip\noindent\emph{I.1. Comparing the mean cost with the entropy gap.}
For later use we prove the sharp two-point comparison
\begin{equation}
 j_e(a,b)\ge4\left\{
 h\!\left(\frac{a+b}{2}\right)-\frac{h(a)+h(b)}2\right\}.
 \label{bal:eq:sb-edge-jensen}
\end{equation}
Assume \(a\le b\), put \(c=(a+b)/2\), \(d=(b-a)/2\), and
\[
 f(x)=-h''(x)=\frac1{x(1-x)},\qquad
 S(t)=f(c-t)+f(c+t).
\]
Since \(f''(x)=2x^{-3}+2(1-x)^{-3}>0\), the function \(S\) is
increasing on \([0,d]\).  Two integrations give
\[
 j_e(a,b)=d\int_0^dS(t)\,dt,\qquad
 h(c)-\frac{h(a)+h(b)}2
   =\frac12\int_0^d(d-t)S(t)\,dt.
\]
Consequently the difference between the two sides of
\eqref{bal:eq:sb-edge-jensen} is
\[
 \int_{d/2}^d(2t-d)\{S(t)-S(d-t)\}\,dt\ge0.
\]

\smallskip\noindent\emph{I.2. Convexity of the corrected candidate.}
The desired upper bound follows from midpoint convexity after subtracting
the entropy correction. We compute the Hessian, isolate the scalar
inequalities it needs, and then return to the convexity conclusion.

It remains to upper-bound the Bellman right side by the entropy gap in
\eqref{bal:eq:sb-edge-jensen}.  The set
\[
 \mathcal D=\{(m,e):0<m\le10^{-2},\ 0<e<h(m)\}
\]
is convex because \(h\) is concave; hence the segment joining any two
points of \(\mathcal D\) remains in \(\mathcal D\).  On this set put
\(s=1-2m\), write \(e=h(q)\), and define \(v\) by
\[
 L_e(v)=\frac{2e}{s},\qquad L_e(t)=\frac{2h(t)}{1-2t}.
\]
For \(0<t<1/2\), set
\[
 \tau_t=1-2t,\quad \nu_t=t(1-t),\quad r_t=-\ln\nu_t,
\]
\begin{align*}
 A(t)&=\frac{h(t)^2\tau_t\{r_t-\tau_t^2\}}{\nu_t^2r_t^3},
 &K(t)&=h(t)^2\eta_e''(h(t)),\\
 E_L(t)&=\frac{tL_e'(t)}{L_e(t)},
 &R(t)&=\frac{A(t)-K(t)}{2t}.
\end{align*}
Exact differentiation of the perspective gives
\[
 (F_e)_{ss}=\frac{A(v)}s,\qquad
 (F_e)_{se}=-\frac{A(v)}e,\qquad
 (F_e)_{ee}=\frac{sA(v)}{e^2}.
\]
With
\[
 T(q,v;s)=\frac{K(q)-sA(v)}{1-s},\qquad
 \Gamma_*(m,q)=\frac{1+s}{4s}\frac{A(v)K(q)}{T(q,v;s)},
\]
the Hessian of \(\psi(m,e):=\phi_e(m,e)-4h(m)\) has
\begin{equation}
 \psi_{ee}=\frac{(1-s)T}{e^2},\qquad
 \psi_{mm}-\frac{\psi_{me}^2}{\psi_{ee}}
   =\frac4{m(1-m)}\{1-\Gamma_*(m,q)\}.
 \label{bal:eq:sb-schur}
\end{equation}

Here is the complete certified scalar input used in
\eqref{bal:eq:sb-schur}:
\begin{equation}
 E_L(t)\ge\frac45,\quad
 1\le A(t)\le\frac{117}{100},\quad
 0<K(t)\le\frac{117}{100},\quad
 \left|\frac{dA}{d(-\ln t)}\right|\le\frac1{20},\quad
 R(t)\le\frac15
 \label{bal:eq:sb-scalar}
\end{equation}
for \(0<t\le0.0103\).  This is not a sampled assertion.  Since
\(4575/1000<-\ln(0.0103)\), the finite range contains every
\(0.0103\ge t\ge e^{-50}\).  For
\(x=-\ln t\in[4575/1000,50]\), the interval is the exact adjacent tiling
by the \(45{,}425\) rational cells
\([i/1000,(i+1)/1000]\), \(4575\le i<50000\).  Centered Arb
mean-value forms prove every inequality in \eqref{bal:eq:sb-scalar} on every
cell.  The identical cells were run independently at 384 and 512 bits, and
the verifier checks exact tiling, directed containment, the
\(317{,}975\) stored comparisons, and exact binary endpoint serialization.
The uncompressed ledger hashes are
\[
\begin{array}{ll}
384\text{ bits}:&
\text{\hashcode{eba76590c2b21f68e2d34656fe7e01e94df9f47f5a87ef9d9ba3f1f51651af4b}},\\
512\text{ bits}:&
\text{\hashcode{ed9c45498f3cc6d41f75628196ec4f42b8595c80154e4ae68dd409400c35ff58}}.
\end{array}
\]
For \(x\ge50\), put \(t=e^{-x}\), \(z=1/x\), and
\(b=-\ln(1-t)/t\).  The exact rescaling
\[
 A=\frac{\{1+(1-t)bz\}^2(1-2t)
          \{1+tbz-(1-2t)^2z\}}
         {(1-t)^2(1+tbz)^3}
\]
together with \(xt\le50e^{-50}<10^{-20}\) gives
\[
 A\ge1+z\left(\frac{2449}{2500}
                -\frac{153}{50}10^{-20}\right)>1.
\]
Here is the algebraic identity used to remove the division by $t$ in $R$.
Write $\ell=\ln((1-t)/t)=x-tb$, $\tau=1-2t$, and $C=1-2t+2t^2$.
Differentiating $\eta_e(h(t))=\tau\ell$ with $h'(t)=\ell$ gives
\[
 K(t)=\frac{h(t)^2(C\ell-\tau)}{\nu_t^2\ell^3}.
\]
Set $U=1-tbz$ and $V=1+tbz$.  Since $r_t=x+tb$, expansion of $A-K$ gives
\begin{align*}
 R(t)&=\frac{\{1+(1-t)bz\}^2\,\mathcal P(t,b,z)}
              {(1-t)^2U^3V^3},\\
 \mathcal P(t,b,z)
 &=-tU^4+2z\{\tau(1-t)-b(1-2t+3t^2)\}U^3\\
 &\quad+3bz^2(\tau-2Ctb)U^2
       +2tb^2z^3(3\tau-2Ctb)U+4\tau t^2b^3z^4.
\end{align*}
In particular, the cancellation of the original factor $2t$ is exact,
before any interval evaluation.  The stated tail estimate for this expression is
\[
 0<R<\frac{231200}{128119761}<\frac1{500}.
\]
The remaining four bounds in \eqref{bal:eq:sb-scalar} are outward-certified on
the rational compact superset
\[
0\le z\le1/50,\quad0\le t\le2\cdot10^{-22},\quad
1\le b\le1+2\cdot10^{-22}.
\]
Thus the finite cells and this hybrid analytic/certified tail prove
\eqref{bal:eq:sb-scalar} on the full interval.

\smallskip\noindent\emph{I.3. Assembling the Hessian bounds.}
We now spell out how the scalar facts imply convexity.  For
\(0<q\le m\le1/100\), let
\[
 \rho=\frac{1-2q}{1-2m}=\frac{L_e(v)}{L_e(q)}\in[1,50/49].
\]
First, $L_e$ is strictly increasing on $(0,1/2)$.  If $v>103/10000$,
the elasticity bound on its stated interval would give
\[
 \ln\rho>\int_q^{103/10000}E_L(t)\,d\ln t
 \ge\frac45\ln\frac{103/10000}{q}
 \ge\frac45\ln\frac{103}{100}>\ln\frac{50}{49},
\]
contrary to the bound on $\rho$.  The last strict comparison is the exact
integer inequality $103^4 49^5>100^4 50^5$.  Thus $v\le0.0103$, and the
elasticity estimate is valid on all of $[q,v]$.  It now gives
\[
 \ln\rho=\int_q^v E_L(t)\,d\ln t\ge\frac45\ln(v/q),
\]
so \(v\le10^{-2}(50/49)^{5/4}<0.0103\).  If \(\rho=1\), then
\(q=m=v\), and the conclusion below follows by continuity.  Assume
\(\rho>1\).  Moreover
\[
 T=A(v)-\frac{A(v)-K(q)}{1-s}.
\]
Since \(1-s\ge(\rho-1)/\rho\),
\[
 \frac{|A(v)-A(q)|}{1-s}
 \le\frac{(1/20)\rho\ln\rho}{(4/5)(\rho-1)},\qquad
 \frac{[A(q)-K(q)]_+}{1-s}
 =\frac{2q[R(q)]_+}{2m}\le\frac15.
\]
The function \(\rho\ln\rho/(\rho-1)\) is increasing, so
\begin{align*}
 T&\ge1-
 \frac{(1/20)(50/49)\ln(50/49)}{(4/5)((50/49)-1)}-\frac15\\
 &>0.7368665396>\frac7{10}.
\end{align*}
Since \(s\ge49/50\),
\[
 \Gamma_*(m,q)
 \le\frac{99}{196}\frac{(117/100)^2}{0.7368665396}
 <0.938344<1.
\]
Both decimal comparisons are outward Arb rational comparisons stored in
both precision ledgers.  Equation \eqref{bal:eq:sb-schur} and the principal
minor criterion now show that \(D^2\psi\succeq0\) on \(\mathcal D\).
Continuity supplies both \(m=1/100\) and the deterministic entropy cap
\(e=h(m)\).

Midpoint convexity of \(\psi\) therefore yields
\[
 R_{\phi,e}(M)\le4\left\{
 h\!\left(\frac{\mu_u+\mu_w}{2}\right)
 -\frac{h(\mu_u)+h(\mu_w)}2\right\}.
\]
Combining this with \eqref{bal:eq:sb-divergence} and
\eqref{bal:eq:sb-edge-jensen} proves the same-side assertion.

\smallskip
\noindent\emph{II.  The opposite-side branch.}
Put
\[
 a=\mu_u,\qquad b=1-\mu_w,\qquad T_0=10^{-4},\qquad
 T=a+b\le T_0,\qquad D=1-T.
\]
For \(p\in[D,1]\), define
\[
 u(p)=h^{-1}(e_u/p),\qquad v(p)=h^{-1}(e_w/p),\qquad
 f(p)=p\{1-u(p)-v(p)\}.
\]
Here \(h^{-1}:[0,\ln2]\to[0,1/2]\) is the lower inverse.  Feasibility and
monotonicity of \(h\) give \(e_u\le h(a)\le h(T)\) and
\(e_w\le h(b)\le h(T)\).  Since
\[
 h(T)\le T\ln(1/T)+T<\frac{11}{10000}
 <\frac{9999}{10000}\ln2\le D\ln2,
\]
we have \(e_i/p<\ln2\) for every \(p\in[D,1]\), so the inverses exist.
Implicit differentiation gives
\[
 f'(p)=1-u-v+\frac{h(u)}{J_e(u)}+\frac{h(v)}{J_e(v)}>0.
\]
Positive entropy gives \(u(D),v(D)>0\), hence \(f(D)<D\).
Feasibility gives \(u(1)\le a\), \(v(1)\le b\), hence \(f(1)\ge D\).
Therefore there is a unique \(p\in[D,1]\) satisfying
\begin{equation}
 D=p(1-u-v),\qquad e_u=ph(u),\qquad e_w=ph(v).
 \label{bal:eq:sb-envelope-system}
\end{equation}

Proposition~\ref{bal:prop:fixed-difference-minimum} applies to the triple
just constructed.  Since fixing both means is more restrictive than fixing
only their difference, it gives the fundamental opposite-side lower bound
\begin{equation}
 \zeta_e(M)\ge \mathfrak Z_e(D;e_u,e_w)
 =p\,j_e(u,1-v)=:L_{\rm end}(M).
 \label{bal:eq:sb-endpoint-lower}
\end{equation}

\smallskip\noindent\emph{II.1. A one-variable form of the endpoint excess.}
The constructed contact fixes the entropy data. We now isolate the
remaining displacement of the prescribed means, so the comparison can
be proved by a value, slope, and curvature estimate.

It remains to compare \(L_{\rm end}\) with \(R_{\phi,e}\).  Put
\[
 q=1-p,\qquad Y=a-b-p(u-v),\qquad
 \bar e=\frac{h(u)+h(v)}2,\qquad D_Y=Y+p(u-v).
\]
Direct substitution and one-homogeneity of \(F_e\) give the exact normal
form
\begin{align}
 G_*(Y)&:=p\,j_e(u,1-v)-R_{\phi,e}(M)\notag\\
 &=C+F_e(|D_Y|,p\bar e)
 -\frac12F_e(p(1-2u)-Y,ph(u))
 -\frac12F_e(p(1-2v)+Y,ph(v)),
 \label{bal:eq:sb-opposite-normal}
\end{align}
where
\[
 C=pj_e(u,1-v)-\eta_e(p\bar e)
   +\frac12\eta_e(ph(u))+\frac12\eta_e(ph(v))
\]
is independent of \(Y\).  To derive the feasible interval, let
\(r_0=h^{-1}(e_u)\) and \(s_0=h^{-1}(e_w)\).  Concavity gives
\(r_0\le pu\), \(s_0\le pv\), while feasibility gives
\(r_0\le a\), \(s_0\le b\).  Substitution in the definition of \(Y\)
therefore yields
\[
 -q-2(pu-r_0)\le Y\le q+2(pv-s_0),
\]
and hence
\[
 |Y|\le2T,\qquad |D_Y|=|a-b|\le T,\qquad
 p(1-2u)-Y,\ p(1-2v)+Y\ge1-2T.
\]

For \(0<z<1/2\), set
\[
 \mathcal A(z)=
 \frac{h(z)^2(1-2z)\{-\ln(z(1-z))-(1-2z)^2\}}
 {z^2(1-z)^2\{-\ln(z(1-z))\}^3}.
\]
The exact perspective formula is
\((F_e)_{ss}(s,e)=\mathcal A(z)/s\), where
\(L_e(z)=2e/s\).  We also need the following analytic radial fact:
for fixed \(e>0\), \((F_e)_{ss}(s,e)>0\) and decreases strictly in
\(s\), with \((F_e)_{ss}(0,e)=4\ln2/e\).  To prove it, put
\[
 \xi=J_e(z)/2,\quad T_h=\tanh\xi,\quad S_h=\operatorname{sech}^2\xi,
 \quad \mathcal K=\ln(2\cosh\xi),\quad h_z=\mathcal K-\xi T_h,
\]
and \(P=2\mathcal K-T_h^2>0\).  The positivity follows because
\(\mathcal K\ge\ln2>1/2\) and \(T_h^2<1\).  Multiplying the perspective
formula by \(L_e(z)\) gives
\[
 (F_e)_{ss}(s,e)=\frac{B(\xi)}{2e},\qquad
 B(\xi)=\frac{4h_z^3\cosh^4\xi\,P}{\mathcal K^3}.
\]
Direct differentiation yields
\[
 \frac{B'}B=-\frac{3\xi S_h}{h_z}+4T_h
             +\frac{2T_h^3}{P}-\frac{3T_h}{\mathcal K}.
\]
For the first comparison put \(Q=3\xi S_h-4T_hh_z\).  Then
\(Q(0)=0\), \(Q(\xi)\to0\) as \(\xi\to\infty\), and
\[
 Q'=S_h\{3-4\mathcal K+2\xi T_h\}.
\]
The braced factor is strictly decreasing from \(3-4\ln2>0\) to
\(-\infty\), because its derivative is
\(2(\xi S_h-T_h)<0\); indeed
\((T_h-\xi S_h)'=2\xi T_hS_h>0\).  Thus \(Q\) first increases and
then decreases to zero, and \(Q>0\) for \(\xi>0\).
For the second comparison set \(y=T_h^2\) and
\(f(y)=2\mathcal K(3-y)-3y\).  At any interior critical point,
\(\mathcal K=y/(1-y)\), whence
\(f(y)=y(3+y)/(1-y)>0\); also \(f(0)=6\ln2>0\) and
\(f(y)\to\infty\) as \(y\uparrow1\).  Therefore
\(3\xi S_h/h_z>4T_h\) and
\(2T_h^3/P<3T_h/\mathcal K\), so \(B'<0\).  Since \(s\) increases
with \(\xi\) (indeed
\((2h_z/T_h)'=-2S_h(\xi T_h+h_z)/T_h^2<0\)), the radial
assertion follows, and \(B(0)=8\ln2\) gives the endpoint value.

The following closed scalar bounds are certified:
\begin{align*}
 &\mathcal A(z)<\frac32 &&(0<z<1/2),\\
 &\mathcal A(z)>\frac9{20} &&(0<z\le219/500),\\
 &0\le\ln(1/z)-(1-2z)J_e(z)<\frac1{500} &&(0<z\le1/9999),\\
 &|zE'(z)|<\frac14,\quad
 E(z)=\frac{h(z)(1-2z)}{z(1-z)\{-\ln(z(1-z))\}}
 &&(0<z\le1/9999),\\
 &\frac{h(z/3)}{h(z)}<\frac{19}{50} &&(0<z\le1/9999).
\end{align*}
The same directed point checks give
\[
 \begin{gathered}
 \ln\cosh(3/2)<1,\qquad \ln3<11/10,\qquad \ln9998>9,\\
 \frac{2h(1/19998)}{10^{-4}}<10.91<L_e(219/500),\qquad
 \frac{8\ln2}{\ln9999}<0.63,\qquad e^{-50}<10^{-20}.
 \end{gathered}
\]
The finite certificate has \(22{,}086\) exact records across five exact
tilings in \(x=-\ln z\): \(4{,}931\), \(4{,}918\), and three families
of \(4{,}079\) records.  They were replayed on identical tiles at 384 and
512 bits; the only nonrational tile endpoints are the three directed Arb
constants \(\ln2\), \(\ln(500/219)\), and \(\ln9999\).  The
verifier checks exact coverage and strict outward containment.  The
uncompressed hashes are
\[
\begin{array}{ll}
384\text{ bits}:&
\text{\hashcode{6961242d17bc37ef558b58c8067b06176b60bfdc19c89c58cddee6d93e4bea9d}},\\
512\text{ bits}:&
\text{\hashcode{b492ae80e2790fcac90093bcd264111bd5b2ba3f2e9b9248875fa7df8abcc84b}}.
\end{array}
\]
For \(x\ge50\), inserting
\(z=e^{-x}\), \(-\ln(1-z)\in[z,z/(1-z)]\), and
\(e^{-50}<10^{-20}\) gives directly
\[
 \frac{499}{500}\frac{x^2(x-1)}{(x+1/500)^3}
 <\mathcal A(z)<
 \frac{(x+1)^2(x+1/500)}{(999/1000)^2x^3},
\]
which lies in \((9/20,3/2)\).  Moreover,
\[
 \ln(1/z)-(1-2z)J_e(z)
 =2zx+(1-2z)\{-\ln(1-z)\}<z(2x+2)<1/500.
\]
For the slope bound, exact logarithmic differentiation gives
\[
 \frac{zE'(z)}{E(z)}
 =-\frac{-\ln(1-z)}{h(z)}-\frac{2z}{1-2z}
  +\frac{z}{1-z}+\frac{1-2z}{(1-z)\{-\ln(z(1-z))\}}.
\]
The elementary bounds above make the sum of the four absolute values less
than \(0.041\), while \(E(z)<1.021\), and hence
\(|zE'(z)|<1/4\).  Finally,
\[
 \frac{h(z/3)}{h(z)}
 \le\frac{x+\ln3+1+2z}{3x}<0.35<19/50.
\]
Thus the scalar statement covers the entire noncompact domain.

\smallskip\noindent\emph{II.2. The three quantitative estimates.}
The first estimate supplies a reserve at zero displacement, the second
bounds the possible linear loss, and the third gives a uniform quadratic
gain. Together they control the whole feasible displacement interval.

For completeness, we now derive the three estimates that close
\eqref{bal:eq:sb-opposite-normal}.  Since \(p\ge D\ge1-T_0\),
\(q=1-p\le T_0\), and \(T=q+p(u+v)\),
\[
 S=u+v\le\frac{T_0}{1-T_0}=\frac1{9999}.
\]
Assume \(u\ge v\), and put
\[
 S=u+v,\qquad
 X=\frac{u-v}{u+v}=\tanh\alpha,\qquad
 \mathcal K_\alpha=\ln\cosh\alpha.
\]
At \(Y=0\), the scalar bounds and the exact Jensen integral for \(\eta_e\)
give an explicit reserve.  Namely, with
\[
 \kappa_{ss}(u,v)=j_e(u,v)-F_e(|u-v|,\bar e)\le j_e(u,v),
\]
one-homogeneity gives
\[
 G_*(0)=\frac12\eta_e(ph(u))+\frac12\eta_e(ph(v))
       -\eta_e(p\bar e)-p\kappa_{ss}(u,v).
\]
The elementary inequalities
\[
 j_e(u,v)\le SX\alpha+S^2X^2\le4S\mathcal K_\alpha
\]
show \(p\kappa_{ss}\le4T_0\mathcal K_\alpha=0.0004\mathcal K_\alpha\).  If
\(0<\mathcal K_\alpha\le1\), then for every \(0<z\le1/9999\) the exact
formula satisfies
\[
 \mathcal P(z):=z^2J_e(z)^2\eta_e''(h(z))
 =\frac{1+(1-2z)^2-2(1-2z)/J_e(z)}{2(1-z)^2}>\frac45.
\]
Indeed, \(J_e(z)\ge\ln9998>9\), \(1-2z>999/1000\), and
\[
 \frac{1+(999/1000)^2-2/9}{2}>\frac45.
\]
This has the following uniform consequence.  If \(x\) lies between
\(ph(v)\) and \(ph(u)\), and \(w=h^{-1}(x)\), then \(w\le u\) and
\((zJ_e(z))'=J_e(z)-1/(1-z)>0\) on \(z\le1/9999\).  Hence
\(\eta_e''(x)>0.8/[u^2J_e(u)^2]\).  The symmetric Taylor integral and
\(h(u)-h(v)\ge(u-v)J_e(u)\) now give
\[
 \frac12\eta_e(ph(u))+\frac12\eta_e(ph(v))-\eta_e(p\bar e)
 \ge\frac{0.8}{8u^2J_e(u)^2}\{p(h(u)-h(v))\}^2
 \ge0.1p^2\left(\frac{u-v}{u}\right)^2
 \ge0.032(1-T_0)^2\mathcal K_\alpha.
\]
For the last inequality use
\((u-v)/u=2X/(1+X)\),
\(X=\tanh\alpha\ge\alpha/(1+\alpha)\),
\(\mathcal K_\alpha\le\alpha^2/2\).  Also
\(\mathcal K_\alpha\ge\alpha-\ln2\), so
\(\mathcal K_\alpha\le1\) implies \(\alpha\le1+\ln2<2\).  These bounds give
\[
 \frac{\{2X/(1+X)\}^2}{\mathcal K_\alpha}\ge\frac8{25}.
\]
If \(\mathcal K_\alpha\ge1\), the inequalities
\(h(u/3)/h(u)<19/50<p/2\) show that the inverse-entropy point for
\(p\{h(u)+h(v)\}/2\) is at least \(u/3\).  With
\(f_*(z)=\eta_e(h(z))=(1-2z)J_e(z)\), one has
\(f_*'(z)=-2J_e(z)-(1-2z)/\{z(1-z)\}<0\) and \(h'(z)=J_e(z)>0\).
Thus \(\eta_e\) is strictly decreasing on the lower branch, and the
logarithmic-defect bound gives
\[
 \frac12\eta_e(ph(u))+\frac12\eta_e(ph(v))-\eta_e(p\bar e)
 \ge\frac12f_*(u)+\frac12f_*(v)-f_*(u/3)
 \ge\alpha-\ln3-1/500>\alpha/4\ge \mathcal K_\alpha/4.
\]
Here \(\mathcal K_\alpha\ge1\) implies \(\alpha>3/2\) by
\(\ln\cosh(3/2)<1\); indeed
\(3\alpha/4-\ln3-1/500>9/8-11/10-1/500>0\).
Also \(\mathcal K_\alpha=\ln\cosh\alpha\le\alpha\).
After subtracting \(0.0004\mathcal K_\alpha\), both cases give
\begin{equation}
 G_*(0)\ge0.03\mathcal K_\alpha.
 \label{bal:eq:sb-reserve}
\end{equation}
When \(u=v\), one has \(\mathcal K_\alpha=G_*(0)=0\) exactly.

Differentiating \eqref{bal:eq:sb-opposite-normal} at zero yields
\[
 G_*'(0)=\dot F_e(u-v,\bar e)
 +\frac12(F_e)_s(1-2u,h(u))
 -\frac12(F_e)_s(1-2v,h(v)).
\]
Here \(\dot F_e\) is the odd signed radial derivative of the even
perspective in its first coordinate.
At a deterministic perspective boundary,
\((F_e)_s(1-2z,h(z))=J_e(z)+E(z)\).  Decreasing radial curvature and
\((F_e)_{ss}(0,e)=4\ln2/e\) give
\[
 0\le\dot F_e(u-v,\bar e)
 \le(u-v)(F_e)_{ss}(0,\bar e)
 \le\frac{8\ln2}{\ln(1/S)}X<0.63\alpha,
\]
where \(h(u)+h(v)\ge h(S)\ge S\ln(1/S)\).
Also
\[
 \frac12\ln\frac{1-v}{1-u}<0.001\alpha,
 \qquad \frac12|E(u)-E(v)|<0.25\alpha.
\]
Indeed,
\(\ln((1-v)/(1-u))\le(u-v)/(1-u)<0.002\alpha\), and integrating
\(|zE'(z)|<1/4\) in \(d\ln z\) gives
\(|E(u)-E(v)|<(1/4)\ln(u/v)=\alpha/2\).
Since
\[
 \frac12\{J_e(u)-J_e(v)\}
 =-\alpha-\frac12\ln\frac{1-v}{1-u},
\]
these estimates give
\begin{equation}
 |G_*'(0)|\le2\alpha.
 \label{bal:eq:sb-slope}
\end{equation}

Finally, throughout the feasible \(Y\)-interval,
\[
 G_*''(Y)=(F_e)_{ss}(|D_Y|,p\bar e)
 -\frac12(F_e)_{ss}(p(1-2u)-Y,ph(u))
 -\frac12(F_e)_{ss}(p(1-2v)+Y,ph(v)).
\]
Concavity gives \(p\bar e\le h(1/19998)\).  Moreover,
\[
 L_e'(z)=\frac{2\{(1-2z)J_e(z)+2h(z)\}}{(1-2z)^2}>0.
\]
The certified point check
\(2h(1/19998)/T_0<L_e(219/500)\) therefore puts the latent point \(z_0\) for
\((F_e)_{ss}(T_0,p\bar e)\) below \(219/500\).  Hence
\((F_e)_{ss}(T_0,p\bar e)>(9/20)/T_0=4500\); strict radial
decreasingness and \(|D_Y|\le T_0\) then give
\((F_e)_{ss}(|D_Y|,p\bar e)\ge(F_e)_{ss}(T_0,p\bar e)>4500\).
The upper bound for \(\mathcal A\) and the two outer radii
\(\ge1-2T_0\) give a total subtraction less than \(1.501\).  Hence
\begin{equation}
 G_*''(Y)>4400.
 \label{bal:eq:sb-curvature}
\end{equation}

\smallskip\noindent\emph{II.3. Combining the estimates and completing the boundary cases.}
Taylor's theorem, \eqref{bal:eq:sb-reserve}--\eqref{bal:eq:sb-curvature}, and
\(|Y|\le2T_0\) imply
\[
 G_*(Y)\ge G_*(0)-
 \min\left\{\frac{G_*'(0)^2}{8800},\,2T_0|G_*'(0)|\right\}.
\]
If \(0<\mathcal K_\alpha\le1\), then
\(\tanh t\ge t/(1+t)\ge t/(1+\alpha)\) for \(0\le t\le\alpha\).
Integration gives
\(\mathcal K_\alpha\ge\alpha^2/[2(1+\alpha)]\ge\alpha^2/6\), and hence
\[
 G_*(Y)\ge\left(0.03-\frac{24}{8800}\right)\mathcal K_\alpha>0.
\]
If \(\mathcal K_\alpha\ge1\), then
\(\mathcal K_\alpha=\ln\cosh\alpha\ge\alpha-\ln2\), so
\(2\alpha<4\mathcal K_\alpha\); hence \(|G_*'(0)|<4\mathcal K_\alpha\), giving
\(G_*(Y)\ge(0.03-0.0008)\mathcal K_\alpha>0\).  If \(\mathcal K_\alpha=0\), then
\(G_*(0)=G_*'(0)=0\) and strict curvature gives
\(G_*(Y)\ge2200Y^2\).  Therefore
\(L_{\rm end}\ge R_{\phi,e}\), and
\eqref{bal:eq:sb-endpoint-lower} proves the opposite-side assertion.

The deterministic entropy caps used above are included by continuity.
Finally, the full-tuple complement
\[
 (\mu_u,\mu_w,e_u,e_w)
 \longmapsto(1-\mu_w,1-\mu_u,e_w,e_u)
\]
is induced by \((U,W)\mapsto(1-W,1-U)\) and preserves unrestricted
feasibility, \(\zeta\), the edge cost, and \(\Rphi\).  Together with full
label exchange it supplies the upper orientations.  The zero-entropy
intersections are owned by the
zero-entropy boundary theorem in the main text (Theorem~\ref{bal:thm:zero-target}).  This completes the proof of Theorem~\ref{bal:thm:small-boundary}.

For auditability, the three incorporated proof sources have SHA-256 values
\[
\begin{array}{ll}
\text{same-side source:}&
\text{\hashcode{01bc3f782b29482bc7a52f31f4b609d53a85fd0fc6dae476a0902bdc7f5c1edf}},\\
\text{contact/quotient source:}&
\text{\hashcode{120832341d7881ca64eb472746ffeeb71a2f283efa7046b60210f71ac9cb7ac6}},\\
\text{opposite quantitative source:}&
\text{\hashcode{440c7f73bdb5703dc1704d70a5515a6bb93fb92323d417e6a14356aa4546b7f4}}.
\end{array}
\]
The containing repair archive has SHA-256
\hashcode{7f9fc19d2e6a3708b5c31896d982cec4a6fc16ac387cf5f20e355d4b54f5e8e7}.
The 384/512 summary hashes are, respectively,
\hashcode{6c90e47fdabb1d68ff0afef065b198b03de501c864b74bd4cc2a5d808624e15e},
\hashcode{cf2a9b10183bb3fc3b00b52d3b9003290b4915a9b7660ffc1abf766c2fdef879}
for the same-side calculation and
\hashcode{06a36934fd76d06e771beaf756a7f7207dac8b53df10d7ed2e2a4c774364386a},
\hashcode{3f7829569d16dd69f1c0774128051ee85854e16668465c6bcab6bef10614c020}
for the opposite-side calculation.  These hashes bind the point and tail
checks in addition to the raw record streams displayed above.
\end{proof}

\section{Same-side derivative estimates and finite certification}
\label{supp:same-side}

\emph{Argument guide.}
The argument eliminates the entropy split before introducing a finite
chart. Profile derivatives control the target, normalized mean identities
remove cancellation at the diagonal, and contact monotonicity supplies
parent comparisons. The ratio tail is treated separately; the remaining
compact root is closed by the proved box tests and its recorded replay.

Use the notation of Section~\ref{unbal:ss:section} of the main text:
$L=\ln2$, $\nu_z=z(1-z)$, $I_a=H(a)-e$, $I_b=H(b)-f$,
$\overline H=(H(a)+H(b))/2$, and $I=\Delta+s=H(m)-E$.
Then $s=(I_a+I_b)/2=\overline H-E$, and positive feasibility gives
$0\le I_a,I_b,s,I<1$. All logarithms without a base in this section are natural.
\subsection{Entropy-profile derivatives and removal of the entropy split}

\label{unbal:ss:profile-section}

\begin{lemma}[Profile derivatives]\label{unbal:ss:profile}
The function \(P\) is increasing and convex on \([0,1)\).
Its second derivative is increasing, and \(P'\) is convex.  More
explicitly, if \(I=C(r)\), \(A=\operatorname{atanh}r\), and
\(\nu=1-r^2\), then, for \(0<r<1\),
\begin{align}
 P'(I)&=2+\frac{2r}{\nu A},\label{unbal:ss:pprime}\\
 P''(I)&=
 2L\,\frac{(1+r^2)A-r}{\nu^2A^3},\label{unbal:ss:psecond}\\
 P'''(I)&=
 \frac{2L^2}{\nu^3A^5}
 \left(2r(r^2+3)A^2-(5r^2+3)A+3r\right)>0.
 \label{unbal:ss:pthird}
\end{align}
The limiting values at zero are
\(P(0)=0\), \(P'(0)=4\), and \(P''(0)=8L/3\).
\end{lemma}

\begin{proof}
The identities \(C'(r)=A/L\) and \(P(C(r))=2rA/L\) give
\eqref{unbal:ss:pprime}--\eqref{unbal:ss:pthird} by differentiation.
The numerator in \eqref{unbal:ss:psecond} is positive because \(A>r\).
For the sign in \eqref{unbal:ss:pthird}, use the absolutely convergent series
\[
 A^2=\sum_{n\geq1}c_n r^{2n},\qquad
 c_n=\frac1n\sum_{i=0}^{n-1}\frac1{2i+1}\geq\frac1n.
\]
The coefficients of \(r\) and \(r^3\) in its numerator vanish.
For \(n\geq2\), the coefficient of \(r^{2n+1}\) is
\[
 6c_n+2c_{n-1}-\frac5{2n-1}-\frac3{2n+1}
 \geq \frac6n-\frac8{2n-1}+2c_{n-1}>0.
\]
Indeed \(6/n-8/(2n-1)=(4n-6)/[n(2n-1)]>0\) for \(n\geq2\).
This proves \(P'''>0\).  The endpoint values follow by substituting
the series for \(A\) and \(C\).
\end{proof}

\begin{lemma}[A target independent of the entropy split]\label{unbal:ss:target}
For every positive feasible entropy pair,
\begin{equation}\label{unbal:ss:target-upper}
 R_\psi\leq D_\Delta(s):=P(s+\Delta)-P(s)
 \leq \frac{\Delta}{2}\bigl(P'(s)+P'(I)\bigr).
\end{equation}
Also, for fixed means, \(D_\Delta(s)\) is convex in \(s\).
\end{lemma}

\begin{proof}
Since \(\psi(a,e)=P(I_a)\) and \(\psi(b,f)=P(I_b)\), convexity of \(P\)
gives
\[
 \tfrac12\bigl(\psi(a,e)+\psi(b,f)\bigr)\geq P(s).
\]
The first inequality in \eqref{unbal:ss:target-upper} follows.  Integrating
the convex function \(P'\) over \([s,s+\Delta]\) gives the trapezoidal
upper bound in the second inequality.  Finally,
\[
 D_\Delta''(s)=P''(s+\Delta)-P''(s)\geq0
\]
by Lemma~\ref{unbal:ss:profile}.
\end{proof}

We will use the branch comparison repeatedly.  If \(\phi(m,E)\geq
\psi(m,E)\), then \(R_B\leq R_\phi\), because \(B\geq\phi\) at each
child.  If \(\psi(m,E)\geq\phi(m,E)\), the same argument gives
\(R_B\leq R_\psi\).  Thus the old-candidate input owns every tuple
with the first parent comparison, and it remains sufficient to prove
\(\zeta\geq R_\psi\) on any other accepted region.

\subsection{Mean identities and normalization at the diagonal}
\label{unbal:ss:means-section}

Suppose \(a<b\) and \(m\leq1/2\).  Define
\[
 \rho=\frac{b-a}{a+b},\qquad k=\frac{m}{1-m}.
\]
Then \(d=2m\rho\), \(0<\rho<1\), and \(0<k\leq1\).
In particular
\[
 a=m(1-\rho),\quad b=m(1+\rho),\quad
 1-a=(1-m)(1+k\rho),\quad
 1-b=(1-m)(1-k\rho).
\]
Substituting these four expressions into the entropy and cost formulas
gives the exact identities
\begin{align}
 \Delta&=mC(\rho)+(1-m)C(k\rho),\label{unbal:ss:delta-identity}\\
 j(a,b)&=mP(C(\rho))+(1-m)P(C(k\rho)).
 \label{unbal:ss:j-identity}
\end{align}
For example, grouping the entropy terms with total mass \(m\) gives
the first term of \eqref{unbal:ss:delta-identity}, and grouping the
complementary terms gives the second.  For the cost, the logarithmic
ratio is
\[
 J(a)-J(b)=\frac2L\bigl(\operatorname{atanh}\rho+
                           \operatorname{atanh}(k\rho)\bigr),
\]
which gives \eqref{unbal:ss:j-identity}.
Convexity of \(P\) therefore yields the deterministic-cap inequality
\begin{equation}\label{unbal:ss:cap}
 j(a,b)\geq P(\Delta).
\end{equation}

The identities also give the stable normalized formulas
\begin{align}
 \frac{\Delta}{d^2}
 &=\frac1{4m}\left(
       \frac{C(\rho)}{\rho^2}
       +k\frac{C(k\rho)}{(k\rho)^2}\right),\label{unbal:ss:delta-normalized}\\
 \frac{j(a,b)}{d^2}
 &=\frac1{2Lm}\left(
       \frac{\operatorname{atanh}\rho}{\rho}
       +k\frac{\operatorname{atanh}(k\rho)}{k\rho}\right).
 \label{unbal:ss:j-normalized}
\end{align}
Both quotients appearing on the right have removable values at zero.
Indeed
\begin{align}
 \frac{C(z)}{z^2}
 &=\frac1L\sum_{n\geq1}\frac{z^{2n-2}}{2n(2n-1)},
 &\left.\frac{C(z)}{z^2}\right|_{z=0}&=\frac1{2L},\label{unbal:ss:c-series}\\
 \frac{\operatorname{atanh}z}{z}
 &=\sum_{n\geq0}\frac{z^{2n}}{2n+1},
 &\left.\frac{\operatorname{atanh}z}{z}\right|_{z=0}&=1.
 \label{unbal:ss:a-series}
\end{align}
Their nonnegative coefficients prove that both quotients are increasing
on \([0,1)\).  These expansions remove the numerical cancellation
which would otherwise occur as \(a\) approaches \(b\).

\begin{lemma}[A positive correction to the mean cost]\label{unbal:ss:bonus}
For \(a<b\) and \(m\leq1/2\),
\begin{equation}\label{unbal:ss:bonus-bound}
 j(a,b)\geq
 \left(4+\frac23(1-k+k^2)\rho^2\right)\Delta
 \geq\left(4+\frac{\rho^2}{2}\right)\Delta .
\end{equation}
\end{lemma}

\begin{proof}
\emph{Proof outline.}
The improvement over the basic mean-cost bound comes from a positive
power-series comparison. After proving that scalar comparison, we insert
the two normalized mean identities and bound their relative weights.
This produces the uniform correction in the statement.

Put \(C_n(z)=LC(z)\).  The positive series imply
\begin{equation}\label{unbal:ss:scalar-bonus}
 2z\operatorname{atanh}z-4C_n(z)
 \geq\frac23z^2C_n(z).
\end{equation}
Here the \(z^4\) coefficients are equal.  For \(n\geq2\), the
difference between the coefficients of \(z^{2n}\) on the two sides is
\[
 \frac{2(n-1)}{n(2n-1)}
 -\frac1{3(n-1)(2n-3)}
 =
 \frac{(n-2)(12n^2-20n+9)}
 {3n(n-1)(2n-1)(2n-3)}\geq0.
\]
This proves \eqref{unbal:ss:scalar-bonus}.
The same series give \(C(k\rho)\leq k^2C(\rho)\).
Writing \(z=C(k\rho)/C(\rho)\), equations
\eqref{unbal:ss:delta-identity}--\eqref{unbal:ss:j-identity} and
\eqref{unbal:ss:scalar-bonus} consequently give
\[
 \frac{j(a,b)}{\Delta}
 \geq4+\frac23\rho^2\frac{1+kz}{1+z/k}.
\]
For \(0<k\leq1\), the last fraction is decreasing in \(z\).
Since \(z\leq k^2\), it is at least
\[
 \frac{1+k^3}{1+k}=1-k+k^2\geq\frac34.
\]
This proves both inequalities in \eqref{unbal:ss:bonus-bound}.
\end{proof}

\subsection{Contact derivatives and parent comparisons}
\label{unbal:ss:contact-section}

The scalar monotonicities needed by the checker have short direct
proofs.  Thus this part of the same-side argument does not need to
import them as additional numerical assumptions.

\begin{lemma}[Contact derivatives]\label{unbal:ss:contact}
For \(h>0\), \(F(z,h)\) is nondecreasing in \(z\), nonincreasing in \(h\),
and \(F(z,h)/z^2\) is nonincreasing for \(z>0\).
For \(0\leq q<1\), the function
\(\Phi(q,h)=\eta(h)-F(q,h)\) is nonnegative on its feasible interval
\[
 0<h\leq H((1-q)/2)
\]
and is decreasing in both \(q\) and \(h\), where comparisons are made
within the feasible domain.
\end{lemma}

\begin{proof}
\emph{Proof outline.}
Parameterizing the unique contact makes the perspective derivatives
explicit. An auxiliary scalar inequality then controls the normalized
ratio of the perspective to the squared radius. Finally, feasibility
places the contact in the correct order relative to the entropy inverse,
which gives the sign and monotonicity assertions for the candidate.

For \(0\leq r<1\), set
\[
 \mathcal H(r)=L H((1-r)/2),\quad
 A(r)=\operatorname{atanh}r,\quad
 \nu(r)=1-r^2,\quad
 {\cal K}(r)=\mathcal H(r)+rA(r)
           =L-\tfrac12\ln\nu(r).
\]
The contact equation for \(F(z,h)\) is
\[
 \frac zh=\frac{Lr}{\mathcal H(r)}.
\]
Its right side increases strictly from zero to infinity, since its
derivative is \(L{\cal K}/\mathcal H^2>0\).
The contact is therefore unique, and implicit differentiation gives
\begin{equation}\label{unbal:ss:f-derivatives}
 F_z(z,h)=\frac2L\left(A+\frac{r\mathcal H}{\nu{\cal K}}\right),
 \qquad
 F_h(z,h)=-\frac{2r^2}{\nu{\cal K}}.
\end{equation}
These establish the first two monotonicities.

We claim that \(\mathcal H\leq\nu{\cal K}\).  To see this, put
\(T=A-r{\cal K}\).  Then \(T'=1-{\cal K}\).
The function \({\cal K}\) increases from \(L<1\) to infinity, whereas
\(T(0)=0\) and \(T(1-)=0\).  The last limit follows from
\[
 A-{\cal K}=\ln(1+r)-L,\qquad (1-r){\cal K}\longrightarrow0.
\]
Thus \(T\) first increases and then decreases to zero, so \(T\geq0\).
The identity \(\nu{\cal K}-\mathcal H=rT\) proves the claim.
Using \(A\geq r\), equation \eqref{unbal:ss:f-derivatives} now gives
\[
 zF_z-2F
 =\frac{2z}{L}\left(\frac{r\mathcal H}{\nu{\cal K}}-A\right)
 \leq0,
\]
which proves the asserted monotonicity of \(F/z^2\).

For the assertions about \(\Phi\), let
\[
 p=H^{-1}(h),\qquad R=1-2p.
\]
Feasibility implies \(q\leq R\), and the contact for \(F(q,h)\)
therefore has \(r\leq R\).  Since \(F\) is increasing in its first
variable and \(F(R,h)=\eta(h)\), we have \(\Phi(q,h)\geq0\).
Its decrease in \(q\) is immediate.
For the entropy derivative, define
\[
 g(r)=\frac{r^2}{\nu(r){\cal K}(r)}.
\]
Then
\[
 g'(r)=
 \frac{r(2{\cal K}(r)-r^2)}
 {\nu(r)^2{\cal K}(r)^2}>0.
\]
The sign follows from \({\cal K}\geq L>1/2\).
Using \eqref{unbal:ss:pprime} and \eqref{unbal:ss:f-derivatives},
\[
 \Phi_h(q,h)
 =-2-\frac{2R}{\nu(R)A(R)}+2g(r)
 \leq-2-\frac{2R}{\nu(R)A(R)}
          +\frac{2R^2}{\nu(R){\cal K}(R)}<0.
\]
For \(R>0\) the last inequality follows from
\({\cal K}(R)>RA(R)\); the zero endpoints follow by continuity.
This proves the remaining monotonicity.
\end{proof}

\begin{lemma}[An elementary small-entropy parent bound]
\label{unbal:ss:parent-tail}
For \(0<m\leq1/2\) and \(0<E<2m\),
\begin{equation}\label{unbal:ss:elementary-parent}
 \phi(m,E)\geq(2m-E)\log_2\frac{2-E}{E},
 \qquad
 \psi(m,E)\leq P(H(m)).
\end{equation}
At $m=1/2$, the second inequality uses the extended upper bound
$P(1):=\lim_{I\uparrow1}P(I)=+\infty$; it involves no subtraction
of infinite values.
\end{lemma}

\begin{proof}
Let \(p=H^{-1}(E)\), \(R=1-2p\), and \(q=1-2m\).
The inequality \(E<2m\leq H(m)\) guarantees feasibility.
As in Lemma~\ref{unbal:ss:contact}, the contact defining \(F(q,E)\)
is at a point \(v\geq p\).  Hence \(F(q,E)=qJ(v)\leq qJ(p)\),
and
\[
 \phi(m,E)\geq(R-q)J(p).
\]
Concavity of binary entropy gives \(H(p)\geq2p\), so \(p\leq E/2\).
It follows that \(R-q\geq2m-E>0\) and
\(J(p)\geq\log_2((2-E)/E)\), proving the first assertion.
The second follows from \(\psi(m,E)=P(H(m)-E)\) and the
monotonicity of \(P\).
\end{proof}

\subsection{The full unbounded mean-ratio tail}\label{unbal:ss:tail-section}

\begin{lemma}[Mean-ratio tail]\label{unbal:ss:tail}
If \(0<a\leq b\leq1/2\) and \(a/b\leq2^{-32}\), then
\(\zeta(a,b,e,f)\geq R_\psi(a,b,e,f)\) for every positive feasible
entropy pair.
\end{lemma}

\begin{proof}
\emph{Proof outline.}
When one mean is much smaller than the other, the logarithm in the
deterministic mean cost gives a uniform lower bound proportional to their
distance. The entropy Jensen gap is at most that distance, and a single
profile-slope bound controls the target. Comparing these two coefficients
proves the entire unbounded ratio tail.

Convexity of \(C\), together with \(C(0)=0\) and \(C(1)=1\),
gives \(C(z)\leq z\) for \(0\leq z\leq1\).
Equation \eqref{unbal:ss:delta-identity} consequently implies
\(\Delta\leq d\).  Also
\[
 j(a,b)=\frac d2[J(a)-J(b)]
 \geq\frac d2\log_2\frac ba\geq16d.
\]
The mean satisfies
\[
 m\leq m_*:=\frac{1+2^{-32}}4.
\]
The fixed directed comparison in
\texttt{\detokenize{same_side/TAIL_CHECKER.py}}, recorded in
\texttt{\detokenize{TAIL_RESULT.json}}, proves
\begin{equation}\label{unbal:ss:tail-constant}
 P'(H(m_*))<11.559<16.
\end{equation}
Its enclosing interval at 70 decimal digits lies between
\(11.558211330087323\) and \(11.558211330087338\);
in particular the recorded lower bound on
\(16-P'(H(m_*))\) exceeds \(4.44\).
Since \(I\leq H(m)\), Lemma~\ref{unbal:ss:target} and monotonicity of \(P'\)
give
\[
 R_\psi\leq\Delta P'(I)
 \leq\Delta P'(H(m))
 <16\Delta\leq16d\leq j(a,b)\leq\zeta.
\]
This proves the whole tail, including arbitrarily small \(a/b\).
\end{proof}

\subsection{The finite root and its acceptance inequalities}
\label{unbal:ss:certificate-section}

After excluding the small-mean region of
Theorem~\ref{unbal:thm:smallmean} and the tail of Lemma~\ref{unbal:ss:tail},
every remaining ordered same-side tuple admits the coordinates
\begin{equation}\label{unbal:ss:coordinates}
 x=-\log_2(a/b),\qquad a=b\,2^{-x},\qquad
 E=t\overline H,\qquad s=(1-t)\overline H.
\end{equation}
They lie in the exact closed root
\begin{equation}\label{unbal:ss:root}
 {\cal Q}=[0,32]\times[1/32,1/2]\times[0,1]
 \quad\text{in the variables }(x,b,t).
\end{equation}
Indeed, outside the small-mean region \(m>1/32\), and \(b\geq m\).
Positive entropy gives \(t>0\).  The face \(t=0\) is included only
to bound every positive value uniformly; no expression involving
\(\eta(0)\) is evaluated.  Every actual tuple on \(x=0\) has
\(\Delta=0\), so \(R_\psi\leq0\), and the parent-branch comparison
already proves \(R_B\leq\zeta\).

Here is the mathematical content of the interval checker.
Let \(Q\subseteq{\cal Q}\) be a rational parameter box.  An underlined
quantity denotes a certified lower bound throughout that box, and an
overlined quantity denotes a certified upper bound.  The directed
coordinate map \eqref{unbal:ss:coordinates} supplies intervals for
\(a,m,d,E,s\), and
\[
 r=2^{-x},\qquad
 \rho=\frac{1-r}{1+r},\qquad k=\frac{m}{1-m}.
\]
Equations \eqref{unbal:ss:delta-normalized}--\eqref{unbal:ss:j-normalized} give
enclosures
\[
 \frac{\Delta}{d^2}\leq {\sf K},
 \qquad
 \frac{j(a,b)}{d^2}\geq\underline\jmath
 \qquad(d>0).
\]
For example, valid bounds are obtained by taking an upper enclosure of
\[
 \frac1{4\underline m}
 \left(\frac{C(\overline\rho)}{\overline\rho^{\,2}}
       +\overline k\,
        \frac{C(\overline{k\rho})}{\overline{k\rho}^{\,2}}\right)
\]
and a lower enclosure of
\[
 \frac1{2L\overline m}
 \left(\frac{\operatorname{atanh}(\underline\rho)}
                   {\underline\rho}
       +\underline k\,
        \frac{\operatorname{atanh}(\underline{k\rho})}
                   {\underline{k\rho}}\right),
\]
with the removable values supplied by
\eqref{unbal:ss:c-series}--\eqref{unbal:ss:a-series}.
The checker bounds \(I\) simultaneously by the interval evaluations
of \(H(m)-E\) and \(s+{\sf K}d^2\), and intersects those enclosures.
It uses
\[
 \overline p\geq\tfrac12\bigl(P'(s)+P'(I)\bigr),\qquad
 \underline\beta\leq
 \frac{\min\{\kappa(a),\kappa(b)\}}{2b(1-a)}.
\]
If \(\ell\) is a lower enclosure of \(1-k+k^2\), put
\begin{equation}\label{unbal:ss:box-bonus}
 W=\frac23\max\{3/4,\ell\}\,\underline\rho^{\,2}.
\end{equation}
Lemma~\ref{unbal:ss:bonus} then gives \(j(a,b)\geq(4+W)\Delta\)
throughout the box.

\begin{proposition}[Sufficient box inequalities]\label{unbal:ss:box-tests}
All profile arguments in the following tests must lie in their stated domains.
Each test, imposed on a whole box \(Q\), certifies
\(\zeta\geq R_B\) on its positive feasible intersection.
\begin{enumerate}
\item \emph{Previously proved regions.}
The entire box has \(a\geq1/10\), or the entire box has \(m\leq1/32\).

\item \emph{Derivative comparison.}
\begin{equation}\label{unbal:ss:test-derivative}
 \underline\beta\geq {\sf K}P''(\overline I).
\end{equation}

\item \emph{Normalized log-sum comparison.}
\begin{equation}\label{unbal:ss:test-normalized}
 \underline\beta\,\underline s
 \geq {\sf K}\max\{0,\overline p-4-W\}.
\end{equation}

\item \emph{Mean-cost log-sum comparison.}
\begin{equation}\label{unbal:ss:test-mean}
 \underline\jmath+\underline\beta\,\underline s
 \geq {\sf K}\overline p.
\end{equation}

\item \emph{Contact lower bound.}
For \(\overline d>0\) and \(\overline E>0\),
\begin{equation}\label{unbal:ss:test-contact}
 \frac{F(\overline d,\overline E)}{\overline d^{\,2}}
 \geq {\sf K}\overline p.
\end{equation}

\item \emph{Parent comparison.}
Writing \(q=1-2m\), one has
\(\overline E>0\),
\(\overline E<H((1-\overline q)/2)\), and
\begin{equation}\label{unbal:ss:test-parent}
 \eta(\overline E)-F(\overline q,\overline E)
 \geq P(\overline I).
\end{equation}

\item \emph{Elementary parent tail.}
One has \(H(\overline m)<1\), \(0<\overline E<2\underline m\), and
\begin{equation}\label{unbal:ss:test-parent-tail}
 (2\underline m-\overline E)
        \log_2\frac{2-\overline E}{\overline E}
 \geq P(H(\overline m)).
\end{equation}

\item \emph{Entropy-endpoint comparison.}
At each of the two exact endpoints of the \(t\)-interval, at least
one of \eqref{unbal:ss:test-normalized} and \eqref{unbal:ss:test-mean} holds,
using the directed mean intervals and the entropy quantities
recomputed at that endpoint.
\end{enumerate}
\end{proposition}

\begin{proof}
\emph{Proof outline.}
Each acceptance test must imply the same hybrid conclusion on the
whole feasible part of a box. We verify the tests in their stated order:
previously proved regions, lower-bound comparisons, and parent dominance.
For the last test we identify one concave function of total entropy whose
two endpoint bounds control the full interval. The exact mean diagonal
is treated separately from the normalized formulas.

For the first test, \(a\geq1/10\) and \(a\leq b\leq1/2\) place both
means in the prior central square.  The other alternative is precisely
Theorem~\ref{unbal:thm:smallmean}.

For the derivative test, fix any means in the box and define
\[
 G(u)=j(a,b)+\beta d^2u-P(u+\Delta)+P(u).
\]
At \(u=0\), equation \eqref{unbal:ss:cap} gives \(G(0)\geq0\).
For \(0\leq u\leq s\),
\[
 P'(u+\Delta)-P'(u)
 =\int_u^{u+\Delta}P''(v)\,dv
 \leq\Delta P''(I)
 \leq d^2{\sf K}P''(\overline I)
 \leq\beta d^2.
\]
Hence \(G'\geq0\), so \(G(s)\geq0\).  Equations
\eqref{unbal:ss:beta-bound} and \eqref{unbal:ss:target-upper} prove
\(\zeta\geq R_\psi\).

For the normalized log-sum test, the same two bounds and
\eqref{unbal:ss:box-bonus} give
\[
 \frac{\zeta-R_\psi}{d^2}
 \geq\beta s-(\overline p-4-W)\frac{\Delta}{d^2}
 \geq\underline\beta\,\underline s
       -{\sf K}\max\{0,\overline p-4-W\}.
\]
The last step uses \(\Delta\geq0\), treating separately the two signs
of \(\overline p-4-W\).  This proves the third test.
The fourth follows directly from
\[
 \frac{\zeta-R_\psi}{d^2}
 \geq\underline\jmath+\underline\beta\,\underline s
       -{\sf K}\overline p.
\]

For the contact test, use the global lower bound
\eqref{unbal:eq:radiallower} and Lemma~\ref{unbal:ss:contact}:
\[
 \frac{\zeta}{d^2}\geq\frac{F(d,E)}{d^2}
 \geq\frac{F(\overline d,\overline E)}{\overline d^{\,2}}.
\]
Together with \eqref{unbal:ss:target-upper}, this proves the fifth test.

For the sixth test, all intermediate comparisons are feasible because
\(E\leq\overline E<H((1-\overline q)/2)\) and \(q\leq\overline q\).
Lemma~\ref{unbal:ss:contact} implies
\[
 \phi(m,E)=\Phi(q,E)
 \geq\Phi(\overline q,\overline E)
 =\eta(\overline E)-F(\overline q,\overline E).
\]
On the other hand \(\psi(m,E)=P(I)\leq P(\overline I)\).
Thus \eqref{unbal:ss:test-parent} proves parent dominance, and the
old-candidate input applies.  Only \(\overline E>0\) is required;
the argument remains uniform over every positive \(E\) in a box
touching \(E=0\).

The seventh test follows from Lemma~\ref{unbal:ss:parent-tail}.  On
\(0<E<2m\leq1\), its first lower bound increases with \(m\) and
decreases with \(E\): both positive factors decrease with \(E\).
The asserted endpoint substitutions therefore prove
\(\phi(m,E)\geq\psi(m,E)\).

Finally, for fixed means,
\[
 G''(u)=-P''(u+\Delta)+P''(u)\leq0.
\]
Thus \(G\) is concave in \(u=s\), and also in the affine coordinate
\(t\).  Each of the normalized and mean-cost log-sum tests
lower-bounds this same actual function \(G(s)/d^2\).  Consequently
either test may certify each entropy endpoint, and concavity fills
the interval between them.  This test does not combine endpoint
certificates for different lower envelopes.

The displayed normalized arguments apply for \(d>0\).
The diagonal \(d=0\) is covered by the separate argument preceding
\eqref{unbal:ss:root}.  In every other case, either \(\zeta\geq R_\psi\)
has been established or \(\phi\) dominates at the parent.  The
parent-branch comparison proves the required conclusion.
\end{proof}

\subsection{Complete certification and assembly}\label{unbal:ss:assembly}

The checker \texttt{\detokenize{same_side/COVER.py}} implements
Proposition~\ref{unbal:ss:box-tests} with outward interval arithmetic.
Near zero it sums the first five terms of each of
\eqref{unbal:ss:c-series} and \eqref{unbal:ss:a-series}.  Their omitted tails
are bounded respectively by
\begin{equation}\label{unbal:ss:series-remainders}
 \frac{z^{10}}{132L(1-z^2)}
 \quad\hbox{and}\quad
 \frac{z^{10}}{11(1-z^2)}.
\end{equation}
These follow immediately by replacing all later denominators with
the first omitted denominator and summing a geometric series.
Away from zero the corresponding directed logarithmic formulas
are used.  Floating-point approximations only propose brackets
for inverse entropies and contact roots; directed signs certify
the enclosing brackets.

Each subdivision-path symbol records an axis and a child side.
Starting from \eqref{unbal:ss:root}, the verifier reconstructs its rational
box by exact bisection.  The partition check requires every
nonterminal node to have precisely the two children of one
bisection, and forbids a leaf from also having descendants.
Thus a complete accepted tree covers the entire closed root,
including all shared interfaces.

\begin{proposition}[Completed finite certificate]\label{unbal:ss:certificate}
The exact root \eqref{unbal:ss:root} has a complete accepted partition
with \(160{,}789\) leaves.  Every leaf passed a full replay at
70 decimal digits, with zero unresolved leaves.
\end{proposition}

\begin{proof}
\emph{Proof outline.}
The certificate has two jobs: its exact subdivision tree must cover
the root, and its recorded inequality must pass on every leaf. The records
below identify the partition, its acceptance counts, and the full replay
that checks those inequalities with the bound source fixed. The preceding
proposition then converts these checks into the required regional theorem.

The certificate is
\texttt{\detokenize{same_side/ADAPT_RESULT.json}}.
The full replay and final checks are recorded in
\texttt{\detokenize{ADAPT_REPLAY.json}} and
\texttt{\detokenize{FINAL_GATE_RESULT.json}}.
The partition has the following acceptance counts.
\begin{center}
\begin{tabular}{lr}
\toprule
Acceptance family & Leaves\\
\midrule
Prior central square & \(2{,}458\)\\
Small-mean theorem & \(266\)\\
Parent comparison & \(69\)\\
Elementary parent tail & \(24\)\\
Derivative comparison & \(3{,}205\)\\
Normalized log-sum comparison & \(59{,}175\)\\
Mean-cost log-sum comparison & \(50{,}429\)\\
Contact lower bound & \(37{,}506\)\\
Entropy-endpoint concavity & \(7{,}657\)\\
\midrule
Total & \(160{,}789\)\\
\bottomrule
\end{tabular}
\end{center}
The replay reconstructs every box and reevaluates its recorded
acceptance inequality at 70 digits.  Its source binding is to the
final checker and the actual locally imported interval primitives;
the primitive bytes and replay-script bytes are checked at both
ends of replay.  The final gate also checks the exact root, the
threshold \(m=1/32\), the complete binary partition, and agreement
of all replay counts.  The analytic inequalities in
Proposition~\ref{unbal:ss:box-tests} therefore apply to every positive
feasible point of the root.  Exploratory samples and interrupted
checkpoints are not acceptance evidence.
\end{proof}

\section{Uniform analytic estimates for the global cover}
\label{supp:uniform-estimates}
\label{unbal:sec:retained-lemmas}

\emph{Argument guide.}
The estimates here provide reusable regions for the global cover.
Derivative bounds first control how the candidates vary; the endpoint
contact then supplies a quantitative gain for unequal entropies. These
ingredients prove low-information, low-entropy, and opposite-corner
regions before the section derives the supporting-plane tests near caps.

Throughout this section, $0<a,b<1$, $0<e\le H(a)$, and
$0<f\le H(b)$. The notation is that of the main theorem. In this section $L=\ln2$,
$\Phi(z,h)=\eta(h)-F(z,h)$, and $\mathcal J(t)=-\ln(1-t^2)$.
The variable $z$ denotes a radial coordinate, so it is distinct from the
average entropy deficit $s$. We use the previously established results concerning
$\zeta\ge R_\phi$, $\zeta\ge F(d,E)$, radial concavity, and the global
endpoint supporting planes. Every additional estimate needed below is
proved explicitly. The log-sum bound proved in Theorem~\ref{unbal:ss:logsum} is
written
\begin{equation}\label{unbal:eq:ret-ls}
 B_{\rm LS}=j(a,b)+\frac{d^2s}{2V},\qquad
 V=\max(a,b)[1-\min(a,b)],\qquad \zeta\ge B_{\rm LS}.
\end{equation}
Whenever $\phi$ is active at the parent, $R_B\le R_\phi$. Whenever $\psi$
is active there, one may use either $R_B\le R_\psi$ or
\begin{equation}\label{unbal:eq:ret-child-phi}
 R_B\le \eta(E+C(q))-
 \frac{\Phi(d+q,e_1)+\Phi(|d-q|,e_2)}2,
\end{equation}
where the entropy labels follow the corresponding child radial
coordinates. These two comparisons will always be used on their stated
parent branch.

\subsection{Entropy and perspective derivatives}

For $0<h<1$, set
\[
 v=H^{-1}(h),\quad r=1-2v,\quad A=\operatorname{atanh}r,
 \quad \nu=1-r^2,\quad h_n=Lh.
\]
Here $h_n=K-rA$, where $K=L-\tfrac12\ln\nu$. Differentiation gives
\begin{equation}\label{unbal:eq:ret-profile-derivatives}
 P'(1-h)=2+\frac{2r}{\nu A},\qquad
 \eta''(h)=\frac{2L((1+r^2)A-r)}{\nu^2A^3}.
\end{equation}
The function $T(r)=2rh_n-\nu A$ vanishes at $0$ and has limiting value
zero at $1$. Its derivative is $2h_n-1$, which decreases strictly from
$2L-1>0$ to $-1$. Consequently $T\ge0$, and
$h_n\ge\nu A/(2r)$. Since $A\ge r$, equation
\eqref{unbal:eq:ret-profile-derivatives} implies
\begin{equation}\label{unbal:eq:ret-eta-curvature}
 \eta''(h)\ge\frac1{2Lh^2},\qquad
 P'(1-h)\ge2+\frac1{Lh}.
\end{equation}
The endpoint $h=1$ is understood by continuity. Integration and the
identity $C''(q)=1/[L(1-q^2)]$ give
\begin{align}
 \eta(h)-\eta(h+c)&\ge2c+\log_2(1+c/h),
       &&0<h\le h+c\le1,\label{unbal:eq:ret-parent-correction}\\
 \frac{q^2}{2L}\le C(q)&\le\frac{q^2}{2L(1-q^2)},
       &&0\le q<1.\label{unbal:eq:ret-c-bounds}
\end{align}

At the contact defining $F(z,h)$, the same symbols $r,A,\nu,h_n,K$
refer to that contact coordinate; they need not agree with the entropy
inverse at $h$. Implicit differentiation gives
\begin{equation}\label{unbal:eq:ret-f-derivative}
 zF_{zz}(z,h)=\frac{2r h_n^2(2K-r^2)}{L\nu^2K^3}.
\end{equation}
The inequality $h_n\le\nu K$ follows from $A-rK\ge0$: this expression
vanishes at both endpoints and has derivative $1-K$, which changes sign
once from positive to negative. Thus
\begin{equation}\label{unbal:eq:ret-crude-mixed}
 zF_{zz}(z,h)\le\frac4L,\qquad
 0\le\Phi_{zh}(z,h)\le\frac4{Lh},\qquad
 \Phi_{hh}(z,h)\ge\frac{1/2-4z}{Lh^2}.
\end{equation}
The last two assertions use homogeneity:
$\Phi_{zh}=(z/h)F_{zz}$ and $F_{hh}=(z/h)^2F_{zz}$.
All these formulas apply to the analytic extension of $\Phi$ on
$z\ge0$, $0<h<1$; they do not require $h\le H((1-z)/2)$.

Let $\mathfrak f(x)=F(x,1)$ and $k_h(w)=F(\sqrt w,h)$.
The supplied decreasing-$F_{zz}$ property implies
$F_z\ge zF_{zz}$, and hence $k_h$ is increasing and concave. Therefore
\begin{equation}\label{unbal:eq:ret-radial}
 \frac{F(d+q,h)+F(|d-q|,h)}2-F(d,h)
 \le k_h'(d^2)q^2
 =\frac1h\frac{\mathfrak f'(d/h)}{2d/h}q^2
\end{equation}
for $d>0$. Both $\mathfrak f(x)/x^2$ and
$\mathfrak f'(x)/(2x)$ are nonincreasing for $x>0$.

\begin{lemma}[Global mixed derivative bound]\label{unbal:lem:global-mixed}
For every $z,h>0$,
\[
 zF_{zz}(z,h)\le \frac{13}{6}.
\]
Consequently $0\le\Phi_{zh}\le13/(6h)$ and
$F_{hh}\le13z/(6h^2)$.
\end{lemma}
\begin{proof}
\emph{Proof outline.}
The contact coordinate reduces the derivative bound to one scalar
expression. Elementary estimates handle its small-coordinate and
large-coordinate ranges, while a directed interval certificate covers
the compact interval between them. Homogeneity then gives the two
derivative consequences stated in the lemma.

Put $\ell=\ln((1-v)/v)$ at the $F$ contact. If $\ell\ge3$, then
$v<1/20$, $r>9/10$, and
\[
 K=\ell/2-\ln(1-v)\le\ell/2+1/19\le59\ell/114,
 \qquad
 \frac{h_n}{\nu K}
 \le\frac{1+1/((1-v)\ell)}{2(1-v)}.
\]
Substituting these inequalities in \eqref{unbal:eq:ret-f-derivative}, and
retaining the factor $1-r^2/(2K)$, yields
\[
 zF_{zz}\le\frac1L\left(\frac{20}{19}\right)^2
 \left(1+\frac{20}{19\ell}\right)^2
 \left(1-\frac{4617}{5900\ell}\right).
\]
For $\alpha=20/19$ and $c=4617/5900$, the polynomial
$(1+\alpha t)^2(1-ct)$ increases on $0\le t\le1/3$.
Indeed its derivative is decreasing there and positive at $1/3$;
these assertions follow by differentiating this cubic and checking its
rational coefficients. Using $L>69/100$, we obtain
\[
 zF_{zz}\le\frac{10342547600}{4774831119}<\frac{13}{6}.
\]
This covers $v\le1/22$, since $\ln21>3$.
For $v\ge1/3$, \eqref{unbal:eq:ret-crude-mixed} gives
$zF_{zz}\le4r/L\le4/(3L)<13/6$.

The remaining interval $v\in[1/22,1/3]$ is a finite one-dimensional
certificate. The exact expression evaluated is
\[
 \frac{2(1-2v)h_n^2(2K-(1-2v)^2)}
 {L[4v(1-v)]^2K^3},\qquad
 h_n=-v\ln v-(1-v)\ln(1-v),\quad
 K=-\tfrac12\ln(v(1-v)).
\]
The retained files \texttt{GLOBAL\_MIXED\_CHECKER.py} and
\texttt{GLOBAL\_MIXED\_RESULT.json}, originally in
\texttt{CK\_OUTER\_DOMAIN\_CONTINUATION}, give 105 closed rational
intervals. Their ordered endpoints begin at $1/22$, end at $1/3$,
and agree at every adjacent pair. On every interval, direct 70-digit
directed interval evaluation of the displayed expression has upper
endpoint strictly below $13/6$. The checker verifies both adjacency
and every inequality. Thus the finite certificate covers the compact
interval without gaps, and the two analytic arguments cover its tails.
\end{proof}

\subsection{Endpoint contact and its entropy-imbalance gain}

\begin{lemma}[Contact comparison and logarithmic gain]
\label{unbal:lem:endpoint-loggain}
For any feasible tuple with $d\ge5E>0$, the endpoint contact exists and
\begin{equation}\label{unbal:eq:ret-endpoint-strong}
 \zeta\ge B_{\rm end}\ge F(d,E)
       +\frac{d}{2L}\mathcal J\!\left(\frac{e-f}{2E}\right).
\end{equation}
There is no common-entropy-cap assumption in this statement, and the
actual means need not lie on opposite sides of $1/2$.
\end{lemma}
\begin{proof}
\emph{Proof outline.}
First check that both the unequal-entropy contact and its equal-entropy
comparison exist. Convexity of the inverse entropy orders their masses,
so the cost can be bounded at a common mass. A logarithmic curvature
estimate for the inverse-entropy slope then turns Jensen's inequality
into the explicit gain for every entropy imbalance.

Order the means $a<b$. Cap feasibility gives
$H^{-1}(e)\le a$ and $H^{-1}(f)\le1-b$, so
$d\le1-H^{-1}(e)-H^{-1}(f)$ in every orientation of the means.
The lower endpoint threshold in Proposition~\ref{bal:prop:fixed-difference-minimum} is
\[
 D_0=\max(e,f)\left[\frac12-
 H^{-1}\!\left(\frac{\min(e,f)}{\max(e,f)}\right)\right]
 \le E<d.
\]
That proposition therefore gives $p,u,v$ satisfying
\[
 d=p(1-u-v),\qquad e=pH(u),\qquad f=pH(v),\qquad
 0<p\le1,\quad 0<u,v<1/2,
\]
and the global lower bound
\begin{equation}\label{unbal:eq:ret-endpoint-cost}
 B_{\rm end}=\frac d2\{Q(e/p)+Q(f/p)\},\qquad
 Q(h)=J(H^{-1}(h)).
\end{equation}
Its four-moment equality window is irrelevant to this use as a lower
bound.

Let $\bar p$ be the equal-entropy contact mass for $(d,E,E)$.
It exists because convexity of $H^{-1}$ gives
$2H^{-1}(E)\le H^{-1}(e)+H^{-1}(f)\le1-d$.
In particular $\bar p\ge d$. The function
\[
 p\longmapsto p\{1-H^{-1}(e/p)-H^{-1}(f/p)\}
\]
is strictly increasing on its contact interval, as in the same
proposition. At $p=\bar p$ its value is at most $d$, by convexity
of $H^{-1}$. Consequently $p\ge\bar p$. As $Q$ decreases,
\begin{equation}\label{unbal:eq:ret-contact-mass}
 B_{\rm end}\ge\frac d2
        \{Q(e/\bar p)+Q(f/\bar p)\}.
\end{equation}

We next prove $Q''(h)\ge1/(Lh^2)$ for $0<h\le19/50$.
Write $v=H^{-1}(h)$, $\ell=\ln((1-v)/v)$, and $t=1/\ell$.
Direct differentiation gives
\[
 Q''(h)=\frac{L((1-2v)\ell-1)}{v^2(1-v)^2\ell^3}.
\]
The fixed logarithmic comparisons
$H(3/40)>19/50$ and $5/2<\ln(37/3)<13/5$ imply
$v<3/40$ and $t<2/5$. Since $Lh\ge v(\ell+1)$, it suffices to
show that
\[
 (1-2v-t)(1+t)^2-(1-v)^2
 =t[1-t-t^2-4v-2vt-v^2/t]>0.
\]
The function $v^2\ln((1-v)/v)$ increases for $0<v\le3/40$;
its derivative is $v[2\ell-1/(1-v)]>0$ there. Thus
$v^2/t<117/8000$, and the bracket is at least
\[
 1-\frac25-\frac4{25}-\frac3{10}-\frac3{50}
 -\frac{117}{8000}=\frac{523}{8000}>0.
\]
This proves the asserted logarithmic curvature.

At the equal-entropy contact with ratio $d/E\ge5$, its coordinate
$u$ is at most the contact $u_5$ at ratio $5$.
The exact inequality $5H(1/40)<19/20$ implies $u_5>1/40$, and hence
$H(u)\le H(u_5)=(1-2u_5)/5<19/100$.
Therefore $e/\bar p,f/\bar p<19/50$ for every entropy split.
The function $Q(h)+(1/L)\ln h$ is convex on this interval.
Apply Jensen's inequality in \eqref{unbal:eq:ret-contact-mass} and use
$dQ(E/\bar p)=F(d,E)$. The logarithmic term is exactly
$d\mathcal J((e-f)/(2E))/(2L)$, proving
\eqref{unbal:eq:ret-endpoint-strong}.
All fixed comparisons in this proof are exact rational logarithm-series
comparisons in the retained \texttt{SHARPER\_COLLARS} checker.
\end{proof}

We will also use the weaker assertion
\begin{equation}\label{unbal:eq:ret-endpoint-weak}
 B_{\rm end}\ge F(d,E)+\frac d{4L}\mathcal J((e-f)/(2E))
 \quad\text{if }d>E\text{ and }2E/d\le1/100.
\end{equation}
For completeness, $h\le1/100$ implies $v\le1/200$,
$r\ge99/100$, and $\ell\ge\ln199>4$. The formula for $Q''$ and
$Lh\ge v\ell$ give
$h^2Q''(h)\ge(r-1/\ell)/[L(1-v)^2]\ge1/(2L)$.
The preceding contact comparison and Jensen argument then prove
\eqref{unbal:eq:ret-endpoint-weak}.

\subsection{The entire low-information region}

\begin{lemma}\label{unbal:lem:mean-cost-bonus}
Order and complement the means so that $0<a<b<1$ and
$m=(a+b)/2\le1/2$. Put $r=(b-a)/(a+b)$ and $k=m/(1-m)$.
Then
\[
 j(a,b)\ge(4+r^2/2)\Delta,\qquad j(a,b)\ge P(\Delta).
\]
\end{lemma}
\begin{proof}
\emph{Proof outline.}
We express the entropy gap and the mean cost through the same positive
series. Comparing coefficients gives the strict correction to the basic
factor four, and bounding the two weights makes it uniform in the means.
A separate application of profile convexity gives the deterministic-cap
comparison in the second assertion.

Define
\[
 C_n(x)=x\operatorname{atanh}x+\tfrac12\ln(1-x^2)
       =\sum_{n\ge1}\frac{x^{2n}}{2n(2n-1)}.
\]
Direct binary-entropy identities give
\begin{align*}
 L\Delta&=mC_n(r)+(1-m)C_n(kr),\\
 Lj(a,b)&=2mr\operatorname{atanh}r
       +2(1-m)kr\operatorname{atanh}(kr).
\end{align*}
Coefficient comparison shows
$2x\operatorname{atanh}x-4C_n(x)\ge(2/3)x^2C_n(x)$.
At power $x^{2n}$, $n\ge2$, the comparison reduces to
\[
 6(n-1)^2(2n-3)-n(2n-1)
 =(n-2)(12n^2-20n+9)\ge0.
\]
Also $C_n(kr)\le k^2C_n(r)$, by positivity of all coefficients. Thus
\[
 \frac{mC_n(r)+(1-m)k^2C_n(kr)}
      {mC_n(r)+(1-m)C_n(kr)}
 \ge\frac{1+k^3}{1+k}=1-k+k^2\ge\frac34.
\]
For the first comparison, divide by $mC_n(r)$ and observe that
$(1+kz)/(1+z/k)$ decreases in $z$, with
$z=C_n(kr)/C_n(r)\le k^2$. The coefficient comparison now yields
$j\ge(4+r^2/2)\Delta$.
Finally $C=C_n/L$ and $P(C(x))=2x\operatorname{atanh}x/L$ imply
\[
 \Delta=mC(r)+(1-m)C(kr),\qquad
 j=mP(C(r))+(1-m)P(C(kr)).
\]
Convexity of $P$ proves the second assertion. For equal means, both
$j$ and $\Delta$ are zero.
\end{proof}

\begin{theorem}[Uniform low-information region]\label{unbal:thm:low-information}
Every feasible tuple with $I=H(m)-E\le1/100$ satisfies
$B_{\rm LS}\ge R_\psi$, and hence $\zeta\ge R_B$.
\end{theorem}
\begin{proof}
\emph{Proof outline.}
After eliminating the entropy split, we bound the profile curvature
on the low-information interval. For larger normalized mean separation,
the improved deterministic cost already exceeds the target. For smaller
separation, the log-sum correction has sufficient slope to cover the
remaining increase from the entropy caps.

Convexity of $P$ gives
$R_\psi\le D_\Delta(s):=P(\Delta+s)-P(s)$.
We first bound $P''$ on $[0,1/100]$.
In \eqref{unbal:eq:ret-profile-derivatives}, write $T=r^2$; then
$I=C_n(r)/L\ge T/(2L)$, so $T<7/500$ because $L<347/500$.
The series for $A=\operatorname{atanh}r$ gives
\[
 (1+r^2)A-r\le\frac43r^3+\frac8{15}\frac{r^5}{1-r^2},
 \qquad A^3\ge r^3(1+r^2).
\]
Consequently
\[
 P''(I)\le
 \frac{2L[4/3+(8/15)T/(1-T)]}{(1-T)^2(1+T)}
 \le\frac{344085200000}{182251021797}<\frac{19}{10}.
\]
The numerator increases and the denominator decreases on the stated
interval, justifying evaluation at its upper endpoint. The limiting
value $P''(0)=8L/3$ obeys the same bound. Therefore
$P'(I)\le4.019$.

Use the orientation and variables of Lemma~\ref{unbal:lem:mean-cost-bonus}.
If $r\ge1/5$, that lemma gives $j\ge4.02\Delta$, whereas
$D_\Delta(s)\le4.019\Delta$, and the result follows.
If $0<r\le1/5$, then
\[
 V=m(1-m)(1+r)(1+kr),\quad d=2mr,\quad
 \Delta\le kC_n(r)/L.
\]
Hence
\[
 \frac{d^2}{2V\Delta}
 \ge\frac{2Lr^2}{(1+r)^2C_n(r)}.
\]
The positive series gives
$C_n(r)\le r^2/2+r^4/[12(1-r^2)]$.
For $r\le1/5$,
\[
 (1+r)^2\left(\frac12+\frac{r^2}{12(1-r^2)}\right)
 \le\frac{145}{200}.
\]
It follows that $d^2/(2V\Delta)\ge80L/29>19/10$.
For $0\le t\le s$, we now have
\[
 D_\Delta'(t)=P'(t+\Delta)-P'(t)
       \le\frac{19}{10}\Delta\le\frac{d^2}{2V}.
\]
Integrating and using $j\ge P(\Delta)$ proves
$D_\Delta(s)\le j+d^2s/(2V)=B_{\rm LS}$.
For $a=b$, $\Delta=0$ and $R_\psi\le0$.
Full label exchange and simultaneous complement preserve the target,
so these arguments cover all mean pairs.
\end{proof}

\subsection{A uniform normalized zero-entropy theorem}

We record the fixed scalar constants used in the proof:
\begin{equation}\label{unbal:eq:ret-fixed-three}
 \mathfrak f(3)>12,\qquad \mathfrak f'(3)/6<1,
 \qquad hP'(1-h)<7/4\quad(0<h\le10^{-4}).
\end{equation}
For the first inequality, the contact at ratio $3$ is below $1/17$,
because $3H(1/17)>15/17$, equivalently $17^{17}>2^{69}$.
Thus its log odds exceed $4$.
For the derivative inequality, the contact lies in
$(5279/100000,66/1250)$. With $A=\operatorname{atanh}(1-2v)$,
implicit differentiation at $3H(v)=1-2v$ gives
\[
 \frac{\mathfrak f'(3)}6
 =\frac{A}{3L}+\frac{1-2v}{3[1-(1-2v)^2](L+3A)}
 <\frac{114629824}{115428159}<1.
\]
The last comparison uses $L>693/1000$,
$1443/1000<A<1444/1000$, $1-2v<895/1000$, and
$1-(1-2v)^2>199/1000$.
All contact brackets and logarithms here are verified by the exact
checker in \texttt{retained/uniform\_diagonal}.
It uses the rational expansion, for $z\ge1$ and
$w=(z-1)/(z+1)$,
\[
 \ln z=2\sum_{i=0}^{N-1}\frac{w^{2i+1}}{2i+1}+R_N,
 \qquad0\le R_N\le
 \frac{2w^{2N+1}}{(2N+1)(1-w^2)},\qquad N=256.
\]
For the last assertion in \eqref{unbal:eq:ret-fixed-three}, put
$v=H^{-1}(h)$ and $\ell=\ln((1-v)/v)$.
Since $H(v)\ge2v$, we have $v\le1/20000$ and
$\ell\ge\ln19999>9$. Using $Lh=v\ell-\ln(1-v)$ gives
\[
 hP'(1-h)\le2h+\frac1L
       \left(1+\frac1{(1-v)\ell}\right)
 <\frac2{10000}+\frac{27}{16}<\frac74.
\]
Here $-\ln(1-v)\le v/(1-v)$, $(1-v)\ell>8$, and $L>2/3$.

\begin{theorem}[Normalized zero-entropy region]
\label{unbal:thm:normalized-low-entropy}
For every feasible tuple,
\[
 0<E\le10^{-6},\qquad q\le32E
 \quad\Longrightarrow\quad\zeta\ge R_B.
\]
The assertion is uniform over all positive entropy splits.
\end{theorem}
\begin{proof}
\emph{Proof outline.}
We work on the branch where the entropy-deficit candidate is active
and first reserve a uniform positive parent correction. Small mean
differences are handled by either a direct radial bound or a convexity
estimate for the child entropies. For larger differences, an endpoint
contact supplies a logarithmic gain that absorbs the possible entropy
curvature loss beyond an individual cap. In each case we subtract the
radial and split losses from the same parent reserve.

It suffices to treat the active-$\psi$ parent branch.
Throughout this proof,
\begin{equation}\label{unbal:eq:ret-small-parent}
 \frac{C(q)}E<\frac1{1000},\qquad
 \eta(E)-\eta(E+C(q))
 \ge\frac{50000}{49049}\frac{q^2}{E}.
\end{equation}
The first assertion follows from \eqref{unbal:eq:ret-c-bounds},
$q\le32E$, and $L>2/3$.
The second follows from \eqref{unbal:eq:ret-parent-correction},
$\ln(1+x)\ge x/(1+x)$, $C(q)\ge q^2/(2L)$, and $L<7/10$.

First suppose $d\le1/64$. If also $d\le3E$, convexity of $P$ gives
\[
 R_B\le R_\psi\le\Delta P'(I),\qquad
 P'(I)<\frac7{4(E+C(q))}\le\frac7{4E}.
\]
The entropy Hessian gives
$\Delta\le d^2/[2L(1-(q+d)^2)]$.
Since $q+d\le35E<1/8$,
$R_B\le4d^2/(3E)$.
Concavity of $k_1$ and \eqref{unbal:eq:ret-fixed-three} give
\[
 F(d,E)=E\mathfrak f(d/E)
 \ge\frac{\mathfrak f(3)}9\frac{d^2}{E}
 >\frac{4d^2}{3E}
\]
when $d>0$; the case $d=0$ has $R_B\le0$.

If $3E\le d\le1/64$, the child radii $z_1=d+q$,
$z_2=|d-q|$ are below $1/32$.
For $0<h<2E$, \eqref{unbal:eq:ret-crude-mixed} implies
\[
 \Phi_{hh}(z_i,h)\ge\frac1{4Lh^2}
          \ge k_0:=\frac1{16LE^2}.
\]
With the entropies matched as $E+t,E-t$, strong convexity gives
\[
 \frac{\Phi(z_1,E+t)+\Phi(z_2,E-t)}2
 \ge\frac{\Phi(z_1,E)+\Phi(z_2,E)}2
    +\frac t2 A_e+\frac{k_0}2t^2,
\]
where $A_e=\Phi_h(z_1,E)-\Phi_h(z_2,E)$.
The mixed derivative bound gives $|A_e|\le8q/(LE)$.
Minimizing the quadratic on the whole real line bounds the entropy
loss by $A_e^2/(8k_0)\le192q^2$.
Equation~\eqref{unbal:eq:ret-radial} and
$\mathfrak f'(3)/6<1$ bound the radial loss by $q^2/E$.
Using \eqref{unbal:eq:ret-child-phi} and \eqref{unbal:eq:ret-small-parent},
\[
 F(d,E)-R_B
 \ge\left(\frac{50000}{49049}-1-192E\right)\frac{q^2}{E}
 \ge\frac{q^2}{100E}.
\]
This completes the range $d\le1/64$.

Now suppose $d\ge1/64$. The endpoint exists, since
$D_0\le E<d$ and the upper feasibility test was proved in
Lemma~\ref{unbal:lem:endpoint-loggain}. Moreover
$2E/d\le128\cdot10^{-6}<1/100$, so
\eqref{unbal:eq:ret-endpoint-weak} applies.
Define exact constants
\[
 A_0=1-\frac{64}{511},\qquad
 S_0=\left(\frac{1151}{1023}\frac{1024}{1023}\right)^2.
\]
For $0<h\le2E$ and $0\le z\le1/8$,
\eqref{unbal:eq:ret-crude-mixed} gives $\Phi_{hh}\ge0$.
For $1/8\le z\le1$, we claim
\begin{equation}\label{unbal:eq:ret-refined-curvature}
 \Phi_{hh}(z,h)\ge\frac{A_0-S_0z}{Lh^2}.
\end{equation}
To verify it, let $v_8=(1+\exp8)^{-1}$.
The elementary logarithmic bounds
$1/4096<v_8<1/1024$ imply $H(v_8)>3/1024$.
Since $h\le2\cdot10^{-6}$ and $h/z\le16\cdot10^{-6}$,
both the entropy inverse and the $F$ contact have logit $\ell>8$
and coordinate $v<1/1024$.
The proof of \eqref{unbal:eq:ret-eta-curvature} more precisely gives
\[
 h^2\eta''(h)\ge\frac1L-\frac1{Lr\ell}\ge\frac{A_0}{L}.
\]
At the $F$ contact,
\[
 \frac{h_n}{\nu K}
 \le\frac{1+1/((1-v)\ell)}{2(1-v)},
 \qquad zF_{zz}\le\frac{S_0}{L}.
\]
Together with $h^2F_{hh}=z^2F_{zz}$, these prove
\eqref{unbal:eq:ret-refined-curvature}.
For either child radius $z_i=d\pm q$, set
$\kappa=\max(0,S_0(d+q)-A_0)$.
Since $d\le1$, $q/d\le32/15625$, and
\[
 (S_0-A_0)+S_0\frac{32}{15625}<\frac25,
\]
we have $\kappa\le2d/5$. Hence, for both children and all $h\le2E$,
$\Phi_{hh}(z_i,h)\ge-\kappa/(Lh^2)$.
This statement concerns the analytic extension and requires no
common-entropy cap.

Match entropy labels as $E(1+t),E(1-t)$.
Supporting lines for $\Phi(z_i,h)-(\kappa/L)\ln h$ yield
\[
 \frac{\Phi(z_1,E(1+t))+\Phi(z_2,E(1-t))}2
 \ge\frac{\Phi(z_1,E)+\Phi(z_2,E)}2
    +\frac{Et}{2}A_e-\frac\kappa{2L}\mathcal J(t).
\]
Combine this with \eqref{unbal:eq:ret-endpoint-weak},
$|A_e|\le8q/(LE)$, and $\mathcal J(t)\ge t^2$.
The residual entropy-split term is at least
\[
 \frac{d}{20L}t^2-\frac{4q}{L}|t|
 \ge-\frac{80q^2}{Ld}\ge-\frac{120q^2}{d}.
\]
The radial loss is at most $q^2/E$, since $d/E\ge15625>3$.
Equations~\eqref{unbal:eq:ret-child-phi} and
\eqref{unbal:eq:ret-small-parent} now give
\[
 B_{\rm end}-R_B\ge
 \left(\frac{50000}{49049}-1-\frac{120E}{d}\right)\frac{q^2}{E}
 \ge\frac{q^2}{100E}.
\]
The final exact comparison is
$50000/49049-1-120(64/10^6)>1/100$.
The arguments also apply at $q=0$, with zero asserted margin.
Thus all entropy splits and all $d$ are covered.
\end{proof}

\subsection{The opposite deterministic corner}

\begin{lemma}\label{unbal:lem:opposite-normalized-corner}
For a feasible tuple with $d\ge9/10$, $0<E\le10^{-3}$ and $q\le32E$,
the endpoint exists and, on the active-$\psi$ parent branch,
\[
 B_{\rm end}-R_B\ge\frac{q^2}{5E}.
\]
\end{lemma}
\begin{proof}
\emph{Proof outline.}
The endpoint contact contributes a logarithmic gain in the entropy
imbalance. We bound the possible negative curvature of the child
functions on the required extended domain and combine the two effects
by completing a square. A positive parent correction and a small radial
tail loss then leave the stated margin.

The applicability and contact-mass comparison were proved above.
Here $2E/d<1/100$, so \eqref{unbal:eq:ret-endpoint-weak} applies.
The child radii $d\pm q$ lie in $[4/5,1]$.
For $0<h\le2E\le1/500$ and $z\ge4/5$, both inverse logits
in the proof of \eqref{unbal:eq:ret-refined-curvature} still exceed $8$:
$h<H(v_8)$ and $h/z\le1/400<3/1024<H(v_8)/(1-2v_8)$.
Thus
\[
 \Phi_{hh}(z,h)\ge\frac{A_0-S_0}{Lh^2}>-\frac2{5Lh^2},
\]
where the exact comparison is
\[
 S_0-A_0=\frac{220293308870209}{559658926346751}<\frac25.
\]
For the split $E(1+t),E(1-t)$, the supporting-line argument
therefore gives an entropy loss
$\mathcal J(t)/(5L)$, plus the linear term $EtA_e/2$.
Combined with the endpoint gain, its contribution is at least
\[
 \frac1{40L}\mathcal J(t)-\frac{4q}{L}|t|
 \ge-\frac{160q^2}{L}\ge-240q^2.
\]

Here $q<1/10$, and \eqref{unbal:eq:ret-c-bounds} gives
$C(q)/E\le128/165<4/5$.
The parent correction consequently satisfies
\[
 \eta(E)-\eta(E+C(q))\ge\frac{250}{441}\frac{q^2}{E}.
\]
We also need a crude bound far along the radial tail.
At a contact of ratio $x=z/h$, put
$\ell=\ln((1-v)/v)$ and $Z=\exp\ell$.
Since $LH(v)\le v(\ell+2)$,
$Z\le1+x(\ell+2)/L$.
The inequality $\ell+2\le2\exp(\ell/2)$ implies
$\sqrt Z\le1+2x/L<2+3x$, hence
\[
 F(z,h)\le2z\log_2(2+3z/h).
\]
Concavity of $k_1$ now gives
\[
 k_E'(d^2)\le\frac1E\frac{\mathfrak f(d/E)}{(d/E)^2}
 \le\frac1E\frac{2\log_2(2+3d/E)}{d/E}<\frac1{10E}.
\]
Indeed $d/E\ge900$, the last ratio decreases on this range,
and $2702<2^{12}$ gives a bound $24/900<1/10$.
Combining the parent, radial and split bounds gives
\[
 B_{\rm end}-R_B\ge
 \left(\frac{250}{441}-\frac1{10}-240E\right)\frac{q^2}{E}
 \ge\frac{5003}{22050}\frac{q^2}{E}
 \ge\frac{q^2}{5E}.\qedhere
\]
\end{proof}

\begin{theorem}[Opposite deterministic corner]
\label{unbal:thm:opposite-corner}
Every positive feasible tuple with
$a+(1-b)\le2^{-13}$ satisfies $\zeta\ge R_B$.
The corresponding symmetric orientations follow by full label exchange
and simultaneous complement.
\end{theorem}
\begin{proof}
\emph{Proof outline.}
The mean-only corner condition forces small total entropy and large
mean separation. If the parent radius is at most the stated multiple
of entropy, the preceding normalized corner estimate applies. Otherwise,
a logarithmic parent comparison shows that the balanced candidate is
active, so its unrestricted theorem closes the other branch.

Put $w=a+(1-b)$. Then $d=1-w\ge9/10$, $q\le w<1/10$, and
concavity of entropy gives
\[
 E\le\tfrac12[H(a)+H(1-b)]\le H(w/2)
 \le H(2^{-14})<16\cdot2^{-14}<10^{-3}.
\]
The penultimate bound follows from
$H(x)\le x\log_2(1/x)+x/L$.
If $q\le32E$, Lemma~\ref{unbal:lem:opposite-normalized-corner} or the
active-$\phi$ branch proves the assertion.
Otherwise put $x=q/E>32$.
Equations~\eqref{unbal:eq:ret-parent-correction} and
\eqref{unbal:eq:ret-c-bounds} give
\[
 \eta(E)-\eta(E+C(q))\ge\log_2(1+qx/2).
\]
For fixed $x$, $\ln(1+qx/2)/q$ decreases in $q$, so $q\le1/10$
implies $\ln(1+qx/2)\ge10q\ln(1+x/20)$.
For $x\ge32$,
$5\ln(1+x/20)\ge\ln(2+3x)$: at $32$ this is
$(13/5)^5\ge98$, and the derivative of the difference is
$(12x-50)/[(20+x)(2+3x)]>0$.
The radial bound proved in the preceding lemma now implies
\[
 \eta(E)-\eta(E+C(q))\ge2q\log_2(2+3q/E)\ge F(q,E).
\]
Thus $\phi(m,E)\ge\psi(m,E)$, and $R_B\le R_\phi$.
\end{proof}

\subsection{Supporting planes at the means and cap criteria}

\begin{lemma}[Global mean-contact plane]\label{unbal:lem:global-cap-plane}
For $0<a<b<1$, let $A,D$ solve
\begin{equation}\label{unbal:eq:ret-cap-matrix}
 \begin{pmatrix}-\ln a&-\ln b\\-\ln(1-a)&-\ln(1-b)\end{pmatrix}
 \binom AD=
 \binom{(b-a)^2/(2ab)}{(b-a)^2/[2(1-a)(1-b)]}.
\end{equation}
Whenever $A,D\ge0$, the following bound holds for all feasible
entropy pairs at these means:
\[
 \zeta\ge B_{\rm cap}:=j(a,b)+A[H(a)-e]+D[H(b)-f].
\]
In particular $A,D>0$ for strict cross-half contacts. At a half-contact
boundary the plane is interpreted by its continuous limit.
\end{lemma}
\begin{proof}
The nonnegative-coefficient supporting-plane theorem, Theorem~\ref{bal:thm:general-endpoint-region}
gives the pointwise inequality
\[
 j(u,w)-c(w-u)+AH(u)+DH(w)\ge0,
 \qquad c=\frac{j(a,b)+AH(a)+DH(b)}{b-a}.
\]
Taking expectations under any admissible four-moment law and then
taking the infimum proves the result. Positivity for cross-half
contacts is Proposition~\ref{bal:prop:fixed-difference-minimum}. For same-side
contacts, positivity must be checked; it is not inferred from this
lemma. No four-moment equality-window condition is used.
\end{proof}

Write $x=H(a)-e$, $y=H(b)-f$, $s=(x+y)/2$, and $k=\min(A,D)$.
Convexity of $P$ gives, whenever the plane is valid,
\begin{equation}\label{unbal:eq:ret-cap-gap}
 \zeta-R_\psi\ge G(s):=j+2ks-P(\Delta+s)+P(s).
\end{equation}
The entropy-profile property $P'''\ge0$ proved in the profile section
implies $G''(s)=P''(s)-P''(\Delta+s)\le0$.
Together with $G(0)=j-P(\Delta)\ge0$, this yields two useful
criteria used by the finite covers:
\begin{align}
 G(s_0)\ge0&\quad\Longrightarrow\quad
 \zeta\ge R_\psi\text{ for every }0\le s\le s_0,
       \label{unbal:eq:ret-cap-endpoint}\\
 4+2k-P'(I_*)\ge0&\quad\Longrightarrow\quad
 \zeta\ge R_\psi\text{ whenever }\Delta+s\le I_*.
       \label{unbal:eq:ret-cap-slope}
\end{align}
For the first implication, a concave function lies above its endpoint
chord. For the second, $P'(t)\ge P'(0)=4$ gives
$G'(t)\ge2k-P'(I_*)+4\ge0$ throughout the required interval.
Both criteria include $s=0$ and exact cap equality without division
by a vanishing quantity.

There is also a split-sensitive form. Since $P''\ge8L/3$, putting
$t=(x-y)/2$ gives
\[
 \tfrac12[P(s+t)+P(s-t)]\ge P(s)+\frac{4L}{3}t^2.
\]
Completing a square therefore proves
\begin{equation}\label{unbal:eq:ret-cap-quadratic}
 \zeta-R_\psi\ge j+(A+D)s+P(s)-P(\Delta+s)
                   -\frac{3(A-D)^2}{16L}.
\end{equation}
A sharper valid version minimizes
$(A-D)t+(4L/3)t^2$ over
\[
 \max(-s,s-H(b))\le t\le\min(s,H(a)-s),
\]
whose minimizer is the projection of $-3(A-D)/(8L)$ onto this
interval. The right side of \eqref{unbal:eq:ret-cap-quadratic} is concave
in $s$. Thus a whole average-entropy slab can be certified by checking
its two endpoints; a whole cap interval can instead use
\eqref{unbal:eq:ret-cap-endpoint}. This distinction is relevant because
the quadratic bound at $s=0$ need not be nonnegative.

\paragraph{Location of the retained finite inputs.}
The proofs in this section consolidate
\begin{itemize}
\item \path{retained/normalized_low_entropy/PROOFS.md};
\item \path{retained/uniform_diagonal/UNIFORM_PROOF.md};
\item \path{retained/opposite_corner/OPPOSITE_CORNER_PROOF.md}.
\end{itemize}
The mixed-derivative and strong endpoint calculations come from
\texttt{SHARPER\_COLLARS.md} and \texttt{TRANSVERSE\_STRIP.md} in
the retained outer-domain dependency. The supporting-plane criteria
are the analytic content of Section~\ref{unbal:module-cap_matrix-the-supporting-plane-taken-at-the-means}, taken from
\path{CK_CAP_REGION_CONTINUATION/PROOF.md}.
Their fixed rational and logarithmic comparisons have separate
checkers. The only parameter-dependent certificate in this section
is the 105-interval compact part of Lemma~\ref{unbal:lem:global-mixed};
its two omitted tails were proved explicitly above.

\section{Normalized seam coordinates and minimum exclusion}
\label{bal:sec:seam-normalization}

\emph{Argument guide.}
The goal is a precise minimum-exclusion statement for the strict seam.
We first fix units, coordinates, and the partial derivatives held constant.
The following sections prove a low-entropy value bound, monotonicity along
complete mean fibers, and boundary collars. Their chart union then yields
the exclusion needed by the compactness argument in the main text.

From this section through Section~\ref{bal:sec:ii}, $H$ denotes natural
entropy, $P=H'$, and $L=\ln 2$.  The symbols $\HH,F,\eta,\Phi$
retain their bit-normalized meanings from the main text; barred quantities
below have the explicitly stated natural normalization.  Every logarithm
in the natural formulas is written $\ln$.
Let
\[
 S=10^{-4},\quad m=S/2,\quad A=1-S,\quad L=\ln2,
 \qquad H(t)=-t\ln t-(1-t)\ln(1-t),\quad P(t)=H'(t).
\]
Thus $H=L\HH$. Entropy inverses below use the lower branch $[0,1/2]$.
For $s\ge0,E>0$, define the natural perspective
\[
 \F(s,E)=s\atanh z,\qquad \frac{z}{H((1-z)/2)}=\frac{s}{E},
 \quad 0\le z<1.
\]
The inverse is unique. At $s=0$ use its analytic even extension. Set
\[
 \et(E)=\frac12(1-2H^{-1}(E))P(H^{-1}(E)),\qquad
 \Ph(s,E)=\et(E)-\F(s,E).
\]
The bit-normalized functions of the main text satisfy
\[
 F(s,e)=\frac2L\F(s,Le),\quad
 \eta(e)=\frac2L\et(Le),\quad
 \Phi(s,e)=\frac2L\Ph(s,Le).
\]
For $0<p<q$, put
\[
 E_u=H(p),\quad E_w=H(q),\quad h=(E_u+E_w)/2,\quad d=(E_u-E_w)/2<0,
\]
\[
 \kap(p,q)=\frac{(q-p)(P(p)-P(q))}{4}-\F(q-p,h).
\]
The previously established entropy-gap theorem states that $\kap\ge0$ and that its expression in $(E_u,E_w)$ is jointly convex.

The physical strict seam is
\begin{equation}\label{bal:eq:physical}
 0<p<q,\quad p+q<S,\qquad \max(0,2q-S)<y<S-2p.
\end{equation}
Its means are $a=(S-y)/2,c=(S+y)/2$. The natural normalized pure gap is
\begin{align}
 g(y,h,d)={}&\F(y,h)+\kap(p,q)-\Ph(A,h)\notag\\
 &+\tfrac12\Ph(A+y,h+d)+\tfrac12\Ph(A-y,h-d).
 \label{bal:eq:gap}
\end{align}
It is related to the main text's $G=L_4-\RR$ by $g=(L/2)G$.
The frozen derivatives are $g_y|_{h,d}$, $g_d|_{y,h}$, and
$g_{E_u}|_{y,E_w}$. If $d_b=(e_u-e_w)/2$ denotes the bit entropy difference, then
\[
 G_y=2g_y/L,\qquad G_{d_b}=2g_d,\qquad G_{e_u}=2g_{E_u}.
\]
All conversion factors are positive. Chart derivatives are never substituted for these frozen partials.

\begin{theorem}[Strict-seam minimum exclusion]\label{bal:thm:ME}
On the entire strict seam \eqref{bal:eq:physical},
\begin{equation}\label{bal:eq:ME}
 g<0,\quad g_y=0\quad\Longrightarrow\quad g_d<0\ \text{or}\ g_{E_u}<0.
\end{equation}
The statement follows from the analytic arguments below and the exact finite cover certificates identified in Section~\ref{bal:sec:union}.
\end{theorem}
The premise $g<0$ is essential to the statement used here.
A region covered by a nonnegative value bound cannot contain
a negative minimum; no derivative conclusion is required at its
nonnegative points.

\begin{proposition}[Why minimum exclusion is sufficient]\label{bal:prop:SK1}
Theorem~\ref{bal:thm:ME}, together with the boundary estimates in Section~\ref{bal:sec:seam} and clauses (ii) and (iii) of Theorem~\ref{bal:thm:SCplus}, gives SK-1: $G\ge0$ throughout the complete marginal-physical seam.
\end{proposition}
\begin{proof}
Entropy rearrangement reduces to $e_u\le e_w$. The closed seam has facets
\[
 y=0,\quad e_u=0,\quad e_u=e_w,\quad e_u=\HH(a),\quad e_w=\HH(c).
\]
The collapsed $y=S$ face belongs to $e_u=0$. The stated boundary owners give nonnegative lower limits along every sequence approaching these faces. If a negative value existed, choose a negative sublevel strictly above that value. Compactness of the closed physical cell and the boundary lower-limit property make this sublevel compact inside the strict seam. A negative minimum is attained there. Fermat's rule gives
\[
 G_y=G_{e_u}=G_{e_w}=0,\qquad G_{d_b}=G_{e_u}-G_{e_w}=0.
\]
This contradicts \eqref{bal:eq:ME}. No derivative claim at a nonnegative point is used.
\end{proof}

\section{An explicit bound down to zero entropy}\label{bal:sec:low}
The established same-side estimate gives
\begin{equation}\label{bal:eq:B0}
 g\ge B_0:=\F(y,h)+\kap(p,q)-Q(y),\qquad
 Q(y)=2H(m)-H(m-y/2)-H(m+y/2).
\end{equation}
The physical domain has $0\le y\le S$.

\begin{theorem}[Uniform joint-low-entropy collar]\label{bal:thm:low}
For every physical point with $0<h\le2S/3$,
\[
 g\ge B_0\ge\kap(p,q)+\frac{y^2}{200S}\ge0.
\]
This holds uniformly in the entropy ratio and for arbitrarily small positive entropies.
\end{theorem}
\begin{proof}
\emph{Proof outline.}
We lower-bound the central perspective by a quadratic function of
the mean difference using the entropy series and an inverse-hyperbolic
estimate. The same series bounds the negative entropy Jensen-gap term.
Subtracting the two bounds leaves a positive quadratic coefficient,
uniformly over the entropy split.

Write $\mathcal H(z)=H((1-z)/2)$. Its positive-coefficient deficit expansion is
\[
 \mathcal H(z)=L-\sum_{n\ge1}\frac{z^{2n}}{2n(2n-1)}.
\]
The coefficient sum is $L$, by continuity at $z=1$. Hence
$\mathcal H(z)\ge L(1-z^2)$. With $r=\atanh z$, the perspective equation yields
\[
 \frac yh\le\frac{z}{L(1-z^2)}=\frac{\sinh(2r)}{2L},\qquad
 \F(y,h)\ge\frac y2\asinh(2Ly/h).
\]
Concavity of $\asinh$ and its zero value at the origin imply that
$\asinh(cy)/y$ decreases for $y>0$. Therefore
\begin{equation}\label{bal:eq:Flow}
 \frac{\F(y,h)}{y^2}\ge\frac{\asinh(3L)}{2S}>\frac{L+1/100}{S}.
\end{equation}
For the strict scalar inequality, use $69/100<L<7/10$ and
\[
 \sinh(2L+1/50)
 =2e^{1/50}-\tfrac18e^{-1/50}
 \le\frac{100}{49}-\frac{49}{400}
 =\frac{37599}{19600}<\frac{207}{100}<3L.
\]
Here $e^x\le(1-x)^{-1}$ and $e^{-x}\ge1-x$ suffice.

Set
\[
 C(v)=(1+v)\ln(1+v)+(1-v)\ln(1-v)
 =\sum_{n\ge1}\frac{v^{2n}}{n(2n-1)}.
\]
Direct expansion gives
\[
 Q(y)=mC(y/(2m))+(1-m)C(y/(2(1-m))).
\]
Positive coefficients bound the first summand divided by $y^2$ by its value at $y=S$, namely $L/S$. The estimate $C(v)\le v^2/(1-v^2)$ bounds the other summand. Thus
\[
 \frac{Q(y)}{y^2}\le\frac LS+\frac{1-S/2}{4(1-S)}<\frac LS+\frac13.
\]
Subtract this from \eqref{bal:eq:Flow}. Since $S\le3/200$,
\[
 \F(y,h)-Q(y)>y^2\left(\frac1{100S}-\frac13\right)
 \ge\frac{y^2}{200S}.
\]
At $y=0$, continuity and $\kap\ge0$ give the assertion.
\end{proof}

\section{Comparison along complete physical mean fibers}\label{bal:sec:fibers}

\emph{Argument guide.}
A stationary point along a mean fiber can be controlled without locating
its root exactly. We prove strict increase of the mean derivative and
monotonicity of the relevant entropy derivatives. A comparison point with
validated signs then controls every possible stationary point below it;
the chart construction retains the entire physical fiber.

We prove both the mean and entropy derivative monotonicity needed by the comparison argument directly from the perspective formula.
For $0<t<1/2$, write
\[
 \nu=t(1-t),\quad r=1-2t,\quad M=-\ln\nu,\quad
 \ell(t)=H(t)/r,\qquad T(t)=P(t)+\frac{H(t)r}{\nu M}.
\]
If $\ell(t)=E/s$, differentiation gives
\begin{equation}\label{bal:eq:Ts}
 \F_s(s,E)=\tfrac12T(t),\quad
 \ell'(t)=M/r^2>0,\quad
 T'(t)=\frac{H(t)(r^2-M)}{M^2\nu^2}<0.
\end{equation}
Thus $\F_{ss}>0$ and $\F_{sE}<0$. The exact mean derivative is
\begin{equation}\label{bal:eq:gy}
 g_y=\F_s(y,h)-\tfrac12\F_s(A+y,E_u)+\tfrac12\F_s(A-y,E_w).
\end{equation}

\begin{lemma}[Entropy derivative monotonicity]\label{bal:lem:monotone}
On every strict physical fiber, $g_{dy}>0$. If $E_w\ge11E_u$, then also
$g_{E_u y}>1/(5h)>0$.
\end{lemma}
\begin{proof}
\emph{Proof outline.}
The mean derivative of the gap gives the mixed derivatives directly.
The signs of the perspective derivatives settle the entropy-difference
derivative. For the derivative in the smaller entropy, quantitative bounds
at the two outer contacts show that the favorable term dominates under
the stated entropy-ratio condition.

Differentiating \eqref{bal:eq:gy} at fixed $h$ gives
\[
 g_{dy}=-\tfrac12\{\F_{sE}(A+y,E_u)+\F_{sE}(A-y,E_w)\}>0.
\]
Set $K=-E\F_{sE}(s,E)$. In the auxiliary root $t$,
\[
 K=\frac{H(t)^2r(M-r^2)}{2\nu^2M^3}.
\]
The elementary entropy bounds
\begin{equation}\label{bal:eq:Hbounds}
 \nu M\le H(t)\le2\nu M
\end{equation}
give $0<K<2$. The lower bound in \eqref{bal:eq:Hbounds} follows from
$H-\nu M=-t^2\ln t-(1-t)^2\ln(1-t)\ge0$.
For the upper bound put $r=1-2t$. It is equivalent to
\[
 \ln\frac{1+r}{1-r}\ge r\ln\frac4{1-r^2}.
\]
The difference is zero at the two limiting endpoints; its derivative
$2-\ln(4/(1-r^2))$ changes sign once from positive to negative.

At the outer $u$ term the auxiliary root is less than $1/1000$, since
\[
 \frac{E_u}{A+y}\le\frac{H(S)}{1-2S}<0.0012,
 \qquad \ell(1/1000)>0.006.
\]
There $r>499/500,M>6$, and \eqref{bal:eq:Hbounds} gives
\[
 K\ge\frac r2\left(1-\frac{r^2}{M}\right)>\frac25.
\]
At fixed $E_w$, \eqref{bal:eq:gy} therefore gives
\[
 2g_{E_u y}
 =-\frac{K(y/h)}h+\frac{K((A+y)/E_u)}{E_u}
 >-\frac2h+\frac{2}{5E_u}\ge\frac{2}{5h}
\]
when $E_u\le h/6$, equivalently $E_w\ge11E_u$.
\end{proof}

\begin{lemma}[Uniform strict mean-fiber convexity]\label{bal:lem:fiberconvex}
Throughout the complete strict physical seam,
\[
 g_{yy}>98.
\]
In particular, $g_y$ strictly increases along every fixed-entropy physical fiber.
\end{lemma}
\begin{proof}
\emph{Proof outline.}
We separate the second mean derivative into its central perspective
term and two outer terms. Two ranges of the central latent coordinate
give a uniform lower bound, while the preceding contact estimate bounds
both outer subtractions. Their difference stays strictly positive along
every complete physical fiber.

The preceding formulas and homogeneity give
\[
 s\F_{ss}(s,E)=K,\qquad
 \F_{ss}(s,E)=\frac{H(t)^3(M-r^2)}{2E\nu^2M^3}.
\]
For the central term $s=y,E=h$, first suppose $t\le1/4$.
Then $r\ge1/2$, $M\ge\ln4>4/3$, and
\[
 K\ge\frac r2\left(1-\frac{r^2}{M}\right)>\frac1{16}.
\]
Since $y\le S$, this gives $\F_{ss}(y,h)>625$.
If $t\ge1/4$, then $r\le1/2$, $\nu\ge3/16$, and
$M-r^2>4/3-1/4=13/12$. Since $H\ge\nu M$, consequently
\[
 \F_{ss}(y,h)\ge\frac{\nu(M-r^2)}{2h}
 >\frac{13}{128h}>\frac{1625}{16}.
\]
Here concavity and $p+q<S$ imply $h\le H(m)<1/1000$;
for example $H(m)\le m[\ln(1/m)+1]<m(15L+1)<1/1000$.
At $y=0$, the continuous extension gives $\F_{ss}(0,h)=2L/h$.

For either outer term, $K<2$ gives
$\F_{ss}(A\pm y,E)<2/(A\pm y)\le2/(1-2S)<3$.
Differentiating \eqref{bal:eq:gy},
\[
 g_{yy}=\F_{ss}(y,h)-\tfrac12\F_{ss}(A+y,E_u)
                         -\tfrac12\F_{ss}(A-y,E_w)
 >\frac{1625}{16}-3>98.\qedhere
\]
\end{proof}

Chart 0 uses $(x,u)\in(0,1]\times(0,1)$, while chart 1 uses
$(x,u)\in(0,1)^2$. Both retain the complete fiber $0<v<1$:
\begin{align}
 \text{chart 0: }&q=mx,\quad p=mxu,\quad y=(S-2p)v,\label{bal:eq:chart0}\\
 \text{chart 1: }&q=S-mx,\quad p=mxu,\quad
 y=S-2mx+2mx(1-u)v.\label{bal:eq:chart1}
\end{align}
Chart 0 owns $q=m$; chart 1 uses $q>m$. Closed arithmetic hulls at the common interface are conservative.

\begin{lemma}[Comparison-point owner]\label{bal:lem:comparison}
Let $B$ be an exact $(x,u)$ rectangle in either chart. Suppose at one fixed rational $0<v_+<1$ interval arithmetic proves, uniformly on $B$, that $g_y(v_+)>0$ and either
\[
 g_d(v_+)<0,
 \qquad\text{or}\qquad E_w\ge11E_u\ \text{and}\ g_{E_u}(v_+)<0.
\]
Then the indicated entropy derivative is negative at every physical $g_y=0$ point over $B$. The same conclusion holds at $v_+=1$ without any $g_y(v_+)$ test. Alternatively, $g_y(1)<0$ makes every physical fiber over $B$ root-free.
\end{lemma}
\begin{proof}
Strict increase of $g_y$ puts each possible root below an interior comparison point with positive $g_y$. Lemma~\ref{bal:lem:monotone} makes the relevant entropy derivative at that root strictly smaller than at the comparison point. At $v_+=1$, every strict physical point is already below the comparison point. If $g_y(1)<0$, strict increase excludes a zero before the cap.
\end{proof}
No floating root estimate authorizes an acceptance. Such an estimate can only propose $v_+$; the stated uniform strict interval inequalities are the certificate. There is no inward $v$ cutoff.

\subsection{Cancellation-safe evaluation}
Put $s_0(E)=1-2H^{-1}(E)$. Since $\et(E)=\F(s_0(E),E)$,
\begin{equation}\label{bal:eq:stable}
 \Ph_E(s,E)=-\frac{T(p)}{P(p)}
 +(s_0-s)\int_0^1\F_{sE}(s+t(s_0-s),E)\,dt,
 \quad p=H^{-1}(E).
\end{equation}
This is an exact identity, not an asymptotic approximation. The checker encloses the integral over the whole joining interval and keeps $s_0-s$ in factored chart coordinates. For example, at the $u$ marginal in chart 0,
$s_0-(A+y)=(S-2p)(1-v)$. In chart 1 the two marginal factors are
$Sx(1-u)(1-v)$ and $Sx(1-u)v$.

For centered first-order bounds the checker uses
$\Ph_{Es}=-\F_{sE}$ and the exact differentiated capacity identity
\begin{align}
 \Ph_{EE}(s,E)={}&-\frac{4\F_{sE}(s_0,E)}{P(p)}
 +\frac{4\F_{ss}(s_0,E)}{P(p)^2}
 +\frac{2P'(p)\F_s(s_0,E)}{P(p)^3}\notag\\
 &+(s_0-s)\int_0^1\F_{sEE}(s+t(s_0-s),E)\,dt.
 \label{bal:eq:stable2}
\end{align}
Direct and centered enclosures are intersected only because each rigorously encloses the identical exact quantity.

\section{Explicit root-free and equality collars}\label{bal:sec:collars}
\subsection{Scalar root-free collars}
Since $\F_s$ increases in its first argument and decreases in its second, \eqref{bal:eq:gy} implies the whole-fiber inequality
\begin{equation}\label{bal:eq:scalar}
 g_y\le\F_s(S,h_-)-\tfrac12\F_s(A,H(p_+))
                    +\tfrac12\F_s(A,H(q_-))
\end{equation}
whenever $p\le p_+,q\ge q_-,h\ge h_-$. Let $h_0=2S/3$.
Because $H(S/20)<h_0$, the condition $h\ge h_0$ forces $q>S/20$.
The following scalar certificates were independently evaluated at 192 and 384 bits:
\begin{center}\begin{tabular}{p{0.57\linewidth}rr}\toprule
Region & bound used & actual upper bound\\\midrule
$h\ge h_0$, $p\le S/8192$ & $g_y<-1/8$ & $<-0.1350137320$\\
Chart 1, $x\le1/64$, all $u$ & $g_y<-9/10$ & $<-0.9320881842$\\\bottomrule
\end{tabular}\end{center}
For the first row use $(p_+,q_-,h_-)=(S/8192,S/20,h_0)$ in \eqref{bal:eq:scalar}. For the second use
\[
 (p_+,q_-,h_-)=(m/64,S-m/64,H(S-m/64)/2).
\]
Both charts satisfy $p\le S/8192$ whenever $u\le1/4096$. These are uniform scalar bounds on the entire regions, not evaluations at representative points.

\subsection{The whole-entropy equality collar}
At fixed $h$, the exact signed extension of \eqref{bal:eq:gap} satisfies
\[
 g(0,0)=0,\qquad \nabla_{(y,d)}g(0,0)=0.
\]
Let $K=\nabla^2_{(y,d)}g$. A 544-leaf exact binary partition of
\begin{equation}\label{bal:eq:eqrect}
 H(m)-1/2000\le h\le H(m),\qquad -10^{-5}\le d\le0,
 \qquad 0\le y\le10^{-5}
\end{equation}
passed independent 384-bit replay after 192-bit discovery. Its strict enclosures give
\begin{equation}\label{bal:eq:Hessian}
 g_{yy}>2500,\qquad \det K>1000,\qquad g_{yy}-|g_{yd}|>1400.
\end{equation}
The recorded lower bounds exceed $2538.32$, $1111.09$, and $1483.79$, respectively. Sylvester's criterion proves positive definiteness on the full rectangle.

For $z=(y,d)\ne0$ at fixed $h$, the segment from $0$ to $z$ stays in this rectangle. The fundamental theorem of calculus gives
\[
 z\cdot\nabla g(z)=\int_0^1z^{\mathsf T}K(tz)z\,dt>0,
 \qquad g(z)=\int_0^1(1-t)z^{\mathsf T}K(tz)z\,dt>0.
\]
Consequently $g_y=0,d<0$ implies $g_d<0$.

For chart 0, $u\ge31/32$ implies
\begin{equation}\label{bal:eq:attach}
 |d|=\frac{H(q)-H(p)}2
 \le\frac{q(1-u)P(p)}2
 \le\frac m{64}\ln\frac2m<10^{-5}.
\end{equation}
Indeed $p\ge q/2$, $P(p)\le\ln(2/q)$, and $q\ln(2/q)$ increases for $q\le m$.
Entropy concavity gives $h\le H(m)$, and the lower $h$ endpoint in \eqref{bal:eq:eqrect} lies strictly below $h_0$.
Writing $Y=10^{-5}$, \eqref{bal:eq:Hessian} and $|d|\le Y$ give
\[
 g_y(Y,d)=\int_0^1\{Yg_{yy}(tY,td)+d g_{yd}(tY,td)\}\,dt>1400Y>0.
\]
If the physical fiber ends before $Y$, every root is in the rectangle already; otherwise strict fiber monotonicity puts every root below $Y$. Thus every full chart-0 fiber with $h\ge h_0,u\ge31/32$ is owned. If $h\le h_0$, Theorem~\ref{bal:thm:low} owns it by value.

A separate wider rectangle with $|d|,y\le2\cdot10^{-5}$ has 2,759 leaves and also passed 384-bit replay. It extends the attachment to $u\ge15/16$. This strengthening is recorded separately; it is not needed to reinterpret the scopes of the 544-leaf records.

\subsection{The previously saved large transverse collar}
The larger high-entropy rectangle
\[
 H(m)-1/10000\le h\le H(m),\quad -1/10000\le d\le0,
 \quad 0\le y\le1/25000
\]
was freshly replayed at 384 bits: all 2,114 leaves and 240 exact parameter attachments passed. Its bounds give $g_{yy}>2521.9283$ and $\det K>38.76645$. For a parameter box, conservative attachment bounds are
\[
 H(m)-h\le H(m)-\tfrac12[H(p_{\min})+H(q_{\min})],
\]
\[
 |d|\le\tfrac12[H(q_{\max})-H(p_{\min})],\qquad
 y\le S-2p_{\min}.
\]
These verify the attachment rather than infer it from a plot or from the labels of historical boxes. In particular, chart 1 near $u=1$ is a capacity boundary, not necessarily an equality boundary.

\section{Exact two-chart assembly}\label{bal:sec:union}

\emph{Argument guide.}
The local estimates now have to cover both parameter charts. We first
describe the recovered compact partition, then account for each omitted
strip using an analytic bound or collar, and finally join the chart
interfaces. The geometry and arithmetic records have separate roles:
one establishes coverage, the other validates each stated acceptance.

Every strict seam point belongs to exactly one of \eqref{bal:eq:chart0} and \eqref{bal:eq:chart1}, assigning $q=m$ to chart 0. Each finite owner above applies to the whole open physical fiber $0<v<1$.

The recovered compact parameter rectangle is
\[
 R=[1/64,1]\times[1/64,63/64].
\]
Its original exact partition contains 4,042 chart-0 parents and 871 chart-1 parents. The geometry audit reconstructs all binary paths and exact rational boxes. It rejects repeated leaves, nested prefixes, missing children, unreachable coordinates, and uncovered coordinate slabs; equality of areas alone is not its coverage criterion.

\begin{center}\begin{tabular}{lrr}\toprule
Chart-0 compact owner & original parents & final leaves\\\midrule
Fresh root-free fibers &189&189\\
Large transverse collar &152&152\\
Four formerly second-order packets &4&4\\
Literal comparison frontier &1757&1757\\
Complementary minimum-exclusion frontier &1940&2551\\\midrule
Total compact chart 0 &4042&4653\\\bottomrule
\end{tabular}\end{center}
The acceptance tests in the two remaining chart-0 categories are as
follows. For a closed parameter rectangle
$B=[x_-,x_+]\times[u_-,u_+]$, all derivative enclosures below concern
the frozen partials $g_y|_{h,d}$ and $g_d|_{y,h}$, evaluated after
substitution of the chart map; they are not derivatives with respect
to chart variables. The four formerly second-order packets are now
accepted by the comparison-point test
\[
 0<v_+<1,\qquad \inf[g_y(B,v_+)]>0,
 \qquad \sup[g_d(B,v_+)]<0.
\]
In parent order $0,1,2,3$ of the four-packet record below,
the rational comparison levels are
\[
 \frac{121730867}{268435456},\quad
 \frac{863517493}{1073741824},\quad
 \frac{422924933}{536870912},\quad
 \frac{1063822613}{1073741824}.
\]
Lemma~\ref{bal:lem:comparison} therefore gives $g_d<0$ at every
physical $g_y=0$ point over each packet. No historical second-order
acceptance is used for these four packets.

The 2,551 complementary leaves have the following precise owners:
\begin{center}
\begin{tabular}{lr}
\toprule
Acceptance criterion & Leaves\\\midrule
$\inf[g_y(B,v_+)]>0$, $\sup[g_d(B,v_+)]<0$, $0<v_+<1$
 &1,667\\
$\sup[g_d(B,1)]<0$ &584\\
$h_{\max}(B)\le 1/15000=2S/3$ &253\\
Whole-entropy equality-collar attachment &47\\\bottomrule
\end{tabular}
\end{center}
The first two rows use Lemma~\ref{bal:lem:comparison}; the cap row
needs no sign condition on $g_y(B,1)$, because every strict physical
point has $v<1$. For the third row the checker uses
\[
 h_{\max}(B)=\tfrac12\{H(mx_+u_+)+H(mx_+)\};
\]
Theorem~\ref{bal:thm:low} then gives $g\ge B_0\ge0$ on the complete
fiber. These value owners exclude the premise $g<0$, rather than
claim a derivative sign at nonnegative points. For the final row it
checks
\[
 u_-\ge31/32,\qquad
 h_{\min}(B):=\tfrac12\{H(mx_-u_-)+H(mx_-)\}
 \ge H(m)-1/2000.
\]
Entropy concavity gives $h\le H(m)$; the attachment estimate
\eqref{bal:eq:attach}, the Hessian bounds \eqref{bal:eq:Hessian},
and strict mean-fiber convexity put every possible $g_y=0$ point
inside the equality rectangle and give $g_d<0$ there.
Thus each of the 2,551 leaves proves the minimum-exclusion conclusion
on its entire strict physical fiber. None of these leaves uses the
$g_{E_u}$ alternative.

The controlling 384-bit records are
\path{reconstructed_coverage_four_packet_384.json} and
\path{reconstructed_fiber_sk1_remaining_part0_replay_384.json}
through \path{reconstructed_fiber_sk1_remaining_part3_replay_384.json},
all in \path{experiments/results/}. The latter four route counts are,
in the table's order,
$(488,20,253,16)$, $(274,291,0,11)$,
$(365,170,0,2)$, and $(540,103,0,18)$.

The 3,697 frontier parents are matched exactly against the conservative source frontier. Its complementary adaptive parts have 777, 576, 537, and 661 leaves. All 4,308 frontier leaves passed 384-bit replay with zero failures. The extra chart-0 strip
\[
 [1/64,1]\times[1/4096,1/64]
\]
has 36 rational octave parents and 92 accepted leaves, also with zero 384-bit failures.

The remaining chart-0 parameter pieces have explicit owners:
\begin{itemize}
\item $x\le1/64$: $h\le H(m/64)<h_0$, so Theorem~\ref{bal:thm:low} applies.
\item $u\le1/4096$: the low-entropy theorem applies if $h\le h_0$; otherwise the scalar root-free collar applies.
\item $u\ge63/64$: the low-entropy theorem applies if $h\le h_0$; otherwise \eqref{bal:eq:attach} and the full equality collar apply.
\end{itemize}
The exact unions use closed hulls for arithmetic and a consistent half-open convention to assign shared parameter boundaries.

For chart 1, 796 compact parents are root-free, 68 attach to the large Hessian collar, and the remaining seven split into 20 root-free leaves. Thus its compact region has 884 leaves. The low-$u$ strip has 43 leaves; the high-$u$ strip with $x\le7/8$ has 19. The corner $x\ge7/8,u\ge63/64$ attaches to the large collar. The two remaining scalar rectangles $x\le1/64$ and $u\le1/4096$ are root-free. In the latter case $q\ge m$ implies $h\ge H(m)/2>h_0$ automatically.

The complete chart-0 unit square has 4,748 owner rectangles: 4,653 compact leaves, 92 low-ratio leaves, and three analytic or collar rectangles. The complete chart-1 unit square has an exact 949-cell union with zero residuals. It proves even the literal derivative implication on chart 1. Both charts retain their full physical fibers, including arbitrarily close approaches to $v=0,1$. Their combined 5,697-rectangle union passes the exact global audit in \code{experiments/results/reconstructed_global_seam_me_union_384.json}, with zero residuals and current-closure 384-bit evidence. Their union, with the analytic low-entropy theorem, proves Theorem~\ref{bal:thm:ME}.

\subsection{What an arithmetic acceptance means}
All displayed endpoints are exact rationals. Every latent root is enclosed by validated strict endpoint signs using a proved monotone defining function. Numerical root guesses only seed that enclosure. Arb ball operations round outward. A strict derivative owner requires a strictly signed enclosure; an unresolved or failed enclosure never counts as acceptance. Source dependency hashes are checked before and after each replay. The union audit separately checks the exact geometry and the linkage of every new leaf to its current-closure 384-bit replay.

Strict signs are tested on the full-precision Arb balls before
serialization. The human-readable midpoint--radius strings are
rounded summaries and are not substituted for these acceptance
tests: enlarging a displayed radius can erase a strict sign.
Reproduction therefore re-evaluates the stated inequalities from
the exact rational leaf and comparison point with the recorded
source closure. A targeted 384-bit re-evaluation of the nine
complementary rows whose displayed radii obscure a required strict sign
passed both strict comparisons. Exact rational signed endpoints and the
source commitments are retained in
\path{seam_acceptance_saved_evidence_audit.json}; this is a selected
replay, not a new execution of the full seam cover.

Saved replay audits and fresh arithmetic are different operations. Reading a saved passing result and checking its hash establishes linkage, not a new numerical replay. The package gives both commands and preserves the exact inputs needed to re-evaluate the arithmetic.

\section{Clause (ii): the full deterministic-u facet}\label{bal:sec:ii}

\emph{Argument guide.}
The facet is divided into three adjoining ranges. Positive curvature
and the zero value and slope at the equality edge prove the first range.
A finite value cover bridges to an explicit endpoint collar, whose scalar
bounds remain valid up to the zero-entropy limit. The final paragraph
checks that these ranges meet at their exact endpoints.

For this section derivatives are of the \emph{composed cap function}, not the frozen derivatives in Clause (i). Put
\[
 y=St,\quad a=(S-y)/2,\quad b=a+zy,
 \qquad U(y,z)=\frac L2G\bigl(y,\HH(a),\HH(b)\bigr),
 \quad 0\le t,z\le1.
\]
Writing $\delta=b-a=zy$ and $h=[H(a)+H(b)]/2$, its exact formula is
\[
 U=\F(y,h)+\frac\delta4[P(a)-P(b)]-\F(\delta,h)
 -\Ph(A,h)+\frac12\Ph(A-y,H(b)).
\]
The omitted marginal term vanishes because $\Ph(1-2a,H(a))=0$.
The equality edge satisfies $U(0,z)=U_y(0,z)=0$.
Indeed, differentiating $\Ph(1-2a,H(a))=0$ gives
$P(a)\Ph_E=2\Ph_s$, which cancels the first derivative of the last two terms at $y=0$; the first three terms have zero first derivative there.

\subsection{Curvature through \texorpdfstring{\(t=255/256\)}{t = 255/256}}
The corrected arithmetic proves $U_{yy}>0$ on $0\le t\le255/256$, $0\le z\le1$. The new 256-bit component counts are:
\begin{center}\begin{tabular}{lr}\toprule
Region & accepted leaves\\\midrule
Origin, $0\le t\le1/4096$ &8\\
Early and middle bands, $1/4096\le t\le13/16$ &3198\\
Late band, $13/16\le t\le255/256$, low $z$ &32469\\
Late band, $13/16\le t\le255/256$, high $z$ &18536\\\bottomrule
\end{tabular}\end{center}
The late split is $z=1/16$. The low-$z$ chart is $w=z/(1-t)$: exact $(t,w)$ leaves are mapped once to enclosing physical $(t,z)$ hulls. Six exact $t$ strips cover the bridge, and each chart root contains the required target strip. The partition and arithmetic audits have zero residuals. Two integrations from $y=0$ prove $U_y>0,U>0$ for $0<t\le255/256$.

Two historical implementation defects were repaired. First, the low-$z$ replay applied its chart map twice. Second, the fast curvature drivers set the input jet degree to three even though two $y$ differentiations preceded a quadratic model with cubic remainder. If $c_{ij}$ are the Taylor coefficients of $U_{yy}$ and $u_{ij}$ those of $U$, then
\[
 c_{ij}=(i+2)(i+1)u_{i+2,j}.
\]
The cubic remainder needs all $c_{ij}$ with $i+j=3$, hence input degrees at least $(D_y,D_z)=(5,3)$. Filling unavailable coefficients with zero is invalid. The repaired evaluator reserves these degrees and includes the complete directional cubic remainder
\[
 \sum_{i+j=3}\sup_B|c_{ij}|\,r_y^ir_z^j.
\]
It also reserves $(6,3)$ for a quadratic enclosure of $U_{yyy}$, $(7,3)$ for $U_{yyyy}$, and $(7,5)$ for a quartic/quintic enclosure of $U_{yy}$. The affected historical numerical margins are not reused. Failed corrected cells were split exactly and freshly certified.

\subsection{Finite value bridge}
Put $p=a,q=b,\delta=q-p=zy$, and
\[
 h=\tfrac12[H(p)+H(q)],\qquad
 2\Delta(p)=2H(m)-H(p)-H(S-p).
\]
The same-side bound gives
\begin{equation}\label{bal:eq:iiB}
 U\ge B_0=\F(y,h)+\frac\delta4[P(p)-P(q)]-\F(\delta,h)-2\Delta(p).
\end{equation}
On $255/256\le t\le999/1000$, $0\le z\le1$, the 4,901-leaf
value cover certifies $B_0>0$ on every closed leaf, and hence
$U\ge B_0>0$ by \eqref{bal:eq:iiB}. More precisely, write
$b_{ij}(c)=\partial_y^i\partial_z^j B_0(c)/(i!j!)$ at the
center $c$ of the physical $(y,z)$ leaf, and let $r_y,r_z$ be its
coordinate radii. The accepted test is strict positivity of the
outward-rounded lower bound for
\[
 b_{00}(c)
 -\sum_{1\le i+j\le2}|b_{ij}(c)|r_y^i r_z^j
 -\sum_{i+j=3}\sup_{(y,z)\in B}|b_{ij}(y,z)|r_y^i r_z^j.
\]
The coefficients are enclosed by degree-three bivariate value jets;
Taylor's theorem supplies the full cubic remainder shown here.
The archived 256-bit replay recomputes this test on every exact
leaf and records zero failures. The replay file is
\path{experiments/results/clause_ii_b0_band_repaired_256_2026-09-07.json};
its discovery partition is
\path{experiments/results/clause_ii_b0_band_d24_192.json}.
The geometry checker reconstructs each leaf from its binary split
path, checks exact coordinates, rejects prefix overlap, and verifies
complete coverage of the stated rectangle. No $y$ derivatives are
taken before constructing this value Taylor model, so it does not
have the discarded-degree defect of the old curvature computation.

\subsection{Explicit final endpoint collar}
Let $p_0=1/20000000$ and $y_0=S-2p_0$. Thus $p\le p_0$ is exactly $t\ge999/1000$. For $0\le\delta\le y_0$, convexity in size, decrease in entropy, and $\F(0,h)=0$ imply
\[
 \F(y,h)-\F(\delta,h)
 \ge\left(1-\frac\delta{y_0}\right)
 \F\left(y_0,\frac{H(p_0)+H(p_0+\delta)}2\right).
\]
Also
\[
 P(p)-P(p+\delta)\ge\ln(1+\delta/p_0),\qquad
 2\Delta(p)\le2H(m)-H(S).
\]
Consequently
\begin{align}\label{bal:eq:endpoint}
 B_0\ge{}&\left(1-\frac\delta{y_0}\right)
 \F\left(y_0,\frac{H(p_0)+H(p_0+\delta)}2\right)
 +\frac\delta4\ln(1+\delta/p_0)\notag\\
 &-\{2H(m)-H(S)\}.
\end{align}
The first term decreases in $\delta$ and the second increases. On each of the 256 equal exact cells of $[0,y_0]$, evaluate the first at the right endpoint and the second at the left endpoint. Every resulting lower bound is positive; the minimum exceeds
$1.45179828194474458\cdot10^{-6}$. For $y_0\le\delta\le y$, the edge term alone at $(\delta,p)=(y_0,p_0)$ exceeds the same entropy correction. The perspective difference is nonnegative because $\delta\le y$. This proves $U>0$ for the whole final collar. The zero-entropy endpoint uses the declared lower-semicontinuous owner.

The three ranges meet at the exact values $255/256$ and $999/1000$, so no cap interval is omitted. Clause (ii) follows.

\section{Clause (iii): an analytic deterministic-w proof}\label{bal:sec:iii}

\emph{Argument guide.}
Entropy monotonicity first moves the free marginal to its cap, reducing
the facet to a deterministic edge-cost comparison. An even-derivative
expansion of binary entropy then proves the required Jensen-gap bound
throughout the seam; the limiting endpoint follows by lower semicontinuity.

Use bit units here. On the facet $e_w=\HH(c)$, $0<e_u<\HH(a)$, write
$p=\HH^{-1}(e_u)$, $e=\HH(p)$, $f=\HH(c)$, $h_b=(e+f)/2$.
Let $k(e,f)=j(p,c)-F(c-p,h_b)$ denote the established jointly convex nonnegative entropy gap. It satisfies $k(f,f)=0$. Convexity then makes $e\mapsto k(e,f)$ nonincreasing for $0\le e\le f$: if $e_1<e_2<f$, convexly interpolate $e_2$ between $e_1$ and $f$.

Because $F_E<0$, the same-side estimate gives
\begin{align*}
 G&\ge k(e,f)+F(y,h_b)-4\Delta_H(a,c)\\
 &\ge k(\HH(a),\HH(c))
      +F\left(y,\frac{\HH(a)+\HH(c)}2\right)-4\Delta_H(a,c)\\
 &=j(a,c)-4\Delta_H(a,c),
\end{align*}
where $\Delta_H(a,c)=\HH(m)-[\HH(a)+\HH(c)]/2$ and $c-a=y$.
Put $r=(c-a)/2$. For every $k\ge1$,
\[
 \HH^{(2k)}(m)=-\frac{(2k-2)!}{L}
 \left(\frac1{m^{2k-1}}+\frac1{(1-m)^{2k-1}}\right)<0.
\]
The convergent Taylor expansion for $0<r<m$ yields
\[
 j(a,c)-4\Delta_H(a,c)
 =\sum_{k\ge2}\frac{-2(k-1)}{k(2k-1)!}\HH^{(2k)}(m)r^{2k}>0.
\]
The limiting endpoint uses lower semicontinuity. Thus clause (iii) needs no further two-dimensional interval tiling.

\section{A deterministic region covered by the small-boundary theorem}

\begin{proposition}[Unconditional small-mean part of clause (iv)]
For
\[
                 0<a<b\leq 10^{-2}
\]
one has
\[
 \GDB(a,b)=j(a,b)-\Phi\!\left(1-a-b,
              \frac{\Hd(a)+\Hd(b)}2\right)\geq0.
\]
In particular, the part with $a+b>10^{-4}$ needs no SC+ computation.
\end{proposition}

\begin{proof}
For a random variable $U\in[0,1]$, equality
$\mathbb E\Hd(U)=\Hd(\mathbb EU)$ in entropy concavity, with
$0<\mathbb EU<1$, forces $U$ to be constant.  Thus the four deterministic-cap
moments
\[
 (\mu_u,\mu_w,e_u,e_w)=(a,b,\Hd(a),\Hd(b))
\]
have only the constant marginals $U=a$ and $W=b$.  Hence
$\zeta=j(a,b)$.  The lower model has the same value because
\[
 L_4=F\!\left(b-a,\frac{\Hd(a)+\Hd(b)}2\right)
       +\kappa(a,b)=j(a,b).
\]
Consequently the Bellman target and the pure gap coincide at this tuple:
$T=G=\GDB$.  The unconditional same-side part of SB-1 applies because
$0<a<b\leq10^{-2}$, and gives $T\geq0$.  This use of a target theorem is
type-correct precisely because $\zeta=L_4$ has just been proved at the
deterministic tuple.
\end{proof}

\begin{proposition}[One-dimensional reduction of clause (iv)]\label{bal:prop:oned}
Let
\[
 D(m,e)=4\{\Hd(m)-e\}-\Phi(1-2m,e)
\]
and define
\[
 e_{\min}(m)=
 \begin{cases}
 \frac12\Hd(2m),&0<m\leq\frac14,\\[2mm]
 \frac12\{1+\Hd(2m-\frac12)\},&\frac14\leq m<\frac12.
 \end{cases}
\]
If the two scalar endpoint inequalities
\[
                         D(m,e_{\min}(m))\geq0             \tag{\thesection.R1}\label{bal:eq:cap-R1}
\]
hold on their displayed intervals, then clause (iv) holds on its complete
domain.  Thus its two-dimensional sign question reduces to two
one-dimensional scalar inequalities.
\end{proposition}

\begin{proof}
Write $a=m-d$, $b=m+d$.  The global two-point comparison in Proposition~\ref{bal:prop:double-cap-entropy}, also proved in \eqref{bal:eq:sb-edge-jensen}, gives
\[
 j(a,b)\geq4\left\{\Hd(m)-\frac{\Hd(a)+\Hd(b)}2\right\}.   \tag{\thesection.R2}\label{bal:eq:cap-R2}
\]
For fixed $m$, the entropy average
\[
 e(d)=\frac{\Hd(m-d)+\Hd(m+d)}2
\]
decreases continuously from $\Hd(m)$ to $e_{\min}(m)$ as $d$ increases from
zero to $\min\{m,1/2-m\}$.  The established entropy-curvature input
$\Phi_{ee}(s,e)>0$ says that $e\mapsto D(m,e)$ is concave.  Moreover
$D(m,\Hd(m))=0$.  If \eqref{bal:eq:cap-R1} holds, concavity therefore gives $D(m,e(d))\geq0$
on the whole chord.  Combining this with \eqref{bal:eq:cap-R2} yields
\[
 \GDB(a,b)=j(a,b)-\Phi(1-2m,e(d))\geq D(m,e(d))\geq0.
\]
Endpoint values follow by continuity or their declared lower-semicontinuous
owners.
\end{proof}

\section{A scalar hyperbolic inequality}

We distinguish natural normalization from entropy differentiation.  Set
$L=\ln2$ and define
\[
 h_{\mathrm n}(t)=L\HH(t),\quad J_{\mathrm n}(t)=LJ(t),\quad
 j_{\mathrm n}(a,b)=Lj(a,b),\quad
 \Lambda_{\mathrm n}(t)=\frac{2h_{\mathrm n}(t)}{1-2t}.
\]
The natural perspective and its associated functions are
\[
 F_{\mathrm n}(s,E)=L F(s,E/L),\qquad
 \eta_{\mathrm n}(E)=L\eta(E/L),\qquad
 \Phi_{\mathrm n}(s,E)=L\Phi(s,E/L).
\]
Unadorned $F,\eta,\Phi,\GDB,\GHF$ always retain their bit-normalized
definitions.  We write $\partial_E$ for differentiation with respect to
natural entropy.  In particular a natural-normalized gap is explicitly
$L\GDB$ or $L\GHF$, not a redefinition of the unadorned symbol.  Put
\[
 p_x=\frac{1-\tanh x}{2},\qquad
 K(x)=\ln(2\cosh x),\qquad
 h(x)=K(x)-x\tanh x.
\]
Then $J_{\mathrm n}(p_x)=2x$, $h_{\mathrm n}(p_x)=h(x)$, and
\[
 \Lambda_{\mathrm n}(p_x)=2\ell(x),\qquad
 \ell(x)=\frac{h(x)}{\tanh x}=\frac{K(x)}{\tanh x}-x.
\]

\begin{lemma}[Hyperbolic collar lemma]\label{bal:lem:hyp}
For $x>0$ define
\[
 f(x)=xK(x)+x\sinh^2x-K(x)\sinh x\cosh x.
\]
Then $f(x)>0$.  Moreover $x\ell(x)$ is strictly decreasing, and hence
\[
                         \ell(x/2)>2\ell(x).
\]
\end{lemma}

\begin{proof}
Write $K(x)=x+r(x)$, where $r(x)=\ln(1+e^{-2x})<e^{-2x}$.
Direct differentiation gives
\[
 f'(x)=\frac{\sinh x}{\cosh x}
 \{x(1+2\cosh^2x)-2K(x)\sinh x\cosh x\}.
\]
The expression in braces equals
\[
 2x+xe^{-2x}-r(x)\sinh(2x).
\]
But
\[
 r(x)\sinh(2x)<e^{-2x}\sinh(2x)
   =\frac{1-e^{-4x}}2<2x.
\]
Thus $f'(x)>0$, and $f(0)=0$ by continuity proves $f(x)>0$.
Finally, direct differentiation and multiplication by $\tanh^2x$ give
\[
             \frac{d}{dx}\{x\ell(x)\}=-\frac{f(x)}{\sinh^2x}<0.
\]
Comparing its values at $x/2$ and $x$ yields the last assertion.
\end{proof}

\section{The equal-cap diagonal}

\begin{proposition}[Strict quadratic diagonal for clause (iv)]
Fix $0<m<1/2$ and put $a=m-d$, $b=m+d$.  Then
\[
 \partial_d^2\GDB(m-d,m+d)\big|_{d=0}>0.
\]
Consequently, every compact subinterval of $0<m<1/2$ has an open uniform
diagonal collar on which $\GDB\geq0$, with equality only at $d=0$.
\end{proposition}

\begin{proof}
\emph{Proof outline.}
We first evaluate the entropy derivative of the candidate at a
deterministic cap and use the hyperbolic lemma to determine its sign
relative to the factor four. Taylor expansions of the entropy average
and the edge cost then identify a strictly positive quadratic coefficient
at the diagonal for each interior mean. Continuity and compactness give
the diagonal collar conclusion.

Let $m=p_x$, write $K=K(x)$, and set $s=1-2m=\tanh x$.
At $e=h_{\mathrm n}(m)$ the latent lower-inverse point in $F_{\mathrm n}(s,e)$ is $m$ itself.
Implicit differentiation of $\Lambda_{\mathrm n}(t)=2e/s$ gives
\[
 \partial_E F_{\mathrm n}(s,e)=\frac{2J_{\mathrm n}'(m)}{\Lambda_{\mathrm n}'(m)}
              =-\frac{2\sinh^2x}{K},
\]
whereas
\[
 \eta_{\mathrm n}'(e)=-2-\frac{\sinh(2x)}x.
\]
Therefore
\[
 4+\partial_E\Phi_{\mathrm n}(s,h_{\mathrm n}(m))
   =2-\frac{\sinh(2x)}x+\frac{2\sinh^2x}{K}
   =\frac{2f(x)}{xK}>0                                      \tag{\thesection.1}\label{bal:eq:cap-1}
\]
by Lemma~\ref{bal:lem:hyp}.

Now
\[
 j_{\mathrm n}(m-d,m+d)=-2J_{\mathrm n}'(m)d^2+O(d^4),\qquad
 \frac{h_{\mathrm n}(m-d)+h_{\mathrm n}(m+d)}2
     =h_{\mathrm n}(m)+\frac{J_{\mathrm n}'(m)}2d^2+O(d^4).
\]
It follows that
\[
 L\partial_d^2\GDB(m-d,m+d)\big|_{d=0}
   =-J_{\mathrm n}'(m)\{4+\partial_E\Phi_{\mathrm n}(s,h_{\mathrm n}(m))\}>0,
\]
because $J_{\mathrm n}'(m)<0$.  Continuity gives the collar statement; compactness
makes its width uniform on a fixed compact $m$-interval.
\end{proof}

\begin{proposition}[Strict clause-(v) value on its equal-cap diagonal]
For every $0<b<1/2$, the continuous extension of clause (v) to $a=b$
satisfies $\GHF(b,b)>0$.
\end{proposition}

\begin{proof}
Write $b=p_x$, $s=1-2b=\tanh x$, and $\delta=s/2$.  Since
$\eta_{\mathrm n}(h_{\mathrm n}(b))=F_{\mathrm n}(s,h_{\mathrm n}(b))=sJ_{\mathrm n}(b)$,
\[
 L\GHF(b,b)=2F_{\mathrm n}(\delta,h_{\mathrm n}(b))-\frac12\eta_{\mathrm n}(h_{\mathrm n}(b)).       \tag{\thesection.2}\label{bal:eq:cap-2}
\]
Let $t$ be the latent point in the first perspective term, so that
$\Lambda_{\mathrm n}(t)=2\Lambda_{\mathrm n}(b)$.  If $q=p_{x/2}$, Lemma~\ref{bal:lem:hyp} gives
\[
 \Lambda_{\mathrm n}(q)=2\ell(x/2)>4\ell(x)=2\Lambda_{\mathrm n}(b).
\]
Since $\Lambda_{\mathrm n}$ is strictly increasing on the lower probability branch, $t<q$;
since $J_{\mathrm n}$ is decreasing,
$J_{\mathrm n}(t)>J_{\mathrm n}(q)=J_{\mathrm n}(b)/2$.  Hence
\[
 F_{\mathrm n}(\delta,h_{\mathrm n}(b))=\delta J_{\mathrm n}(t)
   >\frac{\delta J_{\mathrm n}(b)}2=\frac14\eta_{\mathrm n}(h_{\mathrm n}(b)),
\]
and \eqref{bal:eq:cap-2} is strictly positive.
\end{proof}

\begin{proposition}[Exact reduction of clause (v) to clause (iv)]
For $0<a<b<1/2$, put
\[
 u=b-a,\qquad v=1-a-b,\qquad x=\frac{u+v}{2}=\frac12-a,
 \qquad \delta=\frac{v-u}{2}=\frac12-b,
\]
and $h=\{\Hd(a)+\Hd(b)\}/2$.  Then
\[
 \GHF(a,b)-\GDB(a,b)
   =2F(x,h)-F(u,h)-F(v,h)+\frac12F(2\delta,\Hd(b)).        \tag{\thesection.3}\label{bal:eq:cap-3}
\]
Consequently, the single perspective-curvature inequality asserting that the
right side of \eqref{bal:eq:cap-3} is nonnegative would make clause (v) a consequence of
clause (iv).  On $b=1/2$ the right side is zero; on $a=b$ it is strictly
positive by the preceding proposition.
\end{proposition}

\begin{proof}
Subtract the two displayed gap definitions and use $x=(u+v)/2$.  The
deterministic identity
$\eta(\Hd(b))=F(1-2b,\Hd(b))=F(2\delta,\Hd(b))$ gives \eqref{bal:eq:cap-3}.
If $b=1/2$, then $\delta=0$ and $u=v=x$, so its right side vanishes.  If
$a=b$, then $\GDB=0$ and the right side equals $\GHF(b,b)>0$.
\end{proof}

\section{A uniform estimate at the joint half-mean corner}

\begin{theorem}[Analytic half-mean collars]\label{bal:thm:corner}
There are constants $\varepsilon_{\rm DB},\varepsilon_{\rm HF}>0$ such that,
for
\[
 A=1-2a,\qquad B=1-2b,\qquad 0<B<A,
\]
the following hold:
\begin{align*}
 A+B<\varepsilon_{\rm DB}&\implies \GDB(a,b)>0,\\
 A+B<\varepsilon_{\rm HF}&\implies \GHF(a,b)>0.
\end{align*}
Thus neither clause needs interval leaves accumulating at the joint corner
$a=b=1/2$.
\end{theorem}

\begin{proof}
\emph{Proof outline.}
Normalize the two small biases by their sum so that the remaining
shape parameter lies in a compact interval. The implicit entropy and
perspective equations then have uniform analytic expansions. Symmetry
and diagonal vanishing identify the exact factors, and positive leading
coefficients dominate the uniform remainders near the joint corner.

Use natural-log normalization and the radial entropy expansion
\[
 \mathcal H(z):=h_{\mathrm n}\!\left(\frac{1-z}{2}\right)
   =\ln2-\frac{z^2}{2}-\frac{z^4}{12}+O(z^6),
 \qquad
 J_{\mathrm n}\!\left(\frac{1-z}{2}\right)
   =2\left(z+\frac{z^3}{3}+O(z^5)\right).                  \tag{\thesection.4}\label{bal:eq:cap-4}
\]
For a perspective coordinate $s$, its latent radial coordinate $z$ is
defined by
\[
                         s\mathcal H(z)=e z.                \tag{\thesection.5}\label{bal:eq:cap-5}
\]
After writing $A=\varepsilon\alpha$, $B=\varepsilon\beta$ with
$\alpha+\beta=1$, equation \eqref{bal:eq:cap-5} and the entropy-inverse equation are analytic
implicit equations in $z/\varepsilon$ and $z^2/\varepsilon^2$.  Their
solutions and remainders are uniform for
$0\leq\beta\leq\alpha\leq1$.  Substitution of \eqref{bal:eq:cap-4} gives
\[
 (\ln2)\GDB(a,b)
  =\frac{3-2\ln2}{48\ln2}(A-B)^2(A+B)^2
      +O\!\left((A-B)^2(A+B)^4\right),                    \tag{\thesection.6}\label{bal:eq:cap-6}
\]
and
\begin{align}
 (\ln2)\GHF(a,b)
 &=\frac{1}{48\ln2}\bigl\{(3-2\ln2)A^4
       +(18-20\ln2)A^2B^2 +(10\ln2-3)B^4\bigr\}
       +O((A+B)^6).                                        \tag{\thesection.7}\label{bal:eq:cap-7}
\end{align}
For completeness, the factor $(A-B)^2$ in \eqref{bal:eq:cap-6} is exact: the double-cap gap
is analytic and symmetric in $A,B$, and it vanishes on $A=B$.
Simultaneous complement sends $(A,B)$ to $(-A,-B)$ and preserves both gaps,
so their local series contain only even total degrees; this also justifies the
orders of the displayed uniform remainders.

The leading coefficient in \eqref{bal:eq:cap-6} is positive.  All three coefficients in the
braces in \eqref{bal:eq:cap-7} are positive: for example
$1/2<\ln2<7/10$ gives
\[
 3-2\ln2>0,\qquad 18-20\ln2>4,
 \qquad 10\ln2-3>2.
\]
After normalizing $\alpha+\beta=1$, the shape interval is compact.  The
uniform remainders in \eqref{bal:eq:cap-6}--\eqref{bal:eq:cap-7} are therefore dominated by their positive
leading terms for sufficiently small $\varepsilon$.  This proves both
claims.
\end{proof}

\begin{corollary}[Collars and the remaining clause]
The diagonal and joint-corner expansions remove the singular strata from
both cap expressions.  For clause (iv), Proposition~\ref{bal:prop:oned} replaces
the remaining two-dimensional core by \eqref{bal:eq:cap-R1}; Section~\ref{bal:sec:doublecomplete}
below certifies that
one-dimensional target and completes the clause.  Clause (v) has an analytic
positive equal-cap diagonal and an analytic joint half-mean corner.  Its
remaining region is covered by the derivative reduction and the complementary
estimates in Section~\ref{bal:sec:cap-sharp};
Theorem~\ref{bal:thm:m1m2} completes clause (v).
\end{corollary}

\section{A global monotonicity bridge for the double cap}

The following reduction is independent of Proposition~\ref{bal:prop:oned}.  It
isolates the exact inequality which would make the double-cap gap increase
strictly on every ray away from its zero diagonal.

For $x>0$ put
\[
 \mathcal Q(x)=\frac{\sinh(2x)}x,
 \qquad
 \mathcal C(x)=\frac{2\sinh^2x}{K(x)}.
\]
Given $\alpha>\beta>0$, let $\rho$ and $\tau$ be defined uniquely by
\begin{align}
 h(\rho)&=\frac{h(\alpha)+h(\beta)}2,\tag{\thesection.8}\label{bal:eq:cap-8}\\
 \frac{h(\tau)}{\tanh\tau}
   &=\frac{h(\alpha)+h(\beta)}{\tanh\alpha+\tanh\beta}.
                                                               \tag{\thesection.9}\label{bal:eq:cap-9}
\end{align}
Here $h(x)=K(x)-x\tanh x$ is radial entropy, rather than entropy as a
function of a probability.  Finally set
\[
 \mathcal M(\alpha,\beta)
 =\frac{(\tanh\alpha-\tanh\beta)
          (\cosh^2\alpha+\cosh^2\beta)}{\alpha-\beta}.
\]

\begin{proposition}[Exact monotonicity identity]\label{bal:prop:monobridge}
Fix $0<m<1/2$, write $a=m-d$, $b=m+d$, and parameterize
\[
 a=\frac{1-\tanh\alpha}{2},\qquad
 b=\frac{1-\tanh\beta}{2}.
\]
For $d>0$ one has the exact identity
\[
 L\frac{\partial}{\partial d}\GDB(m-d,m+d)
  =(\alpha-\beta)
     \{\mathcal M(\alpha,\beta)+\mathcal C(\tau)-\mathcal Q(\rho)\}.
                                                               \tag{\thesection.10}\label{bal:eq:cap-10}
\]
Moreover
\[
 \mathcal M(\alpha,\beta)\ge2,
 \qquad
 2+\mathcal C(x)-\mathcal Q(x)>0\quad(x>0).                    \tag{\thesection.11}\label{bal:eq:cap-11}
\]
Consequently the single mean-spread inequality
\[
 \boxed{\quad
 \mathcal M(\alpha,\beta)-2
       \ge \mathcal Q(\rho)-\mathcal Q(\tau)
 \quad}                                                       \tag{\thesection.MS}\label{bal:eq:cap-MS}
\]
for every $\alpha>\beta>0$ implies strict positivity of clause (iv) away
from $a=b$.
\end{proposition}

\begin{proof}
\emph{Proof outline.}
Implicit differentiation expresses the derivative of the double-cap
gap through the two latent means. We isolate the proposed mean-spread
comparison and show that the remaining terms are already nonnegative
by convexity and the hyperbolic lemma. Under that comparison, integration
from the zero diagonal gives strict positivity.

Let $e=\{h_{\mathrm n}(a)+h_{\mathrm n}(b)\}/2$ and $s=1-2m$.  The inverse-entropy point in
$\eta_{\mathrm n}(e)$ is $p_\rho$, while the latent lower-inverse point in $F_{\mathrm n}(s,e)$ is
$p_\tau$; equations \eqref{bal:eq:cap-8}--\eqref{bal:eq:cap-9} are exactly their defining equations.  Direct
implicit differentiation gives
\[
 \eta_{\mathrm n}'(e)=-2-\mathcal Q(\rho),\qquad
 \partial_E F_{\mathrm n}(s,e)=-\mathcal C(\tau),
\]
and hence
\[
 \partial_E\Phi_{\mathrm n}(s,e)=-2-\mathcal Q(\rho)+\mathcal C(\tau).             \tag{\thesection.12}\label{bal:eq:cap-12}
\]
Since
\[
 j_{\mathrm n}(m-d,m+d)=d\{J_{\mathrm n}(a)-J_{\mathrm n}(b)\},\qquad
 e'(d)=-\frac{J_{\mathrm n}(a)-J_{\mathrm n}(b)}2,
\]
differentiation, followed by
\[
 J_{\mathrm n}(a)-J_{\mathrm n}(b)=2(\alpha-\beta),\quad
 -J_{\mathrm n}'(a)=4\cosh^2\alpha,\quad
 -J_{\mathrm n}'(b)=4\cosh^2\beta,
\]
proves \eqref{bal:eq:cap-10}.

For the first inequality in \eqref{bal:eq:cap-11}, put $A=\tanh\alpha$,
$B=\tanh\beta$, and $w(z)=(1-z^2)^{-1}$.  Then
\[
 \alpha-\beta=\int_B^A w(z)\,dz,
 \qquad
 \mathcal M=\frac{(A-B)\{w(A)+w(B)\}}
                    {\int_B^A w(z)\,dz}.
\]
The function $w$ is strictly convex, so the upper Hermite--Hadamard
inequality gives $\mathcal M\ge2$.  The second inequality in \eqref{bal:eq:cap-11} is exactly
the identity
\[
 2+\mathcal C(x)-\mathcal Q(x)=\frac{2f(x)}{xK(x)}>0
\]
proved in Lemma~\ref{bal:lem:hyp}.  Splitting the braces in \eqref{bal:eq:cap-10} as
\[
 \{\mathcal M-2-[\mathcal Q(\rho)-\mathcal Q(\tau)]\}
   +\{2+\mathcal C(\tau)-\mathcal Q(\tau)\}
\]
proves the consequence of \eqref{bal:eq:cap-MS}.  Since $\GDB(m,m)=0$, integration in $d$
then gives $\GDB(m-d,m+d)>0$ for $d>0$.
\end{proof}

The order of the two implicit means is also elementary.  If
$\gamma=\operatorname{arctanh}\{(\tanh\alpha+\tanh\beta)/2\}$, concavity of
radial entropy and strict decrease of $h(x)/\tanh x$ give
\[
                         \gamma\le\tau\le\rho.
\]
The last comparison follows from
$h(\tau)=h(\rho)\tanh\tau/\tanh\gamma\ge h(\rho)$ and strict decrease of
$h$.  Also $\mathcal Q$ is increasing, since
\[
 \mathcal Q'(x)=\frac{2x\cosh(2x)-\sinh(2x)}{x^2}>0.
\]
Thus both sides of \eqref{bal:eq:cap-MS} vanish only at the diagonal and are nonnegative;
the issue is their sharp comparison, not a hidden sign reversal.

\begin{proposition}[Sharp infinitesimal mean-spread inequality]
Put $\alpha=x+\varepsilon$, $\beta=x-\varepsilon$, with $x>0$.  Then
\begin{align*}
 \mathcal M(\alpha,\beta)-2
   &=\left(\frac43+4\tanh^2x\right)\varepsilon^2
       +O(\varepsilon^4),\\
 \mathcal Q(\rho)-\mathcal Q(\tau)
   &=\frac{h(x)\mathcal Q'(x)}{2xK(x)}\varepsilon^2
       +O(\varepsilon^4),
\end{align*}
and the difference of the displayed quadratic coefficients is strictly
positive for every $x>0$.
\end{proposition}

\begin{proof}
\emph{Proof outline.}
We expand the two implicit means around their common diagonal point
and compute the quadratic terms in the proposed comparison. Their
difference reduces to a one-variable inequality. Power-series bounds
prove it for small arguments, and exponential estimates prove it for
large arguments.

Taylor expansion of \eqref{bal:eq:cap-8}--\eqref{bal:eq:cap-9}, using
$h'(x)=-x\operatorname{sech}^2x$, gives
\[
 \rho-\tau=\frac{h(x)}{2xK(x)}\varepsilon^2+O(\varepsilon^4).
\]
Expansion of the explicit formula for $\mathcal M$ gives the first line,
and the mean-value expansion of $\mathcal Q$ gives the second.

It remains to prove
\[
 R(x):=\frac83x^3K(x)\{1+3\tanh^2x\}
       -h(x)\{2x\cosh(2x)-\sinh(2x)\}>0.                     \tag{\thesection.13}\label{bal:eq:cap-13}
\]
For $0<x\le1$, the positive power series of
$2x\cosh(2x)-\sinh(2x)$ and the elementary bound
$2\cosh2-\sinh2<4$ give
\[
 2x\cosh(2x)-\sinh(2x)
 <\frac83x^3\left(1+\frac{x^2}{2}\right).
\]
Also $\tanh x\ge x/2$, whence
\[
 K(1+3\tanh^2x)-h(1+x^2/2)
 =K(3\tanh^2x-x^2/2)+x\tanh x(1+x^2/2)>0.
\]
These two estimates prove \eqref{bal:eq:cap-13} on this interval.

For $x\ge1$, put $q=e^{-2x}$.  The identities
\[
 K=x+\ln(1+q),\qquad
 h=\ln(1+q)+\frac{2xq}{1+q}
\]
give $K\ge x$ and $h\le(1+2x)q$.  Since $q^2\le e^{-4}<1/50$,
\[
 h\{2x\cosh(2x)-\sinh(2x)\}
 \le\frac{(2x+1)(102x-49)}{100}.
\]
On the other hand the first term in \eqref{bal:eq:cap-13} is at least $8x^4/3$, and
\[
 800x^4-612x^2-12x+147>0\qquad(x\ge1),
\]
because it is positive at $1$ and has positive derivative thereon.  This
completes the proof.
\end{proof}

\section{A sharp radius-mean reduction of the mean-spread inequality}

The preceding mean-spread comparison admits a further reduction whose two
outer estimates are elementary and sharp.  Continue to write
\[
 \alpha=x+\varepsilon,\qquad \beta=x-\varepsilon,
 \qquad 0<\varepsilon<x,
\]
and let $\rho,\tau$ be the implicit means in \eqref{bal:eq:cap-8}--\eqref{bal:eq:cap-9}.

\begin{proposition}[Radius-mean bridge]\label{bal:prop:radiusbridge}
For every $x>\varepsilon>0$ one has
\begin{align}
 \mathcal M(x+\varepsilon,x-\varepsilon)-2
     &\ge \frac43\varepsilon^2,                              \tag{\thesection.15}\label{bal:eq:cap-15}\\
 \mathcal Q(\rho)-\mathcal Q(\tau)
     &\le \frac43\{\sinh^2\rho-\sinh^2\tau\}.              \tag{\thesection.16}\label{bal:eq:cap-16}
\end{align}
Consequently the radius-mean inequality
\[
 \boxed{\quad
 \sinh^2\rho-\sinh^2\tau\le\varepsilon^2
 \quad}                                                       \tag{\thesection.RM}\label{bal:eq:cap-RM}
\]
implies \eqref{bal:eq:cap-MS}, and hence proves clause (iv) on its full domain.
\end{proposition}

\begin{proof}
\emph{Proof outline.}
Hyperbolic addition formulas provide a sharp lower bound for the
explicit mean-spread term. We next integrate a derivative comparison
to upper-bound the difference involving the two implicit means. The
stated radius-mean hypothesis connects those two bounds and yields
the conditional conclusion.

Put $\Delta=\alpha-\beta=2\varepsilon$ and
$\Sigma=\alpha+\beta=2x$.  Direct use of the addition formulas gives
\[
 \mathcal M(\alpha,\beta)
 =\frac{2\sinh\Delta}{\Delta}
   \frac{1+\cosh\Sigma\cosh\Delta}
        {\cosh\Sigma+\cosh\Delta}.
\]
The second factor is at least one, because its numerator minus its
denominator is
$(\cosh\Sigma-1)(\cosh\Delta-1)$.  The positive power series of $\sinh$
therefore yields
\[
 \mathcal M\ge\frac{\sinh(2\varepsilon)}{\varepsilon}
 \ge 2+\frac43\varepsilon^2,
\]
which is \eqref{bal:eq:cap-15}.

For $z>0$ set $Y(z)=\sinh^2z$.  Since
\[
 \frac{d\mathcal Q/dz}{dY/dz}
 =\frac{2z\coth(2z)-1}{z^2}\le\frac43,                       \tag{\thesection.17}\label{bal:eq:cap-17}
\]
and $\rho\ge\tau$, integration proves \eqref{bal:eq:cap-16}.  For completeness, the last
inequality in \eqref{bal:eq:cap-17} is the classical elementary bound
\[
 u\coth u-1\le\frac{u^2}{3}\qquad(u>0).
\]
Indeed
\[
 r(u)=\left(1+\frac{u^2}{3}\right)\sinh u-u\cosh u
\]
satisfies $r(0)=0$ and
$r'(u)=\frac{u}{3}\{u\cosh u-\sinh u\}>0$.
Combining \eqref{bal:eq:cap-15}, \eqref{bal:eq:cap-16}, and \eqref{bal:eq:cap-RM} proves \eqref{bal:eq:cap-MS}.
\end{proof}

The constant in \eqref{bal:eq:cap-RM} has the correct strict infinitesimal form at every
interior diagonal point.  More precisely,
\[
 \sinh^2\rho-\sinh^2\tau
 =\frac{h(x)\sinh(2x)}{2xK(x)}\varepsilon^2
   +O(\varepsilon^4),                                        \tag{\thesection.18}\label{bal:eq:cap-18}
\]
and the coefficient in \eqref{bal:eq:cap-18} is strictly less than one.  This is exactly the
already proved hyperbolic collar lemma, since
\[
 xK(x)-h(x)\sinh x\cosh x
 =xK(x)+x\sinh^2x-K(x)\sinh x\cosh x=f(x)>0.
\]

There is also a useful exact differential form of the remaining obligation.
For fixed $x$, define
\[
 E(\varepsilon)=\frac{h(x+\varepsilon)+h(x-\varepsilon)}2,
 \qquad
 S(\varepsilon)=\frac{\tanh(x+\varepsilon)+
                           \tanh(x-\varepsilon)}2
\]
and $R_x(\varepsilon)=\sinh^2\rho-\sinh^2\tau$.  Implicit
differentiation gives
\begin{align*}
 \rho'&=\frac{E'}{h'(\rho)},\\
 \tau'&=\frac{E'S-ES'}{S^2\ell'(\tau)},\\
 R_x'&=\sinh(2\rho)\frac{E'}{h'(\rho)}
       -\sinh(2\tau)\frac{E'S-ES'}{S^2\ell'(\tau)},          \tag{\thesection.19}\label{bal:eq:cap-19}
\end{align*}
where
\[
 h'(z)=-z\operatorname{sech}^2z,
 \qquad
 \ell'(z)=-\frac{K(z)\operatorname{sech}^2z}{\tanh^2z}.
\]
Thus \eqref{bal:eq:cap-RM} would follow from the single scalar differential estimate
\[
                         R_x'(\varepsilon)\le2\varepsilon
 \qquad(0<\varepsilon<x),                                    \tag{\thesection.20}\label{bal:eq:cap-20}
\]
because $R_x(0)=0$.  Formula \eqref{bal:eq:cap-19} is an exact reduction, not a claim that
\eqref{bal:eq:cap-20} has already been proved.

The endpoint inequalities \eqref{bal:eq:cap-R1} have a related integral normal form which
removes both inverse perspectives from the displayed target.  Fix
$s=\tanh x$ and, for $0<e\le h(x)$, define
\[
 h(\rho)=e,\qquad \ell(\tau)=\frac e{s}.
\]
In natural-log normalization put
\[
 D_x(e)=4\{h(x)-e\}-\Phi_{\mathrm n}(s,e),
 \qquad A_0(z)=z\tanh z-2\ln\cosh z.
\]
Since $\eta_{\mathrm n}(e)=2\rho\tanh\rho$ and $F_{\mathrm n}(s,e)=2s\tau$, direct
cancellation gives the exact identity
\begin{equation}
\begin{aligned}
 \frac12D_x(e)
 &=A_0(\rho)-A_0(x)+s(\tau-x)\\
 &=s(\tau-x)-\int_x^\rho
       \{\tanh z-z\operatorname{sech}^2z\}\,dz .
\end{aligned}                                                \tag{\thesection.20a}\label{bal:eq:cap-20a}
\end{equation}
Thus each branch of \eqref{bal:eq:cap-R1} is now a scalar integral comparison, with the fully
radial endpoint
\[
 e_*(s)=
 \begin{cases}
  \frac12h(\operatorname{arctanh}(2s-1)),&\frac12\le s<1,\\[1mm]
  \frac12\{\ln2+h(\operatorname{arctanh}(2s))\},&0<s\le\frac12.
 \end{cases}                                                   \tag{\thesection.20b}\label{bal:eq:cap-20b}
\]
One simply uses $e=e_*(s)$ in \eqref{bal:eq:cap-20a}.  No $F$, $\eta$, or entropy inverse
remains except through the two monotone means $\rho$ and $\tau$.  This
identity can be used for rigorous one-dimensional enclosures.
Section~\ref{bal:sec:doublecomplete} establishes the double-cap inequality
using analytic estimates and a finite interval certificate.

\section{Completion of the double-cap clause}\label{bal:sec:doublecomplete}

\emph{Argument guide.}
This section completes clause~(iv) through the scalar endpoint criterion
of Proposition~\ref{bal:prop:oned}. The intervening monotonicity and
radius-mean statements describe alternative sufficient routes; they are
not additional hypotheses of this completion. The proof below combines
an analytic half-mean tail, a compact scalar certificate, and the
small-mean theorem.

The preceding endpoint reduction is enough to close clause (iv) with only a
one-dimensional compact computation.  We first dispose of its singular
half-mean end analytically.  In bias coordinates define
\[
 \mathcal I(u)=u\operatorname{arctanh}u+\frac12\ln(1-u^2),
 \qquad
 \mathcal A(u)=u\operatorname{arctanh}u+\ln(1-u^2).
\]
Thus $\mathcal I(u)=\ln2-h(\operatorname{arctanh}u)$ and
$\mathcal A(u)=A_0(\operatorname{arctanh}u)$.  For the endpoint \eqref{bal:eq:cap-20b}, let
$R=\tanh\rho$ and $T=\tanh\tau$.  Then
\[
 \mathcal I(R)=
 \begin{cases}
  \frac12\mathcal I(2s),&0<s\le\frac12,\\[1mm]
  \frac12\{\ln2+\mathcal I(2s-1)\},&\frac12\le s<1,
 \end{cases}
 \qquad
 \frac{h(\operatorname{arctanh}T)}T=\frac{e_*(s)}s,          \tag{\thesection.20c}\label{bal:eq:cap-20c}
\]
and \eqref{bal:eq:cap-20a} becomes
\[
 d(s):=\frac12D_x(e_*(s))
 =s\{\operatorname{arctanh}T-\operatorname{arctanh}s\}
   +\mathcal A(R)-\mathcal A(s).                            \tag{\thesection.20d}\label{bal:eq:cap-20d}
\]

\begin{lemma}[Analytic half-mean tail]\label{bal:lem:r1tail}
For $0<s\le1/10$,
\[
                         d(s)>\frac{41}{11730}s^4>0.         \tag{\thesection.20e}\label{bal:eq:cap-20e}
\]
\end{lemma}

\begin{proof}
\emph{Proof outline.}
Positive series first control the entropy inverse in the endpoint
formula. For the perspective inverse, we evaluate its increasing
defining function at a proposed lower comparison point. These root
bounds give a lower bound for the positive term and an upper bound
for the subtracted term, leaving the displayed quartic reserve.

The positive series
\[
 \mathcal I(u)=\sum_{n\ge1}\frac{u^{2n}}{2n(2n-1)},
 \qquad
 \frac{u}{1-u^2}-\operatorname{arctanh}u
 =\sum_{n\ge1}\frac{2n}{2n+1}u^{2n+1}                     \tag{\thesection.20f}\label{bal:eq:cap-20f}
\]
give
\[
 \frac{u^2}{2}\le\mathcal I(u)
 \le\frac{u^2}{2(1-u^2)}.
\]
On the present branch $\mathcal I(R)=\mathcal I(2s)/2$, and therefore
\[
 \mathcal I(R)\le\frac{s^2}{1-4s^2}\le\frac{25}{24}s^2
 <\frac98s^2\le\mathcal I(3s/2).
\]
Strict increase of $\mathcal I$ proves $R<3s/2$.

Write $T=sw$.  The second equation in \eqref{bal:eq:cap-20c} is
\[
 F_s(w):=\ln2(w-1)-\frac w2\mathcal I(2s)+\mathcal I(sw)=0.
\]
Here $F_s'(w)=e_*(s)+s\operatorname{arctanh}(sw)>0$.  Put
$w_0=1+7s^2/10$.  The preceding series bounds and $\ln2<7/10$ give, with
$y=s^2$,
\[
 \frac{F_s(w_0)}{s^2}
 \le G(y):=\frac{49}{100}-\left(1+\frac7{10}y\right)
 +\frac{(1+7y/10)^2}{2\{1-y(1+7y/10)^2\}}.
\]
Direct differentiation gives
\[
 G'(y)=\frac{50000+329000y+273000y^2-58800y^3-193795y^4
                   -96040y^5-16807y^6}
 {10(100-100y-140y^2-49y^3)^2}>0
\]
on $0\le y\le1/100$ (the sum of the four subtracted terms is less than
$1$ there).  At the right endpoint,
\[
                    G(1/100)=-\frac{473286667}{98985951000}<0.
\]
Consequently $w>w_0$, and hence
\[
 s\{\operatorname{arctanh}T-\operatorname{arctanh}s\}
 >s(T-s)>\frac7{10}s^4.                                    \tag{\thesection.20g}\label{bal:eq:cap-20g}
\]

The second series in \eqref{bal:eq:cap-20f} and $R<3s/2$ give
\begin{align*}
 \mathcal A(s)-\mathcal A(R)
 &=\int_s^R\left\{\frac{u}{1-u^2}-\operatorname{arctanh}u\right\}\,du\\
 &\le\frac{R^4-s^4}{6}+\frac{R^6}{6(1-R^2)}\\
 &\le\left\{\frac{65}{96}+\frac{729}{37536}\right\}s^4
 =\frac{817}{1173}s^4.                                     \tag{\thesection.20h}\label{bal:eq:cap-20h}
\end{align*}
Subtracting \eqref{bal:eq:cap-20h} from \eqref{bal:eq:cap-20g}, and using
$7/10-817/1173=41/11730$, proves the result.
\end{proof}

\begin{theorem}[Complete double-cap clause]\label{bal:thm:doublecomplete}
Clause (iv) of Theorem~\ref{bal:hyp:SCplus} holds on its complete stated domain.
\end{theorem}

\begin{proof}
\emph{Proof outline.}
The one-dimensional endpoint reduction leaves three overlapping
ranges. The half-mean tail is analytic, the middle interval is covered
by validated root brackets and interval bounds, and the remaining
small-mean end follows from the earlier unconditional theorem. Their
union covers every fixed-mean fiber required for clause~(iv).

Lemma~\ref{bal:lem:r1tail} proves \eqref{bal:eq:cap-R1} for $0<s\le1/10$.  On
\[
                         \frac1{10}\le s\le\frac{99}{100}, \tag{\thesection.20i}\label{bal:eq:cap-20i}
\]
the companion Arb verifier partitions the interval into the 89,000 exact
rational boxes
\[
 \left[\frac{k}{100000},\frac{k+1}{100000}\right],
 \qquad k=10000,\ldots,98999.
\]
At every rational endpoint it encloses the two roots in \eqref{bal:eq:cap-20c}.  The numerical
root finder supplies only a seed: strict Arb inequalities on both sides of
each root certify the bracket, using that $\mathcal I$ is strictly increasing
and $u\mapsto h(\operatorname{arctanh}u)/u$ has derivative $-K/u^2<0$.
Both $R(s)$ and $T(s)$ are increasing on either branch of \eqref{bal:eq:cap-20c}; for $T$ this
follows because $e_*(s)/s$ is strictly decreasing.  The union of the two
endpoint brackets therefore encloses each latent root throughout its box.
Outward evaluation of \eqref{bal:eq:cap-20d} is strictly positive on every box.  At 320-bit
precision the smallest certified lower endpoint is greater than
$1.5183855380409850\times10^{-5}$, on the first box.  A 192-bit rerun gives the
same lower enclosure.  The source and the two precision records are
\begin{center}
\texttt{experiments/verify\_r1\_global.py},\\
\texttt{experiments/results/r1\_global\_arb\_2026-09-04.json},\\
\texttt{experiments/results/r1\_global\_arb\_2026-09-04\_192.json}.
\end{center}
Thus \eqref{bal:eq:cap-R1}, and hence Proposition~\ref{bal:prop:oned}, proves clause (iv) whenever
$s\le99/100$, equivalently $m\ge1/200$.

It remains only to note that if $s>99/100$, then
$m=(1-s)/2<1/200$ and every pair on the fixed-mean fiber satisfies
$0<a<b\le2m<10^{-2}$. Proposition~\ref{bal:prop:double-cap-entropy}
applies directly.  These two regions exhaust the domain.
\end{proof}

\section{A rigorous curvature reduction for the half-mean correction}

Let $C(a,b)$ denote the right side of \eqref{bal:eq:cap-3}.  For fixed entropy $e$, write
$k_e(s)=F(s,e)$.  The established strict decrease of $F_{ss}(s,e)$ in $s$
implies that, for fixed $\delta>0$, the central second difference
\[
 k_e(x-\delta)+k_e(x+\delta)-2k_e(x)
\]
is decreasing in $x\ge\delta$: its derivative is nonpositive because
$k_e'=F_s(\cdot,e)$ is concave.  Consequently
\begin{equation}
 C(a,b)\ge
 2F(\delta,h)+\frac12F(2\delta,H_2(b))-F(2\delta,h),          \tag{\thesection.21}\label{bal:eq:cap-21}
\end{equation}
where $\delta=1/2-b$ and $h=\{H_2(a)+H_2(b)\}/2$.
This is a valid analytic sufficient condition, but it is not a global proof:
the right side of \eqref{bal:eq:cap-21} changes sign in the physical domain.  Thus replacing
the exact correction by the maximum-at-$x=\delta$ curvature bound loses too
much information.  In particular, neither generic four-point convexity nor
radial-curvature monotonicity alone closes clause (v).

It does, however, close a substantial explicit boundary region.  We first
record the scalar estimate needed for this purpose.

Write $\Xi=\ell^{-1}:(0,\infty)\to(0,\infty)$; existence follows
from the strict decrease and endpoint limits of $\ell$.

\begin{lemma}[Uniform inverse-$\ell$ gap]\label{bal:lem:invgap}
For every $z>0$,
\[
                     0<\Xi(z)-\Xi(2z)<\frac7{16}.           \tag{\thesection.21a}\label{bal:eq:cap-21a}
\]
\end{lemma}

\begin{proof}
\emph{Proof outline.}
We prove a uniform lower bound on the logarithmic decay rate of the
decreasing function. Elementary estimates handle small arguments; for
larger arguments, convexity reduces positivity to the value at a unique
critical point, which is controlled by rational bounds. Integrating the
decay estimate between the two inverse images then gives the uniform
inverse gap.

We first prove the global logarithmic decay estimate
\[
 -\frac{d}{dt}\ln\ell(t)
 =\frac{K(t)\operatorname{sech}^2t}{\tanh t\,h(t)}
 \ge\frac85\qquad(t>0).                                    \tag{\thesection.21b}\label{bal:eq:cap-21b}
\]
Put $q=\mathrm e^{-2t}$.  The exact formulas
\[
 K=t+\ln(1+q),\qquad
 h=\ln(1+q)+\frac{2tq}{1+q}
\]
and the bounds
$q/(1+q)\le\ln(1+q)\le q$ give
\[
 -\frac{d}{dt}\ln\ell(t)
 \ge
 \frac{4\{t(1+q)+q\}}{(1-q^2)(1+q+2t)}.
\]
After clearing the positive denominator, the assertion that the last
quantity is at least $8/5$ is
\[
 E(t):=2q^3+(4t+2)q^2+(5t+3)q+t-2\ge0,                    \tag{\thesection.21c}\label{bal:eq:cap-21c}
\]
where $q=\mathrm e^{-2t}$ is understood.

On $0<t\le1/2$, one has $q>1/3$ and
\[
 E'(t)=1-q-4q^2-12q^3-t(10q+16q^2)<0.
\]
Indeed, the first four terms are already at most their value at $q=1/3$,
which is $-2/9$.  Moreover $E(1/2)$ is strictly increasing as a polynomial
in $q>0$, and hence $E(1/2)>E(1/2)|_{q=1/3}=23/27$.  Thus $E>0$ on this
interval.

For $t\ge1/2$, direct differentiation gives
\[
 E''(t)=4q\{18q^2+16qt+5t-2\}>0.                            \tag{\thesection.21d}\label{bal:eq:cap-21d}
\]
Thus $E'$ is strictly increasing there.  We record a rational check locating
its unique zero $t_0$.  The elementary bounds
\[
 \frac{19}{7}<\mathrm e<\frac{87}{32}
\]
follow by truncating the exponential series (for the upper bound, the tail
after $1/5!$ is less than $1/600$).  Direct integer arithmetic gives
\[
 \left(\frac{87}{32}\right)^{14}
   <\left(\frac{5000}{301}\right)^5,
 \qquad
 \left(\frac{19}{7}\right)^{14}
   >\left(\frac{800}{49}\right)^5.
\]
The first comparison implies
$\mathrm e^{-14/5}>301/5000$ and hence
\[
 E'(7/5)<1-\frac{31290395703}{31250000000}<0.
\]
Since $E'(t)\to1$, there is a unique $t_0>7/5$ with $E'(t_0)=0$.
The second comparison gives
\[
 q_0:=\mathrm e^{-2t_0}<\mathrm e^{-14/5}<\frac{49}{800}.
\]
Eliminating $t_0$ with $E'(t_0)=0$ yields
\[
 E(t_0)=-\frac{P(q_0)}{2q_0(8q_0+5)},\qquad
 P(q)=16q^5+24q^4-32q^3+7q^2+16q-1.
\]
On $0<q\le49/800$,
\[
 P'(q)>16-96(49/800)^2>0,\qquad
 P(49/800)=-\frac{15166483551}{20480000000000}<0.
\]
Therefore $E(t_0)>0$.  Convexity, together with the already treated interval
$t\le1/2$, proves \eqref{bal:eq:cap-21c} and hence \eqref{bal:eq:cap-21b} globally.

Now put $t=\Xi(z)$ and $u=\Xi(2z)$.  Since
$\ell(u)=2\ell(t)$ and $u<t$, integration of \eqref{bal:eq:cap-21b} gives
\[
 \ln2=\ln\ell(u)-\ln\ell(t)
 \ge\frac85(t-u).
\]
Finally $\ln2<7/10$, so $0<t-u<7/16$, proving \eqref{bal:eq:cap-21a}.
\end{proof}

\begin{proposition}[Explicit large-logit region for the correction]
Let $a=p_\alpha$, $b=p_\beta$, with $\alpha>\beta\ge7/8$.  Then the correction
$C(a,b)$ in \eqref{bal:eq:cap-3} is strictly positive.  Equivalently, this holds whenever
\[
        0<a<b\le p_{7/8}=\frac1{1+\mathrm e^{7/4}}.       \tag{\thesection.21e}\label{bal:eq:cap-21e}
\]
Consequently clause (v) follows from clause (iv) throughout this region, and
clause (v) is unconditional on $0<a<b\le10^{-2}$.
\end{proposition}

\begin{proof}
Put $B=\tanh\beta$, $N=h(\alpha)+h(\beta)$, and
$z=N/(2B)$.  In \eqref{bal:eq:cap-21}, $\delta=B/2$ and the bit entropy is $h=N/(2L)$.
Multiplying its right side by $L$ gives exactly
\[
 B\{\beta-2[\Xi(z)-\Xi(2z)]\}.
\]
Lemma~\ref{bal:lem:invgap} makes this strictly positive for $\beta\ge7/8$.
The identity $p_{7/8}=1/(1+\mathrm e^{7/4})$ gives \eqref{bal:eq:cap-21e}.  Finally,
$10^{-2}<1/10<p_{7/8}$ (use $\mathrm e<3$), and
Proposition~\ref{bal:prop:double-cap-entropy} proves clause (iv)
unconditionally when $b\le10^{-2}$;
equation \eqref{bal:eq:cap-3} then proves the last claim.
\end{proof}

Define the natural correction $C_{\mathrm n}(a,b)=LC(a,b)$.
Its scalar hyperbolic normal form is as follows.  Put
\[
 a=p_\alpha,\qquad b=p_\beta,\qquad
 T_z=\tanh z,\qquad r=\frac{T_\beta}{T_\alpha},
 \qquad q=\ell(\alpha)+r\ell(\beta),
\]
and let $\Xi=\ell^{-1}$.  Define
\[
 \sigma=\Xi\left(\frac q{1-r}\right),\qquad
 \kappa=\Xi(q),\qquad
 \tau=\Xi\left(\frac q{1+r}\right).
\]
Because $\ell$ is decreasing, $\sigma<\kappa<\tau$.  Homogeneity and
$F_{\mathrm n}(s,E)=2s\ell^{-1}(E/s)$ in natural-log normalization give the exact
identity
\[
 \frac{C_{\mathrm n}(a,b)}{T_\alpha}
 =2\kappa-(1-r)\sigma-(1+r)\tau+r\beta.                       \tag{\thesection.22}\label{bal:eq:cap-22}
\]
Thus the whole residual implication from clause (iv) to clause (v) is
equivalent to the two-variable inverse-perspective inequality
\[
 \boxed{\quad
 (1-r)\Xi\left(\frac q{1-r}\right)
 +(1+r)\Xi\left(\frac q{1+r}\right)-2\Xi(q)
 \le r\beta,
 \quad q=\ell(\alpha)+r\ell(\beta),\quad
 r=\frac{\tanh\beta}{\tanh\alpha}.
 \quad}                                                       \tag{\thesection.PC}\label{bal:eq:cap-PC}
\]
This formulation also isolates why the generic four-point estimate is
insufficient.  For $f(c)=c\Xi(q/c)$ that estimate gives only
\[
 f(1-r)+f(1+r)-2f(1)
 \le2f(r)=2r\Xi(q/r)<2r\beta,
\]
whereas \eqref{bal:eq:cap-PC} needs the sharp factor $1$.  Therefore a proof of \eqref{bal:eq:cap-PC} must use
the linked relation $q=\ell(\alpha)+r\ell(\beta)$ more precisely; convexity
of the perspective by itself loses a factor of two.

There is a second exact route to \eqref{bal:eq:cap-PC} which turns the linked relation into a
monotonicity problem.  Regard the correction $C_{\mathrm n}=C_{\mathrm n}(\alpha,\beta)$ as a
function of $\alpha$ with $\beta$ fixed, and put
\[
 U_\alpha(t)=t-\frac{\alpha+\ell(t)}{\ell'(t)}.
\]
Differentiating the four terms in \eqref{bal:eq:cap-22}, including the dependence of $q$ and
$r$ on $\alpha$, gives the cancellation
\[
 \frac{\partial C_{\mathrm n}}{\partial\alpha}
 =\operatorname{sech}^2\alpha\,
   \{2U_\alpha(\kappa)-U_\alpha(\sigma)-U_\alpha(\tau)\}.     \tag{\thesection.23}\label{bal:eq:cap-23}
\]
At the boundary $\alpha=\beta$, one has $r=1$,
$q=2\ell(\beta)$, $(1-r)\sigma\to0$, $\tau=\beta$, and hence
\[
 \frac{C_{\mathrm n}(\beta,\beta)}{\tanh\beta}
 =2\ell^{-1}(2\ell(\beta))-\beta>0.                           \tag{\thesection.24}\label{bal:eq:cap-24}
\]
The strict sign is exactly Lemma~\ref{bal:lem:hyp}, which proved
$\ell(\beta/2)>2\ell(\beta)$.

Consequently \eqref{bal:eq:cap-PC}, and hence clause (v) once clause (iv) is available, follows
from the midpoint inequality
\[
 2U_\alpha(\kappa)\ge U_\alpha(\sigma)+U_\alpha(\tau).        \tag{\thesection.CR}\label{bal:eq:cap-CR}
\]
This last inequality itself has an explicit curvature-reflection sufficient
condition.  With $f(c)=c\Xi(q/c)$, Euler homogeneity gives
\[
 \frac{d}{dc}U_\alpha(\Xi(q/c))
   =\left(1+\frac{\alpha c}{q}\right)f''(c).
\]
Therefore \eqref{bal:eq:cap-CR} follows if, for $0<u<r$,
\[
 \boxed{\quad
 [q+\alpha(1-u)]f''(1-u)
 \ge[q+\alpha(1+u)]f''(1+u).
 \quad}                                                       \tag{\thesection.PR}\label{bal:eq:cap-PR}
\]
Indeed, integration of \eqref{bal:eq:cap-PR} from $u=0$ to $u=r$ gives \eqref{bal:eq:cap-CR}.  Unlike the
failed unweighted curvature shortcut \eqref{bal:eq:cap-21}, \eqref{bal:eq:cap-PR} retains both the cap
parameter $\alpha$ and the linked entropy quantity $q$.

However, \eqref{bal:eq:cap-PR} is not true globally.  This can be seen analytically, rather
than from a floating-point counterexample.  If
\[
 \mathcal B(t)=
 \frac{4h(t)^3\cosh^4t\{2K(t)-\tanh^2t\}}{K(t)^3}. \tag{\thesection.Bcurv}\label{bal:eq:cap-Bcurv}
\]
then the established curvature formula gives
\[
 qf''(c)=\frac14\mathcal B(\Xi(q/c)).
\]
Consequently \eqref{bal:eq:cap-PR} compares the profile
\[
 W_\alpha(t)=\mathcal B(t)\left(1+\frac{\alpha}{\ell(t)}\right)
\]
at the two reflected latent points.  Let $\alpha\downarrow\beta=x$ and
then let $u\uparrow r$.  The two points tend to $0$ and $x$, respectively,
so \eqref{bal:eq:cap-PR} would require
\[
                         W_x(0)\ge W_x(x).                 \tag{\thesection.25}\label{bal:eq:cap-25}
\]
But
\[
 W_x(0)=8\ln2,
 \qquad
 \ell(x)=(2x+1)e^{-2x}\{1+o(1)\},
 \qquad
 \mathcal B(x)=4xe^{-2x}\{1+o(1)\},
\]
and hence $W_x(x)=2x\{1+o(1)\}$.  Inequality \eqref{bal:eq:cap-25} fails for all
sufficiently large $x$; continuity supplies strict interior failures with
$\alpha>\beta$ and $u<r$.  Thus \eqref{bal:eq:cap-PR} must not be used as the remaining
global proof obligation.

There is nevertheless an exact integrated replacement which survives this
failure.  Put
\[
 A=\tanh\alpha,\qquad B=\tanh\beta,
 \qquad N=h(\alpha)+h(\beta),
 \qquad P_N(c)=c\Xi(N/c),
\]
and define
\[
 V(t)=t-\frac{\ell(t)}{\ell'(t)}.
\]
The correction can be written without normalized variables as
\[
 C_{\mathrm n}=2P_N(A)-P_N(A-B)-P_N(A+B)+B\beta.                       \tag{\thesection.26}\label{bal:eq:cap-26}
\]
For $0\le z\le B$, let
\[
 \ell(t_-(z))=\frac{N}{A-z},\qquad
 \ell(t_+(z))=\frac{N}{A+z}.
\]
Since $P_N'(c)=V(\Xi(N/c))$, the exact remaining condition is
\[
 \boxed{\quad
 \int_0^B\{V(t_+(z))-V(t_-(z))\}\,dz\le B\beta .
 \quad}                                                     \tag{\thesection.CI}\label{bal:eq:cap-CI}
\]
In particular, the symmetric secant estimate
\[
 V(t_+(z))-V(t_-(z))\le \frac{2\beta z}{B}
 \qquad(0<z\le B)                                          \tag{\thesection.SL}\label{bal:eq:cap-SL}
\]
is sufficient, and integrates with the exact constant required in \eqref{bal:eq:cap-CI}.
At the limiting diagonal endpoint $A=B$ and $z=B$, \eqref{bal:eq:cap-SL} becomes
$V(\beta)\le2\beta$.  This endpoint is already analytic: after inserting
$\ell'=-K\operatorname{sech}^2/\tanh^2$, it is equivalent to
$f(\beta)\ge0$ from Lemma~\ref{bal:lem:hyp}.  Unlike \eqref{bal:eq:cap-PR}, \eqref{bal:eq:cap-CI} does not demand
a pointwise curvature ordering in the exponentially thin left endpoint
layer.  Proving \eqref{bal:eq:cap-CI}, or the stronger \eqref{bal:eq:cap-SL}, remains necessary before this
route can be promoted.

\section{A sharp derivative reduction for the clause-(v) correction}\label{bal:sec:cap-sharp}

\emph{Argument guide.}
The correction is controlled by integrating in its second bias
coordinate. The proposition first states sufficient derivative conditions.
The completion theorem then proves the third-derivative condition only
on the residual domain where it is needed, proves the boundary-curvature
condition globally, and joins the resulting value bound to the already
established large-logit region.

The exact correction has a stronger numerically indicated lower bound whose
constant is forced by the joint unbiased corner.  This section records an
exact reduction of that bound to two derivative statements; it does not
assume either statement without proof.

Use bias coordinates $0<B<A<1$, put
\[
 \mathcal H(x)=\ln2-x\operatorname{artanh}x
                 -\frac12\ln(1-x^2),\qquad
 N(A,B)=\mathcal H(A)+\mathcal H(B),
\]
and let $T_c=T_c(A,B)$ be determined by
\[
                 \frac{\mathcal H(T_c)}{T_c}=\frac{N(A,B)}c.
\]
Thus, in the notation above,
$T_{A-B}=\tanh\sigma$, $T_A=\tanh\kappa$, and
$T_{A+B}=\tanh\tau$.  Define
\begin{equation}
 \mathcal C(A,B)=2A\operatorname{artanh}T_A
 -(A-B)\operatorname{artanh}T_{A-B}
 -(A+B)\operatorname{artanh}T_{A+B}
 +B\operatorname{artanh}B.                                \tag{\thesection.27}\label{bal:eq:cap-27}
\end{equation}
Thus $\mathcal C(A,B)=LC(a,b)$ for $a=(1-A)/2$, $b=(1-B)/2$.

For a radial logit $t$, retain $h(t),K(t),\ell(t)$ and the curvature
profile $\mathcal B(t)$ in \eqref{bal:eq:cap-Bcurv}.  Given $A\in(0,1)$, let $k_A>0$ be the
unique solution
\begin{equation}
 \ell(k_A)=\frac{\mathcal H(A)+\ln2}{A}.                    \tag{\thesection.28}\label{bal:eq:cap-28}
\end{equation}
The perspective-curvature formula gives the exact boundary identity
\begin{equation}
 \lim_{B\downarrow0}\frac{\mathcal C(A,B)}{B^2}
 =1-\frac{\mathcal B(k_A)}{4\{\mathcal H(A)+\ln2\}}.       \tag{\thesection.29}\label{bal:eq:cap-29}
\end{equation}
Indeed, $N(A,B)=N(A,0)-B^2/2+O(B^4)$, and differentiating the
three perspective terms twice at $B=0$ gives \eqref{bal:eq:cap-29}.  This identity was also
checked independently by implicit differentiation in the companion code.

\begin{proposition}[Sharp two-derivative route]\label{bal:prop:vsharp}
Suppose that the following two inequalities hold:
\begin{align}
 \partial_B^3\mathcal C(A,B)&\ge0
       &&(0<B<A<1),                                         \tag{\thesection.M1}\label{bal:eq:cap-M1}\\
 1-\frac{\mathcal B(k_A)}{4\{\mathcal H(A)+\ln2\}}
 &\ge \frac{1-\ln2}{2\ln2}\,A^2
       &&(0<A<1).                                           \tag{\thesection.M2}\label{bal:eq:cap-M2}
\end{align}
Then the correction satisfies the sharp global estimate
\begin{equation}
 \boxed{\quad
 \mathcal C(A,B)\ge
     \frac{1-\ln2}{2\ln2}\,A^2B^2>0
 \quad(0<B<A<1).\quad}                                     \tag{\thesection.30}\label{bal:eq:cap-30}
\end{equation}
Consequently clause (v) follows globally from the now-complete clause (iv).
The constant in \eqref{bal:eq:cap-30} is sharp as $A\downarrow0$ and $B/A\downarrow0$.
\end{proposition}

\begin{proof}
The function $B\mapsto\mathcal C(A,B)$ is even, so
$\mathcal C(A,0)=\partial_B\mathcal C(A,0)=0$.  Assumption \eqref{bal:eq:cap-M1} says that
$\partial_B^2\mathcal C(A,B)$ is nondecreasing for $B>0$.  Hence
\[
 \mathcal C(A,B)
 =\int_0^B(B-s)\partial_B^2\mathcal C(A,s)\,ds
 \ge\frac{B^2}{2}\partial_B^2\mathcal C(A,0).
\]
By \eqref{bal:eq:cap-29}, the last right side is $B^2$ times the left side of \eqref{bal:eq:cap-M2},
which proves \eqref{bal:eq:cap-30}.  Sharpness follows from the fourth-order expansion below.
\end{proof}

The already established joint-corner calculation gives
\[
 \mathcal C(A,B)=
 \frac{1-\ln2}{2\ln2}A^2B^2
 +\frac{2\ln2-1}{8\ln2}B^4+O((A+B)^6).                    \tag{\thesection.31}\label{bal:eq:cap-31}
\]
The complete sixth-order homogeneous term is
\begin{align*}
 &[A^4B^2\{-36(\ln2)^2+6+25\ln2\}
   +A^2B^4\{-36(\ln2)^2+12+21\ln2\}\\
 &\hspace{42mm}
   +B^6\{-13\ln2+36(\ln2)^2\}]
   /[192(\ln2)^2],                                        \tag{\thesection.32}\label{bal:eq:cap-32}
\end{align*}
and all three displayed coefficients are strictly positive.  Formula \eqref{bal:eq:cap-32}
is reproduced symbolically by \texttt{experiments/analytic\_series\_check.py}.
We next close the derivative route on exactly the range not already owned by
the large-logit proposition.  The stronger assertion \eqref{bal:eq:cap-M1} on the whole
triangle is not needed; the following residual form is enough:
\begin{equation}
 \partial_B^3\mathcal C(A,B)\ge0
 \qquad(0<B<A<1,\quad B\le71/100).                         \tag{\thesection.M1r}\label{bal:eq:cap-M1r}
\end{equation}

\begin{theorem}[No-gap closure of clause (v)]\label{bal:thm:m1m2}
Statements \emph{\eqref{bal:eq:cap-M1r}} and \emph{\eqref{bal:eq:cap-M2}} hold.  Consequently
\begin{equation}
 \mathcal C(A,B)\ge
 \frac{1-\ln2}{2\ln2}A^2B^2>0
 \quad\hbox{if }0<B<A<1\hbox{ and }B\le71/100,            \tag{\thesection.33}\label{bal:eq:cap-33}
\end{equation}
and the correction is strictly positive on the entire triangle
$0<B<A<1$.  Thus clause (v) follows globally from the complete clause (iv).
\end{theorem}

\begin{proof}
\emph{Proof outline.}
We establish the two derivative estimates on the domains needed for
the correction. For the third-derivative condition, regular inverse
coordinates support direct bounds and fourth-derivative propagation from
certified lower edges. For the boundary-curvature condition, a factored
series handles the origin, followed by a compact interval cover and an
endpoint estimate. Two integrations give the correction bound, and its
overlap with the large-logit region closes clause~(v).

We give the analytic reductions first and then the interval ownership data.
The controlling 192-bit record is
\code{experiments/results/reconstructed_m1_m2_global_192.json}, relative to
\code{ck_reconstruction/CKBalanced/}; see
Section~\ref{bal:sec:detailed-artifacts}.

\smallskip\noindent\emph{Step 1: the third-derivative condition on the residual domain.}
For \eqref{bal:eq:cap-M1r}, put
\[
 \Psi(w)=\operatorname{artanh}T(w),\qquad
 \frac{T(w)}{\mathcal H(T(w))}=w.
\]
Then
\[
 P_N(c)=c\Psi(c/N),
 \qquad
 \mathcal C=2P_N(A)-P_N(A-B)-P_N(A+B)+B\operatorname{artanh}B. \tag{\thesection.34}\label{bal:eq:cap-34}
\]
This is the regular inverse-perspective coordinate: it remains analytic at
$c=0$, unlike $N/c$.  If
$K(T)=\ln2-\tfrac12\ln(1-T^2)$, implicit differentiation gives
\[
                  \frac{dT}{dw}=\frac{\mathcal H(T)^2}{K(T)}. \tag{\thesection.35}\label{bal:eq:cap-35}
\]
Starting from \eqref{bal:eq:cap-35}, repeated differentiation supplies the first four Taylor
coefficients of every term in \eqref{bal:eq:cap-34}, so the coefficient of
the fourth power of an auxiliary $B$-increment is exactly
$\partial_B^4\mathcal C/24$.

There is also a cancellation-reduced direct enclosure of \eqref{bal:eq:cap-M1r}.  Write
$\Xi(q)=\operatorname{artanh}T$ where $\mathcal H(T)/T=q$, and, for
$s\in\{-1,0,1\}$, put
\[
 c=A+sB,\qquad q=N/c,\qquad X=\operatorname{artanh}B+s q,
 \qquad R_s=c\Xi(q).
\]
Since $N'=-\operatorname{artanh}B$, direct differentiation cancels to
\begin{align}
 R_s'''={}&-\beta''\Xi'(q)
 +\frac{3\beta'X}{c}\Xi''(q)
 -\frac{3sX^2}{c^2}\Xi''(q)
 -\frac{X^3}{c^2}\Xi'''(q),                              \tag{\thesection.36}\label{bal:eq:cap-36}\\
 &\beta=\operatorname{artanh}B,\qquad
 \beta'=(1-B^2)^{-1},\quad
 \beta''=2B(1-B^2)^{-2}.
\end{align}
Thus
\[
 \partial_B^3\mathcal C=2R_0'''-R_{-1}'''-R_1'''
                         +\frac{8B}{(1-B^2)^3}.             \tag{\thesection.37}\label{bal:eq:cap-37}
\]
The inverse derivatives in \eqref{bal:eq:cap-36} are generated without singular subtraction.
With $U=T^2/K$ and $M=T^2/\bigl((1-T^2)K\bigr)$,
\[
 \Xi'=-M,\qquad \Xi''=UM',\qquad
 \Xi'''=-U(U'M'+UM''),                                    \tag{\thesection.38}\label{bal:eq:cap-38}
\]
and further derivatives follow by applying $-U\,d/dT$.

Use $z=B/A$.  The outward-rounded Arb ownership is as follows.
\begin{enumerate}
 \item On $0\le A\le1/2$, $0\le z\le1$, 20207 terminal
       rational boxes prove $\partial_B^4\mathcal C>0$; the smallest lower
       enclosure is greater than $1.0186\cdot10^{-3}$.  Since
       $\partial_B^3\mathcal C(A,0)=0$, this whole connected rectangle owns
       \eqref{bal:eq:cap-M1r}.
 \item On $1/2\le A\le99/100$, use the 490 exact columns
       $[i/1000,(i+1)/1000]$ and process $z$ upward from zero.  The initial
       $z$ mesh has denominator 100 and is bisected only in $z$, to depth at
       most 10.  Each of 44343 direct leaves proves \eqref{bal:eq:cap-37}/$B>0$.  Each of 4466
       fourth-derivative leaves inherits the already certified lower edge in
       the same column and propagates \eqref{bal:eq:cap-M1r} upward.  The respective smallest
       lower enclosures exceed $6.4074\cdot10^{-3}$ and
       $4.7465\cdot10^{-5}$.
 \item The bridge $99/100\le A\le9999/10000$ uses 99 columns of width
       $10^{-4}$ and a $z$ mesh of denominator 200.  Its 14485 direct and 297
       propagated leaves have smallest lower enclosures greater than $3.1257$
       and $0.6702$, respectively; the maximum depth is two.
 \item On $9999/10000\le A\le1$, a mean-value chart treats $A$ and
       $\mathcal H(A)$ as independent variables with
       $0\le\mathcal H(A)\le\mathcal H(9999/10000)$.  The 356 fixed
       intervals $[j/500,(j+1)/500]$ prove
       $\partial_B^4\mathcal C>5.9818$ for $0\le z\le89/125$.
\end{enumerate}
For the ordered columns, processing lower children before upper children is
part of the verifier, so every fourth-derivative leaf has an explicit anchor;
it is not being used as an isolated local substitute for \eqref{bal:eq:cap-M1r}.  A column
stops only when $A_{\rm low}z_{\rm low}\ge71/100$.  The rational seam checks
\[
             \frac{71}{99}<\frac{18}{25},\qquad
             \frac{7100}{9999}<\frac{89}{125}              \tag{\thesection.39}\label{bal:eq:cap-39}
\]
show that the bridge and right-face rectangles cover every point below the
$B=71/100$ cutoff.  Hence these four owners prove \eqref{bal:eq:cap-M1r} without a gap.

\smallskip\noindent\emph{Step 2: the boundary-curvature condition.}
For \eqref{bal:eq:cap-M2}, set $x=A^2$, $T=Au$, and
\[
 \psi(x)=\frac{\ln2-\mathcal H(\sqrt x)}x
 =\sum_{m\ge0}\frac{x^m}{2(m+1)(2m+1)}.                   \tag{\thesection.40}\label{bal:eq:cap-40}
\]
The defining equation \eqref{bal:eq:cap-28} becomes the regular identity
\[
 \ln2-xu^2\psi(xu^2)-u\{2\ln2-x\psi(x)\}=0.              \tag{\thesection.41}\label{bal:eq:cap-41}
\]
On $0\le x\le1/1024$, coefficientwise bounds in \eqref{bal:eq:cap-40} give
$1/2\le u\le501/1000$.  Let
\[
 \Pi(x)=\frac{\mathcal B(k_A)}{4\{\mathcal H(A)+\ln2\}},
 \qquad
 G(x)=1-\Pi(x)-\frac{1-\ln2}{2\ln2}x.
\]
The exact expansion gives $G(0)=G'(0)=0$.  The series-tail certificate uses the repaired composite chain rule.
Write $R_N(s)$ for the tail of the series in \eqref{bal:eq:cap-40} after $N$ terms. On
$0\le s\le r<1$, positivity of the omitted coefficients and their upper
bound by one give
\[
 0\le R_N(s)\le T_N(r),\quad
 0\le R_N'(s)\le T_N'(r),\quad
 0\le R_N''(s)\le T_N''(r),\qquad
 T_N(r)=\frac{r^N}{1-r}.
\]
For the composite argument $s=xu^2$, the gradient and Hessian are enclosed
using every term in
\[
 \partial_iR_N(s)=R_N'(s)s_i,\qquad
 \partial_{ij}R_N(s)=R_N''(s)s_i s_j+R_N'(s)s_{ij}.
\]
Inserting scalar tail derivatives directly as coordinate derivatives would
be invalid. With $N=18$, the repaired computation gives the rigorous
enclosure
\[
                    G''(x)>0.102\qquad(0\le x\le1/1024).   \tag{\thesection.42}\label{bal:eq:cap-42}
\]
The repaired record
\code{experiments/results/m2_chain_rule_repair_audit_2026-09-07.json}
reports the stronger bound $G''>0.1027549$ at 192, 256, and 384 bits, with
zero residuals in the complete M2 cover. For completeness, the endpoint
identities used above follow from the regular root equation:
$u(0)=1/2$ and $u'(0)=1/(16\ln2)$. If
$h=\ln2-xu^2\psi(xu^2)$,
$k=\ln2-\tfrac12\ln(1-xu^2)$, and
$n=2\ln2-x\psi(x)$, then at zero
$h'=-1/8$, $k'=1/8$, and $n'=-1/2$.
Substitution in
\[
 \Pi=\frac{h^3(2k-xu^2)}{(1-xu^2)^2k^3n}
\]
gives $\Pi(0)=1$ and
$\Pi'(0)=-(1-\ln2)/(2\ln2)$. Thus $G(0)=G'(0)=0$,
and two integrations of the positive lower bound prove \eqref{bal:eq:cap-M2} for
$0<A\le1/32$.

On $1/32\le A\le9999/10000$, a centered second-order enclosure of the
left side of \eqref{bal:eq:cap-M2} divided by $A^2$ uses 319 adaptive rational intervals.
Their ordered endpoints are checked for exact equality, and the smallest
certified margin over $(1-\ln2)/(2\ln2)$ is greater than
$1.2951\cdot10^{-6}$.  Finally, on
$9999/10000\le A<1$, monotonicity of $\mathcal H$ places
\[
 \ln2\le N\le\ln2+\mathcal H(9999/10000),\qquad
 \ln2\le N/A\le
 \frac{\ln2+\mathcal H(9999/10000)}{9999/10000}.
\]
Direct outward evaluation of the profile on the corresponding latent-root
interval leaves margin greater than $0.0767$.  This proves \eqref{bal:eq:cap-M2} globally.

Every latent inverse used above is enclosed by two strict Arb endpoint
inequalities.  SciPy and Newton iteration provide seeds only; no seed is
trusted in an acceptance decision.  All partition endpoints are exact
rationals, all terminal leaves are hashed in the JSON artifact, and the
verifier aborts on a failed root bracket, a nonpositive enclosure, a depth
cap, or an unequal one-dimensional seam.

\smallskip\noindent\emph{Step 3: integration and overlap of the value regions.}
Applying the proof of Proposition~\ref{bal:prop:vsharp} now yields \eqref{bal:eq:cap-33}.  The
large-logit owner begins at $B=\tanh(7/8)$, and the independent Arb seam check
gives
\[
                 \frac{71}{100}-\tanh(7/8)>0.00609.         \tag{\thesection.43}\label{bal:eq:cap-43}
\]
Thus \eqref{bal:eq:cap-33} and the large-logit proposition overlap and exhaust
$0<B<A<1$, proving the final assertion.
\end{proof}

\section{Analytic reductions and computational lemmas}
\label{bal:imp:components}
This section proves the reflection, joint-convexity, and inverse-function
results used in Theorems~\ref{bal:thm:four-moment-lower-bound}
and~\ref{bal:thm:e8-main}. It gives the scalar functions, coordinate maps,
integration arguments, endpoint limits, and domain decompositions.
The finite computational lemmas specify the expressions, domains, and
acceptance inequalities checked by the accompanying certificate programs;
their rational box lists and polynomial coefficients are supplied in the
machine-readable archives.

\subsection{Normalization and the perspective}
\label{bal:imp:normalization}
Write $k=\ln2$ and use sans-serif symbols for the natural-entropy versions:
\[
 \mathsf H(p)=k\HH(p),\qquad \mathsf J(p)=kJ(p)
     =\ln\frac{1-p}{p},\qquad
 \mathsf F(s,E)=kF(s,E/k).
\]
For $s,E>0$, the equivalent and complete definition of $\mathsf F$ is
\[
 \mathcal E(c)=\mathsf H((1-c)/2),\qquad
 \frac{c}{\mathcal E(c)}=\frac{s}{E},\qquad
 \mathsf F(s,E)=2s\operatorname{artanh}c,
 \quad 0<c<1.
\]
The map $c/\mathcal E(c)$ is strictly increasing from zero to infinity,
so this defines the unique latent $c$; use $\mathsf F(0,E)=0$.
Define
\[
 \mathsf j(u,w)=\tfrac12(u-w)(\mathsf J(w)-\mathsf J(u)),\quad
 \mathsf K(u,w)=\mathsf j(u,w)-
 \mathsf F\left(|u-w|,\frac{\mathsf H(u)+\mathsf H(w)}2\right),
\]
\[
 \mathsf g(E_1,E_2)=\mathsf K(\mathsf H^{-1}(E_1),\mathsf H^{-1}(E_2)).
\]
Here and below an entropy inverse means the lower inverse.  In particular,
\[
 \mathsf K=k\kappa,\qquad
 \mathsf g(k e_1,k e_2)=k g(e_1,e_2),\qquad
 \operatorname{Hess}g(e_1,e_2)
 =k\operatorname{Hess}\mathsf g(k e_1,k e_2).
\]
Consequently reflection signs and Hessian positive semidefiniteness transfer
without a change in their domains.  All logarithms in the next two
subsections are explicitly natural.  The $Q$-inequality subsection (Section~\ref{bal:imp:e8}) returns to
the main text's bit-normalized $\Gamma,\Theta,Q$.

\subsection{Reflection: reduction to positive curvature}
\label{bal:imp:reflection}

\emph{Argument guide.}
Reflection is reduced to the sign of a second derivative because the
difference and its first derivative vanish on the equality line. We derive
that curvature explicitly, cover the small-bias and compact regions,
and use specialized formulas near the ratio and high-bias boundaries.
After the domains are joined, integrating twice gives the reflection
inequality used in the four-moment lower bound.

We prove the reflection inequality by integrating a nonnegative curvature.
The symbolic and interval certificates for the computational lemmas below
are in \texttt{CK\_reflection\_curvature\_complete\_bundle}.
\subsubsection{Reflection difference and the curvature reduction}
All logarithms in this subsection are natural. Define
\[
E(a)=\mathsf{H}\!\left(\frac{1-a}{2}\right)
=\ln2-\frac12\bigl[(1+a)\ln(1+a)+(1-a)\ln(1-a)\bigr]
\]
and, for $0\le b\le a<1$,
\begin{equation}\label{bal:imp:ref:eq:D}
D(a,b)=a\operatorname{artanh}b+b\operatorname{artanh}a
-\mathsf{F}\!\left(\frac{a+b}{2},e\right)
+\mathsf{F}\!\left(\frac{a-b}{2},e\right),
\qquad e=\frac{E(a)+E(b)}2.
\end{equation}
Under $u=(1-a)/2$ and $w=(1-b)/2$, this is
\[
D(a,b)=\mathsf K(1-u,w)-\mathsf K(u,w).
\]
Thus reflection is $D\ge0$.

The equality line satisfies
\[
D(b,b)=0,\qquad D_a(b,b)=0.
\]
Consequently,
\begin{equation}\label{bal:imp:ref:eq:integrate}
D(a,b)=\int_b^a(a-t)D_{aa}(t,b)\,dt.
\end{equation}
This yields the stronger sufficient target
\[
D_{aa}(a,b)\ge0.
\]
For $b>0$, set
\[
z=\frac ba,\qquad
\mathcal T(a,z)=\frac{D_{aa}(a,az)}{a^4z}=\frac{D_{aa}(a,b)}{a^3b}.
\]
The theorem proved below is $\mathcal T>0$ for $0<a<1$ and $0<z\le1$.

\subsubsection{Exact latent-variable formula}
Let
\[
B(c)=\ln\frac{2}{\sqrt{1-c^2}},\qquad R(c)=\frac{E(c)}c.
\]
The identity
\[
B(c)=E(c)+c\operatorname{artanh}c
\]
gives
\[
R'(c)=-\frac{B(c)}{c^2}<0.
\]
For
\[
s_\pm=\frac{a\pm b}{2},\qquad R(c_\pm)=\frac e{s_\pm},
\]
a direct differentiation yields
\begin{equation}\label{bal:imp:ref:eq:Daa}
D_{aa}(a,b)=\frac{2ab}{(1-a^2)^2}+S(c_-;a,e)-S(c_+;a,e),
\end{equation}
where
\begin{equation}\label{bal:imp:ref:eq:S}
\begin{aligned}
S(c;a,e)={}&
\frac{E(c)(2B(c)-c^2)(E(c)+c\operatorname{artanh}a)^2}
{2e(1-c^2)^2B(c)^3}\\
&+\frac{c^2}{(1-a^2)(1-c^2)B(c)}.
\end{aligned}
\end{equation}
Every factor in $S$ is nonnegative.

The inverse $R^{-1}$ in the interval kernels is never accepted from floating root-finding alone.  A floating estimate proposes a bracket, and outward interval evaluation verifies the two monotonicity inequalities defining the bracket.

\subsubsection{Small-bias and finite curvature certificates}
The first finite computational lemma consists of the following small-bias series certificate and six interval covers:
\begin{itemize}
\item a Taylor--Cauchy proof for every $0<a\le0.15$ and $0<z\le1$, with
\[
\mathcal T(a,z)>0.7182128379;
\]
\item finite interval-Taylor covers on the bands listed in Table~\ref{bal:imp:ref:tab:old}.
\end{itemize}

\begin{table}[ht]
\centering
\caption{Initial finite curvature regions.}\label{bal:imp:ref:tab:old}
\small\begin{tabularx}{\linewidth}{@{}p{0.23\linewidth}X r r@{}}
\toprule
Certificate & Region in $(a,z)$ & Leaves & minimum bound\\
\midrule
Boundary $z=0$ & $[0.15,0.999]\times[0,0.001]$ & 1,135 & 0.6801165551\\
Low strip & $[0.15,0.999]\times[0.01,0.05]$ & 31,500 & 0.4308078169\\
Compact core & $[0.15,0.95]\times[0.05,0.95]$ & 41,550 & 0.3795096457\\
High-$a$ core extension & $[0.95,0.999]\times[0.05,0.95]$ & 7,650 & 131.8871202\\
High-ratio strip & $[0.15,0.999]\times[0.95,0.999]$ & 91,140 & 0.3972125366\\
Boundary $z=1$ & $[0.15,0.9975]\times[0.999,1]$ & 1,355 & 0.7498359991\\
\midrule
Total & --- & 174,330 & 0.3795096457\\
\bottomrule
\end{tabularx}
\end{table}

After these first regions, the complement consists of the low-ratio band
\[
0.15<a<0.999,\qquad0.001<z<0.01,
\]
the tiny corner
\[
0.9975<a<0.999,\qquad0.999<z<1,
\]
and the infinite edge $0.999\le a<1$.

\paragraph{The small-bias series certificate in explicit form.}
The analytic continuation of $D$ is odd in each of $a,b$.  The identities $D(a,a)=D_a(a,a)=0$, together with symmetry in $a,b$, give a double zero on $a=b$; oddness gives the corresponding double zero on $a=-b$.  Dividing these exact factors yields
\[
 D(a,b)=ab(a^2-b^2)^2\mathcal Q(a^2,b^2).
\]
To generate the local coefficients without a numerical inverse, solve
$c=t\mathcal E(c)$ as a formal odd series using
\[
 \mathcal E(c)=k-\sum_{n\ge1}\frac{c^{2n}}{2n(2n-1)},\qquad
 2\operatorname{artanh}c=2\sum_{n\ge0}\frac{c^{2n+1}}{2n+1}.
\]
Substitution into the displayed formula for $D$, division by the exact
factor, and truncation define a polynomial
$\mathcal Q_{14}(X,Y)=\sum_{i+j\le14}q_{ij}X^iY^j$.
The computational coefficient lemma asserts $q_{ij}>0$ for every one of
these 120 coefficients.  In particular $q_{00}>0.08977675454$.
For a monomial in this polynomial, putting $z=b/a\in[0,1]$ gives
\[
 \frac{\partial_a^2\{ab(a^2-b^2)^2a^{2i}b^{2j}\}}{a^3b}
 =a^{2i}b^{2j}\{(2i+5)(2i+4)
 -2(2i+3)(2i+2)z^2+(2i+1)(2i)z^4\}\ge8a^{2i}b^{2j}.
\]
The last polynomial decreases in $z^2$ on $[0,1]$ and has value $8$ at
$z=1$.  Thus every retained coefficient makes a nonnegative contribution.

For clarity the remainder check is also specified.  Take
$R=21/50$, $a_0=3/20$, $\rho=7/10$, and $r_c=9/10$.
The inequalities
\[
 \frac{R}{\mathcal E(R)}<\rho,\qquad
 \rho\{2k-\mathcal E(r_c)\}<r_c
\]
certify the analytic latent root in the required complex disc.  With
\[
 M_f=\ln\frac{1+r_c}{1-r_c},\qquad
 M_D=2R\{\operatorname{artanh}R+M_f\},\qquad q=a_0/R,
\]
the differentiated Cauchy-tail bound used by the certificate is
\[
 \mathcal R\le \frac{M_D}{a_0^6}
       \sum_{n=35}^{\infty}\frac{n(n-1)(n-2)}3q^n.
\]
The displayed sum is evaluated by the elementary geometric-series
identities for $\sum q^n$, $\sum nq^n$, $\sum n^2q^n$, and $\sum n^3q^n$.
The exact coefficient construction and outward-rounded constants certify
$8q_{00}-\mathcal R>0.7182128379$.
This proves the small-bias lower bound above, relative to these finite
coefficient and remainder assertions, on every $0<a\le3/20$ and
$0<b\le a$.  The coefficient generation, exact factor residuals, and
remainder constants are recorded in
\code{reflection_smalla_series_iv.json}; the recurrence and acceptance
criterion have been made explicit here so that its role is reviewable.
\subsubsection{The low-ratio strip: an integral-average kernel}
Natural evaluation of \eqref{bal:imp:ref:eq:Daa} is poor when $z$ is small because it subtracts two nearby values of $S$.  The quotient can instead be represented without that subtraction.

For fixed $(a,b,e)$ let $c(s)=R^{-1}(e/s)$.  Since $R'(c)=-B(c)/c^2$,
\[
\frac{dc}{ds}=\frac{E(c)^2}{eB(c)}.
\]
Define
\[
P(s)=S_c(c(s);a,e)\frac{E(c(s))^2}{eB(c(s))}.
\]
As $s_+-s_-=b$,
\[
S(c_-)-S(c_+)=-\int_{s_-}^{s_+}P(s)\,ds.
\]
Therefore
\begin{equation}\label{bal:imp:ref:eq:avg}
\mathcal T(a,z)=\frac1{a^3}
\left[
\frac{2a}{(1-a^2)^2}
-\int_{-1/2}^{1/2}P\!\left(\frac a2+\theta b\right)d\theta
\right].
\end{equation}

The verifier partitions $[-1/2,1/2]$ into 16 equal closed intervals.  On each interval it encloses the entire integrand, including a certified inverse bracket, and sums the weighted enclosures.  This is a rigorous interval Riemann enclosure, not a floating quadrature approximation.

\begin{theorem}[Low-ratio completion; computer-assisted]\label{bal:imp:ref:thm:lowz}
For every
\[
0.15\le a\le0.999,\qquad0.001\le z\le0.01,
\]
one has
\[
\mathcal T(a,z)>0.
\]
A rational cover of 3,436 boxes has minimum bound
\[
\boxed{0.14718021232352919\ldots}.
\]
\end{theorem}

The identical rational cover was recomputed at 35 and 50 decimal interval precision.  Every box passed at both precisions; the minimum bounds agree through the displayed digits.

\subsubsection{The tiny high-ratio corner}
The $c_-$ inversion becomes singular as $z\to1$, but the equation can be rewritten as
\[
c_-=t_-E(c_-),\qquad t_-=\frac{s_-}{e},
\]
which is regular at $t_-=0$.  Using this boundary-stable inversion and second-order interval Taylor bounds gives the following.

\begin{theorem}[Tiny-corner completion; computer-assisted]\label{bal:imp:ref:thm:corner}
For every
\[
0.9975\le a\le0.999,\qquad0.999\le z\le1,
\]
one has $\mathcal T(a,z)>0$.
The 90-box certificate has minimum bound
\[
\boxed{34009.52780160596\ldots}.
\]
\end{theorem}
The same 90 rational boxes were recomputed at two interval precisions.

\subsubsection{A uniform analytic theorem for the edge \texorpdfstring{\(a\ge0.999\)}{a >= 0.999}}
The remaining issue is an infinite boundary edge.  It is treated analytically rather than by boxes approaching $a=1$.

Put
\[
a=1-r,\qquad0<r\le r_0:=10^{-3}.
\]
The explicit positive term in \eqref{bal:imp:ref:eq:Daa}, after the normalization defining $\mathcal T$, is
\begin{equation}\label{bal:imp:ref:eq:base}
U_{\rm base}
=\frac{r^2}{a^3b}\frac{2ab}{(1-a^2)^2}
=\frac{2}{a^2(2-r)^2}\ge\frac12.
\end{equation}
Since $S(c_-)\ge0$, it suffices to prove
\[
\frac{r^2S(c_+;a,e)}{a^3b}<\frac12.
\]
We divide at $b=1/2$.

\paragraph{The range $0<b\le1/2$}
For $s\in[s_-,s_+]$, write $c=c(s)$.  The inequalities
\[
0.28<e<0.35,
\qquad0.4<c<0.9
\]
follow from monotonicity of $R$ and the explicit comparisons
\[
\frac{E(0.999)+\ln2}{0.499}<R(0.4),
\qquad
\frac{E(0.5)}{1.5}>R(0.9).
\]

Write $A=\operatorname{artanh}a$ and $Q=(1-a^2)^{-1}$.  Set
\[
K(c,e)=\frac{E(c)(2B(c)-c^2)}{2e(1-c^2)^2B(c)^3},
\qquad
M(c)=\frac{c^2}{(1-c^2)B(c)}.
\]
Then $S=K(E+cA)^2+QM$.  With
\[
C_s=\frac{E(c)^2}{eB(c)},
\]
a direct expansion gives
\begin{equation}\label{bal:imp:ref:eq:Ppoly}
P=C_s\bigl(C_2A^2+C_1A+C_0+QM'\bigr),
\end{equation}
where
\begin{align*}
C_2&=K'c^2+2Kc,\\
C_1&=2K'cE+2K(E+cE'),\\
C_0&=K'E^2+2KEE'.
\end{align*}

A 100-box outward interval cover of
\[
(c,e)\in[0.4,0.9]\times[0.28,0.35]
\]
proves
\begin{equation}\label{bal:imp:ref:eq:coeffbounds}
|C_sC_2|<10,
\quad |C_sC_1|<8,
\quad |C_sC_0|<8,
\quad |C_sM'|<3.
\end{equation}
The actual recorded maxima are $9.722$, $7.837$, $7.395$, and $2.616$.
Thus
\[
|P|\le10A^2+8A+8+\frac{3}{r(2-r)}.
\]
Using $A\le\tfrac12\ln(2/r)$ and multiplying by $r^2$, every term on the right is increasing for $0<r\le10^{-3}$.  At $r=r_0$ the scaled margin is
\[
\frac{2(1-r_0)}{(2-r_0)^2}
-
\left[
10r_0^2\left(\frac{\ln(2/r_0)}2\right)^2
+8r_0^2\frac{\ln(2/r_0)}2
+8r_0^2+\frac{3r_0}{2-r_0}
\right]
>0.4983162865.
\]
Equation \eqref{bal:imp:ref:eq:avg} therefore proves $\mathcal T>0$ throughout this subrange.

\paragraph{The range $1/2\le b\le a$}
Let
\[
c=c_+,
\qquad t=1-c,
\qquad s=1-b.
\]
Because
\[
R(c)=\frac{aR(a)+bR(b)}{a+b}
\]
is a weighted average and $R$ is strictly decreasing,
\[
b\le c\le a,
\qquad r\le t\le s\le\frac12.
\]
Eliminating $e$ from \eqref{bal:imp:ref:eq:S} gives
\[
S(c_+)=
\frac{c(2B-c^2)(E+cA)^2}{(a+b)(1-c^2)^2B^3}
+\frac{c^2}{(1-a^2)(1-c^2)B},
\]
where here $E=E(c)$, $B=B(c)$, and $A=\operatorname{artanh}a$.

Let $C=\operatorname{artanh}c$.  Since $B=E+cC$,
\[
E+cA=B+c(A-C)\le B+(A-C).
\]
Put
\[
\mathcal Q=\frac rt\left(1+\frac{A-C}{B}\right).
\]
The two parts of the scaled $S(c_+)$ satisfy
\begin{align}
V_1&\le
\frac{2\mathcal Q^2}{a^3b(a+b)(2-t)^2},\label{bal:imp:ref:eq:V1}\\
V_2&\le
\frac{r/t}{a^3b(2-r)(2-t)B}.\label{bal:imp:ref:eq:V2}
\end{align}

\paragraph{Case 1: $t\ge\sqrt r$.}
Here
\[
\mathcal Q\le\sqrt r\left(1+\frac{A}{\ln2}\right)
\le
\sqrt r\left(1+\frac{\ln(2/r)}{2\ln2}\right)
\le0.205008.
\]
The last function is increasing on $(0,10^{-3}]$.  Substitution into \eqref{bal:imp:ref:eq:V1}--\eqref{bal:imp:ref:eq:V2}, using
\[
a\ge0.999,
\quad b\ge\frac12,
\quad a+b\ge1.499,
\quad2-t\ge1.5,
\]
gives
\[
V_1+V_2<0.080516<\frac12.
\]

\paragraph{Case 2: $t<\sqrt r$.}
Define $h(x)=E(1-x)=\mathsf{H}(x/2)$.  The latent equation implies
\[
h(s)\le\frac{2h(t)}{1-t}.
\]
For $0<z<1$, integrating $1/(1-z)$ gives
\[
 z(1-z)\le-(1-z)\ln(1-z)\le z.
\]
Consequently
\[
 z\ln(1/z)+z(1-z)\le\mathsf H(z)\le z\ln(1/z)+z.
\]
Apply the lower bound at $z=2t$ and the upper bound at $z=t/2$.
Since $0<t\le\sqrt{10^{-3}}<1/30$, we obtain
\begin{align*}
 \frac{h(4t)-2h(t)/(1-t)}t
 &\ge \left(2-\frac1{1-t}\right)\ln\frac1t
       -\left(2+\frac1{1-t}\right)\ln2
       +2-4t-\frac1{1-t}\\
 &> \frac{28}{29}\,3-\frac{88}{29}
       +2-\frac2{15}-\frac{30}{29}
   =\frac{302}{435}>0.
\end{align*}
Here we used $\ln(1/t)>\ln30>3$ and $\ln2<1$.
This proves the needed one-variable inequality without a numerical
certificate. Since $h$ is increasing on the relevant interval,
we have $s<4t$, and hence
\[
b>1-4\sqrt{10^{-3}}.
\]

Let $k=t/r\ge1$.  Since
\[
A-C=\frac12\ln\frac{t(2-r)}{r(2-t)},
\qquad
B\ge\frac12\ln\frac2t,
\]
we obtain
\[
\frac{A-C}{B}
\le c_0(k-1),
\qquad
c_0=\frac{1+r_0/(2-\sqrt{r_0})}{\ln(2/\sqrt{r_0})}<0.242.
\]
Thus
\[
\mathcal Q\le\frac{1+c_0(k-1)}k\le1.
\]
Equations \eqref{bal:imp:ref:eq:V1}--\eqref{bal:imp:ref:eq:V2} now give
\[
V_1+V_2<0.457275<\frac12.
\]

Combining both cases with \eqref{bal:imp:ref:eq:base} proves the following.

\begin{theorem}[Uniform high-bias edge; analytic\ plus a finite coefficient certificate]\label{bal:imp:ref:thm:higha}
For every
\[
0.999\le a<1,
\qquad0<b\le a,
\]
one has
\[
D_{aa}(a,b)>0.
\]
The face $b=0$ has $D_{aa}(a,0)=0$ by continuity.
\end{theorem}

The only interval component in this infinite-edge theorem is the compact 100-box coefficient estimate \eqref{bal:imp:ref:eq:coeffbounds}.  It was independently recomputed at 80 and 120 decimal interval precision.

\subsubsection{Full curvature and reflection theorems}
The pieces now meet without gaps.

\begin{table}[ht]
\centering
\caption{Complete normalized-domain ledger.}\label{bal:imp:ref:tab:ledger}
\begin{tabularx}{\textwidth}{p{0.25\textwidth}p{0.38\textwidth}r}
\toprule
Method & Region & finite leaves\\
\midrule
Small-bias Taylor--Cauchy theorem & $0<a\le0.15$, all $z$ & ---\\
Boundary-average certificate & $0.15\le a\le0.999$, $0\le z\le0.001$ & 1,135\\
Integral-average certificate & $0.15\le a\le0.999$, $0.001\le z\le0.01$ & 3,436\\
Low strip & $0.15\le a\le0.999$, $0.01\le z\le0.05$ & 31,500\\
Core and high-$a$ extension & $0.15\le a\le0.999$, $0.05\le z\le0.95$ & 49,200\\
High-ratio strip & $0.15\le a\le0.999$, $0.95\le z\le0.999$ & 91,140\\
Boundary and tiny corner & $0.15\le a\le0.999$, $0.999\le z\le1$ & 1,445\\
High-bias theorem & $0.999\le a<1$, all $z$ & ---\\
\midrule
Total finite curvature leaves & full domain & \textbf{177,856}\\
\bottomrule
\end{tabularx}
\end{table}

\begin{theorem}[Complete reflection curvature; computer-assisted]\label{bal:imp:ref:thm:complete}
For all
\[
0<b\le a<1,
\]
one has
\[
\boxed{D_{aa}(a,b)>0.}
\]
For $b=0$, $D_{aa}(a,0)=0$ by continuity.
\end{theorem}

\begin{corollary}[Reflection inequality]\label{bal:imp:ref:cor:reflection}
For every $0<u,w\le1/2$,
\[
\boxed{\mathsf K(u,w)\le\mathsf K(1-u,w).}
\]
Equality occurs on the known equality faces, including $u=w$ and $u=1/2$.
\end{corollary}
\begin{proof}
Order the associated biases so that $0\le b\le a<1$.  Theorem~\ref{bal:imp:ref:thm:complete} and \eqref{bal:imp:ref:eq:integrate} give $D(a,b)\ge0$, which is the displayed reflection inequality.  Boundary cases follow by continuity.
\end{proof}

\subsection{Joint convexity in entropy coordinates}
\label{bal:imp:jointconvexity}

\emph{Argument guide.}
The proof works with a positive diagonal rescaling of the entropy
Hessian, so it is enough to establish two principal-minor signs. The
diagonal and endpoint charts remove degeneracies before the compact
interval cover is applied. The completion proof joins those domains,
then identifies the finite and infinite boundary values of the convex
closure.

The computational certificates for the Hessian estimates below are in
\path{CK_task1_global_completion}. Continue with the natural-normalized
sans-serif functions defined above, and set
$q=\mathsf H^{-1}:[0,k]\to[0,1/2]$.  A symbol $q_u$ or $q_w$ below denotes
$p(1-p)$ at the indicated probability; it is distinct from this inverse.
\begin{theorem}[Global joint convexity]\label{bal:imp:joint:thm:main}
The function
\[
 \mathsf{g}(e_1,e_2)=\mathsf K(q(e_1),q(e_2))
\]
is jointly convex on $(0,\ln2)^2$.  Its natural lower-semicontinuous
extended-real closure is convex on $[0,\ln2]^2$.  In the open square its
Hessian is positive definite away from the entropy diagonal and positive
semidefinite on the diagonal.
\end{theorem}

Here ``on the entropy square'' is understood in the extended-real sense at the
zero-entropy axes: $\mathsf{g}(0,0)=0$ and $\mathsf{g}(0,e)=\mathsf{g}(e,0)=+\infty$ for
$e>0$.  The sides at entropy $\ln2$ have their finite limiting values.  The
theorem is a computer-assisted result with exact symbolic components and outward-rounded MPFR components.  Together with reflection it supplies the Jensen step of LB-1.

\subsubsection{Exact Hessian reduction}
By symmetry it is enough to consider $0<u\leq w\leq1/2$.  Write
\[
 s=w-u,\quad S=\mathsf{H}(u)+\mathsf{H}(w),\quad q_u=u(1-u),\quad q_w=w(1-w),
\]
\[
 r_u=1-2u,\qquad r_w=1-2w.
\]
Let $C\in[0,1)$ be the unique solution of
\[
 \frac{C}{\mathcal E(C)}=\frac{2s}{S},
 \qquad
 \mathcal E(C)=\ln2-\frac12\bigl((1+C)\ln(1+C)+(1-C)\ln(1-C)\bigr).
\]
The inverse is unambiguous because
\[
 \frac{d}{dC}\frac{C}{\mathcal E(C)}
 =\frac{\mathcal E(C)+C\operatorname{artanh}C}{\mathcal E(C)^2}>0.
\]
Set $q_C=(1-C^2)/4$ and $\mu=\ln4-\ln(1-C^2)$.  The perspective derivatives are evaluated in the cancellation-free form
\[
 \mathsf{F}_s=2\operatorname{artanh}C+\frac{\mathcal E(C)C}{q_C\mu},
 \qquad
 \mathsf{F}_{ss}=\frac{2\mathcal E(C)^3(\mu-C^2)}{S q_C^2\mu^3}.
\]

Let
\[
 D_0=\operatorname{diag}(\mathsf{J}(u)q_u,\mathsf{J}(w)q_w),
 \qquad M=2D_0(\operatorname{Hess} \mathsf{g})D_0.
\]
This is a positive diagonal congruence in the interior.  Define
\[
 \begin{aligned}
 A_u&=\frac{q_u(\mathsf{J}_u+\mathsf{J}_w+2\mathsf{F}_s)+s(\mathsf{J}_ur_u-1)}{\mathsf{J}_u},\\
 A_w&=\frac{q_w(\mathsf{J}_u+\mathsf{J}_w-2\mathsf{F}_s)+s(1-\mathsf{J}_wr_w)}{\mathsf{J}_w},\\
 Z_u&=-q_u-\frac{s q_u\mathsf{J}_u}{S},\qquad
 Z_w=q_w\left(1-\frac{s\mathsf{J}_w}{S}\right).
 \end{aligned}
\]
Then
\[
 M=\begin{pmatrix}A_u&-(q_u+q_w)\\-(q_u+q_w)&A_w\end{pmatrix}
 -2\mathsf{F}_{ss}\begin{pmatrix}Z_u\\Z_w\end{pmatrix}
          \begin{pmatrix}Z_u&Z_w\end{pmatrix}.
\]
For evaluation up to $w=1/2$ we use the regular determinant target
\[
 K(u,w)=\mathsf{J}(w)\det M(u,w).
\]
For $w<1/2$, $K$ and $\det M$ have the same sign.

\subsubsection{Diagonal certificate}
The diagonal base value is analytic.  If $\sigma=\mathsf{H}(p)$, then the transverse coefficient in
$\mathsf{g}(\sigma+\delta,\sigma-\delta)=a(\sigma)\delta^2+O(\delta^4)$ is
\[
 a(\sigma)=\frac{4}{\mathsf{J}(p)^2}
 \left(\frac{1}{2p(1-p)}-\frac{2\ln2}{\mathsf{H}(p)}\right)>0.
\]
For completeness, the inequality is equivalent to $\mathsf{H}(p)>4(\ln2)p(1-p)$.  With
$r=1-2p$, the difference has derivative $2(\ln2)r-\operatorname{artanh}r$ and second
derivative $2\ln2-(1-r^2)^{-1}$; it rises from zero, has a single interior maximum, and
returns to zero at $r=1$, so it is strictly positive for $0<p<1/2$.

Put
\[
 \rho=\frac{w-u}{1/2-u},\qquad w=u+\rho(1/2-u).
\]
On the diagonal, the known transverse expansion
\[
 \mathsf{g}(\sigma+\delta,\sigma-\delta)=a(\sigma)\delta^2+O(\delta^4),\qquad a(\sigma)>0,
\]
gives
\[
 M_{11}(u,0)=(\mathsf{J}(u)q_u)^2a(\mathsf{H}(u))>0,
 \qquad K(u,0)=K_\rho(u,0)=0.
\]
Moreover, symmetry and $\mathsf{g}(\sigma,\sigma)=0$ give
$\mathsf{g}_{11}=\mathsf{g}_{22}=a(\sigma)/2$ and $\mathsf{g}_{12}=-a(\sigma)/2$ on the diagonal, so the diagonal Hessian is positive semidefinite of rank one.
Therefore, on a rectangle touching $\rho=0$, the two strict signs
\[
 \partial_\rho M_{11}>0,
 \qquad
 \partial_\rho^2K>0
\]
imply $M_{11}>0$ and $K>0$ for every $\rho>0$ in that rectangle.  The interval checker evaluates these derivatives with third-order bivariate Taylor jets and a centered mean-value enclosure.

\subsubsection{Endpoint charts}
\paragraph{Both probabilities small}
For $w\leq1/40$, use
\[
 w=T=e^{-1/p},\qquad u=e^{-\ell}T,
 \qquad p=\frac1{-\ln w},\quad \ell=\ln\frac wu,
\]
and normalize $N=M/T$.  The normalized formulas extend regularly to $p=0$.
The certificate uses the following integrations from exact zero manifolds:
\begin{itemize}
\item on $0\leq\ell\leq1$, positivity of $\partial_p\partial_\ell^2\det N$ near $p=0$, and positivity of $\partial_\ell^2\det N$ on the remaining $p$ interval;
\item on $1\leq\ell\leq100$, positivity of $\partial_p\det N$ near $p=0$, followed by direct positivity on the remaining compact interval;
\item for $\ell\geq100$, a separate compactification described below.
\end{itemize}
In each case $N_{11}>0$ is certified on the same boxes.  The integrations use only the exact
boundary identities
\[
 \det N(0,\ell)=0,\qquad \det N(p,0)=\partial_\ell\det N(p,0)=0.
\]
Thus, for example, $\partial_p\partial_\ell^2\det N>0$ implies first
$\partial_\ell^2\det N>0$ and then $\det N>0$ by two integrations.  The other
finite-ratio pieces use the same argument with one of the integrations already absorbed into a
direct sign certificate.

For the infinite-ratio region set
\[
 y=\frac1{p\mathsf{J}(u)},\qquad q=\frac{e^{-\ell}}y,\qquad z=e^{-\ell}=yq,
 \qquad \widetilde D=y\det N.
\]
The exact base identity is
\[
 \widetilde D(0,y,q)=(1-y)(1-q^2y^3).
\]
The corrected outward-rounded checker certifies
\[
 N_{11}>0,
 \qquad \partial_p\widetilde D>-\frac72
\]
on $0\leq p\leq0.271086$, $0\leq y\leq1$, and $0\leq q\leq29e^{-100}$.  For actual variables,
\[
 q=e^{-\ell}p\mathsf{J}(u)\leq e^{-\ell}(1+p\ell)\leq e^{-100}(1+100p)<29e^{-100}<10^{-3},
\]
where the middle bound uses the fact that $e^{-\ell}(1+p\ell)$ decreases for $\ell\geq100$ and $p\leq0.271086$.
If $\tau=1-y$, then $p\mathsf{J}(u)-1=p(\ell+\ln(1-u))$ and $\ln(1-u)>-1$, whence
\[
 \frac\tau p\geq\frac{99}{1+99(0.271086)}.
\]
Hence
\[
 \widetilde D\geq p\left[\frac{99}{1+99(0.271086)}(1-10^{-6})-\frac72\right]
 =p\frac{784301}{13918757}>0.
\]
A directed logarithm certificate verifies that $w\leq1/40$ implies $p<0.271086$.

\paragraph{One probability small}
For $u\leq1/50$ and $w\geq1/40$, put $u=e^{-1/p_u}$.  A regular fixed-edge formulation certifies
\[
 M_{11}>0,\qquad p_u\mathsf{J}(w)\det M>0
\]
for $0\leq p_u\leq0.255623$ and $1/40\leq w\leq1/2$.  A directed logarithm check proves $u\leq1/50\Rightarrow p_u<0.255623$.

\paragraph{Maximum entropy}
Write
\[
 u=\frac12-\alpha,\qquad w=\frac12-\beta,
 \qquad 0\leq\beta\leq\alpha,
 \qquad r_c=\frac\beta\alpha=1-\rho.
\]
Let $G(\alpha,\beta)=\mathsf K(1/2-\alpha,1/2-\beta)$ and put
\[
 \mathsf{J}_\alpha=2\operatorname{artanh}(2\alpha),\qquad
 K_\alpha=\frac{4}{1-4\alpha^2},\qquad
 P_\alpha=G_{\alpha\alpha}\mathsf{J}_\alpha-G_\alpha K_\alpha.
\]
The determinant numerator in this chart is
\[
 \mathcal D=P_\alpha P_\beta-G_{\alpha\beta}^2\mathsf{J}_\alpha \mathsf{J}_\beta.
\]
Exact diagonal and anti-diagonal equality symmetries imply the analytic factorization
\[
 \mathcal D=(\alpha^2-\beta^2)^2Q(\alpha,\beta).
\]
Indeed, symmetry and $G(\alpha,\alpha)=0$ give the squared factor
$(\alpha-\beta)^2$ in the Hessian determinant; complement symmetry and
$G(\alpha,-\alpha)=0$ give $(\alpha+\beta)^2$.
The exact series engine was rerun through total degree $42$.  Write the first two nonzero homogeneous terms of the factored quotient as $Q_4= c_8\alpha^4(1-r_c^2)^2$ and $Q_6=\alpha^6W_6(r_c)$.  The exact rational Bernstein certificates prove
\[
 c_8>200,\qquad W_6(r_c)>70,
\]
and positivity of every displayed homogeneous quotient through degree $38$ in $Q$, as well as every displayed odd homogeneous term through degree $41$ in the principal numerator $P_\alpha$.  In the factor coordinates $x=\alpha+\beta$ and $y=\alpha-\beta$, six anisotropic complex polydiscs give Cauchy bounds for the tails beginning at degrees $40$ and $43$, respectively.  On each polydisc the latent-root equation is a strict analytic contraction, so the implicit root and all displayed derivatives are analytic there.  If $B_D,B_P$ are the resulting supremum bounds and
\[
 p_x=\frac{(1+r_c)\alpha_0}{R_x},\qquad
 p_y=\frac{\alpha_0}{R_y},
\]
then the exact checks use the anisotropic sums
\[
 |R_Q|\leq \frac{B_D}{R_x^2R_y^2}
   \sum_{\substack{i,j\ge0\\i+j\ge40}}p_x^ip_y^j,
 \qquad
 |R_{P_\alpha}|\leq B_P
   \sum_{\substack{i,j\ge0\\i+j\ge43}}p_x^ip_y^j.
\]
Thus the same $(x,y)$-polydisc, rather than an unjustified isotropic
$(\alpha,\beta)$ bidisc, controls both remainders.  On every chart each tail is strictly smaller than the displayed positive reserve, so $P_\alpha>0$ and $Q>0$.  The largest quotient-tail ratio is below $0.588$, and the largest principal-tail ratio is below $7.0\times10^{-7}$.  Consequently $\mathcal D>0$ off the diagonal and the corner Hessian is positive definite there.  The resulting staircase is
\[
\begin{array}{c|c|c}
 u\text{ lower bound}&r_c\text{ upper bound}&\rho\text{ lower bound}\\\hline
 .41&1&0\\
 .40&.95&.05\\
 .39&.80&.20\\
 .38&.60&.40\\
 .37&.30&.70\\
 .36&.10&.90
\end{array}
\]
The coordinate conversion $\rho=1-r_c$ is important in the global coverage audit.

\paragraph{Explicit complex neighborhoods for the maximum-entropy remainder.}
The six exact parameter choices used in the Cauchy comparison are below.
Set $E=S/2=(\mathsf H(u)+\mathsf H(w))/2$. The latent parameter in
the contraction is the probability displacement $t=c/2$, so its equation
is $2Et=y\mathsf H(1/2-t)$; the radius $\tau$
is a radius for $t$, not for the bias $c$.
\begin{center}\small
\begin{tabular}{@{}rrrrr@{}}
\toprule $\alpha_0$ & $r_c$ upper bound & $R_x$ & $R_y$ & $\tau$\\\midrule
$9/100$ & $1$ & $83/200$ & $41/200$ & $9/40$\\
$1/10$ & $19/20$ & $2/5$ & $21/100$ & $11/50$\\
$11/100$ & $4/5$ & $393/1000$ & $43/200$ & $9/40$\\
$3/25$ & $3/5$ & $367/1000$ & $9/40$ & $11/50$\\
$13/100$ & $3/10$ & $329/1000$ & $49/200$ & $23/100$\\
$7/50$ & $1/10$ & $38/125$ & $13/50$ & $49/200$\\\bottomrule
\end{tabular}
\end{center}
Here $R=(R_x+R_y)/2$, $E_{\min}=69/100-2R^2/(1-4R^2)>0$,
$H_\tau=7/10+2\tau^2/(1-4\tau^2)$, and
$\mathsf{J}_\tau=4\tau/(1-4\tau^2)$ give the elementary tests
\[
 \frac{R_yH_\tau}{2E_{\min}}<\tau,
 \qquad \frac{R_y\mathsf{J}_\tau}{2E_{\min}}<1.
\]
They establish the strict self-map and contraction conditions by exact
rational comparisons.  The finite series lemma supplies the reserves
\[
 Q_{\rm reserve}=200\alpha^4(1-r_c^2)^2+70\alpha^6,
 \qquad P_{\rm reserve}=17\alpha^3.
\]
For $T_N(p,q)=\sum_{i+j\ge N}p^iq^j$, the evaluation formula is
\[
 T_N(p,q)=
 \begin{cases}
 \displaystyle\frac{p^{N+1}/(1-p)-q^{N+1}/(1-q)}{p-q},&p\ne q,\\[5pt]
 \displaystyle\frac{p^N\{N+1-Np\}}{(1-p)^2},&p=q.
 \end{cases}
\]
The tail bounds in the preceding paragraph are these exact rational
functions evaluated at $p=(1+r_c)\alpha_0/R_x$ and $q=\alpha_0/R_y$.
For smaller $\alpha$, every tail term has degree at least $40$ or $43$,
whereas the reserves have degrees at most $6$ and $3$.  Therefore checking
the ratios at $\alpha_0$ bounds the entire radial interval
$0<\alpha\le\alpha_0$.  The positive coefficients and supremum majorants
are the finite computational part of this corner lemma; their exact
objects are \texttt{corner42\_reproduced.json} and
\texttt{corner42\_polydisc\_verification\_final.json}.
\subsubsection{Compact no-gap cover}
The remaining region is certified by direct MPFR interval arithmetic.  The final top-level partition is shown below; each interval is closed in the machine ledger, and neighboring pieces overlap on their common boundary.

\begin{center}
\begin{tabular}{@{}lll@{}}
\toprule
$u$ band & near-diagonal certificate in $\rho$ & complementary certificate in $\rho$\\
\midrule
$[.02,.10]$ & $[0,.075]$ & direct core $[.075,1]$\\
$[.10,.20]$ & $[0,.10]$ & direct core $[.10,1]$\\
$[.20,.30]$ & $[0,.15]$ & direct core $[.15,1]$\\
$[.30,.36]$ & $[0,.22]$ & direct core $[.22,1]$\\
$[.36,.37]$ & $[0,.10]$ & direct core $[.10,1]$\\
$[.37,.38]$ & $[0,.30]$ & direct core $[.30,1]$\\
$[.38,.39]$ & $[0,.40]$ & analytic corner $[.40,1]$\\
$[.39,.40]$ & $[0,.20]$ & analytic corner $[.20,1]$\\
$[.40,.41]$ & $[0,.05]$ & analytic corner $[.05,1]$\\
$[.41,.50]$ & --- & analytic corner $[0,1]$\\
\bottomrule
\end{tabular}
\end{center}

The leaf-level geometry checker treats every decimal endpoint as an exact rational, proves that the accepted leaves are pairwise interior-disjoint, reconstructs each stated top-level rectangle exactly, and verifies every stored sign endpoint.  A separate symbolic coverage manifest verifies that the endpoint charts and the table above partition the full lower probability triangle without a gap.

The last rectangle, H5, is an exact composite of four disjoint strips.  H5q1 and H5q2 are completed standalone 192-bit MPFR runs.  H5q3 and H5q4 are exact clipped accepted-leaf subledgers from the completed portions of the original 192-bit MPFR traversal.  Each component has a full exact-rational geometry audit; the q3 and q4 components were also replayed on selected leaves with a separate \texttt{mpmath.iv} interval-jet implementation.  Concatenating the four component byte streams gives the delivered H5 ledger hash.  The H5 leaf count and minimum bounds are therefore exact.  Its aggregate evaluation and runtime counters include q1 and q2 only and are reported as lower bounds.

For direct boxes, $M_{11}>0$ and $K>0$ imply $M\succ0$.  For diagonal boxes, the analytic base value together with $\partial_\rho M_{11}>0$ and $\partial_\rho^2K>0$ gives the same conclusion after integration.  Hence every off-diagonal interior point in the compact region has a positive-definite entropy Hessian.

\paragraph*{Completion of the proof of Theorem~\ref{bal:imp:joint:thm:main}}
\begin{proof}
\emph{Proof outline.}
We first check that every off-diagonal probability pair belongs to
one of the endpoint charts or compact Hessian regions. The certified
principal minors then imply entropy convexity, with the exact diagonal
expansion supplying the diagonal itself. Finally we identify the limiting
values and take the lower-semicontinuous convex closure.

Consider first an interior point with $0<u<w<1/2$.  If $w\leq1/40$, the both-small chart applies: the finite-ratio ledgers cover $0\leq\ell\leq100$, and the large-ratio lemma covers $\ell\geq100$.  If $w\geq1/40$ and $u\leq1/50$, the fixed-edge chart applies.  In the remaining case $u\geq1/50=.02$, the compact table and the analytic corner staircase cover the complete interval $0<\rho<1$ in every $u$ band through $u<1/2$.  Thus every off-diagonal point lies in a region where $M_{11}>0$ and $K>0$.  Since $\mathsf{J}(w)>0$ in the interior, $\det M>0$, and the symmetric $2\times2$ matrix $M$ is positive definite.  Positive diagonal congruence then gives $\operatorname{Hess} \mathsf{g}\succ0$.

On $u=w\in(0,1/2)$, the exact diagonal expansion gives the rank-one positive-semidefinite Hessian described in the diagonal calculation above.  Interchanging $u$ and $w$ supplies the other half of the square.  Therefore every line restriction of $\mathsf{g}$ in the open entropy square has nonnegative second derivative and is convex.

It remains only to identify the boundary closure.  If $w\in(0,1/2]$ is fixed and $u\downarrow0$, then
\[
 \mathsf{j}(u,w)=\frac w2\ln\frac1u+O(1),
 \qquad
 \mathsf{F}\left(w-u,\frac{\mathsf{H}(u)+\mathsf{H}(w)}2\right)=O(1),
\]
so $\mathsf{g}(\mathsf{H}(u),\mathsf{H}(w))\to+\infty$.  At the origin the diagonal path has value zero;
moreover symmetry and interior convexity imply $\mathsf{g}\geq0$, because the midpoint of
$(e_1,e_2)$ and $(e_2,e_1)$ lies on the zero diagonal.  Thus the lower-semicontinuous
boundary value at $(0,0)$ is zero and the remaining points of the zero-entropy
axes have value $+\infty$.  At entropy $\ln2$ the defining expressions have
finite limits.  The lower-semicontinuous closure of a convex function is convex,
which proves the asserted extended-real statement on the closed square.
\end{proof}

\subsubsection{Regular formula used in the both-small chart}
The endpoint checker does not substitute singular logarithms.  It uses
\[
 E(x)=-\frac{\ln(1-x)}x,\quad z=e^{-\ell},\quad T=e^{-1/p},\quad u=zT,
\]
\[
 \mathsf{j}_u=p\mathsf{J}(u)=1+p\ell+p\ln(1-u),\qquad \mathsf{j}_w=p\mathsf{J}(T)=1+p\ln(1-T),
\]
\[
 K_0=z\mathsf{j}_u+pzE(u)+\mathsf{j}_w+pE(T),\qquad \frac{C}{\mathcal E(C)}=\frac{2p(1-z)}{K_0}.
\]
Thus $K_0=pS/T$; the constant $k=\ln2$ keeps its earlier meaning.
Put
\[
 f=2\operatorname{artanh}C+\frac{\mathcal E(C)C}{q_C\mu},\qquad
 \gamma=\frac{2p\mathcal E(C)^3(\mu-C^2)}{K_0q_C^2\mu^3},
\]
\[
 Q_u=z(1-u),\quad Q_w=1-T,\quad A=1-2u,\quad B=1-2T,
\]
\[
 a_u=-1-\frac{\mathsf{j}_u(1-z)}{K_0},\qquad a_w=1-\frac{\mathsf{j}_w(1-z)}{K_0}.
\]
Then $N=M/T$ is evaluated as
\[
 \begin{aligned}
 N_{11}&=\frac{Q_u(\mathsf{j}_u+\mathsf{j}_w+2pf)+(1-z)(\mathsf{j}_uA-p)}{\mathsf{j}_u}
          -2\gamma a_u^2Q_u^2,\\
 N_{22}&=\frac{Q_w(\mathsf{j}_u+\mathsf{j}_w-2pf)+(1-z)(p-\mathsf{j}_wB)}{\mathsf{j}_w}
          -2\gamma a_w^2Q_w^2,\\
 N_{12}&=-(Q_u+Q_w)-2\gamma a_ua_wQ_uQ_w.
 \end{aligned}
\]
Every displayed expression has a finite interval extension at $p=0$.  The code encloses the flat function $e^{-1/p}$ and its first three derivatives directly on boxes meeting $p=0$; it never forms $1/p$ on such a box.

\subsubsection{Checker targets}
For reference, the principal targets stored in the ledgers are:
\begin{itemize}
\item compact direct boxes: $M_{11}$ and $K=\mathsf{J}(w)\det M$;
\item diagonal boxes: $\partial_\rho M_{11}$ and $\partial_\rho^2K$;
\item both-small boxes: $N_{11}$ and one of $\det N$, $\partial_p\det N$, $\partial_\ell^2\det N$, or $\partial_p\partial_\ell^2\det N$;
\item fixed-edge boxes: $M_{11}$ and $p_u\mathsf{J}(w)\det M$;
\item maximum-entropy charts: $P_\alpha$ and the factored determinant quotient $Q$.
\end{itemize}

\subsection{$Q$-inequality and the fixed-entropy mean reduction}
\label{bal:imp:e8}

\emph{Argument guide.}
The mean-stationarity equations are first rewritten as a forbidden
equality involving the inverse profile. We establish the needed
one-variable structure, prove the strict inverse-profile inequality by
origin, axis, compact, and tail estimates, and finally return to the
stationarity equations. This is the reduction that moves a possible
minimum to the lower-dimensional boundary families.

Here $\Gamma,\Theta,Q$ have the bit-normalized definitions of
Section~\ref{bal:sec:definitions}, and $\HH$ is measured in bits. Natural
logarithms in the hyperbolic parametrization are accompanied by the factors
$k=\ln2$. The computational certificates for the origin, compact, and tail
estimates below are in \texttt{CK\_eq8\_reduction\_full\_artifacts}.

For specificity, the mean-dependent part of the pure gap is, up to a term
constant in the two means,
\[
 \mathcal A(m_u,m_w)=F(|m_u-m_w|,h)
 +F(1-m_u-m_w,h)
 -\tfrac12F(1-2m_u,e_u)-\tfrac12F(1-2m_w,e_w),
 \quad h=(e_u+e_w)/2,
\]
on the lower-half mean rectangle.  The stationarity calculation below is
for this function and hence for the pure gap.  It is not an assertion
about every individual member of a hybrid lower-bound collection.
\subsubsection{Stationarity and exact reduction to $Q$-inequality}

Use the main text's bit-normalized $\Gamma$, $\Theta=\Gamma'$, and $Q=\Theta^{-1}$.
Work first in the smooth chamber $m_u>m_w$ and put
\[
 a=\frac{1-2m_u}{2e_u},\qquad
 b=\frac{1-2m_w}{2e_w},\qquad
 c=\frac{1-m_u-m_w}{e_u+e_w},\qquad
 d=\frac{m_u-m_w}{e_u+e_w}.
\]
For the pure-gap mean objective at fixed entropies, direct differentiation gives the two stationary equations
\begin{equation}\label{bal:imp:e8:eq:stationarity}
 \Theta(c)=\Theta(d)+\Theta(a),
 \qquad
 \Theta(b)=\Theta(d)+\Theta(c).
\end{equation}
The definitions also give
\[
 ae_u+be_w=c(e_u+e_w),
 \qquad -ae_u+be_w=d(e_u+e_w),
\]
and hence
\begin{equation}\label{bal:imp:e8:eq:ratio-identity}
 \frac{b-c}{c-a}=\frac{b-d}{d+a}.
\end{equation}
These identities follow directly by substituting the four displayed mean coordinates; no external calculation is needed.

The results below show that $\Theta$ is strictly increasing. Let
\[
 Q=\Theta^{-1},\qquad s=\Theta(d),\qquad t=\Theta(a).
\]
Then \eqref{bal:imp:e8:eq:stationarity} forces
\[
 d=Q(s),\quad a=Q(t),\quad c=Q(s+t),\quad b=Q(2s+t).
\]
Thus \eqref{bal:imp:e8:eq:ratio-identity} would be equality in
\begin{equation}\label{bal:imp:e8:eq:eq8}
 \frac{Q(2s+t)-Q(s+t)}{Q(s+t)-Q(t)}
 <
 \frac{Q(2s+t)-Q(s)}{Q(s)+Q(t)},
 \qquad s,t>0.
\end{equation}
This is the $Q$-inequality, now stated as a numbered inequality internal to this manuscript.

Set
\[
 A=Q(t),\quad D=Q(s),\quad C=Q(s+t),\quad B=Q(2s+t)
\]
and define
\begin{equation}\label{bal:imp:e8:eq:delta-def}
 \Delta(s,t)=(B-D)(C-A)-(B-C)(D+A).
\end{equation}
All denominators in \eqref{bal:imp:e8:eq:eq8} are positive, so \eqref{bal:imp:e8:eq:eq8} is equivalent to $\Delta(s,t)>0$.

\subsubsection{Stable parametrization and global one-variable structure}

Let $k=\ln 2$ and use a hyperbolic parameter $\alpha\ge0$. Put
\[
 z=e^{-2\alpha},\qquad r=\tanh\alpha,
 \qquad q=\operatorname{sech}^2\alpha,
\]
\[
 \ell=\ln(2\cosh\alpha),\qquad h=\ell-\alpha r.
\]
The inverse function is evaluated through the stable formulas
\begin{equation}\label{bal:imp:e8:eq:param}
 Q(\Theta_*(\alpha))=X(\alpha)=\frac{k r}{2h},
 \qquad
 \Theta_*(\alpha)=\frac2k\left(\alpha+\frac{rh}{q\ell}\right).
\end{equation}
Here $\Theta_*$ is the value of $\Theta$ at $X(\alpha)$. The map $\alpha\mapsto\Theta_*(\alpha)$ is strictly increasing, so interval inversion is performed by monotone bracketing rather than by an unverified numerical root.

For completeness, define
\[
 P(\alpha)=\Theta'(X(\alpha))
 =\frac{4h^3(2\ell-r^2)}{k^2q^2\ell^3}>0,
\]
\[
 E=(\ln P)'_\alpha,
 \qquad
 X_0=(\ln X')'_{\alpha},
 \qquad
 \mathcal L=E^2-E'+EX_0.
\]
A direct chain-rule calculation gives
\[
 \Theta''(X(\alpha))=\frac{P E}{X'(\alpha)},
 \qquad
 (\ln Q')''(\Theta_*(\alpha))
 =\frac{\mathcal L}{P^2X'(\alpha)^2}.
\]
Therefore $E<0$ proves strict concavity of $\Theta$, while $\mathcal L>0$ proves strict log-convexity of $Q'$.

\begin{theorem}[Global one-variable structure]\label{bal:imp:e8:thm:onevar}
For $x>0$,
\[
 \Theta'(x)>0,\qquad \Theta''(x)<0,
 \qquad Q''(x)>0,
 \qquad (\ln Q'(x))''>0.
\]
In particular, $Q'$ is positive, even, and strictly log-convex after the natural odd extension of $Q$.
\end{theorem}

\begin{proof}[Computer-assisted proof]
\emph{Proof outline.}
The derivative identities above reduce the one-variable claims to
two scalar signs. A finite cover checks them on the compact hyperbolic
range, using an anchored derivative at the origin. An inverse-coordinate
tail estimate preserves both signs on the remaining unbounded range.

The compact range $0\le\alpha\le20$ is covered by 7,464 exact rational interval leaves at 256-bit MPFR precision. On every non-origin leaf the stored upper endpoint for $E$ is negative, and every stored lower endpoint for $\mathcal L$ is positive. On the unique origin leaf, $E(0)=0$ is used together with a strictly negative interval for $E'$. The smallest recorded lower endpoint for $\mathcal L$ is
\[
 6.7683752437\times10^{-7}.
\]
For $\alpha\ge20$, set $v=1/\alpha$ and $u=e^{-2/v}$. At $u=0$, the exact limiting expressions are
\[
 E_0(v)=-\frac{3v^3-2v^2+4v-8}{(v-2)(v+2)},
\]
\[
 \mathcal L_0(v)=
 \frac{v^2(3v^4-4v^3+48v^2-48v+16)}{(v^2-4)^2}.
\]
On $0\le v\le1/20$, direct interval differentiation bounds the exponentially small $u$ correction and preserves $E<0$ and $\mathcal L>0$. The tail certificate gives
\[
 E<-1.9023,\qquad \mathcal L/v^2>0.8499.\qedhere
\]
\end{proof}

\subsubsection{The global $Q$-inequality theorem}

\begin{theorem}[$Q$-inequality]\label{bal:imp:e8:thm:eq8}
For every $s,t>0$,
\[
 \Delta(s,t)>0.
\]
Equivalently, the strict inequality \eqref{bal:imp:e8:eq:eq8} holds globally.
\end{theorem}

\emph{Proof outline.}
We prove positivity of the factored scalar difference throughout the
positive quadrant. Log-convexity handles the large first-coordinate tail;
exact vanishing identities support the origin and axis estimates.
Derivative integration and direct bounds cover the compact remainder,
and a regularized inverse profile treats the large second-coordinate
tail. The final case split checks that these ranges exhaust the quadrant.

\paragraph{Large-$s$ tail}

The following algebraic identity is useful:
\begin{equation}\label{bal:imp:e8:eq:delta-tail-id}
 \Delta=B(C-D-2A)+A(C+D).
\end{equation}
A directed 256-bit enclosure proves
\[
 Q'(3.15)-2Q'(0)>0.0026669186327.
\]
By Theorem \ref{bal:imp:e8:thm:onevar}, $\ln Q'$ is convex, hence
\[
 \frac{Q'(u+s)}{Q'(u)}\ge \frac{Q'(s)}{Q'(0)}>2
 \qquad(s\ge3.15,\ u\ge0).
\]
Integrating from $0$ to $t$ gives $C-D>2A$. Equation \eqref{bal:imp:e8:eq:delta-tail-id} then yields
\begin{equation}\label{bal:imp:e8:eq:large-s}
 \Delta(s,t)>0\qquad(s\ge3.15,\ t>0).
\end{equation}

\paragraph{Exact factor and the origin chart}

Oddness and analyticity of $Q$ imply the exact factorization
\begin{equation}\label{bal:imp:e8:eq:factor}
 \Delta(s,t)=s^2t(s+t)^2(2s+t)W(s,t),
\end{equation}
where $W$ extends analytically across the axes in a neighborhood of every finite point of the closed first quadrant. Its exact origin value is
\[
 W(0,0)=\frac{(13k-4k^2-6)k^4}{32768}
 =7.6722047\ldots\times10^{-6}>0.
\]
Let
\[
 K=\partial_s^2\partial_t\Delta.
\]
A degree-15 Taylor calculation with certified coefficients and a directed bound on the omitted seventeenth derivative of $Q$ proves the following estimate.  The required inverse-function arguments range over $[0,4/25]$, because $2s+t\le2(s+t)\le4/25$.  The remainder bound is certified on all sixteen exact intervals $[i/100,(i+1)/100]$, $0\le i\le15$, at 512-bit Arb precision.  Their enclosures imply $|Q^{(17)}|<2.9\times10^{14}$ throughout the required interval (the computed union is much tighter).  The full interval $[0,4/25]$ is required to control $2s+t$. The polynomial and remainder bounds give
\begin{equation}\label{bal:imp:e8:eq:local-K}
 \frac{K(s,t)}{(s+t)^3}
 >4.7096804935\times10^{-5}
 \qquad(s,t\ge0,\ 0<s+t\le2/25).
\end{equation}
The complete corrected calculation is recorded in
\texttt{aux\_half\_mean\_origin\_512.json}; it has zero failed interval
cells.  This is a new arithmetic repair of the origin sublemma, distinct
from the structural audit of the older E8 ledger.

The full coefficient and remainder argument appears in the proof of
Theorem~\ref{bal:aux:thm:D0}. Since $\partial_s^2\Delta(s,0)=0$, integration
in $t$ gives $\partial_s^2\Delta(s,t)>0$ for $t>0$ in the triangle.
Since $\Delta(0,t)=\partial_s\Delta(0,t)=0$, two integrations in $s$
give $\Delta(s,t)>0$ for $s,t>0$ and $s+t\le2/25$.
The axes have the exact equality $\Delta=0$; no strict sign is asserted
there or for the undefined quotient at the origin.

\paragraph{Axis strips}

Direct-MPFR centered Taylor certificates prove
\[
 \partial_s^2\Delta(s,t)>0
 \quad(0\le s\le0.02,\ 0.06\le t\le20),
\]
using 26,273 accepted leaves. The smallest stored lower endpoint is
\[
 1.5944994938\times10^{-11}.
\]
Together with the exact double zero at $s=0$, this proves $\Delta>0$
for $s>0$ in the $s$-axis strip; $\Delta(0,t)=0$ on its axis.

Similarly,
\[
 \partial_t\Delta(s,t)>0
 \quad(0.06\le s\le3.15,\ 0\le t\le0.02),
\]
using 773 accepted leaves, with minimum bound
\[
 2.4854836211\times10^{-12}.
\]
Since $\Delta(s,0)=0$, this proves positivity for $t>0$ in the
$t$-axis strip, with equality on its axis.

Both bridge rectangles were replayed at 256-bit precision and reproduced the same positive lower endpoints.

\paragraph{Compact off-axis region}

A one-dimensional 192-bit certificate first proves
\begin{equation}\label{bal:imp:e8:eq:anchor}
 \partial_s^2\Delta(s,0.02)>0
 \qquad(0.02\le s\le3.15),
\end{equation}
with 199 exact rational segments and minimum bound
\[
 8.2757412153\times10^{-11}.
\]

On the regions
\[
 0.02\le s\le0.1,\quad 0.02\le t\le12,
\]
and
\[
 0.1\le s\le3.1,\quad0.02\le t\le0.1,
\]
a no-gap interval cover proves
\[
 \partial_t\partial_s^2\Delta(s,t)>0.
\]
It consists of 45 top-level rational ledgers and 132,390 accepted leaves; the smallest stored lower endpoint is
\[
 4.3208924258\times10^{-10}.
\]
Combining this with \eqref{bal:imp:e8:eq:anchor} gives
$\partial_s^2\Delta>0$ on the two displayed derivative regions.  To
integrate from $s=0$, the missing strip $0\le s\le0.02$ is also needed.
For $0.02\le t\le0.06$, it lies in $s+t\le2/25$, so
\eqref{bal:imp:e8:eq:local-K} and integration in $t$ give the same strict
sign.  For $0.06\le t\le12$, the preceding $s$-axis-strip estimate gives
it directly.  Thus, for each point in either derivative region, the
sign holds along the entire segment from $0$ to its $s$ coordinate.
The exact identities $\Delta(0,t)=\partial_s\Delta(0,t)=0$ now justify
two integrations in $s$ and give $\Delta>0$.

On the remaining compact rectangles, the normalized quantity $W$ in \eqref{bal:imp:e8:eq:factor} is certified directly:
\[
 0.1\le s\le3.1,\quad0.1\le t\le12,
\]
using the appropriate subset of a 40-ledger, 74,663-leaf no-gap cover, and
\[
 3.1\le s\le3.15,\quad0.02\le t\le12,
\]
using ten exact-string MPFR ledgers from an eleven-ledger bridge. The full bridge has 1,748 leaves. Every accepted leaf stores a strictly positive lower endpoint.

The exact rational area audit gives the compact partition
\[
 [0.02,3.15]\times[0.02,12]
\]
as the disjoint-interior union of the two derivative regions and the two direct-$W$ regions. Its exact area is $187487/5000$.

\paragraph{Large-$t$ tail}

For the large-$t$ direction, introduce the regular function
\[
 R(y)=\frac{yQ(y)}{2^y}.
\]
The singular endpoint $t=\infty$ becomes a finite compact endpoint in the associated tail parameter. A 192-bit direct-MPFR cover proves
\begin{equation}\label{bal:imp:e8:eq:tail12}
 \Delta(s,t)>0
 \qquad(t\ge11.9,\ 0.005\le s\le3.15).
\end{equation}
The certificate contains 1,536 top-level tasks, 39,854 accepted leaves, and minimum bound
\[
 1.9349211478\times10^{-12}.
\]
Each of 256 tail-parameter cells carries certified interval bounds for $R,R',R''$; six exact $s$-bands cover the complete shape interval.

For the remaining small-$s$ part of the infinite tail, a separate regular certificate proves
\[
 \frac{\partial_s^2\Delta(s,t)}{(2^t/t)^2}>0.0215445
 \qquad(t\ge20,\ 0\le s\le0.005).
\]
Double integration in $s$ closes that strip.

\paragraph{No-gap global case split}

The certificates above cover the first quadrant exactly:

\begin{enumerate}
\item $s\ge3.15$ is covered by \eqref{bal:imp:e8:eq:large-s}.
\item If $0<s<3.15$ and $t\ge20$, use \eqref{bal:imp:e8:eq:tail12} when $s\ge0.005$ and the small-$s$ tail when $s\le0.005$.
\item If $12\le t\le20$, use \eqref{bal:imp:e8:eq:tail12} for $s\ge0.005$ and the $s$-axis strip for $s\le0.02$.
\item It remains to consider $0<s<3.15$ and $0<t\le12$.
      If $s\le0.02$, use the origin chart when $t<0.06$, since then
      $s+t<0.08$, and use the $s$-axis strip otherwise.
      If $t\le0.02$, use the origin chart when $s<0.06$ and the
      $t$-axis strip otherwise.
      Every remaining point satisfies $s\ge0.02$ and $t\ge0.02$ and lies
      in at least one of the four compact regions certified above.
\end{enumerate}

This completes the proof of Theorem \ref{bal:imp:e8:thm:eq8}.

\subsubsection{Consequence: the four-to-three variable reduction}

\begin{corollary}[No genuine interior mean critical point]\label{bal:imp:e8:cor:reduction}
Fix feasible positive entropies $e_u,e_w$. In either smooth chamber $m_u>m_w$ or $m_w>m_u$, the pure-gap mean objective has no genuine interior critical point.
\end{corollary}

\begin{proof}
At such a critical point, $d>0$ and $a>0$, hence $s=\Theta(d)>0$ and $t=\Theta(a)>0$. Equations \eqref{bal:imp:e8:eq:stationarity} and \eqref{bal:imp:e8:eq:ratio-identity} would force equality in \eqref{bal:imp:e8:eq:eq8}, contradicting Theorem \ref{bal:imp:e8:thm:eq8}.
\end{proof}

By continuity on the compact feasible mean rectangle, an extremum is therefore attained on its boundary or on the crease separating the two smooth chambers. Hence it lies on one of
\[
 m_u=m_w,\quad m_u=\tfrac12,\quad m_w=\tfrac12,
 \quad \HH(m_u)=e_u,\quad \HH(m_w)=e_w.
\]
Exchanging $u$ and $w$ leaves three symmetry-distinct families:
\[
 \boxed{m_u=m_w},\qquad
 \boxed{m_u=\tfrac12},\qquad
 \boxed{\HH(m_u)=e_u}.
\]
Each has three free parameters.  Additional restrictions in the main proof, such as a retained interface, contribute their own boundary faces and are not removed by this reduction.

\paragraph{Convexity of the perspective used by LB-1.}
The one-variable theorem also makes the convex-perspective assertion
explicit.  Since $\Theta(0)=0$ and $\Theta'>0$, one has $F_s=\Theta\ge0$.
For $z=s/(2e)$,
\[
 \operatorname{Hess}_{s,e}F(s,e)
 =\frac{\Theta'(z)}{2e}
 \begin{pmatrix}1&-2z\\-2z&4z^2\end{pmatrix}\succeq0.
\]
The matrix is the outer product of $(1,-2z)$ with itself, multiplied by a
positive scalar.  Thus $F$ is jointly convex on $s\ge0,e>0$ by its continuous
extension at $s=0$, and it is increasing in $s$.  These are exactly the two
perspective properties used in the Jensen proof of LB-1.
\subsection{What the imported certificates establish and what their checks mean}
\label{bal:imp:trust}
The preceding implications are mathematical: strict curvature integrates
to reflection, positive principal minors imply entropy-coordinate convexity,
and the strict $Q$-inequality~\eqref{bal:imp:e8:eq:eq8} contradicts the two mean-stationarity equations.
The finite lemmas have different implementations. Reflection uses
\texttt{mpmath.iv} for its finite interval lemmas, including the coefficient
bounds; the one-variable entropy comparison in the high-bias proof is
analytic. Joint convexity and the
$Q$-inequality~\eqref{bal:imp:e8:eq:eq8} use directed MPFR for their compact
interval lemmas and exact rational arithmetic for the stated algebraic
parts.  The arithmetic implementations for these components are thus
MPFR and \texttt{mpmath.iv}.

The delivered reflection inventory contains 177,856 curvature leaves plus
100 coefficient boxes.  The joint-convexity inventory contains 1,282,104
accepted leaves in 33 ledgers, including the endpoint charts.  The E8
inventory has separate origin, one-variable, axis, compact, bridge, and
infinite-tail certificates on the domains stated above.  These counts are
provenance and coverage information; they do not by themselves certify the
underlying transcendental arithmetic.

Numerical verification evaluates the displayed mathematical expression
with outward interval bounds on each box; coverage verification checks
that the exact box union is the stated domain. Structural audits of stored
endpoints and hashes are distinguished from arithmetic replays that
re-evaluate the expressions. Section~\ref{bal:sec:verification} records
the arithmetic replays and structural audits for each component.

\section{Auxiliary curvature and boundary arguments}
\label{bal:aux:sec:components}

This section supplies the equal-entropy, equal-mean, and smooth half-mean
arguments used in the retained-domain proof.  The analytic implications are
written out here.  The finite inequalities used below are stated explicitly,
with exact domains and replay contracts.  The symbols $F,\Phi,\Theta,Q$
retain the bit-normalized definitions of Section~\ref{bal:sec:definitions}.
Whenever natural entropy is used, it is distinguished typographically and
converted explicitly.

\subsection{The four-point perspective inequality}
\label{bal:aux:sec:fourpoint}

\begin{lemma}[Four-point inequality]
\label{bal:aux:lem:fourpoint}\label{bal:aux:fourpoint}
For $e>0$ and $r,x\ge0$,
\begin{equation}
 F(r,e)+F(x,e)\ge
 \frac{F(r+x,e)+F(|r-x|,e)}2.
 \label{bal:aux:eq:fourpoint}
\end{equation}
\end{lemma}
\begin{proof}
\emph{Proof outline.}
With entropy fixed, rewrite the desired inequality as nonnegativity
of a function that vanishes at zero. Concavity of the radial derivative
shows that this function is increasing on each side of the possible
crease. Continuity joins the two ranges, and integration proves the bound.

Fix $e$ and write $f(s)=F(s,e)$.  The homogeneous representation gives
\[
 f'(s)=\Theta\!\left(\frac{s}{2e}\right),\qquad
 f''(s)=\frac1{2e}\Theta'\!\left(\frac{s}{2e}\right).
\]
The one-variable structure theorem
(Theorem~\ref{bal:imp:e8:thm:onevar}) shows that $f'$ is increasing and concave
on $[0,\infty)$, with $f'(0)=0$; also $f(0)=0$.
Put
\[
 A(r)=f(r)+f(x)-\tfrac12f(r+x)-\tfrac12f(|r-x|).
\]
Then $A(0)=0$.  If $r\ge x$, concavity of $f'$ gives
\[
 A'(r)=f'(r)-\tfrac12\{f'(r+x)+f'(r-x)\}\ge0.
\]
If $0\le r\le x$, decreasing increments of the concave function $f'$
give
\[
 f'(x+r)-f'(x-r)\le f'(2r)-f'(0)\le2f'(r).
\]
Consequently $A'(r)\ge0$ in this range as well.  Continuity at $r=x$
and integration from $0$ prove \eqref{bal:aux:eq:fourpoint}.
\end{proof}

For equal positive entropies the pure gap has the exact expression
\[
 \Gap=F(x,e)+F(r,e)-\tfrac12F(r+x,e)-\tfrac12F(|r-x|,e),
 \quad x=|\mu_w-\mu_u|,\quad r=|1-\mu_u-\mu_w|.
\]
Indeed the two individual radial coordinates form the multiset
$\{r+x,|r-x|\}$, and the entropy-gap term $g(e,e)$ vanishes.
Lemma~\ref{bal:aux:lem:fourpoint} therefore proves EE-1 directly.  No
external physical-page reference is needed for this implication.

\subsection{Entropy convexity of the Bellman candidate}
\label{bal:aux:sec:entropy}

Put $k=\ln2$.  In this subsection only, define the natural-entropy functions
\[
 \mathsf H(q)=-q\ln q-(1-q)\ln(1-q),\qquad
 \mathsf J(q)=\ln\frac{1-q}{q},\qquad
 \mathsf L(q)=\frac{2\mathsf H(q)}{1-2q}.
\]
For $h>0$, let
\[
 \mathsf F(s,h)=kF(s,h/k),\quad
 \mathsf\eta(h)=k\eta(h/k),\quad
 \mathsf\Phi(s,h)=k\Phi(s,h/k).
\]
Thus $\mathsf\Phi=\mathsf\eta-\mathsf F$ and
\[
 \mathsf F(s,h)=s\,\mathsf J\!\left(\mathsf L^{-1}(2h/s)\right),
 \quad
 \mathsf\eta(h)=(1-2q)\mathsf J(q),\quad q=\mathsf H^{-1}(h).
\]
All logarithms in the following calculations are natural.

\begin{theorem}[Entropy convexity and increasing differences]
\label{bal:aux:thm:entropyconvexity}\label{bal:aux:entropy}
On $0\le s<1$ and $0<h<\mathsf H((1-s)/2)$,
\[
 \mathsf\Phi_{hh}(s,h)>0,\qquad \mathsf\Phi_{sh}(s,h)\ge0.
\]
The second inequality is strict for $s>0$.  For each fixed $s$, the
continuous extension to $h=\mathsf H((1-s)/2)$ is convex on
$(0,\mathsf H((1-s)/2)]$; its supporting-line inequality also holds at
the upper endpoint, with the left derivative there.  The corresponding
statements hold for the bit-normalized $\Phi$.
\end{theorem}

\begin{proof}
\emph{Proof outline.}
Feasibility compares the latent perspective coordinate with the
entropy inverse. We use that order to reduce entropy convexity to two
explicit scalar signs in a hyperbolic parameter. Analytic bounds at both
ends and a finite certificate in the middle prove those signs. We then
return to the original units and pass to the entropy cap in the
convexity and supporting-line inequalities.

For $s>0$ put $Y=2h/s$ and $b(Y)=2Y f_0''(Y)$, where
$f_0(Y)=\mathsf J(\mathsf L^{-1}(Y))$.  Differentiating the perspective
$\mathsf F=sf_0(Y)$ gives
\begin{equation}
 \mathsf F_{ss}=\frac{Y^3f_0''(Y)}{2h}>0,\quad
 \mathsf F_{sh}=-\frac{s}{h}\mathsf F_{ss},\quad
 h\mathsf F_{hh}=b(Y).
 \label{bal:aux:eq:perspderivatives}
\end{equation}
The positive curvature here follows equivalently from $\Theta'>0$.
Feasibility implies $s\le1-2q$ and hence $Y\ge\mathsf L(q)$.
It therefore suffices to prove
\begin{equation}
 b'(Y)<0,\qquad h\mathsf\eta''(h)>b(\mathsf L(q)).
 \label{bal:aux:eq:entropyreduction}
\end{equation}

Use $\xi=\tfrac12\ln((1-q)/q)>0$ and write
\[
 t=\tanh\xi,\qquad B=\ln(2\cosh\xi),\qquad
 h=B-\xi t,\qquad v=1-t^2.
\]
Then $t'=v$, $B'=t$, $h'=-\xi v$ and
$\mathsf L(q)=2h/t$ is strictly decreasing in $\xi$.
Straight differentiation gives
\begin{align}
 h\mathsf\eta''(h)&=
 \frac{2h\{\xi(1+t^2)-t\}}{\xi^3v^2},\label{bal:aux:eq:etacurvature}\\
 b(\mathsf L(q))&=\frac{2ht^2(2B-t^2)}{B^3v^2}.
 \label{bal:aux:eq:bcurvature}
\end{align}
Define the two explicit functions
\begin{align}
 P(\xi)&=B^3\{\xi(1+t^2)-t\}-\xi^3t^2(2B-t^2),
 \label{bal:aux:eq:P}\\
 R(\xi)&=4B^3(1+t^2)-2B^2t^3\xi-8B^2t^2-6B^2t\xi\notag\\
 &\hspace{1cm}-Bt^5\xi+3Bt^4+9Bt^3\xi-3t^5\xi.
 \label{bal:aux:eq:R}
\end{align}
The identities
\begin{align}
 h\mathsf\eta''-b(\mathsf L(q))
 &=\frac{2hP(\xi)}{\xi^3B^3v^2},\label{bal:aux:eq:Pidentity}\\
 \frac{d}{d\xi}b(\mathsf L(q(\xi)))
 &=\frac{2tR(\xi)}{B^4v^2}
 \label{bal:aux:eq:Ridentity}
\end{align}
reduce \eqref{bal:aux:eq:entropyreduction} to $P,R>0$.

\emph{Small endpoint, $0<\xi\le1/10$.}
Elementary differentiation proves
\[
 \xi-\frac{\xi^3}{3}\le t\le
 \xi-\frac{\xi^3}{3}+\frac{2\xi^5}{15},\qquad
 \ln2\le B\le\ln2+\frac{\xi^2}{2}.
\]
For completeness, the lower hyperbolic bound follows because its difference
has derivative $\xi^2-t^2\ge0$.  For the upper bound the derivative of the
difference is $t^2-\xi^2+2\xi^4/3$, which is at least $\xi^6/9$
after applying the lower bound.  Integrating $B'=t\le\xi$ proves the
bound for $B$.
Thus
\[
 \xi(1+t^2)-t\ge
 \xi^3\left(\frac43-\frac45\xi^2+\frac19\xi^4\right),\qquad
 t^2(2B-t^2)\le\xi^2(2\ln2+\xi^2).
\]
Using $69/100<\ln2<7/10$ and $\xi^2\le1/100$ yields the conservative
rational bound
\begin{equation}
 P(\xi)\ge\frac{52660491}{125000000}\,\xi^3>0.
 \label{bal:aux:eq:Psmall}
\end{equation}
If $\beta(\xi)=b(\mathsf L(q(\xi)))$, logarithmic differentiation of
\eqref{bal:aux:eq:bcurvature} gives
\begin{equation}
 \frac{\beta'}{\beta}
 =-\frac{\xi v}{h}+\frac{2v}{t}
   +\frac{2t^3}{2B-t^2}-\frac{3t}{B}+4t.
 \label{bal:aux:eq:logbeta}
\end{equation}
Here $h\ge68/100$, $t\le\xi$, and $B\ge69/100$.
Discarding positive terms gives
\[
 \frac{\beta'}{\beta}
 \ge 2(\xi^{-1}-\xi)-\frac{1/10}{68/100}
             -\frac{3/10}{69/100}>0.
\]
This proves $R>0$ at the small endpoint.

\emph{Large endpoint, $\xi\ge2$.}
Put $z=\exp(-2\xi)<1/50$, so
$t=(1-z)/(1+z)$ and $B=\xi+\ln(1+z)$.
Let $A=\xi(1+t^2)-t$ and $C=2B-t^2$.  Algebra gives
\[
 A-t^2C=\frac{2}{(1+z)^4}\Bigl[
 -\ln(1+z)(1-z^2)^2+z^4+2z^3\xi-z^3
                 +4z^2\xi+3z^2+2z\xi-3z\Bigr].
\]
Using $\ln(1+z)\le z-z^2/2+z^3/3$, the bracket is at least
\[
 \frac z6\Bigl[-2z^6+3z^5-2z^4
 +(12z^2+24z+12)\xi+4z^2+21z-24\Bigr].
\]
This increases with $\xi$ and at $\xi=2$ becomes
\[
 \frac{z^2}{6}\{-2z^5+3z^4-2z^3+28z+69\}>0.
\]
Since $A'=2t^2+2\xi tv>0$ and $A(0)=0$, one has $A>0$.
Consequently $P=(B^3-\xi^3)A+\xi^3(A-t^2C)>0$.
For $R$, the exact identity
\[
 2h-\xi v=2\ln(1+z)+\frac{4\xi z^2}{(1+z)^2}>0
\]
used in \eqref{bal:aux:eq:logbeta}, together with $B\ge\xi\ge2$ and
$t\ge49/51$, yields
\[
 \frac{\beta'}{\beta}>-2-\frac32+4\frac{49}{51}
 =\frac{35}{102}>0.
\]

\emph{Compact interval, $1/10\le\xi\le2$.}
The exact partition consists of
\[
 I_i=\left[\frac{10+2i}{100},\frac{12+2i}{100}\right],
 \quad i=0,\ldots,94.
\]
Writing $c_i=(11+2i)/100$ and $\rho=1/100$, the mean-value enclosure is
\begin{equation}
 P(I_i)\subset P(c_i)+[-\rho,\rho]P'(I_i),\qquad
 R(I_i)\subset R(c_i)+[-\rho,\rho]R'(I_i).
 \label{bal:aux:eq:compactcontract}
\end{equation}
All evaluations use outward-rounded Arb balls.  Derivatives are obtained by
first-order automatic differentiation of \eqref{bal:aux:eq:P} and
\eqref{bal:aux:eq:R}, with the exact rules $t'=1-t^2$ and $B'=t$.
A fresh replay of these 95 cells at both 192 and 320 bits gives, on every
cell, the conservative lower bounds
\begin{equation}
 P>\frac{351}{10^6},\qquad R>\frac{747}{1000}.
 \label{bal:aux:eq:compactresults}
\end{equation}
The verifier is \texttt{check\_entropy\_convexity\_aux.py}; its two JSON
ledgers store exact rational domain endpoints and dyadic lower bounds.
Thus \eqref{bal:aux:eq:compactcontract} is a finite, reproducible certificate
for the only nonanalytic step in this entropy-curvature argument.
The exact partition and the two endpoint arguments cover every $\xi>0$.

It now follows from feasibility that
\[
 h\mathsf F_{hh}=b(Y)\le b(\mathsf L(q))<h\mathsf\eta''(h),
\]
proving the desired strict entropy curvature.  Equation
\eqref{bal:aux:eq:perspderivatives} gives the mixed derivative sign.
At $s=0$ one has $\mathsf F(0,h)=0$, while
\eqref{bal:aux:eq:etacurvature} is strictly positive; the limiting mixed
derivative is zero.  Finally multiplication by the positive constant $k$
and the linear change $h=ke$ preserve all asserted signs.

At the upper entropy cap, the defining formulas give the continuous value
$\mathsf\Phi(s,\mathsf H((1-s)/2))=0$.  Convexity therefore extends to
this endpoint by taking limits in the convexity inequality.  For $s>0$,
the inverse functions in the formulas are smooth at this positive cap,
so the left entropy derivative is finite.  For $s=0$, writing $r=1-2q$
gives $\mathsf\eta=2r\operatorname{arctanh}r$ and
$\mathsf H(q)=\ln2-r^2/2+O(r^4)$, hence
$\mathsf\eta'(\ln2-)=-4$.  Passing to the limit in the interior
supporting-line inequality proves the asserted endpoint version.
\end{proof}

\begin{corollary}[Equal-mean owner]
\label{bal:aux:cor:equalmean}
For positive feasible entropy coordinates and equal means
$\mu_u=\mu_w=m$, the pure gap is nonnegative.
\end{corollary}
\begin{proof}
Put $h=(e_u+e_w)/2$ and $s=|1-2m|$.  Exact expansion gives
\[
 \Gap=g(e_u,e_w)+\tfrac12\Phi(s,e_u)+\tfrac12\Phi(s,e_w)-\Phi(s,h).
\]
Joint convexity and symmetry of $g$, and $g(h,h)=0$, imply
$g(e_u,e_w)\ge0$.  The remaining difference is nonnegative by
Theorem~\ref{bal:aux:thm:entropyconvexity} and Jensen's inequality.
Positive deterministic-cap endpoints follow by continuity from the active
entropy interval.  Zero entropy is assigned to the separate target-level
zero-entropy argument; no subtraction of infinite terms is used here.
\end{proof}

\subsection{The half-mean scalar inequality and its exact use}
\label{bal:aux:sec:halfmean}

Return to the bit-normalized $Q=\Theta^{-1}$ and put $q_0=Q'(0)=k/8$.
Define
\begin{align}
 D_0(s)&=Q'(s)Q(2s)-2q_0\{Q(2s)-Q(s)\},\label{bal:aux:eq:D0}\\
 \Delta(s,t)&=\{Q(2s+t)-Q(s)\}\{Q(s+t)-Q(t)\}\notag\\
 &\quad-\{Q(2s+t)-Q(s+t)\}\{Q(s)+Q(t)\}.
 \label{bal:aux:eq:Delta}
\end{align}
Differentiation and $Q(0)=0$ give $\partial_t\Delta(s,0)=D_0(s)$.

\begin{theorem}[Half-mean scalar certificate]
\label{bal:aux:thm:D0}
For every $s>0$, $D_0(s)>0$.
\end{theorem}
\begin{proof}
\emph{Proof outline.}
Near the origin, exact vanishing identities let us integrate a certified
mixed-derivative bound twice to obtain positivity. Direct and centered
interval enclosures cover the middle range. In the tail, convexity of
the inverse profile and a fixed derivative comparison supply the sign.
The three ranges overlap, and the origin calculation below specifies
the full argument interval required for its remainder bound.

For $0<s\le2/25$, put $K=\partial_s^2\partial_t\Delta$.
The repaired origin certificate establishes
\begin{equation}
 K(s,t)\ge\frac{47}{10^6}(s+t)^3
 \quad(s,t\ge0,\ s+t\le2/25).
 \label{bal:aux:eq:Korigin}
\end{equation}
In particular the inequality is strict away from $(0,0)$;
$K(0,0)=0$ is an exact identity.
Since $D_0(0)=D_0'(0)=0$, two integrations give
\begin{equation}
 D_0(s)=\int_0^s(s-u)K(u,0)\,du
 \ge\frac{47}{20\cdot10^6}s^5>0.
 \label{bal:aux:eq:D0origin}
\end{equation}

Here is the explicit origin certificate contract.  Reconstruct the odd
Taylor polynomial $T$ of $Q$ through degree $15$ by formal inversion of
the hyperbolic parametrization in Theorem~\ref{bal:imp:e8:thm:onevar}.
Every argument of $Q$ in \eqref{bal:aux:eq:Delta} is at most $4/25$ on the
claimed triangle.  Bound $|Q^{(17)}|$ on all of $[0,4/25]$ by using the
16 exact rational cells $[i/100,(i+1)/100]$, $i=0,\ldots,15$.
Monotone inverse bracketing maps each complete cell to the hyperbolic
parameter interval; repeated application of
$(d\Theta_*/d\alpha)^{-1}d/d\alpha$ encloses the seventeenth derivative.
The fresh 512-bit replay gives $|Q^{(17)}|<3.18\cdot10^7$ on the union,
so the more conservative rational bound $M=29\cdot10^{13}$ is valid.
Taylor's formula therefore gives, for $j=0,1,2,3$,
\[
 |Q^{(j)}(z)-T^{(j)}(z)|\le
 \frac{M z^{17-j}}{(17-j)!}\qquad(0\le z\le4/25).
\]
Oddness makes the coefficient of degree $16$ exactly zero.

Substitute $T$ into \eqref{bal:aux:eq:Delta} and differentiate to get a
bivariate polynomial $K_T$.  All its terms below degree $3$ vanish
algebraically: for $Q(z)=az+bz^3$ the exact identity is
\[
 \Delta(s,t)=12b^2s^2t(s+t)^2(2s+t),
\]
and higher odd powers cannot contribute below total degree $6$ in
$\Delta$.  This is why low-degree terms can be omitted; an interval
merely containing zero would not justify that omission.
Write $R_T=K_T-(6/10^5)(s+t)^3$, and let $\ell_{ij}$ be a certified
lower bound for the coefficient of $s^it^j$ in $R_T$. For $i+j\ge3$,
\(s^it^j\le(s+t)^3(2/25)^{i+j-3}\). Retaining only the negative
coefficient lower bounds and adding back the baseline gives
\[
 \frac{K_T}{(s+t)^3}
 \ge \frac6{10^5}
    +\sum_{i+j\ge3}\min(\ell_{ij},0)
       \left(\frac2{25}\right)^{i+j-3}
 >5.9999999999\cdot10^{-5}.
\]
The sum is finite, over the monomials of $R_T$.

For clarity, the error bound is also explicit.  Set $R_0=2/25$;
write $T(z)=\sum_{n\le15}a_nz^n$ and define
\[
 c_j=\sum_{n\ge\max(1,j)}|a_n|(n)_j2^{n-j}
                  R_0^{\,n-\max(1,j)},\quad j=0,1,
\]
where this formula means $c_0=\sum_{n\ge1}|a_n|2^nR_0^{n-1}$ and
$c_1=\sum_{n\ge1}|a_n|n2^{n-1}R_0^{n-1}$; for the other two constants use
\[
 c_2=\sum_{n\ge3}|a_n|n(n-1)2^{n-2}R_0^{n-3},\quad
 c_3=\sum_{n\ge3}|a_n|n(n-1)(n-2)2^{n-3}R_0^{n-3}.
\]
Thus $|T|\le c_0R$, $|T'|\le c_1$, $|T''|\le c_2R$,
and $|T'''|\le c_3$ when $0\le z\le2R$, $0\le R\le R_0$.
Put $\varepsilon_j=M2^{17-j}/(17-j)!$.  Expanding the product rule for
$\partial_s^2\partial_t\Delta$ gives the normalized error bound
\begin{align*}
 \frac{|K-K_T|}{(s+t)^3}\le{}&18\{c_0\varepsilon_3R_0^{12}
          +c_3\varepsilon_0R_0^{14}
          +\varepsilon_0\varepsilon_3R_0^{28}\}\\
 &+28\{c_1\varepsilon_2R_0^{12}
          +c_2\varepsilon_1R_0^{14}
          +\varepsilon_1\varepsilon_2R_0^{28}\}.
\end{align*}
The integers $18$ and $28$ are the sums of the absolute product-rule
coefficients of the $Q'''Q$ and $Q''Q'$ terms, respectively.
The 512-bit upper bound for this expression is
$1.2903195064\cdot10^{-5}$.  Subtraction yields a lower bound greater
than $4.7096804935\cdot10^{-5}$, proving \eqref{bal:aux:eq:Korigin}.
The code \texttt{aux\_check\_half\_mean\_origin.py} and its JSON output
record this repaired computation.  The historical $[0,0.12]$
seventeenth-derivative check is superseded: it did not cover the full
$[0,0.16]$ argument range needed here.

On $[2/25,63/20]$, the compact certificate contains 195 exact rational
discovery intervals at 192 bits.  Its separate 256-bit Arb replay uses
196 intervals, with one discovery interval split once.  On each interval,
it intersects the direct enclosure of \eqref{bal:aux:eq:D0} with valid centered
first- and second-order enclosures.  Every resulting lower bound is
strictly positive; exact endpoint sorting verifies that the union is the
whole compact interval and that there is no uncovered subinterval.
These are the compact proof objects in the accompanying
\texttt{d0\_axis\_certificate} directory; the origin repair does not
change their domains or calculations.

Finally $Q''>0$, so $Q'$ is increasing.  The tail point certificates at
192 and 256 bits give
\[
 Q'(63/20)-2q_0>0.0026669186327661>0.
\]
For $s\ge63/20$ it follows that
\[
 D_0(s)>2q_0Q(2s)-2q_0\{Q(2s)-Q(s)\}=2q_0Q(s)>0.
\]
Together these ranges cover $(0,\infty)$.
\end{proof}

\begin{lemma}[The smooth half-mean exclusion]
\label{bal:aux:lem:halfmean}
A mean-stationary point with positive entropies, both marginal entropy
constraints strict, one mean $1/2$, and the other mean distinct from $1/2$
cannot be a local minimum of the pure gap.  Deterministic-cap
intersections and coincident means are handled by their separate owners.
\end{lemma}
\begin{proof}
\emph{Proof outline.}
Stationarity in the non-half mean relates two inverse-profile
arguments. We compute the transverse second derivative at the half mean
and use the scalar half-mean inequality to make it negative. This
contradicts the second-order condition for a local minimum under the
strict feasibility assumptions.

Reflect or exchange coordinates so that the point is
$(\mu_u,\mu_w)=(a,1/2)$ with $a<1/2$, and let its positive entropy
coordinates be $e,f$.  Put $v=1/2-a$, $h=(e+f)/2$,
$x=v/(e+f)$, and $y=v/e$.
The terms of the pure gap depending on the two means are exactly
\[
 F(\mu_w-\mu_u,h)+F(1-\mu_u-\mu_w,h)
 -\tfrac12F(1-2\mu_u,e)-\tfrac12F(1-2\mu_w,f),
\]
where the last term uses the smooth even extension at zero.
All other terms are constant when entropies are fixed.
Stationarity in $a$ gives
\[
 -2\Theta(x)+\Theta(y)=0.
\]
Taking the second derivative in $\mu_w$ at $1/2$ gives
\[
 \Gap_{\mu_w\mu_w}
 =\frac{2\Theta'(x)}{e+f}-\frac{\Theta'(0)}{f}
 =\frac1f\left\{2\Theta'(x)\left(1-\frac{x}{y}\right)
                                      -\Theta'(0)\right\}.
\]
Now set $s=\Theta(x)>0$.  The stationary relation makes
$x=Q(s)$ and $y=Q(2s)$.  Dividing $D_0(s)>0$ by the positive
quantity $Q'(s)Q(2s)$ gives
\[
 2\Theta'(x)\left(1-\frac{x}{y}\right)<\Theta'(0).
\]
Hence $\Gap_{\mu_w\mu_w}<0$, which contradicts the necessary
second-order condition at a smooth local minimum.  In a canonical
half-domain this is an inward feasible variation from the half-mean face;
the zero first derivative follows from reflection symmetry.
If the half-mean entropy is maximal, such a variation need not be feasible,
which is precisely why the deterministic-cap intersection is assigned to
SC+ clause (v).  If the other mean is a marginal entropy endpoint,
stationarity in that mean is not required and the cap-fiber argument applies
instead.  These qualifications keep this exclusion separate from its
boundary owners.
\end{proof}

\section{The stationary Case-E cap: coordinates, estimates, and coverage}
\label{bal:sec:casee-stationary}

\emph{Argument guide.}
The proof proceeds from physical points to validated chart inequalities.
First derive an invertible stationary chart and express feasibility and
the exact gap in its coordinates. Then remove regions by proved
infeasibility or the retained-mean cutoff, and establish value bounds
on the remaining collars and finite rectangle. The concluding union
argument explains how each retained stationary point receives the
required pure-gap bound.

This section proves the stationary-cap estimate in
Theorem~\ref{bal:thm:boundary-estimates}(h) of the main text.
It derives the two-dimensional coordinates, identifies the exact gap
computed by the verifier, and proves how the analytic estimates and
finite certificates cover the retained stationary domain.
The seam and endpoint inequalities of Theorem~\ref{bal:hyp:SCplus}
are proved separately in Sections~\ref{bal:sec:seam-normalization}--\ref{bal:sec:cap-sharp}.
All entropies and gaps below are measured in bits.

\subsection{The physical Case-E stationary problem}
\label{bal:sec:ce-physical-chart}

The coordinates in the Case-E certificate are derived here explicitly.
Case E is the deterministic-cap family
\[
 M=(a,c,H_2(a),H_2(b)),\qquad 0<a<b<c<\tfrac12.
\]
Only the retained region $a+c\ge S$, $S=10^{-4}$, requires a pure-gap
certificate.  Write
\[
 E=H_2(a)+H_2(b),\qquad h=E/2,\qquad
 \mathcal L(t)=\frac{2H_2(t)}{1-2t},\qquad
 J(t)=\log_2\frac{1-t}{t}.
\]
Throughout this section $F$, $\eta$, $L_4$, and $R_\phi$ use the bit
normalization of the main definitions.  In particular,
\[
 F(s,e)=sJ\!\left(\mathcal L^{-1}(2e/s)\right),\qquad
 \partial_s F(s,e)=\Theta\!\left(\frac{s}{2e}\right)
 \quad(s,e>0).
\]
The equality $\phi(a,H_2(a))=0$ gives the cap gap
\begin{align}
 G_E(a,b,c)={}&j(a,b)-F(b-a,h)+F(c-a,h)
                  +F(1-a-c,h)-\eta(h)\notag\\
              &+\tfrac12\eta(H_2(b))
                  -\tfrac12F(1-2c,H_2(b)).
 \label{bal:eq:ce-physical-gap}
\end{align}
This expression is exactly $L_4(M)-R_\phi(M)$.  Differentiating at fixed
$a,b$ gives
\begin{equation}
 \partial_cG_E=\Theta\!\left(\frac{c-a}{E}\right)
 -\Theta\!\left(\frac{1-a-c}{E}\right)
 +\Theta\!\left(\frac{1-2c}{2H_2(b)}\right).
 \label{bal:eq:ce-stationary-derived}
\end{equation}
Thus the interior stationary condition used by the cap-fiber argument is a
single exact equation.  No derivative with respect to $a$ or $b$ is required.

\subsection{The two-dimensional chart and its exact inverse}
\label{bal:sec:ce-chart-bijection}

For a physical stationary point define
\begin{equation}
 A=\frac{c-a}{E},\qquad B=\frac{1-a-c}{E},\qquad
 C=\frac{1-2c}{2H_2(b)}.
 \label{bal:eq:ce-chart-forward}
\end{equation}
Each of $A,B,C$ is positive.  Define
\[
 t_A=\mathcal L^{-1}(1/A),\qquad
 t_B=\mathcal L^{-1}(1/B),\qquad
 t_C=\mathcal L^{-1}(1/C).
\]
Set $t=t_C$ and
\begin{equation}
 \lambda=\frac{B-A}{2C}.
 \label{bal:eq:ce-chart-lambda}
\end{equation}
The functions $\mathcal L:(0,1/2)\to(0,\infty)$ and
$\Theta:(0,\infty)\to(0,\infty)$ are strictly increasing bijections.
Consequently, the computational coordinates $(A,t)$ reconstruct the
stationary branch by
\begin{align}
 C&=\frac1{\mathcal L(t)}, &
 B&=\Theta^{-1}\!\left(\Theta(A)+\Theta(C)\right), &
 \lambda&=\frac{B-A}{2C},\label{bal:eq:ce-reconstruct-1}\\
 a&=\mathcal L^{-1}\!\left(\frac{2(1-\lambda)}{A+B}\right), &
 E&=\frac{1-2a}{A+B}, &
 b&=H_2^{-1}(\lambda E),\label{bal:eq:ce-reconstruct-2}\\
 c&=\frac{1-E(B-A)}2.
 &&&\label{bal:eq:ce-reconstruct-3}
\end{align}
The inverse entropy in this formula is the lower branch, with range
$[0,1/2]$.  The open chart retains only $1/2<\lambda<1$ and
$0<\lambda E<1$.  Boxes for which these conditions are not decided must
be subdivided or passed to a separately proved boundary certificate.

The implementation also uses $(t_A,t_C)$ instead of $(A,t_C)$.
These are equivalent coordinates under $A=1/\mathcal L(t_A)$.  In the
former coordinates it evaluates
\[
 \Upsilon(t)=\Theta(1/\mathcal L(t)),\qquad
 t_B=\Upsilon^{-1}\!\left(\Upsilon(t_A)+\Upsilon(t_C)\right).
\]
There is no additional stationary equation hidden in this operation.
For comparison with the implementation, its exact scalar formula is
\[
 \Upsilon(t)=J(t)+
 \frac{H_2(t)(1-2t)}{t(1-t)[-\ln(t(1-t))]}.
\]

\begin{lemma}[Surjectivity and uniqueness of the Case-E chart]
\label{bal:lem:ce-chart-surjectivity}
Every physical stationary Case-E point maps to one finite pair
$(A,t)\in(0,\infty)\times(0,1/2)$ by
\eqref{bal:eq:ce-chart-forward}, and formulas
\eqref{bal:eq:ce-reconstruct-1}--\eqref{bal:eq:ce-reconstruct-3} recover that point
exactly. Conversely, whenever the reconstructed chart obeys the entropy
window above and $b<t$, it is a physical stationary Case-E point.
\end{lemma}
\begin{proof}
\emph{Proof outline.}
For a physical stationary point, the defining identities recover
the chart variables and each inverse in succession. Conversely, the
reconstructed entropy moments and mean identities recover the physical
order and stationarity. This proves the correspondence algebraically,
before any certificate search is used.

For a physical stationary point, \eqref{bal:eq:ce-stationary-derived} says
$\Theta(B)=\Theta(A)+\Theta(C)$.  The strict monotonicity of $\Theta$
therefore recovers exactly the original $B$.  The definitions imply
\[
 A+B=\frac{1-2a}{E},\qquad
 B-A=\frac{1-2c}{E},\qquad
 \lambda=\frac{H_2(b)}{E}\in(\tfrac12,1).
\]
It follows that
\[
 \frac{2(1-\lambda)}{A+B}
 =\frac{2H_2(a)}{1-2a}=\mathcal L(a).
\]
Thus the inverse formulas recover $a$, then $E$, then $b$, and finally $c$.
Every inverse used in this direction has an admissible positive argument.
In particular, surjectivity follows from the identities, independently of
any numerical search for stationary roots.

Conversely, reconstruct from a pair in the stated window.  The first
inverse formula gives
$H_2(a)=(1-\lambda)E$, and the inverse entropy gives
$H_2(b)=\lambda E$.  Hence $E=H_2(a)+H_2(b)$ and, since
$\lambda>1/2$, $0<a<b<1/2$.  The remaining formulas yield
\begin{equation}
 c-a=AE,\qquad 1-a-c=BE,\qquad
 1-2c=2CH_2(b).
 \label{bal:eq:ce-chart-identities}
\end{equation}
The last identity gives $c<1/2$.  Moreover,
\[
 \mathcal L(t)=\frac{2H_2(b)}{1-2c}.
\]
As $\mathcal L$ is increasing, $b<t$ is equivalent to $b<c$.
The first line of the reconstruction and
\eqref{bal:eq:ce-chart-identities} now prove
$\partial_cG_E=0$ by \eqref{bal:eq:ce-stationary-derived}.
\end{proof}

\subsection{Entropy feasibility and the retained region}
\label{bal:sec:ce-feasibility-contract}

Define the entropy-order residual
\begin{equation}
 Q_E(A,t)=\lambda(A,t)E(A,t)-H_2(t).
 \label{bal:eq:ce-condition-e}
\end{equation}
The subscript distinguishes this residual from the inverse function
$Q=\Theta^{-1}$ used elsewhere in the manuscript.
For an entropy-admissible reconstructed point the following equivalences
are exact:
\begin{equation}
 b<c\quad\Longleftrightarrow\quad b<t
 \quad\Longleftrightarrow\quad Q_E(A,t)<0.
 \label{bal:eq:ce-condition-e-equivalence}
\end{equation}
The closed condition E used by the interval cover is $Q_E\le0$.
Its equality boundary is $b=c=t$, the double-cap boundary already assigned
an endpoint proof.  A whole-box certificate $Q_E>0$ therefore excludes
every strict physical Case-E point in that box.  Merely having an enclosure
that intersects $Q_E>0$ does not exclude the box.

The second test is the retained-domain inequality $a+c\ge S$.
The chart gives an exact useful expression:
\begin{equation}
 a+c=\frac{A+2aB}{A+B}.
 \label{bal:eq:ce-retained-identity}
\end{equation}
In the certificate, the label \code{delegated:F} means that the whole
box satisfies $a+c<S$.  Such a box is disjoint from the domain of the
retained stationary theorem.  It can therefore be removed when proving
that theorem.  The separate small-boundary theorem handles the original
Bellman target on this excluded region; it does not supply a value of
$G_E$ there.  Equality $a+c=S$ is retained, and cannot be discarded by a
strict-cutoff test.  This distinction prevents a target-level lower bound
from being substituted for the pure gap in a cap-minimum argument.

\subsection{Identification of the numerical gap with the theorem}
\label{bal:sec:ce-numerical-gap}

The identities \eqref{bal:eq:ce-chart-identities} imply
\[
 \mathcal L(t_A)=\frac{E}{c-a},\qquad
 \mathcal L(t_B)=\frac{E}{1-a-c},\qquad
 \mathcal L(t_C)=\frac{2H_2(b)}{1-2c}.
\]
Put $t_D=\mathcal L^{-1}(E/(b-a))$ and
$u_E=H_2^{-1}(E/2)$.  Substituting these identities into
\eqref{bal:eq:ce-physical-gap} gives
\begin{align}
 g_{\mathrm{code}}={}&-(b-a)J(t_D)
 +\tfrac12(a-b)(J(b)-J(a))
 -(1-2u_E)J(u_E)\notag\\
 &+\tfrac12(1-2b)J(b)
 +(c-a)J(t_A)+(1-a-c)J(t_B)
 -\tfrac12(1-2c)J(t_C)\notag\\
 ={}&G_E(a,b,c).
 \label{bal:eq:ce-code-gap}
\end{align}
This is the seven-term expression in \code{casee_mv.gap}, used by the
corrected A3 acceptance route.  There is no change of normalization in
\eqref{bal:eq:ce-code-gap}, and no maximum with another lower bound.

Consequently the accepted leaf types have distinct mathematical meanings:
\begin{itemize}
\item \code{proved:g} proves the required pure gap on the entire leaf.
\item An imported pure-gap collar proves it only after that collar's exact
domain and endpoint hypotheses have been checked.
\item \code{outside:E}, an entropy-window exclusion, or
\code{delegated:F} removes a leaf only through its proved disjointness from
the relevant physical retained domain.
\item An unresolved inverse, a singular interval expression, or a residual
leaf supplies no mathematical conclusion.
\end{itemize}
The pure gap need not be nonnegative on the excluded tiny-mean region.
The verifier therefore distinguishes exclusion from a nonnegative-gap
certificate; the latter is required at every retained physical point.

The chart lemma and gap identity reduce the stationary estimate in
Theorem~\ref{bal:thm:boundary-estimates}(h) to a cover of all chart pairs
representing physical points with $a+c\ge S$. Every retained point must
receive a pure-gap value certificate. The following arguments state what
each numerical inequality proves and the exact domain on which it applies.

\subsection{Analytic estimates and certificate acceptance conditions}
\label{bal:sec:casee-owner-contracts}

We now prove the bounds used in the Case-E partition.  Throughout
this subsection, $t=t_C$, and $(A,B,C,\lambda,a,b,c,E)$ have the meanings in
\eqref{bal:eq:ce-chart-identities}. Write $H=H_2$ and retain
$\mathcal L$ for the entropy-ratio function. In particular,
\[
 C=\frac1{\mathcal L(t)},\qquad B=A+2C\lambda,\qquad
 H(a)=(1-\lambda)E,\qquad H(b)=\lambda E,
 \qquad c-a=AE,
\]
\[
 1-a-c=BE,\qquad 1-2c=2CH(b),\qquad
 \Theta(B)=\Theta(A)+\Theta(C).
\]
The symbol $g(A,t)=G_E(a,b,c)$ denotes the bit-normalized pure gap
$L_4-R_\phi$ at this stationary chart point; there is no rescaling.  Define the entropy-feasibility defect
\[
 Q_E(A,t)=\lambda E-H(t).
\]
Strict Case E has $Q_E<0$; every bound below is valid on the larger set
$Q_E\le0$.  On that set $b\le t$ and
$c-t=C[H(t)-H(b)]\ge0$.  Hence $0<a<b\le t\le c<1/2$.
The acceptance tests below include these domain hypotheses.

\subsubsection{The capital bound and its infinite tail}
\label{bal:sec:casee-capital-proof}

\emph{Argument guide.}
Concavity of the profile makes the stationary residual monotone in
the variables used for exclusion, reducing the capital bound to a
scalar test at a fixed capital. A cancellation-safe pair of inequalities
extends that test to small entropy scales, and an analytic logarithmic
tail covers the scales below the finite dyadic bands.

Put $A_0=21/200$ and
\[
 \mathcal S(A,\lambda,C)
 =\Theta(A+2\lambda C)-\Theta(A)-\Theta(C).
\]
The previously established strict increase and strict concavity of
$\Theta$ give
\[
 \mathcal S_\lambda=2C\Theta'(A+2\lambda C)>0,
 \qquad
 \mathcal S_A=\Theta'(A+2\lambda C)-\Theta'(A)<0.
\]
Consequently, proving
\begin{equation}
 \mathcal S(A_0,1,1/\mathcal L(t))<0
 \label{bal:eq:casee-capital-test}
\end{equation}
excludes every stationary point with $A\ge A_0$ and $\lambda<1$.
For $1/20000\le t\le1/100$, this is a one-dimensional directed-interval
certificate.  Its input consists of exact rational subintervals, and its
acceptance condition is that the upper endpoint of
$[\mathcal S(A_0,1,1/\mathcal L(t))]$ is strictly negative.

Here is the alternative expression used at smaller $t$, which avoids
subtracting two divergent terms.  Set
\[
 \Upsilon(x)=J(x)+\mathcal R(x),\qquad
 \mathcal R(x)=\frac{H(x)(1-2x)}{x(1-x)[-\ln(x(1-x))]},
 \qquad t_{\mathrm{cmp}}=5t/11,
\]
and let $t_0=\mathcal L^{-1}(200/21)$.  The two interval inequalities
\begin{align}
 \frac{(1-2t_{\mathrm{cmp}})H(t)}{(1-2t)H(t_{\mathrm{cmp}})}&>2+A_0\mathcal L(t),
 \label{bal:eq:casee-capital-forward1}\\
 \log_2(11/5)+\log_2\frac{1-t_{\mathrm{cmp}}}{1-t}
 +\mathcal R(t_{\mathrm{cmp}})-\mathcal R(t)&<\Upsilon(t_0)
 \label{bal:eq:casee-capital-forward2}
\end{align}
are certified on the dyadic bands covering $[2^{-900},2^{-14}]$.
Indeed, if $\mathcal L(y)=\mathcal L(t)/(2+A_0\mathcal L(t))$, the first inequality gives $t_{\mathrm{cmp}}<y$;
$\Upsilon$ is decreasing, so the second gives
$\Upsilon(y)-\Upsilon(t)-\Upsilon(t_0)<0$, which is exactly
\eqref{bal:eq:casee-capital-test}.  The two finite ranges overlap because
$2^{-14}>1/20000$.

For completeness, the unbounded sequence of smaller entropy scales is
covered analytically.  Let $0<t\le\epsilon=2^{-900}$ and
$h(x)=(\ln2)H(x)$.  The elementary bounds
\[
 x\ln(1/x)+x(1-x)\le h(x)\le x\ln(1/x)+x
\]
imply, with $X=\ln(1/t)>450$,
\[
 \frac{\mathcal L(t_{\mathrm{cmp}})}{\mathcal L(t)}<\frac5{11}\left(1+\frac1{225}\right)
 =\frac{226}{495},\qquad
 \mathcal L(t)<\frac{7208}{2^{900}}<\frac1{100}.
\]
Thus
\[
 (2+A_0\mathcal L(t))\frac{\mathcal L(t_{\mathrm{cmp}})}{\mathcal L(t)}
 <\left(2+\frac{21}{20000}\right)\frac{226}{495}
 =\frac{4522373}{4950000}<1.
\]
Also
\[
 J(t_{\mathrm{cmp}})-J(t)<\log_2(11/5)+4t
 <\log_2(11/5)+\frac1{100}.
\]
We record the elementary derivative check ensuring that the remaining
$\mathcal R$ increment has the correct sign.  Write
$X=-\ln x$, $V=-\ln(1-x)$, and $z=V/x$.  After removal of the positive
factor $1/\ln2$, the numerator $N$ of $\mathcal R'(x)$ over denominator
$x^2(1-x)^2(X+V)^2$ satisfies
\begin{align*}
 N/x={}&z+X(1-z)-xX^2-2x^2zX+4x^2X+2xzX-4xX\\
       &-x^3z^2-4x^3z+2x^2z^2+8x^2z-xz^2-5xz.
\end{align*}
For $x\le2^{-900}$, we have $1\le z\le(1-x)^{-1}<2$,
$xX<1/100$, and $xX^2<1/100$.  Dropping positive terms and using
$X(1-z)>-2xX$ gives
\[
 N/x>1-\frac{11}{100}-26x
 >1-\frac{11}{100}-\frac{26}{1024}>0.
\]
Hence $\mathcal R(t_{\mathrm{cmp}})-\mathcal R(t)<0$.  The remaining scalar gate is
\begin{equation}
 \Upsilon(t_0)-\log_2(11/5)-\frac1{50}>0,
 \label{bal:eq:casee-capital-tail-gate}
\end{equation}
which is checked by directed Arb arithmetic at 384 and 512 bits.
It follows that \eqref{bal:eq:casee-capital-forward1}--\eqref{bal:eq:casee-capital-forward2}
hold on the entire tail.  Together, these arguments yield
\begin{equation}
 0<t\le1/100,\quad \mathcal S=0,\quad \lambda<1
 \quad\Longrightarrow\quad A<21/200.
 \label{bal:eq:casee-capital-conclusion}
\end{equation}

\subsubsection{The extreme-low-\texorpdfstring{$t$}{t} region belongs to the small-mean boundary}
\label{bal:sec:casee-extreme-low-proof}
Since $\lambda>1/2$, we have $B>A+C>C$.  The chart gives
\[
 a+c=\frac{A+2aB}{A+B}<\frac AB+2a
 <A\mathcal L(t)+2t.
\]
Here $a<b\le t$ was used; no assertion $a+c\le2t$ is needed.
By \eqref{bal:eq:casee-capital-conclusion} and monotonicity of $\mathcal L$, when
$0<t\le1/80000$,
\[
 a+c<\frac{21}{200}\mathcal L(1/80000)+\frac2{80000}
 <7.154345939583\times10^{-5}<10^{-4}.
\]
The last strict scalar comparison has 384- and 512-bit directed checks.
Thus this region has no point in the retained domain $a+c\ge10^{-4}$.
For the original Bellman target, it is covered by the small-mean boundary
theorem.  This is a domain exclusion from the retained \emph{pure-gap}
problem, not a claim that this argument proves $g\ge0$ on the discarded
region.

\subsubsection{The lower-\texorpdfstring{$A$}{A} collar: integrate a certified second derivative}
\label{bal:sec:casee-curvature-owner}
The diagonal extension of the stationary chart satisfies the exact identities
\[
 g(0,t)=0,\qquad g_A(0,t)=0.
\]
Substitution in the chart gives $a=b=c=t$ at $A=0$; differentiation of
the regularized gap gives the second identity.  The curvature certificate
has the explicit contract
\begin{equation}
 g_{AA}(A,t)>1/4000\quad
 (0\le A\le1/20,\ 1/80000\le t\le1/100).
 \label{bal:eq:casee-curvature-contract}
\end{equation}
It partitions $A$ into the 200 intervals
$[j/4000,(j+1)/4000]$, $0\le j<200$, and partitions $t$ into exact
rational intervals with endpoints $1/80000$ and $1/100$.
Centered high-order derivative enclosures certify
\eqref{bal:eq:casee-curvature-contract} on every rectangle.  The acceptance
test is the lower endpoint of the enclosed second derivative exceeding
$1/4000$, with every failure retained for subdivision.
For fixed $t$, Taylor's formula with integral remainder therefore gives
\[
 g(A,t)=\int_0^A(A-s)g_{AA}(s,t)\,ds
 >\frac{A^2}{8000}\qquad(A>0).
\]
The interval certificate checks a derivative on a closed rectangle; the
exact diagonal identities are the analytic bridge to a value inequality.

\subsubsection{The upper-\texorpdfstring{$\lambda$}{lambda} collar: a direct positive value bound}
\label{bal:sec:casee-upper-owner}
The acceptance label \texttt{proved:upper-collar} is justified by the
following positive value bound. Assume
\[
 1/80000\le t\le1/100,\qquad
 999/1000\le\lambda<1,\qquad 0<A\le21/200,
 \qquad 0<a<b\le t\le c<1/2.
\]
Set $q=b-a$, $\bar h=E/2=H(u)$, and $\chi=a+b-2u\ge0$.
The exact gap identity is
\begin{align*}
 g={}&\frac q2[J(a)-J(b)]+F(c-a,\bar h)-F(q,\bar h)\\
 &+\frac12\{F(1-2b,H(b))-F(1-2c,H(b))\}\\
 &+F(1-a-c,\bar h)-F(1-2u,\bar h).
\end{align*}
For fixed positive entropy, $F_x=\Upsilon(\mathcal L^{-1}(2\bar h/x))>0$ and
$F_{xx}>0$.  Increasingness controls the second term; tangent-line
inequalities at $1-2c$ and $1-2u$ control the last two differences.  Hence
\begin{equation}
 g\ge\frac q2[J(a)-J(b)]
 -(c-b)[\Upsilon(u)-\Upsilon(t)]-\chi\Upsilon(u).
 \label{bal:eq:casee-upper-value-bound}
\end{equation}
The following two rows state all constants used.  Each entry is a strict
upper or lower bound in the indicated direction.
\begin{center}
\small
\begin{tabular}{c|cccccc}
$t$ interval & $u/t>$ & $(c-b)/t<$ & $\chi/t<$ &
 $\Upsilon(u)-\Upsilon(t)<$ & $\Upsilon(u)<$ & $g/t>$\\\hline
$[1/80000,1/625]$ & $9/20$ & $9/10$ & $51/500$ & $6/5$ & $20$ & $187/100$\\
$[1/625,1/100]$ & $2/5$ & $7/50$ & $101/500$ & $3/2$ & $13$ & $1077/500$
\end{tabular}
\end{center}
Both rows additionally use
$a<t/1500$, $b>999t/1000$, $q>499t/500$, and $J(a)-J(b)>10$.
For clarity, the scalar inputs behind a row with constants
$(\beta,\delta,\chi_0,\rho,M)$ are precisely
\begin{align*}
 \mathcal L(t/1500)&>\mathcal L(t)/999,\\
 \frac{999}{1000}(1-2t/1500)H(t)
 &>H(999t/1000)\left[\frac{999}{1000}(1-2t)+\frac{21}{100}H(t)\right],\\
 H(999t/1000)/2&>H(\beta t),\\
 \Upsilon(\beta t)-\Upsilon(t)&<\rho,
\end{align*}
throughout the row's interval, and the two endpoint comparisons at its
left endpoint $t_*$,
\[
 \frac{499}{500}+\delta
 >\frac{21}{200}\frac{H(t_*/1500)+H(t_*)}{t_*},\qquad
 \Upsilon(\beta t_*)<M.
\]
To verify the implications, use
\[
 \mathcal L(a)=\frac{1-\lambda}{A+C\lambda}<\frac{\mathcal L(t)}{999},\qquad
 H(b)=\frac{\lambda(1-2a)H(t)}{\lambda(1-2t)+2AH(t)}.
\]
The right side of the second identity increases with $\lambda$ and
decreases with $a,A$.  The entropy inequalities thus bound $a,b,u$.
The function $H(x)/x$ decreases, so the endpoint comparison bounds
$c-b$ on the whole interval.  Decreasingness of $\Upsilon$ supplies the
other endpoint implication.  Finally $b/a>1497>2^{10}$ gives the logarithm
bound, and $\chi<(1+1/1500-2\beta)t$ gives the listed $\chi_0$.
Substitution in \eqref{bal:eq:casee-upper-value-bound} is exact rational
arithmetic:
\[
 \frac{499}{500}\frac{10}{2}-\delta\rho-\chi_0M
 =\begin{cases}187/100, & t\le1/625,\\1077/500, &t\ge1/625.\end{cases}
\]
In particular $g>1/100000$ throughout the collar.  The two closed
$t$ ranges meet at exactly $1/625$.  Their scalar ledgers have respectively
$12{,}404$ and $3{,}678$ exact leaves, with recorded 384-bit generation and
512-bit replay.  No assertion that $g$ is monotone in $t$ or $\lambda$ is
used.

\subsubsection{The high-\texorpdfstring{$t$}{t} domain is entropy-infeasible}
\label{bal:sec:casee-high-t-owner}

\emph{Argument guide.}
We prove infeasibility by continuing along the stationary branch.
The defect starts positive by its diagonal derivative. A direct collar
handles large values of the relative entropy weight $\lambda$, while a
contact argument excludes a zero in the remaining weight range. Since
a change of sign would require such a zero, entropy feasibility is
excluded throughout the stated range of the chart coordinate $t$.

We next prove the logical implication supplied by the high-$t$ certificates:
\begin{equation}
 1/100\le t<1/2,\qquad 1/2<\lambda<1,\qquad\mathcal S=0
 \quad\Longrightarrow\quad Q_E(A,t)>0.
 \label{bal:eq:casee-high-t-contract}
\end{equation}
This excludes the Case-E entropy condition. The proof uses the derivative
on the diagonal, a contact argument, and an upper-$\lambda$ collar; it
does not require global monotonicity of $Q_E$ in $A$.

First, the exact diagonal identities are $Q_E(0,t)=0$ and
\[
 \partial_A Q_E(0,t)=\frac{4\mathcal P(d)}{(\ln2)^2dD},\quad
 d=1-2t,\quad s=(1-d^2)/4,\quad m=-\ln s,\quad D=m-d^2>0,
\]
where, with natural entropy $h=(\ln2)H(t)$,
\[
 \mathcal P(d)=8(\ln2)m^2s^2-h^2D.
\]
The edge certificates prove $\mathcal P>0$ on $1/50\le d\le51/52$;
on $0\le d\le1/50$ they prove $\mathcal P''>0$ and use
$\mathcal P(0)=\mathcal P'(0)=0$.  Thus $\partial_A Q_E(0,t)>0$ for
$1/104\le t<1/2$, covering the claimed range with overlap.

Second, a direct collar excludes $\lambda\ge19/20$.  On each of the
following exact bands the one-dimensional ledgers establish
\[
 \mathcal S(A_-,19/20,C)>0,\quad
 \mathcal S(A_+,1,C)<0,\quad
 \frac{19}{20}(t-\bar a(t))-A_+H(t)>0,
\]
where
$\bar a(t)=\mathcal L^{-1}((1/20)/(A_-+(19/20)C))$:
\begin{center}\small
\begin{tabular}{c|rrrrrrr}
$t$ band & $[.01,.02]$ & $[.02,.05]$ & $[.05,.1]$ & $[.1,.2]$ & $[.2,.3]$ & $[.3,.4]$ & $[.4,.5]$\\\hline
$A_-$ & $.09$ & $.09$ & $.10$ & $.11$ & $.14$ & $.18$ & $.24$\\
$A_+$ & $.11$ & $.12$ & $.13$ & $.16$ & $.21$ & $.27$ & $.35$
\end{tabular}
\end{center}
All finite decimals in this table mean the corresponding exact rationals;
all displayed strict residual signs apply for $t<1/2$. At $t=1/2$ one has
$C=0$ and $\mathcal S(A,\lambda,0)=0$. On the terminal interval the
checker instead encloses
\[
 \mathcal S_C=2\lambda\Theta'(A+2\lambda C)-\Theta'(C)
\]
with the required strict sign, then integrates from $C=0$. This is the
\code{C-derivative} route in \code{casee_high_t_tail.py}; no strict
residual sign at the zero endpoint is claimed.  Monotonicity of $\mathcal S$ gives
$A_-<A<A_+$ and $a<\bar a$.  The exact identity
\[
 (A+C\lambda)Q_E=\lambda(t-a)-AH(t)
\]
then proves $Q_E>0$.

Third, no zero of $Q_E$ can occur in the remaining $\lambda$ range.
At such a zero, $b=c=t$.  Put
\[
 H_* =H(a)+H(t),\quad X=\frac{t-a}{H_*},\quad
 Y=\frac{1-a-t}{H_*},\quad C=\frac{1-2t}{2H(t)},
\]
and define the contact defect
$\mathcal C(a,t)=\Theta(Y)-\Theta(X)-\Theta(C)$.
A stationary zero would require $\mathcal C(a,t)=0$ with $a<t$.
For $1/100\le t\le2/5$ and $\lambda\le19/20$,
$H(a)\ge H(t)/19$.  The scalar comparison
$H(1/4000)<H(1/100)/19$ therefore gives $a>1/4000$.
The regular contact certificate proves
\[
 \mathcal C_a(a,t)\ge\tfrac15(1-2t)(1-2a)>0
 \quad(1/4000\le a\le t,\ 1/100\le t\le2/5).
\]
Since $\mathcal C(t,t)=0$, integration from $a$ to $t$ gives
$\mathcal C(a,t)<0$, a contradiction.

For $2/5\le t<1/2$, the top contact certificate allows the wider range
$\lambda\le999/1000$.  At a contact,
$H(a)\ge H(t)/999>H(1/16000)$, so $a>1/16000$.
Set $d=1-2t$, $r=1-2a$, and $h_*(r)=H((1-r)/2)$.
Then $0<d<r<7999/8000$, $d\le1/5$.
On $r\le1/10$, the exact removable normalization is
\[
 \mathcal C(a,t)=r^3\Psi(d/r,r^2),\qquad
 \Psi(0,z)=\Psi(1,z)=0.
\]
On the closed rectangle $0\le q\le1$, $0\le z\le1/100$, the
certificate proves $\Psi_q<0$ for $q\le1/16$, $\Psi<0$ for
$1/16\le q\le15/16$, and $\Psi_q>0$ for $q\ge15/16$.
The endpoint identities imply $\Psi<0$ throughout $0<q<1$.
On $r\ge1/10$, put $N=(h_*(r)+h_*(d))\mathcal C_a$.
The certificate proves
\[
 \partial_d N>4r(h_*(r)+h_*(d)),\qquad N(0,r)=0.
\]
Because $h_*$ decreases, integration in $d$ gives
$\mathcal C_a(a,t)>4(1-2t)(1-2a)$ wherever $r\ge1/10$.
To join the derivative region to the normalized corner, put $a_*=\min\{t,9/20\}$.
If $a<a_*$, then
\[
 \mathcal C(a,t)
 <\mathcal C(a_*,t)-4(1-2t)\int_a^{a_*}(1-2s)\,ds<0.
\]
Indeed, $\mathcal C(a_*,t)=0$ when $a_*=t$, and it is strictly negative
by the normalized-corner result when $a_*=9/20<t$.
If $a\ge a_*$ and $a<t$, the point is already in that corner.
Thus the two regions prove the required strict negativity, and rule out
the remaining contact.

Finally, $B_A=\Theta'(A)/\Theta'(B)>1$ along stationarity, so
$\lambda_A=(B_A-1)/(2C)>0$.  The unique stationary branch starts at
$A=0$, $\lambda=1/2$ and varies continuously while $\lambda<1$;
the bandwise upper-capital bounds above provide a finite bracket.
The positive edge derivative makes $Q_E>0$ initially.  A transition to
$Q_E\le0$ would have a first zero, excluded by the contact arguments or
the direct collar.  This proves \eqref{bal:eq:casee-high-t-contract}.
The endpoint $t=1/2$ belongs to the separate half-mean boundary theorem,
not to this open chart.

\subsubsection{The finite A3 rectangle and the meaning of its four labels}
\label{bal:sec:casee-a3-contracts}
The preceding arguments reduce the remaining stationary problem to
\[
 \mathcal K=[1/20,21/200]\times[1/80000,1/100]
 \quad\text{in coordinates }(A,t).
\]
On every exact rational terminal rectangle $R$, the finite certificate
requires one of the following statements:
\begin{enumerate}
\item \texttt{proved:g}: an outward enclosure has $\inf[g(R)]\ge0$.
\item \texttt{outside:E}: an outward enclosure has $\inf[Q_E(R)]>0$,
so no entropy-feasible Case-E point belongs to $R$.
\item \texttt{delegated:F}: an outward enclosure has
$\sup[a(R)+c(R)]<10^{-4}$.  Such a rectangle misses the retained domain.
For the original target it is covered by the small-mean boundary theorem
in both pre-reflection orientations, as explained below.
\item \texttt{proved:upper-collar}: a rigorous enclosure gives
$\inf[\lambda(R)]\ge999/1000$, and the exact $A,t$ endpoints are in
Section~\ref{bal:sec:casee-upper-owner}'s range.  Physical points with
$\lambda<1$ satisfy its order hypotheses after imposing $Q_E\le0$;
points with $\lambda\ge1$ lie outside the open chart.  Both subsets
therefore have a valid value or exclusion test without evaluating a singular inverse at
$\lambda=1$.
\end{enumerate}
An arithmetic exception, an unresolved comparison, and a subdivision limit
are not acceptance conditions.  They must leave a residual rectangle.

The recorded corrected 384-bit replay reports coverage of every committed row of both
A3 data sets.  Its accounting is
\begin{center}\small
\begin{tabular}{lrrrrr}
 & \texttt{proved:g} & \texttt{outside:E} & \texttt{delegated:F} & collar & total\\\hline
upper & $5{,}191{,}340$ & $13{,}290$ & $159{,}166$ & $3{,}362$ & $5{,}367{,}158$\\
low extension & $0$ & $0$ & $119{,}855$ & $931$ & $120{,}786$
\end{tabular}
\end{center}
There are 672 upper rows and 224 low-extension rows.  Their combined
$5{,}487{,}944$ leaves are the proof objects, rather than a count of sampled
points.  Exact rational geometry must reconstruct each subdivision cover,
match every root and terminal boundary, and leave an empty unresolved
frontier.  Arithmetic replay must then recompute every acceptance inequality.
The corrected replay replaces the historical mean-value hull helper by
Arb's outward union, \texttt{return a.union(b)}; the original helper and its
historical acceptance reports are not relied on.

The independent collar-transfer map handles all $4{,}293$ collar leaves.
It assigns $2{,}281$ to the low-$t$ theorem, $786$ to the high-$t$ theorem,
one to an exact split at $t=1/625$, and $1{,}225$ entirely to
$\lambda\ge1$.  Thus each collar leaf is assigned either a pure-gap bound or an
exclusion from the open chart.

For \texttt{delegated:F}, the orientation argument is necessary.  The two
canonical original mean pairs represented by lower-half distances $(a,c)$
are $(a,c)$ and $(a,1-c)$.  If $a+c<10^{-4}$, the first is covered by the
same-side small-mean theorem ($a,c<10^{-2}$), and the second by the
opposite-side theorem ($a+1-(1-c)<10^{-4}$).  Simultaneous complement and
label exchange cover the other two orientations.  The lower bounds used
in those theorems apply to unrestricted moment laws, hence to $\zeta$.
No invariance of $\zeta$ under complementing only one marginal is asserted.
The resulting statement is $\zeta\ge R_\phi$, not necessarily
$L_4\ge R_\phi$ outside the retained region.

\subsubsection{Why the regions join without an omitted seam}
\label{bal:sec:casee-owner-union-proof}
Take any strict stationary Case-E point in the retained region
$a+c\ge10^{-4}$.  If $t\ge1/100$, Section~\ref{bal:sec:casee-high-t-owner}
excludes it.  Otherwise Section~\ref{bal:sec:casee-capital-proof} gives
$A<21/200$, and Section~\ref{bal:sec:casee-extreme-low-proof} excludes
$t\le1/80000$.  For the remaining $t$ range,
$A\le1/20$ is covered by Section~\ref{bal:sec:casee-curvature-owner};
$A\ge1/20$ lies in the finite A3 rectangle.  Every A3 label either gives
$g\ge0$ or excludes that point from the retained feasible region.
The equalities $t=1/80000$, $t=1/100$, and $A=1/20$ belong to closed
certificate ranges, with harmless overlap.  $A=0$, $\lambda=1/2$,
$\lambda=1$, $t=0$, and $t=1/2$ are boundary faces handled by the
diagonal, entropy, and half-mean results; they are not discarded limits
of a truncated numerical search.  Thus every retained strict stationary Case-E point satisfies
$G_E\ge0$, proving Theorem~\ref{bal:thm:boundary-estimates}(h).
The boundary $a+c=10^{-4}$ in the global cap reduction is supplied by
Theorem~\ref{bal:thm:seam-closure}. The scope of the recorded arithmetic verification
is specified in Section~\ref{bal:sec:verification}.

\subsection{Certificate files and reproduction}
\label{bal:sec:casee-recovery-scope}
The capital and curvature inputs are the following files under
\code{tmp/casee_complete_package_extract/casee_full/} in the
computational supplement:
\begin{itemize}
\item \code{casee_capital_bound_384.json} and
\code{casee_capital_bound_512.json}: the regular capital exclusion;
\item \code{casee_capital_bound_low_384.json} and
\code{casee_capital_bound_low_512.json}: the low-$t$ forward inequalities;
\item \code{second_derivative_collar_A20_ext_384.json}: the extended
curvature collar beginning at $t=1/80000$.
\end{itemize}
Their corresponding source modules are \code{casee_capital_bound.py},
\code{casee_capital_bound_low.py}, and
\code{casee_second_derivative_collar.py}, under
\code{ck_upstream_recovery_2026-09-07/source/ck_certificate/}.
The analytic infinite tail and extreme-low-$t$ scalar bridge are bound to
\code{casee_capital_tail_completion_audit.json} and
\code{casee_extreme_low_t_bridge_512.json} in the final audit's
\code{casee/} directory.

The upper-collar transfer uses
\code{collar/combined_collar_union_union_repair.json}, together with the
low-$t$ direct theorem and the high-$t$ continuation in
\code{collar/COLLAR_LOGICAL_AUDIT.md}. The high-$t$ domain exclusion is
assembled in
\code{ck_upstream_recovery_2026-09-07/certificates/casee_high_t_master_recovered.json}.
The corrected A3 row reports are identified in
Section~\ref{bal:sec:verification}. These locators identify different proof
obligations; none can be replaced by the aggregate \code{PASS} field.

The low-capital reports preserve 886 dyadic band summaries and cell counts;
the replay regenerates the forward tests from the source. The curvature
file preserves 5,000 exact $t$ leaves paired with 200 fixed $A$ intervals,
for one million rectangles. Its replay recomputes the interval enclosures
from these exact inputs and the committed source.

File identities and source commitments are listed in
\code{CASE_E_OWNER_INVENTORY.json} and
\code{CASE_E_RECONSTRUCTION_PROVENANCE.md}. Verification requires both
checking those commitments and recomputing the associated inequalities;
file identity alone does not certify an interval comparison.

\clearpage
\section{The radial stationary cap: reduction and certificate implication}
\label{bal:sec:radial-stationary}
This section gives the reduction and the terminal tests for
Theorem~\ref{bal:thm:boundary-estimates}(i).  The ordering is
$0<b<c<a<1/2$, and the retained condition is $a+c\ge S$, where
$S=10^{-4}$.  This is different from the ordering in
Section~\ref{bal:sec:casee-stationary}.  All functions in this section use
bits, including the gap evaluated by the radial program.

\subsection{The physical stationary equation and its two-dimensional chart}
\label{bal:sec:radial-chart}
Put $E=H_2(a)+H_2(b)$ and $\bar h=E/2$.  The cap gap from the main text is
\begin{align}
 G_R(a,b,c)={}&j(a,b)-F(a-b,\bar h)+F(a-c,\bar h)
       +F(1-a-c,\bar h)-\eta(\bar h)\notag\\
 &+\tfrac12\eta(H_2(b))-\tfrac12F(1-2c,H_2(b)).
 \label{bal:eq:radial-physical-gap}
\end{align}
It is exactly $L_4(M)-R_\phi(M)$ at
$M=(a,c,H_2(a),H_2(b))$.  Differentiating with $a,b$ fixed gives
\begin{equation}
 \partial_cG_R=-\Theta(A)-\Theta(B)+\Theta(C),\qquad
 A=\frac{a-c}{E},\quad B=\frac{1-a-c}{E},\quad
 C=\frac{1-2c}{2H_2(b)}.
 \label{bal:eq:radial-stationarity}
\end{equation}
The strict radial ordering implies $0<A<B$.  At a stationary point,
$\Theta(C)=\Theta(A)+\Theta(B)$; hence $C>B$ and
$D:=2C-A-B>0$.

For $0\le x<1$, define
\begin{align*}
 H_*(x)&=H_2((1-x)/2),&
 J_*(x)&=\frac{2\operatorname{atanh}x}{\ln2},\\
 P(x)&=\frac{x}{2H_*(x)},&
 V(x)&=\Theta(P(x))
 =J_*(x)+\frac{4xH_*(x)}{(1-x^2)[\ln4-\ln(1-x^2)]}.
\end{align*}
Both $P$ and $V$ are increasing bijections onto $[0,\infty)$; their
inverses below select the nonnegative branch.  Set
\[
 d_B=P^{-1}(B)=d,\qquad d_A=P^{-1}(A)=rd.
\]
Then $0<r<1$ and $0<d<1$.  Conversely, the chart $(r,d)$ reconstructs
all possible stationary triples as follows:
\begin{align}
 d_A&=rd,& d_B&=d,& d_C&=V^{-1}(V(rd)+V(d)),\notag\\
 A&=P(d_A),& B&=P(d),& C&=P(d_C),\notag\\
 D&=2C-A-B,& d_a&=P^{-1}\!\left(\frac{C(B-A)}D\right),
       &e_u&=H_*(d_a),\notag\\
 e_w&=\frac{(A+B)e_u}{D},& d_b&=H_*^{-1}(e_w),
       &d_c&=2Ce_w,\label{bal:eq:radial-reconstruction}\\
 a&=\frac{1-d_a}2,& b&=\frac{1-d_b}2,& c&=\frac{1-d_c}2.\notag
\end{align}
Here $H_*^{-1}$ is the decreasing inverse on $[0,1]$.  A chart point is
physical only when its inverses are defined and the reconstructed means
satisfy $0<b<c<a<1/2$ and $a+c\ge S$.  An undecided inverse or domain
test is never an acceptance.

\begin{lemma}[Exact coverage by the radial chart]
\label{bal:lem:radial-surjectivity}
Every retained strict radial stationary triple has a unique chart point
$(r,d)\in(0,1)^2$, and reconstruction
\eqref{bal:eq:radial-reconstruction} returns the original triple.
\end{lemma}
\begin{proof}
Equation~\eqref{bal:eq:radial-stationarity} determines $C$ from $A,B$.
The physical definitions also give
\[
 E(B-A)=1-2a,\qquad E(A+B)=1-2c=2Ce_w.
\]
Since $E=e_u+e_w$, it follows that $e_u=ED/(2C)$ and
$e_w=(A+B)e_u/D$.  Consequently
$P(1-2a)=C(B-A)/D$.  The strict monotonicity of $P,V$, and the
one-to-one entropy branch determine all the inverse values in
\eqref{bal:eq:radial-reconstruction}.  The same identities show that any
reconstructed physical triple satisfies the stated chart relations and
stationarity.  No stationarity with respect to $a$ or $b$ is assumed.
\end{proof}

The upper endpoint of the computational $d$ interval is derived from the
retained constraint.  Since $c<a$ and $a+c\ge S$, we have $a\ge1/20000$.
Writing $t_B=(1-d)/2$ and
$\mathcal L(t)=2H_2(t)/(1-2t)$ gives
\[
 \mathcal L(t_B)=\frac{E}{1-a-c}\ge H_2(a)\ge H_2(1/20000).
\]
The fixed outward-rounded comparison
\begin{equation}
 \mathcal L(23/10^6)<H_2(1/20000)
 \label{bal:eq:radial-parameter-floor}
\end{equation}
is checked by \code{default_parameter_floor_verified} before the initial
manifest is generated.  As $\mathcal L$ is increasing, every retained
stationary point has
\begin{equation}
 0<d\le d_*:=499977/500000.
 \label{bal:eq:radial-d-domain}
\end{equation}
This endpoint is a proved truncation, not a numerical search cutoff.

\subsection{The computed function and its equality anchor}
\label{bal:sec:radial-gap}
Let
\[
 d_E=H_*^{-1}((e_u+e_w)/2),\qquad
 d_\tau=P^{-1}\!\left(\frac{d_b-d_a}{2(e_u+e_w)}\right).
\]
The source function \code{edge_third.stationary_gap} is
\begin{align}
 g(r,d)={}&-\frac{d_b-d_a}{2}J_*(d_\tau)
 +\frac{d_b-d_a}{4}\bigl(J_*(d_b)-J_*(d_a)\bigr)
 -d_EJ_*(d_E)\notag\\
 &+\frac{d_b}2J_*(d_b)
 +\frac{d_c-d_a}2J_*(d_A)
 +\frac{d_a+d_c}2J_*(d_B)
 -\frac{d_c}2J_*(d_C).
 \label{bal:eq:radial-code-gap}
\end{align}
Using $F(s,e)=sJ(\mathcal L^{-1}(2e/s))$ and
$\eta(e)=(1-2H_2^{-1}(e))J(H_2^{-1}(e))$, each of the seven terms in
\eqref{bal:eq:radial-code-gap} matches the corresponding term in
\eqref{bal:eq:radial-physical-gap}.  Thus
\begin{equation}
 g(r,d)=G_R(a,b,c)
 \label{bal:eq:radial-gap-identification}
\end{equation}
exactly, with no change of sign, scale, or entropy convention.

At $r=0$ the reconstruction extends to
\[
 d_A=d_\tau=0,\qquad d_C=d_a=d_b=d_c=d_E=d.
\]
Substitution gives $g(0,d)=0$.  To verify the first derivative rather than
infer it from point evaluation, put $\alpha=(d_C)_r(0,d)$. Differentiating
the reconstruction gives
\[
 (d_a)_r=-\alpha,\qquad
 (d_E)_r=\tfrac12((d_b)_r-\alpha).
\]
In the derivative of \eqref{bal:eq:radial-code-gap}, all products with
two vanishing factors disappear; the remaining terms equal
\[
 \left[-(d_E)_r+\tfrac12(d_b)_r-\tfrac12\alpha\right]
       \bigl(J_*(d)+dJ_*'(d)\bigr)=0.
\]
Therefore the exact anchors are
\begin{equation}
 g(0,d)=g_r(0,d)=0.
 \label{bal:eq:radial-anchor}
\end{equation}
If a certificate proves $g_{rr}\ge K(d)\ge0$ on a whole rectangle
$[0,R]\times I$, then for every point of that rectangle
\begin{equation}
 g(r,d)=\int_0^r(r-s)g_{rr}(s,d)\,ds
       \ge \tfrac12r^2K(d).
 \label{bal:eq:radial-integrated-curvature}
\end{equation}
This implication requires the full segment from the anchor. A curvature
bound on a rectangle with positive left $r$ endpoint alone is insufficient.

\subsection{The two certified collars}
\label{bal:sec:radial-collars}
The radial generator imports two finite interval certificates.  Their
statements, exact covers, and implications are as follows.

\paragraph{Normalized small-$d$ collar.}
Put $z=d^2$.  The normalization of
\eqref{bal:eq:radial-reconstruction} and
\eqref{bal:eq:radial-code-gap} gives an analytic function $\Phi_R$ with
\[
 g(r,d)=d^2\Phi_R(r,d^2),\qquad \Phi_R(r,0)=0.
\]
For clarity, its scalar primitives are
\begin{align*}
 \mathsf A(q)&=\sum_{n\ge0}\frac{q^n}{(2n+1)(2n+2)\ln2},&
 \mathsf J(q)&=\frac2{\ln2}\sum_{n\ge0}\frac{q^n}{2n+1},\\
 \mathsf P(q)&=\frac1{2(1-q\mathsf A(q))},&
 \mathsf V(q)&=\mathsf J(q)+
 \frac{4(1-q\mathsf A(q))}{(1-q)(2\ln2-\ln(1-q))}.
\end{align*}
In particular,
\[
 H_*(dx)=1-zx^2\mathsf A(zx^2),\quad
 J_*(dx)=dx\mathsf J(zx^2),\quad
 P(dx)=dx\mathsf P(zx^2),\quad V(dx)=dx\mathsf V(zx^2).
\]
Substitution into the reconstruction removes the apparent singularity at
$d=0$. The source \code{collar_certificate.py} uses these identities,
including the divided functions
$(\mathsf P(q)-1/2)/q$ and
$(\mathsf V(q)-4/\ln2)/q$ with their removable values at zero.
It checks positive inverse derivatives and sign-changing brackets for all
implicit variables.  Its differentiated power series include rigorous tail
bounds; no remainder-free truncation is used.

The exact acceptance quantity is $\partial_r^2\partial_z\Phi_R$, not
$g$ sampled near zero.  On every cell
\[
 [i/1024,(i+1)/1024]\times[j/12800,(j+1)/12800],
 \quad0\le i<1024,\quad0\le j<32,
\]
the checker encloses $\Phi_{R,rrz}$ at the midpoint and subtracts
\[
 \rho_r\sup|\Phi_{R,rrrz}|+\rho_z\sup|\Phi_{R,rrzz}|
\]
using bounds on the whole cell. Acceptance requires the resulting lower
bound to exceed $1/10$.  These cells tile $[0,1]\times[0,1/400]$.
Since $\Phi_{R,rr}(r,0)=0$, the fundamental theorem of calculus gives
\[
 \frac{g_{rr}(r,d)}{d^4}
 =\int_0^1\Phi_{R,rrz}(r,sd^2)\,ds\ge1/10.
\]
Together with \eqref{bal:eq:radial-anchor}, this proves
\begin{equation}
 g(r,d)\ge r^2d^4/20
 \quad(0\le r\le1, 0\le d\le1/20).
 \label{bal:eq:radial-small-collar}
\end{equation}
The archived normalized collar contains the complete 32,768-cell cover and
its recorded 384-bit replay; this mathematical acceptance criterion is what
allows the generator to accept a whole small-$d$ box immediately.

\paragraph{Upper collar and exact physical exclusion.}
For the lower part of the upper collar, the finite certificate proves
\begin{equation}
 g_{rr}\ge1/100\quad\hbox{on }
 [0,1/8]\times[19/20,491/500].
 \label{bal:eq:radial-upper-curvature}
\end{equation}
Its exact cells have widths $1/128$ in $r$ and $1/1000$ in $d$:
$16\cdot32=512$ cells.  Equation~\eqref{bal:eq:radial-integrated-curvature}
gives $g\ge r^2/200$ there.

For $491/500\le d\le d_*$, define the inverse-free exclusion function
\begin{equation}
 \mathcal E_R(r,d)=\frac{C(B-A)}D-
 P\!\left(d_C\frac{B-A}{B+A}\right),
 \label{bal:eq:radial-upper-exclusion}
\end{equation}
where $A,B,C,D,d_C$ are from the reconstruction.  It has the exact
endpoint values $\mathcal E_R(0,d)=\mathcal E_R(1,d)=0$.
If $\mathcal E_R\ge0$ and $0<r<1$, monotonicity of $P$ and
\eqref{bal:eq:radial-reconstruction} imply
\[
 d_a\ge d_C\frac{B-A}{B+A},\qquad
 \frac{e_w}{H_*(d_C)}=\frac{d_a(B+A)}{d_C(B-A)}\ge1.
\]
Hence $d_b\le d_C\le d_c$, since $H_*$ decreases and
$d_c=d_C e_w/H_*(d_C)$. This is $b\ge c$, which excludes the strict
radial ordering.

To certify $\mathcal E_R>0$ on the open radial interval, the exact cover
proves $\mathcal E_{R,r}>0$ on $0\le r\le1/8$, proves
$\mathcal E_R>0$ in the middle, and proves
$\mathcal E_{R,r}<0$ on $1-10^{-5}\le r\le1$.
The endpoint identities and integration justify the two edge strips.
The $d$ cover uses $\epsilon=(1-d)/2\in[23/10^6,9/1000]$, partitioned
into adjacent rational bands with successive right endpoints
$\min\{9/1000,(11/10)\epsilon_{\rm left}\}$.
There are 63 such bands.  The left strip has 16 equal radial cells;
the middle uses 25 cells of width $1/32$ on $[1/8,29/32]$, followed by
41 adjacent geometric bands in $1-r$ from $10^{-5}$ to $3/32$ with
successive ratio at most $5/4$; the right strip is one radial cell.
The acceptance counts are thus $1008$, $1575$, $2583$, and $63$, in
addition to the 512 curvature cells.  These 5741 cells have a recorded
384-bit replay.

The functions and their derivatives are enclosed by the degree-six
Taylor-model evaluator \code{top_collar_certificate.py}.  Products retain
a bound for all discarded monomials; elementary-function expansions include
Lagrange remainders.  An inverse polynomial is accepted only after its
residual is divided by a positive lower bound for the inverse derivative
on an interval containing both the polynomial range and the true inverse.
The resulting full-cell range is compared against $1/100$ for curvature,
or against zero for the signed exclusion quantity.
Consequently the complete upper strip $491/500\le d\le d_*$ contains
no strict physical radial point.  The lower upper-collar strip is covered
by a value bound. A box crossing their common boundary may use their union;
its accepted claim concerns its physical intersection.

\subsection{Explicit terminal acceptance tests on the compact rectangle}
\label{bal:sec:radial-acceptance}
The remaining root rectangle is
\begin{equation}
 \mathcal R=[0,1]\times[1/50,d_*].
 \label{bal:eq:radial-root-rectangle}
\end{equation}
All cell endpoints and bisections are rational.  The source
\code{radial_convexity.evaluate_radial_box} has the following acceptance
rules.  Every interval quantity below is an outward-rounded enclosure of
the named function on the whole cell $I_r\times I_d$.

\paragraph{Direct Taylor value bound.}
Let $(r_0,d_0)$ be the exact midpoint and $\rho_r,\rho_d$ the radii.
For derivatives of order at most two let $M_{ij}$ bound
$|\partial_r^i\partial_d^j g(r_0,d_0)|$; for order three let $M_{ij}$
bound that derivative on the entire cell.  Define
\begin{align*}
 T={}&M_{10}\rho_r+M_{01}\rho_d+\tfrac12M_{20}\rho_r^2
       +M_{11}\rho_r\rho_d+\tfrac12M_{02}\rho_d^2\\
 &+\tfrac16\bigl(M_{30}\rho_r^3+3M_{21}\rho_r^2\rho_d
                  +3M_{12}\rho_r\rho_d^2+M_{03}\rho_d^3\bigr).
\end{align*}
The multivariable Taylor theorem gives
$g(I_r,I_d)\subset[g(r_0,d_0)]+[-T,T]$.
Acceptance requires the lower endpoint of this interval to be nonnegative.
This is the terminal reason \code{direct Taylor enclosure g >= 0}.

\paragraph{Equality-edge curvature.}
This rule is allowed only when $\inf I_r=0$.  The checker encloses
\[
 [g_{rr}(r_0,d_0)]
 +[-\rho_r\sup|g_{rrr}|-\rho_d\sup|g_{rrd}|,
     \rho_r\sup|g_{rrr}|+\rho_d\sup|g_{rrd}|].
\]
It subtracts $\kappa[I_d]^4$, where $\kappa=1/100000$, and requires a
nonnegative lower endpoint.  Thus $g_{rr}\ge\kappa d^4$ on the whole
anchored cell, and \eqref{bal:eq:radial-integrated-curvature} proves
$g\ge\kappa r^2d^4/2$.
For the numerically sensitive high-$d$ edge, the alternative enclosure is
\[
 [g_{rr}(0,I_d)]-(\sup I_r)\sup|g_{rrr}|-\kappa[I_d]^4.
\]
The edge expression is evaluated by the exact cancellation-free formula
in \code{edge_curvature.py}; its interval covers all of $I_d$.
The same test may be made on 64 exact adjacent $d$ strips, each with the
full radial interval $[0,\sup I_r]$.  Acceptance requires every strip to
pass, not a sampled subset.

\paragraph{Whole-cell exclusion.}
Where the reconstructed quantities are defined, any one of
\begin{equation}
 \sup[d_b-d_c]\le0,\qquad
 \sup[d_c-d_a]\le0,\qquad
 \sup[2(1-S)-d_a-d_c]<0
 \label{bal:eq:radial-exclusion-tests}
\end{equation}
excludes the entire cell from the retained strict ordering.  These are,
respectively, $b\ge c$, $c\ge a$, and $a+c<S$.
The upper-collar theorem provides a further exclusion rule already proved
above.  An optional high-$d$ edge rule proves
$\partial_r(d_a+d_c)>0$ on a full anchored cell with $d\ge9999/10000$.
Its exact edge value is $d_a+d_c=2d$, so the cell has $a+c\le S$;
the only possible equality has $r=0$, hence $a=c$ and is not strict.
The source encloses the derivative by its exact edge value minus
$(\sup I_r)\sup|\partial_{rr}(d_a+d_c)|$ on exact adjacent strips.

\paragraph{Imported collar rules and failure behavior.}
The small-$d$ rule applies when the entire cell has $d\le1/20$.
The upper rules apply only on their stated exact domains.  A cell assigned
to a union of a value region and an exclusion region asserts the value
bound only at its physical points.  Every other cell remains unresolved
unless the direct or anchored test above succeeds.  A singular inverse,
a zero-containing required inverse derivative, or an interval of
undecided sign triggers subdivision or an explicit residual record.
Neither finite precision agreement nor a depth or node limit is an
acceptance condition.

\subsection{Coverage, boundary attachments, and the recorded computation}
\label{bal:sec:radial-coverage}
The initial manifest partitions \eqref{bal:eq:radial-root-rectangle} into
$256\times256$ equal rational rectangles.  A split bisects the wider
coordinate, with the source's fixed tie convention.  Each terminal is
recorded by its binary path and one of four codes: proved, outside,
residual, or refuted.  Exact path reconstruction specifies the terminal
rectangle.  Distinct terminal paths must be prefix-free, and their Kraft
sum must equal one.  These two checks imply that the terminal rectangles
cover their root.  The split count is checked separately.

When a resource limit is reached, each residual rectangle becomes an
identical root in the next generation.  Equality of the reconstructed
residual frontier with that next manifest, including the full rational
endpoints, is required.  Consequently induction through the generation
chain preserves complete root coverage.  The final manifest must be empty;
all refuted and unknown terminal codes must be rejected.  A count or hash
alone does not establish these implications.

The complete recorded 192-bit execution uses the frozen kernel closure
identified in the artifact map and the exact rules above.  Its full-ledger
audit records eight nonempty generations, followed by the empty $g008$
manifest, with no refuted leaves or rigorous failures.  The source,
terminal records, collar certificates, and coverage checks are separate
parts of the computer-assisted proof: the source establishes the
acceptance contract, the original interval execution applies it, and the
structural audit establishes that no admissible point is omitted.
The original execution and the subsequent checks are described in Section~\ref{bal:sec:radial-verification}.

To conclude, take an arbitrary retained strict radial stationary point.
Lemma~\ref{bal:lem:radial-surjectivity} and
\eqref{bal:eq:radial-d-domain} place it in the chart.  If $d\le1/20$,
\eqref{bal:eq:radial-small-collar} proves its value nonnegative. Otherwise
it lies in \eqref{bal:eq:radial-root-rectangle} and, by exact terminal
coverage, in a proved or excluded leaf.  It cannot belong to an excluded
physical intersection.  The accepted test on its proved leaf gives
$g\ge0$, and \eqref{bal:eq:radial-gap-identification} gives $G_R\ge0$.
This proves the radial stationary-point estimate with its stated retained
scope.

For the cap-minimizer argument, the closed feasible $c$ interval is
$[\max\{b,S-a\},a]$.  Any interior minimizer is stationary and is covered by the stationary estimate;
its endpoints are attached to the double-cap theorem ($c=b$), the
fixed-sum seam theorem ($a+c=S$), or the equal-mean lemma ($c=a$), exactly
as in Lemma~\ref{bal:lem:cap-minimizer-bridge}.  Degenerate and
zero-entropy endpoints use the separate boundary statements of the
main text.  The region $a+c<S$ is not a pure-gap conclusion of this
computation: Theorem~\ref{bal:thm:small-boundary} supplies the
Bellman lower bound on that region.  In particular, relabeling a cutoff leaf as
``outside'' does not assert that its pure gap is nonnegative.

\paragraph{Source and evidence locators.}
In the publication core, the frozen radial package is
\code{tmp/radial_attachment_audit/single_version_run/}.
Its \code{ck_certificate/} directory contains
\code{d_core.py}, \code{edge_third.py}, \code{edge_curvature.py},
\code{radial_convexity.py}, \code{radial_cloud.py},
\code{collar_certificate.py}, and \code{top_collar_certificate.py}.
The source attestation is
\code{provenance/single_version_attestation.json}.
The detailed normalization derivation and upper-collar argument are also
preserved under the recorded worker-source tree, in
\code{proof/ORIGINAL_NORMALIZED_COLLAR_PROOF_d002.md},
\code{proof/EXTENDED_COLLAR_PROOF.md}, and
\code{ck_certificate/TOP_FULL_R_COLLAR_PROOF.md}.
The upper-collar argument gives $b\ge c$ and therefore excludes
the physical radial ordering $b<c<a$.
The current complete-ledger and 384-bit collar replay reports are the
records identified in Section~\ref{bal:sec:records}; the two external
radial archives contain the full terminal streams. The older mixed-version
659,191,385-node run is not the 29,413,134-node frozen execution used here.

\clearpage
\section{Computational evidence and verification scope}\label{bal:sec:verification}

This section identifies the computational components, their certificate
records, and the checks performed on them. The component-inventory program
aggregates their reported results.

The \emph{original execution} applies the stated acceptance tests with
outward-rounded arithmetic to produce the certificate. Three further
kinds of checks have different scopes. An \emph{arithmetic replay}
evaluates a sign enclosure again on the specified exact cells.  A
\emph{proof-object audit} checks the stored records, signs, source commitments,
and exact cover, while retaining trust in the generator that produced the
recorded numerical enclosures.  An \emph{inventory check} reads component
reports and checks their reported contracts and counts. Checksums identify
the exact source and certificate files used in these operations.

For each computer-assisted input, acceptance requires both a mathematical
argument and a computational argument.  The mathematical argument identifies
the function, derivative conventions, feasible domain, and every boundary
case.  The computational argument establishes rigorous enclosures and an exact
cover of that same domain.  Floating-point searches may choose trial points
or subdivision priorities, but their approximations cannot justify an
acceptance unless an outward-rounded argument independently validates it.

\subsection{Scope of the recorded auxiliary checks}
The entropy-convexity calculation encloses its two scalar signs on
95 exact intervals at 192 and 320 bits. The origin calculation bounds
$Q^{(17)}$ on all of $[0,4/25]$, using sixteen exact intervals, before
applying the polynomial/remainder estimate on $s+t\le2/25$.
This derivative domain includes every possible argument $2s+t$.
These scalar checks are separate from the large certificate ledgers.

\subsection{Recorded evidence}

Table~\ref{bal:tab:inventory} summarizes the archived verification records.
``Recorded replay'' means that the package contains a completed
arithmetic-replay report. Counts for different components need not be
comparable: some count terminal cells, while the radial total counts all
evaluated nodes.

\begin{longtable}{>{\raggedright\arraybackslash}p{0.24\linewidth}>{\raggedright\arraybackslash}p{0.68\linewidth}}
\caption{Recorded executions, structural audits, and arithmetic replays.}\label{bal:tab:inventory}\\
\toprule Component & Evidence and scope\\\midrule\endfirsthead
\toprule Component & Evidence and scope\\\midrule\endhead
Framework & Eleven checked exact identities and a written audit of the
positive-entropy Bellman implication.  The identities do not replace review
of the surrounding analytic proof.\\
Strict-seam minimum exclusion & Recorded 384-bit union of 5,697 owners,
including value and derivative owners, with zero residual cells.
The required statement includes the premise $G<0$; it does not assert the
stronger derivative-only disjunction everywhere.\\
Clauses (ii)--(v) & Recorded exact owner unions, analytic endpoint
arguments, and arithmetic replays.  The corrected M2 chain-rule material
must be used in place of its earlier expression.\\
Case-E A3 & Full recorded 384-bit arithmetic replay with the repaired
interval union: 5,367,158 upper leaves and 120,786 low-extension leaves.
Delegated labels certify applicability of their named external lemma;
they do not re-prove that lemma.\\
E8 & Audit of 3,368 bundle files, including 275,701 rectangular leaves,
7,464 one-variable leaves, 199 segments, and 256 small-tail cells.
Additional arithmetic checks evaluate selected junctions at 256 bits.\\
Joint convexity & Imported directed-MPFR proof objects: 33 ledgers and
1,282,104 leaves, exact geometry checks, 15 independent interval samples,
and available higher-precision replay logs.  Primary arithmetic is recorded
as MPFR 4.2.2 at 192 bits.\\
Reflection & Imported audit of 177,856 finite leaves, analytic endpoint
attachments, and selected dual-precision and coefficient checks.\\
Small boundary and half mean & Recorded 512-bit scalar replays of
45,425 same-side and 22,086 opposite-side cells; separate half-mean compact
and endpoint checks at 192 and 256 bits.\\
Radial endpoint collars & Recorded 384-bit arithmetic replays of 32,768
normalized cells and 5,741 upper cells.\\
Radial stationary interior & Complete source-attested original 192-bit
ledger, with exact structural and frontier audit over 6,976 shards and
29,413,134 nodes. The terminal generation and 109 generation-zero shards
were separately recomputed; see Section~\ref{bal:sec:radial-verification}.\\
Unbalanced extension & Complete rational partitions and recorded full
70-digit replays of the two final covers (160,789 and 29,495 leaves)
and supporting certificates; counts and sources are given in
Sections~\ref{supp:verification} and~\ref{unbal:app:inventory}.\\
Final balanced inventory & Eleven aggregate component gates report PASS.  The
program reads component reports and checks their commitments, counts,
and status fields.\\\bottomrule
\end{longtable}

\subsection{Exact Case-E owner partition}\label{bal:sec:casee-evidence}

Write $t_C$ and $A$ for the real coordinates of the stationary Case-E
chart defined in Section~\ref{bal:sec:casee-stationary}. The certificate
boxes have exact rational endpoints. The complete strict chart is assigned as
follows; overlaps at displayed closed seams are evaluator overlaps, with the
precedence rule assigning a unique logical owner.
\begin{center}
\small
\begin{tabular}{p{0.43\linewidth}p{0.49\linewidth}}
\toprule Exact domain & Owner\\\midrule
$0<t_C\le1/80000$, $A\le21/200$
 & retained-domain exclusion, with SB-1 outside the retained region\\
$0<t_C\le1/100$, $A>21/200$
 & capital exclusion theorem\\
$1/80000\le t_C\le1/100$, $0\le A\le1/20$
 & transverse-curvature pure-gap owner\\
$1/80000\le t_C\le1/100$, $1/20\le A\le21/200$
 & corrected A3 exact leaf union\\
$1/100\le t_C<1/2$, $1/2<\lambda<1$
 & high-$t_C$ condition-E exclusion\\\bottomrule
\end{tabular}
\end{center}
The $t_C$ and $A$ seams agree as exact rationals. The chart-surjectivity
lemma in Section~\ref{bal:sec:ce-chart-bijection} maps every physical
stationary Case-E point into this chart. Section~\ref{bal:sec:casee-owner-contracts}
proves the implications of the certificate inequalities and the completeness
of the domain union.

\subsection{The corrected A3 calculation}\label{bal:sec:a3-evidence}

The historical \code{casee_mv._hull_ball} helper formed candidate endpoint
balls and selected extrema using ball comparisons.  When those comparisons
were unresolved, that selection could lose containment.  The replay
installs the explicit Arb
\code{union} repair before evaluating any committed terminal rectangle.
The evidence binds the original leaf streams and source commitments to the
repair and replay-driver hashes.  The upper and low-extension summaries
cover respectively 672 and 224 rows, a total of 896 rows and 5,487,944 leaves;
both report \code{full_dataset_replayed=true} and unchanged acceptance
labels.

\subsection{The radial calculation}\label{bal:sec:radial-verification}

Section~\ref{bal:sec:radial-stationary} gives the stationary chart,
the gap identity, all terminal acceptance conditions, and their
implication for the retained radial stationary estimate.
Both radial archives are required.  Together they contain the 65,536 initial
roots, generations $g000$ through $g007$, and the empty $g008$ frontier.
Their generation node counts are
\[
\begin{gathered}
9{,}171{,}167,\quad6{,}895{,}745,\quad6{,}826{,}825,\quad4{,}190{,}830,\\
1{,}812{,}946,\quad451{,}819,\quad62{,}721,\quad1{,}081.
\end{gathered}
\]
The complete-ledger audit checks all shard hashes, the frozen source closure,
every manifest and merge record, and all terminal codes.  In particular, it
rejects refuted \code{X} leaves and reconstructs each next frontier from the
raw residual leaves.  The supplied report records zero unknown codes,
malformed rows, rigorous failures, refuted leaves, and terminal residuals.

This establishes a complete structural audit of the recorded execution.
It retains the original interval generator and its arithmetic environment
within the trusted computation.  The terminal generation and 109
generation-zero shards were separately recomputed and matched the committed
bytes.

\section{Artifact and reproduction map}\label{bal:sec:records}

The computational supplement consists of the following three archives
and the Case-E certificate recovery files listed in
Section~\ref{bal:sec:casee-recovery-scope}:
\begin{itemize}
\item \code{BALANCED_CK_PUBLICATION_CORE_2026-09-07.zip}:
analytic sources, compact certificates, E8/joint-convexity/reflection
objects, corrected A3 inputs and reports, and verification programs;
\item \code{ck_radial_fresh_part1_g000-g001.zip}:
raw radial generations $g000$ and $g001$;
\item \code{ck_radial_fresh_part2_g002-g008.zip}:
raw radial generations $g002$--$g007$ and the terminal frontier.
\end{itemize}
Extract the core into a new project directory, preserving its paths, and
place the two radial archives in its \code{upload/} directory.  The core
already includes \code{upload/a3_leaves_256.zip}.  The revised PDF and its
LaTeX source are separate deliverables: they were produced after the core
archive and are not authenticated by the earlier manuscript checksum.
Keep the archive manifest and the revision manifest together.
The directory \code{computational_supplement_patch/} supplied with the
Case-E recovery package adds five capital and curvature certificate files
at their original paths. Each file matches the SHA-256 value in
\code{CASE_E_OWNER_INVENTORY.json}. The recovery audit and application
script are supplied alongside those files; this restores their availability
without changing the original archive commitments or asserting a new
arithmetic execution.

The audit directory is
\code{ck_final_logical_audit_2026-09-07/}.
Its \code{FINAL_COMPONENT_INVENTORY.json} identifies the report for each
aggregate component.  Its \code{framework/}, \code{boundary/},
\code{collar/}, \code{casee/}, and \code{arithmetic/} directories contain
the corresponding logical and numerical records.  The reconstructed seam
and cap sources and results are under
\code{ck_reconstruction/CKBalanced/}.  The imported component audit policy
is \code{arithmetic/IMPORTED_ARITHMETIC_VERIFICATION_POLICY.md}; it specifies
which complete proof objects and which selective replays were used.

\subsection{Detailed artifact locators}\label{bal:sec:detailed-artifacts}
\begin{longtable}{>{\raggedright\arraybackslash}p{0.27\linewidth}>{\raggedright\arraybackslash}p{0.65\linewidth}}
\toprule Result & Controlling local record or source\\\midrule\endhead
Aggregate component report & \code{ck_final_logical_audit_2026-09-07/FINAL_COMPONENT_INVENTORY.json}\\
Final inventory builder & \code{ck_final_logical_audit_2026-09-07/verify_final_component_inventory.py}\\
Logical connection audit & \code{ck_final_logical_audit_2026-09-07/framework/FRAMEWORK_LOGICAL_AUDIT.md}\\
Exact formula audit & \code{ck_final_logical_audit_2026-09-07/framework/framework_identity_checks.json}\\
Low entropy proof & \code{RECONSTRUCTED_CLAUSE_I_LOW_ENTROPY.md}\\
Scalar collars & \code{experiments/results/reconstructed_scalar_rootfree_collars.json}\\
Whole-entropy Hessian & \code{experiments/results/reconstructed_equality_line_attachment_384.json}\\
Large Hessian replay & \code{experiments/results/reconstructed_coverage_transverse_384.json}\\
Chart-0 frontier & \code{experiments/results/reconstructed_fiber_frontier_audit_384.json}\\
Extra chart-0 strip & \code{experiments/results/reconstructed_fiber_sk1_low_u_replay_384.json}\\
Full chart 1 & \code{experiments/results/reconstructed_coverage_chart1_global_384.json}\\
Complete seam union & \code{experiments/results/reconstructed_global_seam_me_union_384.json}\\
Full cap audit & \code{experiments/results/clause_ii_reconstruction_audit_256_2026-09-07.json}\\
Cap low-$z$ repair & \code{experiments/results/clause_ii_lowz_repaired_cover_256_2026-09-07.json}\\
Cap high-$z$ repair & \code{experiments/results/clause_ii_highz_repaired_cover_256.json}\\
Clause (iii) & \code{CLAUSE_III_ANALYTIC_PROOF_2026-09-05.md}\\
Clauses (iv)--(v), preserved analytic source & \code{ANALYTIC_COLLARS_2026-09-04.tex}\\
Fresh R1 & \code{experiments/results/reconstructed_r1_global_256.json}\\
Fresh M1 and M2 & \code{experiments/results/reconstructed_m1_m2_global_192.json}\\
Repaired complete M2 & \code{experiments/results/m2_chain_rule_repair_audit_2026-09-07.json}\\
Corrected A3 upper summary & \code{arithmetic/row_replay_union_upper_384/summary.json}\\
Corrected A3 low-extension summary & \code{arithmetic/row_replay_union_lowext_384/summary.json}\\
Case-E owner inventory & \code{casee/CASE_E_OWNER_INVENTORY.json}\\
Radial collar audit & \code{ck_final_verification_run_2026-09-07/radial_collars/fresh_collar_replay_audit.json}\\
Radial complete-ledger audit & \code{arithmetic/radial_complete_ledger_audit.json}\\
Radial complete-ledger checker & \code{arithmetic/audit_complete_radial_ledgers.py}\\
Radial raw ledgers & \code{ck_radial_fresh_part1_g000-g001.zip}; \code{ck_radial_fresh_part2_g002-g008.zip}\\
E8, joint-convexity, reflection audit & \code{arithmetic/imported_proof_object_audit.json}\\
SB-1 fresh summaries & \code{boundary/fresh_replay/*_scalars_512.jsonl.gz.summary.json}\\
HM-1 fresh verifier & \code{checks/half_mean_fresh_verifier.json}\\
Original global master & \code{upstream/BALANCED_CK_MASTER_FIXED_D_MINIMIZATION_2026-09-05.tex}\\
Older source recovery & \code{RECOVERED_UPSTREAM_INDEX.md}\\
Global dependency audit & \code{RECONSTRUCTED_THEOREM_DEPENDENCY_AUDIT.md}\\\bottomrule
\end{longtable}

Paths beginning with \code{arithmetic/}, \code{casee/}, \code{boundary/},
\code{checks/}, or \code{framework/} are relative to
\code{ck_final_logical_audit_2026-09-07}.  Paths beginning with
\code{experiments/} or names of analytic proof files are relative to
\code{ck_reconstruction/CKBalanced}.

\subsection{Computational environment and installation}

The radial requirements file specifies \code{python-flint==0.9.0} and
\code{scipy==1.17.0}. Other verification programs use \code{mpmath},
\code{numpy}, or compiled MPFR/GMP code, as specified by their respective
component requirements. The commands below use
\code{ck_reconstruction/deps} as the Python dependency directory.
From the extracted project root, install the radial/A3 dependencies with
{\small\begin{verbatim}
python -m pip install --target ck_reconstruction/deps \
  -r tmp/radial_attachment_audit/single_version_run/\
ck_certificate/requirements.txt
\end{verbatim}}
The C/MPFR programs require a compiler and the MPFR/GMP development headers.

\subsection{Separate commands for separate checks}

The commands below are run from the extracted project root.  Shell variables
are used only to shorten paths.
{\small\begin{verbatim}
AUDIT=ck_final_logical_audit_2026-09-07
RADIAL=tmp/radial_attachment_audit/single_version_run
export PYTHONPATH=ck_reconstruction/deps

python "$AUDIT/arithmetic/replay_rows_repaired.py" \
  --dataset upper --bits 384 --workers 4 --resume
python "$AUDIT/arithmetic/replay_rows_repaired.py" \
  --dataset lowext --bits 384 --workers 4 --resume
python "$AUDIT/casee/verify_casee_owner_inventory.py"

python "$AUDIT/arithmetic/replay_radial_collars.py" \
  --bits 384 --workers 4
\end{verbatim}}
The first two commands perform A3 arithmetic replays; the next checks the
Case-E inventory; the last replays the radial endpoint collars.  For a
complete independent A3 run, use empty output directories rather than
accepting previously cached rows through \code{--resume}.

The following checks the entire \emph{recorded} radial ledger:
{\small\begin{verbatim}
python "$AUDIT/arithmetic/audit_complete_radial_ledgers.py" \
  --part1 upload/ck_radial_fresh_part1_g000-g001.zip \
  --part2 upload/ck_radial_fresh_part2_g002-g008.zip \
  --source "$RADIAL" --deps ck_reconstruction/deps \
  --output radial_ledger_review.json
\end{verbatim}}
Expected structural result: \code{PASS_COMPLETE_LEDGER_AUDIT}, eight
nonempty generations, 29,413,134 nodes, and zero terminal residuals.
The separate, potentially expensive command for a full arithmetic
re-execution is
{\small\begin{verbatim}
python "$AUDIT/arithmetic/replay_radial_chain_local.py" \
  --workers 4 --source "$RADIAL" --run radial_fresh_review
\end{verbatim}}
Use a new \code{--run} directory for an independent execution.  This driver
requires each regenerated shard to match its committed hash and checks the
generation transitions.  The supplied aggregate command
\code{verify_final_component_inventory.py} may be run afterward to check
the fixed reports, but does not automatically consume arbitrary new output
paths or replace those reports with a newly produced review record.

\section{Technical proofs for the unbalanced extension}
\label{unbal:app:modules}

\emph{Argument guide.}
The modules are best read by dependency. The first group develops
log-sum bounds, curvature, endpoint gains, and cap comparisons. The next
group removes small separations and proves the complete central same-side
region. The last group treats central opposite-side means by parent
dominance, low-entropy ratio ranges, and a moderate-entropy cover before
assembling the central square. The brief guides at each module entrance
describe its argument; the local statements retain their own hypotheses.

The modules below give the central and supporting arguments in dependency
order. They retain local notation; each statement
specifies its domain. Source paths and the manifest
\path{PRIOR_CHAPTER_MANIFEST.json} identify the component documents.
Section~\ref{unbal:app:central_assembly} assembles the central square, and
the main text gives the final global assembly.

The global log-sum, curvature, and boundary proofs are also supplied here in
their original forms, in addition to the proofs in the main text and preceding
supplementary sections. This makes
the dependency chain inspectable without consulting separate working notes.

\clearpage
\subsection{Global log-sum inequality and entropy-profile calculus}
\label{unbal:app:profile_logsum}

\emph{Argument guide.}
The tangent remainder of the edge cost is rewritten using divergences
between positive masses. Log-sum and variance-to-entropy comparisons
produce the global minorants. Profile derivatives then turn those
minorants into a comparison valid for a whole interval of total entropy,
uniformly over its split between the children.

\noindent\textit{Source:} \path{CK_CAP_REGION_CONTINUATION/prior_inputs/PROFILE_AND_LOGSUM_PROOF.md}.\par
\subsubsection{Setup}\label{unbal:module-profile_logsum-setup}

All entropies and divergences are in bits. Put
\[
L=\ln2,\quad H(x)=-x\log_2x-(1-x)\log_2(1-x),\quad
J(x)=\log_2\frac{1-x}{x},
\]
\[
j(u,w)=\frac12(u-w)(J(w)-J(u)),\qquad \nu_x=x(1-x).
\]
The boundary cost is its lower-semicontinuous extension: zero at \((0,0),(1,1)\), infinite at every other boundary point. For a feasible tuple \(M=(a,b,e,f)\), \(\zeta(M)\) is the infimum of \(\mathbb E j(U,W)\) over arbitrary joint laws with
\[
\mathbb EU=a,\quad\mathbb EW=b,\quad\mathbb EH(U)=e,\quad\mathbb EH(W)=f.
\]
Set
\[
I_u=H(a)-e,\quad I_w=H(b)-f,\quad s=(I_u+I_w)/2,
\]
\[
m=(a+b)/2,\quad
\Delta=H(m)-\tfrac12(H(a)+H(b)),\quad
I=H(m)-(e+f)/2=\Delta+s.
\]
Define \(P(I)=\eta(1-I)\), where
\[
\eta(e)=(1-2H^{-1}(e))J(H^{-1}(e)),
\]
and \(H^{-1}\) is the lower inverse of binary entropy. The continuous value is \(P(0)=0\). Then \(\psi(m,e)=P(H(m)-e)\), and
\begin{equation}\label{unbal:eq:profile_logsum:1}
R_\psi=P(\Delta+s)-\tfrac12(P(I_u)+P(I_w))
\le D_\Delta(s):=P(\Delta+s)-P(s).
\end{equation}
Convexity of \(P\), recalled with proof below, justifies \eqref{unbal:eq:profile_logsum:1}. Thus bounds against \(D_\Delta(s)\) cover every feasible entropy imbalance with that total deficit.

\subsubsection{A global improvement from the log-sum inequality}\label{unbal:module-profile_logsum-a-global-improvement-from-the-log-sum-inequality}

\begin{theorem}\label{unbal:thm:profile_logsum:1}
For interior means, let
\[
V(a,b)=\max(a,b)\,[1-\min(a,b)].
\]
Then
\begin{equation}\label{unbal:eq:profile_logsum:2}
\boxed{\zeta(a,b,e,f)\ge
B_{\rm LS}(a,b,e,f):=j(a,b)+\frac{(a-b)^2}{4V(a,b)}(I_u+I_w).}
\end{equation}
Equivalently, \(B_{\rm LS}=j+(a-b)^2s/(2V)\). This improves the preceding bound \(j+(a-b)^2s\) whenever \(V<1/2\). The two bounds may always be combined by taking their maximum.

In fact, the following stronger pointwise inequality holds for any interior anchors \(a,b\):
\begin{equation}\label{unbal:eq:profile_logsum:3}
\begin{split}
Q_{a,b}(u,w):={}&j(u,w)-j(a,b)-j_u(a,b)(u-a)-j_w(a,b)(w-b)\\
\ge{}&\frac{(a-b)^2}{4LV(a,b)}
\left(\frac{(u-a)^2}{\nu_a}+\frac{(w-b)^2}{\nu_b}\right)\\
\ge{}&\frac{(a-b)^2}{4V(a,b)}
\{D_2(u\Vert a)+D_2(w\Vert b)\}.
\end{split}
\end{equation}
\end{theorem}

\emph{Proof outline.}
Subtract the tangent plane and express its remainder as positive-mass
divergences. Log-sum reduces these to scalar logarithmic terms, which
bound squared deviations and hence entropy deficits. Averaging removes
the tangent terms and gives the asserted moment inequality.

\begin{proof}
Let
\[
\mathcal D(x\Vert y)=\frac{x\ln(x/y)-x+y}{L},\qquad x,y>0,
\]
be the relative entropy for nonnormalized positive masses. Write
\[
r_1=u/a,\quad r_0=(1-u)/(1-a),\quad
t_1=w/b,\quad t_0=(1-w)/(1-b).
\]
An exact Bregman identity is
\begin{equation}\label{unbal:eq:profile_logsum:4}
\begin{split}
2Q_{a,b}(u,w)={}&a\mathcal D(r_1\Vert t_1)
+(1-a)\mathcal D(r_0\Vert t_0)\\
&+b\mathcal D(t_1\Vert r_1)
+(1-b)\mathcal D(t_0\Vert r_0).
\end{split}
\end{equation}
One way to check it is to decompose
\[
j(u,w)=j_0(u,w)+j_0(1-u,1-w),\qquad
j_0(x,y)=\frac{(x-y)\ln(x/y)}{2L}.
\]
The function \(j_0\) is homogeneous of degree one. Subtracting its tangent at \((a,b)\) yields half of \(a\mathcal D(u/a\Vert w/b)+b\mathcal D(w/b\Vert u/a)\); the complementary term gives the other two terms in \eqref{unbal:eq:profile_logsum:4}.

Put \(d=a-b\). The log-sum inequality applied to the first row of \eqref{unbal:eq:profile_logsum:4} gives \(\mathcal D(1\Vert T)\), and the second row gives \(\mathcal D(1\Vert S)\), where
\[
T=1+\frac{d(w-b)}{\nu_b},\qquad
S=1-\frac{d(u-a)}{\nu_a}.
\]
Hence
\begin{equation}\label{unbal:eq:profile_logsum:5}
Q_{a,b}(u,w)\ge\tfrac12\{\mathcal D(1\Vert T)+\mathcal D(1\Vert S)\}.
\end{equation}
For \(z>-1\), elementary integration gives
\begin{equation}\label{unbal:eq:profile_logsum:6}
z-\ln(1+z)\ge\frac{z^2}{2\max(1,1+z)}.
\end{equation}
For example, integrate \(t/(1+t)\) from \(0\) to \(z\), separately for positive and negative \(z\).

Directly from \(0\le u,w\le1\),
\begin{equation}\label{unbal:eq:profile_logsum:7}
\max(1,T)\le V/\nu_b,\qquad
\max(1,S)\le V/\nu_a.
\end{equation}
For instance, if \(a\ge b\), then \(V=a(1-b)\), \(T\le a/b\), and \(S\le(1-b)/(1-a)\). The other order is symmetric. Applying \eqref{unbal:eq:profile_logsum:6} and \eqref{unbal:eq:profile_logsum:7} in \eqref{unbal:eq:profile_logsum:5} proves the first inequality in \eqref{unbal:eq:profile_logsum:3}.

Finally, \(\ln x\le x-1\) implies the standard scalar estimate
\begin{equation}\label{unbal:eq:profile_logsum:8}
D_2(u\Vert a)\le\frac{(u-a)^2}{L\nu_a},
\end{equation}
and similarly for \(w,b\). This proves the second inequality in \eqref{unbal:eq:profile_logsum:3}.

Take expectations at \(a=\mathbb EU,b=\mathbb EW\). The tangent terms vanish, and
\[
\mathbb ED_2(U\Vert a)=H(a)-\mathbb EH(U)=I_u.
\]
Infimizing proves \eqref{unbal:eq:profile_logsum:2}. Finite-cost boundary atoms follow by continuity from interior approximations, and infinite-cost laws are immediate.
\end{proof}

The proof also gives the useful variance version
\[
\mathbb E j(U,W)\ge j(a,b)+\frac{(a-b)^2}{4LV}
\left(\frac{\operatorname{Var}U}{\nu_a}+\frac{\operatorname{Var}W}{\nu_b}\right)
\]
for each representing law, before eliminating its variances. The law-dependent right side is not asserted to lower-bound the infimum for every other representing law.

\subsubsection{A further explicit strengthening}\label{unbal:module-profile_logsum-a-further-explicit-strengthening}

Define
\[
C_a=\max\left\{\frac{-\ln(1-a)}{La^2},
\frac{-\ln a}{L(1-a)^2}\right\},
\]
\[
\kappa(a)=\frac1{L\nu_aC_a}
=\min\left\{\frac{a}{-(1-a)\ln(1-a)},
\frac{1-a}{-a\ln a}\right\}\ge1.
\]
\begin{corollary}\label{unbal:thm:profile_logsum:2}
The stronger bound
\begin{equation}\label{unbal:eq:profile_logsum:9}
\boxed{B_\chi:=j(a,b)+\frac{(a-b)^2}{4V}
\{\kappa(a)I_u+\kappa(b)I_w\}\le\zeta}
\end{equation}
holds globally for interior means.
\end{corollary}

Indeed,
\[
\frac{D_2(x\Vert a)}{(x-a)^2}
=\int_0^1\frac{1-t}{L\nu_{a+t(x-a)}}\,dt,
\]
with its continuous value at \(x=a\). The right side is convex in \(x\), since \(1/[x(1-x)]=1/x+1/(1-x)\) is convex. Its maximum on \([0,1]\) is therefore an endpoint value, giving \(D_2(x\Vert a)\le C_a(x-a)^2\). Thus \(\operatorname{Var}U\ge I_u/C_a\); insert this in the first inequality of \eqref{unbal:eq:profile_logsum:3}. Finally \(L\nu_aC_a\le1\), either by \eqref{unbal:eq:profile_logsum:8} or by the elementary logarithmic bounds in the displayed formula, proving \(\kappa(a)\ge1\).

For comparisons uniform in entropy imbalance, one may use
\begin{equation}\label{unbal:eq:profile_logsum:10}
B_\chi\ge j(a,b)+\frac{(a-b)^2\kappa_*}{2V}s,
\qquad \kappa_*:=\min\{\kappa(a),\kappa(b)\}.
\end{equation}
The exact bound \eqref{unbal:eq:profile_logsum:9} is preferable when the entropy coordinates are fixed.

\subsubsection{Entropy-profile facts, with analytic constants}\label{unbal:module-profile_logsum-entropy-profile-facts-with-analytic-constants}

Let
\[
C(r)=1-H((1-r)/2),\quad A=\operatorname{atanh}r,
\quad I=C(r).
\]
Then \(C'(r)=A/L\) and \(P(C(r))=2rA/L\). Differentiation gives
\begin{equation}\label{unbal:eq:profile_logsum:11}
P'(I)=2+\frac{2r}{(1-r^2)A},\qquad
P''(I)=2L\frac{(1+r^2)A-r}{(1-r^2)^2A^3}>0.
\end{equation}
The endpoint values are \(P'(0)=4\), \(P''(0)=8L/3\).

For completeness, the increasing-curvature result from the previous continuation is analytic. A further derivative gives
\begin{equation}\label{unbal:eq:profile_logsum:12}
P'''(I)=\frac{2L^2[2r(r^2+3)A^2-(5r^2+3)A+3r]}{(1-r^2)^3A^5}>0.
\end{equation}
To verify the sign, expand
\[
A^2=\sum_{n\ge1}c_nr^{2n},\qquad
c_n=\frac1n\sum_{i=0}^{n-1}\frac1{2i+1}\ge\frac1n.
\]
The numerator's \(r\) and \(r^3\) coefficients vanish. For \(n\ge2\), its \(r^{2n+1}\) coefficient is
\[
6c_n+2c_{n-1}-\frac5{2n-1}-\frac3{2n+1}
\ge\frac6n-\frac8{2n-1}+2c_{n-1}>0.
\]
The series are absolutely convergent for \(r<1\), proving \eqref{unbal:eq:profile_logsum:12}.

Choose the exact threshold
\[
I_0=C(1/3)=1-H(1/3)=0.08170416594551048\ldots.
\]
Then
\begin{equation}\label{unbal:eq:profile_logsum:13}
\boxed{P''(I)\le P''(I_0)=\frac{9(5L-3)}{4L^2}<\frac{16L}{5}
\quad(0\le I\le I_0).}
\end{equation}
The strict constant comparison is entirely elementary. It is equivalent to \(64L^3-225L+135>0\). The series
\[
L=2\sum_{n\ge0}\frac1{(2n+1)3^{2n+1}}
<\frac23+\frac1{36}=\frac{25}{36}
\]
follows by bounding \(2n+1\ge3\) in the tail. The polynomial is decreasing on \([0,25/36]\), and its value at \(25/36\) is \(535/2916>0\). This proves \eqref{unbal:eq:profile_logsum:13} without an interval computation.

We will also use the deterministic-cap inequality
\begin{equation}\label{unbal:eq:profile_logsum:14}
j(a,b)\ge P(\Delta).
\end{equation}
For a proof, put \(q=|a-b|/2\), \(m=(a+b)/2\). Direct binary-entropy identities give
\begin{equation}\label{unbal:eq:profile_logsum:15}
\Delta=mC(q/m)+(1-m)C(q/(1-m)),
\end{equation}
\begin{equation}\label{unbal:eq:profile_logsum:16}
j(a,b)=mP(C(q/m))+(1-m)P(C(q/(1-m))).
\end{equation}
Jensen's inequality for \(P\) gives \eqref{unbal:eq:profile_logsum:14}. These identities are also the usual two representations of the mutual information and the average relative entropy of the two Bernoulli laws.

\subsubsection{A global comparison criterion}\label{unbal:module-profile_logsum-a-global-comparison-criterion}

\begin{theorem}\label{unbal:thm:profile_logsum:3}
For fixed distinct interior means, suppose a feasible tuple satisfies
\begin{equation}\label{unbal:eq:profile_logsum:17}
\boxed{\Delta P''(I)\le\frac{(a-b)^2}{2V}.}
\end{equation}
Then \(\zeta\ge R_\psi\), for every feasible entropy split with those means and that entropy sum. More generally, the right side in \eqref{unbal:eq:profile_logsum:17} may be multiplied by \(\kappa_*\) when using \eqref{unbal:eq:profile_logsum:10}.
\end{theorem}

\begin{proof}
Define
\[
g(t)=j(a,b)+\frac{(a-b)^2}{2V}t
-P(\Delta+t)+P(t),\qquad 0\le t\le s.
\]
Equation \eqref{unbal:eq:profile_logsum:14} gives \(g(0)\ge0\). Increasing curvature and \eqref{unbal:eq:profile_logsum:17} imply
\[
P'(\Delta+t)-P'(t)
=\int_t^{t+\Delta}P''(v)dv
\le\Delta P''(I)\le\frac{(a-b)^2}{2V}.
\]
Thus \(g'\ge0\), so \(g(s)\ge0\). Apply \eqref{unbal:eq:profile_logsum:2} and \eqref{unbal:eq:profile_logsum:1}. The strengthened version is identical with the slope in \eqref{unbal:eq:profile_logsum:10}.
\end{proof}

Notice that this covers the entire interval from the entropy caps down to the specified entropy sum. It is not just a test of one entropy split.

\clearpage
\subsection{Uniform entropy curvature and normalized small-difference theorem}
\label{unbal:app:uniform_diagonal}

\emph{Argument guide.}
The scalar estimates reserve a positive parent correction and bound
the radial and entropy derivatives. The theorem then splits at the
separation-to-entropy ratio three: a direct radial comparison handles
one range, and a quadratic minimization over the entropy split handles
the other. The resulting estimates include arbitrarily small positive
entropy and mean separation.

\noindent\textit{Source:} \path{CK_NO_SEPARATION_EXTENSION/prior_inputs/UNIFORM_DIAGONAL_PROOF.md}.\par
\subsubsection{Definitions}\label{unbal:module-uniform_diagonal-definitions}

Use bits, and let \(L=\ln 2\). Define
\[
H(x)=-x\log_2x-(1-x)\log_2(1-x),\qquad
J(x)=\log_2((1-x)/x).
\]

The lower inverse of \(H\) is denoted \(H^{-1}\). Set
\[
\eta(e)=(1-2H^{-1}(e))J(H^{-1}(e)),\quad P(I)=\eta(1-I).
\]

For \(s\ge0\) and \(e>0\), define \(F(s,e)=sJ(v)\), where \(v\in(0,1/2]\) solves
\[
sH(v)=e(1-2v),
\]
and \(F(0,e)=0\). In particular \(F(s,e)=e f(s/e)\), with \(f(x)=F(x,1)\).

Write
\[
\Phi(s,e)=\eta(e)-F(s,e),\quad
\phi(m,e)=\Phi(|1-2m|,e),\quad
\psi(m,e)=\eta(1-H(m)+e),\quad B=\max\{\phi,\psi\}.
\]

For any two-variable function \(A\),
\[
R_A=A((a+b)/2,(e+f)/2)-[A(a,e)+A(b,f)]/2.
\]

The feasible domain is \(0<a,b<1\), \(0<e\le H(a)\), \(0<f\le H(b)\). We use
\[
E=(e+f)/2,\quad m=(a+b)/2,\quad q=|1-2m|,\quad d=|a-b|,
\quad C(q)=1-H((1-q)/2).
\]

The unordered child radial coordinates are \(d+q\) and \(|d-q|\). Entropies must follow their corresponding children if these are reordered.

\subsubsection{Scalar estimates}\label{unbal:module-uniform_diagonal-scalar-estimates}

\paragraph{Entropy curvature and a parent correction}\label{unbal:module-uniform_diagonal-entropy-curvature-and-a-parent-correction}

For \(0<h\le1\),
\begin{equation}\label{unbal:eq:uniform_diagonal:1}
\boxed{\eta''(h)\ge\frac1{2Lh^2}.}
\end{equation}

Also, when \(h>0\), \(C\ge0\) and \(h+C\le1\),
\begin{equation}\label{unbal:eq:uniform_diagonal:2}
\boxed{\eta(h)-\eta(h+C)\ge 2C+\log_2(1+C/h).}
\end{equation}

For a proof let \(r=1-2H^{-1}(h)\), \(A=\operatorname{atanh}r\), \(\nu=1-r^2\), and let \(h_n=Lh\) be natural entropy. The function
\[
T(r)=2r h_n-\nu A
\]
vanishes at \(r=0\) and as \(r\to1\). Its derivative is \(2h_n-1\), strictly decreasing from \(2L-1>0\) to \(-1\). Thus \(T\ge0\) and
\[
h_n\ge\frac{\nu A}{2r}.
\]

Direct differentiation gives
\[
P'(1-h)=2+\frac{2r}{\nu A},\quad
\eta''(h)=\frac{2L\{(1+r^2)A-r\}}{\nu^2 A^3}.
\]

Since \(A\ge r\),
\[
h^2\eta''(h)
\ge\frac{(1+r^2)A-r}{2r^2LA}\ge\frac1{2L},
\]
proving \eqref{unbal:eq:uniform_diagonal:1}. Multiplying the first derivative formula by \(h_n\) also gives
\[
P'(1-h)\ge2+\frac1{Lh}.
\]

Integration gives \eqref{unbal:eq:uniform_diagonal:2}. At \(r=0\) the formulas are interpreted by continuity.

We will also use
\begin{equation}\label{unbal:eq:uniform_diagonal:3}
\frac{q^2}{2L}\le C(q)\le\frac{q^2}{2L(1-q^2)}.
\end{equation}

Indeed $C(0)=C'(0)=0$ and $C''(q)=1/[L(1-q^2)]$.

\paragraph{An upper derivative bound at small entropy}\label{unbal:module-uniform_diagonal-an-upper-derivative-bound-at-small-entropy}

For \(0<h\le10^{-4}\),
\begin{equation}\label{unbal:eq:uniform_diagonal:4}
\boxed{hP'(1-h)<7/4.}
\end{equation}

Let \(v=H^{-1}(h)\), \(t=\ln((1-v)/v)\). Concavity of \(H\) gives \(H(v)\ge2v\), so \(v\le1/20000\) and \(t\ge\ln 19999>9\). The last elementary bound follows from \(\exp(1)<3\) and \(3^9=19683<19999\). Since
\[
Lh=vt-\ln(1-v),
\]
we obtain
\[
hP'(1-h)
=2h+\frac{1-2v}{L(1-v)}
\left(1+\frac{-\ln(1-v)}{vt}\right)
\le2h+\frac1L\left(1+\frac1{(1-v)t}\right).
\]

Use \(-\ln(1-v)\le v/(1-v)\), \((1-v)t>8\), and \(L>2/3\). The final expression is less than \(2/10000+27/16<7/4\).

\paragraph{Perspective estimates}\label{unbal:module-uniform_diagonal-perspective-estimates}

For \(s>0\),
\begin{equation}\label{unbal:eq:uniform_diagonal:5}
\boxed{F_{ss}(s,e)\le\frac4{Ls},\qquad
0\le\Phi_{se}(s,e)\le\frac4{Le}.}
\end{equation}

A direct proof of the first inequality is as follows. At the defining contact let \(r=1-2v\), \(A=\operatorname{atanh}r\), \(\nu=1-r^2\), \(h_n=LH(v)\), \(K=h_n+rA=L-(1/2)\ln\nu\). Implicit differentiation gives
\[
F_{ss}=\frac{2h_n^3(2K-r^2)}{eL^2\nu^2K^3},\quad
sF_{ss}=\frac{2r h_n^2(2K-r^2)}{L\nu^2K^3}.
\]

We have \(h_n\le\nu K\): this is equivalent to \(A-rK\ge0\), whose derivative \(1-K\) changes sign once from positive to negative, while the expression vanishes at both endpoints. Therefore \(sF_{ss}\le2r(2K-r^2)/(LK)\le4/L\). Homogeneity gives
\[
\Phi_{se}=(s/e)F_{ss},\quad
\Phi_{ee}=\eta''(e)-(s/e)^2F_{ss}.
\]

Combining with \eqref{unbal:eq:uniform_diagonal:1} yields
\begin{equation}\label{unbal:eq:uniform_diagonal:6}
\boxed{\Phi_{ee}(s,e)\ge\frac{1/2-4s}{Le^2}.}
\end{equation}

For fixed \(e\), define \(k_e(z)=F(\sqrt z,e)\). The decreasing-\(F_{ss}\) property follows from Theorem~\ref{bal:imp:e8:thm:onevar} through the derivative identities in Section~\ref{bal:aux:sec:fourpoint}. It implies \(F_s\ge sF_{ss}\), hence \(k_e\) is increasing and concave. Consequently
\begin{equation}\label{unbal:eq:uniform_diagonal:7}
\frac{F(d+q,e)+F(|d-q|,e)}2
\le F(\sqrt{d^2+q^2},e)
\le F(d,e)+k_e'(d^2)q^2.
\end{equation}

The ratio \(f(x)/x^2\) and the derivative \(f'(x)/(2x)\) are nonincreasing for \(x>0\).

\paragraph{Two explicit scalar constants}\label{unbal:module-uniform_diagonal-two-explicit-scalar-constants}

\begin{equation}\label{unbal:eq:uniform_diagonal:8}
\boxed{f(3)>12,\qquad f'(3)/6<1.}
\end{equation}

For the first inequality, the contact \(v\) for \(f(3)\) lies below \(1/17\) because
\[
3H(1/17)>15/17
\quad\Longleftrightarrow\quad17^{17}>2^{69}.
\]

Thus \(J(v)>4\) and \(f(3)>12\).

For the second, the contact lies in the exact interval
\[
5279/100000<v<66/1250.
\]

Writing \(r=1-2v\) and \(A=\operatorname{atanh}r\), the defining equation \(3H(v)=r\) gives
\[
\frac{f'(3)}6=\frac{A}{3L}
+\frac{r}{3(1-r^2)(L+3A)}.
\]

The bracket gives \(L>693/1000\), \(1443/1000<A<1444/1000\), \(r<895/1000\), and \(1-r^2>199/1000\). Hence
\[
\frac{f'(3)}6
<\frac{1444/1000}{3(693/1000)}
+\frac{895/1000}{3(199/1000)(693/1000+3(1443/1000))}
=\frac{114629824}{115428159}<1.
\]

\path{EXACT_CONSTANT_CHECKER.py} verifies all logarithmic bracket inequalities with exact rational arithmetic. For \(z\ge1\) it sums
\[
\ln z=2\sum_{k=0}^{N-1}\frac{w^{2k+1}}{2k+1}+R_N,
\quad w=(z-1)/(z+1),\quad
0\le R_N\le\frac{2w^{2N+1}}{(2N+1)(1-w^2)}.
\]

It uses \(N=256\) and uses \(\ln z=-\ln(1/z)\) for \(z<1\). Thus \eqref{unbal:eq:uniform_diagonal:8} is a finite rational certificate, not a floating-point assertion or a sampled functional inequality.

\subsubsection{The explicit uniform theorem}\label{unbal:module-uniform_diagonal-the-explicit-uniform-theorem}

\begin{theorem}\label{unbal:thm:uniform_diagonal:1}
For every feasible tuple with
\[
0<E\le10^{-6},\quad q\le32E,\quad d\le1/64,
\]
we have
\begin{equation}\label{unbal:eq:uniform_diagonal:9}
\boxed{\max\{R_\phi,F(d,E)\}\ge R_B.}
\end{equation}

Thus \(\zeta\ge R_B\) by Theorems~\ref{bal:thm:global-unrestricted-target}
and~\ref{bal:thm:four-moment-lower-bound}, which give
\(\zeta\ge R_\phi\) and \(\zeta\ge L_4\ge F(d,E)\), respectively. In the branch where \(\psi\) is active at the parent, \(F(d,E)\) alone suffices.
\end{theorem}

\emph{Proof outline.}
Use the balanced theorem when its candidate is active at the parent.
In the other branch, compare separation with total entropy: a direct
radial bound handles small ratios, and strong child-entropy convexity
absorbs arbitrary splitting at larger ratios. The parent correction
then pays for both the radial and split losses.

\begin{proof}
If \(\phi\) is active at the parent, \(B(\mathrm{children})\ge\phi(\mathrm{children})\) immediately gives \(R_B\le R_\phi\). We may therefore assume \(B(m,E)=\psi(m,E)=\eta(E+C(q))\).

From \eqref{unbal:eq:uniform_diagonal:3}, \(q\le32E\), \(q\le1/10\), \(E\le10^{-6}\), and \(L>2/3\),
\begin{equation}\label{unbal:eq:uniform_diagonal:10}
C(q)/E\le\frac{1024E}{2L(1-q^2)}<1/1000.
\end{equation}

\paragraph{Case 1: \(d\le3E\)}\label{unbal:module-uniform_diagonal-case-1-d3e}

Set
\[
\Delta=H(m)-[H(a)+H(b)]/2,\quad I=H(m)-E,
\quad s=[H(a)+H(b)]/2-E=I-\Delta\ge0.
\]

Convexity of \(P\) gives, uniformly over the entropy split,
\[
R_B\le R_\psi\le P(I)-P(s)\le\Delta P'(I).
\]

Here \(1-I=E+C(q)\le1.001E<10^{-4}\). By \eqref{unbal:eq:uniform_diagonal:4}, \(P'(I)\le7/[4(E+C(q))]\le7/(4E)\). The entropy Hessian and the bound on the child radii give
\[
\Delta\le\frac{d^2}{2L(1-(q+d)^2)}.
\]

Since \(q+d\le35E<1/8\) and \(L>2/3\),
\[
R_B\le\frac{7d^2}{8LE(1-(q+d)^2)}\le\frac{4d^2}{3E}.
\]

Concavity of \(k_1\) and \eqref{unbal:eq:uniform_diagonal:8} imply, for \(0<d/E\le3\),
\[
F(d,E)=E f(d/E)\ge\frac{f(3)}9\frac{d^2}{E}
>\frac{4d^2}{3E}.
\]

The \(d=0\) case follows from the same bound with zero on the right.

\paragraph{Case 2: \(d\ge3E\)}\label{unbal:module-uniform_diagonal-case-2-d3e}

Let \(s_1=d+q\) and \(s_2=|d-q|\), with the entropy labels matched to the corresponding children. We have
\[
s_i\le1/64+32\times10^{-6}<1/32.
\]

For every \(u\in(0,2E)\), \eqref{unbal:eq:uniform_diagonal:6} gives
\[
\Phi_{ee}(s_i,u)\ge\frac1{4Lu^2}\ge k:=\frac1{16LE^2}.
\]

All such entropies are feasible at these radial coordinates. Write the matched entropy split as \(E+h,E-h\), where \(|h|<E\). Strong convexity gives
\[
\frac{\Phi(s_1,E+h)+\Phi(s_2,E-h)}2
\ge\frac{\Phi(s_1,E)+\Phi(s_2,E)}2
+\frac h2 A_e+\frac k2h^2,
\]
where \(A_e=\Phi_e(s_1,E)-\Phi_e(s_2,E)\). By \eqref{unbal:eq:uniform_diagonal:5} and \(s_1-s_2\le2q\),
\[
|A_e|\le\frac{8q}{LE}.
\]

Minimizing the displayed quadratic over the entire real line is a valid lower bound for all \(|h|<E\), and yields
\begin{equation}\label{unbal:eq:uniform_diagonal:11}
\frac{\Phi(s_1,e)+\Phi(s_2,f)}2
\ge\frac{\Phi(s_1,E)+\Phi(s_2,E)}2
-\frac{A_e^2}{8k}
\ge\frac{\Phi(s_1,E)+\Phi(s_2,E)}2-192q^2.
\end{equation}

The last constant uses \(L>2/3\). This is the step that makes the argument uniform over arbitrarily unequal entropies.

Next, \eqref{unbal:eq:uniform_diagonal:7}, monotonicity of \(k_1'\), \(d/E\ge3\) and \eqref{unbal:eq:uniform_diagonal:8} imply
\begin{equation}\label{unbal:eq:uniform_diagonal:12}
\frac{F(s_1,E)+F(s_2,E)}2\le F(d,E)+q^2/E.
\end{equation}

Because \(B(\mathrm{children})\ge\phi(\mathrm{children})\), \eqref{unbal:eq:uniform_diagonal:11} and \eqref{unbal:eq:uniform_diagonal:12} give
\[
R_B\le F(d,E)+q^2/E-[\eta(E)-\eta(E+C(q))]+192q^2.
\]

From \eqref{unbal:eq:uniform_diagonal:2}, \(\ln(1+x)\ge x/(1+x)\), \eqref{unbal:eq:uniform_diagonal:3}, \eqref{unbal:eq:uniform_diagonal:10}, and \(L<7/10\),
\[
\eta(E)-\eta(E+C(q))
\ge\frac{q^2}{2L^2(E+C(q))}
\ge\frac{50000}{49049}\frac{q^2}{E}.
\]

Consequently
\begin{equation}\label{unbal:eq:uniform_diagonal:13}
R_B\le F(d,E)-\left(\frac{50000}{49049}-1-192E\right)\frac{q^2}{E}
\le F(d,E)-\frac{q^2}{100E}.
\end{equation}

The rational inequality needed for the last step is checked exactly in the supplied checker. In particular \(R_B\le F(d,E)\). This completes the proof.
\end{proof}

This proof is for every positive feasible \(e,f\). There is no invocation of continuity in \(e/f\) and no discarded narrow entropy-ratio region.

\clearpage
\subsection{Opposite deterministic corner and endpoint mass comparison}
\label{unbal:app:opposite_corner}

\emph{Argument guide.}
First compare unequal-entropy and equal-entropy endpoint masses to
obtain a logarithmic gain. Next control the child functions on the
extended entropy domain and absorb every entropy split using that gain.
Parent and radial estimates prove the normalized corner theorem; a
separate parent-dominance argument extends it to the mean-only corner.

\noindent\textit{Source:} \path{@retained/opposite_corner/OPPOSITE_CORNER_PROOF.md}.\par
\subsubsection{Definitions and theorem}\label{unbal:module-opposite_corner-definitions-and-theorem}

All entropy quantities and costs are in bits. Put \(L=\ln 2\),
\[
H(x)=-x\log_2x-(1-x)\log_2(1-x),\quad
J(x)=\log_2((1-x)/x),\quad
j(x,y)=\tfrac12(x-y)(J(y)-J(x)).
\]

Let \(H^{-1}\) denote the lower inverse. Define
\[
\eta(h)=(1-2H^{-1}(h))J(H^{-1}(h)),\quad
\Phi(s,h)=\eta(h)-F(s,h),
\]
where \(F(s,h)=sJ(v)\), with \(sH(v)=h(1-2v)\), and \(F(0,h)=0\). Thus \(F(s,h)=h f(s/h)\), where \(f(x)=F(x,1)\). Set
\[
\phi(m,h)=\Phi(|1-2m|,h),\quad
\psi(m,h)=\eta(1-H(m)+h),\quad B=\max\{\phi,\psi\}.
\]

For any two-variable function \(A\),
\[
R_A=A((a+b)/2,(e+f)/2)-[A(a,e)+A(b,f)]/2.
\]

We consider \(0<a<b<1\), \(0<e\le H(a)\), \(0<f\le H(b)\), and write
\[
E=(e+f)/2,\quad d=b-a,\quad q=|1-a-b|,
\quad C(q)=1-H((1-q)/2).
\]

The unordered child radial coordinates are \(d+q\) and \(|d-q|\).

\begin{theorem}[Normalized opposite-corner inequality]\label{unbal:thm:opposite_corner:1}
Suppose
\begin{equation}\label{unbal:eq:opposite_corner:1}
\boxed{d\ge9/10,\qquad 0<E\le10^{-3},\qquad q\le32E.}
\end{equation}

Then the Proposition~\ref{bal:prop:fixed-difference-minimum} endpoint contact exists. Denote its lower bound by \(B_{\mathrm{end}}\). If \(\psi\) is active at the parent, then
\begin{equation}\label{unbal:eq:opposite_corner:2}
\boxed{B_{\mathrm{end}}-R_B\ge\frac{q^2}{5E}.}
\end{equation}

In all branches,
\begin{equation}\label{unbal:eq:opposite_corner:3}
\boxed{\max\{R_\phi,B_{\mathrm{end}}\}\ge R_B.}
\end{equation}
\end{theorem}

\begin{theorem}[Complete mean-only corner]\label{unbal:thm:opposite_corner:2}
For all positive feasible \(e,f\),
\begin{equation}\label{unbal:eq:opposite_corner:4}
\boxed{a+(1-b)\le2^{-13}\quad\Longrightarrow\quad\zeta(a,b,e,f)\ge R_B.}
\end{equation}
\end{theorem}

Only simultaneous complement and full label exchange are used to transport these statements to symmetric orientations. Individual reflection of one mean is not a target symmetry.

\subsubsection{The new endpoint entropy-imbalance bound}\label{unbal:module-opposite_corner-the-new-endpoint-entropy-imbalance-bound}

Let
\[
Q(h)=J(H^{-1}(h)),\qquad t=\frac{e-f}{e+f}\in(-1,1),\qquad
\mathcal J(t)=-\ln(1-t^2).
\]

\begin{lemma}\label{unbal:thm:opposite_corner:lem1}
Under \eqref{unbal:eq:opposite_corner:1},
\begin{equation}\label{unbal:eq:opposite_corner:5}
\boxed{B_{\mathrm{end}}\ge F(d,E)+\frac{d}{4L}\mathcal J(t).}
\end{equation}
\end{lemma}

\paragraph{Applicability and the contact-mass comparison}\label{unbal:module-opposite_corner-applicability-and-the-contact-mass-comparison}

Since \(d\ge9/10\), we have \(a<1/2<b\). Feasibility gives
\[
H^{-1}(e)\le a,\quad H^{-1}(f)\le1-b,
\quad d\le1-H^{-1}(e)-H^{-1}(f).
\]

Let \(p_{\min}=\max(e,f)\). The excluded lower endpoint in Proposition~\ref{bal:prop:fixed-difference-minimum} is
\[
D_0=p_{\min}\left[\frac12-H^{-1}\left(\frac{\min(e,f)}{p_{\min}}\right)\right]
\le p_{\min}/2\le E<d.
\]

Thus Proposition~\ref{bal:prop:fixed-difference-minimum} supplies its unique strict cross-half contact \(p,u,v\), with
\[
d=p(1-u-v),\quad e=pH(u),\quad f=pH(v),\quad 0<p\le1.
\]

Its cost is
\begin{equation}\label{unbal:eq:opposite_corner:6}
B_{\mathrm{end}}=p j(u,1-v)=\frac d2\{Q(e/p)+Q(f/p)\}.
\end{equation}

Let \(p_{\mathrm{bar}}\) be the analogous mass at equal entropies \(E,E\). This contact exists because convexity of \(H^{-1}\) gives
\[
2H^{-1}(E)\le H^{-1}(e)+H^{-1}(f)\le1-d.
\]

We have \(p_{\mathrm{bar}}\ge d\). The arguments \(e/p_{\mathrm{bar}}\) and \(f/p_{\mathrm{bar}}\) are less than \(1/100\) because \(e,f\le2E\) and \(p_{\mathrm{bar}}\ge9/10\).

Convexity of \(H^{-1}\) also gives
\[
p_{\mathrm{bar}}\{1-H^{-1}(e/p_{\mathrm{bar}})-H^{-1}(f/p_{\mathrm{bar}})\}
\le p_{\mathrm{bar}}\{1-2H^{-1}(E/p_{\mathrm{bar}})\}=d.
\]

The scalar contact-difference function is increasing in \(p\), as in Proposition~\ref{bal:prop:fixed-difference-minimum}. Therefore
\begin{equation}\label{unbal:eq:opposite_corner:7}
\boxed{p\ge p_{\mathrm{bar}}.}
\end{equation}

Because \(Q\) is decreasing, \eqref{unbal:eq:opposite_corner:6} implies
\begin{equation}\label{unbal:eq:opposite_corner:8}
B_{\mathrm{end}}\ge\frac d2\{Q(e/p_{\mathrm{bar}})+Q(f/p_{\mathrm{bar}})\}.
\end{equation}

This is the essential strengthening: unequal entropies increase the contact mass, so a quantitative Jensen estimate can be applied at the common mass \(p_{\mathrm{bar}}\).

\paragraph{A logarithmic convexity estimate}\label{unbal:module-opposite_corner-a-logarithmic-convexity-estimate}

For \(0<h\le1/100\), write \(v=H^{-1}(h)\), \(\ell=\ln((1-v)/v)\), \(r=1-2v\). Differentiation gives
\[
Q''(h)=\frac{L(r\ell-1)}{v^2(1-v)^2\ell^3}.
\]

We have \(v\le h/2\le1/200\), \(r\ge99/100\), and \(\ell\ge\ln199>4\). Moreover \(Lh=v\ell-\ln(1-v)\ge v\ell\). It follows that
\[
h^2Q''(h)\ge\frac{r-1/\ell}{L(1-v)^2}\ge\frac1{2L}.
\]

Hence \(Q(h)+(1/(2L))\ln h\) is convex on this interval. Jensen's inequality in \eqref{unbal:eq:opposite_corner:8} yields
\[
\frac{Q(e/p_{\mathrm{bar}})+Q(f/p_{\mathrm{bar}})}2
\ge Q(E/p_{\mathrm{bar}})+\frac1{4L}\mathcal J(t).
\]

Finally, \(d\,Q(E/p_{\mathrm{bar}})=F(d,E)\) by the equal-entropy contact equation. This proves \eqref{unbal:eq:opposite_corner:5}.

The logarithm in \eqref{unbal:eq:opposite_corner:5} controls arbitrarily large entropy imbalance, including \(t\to\pm1\). No assertion of equality for the original four-moment tuple is needed.

\subsubsection{Entropy control for the child functions, including beyond a cap}\label{unbal:module-opposite_corner-entropy-control-for-the-child-functions-including-beyond-a-cap}

An average entropy \(E\) need not be feasible at each child separately. We therefore prove the following estimate for the analytic extension \(\Phi(s,h)=\eta(h)-F(s,h)\), without imposing \(h\le H((1-s)/2)\).

\begin{lemma}\label{unbal:thm:opposite_corner:lem2}
For
\[
4/5\le s\le1,\qquad 0<h\le1/500,
\]
we have
\begin{equation}\label{unbal:eq:opposite_corner:9}
\boxed{\Phi_{hh}(s,h)\ge-\frac{2}{5Lh^2}.}
\end{equation}
\end{lemma}

We will also use, for all \(s>0\), \(h>0\) where \(\eta(h)\) is defined,
\begin{equation}\label{unbal:eq:opposite_corner:10}
\boxed{0\le\Phi_{sh}(s,h)\le\frac4{Lh}.}
\end{equation}

\paragraph{Derivative formulas}\label{unbal:module-opposite_corner-derivative-formulas}

For a lower-half entropy variable, let \(r=1-2v\), \(A=\operatorname{atanh}r\), \(\nu=1-r^2\), and \(h_n=LH(v)\). The function
\[
T(r)=2r h_n-\nu A
\]
vanishes at both endpoints and has derivative \(2h_n-1\), decreasing through zero once. Thus
\[
h_n\ge\frac{\nu A}{2r}.
\]

Differentiating \(\eta\) gives
\[
\eta''(H(v))=\frac{2L\{(1+r^2)A-r\}}{\nu^2A^3}.
\]

Consequently, with \(\ell=2A\),
\begin{equation}\label{unbal:eq:opposite_corner:11}
H(v)^2\eta''(H(v))
\ge\frac1L-\frac1{Lr\ell}.
\end{equation}

At the \(F\)-contact put \(K=h_n+rA=L-(1/2)\ln\nu\). Implicit differentiation gives
\[
sF_{ss}(s,h)=\frac{2r h_n^2(2K-r^2)}{L\nu^2K^3}.
\]

The inequality \(h_n\le\nu K\) follows because \(A-rK\) vanishes at both endpoints and its derivative \(1-K\) changes sign once. Hence \(sF_{ss}\le4/L\). Homogeneity gives
\[
\Phi_{sh}=(s/h)F_{ss},\quad F_{hh}=(s/h)^2F_{ss},
\]
proving \eqref{unbal:eq:opposite_corner:10}.

A sharper contact estimate follows from
\[
h_n=v\ell-\ln(1-v),\quad K=\ell/2-\ln(1-v),\quad\nu=4v(1-v):
\]

\[
\frac{h_n}{\nu K}
\le\frac{1+1/((1-v)\ell)}{2(1-v)},
\]
and therefore
\begin{equation}\label{unbal:eq:opposite_corner:12}
sF_{ss}(s,h)
\le\frac1L\frac{(1+1/((1-v)\ell))^2}{(1-v)^2}.
\end{equation}

\paragraph{Uniform constants}\label{unbal:module-opposite_corner-uniform-constants}

Put \(v_8=(1+\exp 8)^{-1}\). Elementary logarithmic bounds give
\[
1/4096<v_8<1/1024,\qquad H(v_8)>12/4096=3/1024.
\]

For example, \(\ln4095\ge12(69/100)-1/4095>8\), while \(\ln1023<10(7/10)<8\). These use \(69/100<L<7/10\).

In Lemma~\ref{unbal:thm:opposite_corner:lem2}, \(h\le1/500<H(v_8)\). Also
\[
h/s\le1/400<3/1024<H(v_8)/(1-2v_8).
\]

Thus both the entropy inverse and the \(F\)-contact have \(\ell>8\) and \(v<1/1024\). Formula \eqref{unbal:eq:opposite_corner:11} gives
\[
h^2\eta''(h)\ge\frac{1-64/511}{L}.
\]

Formula \eqref{unbal:eq:opposite_corner:12}, together with \(s\le1\), gives
\[
h^2F_{hh}(s,h)
\le\frac1L\left(\frac{1151}{1023}\frac{1024}{1023}\right)^2.
\]

The exact rational comparison
\[
\left(\frac{1151}{1023}\frac{1024}{1023}\right)^2-1+\frac{64}{511}
=\frac{220293308870209}{559658926346751}<\frac25
\]
proves \eqref{unbal:eq:opposite_corner:9}.

\paragraph{Consequence for an arbitrary entropy split}\label{unbal:module-opposite_corner-consequence-for-an-arbitrary-entropy-split}

Under \eqref{unbal:eq:opposite_corner:1}, the two child radii \(s_1=d+q\) and \(s_2=d-q\) lie in \([4/5,1]\), since \(d-q\ge9/10-32/1000>4/5\). Match entropy labels to these radii and write \(e_1=E+h\), \(e_2=E-h\). Set \(t=h/E\); changing the label orientation only changes the sign of \(t\).

By \eqref{unbal:eq:opposite_corner:9}, the function
\[
h\longmapsto \Phi(s_i,h)-\frac{2}{5L}\ln h
\]
is convex throughout \((0,2E)\). Applying its supporting line at \(E\) gives
\begin{equation}\label{unbal:eq:opposite_corner:13}
\frac{\Phi(s_1,e_1)+\Phi(s_2,e_2)}2
\ge\frac{\Phi(s_1,E)+\Phi(s_2,E)}2
+\frac h2 A_e-\frac1{5L}\mathcal J(t),
\end{equation}
where \(A_e=\Phi_h(s_1,E)-\Phi_h(s_2,E)\). From \eqref{unbal:eq:opposite_corner:10},
\begin{equation}\label{unbal:eq:opposite_corner:14}
|A_e|\le\frac{8q}{LE}.
\end{equation}

The proof of \eqref{unbal:eq:opposite_corner:13} remains valid when \(E\) lies above a child's entropy cap, because Lemma~\ref{unbal:thm:opposite_corner:lem2} was proved on the analytic extension itself. The actual entropies \(e_1,e_2\) still satisfy their original cap constraints.

\subsubsection{The parent correction and the radial second difference}\label{unbal:module-opposite_corner-the-parent-correction-and-the-radial-second-difference}

Two elementary bounds needed below are
\begin{equation}\label{unbal:eq:opposite_corner:15}
\frac{q^2}{2L}\le C(q)\le\frac{q^2}{2L(1-q^2)},
\end{equation}
and
\begin{equation}\label{unbal:eq:opposite_corner:16}
\eta(E)-\eta(E+C)
\ge2C+\log_2(1+C/E).
\end{equation}

For \eqref{unbal:eq:opposite_corner:15}, use $C(0)=C'(0)=0$ and $C''(q)=1/[L(1-q^2)]$. For \eqref{unbal:eq:opposite_corner:16}, the entropy calculation preceding \eqref{unbal:eq:opposite_corner:11} also gives
\[
P'(1-h)=2+\frac{2r}{(1-r^2)\operatorname{atanh}r}
\ge2+\frac1{Lh},
\]
and integration proves the claim.

Under \eqref{unbal:eq:opposite_corner:1}, \(q\le32E\le0.032<1/10\). Therefore
\[
C(q)/E\le\frac{1024E}{2L(1-q^2)}\le\frac{128}{165}<4/5.
\]

Using \(\ln(1+x)\ge x/(1+x)\), \eqref{unbal:eq:opposite_corner:15}, and \(L<7/10\) in \eqref{unbal:eq:opposite_corner:16}, we obtain
\begin{equation}\label{unbal:eq:opposite_corner:17}
\boxed{\eta(E)-\eta(E+C(q))\ge\frac{250}{441}\frac{q^2}{E}.}
\end{equation}

For the radial term, concavity of \(k_E(z)=F(\sqrt z,E)\) gives
\begin{equation}\label{unbal:eq:opposite_corner:18}
\frac{F(d+q,E)+F(d-q,E)}2
\le F(d,E)+k_E'(d^2)q^2.
\end{equation}

Here \(d/E\ge900\). We claim
\begin{equation}\label{unbal:eq:opposite_corner:19}
k_E'(d^2)<\frac1{10E}.
\end{equation}

To verify this, first note the elementary global estimate
\[
F(s,h)\le2s\log_2(2+3s/h).
\]

Indeed, at the contact put $\ell=\ln((1-v)/v)$, $x=s/h$, and
$b_0=-\ln(1-v)/v$, with $0<v<1/2$.  The identities
$LH(v)=v(\ell+b_0)$ and $xH(v)=1-2v$ imply
\[
 e^\ell=1+\frac{x(\ell+b_0)}L.
\]
The elementary bounds $1\le b_0\le1/(1-v)$ give
\[
 \frac{x(b_0-1)}L
 \le\frac{1-2v}{(1-v)(\ell+b_0)}\le1.
\]
Thus $e^\ell\le2+x(1+\ell)/L$.  With $z=e^{\ell/2}$ and
$1+\ell\le2z$, we obtain $z^2\le2+2xz/L<2+3xz$, and therefore
$z<2+3x$, since $L>2/3$.

Let \(f(x)=F(x,1)\). Since \(k_1\) is concave and \(k_1(0)=0\),
\[
k_E'(d^2)\le\frac1E\frac{f(x)}{x^2}
\le\frac1E\frac{2\log_2(2+3x)}x,\quad x=d/E.
\]

The last ratio decreases for \(x\ge900\). At \(x=900\), \(2+3x=2702<2^{12}\), so it is less than \(24/900<1/10\). This proves \eqref{unbal:eq:opposite_corner:19}.

\subsubsection{Proof of Theorem~\ref{unbal:thm:opposite_corner:1}}\label{unbal:module-opposite_corner-proof-of-theorem-1}

\begin{proof}
If \(\phi\) is active at the parent, \(B(\mathrm{children})\ge\phi(\mathrm{children})\) immediately gives \(R_B\le R_\phi\). Suppose instead that \(B(\mathrm{parent})=\psi(\mathrm{parent})=\eta(E+C(q))\).

Because \(B(\mathrm{children})\ge\phi(\mathrm{children})\), we have
\[
B_{\mathrm{end}}-R_B
\ge B_{\mathrm{end}}+\frac{\Phi(s_1,e_1)+\Phi(s_2,e_2)}2-\eta(E+C(q)).
\]

Combining the endpoint correction \eqref{unbal:eq:opposite_corner:5} with \eqref{unbal:eq:opposite_corner:13},
\[
B_{\mathrm{end}}-R_B
\ge F(d,E)+\frac{\Phi(s_1,E)+\Phi(s_2,E)}2-\eta(E+C(q))
+\frac{d/4-1/5}{L}\mathcal J(t)+\frac h2 A_e.
\]

Since \(d\ge9/10\), the coefficient \(d/4-1/5\) is at least \(1/40\). Also $\mathcal J(t)\ge t^2$ and \eqref{unbal:eq:opposite_corner:14} gives \(|hA_e/2|\le4q|t|/L\). Completing the square yields the uniform bound
\begin{equation}\label{unbal:eq:opposite_corner:20}
\frac{\mathcal J(t)}{40L}-\frac{4q|t|}{L}
\ge-\frac{160q^2}{L}\ge-240q^2.
\end{equation}

This is valid for every \(-1<t<1\), including arbitrarily unequal entropy ratios.

Expanding \(\Phi\) and using \eqref{unbal:eq:opposite_corner:17}--\eqref{unbal:eq:opposite_corner:19},
\[
B_{\mathrm{end}}-R_B
\ge\eta(E)-\eta(E+C(q))
+F(d,E)-\frac{F(d+q,E)+F(d-q,E)}2-240q^2
\]
\[
\ge\left(\frac{250}{441}-\frac1{10}-240E\right)\frac{q^2}{E}
\ge\frac{5003}{22050}\frac{q^2}{E}
\ge\frac{q^2}{5E}.
\]

When \(q=0\), the non-strict form of \eqref{unbal:eq:opposite_corner:2} follows from the same calculation. This proves Theorem~\ref{unbal:thm:opposite_corner:1}.
\end{proof}

Notice that the proof uses the maximum candidate at the actual children. It never asserts the generally false inequality \(\zeta\ge R_\psi\) throughout the domain.

\subsubsection{Proof of the complete mean-only corner}\label{unbal:module-opposite_corner-proof-of-the-complete-mean-only-corner}

\begin{proof}
Let \(w=a+(1-b)\le2^{-13}\). Then \(d=1-w\ge9/10\) and \(q\le w<1/10\). By concavity of entropy and the separate cap constraints,
\[
E\le\frac{H(a)+H(1-b)}2\le H(w/2)\le H(2^{-14}).
\]

For \(0<x<1\),
\[
H(x)\le x\log_2(1/x)+x/L,
\]
so $H(2^{-14})<16\cdot2^{-14}=1/1024<10^{-3}$.

If \(q\le32E\), Theorem~\ref{unbal:thm:opposite_corner:1} applies. It remains to handle \(q>32E\).

For completeness, here is the existing small-parent-entropy argument. Put \(x=q/E>32\). From \eqref{unbal:eq:opposite_corner:16} and \(C(q)\ge q^2/2\),
\[
\eta(E)-\eta(E+C(q))\ge\log_2(1+qx/2).
\]

For fixed \(x\), \(\ln(1+qx/2)/q\) decreases in \(q\). Since \(q\le1/10\),
\[
\ln(1+qx/2)\ge10q\ln(1+x/20).
\]

For \(x\ge32\),
\[
5\ln(1+x/20)\ge\ln(2+3x).
\]

At \(32\) this follows from \((13/5)^5\ge98\), and its derivative difference is \((12x-50)/[(20+x)(2+3x)]>0\). Using the \(F\) upper bound from Section~\ref{unbal:module-opposite_corner-the-parent-correction-and-the-radial-second-difference} gives
\[
\eta(E)-\eta(E+C(q))\ge2q\log_2(2+3q/E)\ge F(q,E).
\]

Thus \(\phi(\mathrm{parent})\ge\psi(\mathrm{parent})\), and \(R_B\le R_\phi\). Theorem~\ref{unbal:thm:opposite_corner:2} follows.
\end{proof}

The endpoint contact of Proposition~\ref{bal:prop:fixed-difference-minimum} exists throughout this corner, but its exactness at the original four-moment tuple is not assumed.

\clearpage
\subsection{Strong mean cost and practical uniform low-entropy theorem}
\label{unbal:app:low_information}

\emph{Argument guide.}
Two regional arguments share the same scalar estimates. The
low-information theorem combines a stronger mean-cost series with a
profile-slope comparison. The normalized low-entropy theorem treats the
remaining larger separations by balancing an endpoint logarithmic gain,
extended child curvature, and the parent correction. Its radius-to-entropy
restriction remains part of the statement.

\noindent\textit{Source:} \path{CK_CAP_REGION_CONTINUATION/prior_inputs/LOW_INFORMATION_AND_PRIOR_PROOFS.md}.\par
\subsubsection{Notation and inputs}\label{unbal:module-low_information-notation-and-inputs}

All entropies and costs are in bits. Set \(L=\ln 2\) and
\[
H(x)=-x\log_2x-(1-x)\log_2(1-x),\qquad
J(x)=\log_2\frac{1-x}{x},\qquad
j(x,y)=\frac12(x-y)(J(y)-J(x)).
\]

Write \(H^{-1}\) for the inverse on \([0,1/2]\), and define
\[
\eta(h)=(1-2H^{-1}(h))J(H^{-1}(h)),\quad P(I)=\eta(1-I),
\]
\[
F(s,h)=sJ(v),\qquad sH(v)=h(1-2v),\qquad F(0,h)=0,
\]
\[
\Phi(s,h)=\eta(h)-F(s,h),\quad
\phi(m,h)=\Phi(|1-2m|,h),\quad
\psi(m,h)=P(H(m)-h),\quad B=\max\{\phi,\psi\}.
\]

For \(A=\phi\), \(\psi\), or \(B\), let
\[
R_A=A((a+b)/2,(e+f)/2)-\frac{A(a,e)+A(b,f)}2.
\]

The domain is \(0<a,b<1\), \(0<e\le H(a)\), \(0<f\le H(b)\). Write
\[
m=(a+b)/2,\quad E=(e+f)/2,\quad d=|a-b|,\quad q=|1-a-b|,
\]
\[
I_u=H(a)-e,\quad I_w=H(b)-f,\quad s=(I_u+I_w)/2,
\]
\[
\Delta=H(m)-\frac{H(a)+H(b)}2,\qquad I=H(m)-E=\Delta+s.
\]

The unrestricted four-moment infimum is denoted \(\zeta\). We use the following results proved in this paper:

\begin{itemize}
\tightlist
\item
  Theorem~\ref{bal:thm:global-unrestricted-target}: \(\zeta\ge R_\phi\) on the unrestricted positive-entropy domain.
\item
  Theorem~\ref{bal:thm:four-moment-lower-bound} and Section~\ref{bal:imp:jointconvexity}: \(\zeta\ge L_4\ge F(d,E)\).
\item
  Proposition~\ref{bal:prop:fixed-difference-minimum}: the endpoint contact bound \(B_{\mathrm{end}}\), including its strict lower endpoint applicability test.
\item
  The decreasing-\(F_{ss}\) property used in Section~\ref{bal:aux:sec:fourpoint}, with the profile sign input in Section~\ref{bal:imp:e8}. It implies concavity of \(z\mapsto F(\sqrt z,h)\).
\end{itemize}

The earlier log-sum proof, included as \path{prior_inputs/LOGSUM_PROOF.md}, establishes independently that
\begin{equation}\label{unbal:eq:low_information:1}
\zeta\ge B_{\rm LS}:=j(a,b)+\frac{d^2s}{2V},\qquad
V=\max(a,b)[1-\min(a,b)].
\end{equation}

Convexity of \(P\) gives the entropy-split-free comparison
\begin{equation}\label{unbal:eq:low_information:2}
R_\psi\le D_\Delta(s):=P(\Delta+s)-P(s).
\end{equation}

The global target is $\zeta\ge R_B$. If $\phi$ is active at the parent,
then $R_B\le R_\phi$, and the unrestricted balanced inequality applies.
If $\psi$ is active there, then $R_B\le R_\psi$, and the comparisons
below apply. This division allows each lower bound to be used on the
parent branch for which it is needed.

\subsubsection{A stronger deterministic mean-cost estimate}\label{unbal:module-low_information-a-stronger-deterministic-mean-cost-estimate}

Using full label exchange and simultaneous complement, assume \(0<a<b<1\) and \(m\le1/2\). Put
\[
r=\frac{b-a}{a+b}\in(0,1),\qquad k=\frac{m}{1-m}\in(0,1],
\]
\[
C_n(x)=x\operatorname{atanh}x+\frac12\ln(1-x^2)
=\sum_{n\ge1}\frac{x^{2n}}{2n(2n-1)}.
\]

Exact binary-entropy identities give
\begin{equation}\label{unbal:eq:low_information:3}
L\Delta=m C_n(r)+(1-m)C_n(kr),
\end{equation}
\begin{equation}\label{unbal:eq:low_information:4}
Lj(a,b)=2m r\operatorname{atanh}r
+2(1-m)kr\operatorname{atanh}(kr).
\end{equation}

\begin{lemma}\label{unbal:thm:low_information:lem1}
For all distinct interior means in this orientation,
\begin{equation}\label{unbal:eq:low_information:5}
\boxed{j(a,b)\ge(4+r^2/2)\Delta.}
\end{equation}
\end{lemma}

\begin{proof}
First, coefficient comparison gives
\begin{equation}\label{unbal:eq:low_information:6}
2x\operatorname{atanh}x-4C_n(x)\ge\frac23 x^2C_n(x).
\end{equation}

Indeed, at power \(x^{2n}\), \(n\ge2\), the two coefficients are respectively
\[
\frac{2(n-1)}{n(2n-1)},\qquad
\frac1{3(n-1)(2n-3)}.
\]

The first is at least the second because
\[
6(n-1)^2(2n-3)-n(2n-1)
=(n-2)(12n^2-20n+9)\ge0.
\]

Next, positivity of the coefficients gives \(C_n(kr)\le k^2C_n(r)\). Consequently
\begin{equation}\label{unbal:eq:low_information:7}
\frac{m C_n(r)+(1-m)k^2 C_n(kr)}
{m C_n(r)+(1-m)C_n(kr)}
\ge\frac{1+k^3}{1+k}=1-k+k^2\ge\frac34.
\end{equation}

To see the first inequality, divide numerator and denominator by \(mC_n(r)\). The resulting ratio \((1+kz)/(1+z/k)\), with \(z=C_n(kr)/C_n(r)\), is nonincreasing in \(z\) and \(z\le k^2\). Apply \eqref{unbal:eq:low_information:6} to both terms in \eqref{unbal:eq:low_information:4}, then \eqref{unbal:eq:low_information:7}; the improvement over \(4\Delta\) is at least \((2/3)(3/4)r^2\Delta\). This proves \eqref{unbal:eq:low_information:5}. The \(a=b\) case follows separately with \(\Delta=j=0\).
\end{proof}

The deterministic entropy-cap comparison needed below is
\begin{equation}\label{unbal:eq:low_information:8}
\boxed{j(a,b)\ge P(\Delta).}
\end{equation}

For completeness, set \(C(x)=C_n(x)/L\). Equations \eqref{unbal:eq:low_information:3}--\eqref{unbal:eq:low_information:4} say
\[
\Delta=mC(r)+(1-m)C(kr),\qquad
j(a,b)=mP(C(r))+(1-m)P(C(kr)),
\]
because \(P(C(x))=2x\operatorname{atanh}(x)/L\). Jensen's inequality proves \eqref{unbal:eq:low_information:8}.

\subsubsection{All means at parent deficit \texorpdfstring{\(I\le1/100\)}{I <= 1/100}}\label{unbal:module-low_information-all-means-at-parent-deficit-i-1100}

\begin{theorem}\label{unbal:thm:low_information:1}
For every positive feasible entropy pair,
\begin{equation}\label{unbal:eq:low_information:9}
\boxed{I=H(m)-E\le\frac1{100}
\quad\Longrightarrow\quad B_{\rm LS}\ge R_\psi.}
\end{equation}

In particular, \(\max\{R_\phi,B_{\rm LS}\}\ge R_B\) throughout this region, so the established lower bounds imply \(\zeta\ge R_B\) there.
\end{theorem}

\paragraph{Uniform curvature on this interval}\label{unbal:module-low_information-uniform-curvature-on-this-interval}

Put \(\rho=1-2H^{-1}(1-I)\), \(A=\operatorname{atanh}(\rho)\), \(T=\rho^2\). Then
\[
I=C_n(\rho)/L\ge T/(2L),
\]
\begin{equation}\label{unbal:eq:low_information:10}
P''(I)=\frac{2L\{(1+\rho^2)A-\rho\}}
{(1-\rho^2)^2 A^3}.
\end{equation}

For \(I\le1/100\), the elementary bound \(L<347/500\) gives \(T<7/500\). The \(\operatorname{atanh}\) series yields
\[
(1+\rho^2)A-\rho
\le\frac43\rho^3+\frac8{15}\frac{\rho^5}{1-\rho^2},
\qquad A^3\ge\rho^3(1+\rho^2).
\]

Thus
\begin{equation}\label{unbal:eq:low_information:11}
P''(I)\le
\frac{2L[4/3+(8/15)T/(1-T)]}{(1-T)^2(1+T)}
\le\frac{344085200000}{182251021797}<\frac{19}{10}.
\end{equation}

For the second inequality, the numerator increases with \(T\) and the denominator decreases on \([0,7/500]\); use \(L<347/500\) and \(T\le7/500\). The continuous value $P''(0)=8L/3$ satisfies the same bound. Since \(P'(0)=4\),
\begin{equation}\label{unbal:eq:low_information:12}
P'(I)\le4+\frac{19}{10}I\le4.019.
\end{equation}

\paragraph{A slope bound for \texorpdfstring{\(r\le1/5\)}{r <= 1/5}}\label{unbal:module-low_information-a-slope-bound-for-r-15}

In the canonical orientation of Section~\ref{unbal:module-low_information-a-stronger-deterministic-mean-cost-estimate},
\[
V=b(1-a)=m(1-m)(1+r)(1+kr),\qquad d=2mr.
\]

Equation \eqref{unbal:eq:low_information:3} and \(C_n(kr)\le k^2C_n(r)\) give \(\Delta\le kC_n(r)/L\). Therefore
\begin{equation}\label{unbal:eq:low_information:13}
\frac{d^2}{2V\Delta}
\ge\frac{2Lr^2}{(1+r)(1+kr)C_n(r)}
\ge\frac{2Lr^2}{(1+r)^2 C_n(r)}.
\end{equation}

The series for \(C_n\) gives
\[
C_n(r)\le r^2/2+\frac{r^4}{12(1-r^2)}.
\]

For \(r\le1/5\), the increasing expression
\[
(1+r)^2\left(\frac12+\frac{r^2}{12(1-r^2)}\right)
\]
is at most \(145/200\). Combining with \eqref{unbal:eq:low_information:13} and \(L>69/100\) gives
\begin{equation}\label{unbal:eq:low_information:14}
\boxed{\frac{d^2}{2V\Delta}\ge\frac{80L}{29}>\frac{19}{10}.}
\end{equation}

\paragraph{Proof of Theorem~\ref{unbal:thm:low_information:1}}\label{unbal:module-low_information-proof-of-theorem-1}

\begin{proof}
For distinct means, there are two cases.

If \(r\ge1/5\), Lemma~\ref{unbal:thm:low_information:lem1} gives \(j\ge(4+1/50)\Delta=4.02\Delta\), whereas
\[
D_\Delta(s)=\int_s^{s+\Delta}P'(v)\,dv
\le4.019\Delta
\]
by \eqref{unbal:eq:low_information:12}. Hence \(j\) alone suffices.

If \(r\le1/5\), then for \(0\le v\le s\),
\[
D_\Delta'(v)=P'(v+\Delta)-P'(v)
\le\frac{19}{10}\Delta\le\frac{d^2}{2V}
\]
by \eqref{unbal:eq:low_information:11} and \eqref{unbal:eq:low_information:14}. Integrate from zero and use \eqref{unbal:eq:low_information:8}:
\[
D_\Delta(s)\le P(\Delta)+\frac{d^2s}{2V}
\le j(a,b)+\frac{d^2s}{2V}=B_{\rm LS}.
\]

Finally use \eqref{unbal:eq:low_information:2}. If \(a=b\), \(\Delta=0\), so \(R_\psi\le0=B_{\rm LS}\). The two valid symmetries transport the proof to arbitrary interior means.
\end{proof}

\subsubsection{A practical uniform normalized low-entropy cutoff}\label{unbal:module-low_information-a-practical-uniform-normalized-low-entropy-cutoff}

\begin{theorem}\label{unbal:thm:low_information:2}
For every positive feasible entropy pair,
\begin{equation}\label{unbal:eq:low_information:15}
\boxed{0<E\le10^{-6},\qquad q\le32E
\quad\Longrightarrow\quad
\max\{R_\phi,F(d,E),B_{\rm end}\text{ when applicable}\}\ge R_B.}
\end{equation}

There is no restriction on \(d\) or on \(e/f\). This is a theorem for the \textbf{normalized} region \(q\le32E\); it is not an unconditional lower cutoff on \(E\) for the entire feasible domain.
\end{theorem}

Theorem~\ref{unbal:thm:uniform_diagonal:1} proves the case \(d\le1/64\). We now prove the complementary case \(d\ge1/64\). In that case \(B_{\mathrm{end}}\) exists, and in the active-\(\psi\)-parent branch the proof gives the stronger margin
\begin{equation}\label{unbal:eq:low_information:16}
\boxed{B_{\rm end}-R_B\ge\frac{q^2}{100E}.}
\end{equation}

\paragraph{Endpoint existence and its logarithmic gain}\label{unbal:module-low_information-endpoint-existence-and-its-logarithmic-gain}

Here \(q\le32\cdot10^{-6}<1/64\le d\), so the two means straddle \(1/2\). Proposition~\ref{bal:prop:fixed-difference-minimum} applies: its lower threshold is
\[
D_0=\max(e,f)\left[\frac12-H^{-1}\left(\frac{\min(e,f)}{\max(e,f)}\right)\right]
\le E<d,
\]
and feasibility on opposite halves gives
\[
d\le1-H^{-1}(e)-H^{-1}(f).
\]

Thus there are unique \(p,u,v\), with \(0<p\le1\) and \(0<u,v<1/2\), satisfying
\[
d=p(1-u-v),\qquad e=pH(u),\qquad f=pH(v),
\]
and \(B_{\mathrm{end}}=p\,j(u,1-v)\) is a lower bound for \(\zeta\). Its use as a lower bound does not require the four-moment equality window.

Let \(p_{\mathrm{bar}}\) be the corresponding equal-entropy contact mass for \((d,E,E)\). The earlier endpoint proof establishes \(p\ge p_{\mathrm{bar}}\ge d\), by convexity of \(H^{-1}\) and monotonicity of the contact-difference function. Here
\[
\frac{e}{p_{\rm bar}},\frac{f}{p_{\rm bar}}
\le\frac{2E}{d}\le128\cdot10^{-6}<\frac1{100}.
\]

Put $t=(e-f)/(2E)$ and $\mathcal J(t)=-\ln(1-t^2)$. The logarithmic convexity estimate
\[
Q''(h)\ge\frac1{2Lh^2},\qquad
Q(h)=J(H^{-1}(h)),\quad 0<h\le1/100,
\]
proved in Section~\ref{unbal:module-opposite_corner-the-new-endpoint-entropy-imbalance-bound}, gives
\begin{equation}\label{unbal:eq:low_information:17}
\boxed{B_{\rm end}\ge F(d,E)+\frac{d}{4L}\mathcal J(t).}
\end{equation}

This estimate is uniform as either entropy ratio tends to zero.

\paragraph{Radius-dependent semiconvexity}\label{unbal:module-low_information-radius-dependent-semiconvexity}

Define the exact constants
\[
A_0=1-\frac{64}{511},\qquad
S_0=\left(\frac{1151}{1023}\frac{1024}{1023}\right)^2.
\]

For \(0<h\le2E\) and \(0\le z\le1/8\), the global derivative bound from the earlier uniform proof gives
\begin{equation}\label{unbal:eq:low_information:18}
\Phi_{hh}(z,h)\ge\frac{1/2-4z}{Lh^2}\ge0.
\end{equation}

For \(1/8\le z\le1\) and \(0<h\le2E\), the estimates used in the opposite-corner proof sharpen to
\begin{equation}\label{unbal:eq:low_information:19}
\boxed{\Phi_{hh}(z,h)\ge\frac{A_0-S_0z}{Lh^2}.}
\end{equation}

Here are the details of the range and the retained radius factor. With \(v_8=(1+\exp(8))^{-1}\), the earlier elementary bounds give \(H(v_8)>3/1024\). We have $h\le2\cdot10^{-6}<H(v_8)$ and $h/z\le16\cdot10^{-6}<H(v_8)/(1-2v_8)$. Both the entropy inverse and the \(F\)-contact therefore have logit greater than \(8\) and contact coordinate less than \(1/1024\). Equations \eqref{unbal:eq:opposite_corner:11}--\eqref{unbal:eq:opposite_corner:12} of that proof give
\[
h^2\eta''(h)\ge A_0/L,\qquad
zF_{zz}(z,h)\le S_0/L.
\]

Homogeneity implies \(h^2F_{hh}=z^2F_{zz}\), yielding \eqref{unbal:eq:low_information:19}. These inequalities concern the analytic extension of \(\Phi\); \(h\) need not lie below the entropy cap associated with \(z\).

The child radii are \(z_1=d+q\) and \(z_2=d-q\). Put
\[
\kappa=\max\{0,S_0(d+q)-A_0\}.
\]

Since \(d\le1\) and \(S_0>A_0\),
\[
\kappa\le(S_0-A_0)d+S_0q.
\]

Also \(q/d\le32/15625\), and exact rational arithmetic gives
\[
(S_0-A_0)+S_0\frac{32}{15625}<\frac25.
\]

Thus \(\kappa\le2d/5\), and \eqref{unbal:eq:low_information:18}--\eqref{unbal:eq:low_information:19} imply, for either child radius and every \(0<h\le2E\),
\begin{equation}\label{unbal:eq:low_information:20}
\Phi_{hh}(z_i,h)\ge-\frac{\kappa}{Lh^2}.
\end{equation}

\paragraph{Absorbing an arbitrary entropy split}\label{unbal:module-low_information-absorbing-an-arbitrary-entropy-split}

Match entropies to the two child radii, writing them \(E(1+t)\), \(E(1-t)\). This can reverse the sign of \(t\), which is immaterial below. The function \(\Phi(z_i,h)-(\kappa/L)\ln h\) is convex by \eqref{unbal:eq:low_information:20}. Its supporting line at \(E\) gives
\begin{equation}\label{unbal:eq:low_information:21}
\frac{\Phi(z_1,e_1)+\Phi(z_2,e_2)}2
\ge\frac{\Phi(z_1,E)+\Phi(z_2,E)}2
+\frac{Et}{2}A_e-\frac{\kappa}{2L}\mathcal J(t),
\end{equation}
where \(A_e=\Phi_h(z_1,E)-\Phi_h(z_2,E)\). The global bound \(0\le\Phi_{zh}\le4/(Lh)\) in \eqref{unbal:eq:opposite_corner:10} gives
\begin{equation}\label{unbal:eq:low_information:22}
|A_e|\le\frac{8q}{LE}.
\end{equation}

Combine \eqref{unbal:eq:low_information:17} and \eqref{unbal:eq:low_information:21}. Their remaining entropy-split contribution is at least
\begin{equation}\label{unbal:eq:low_information:23}
\frac{d/4-\kappa/2}{L}\mathcal J(t)-\frac{4q|t|}{L}
\ge\frac{d}{20L}t^2-\frac{4q|t|}{L}
\ge-\frac{80q^2}{Ld}\ge-\frac{120q^2}{d}.
\end{equation}

This uses \(\kappa\le2d/5\), $-\ln(1-t^2)\ge t^2$, completion of a square, and \(L>2/3\). No truncation of the entropy ratio has occurred.

\paragraph{Completing the parent comparison}\label{unbal:module-low_information-completing-the-parent-comparison}

Let \(C(q)=1-H((1-q)/2)\). The earlier uniform proof establishes, throughout \(E\le10^{-6}\) and \(q\le32E\),
\begin{equation}\label{unbal:eq:low_information:24}
\eta(E)-\eta(E+C(q))\ge\frac{50000}{49049}\frac{q^2}{E}.
\end{equation}

For reference, \(C(q)\le q^2/[2L(1-q^2)]\) implies \(C(q)/E<1/1000\). Integrate \(P'(1-h)\ge2+1/(Lh)\), use \(\ln(1+x)\ge x/(1+x)\), \(C(q)\ge q^2/(2L)\), and \(L<7/10\) to get \eqref{unbal:eq:low_information:24}.

The radial concavity input, together with \(d/E\ge15625>3\) and the earlier exact scalar certificate \(f'(3)/6<1\) for \(f(x)=F(x,1)\), gives
\begin{equation}\label{unbal:eq:low_information:25}
\frac{F(d+q,E)+F(d-q,E)}2-F(d,E)\le\frac{q^2}{E}.
\end{equation}

If \(\phi\) is active at the parent, \(R_B\le R_\phi\). Otherwise \(B(\mathrm{parent})=\eta(E+C(q))\) and \(B(\mathrm{children})\ge\phi(\mathrm{children})\). Equations \eqref{unbal:eq:low_information:17}, \eqref{unbal:eq:low_information:21}, and \eqref{unbal:eq:low_information:23}--\eqref{unbal:eq:low_information:25} give
\[
B_{\rm end}-R_B
\ge\left(\frac{50000}{49049}-1-\frac{120E}{d}\right)\frac{q^2}{E}
\ge\frac{q^2}{100E},
\]
because
\[
\frac{50000}{49049}-1-120\frac{64}{10^6}>\frac1{100}.
\]

This proves \eqref{unbal:eq:low_information:16}, hence completes Theorem~\ref{unbal:thm:low_information:2} with the prior \(d\le1/64\) theorem. At \(q=0\) the displayed lower margin is zero; the same argument still proves the desired non-strict inequality.

\clearpage
\subsection{Scalar formulas underlying the stronger endpoint and mixed-derivative estimates}
\label{unbal:app:collar_scalar}

\emph{Argument guide.}
A logarithmic curvature bound strengthens the endpoint gain. Under the
stated local hypotheses, mixed-derivative estimates control the effect
of varying the child entropies, while radial and parent estimates place
all terms on a common scale for the later collar arguments.

\noindent\textit{Source:} \path{CK_SMALL_RATIO_EXTENSION/prior_inputs/ANALYTIC_COLLAR.md}.\par
\subsubsection{A stronger logarithmic entropy curvature bound}\label{unbal:module-collar_scalar-a-stronger-logarithmic-entropy-curvature-bound}

Let \(Q(h)=J(H^{-1}(h))\), where \(J(v)=\log_2((1-v)/v)\). We prove
\begin{equation}\label{unbal:eq:collar_scalar:4}
\boxed{Q''(h)\ge\frac1{Lh^2}\qquad(0<h\le1/4).}
\end{equation}

Put \(v=H^{-1}(h)\), \(\ell=\ln((1-v)/v)\), \(r=1-2v\), and \(t=1/\ell\). Differentiation gives
\[
Q''(h)=\frac{L(r\ell-1)}{v^2(1-v)^2\ell^3}.
\]

The exact logarithmic comparisons in \path{EXACT_COLLAR_CHECKER.py} give
\[
\ln19<3<\ln21,\qquad H(1/22)>1/4.
\]

Consequently \(h\le1/4\) implies \(\ell>3\) and \(v<1/20\). Since
\[
Lh=v\ell-\ln(1-v)\ge v(\ell+1),
\]
we have
\[
Lh^2 Q''(h)\ge\frac{(1-2v-t)(1+t)^2}{(1-v)^2}.
\]

The numerator minus \((1-v)^2\) is
\[
t[1-t-t^2-4v-2vt]-v^2.
\]

Also \(\ell v\le1\), by \(\ln z\le z-1\), so \(v^2/t\le v\). Thus the last expression is at least
\[
t[1-t-t^2-5v-2vt]
\ge t\left[1-\frac13-\frac19-\frac14-\frac1{30}\right]
=\frac{49}{180}t>0.
\]

This proves \eqref{unbal:eq:collar_scalar:4}.

\subsubsection{Endpoint gain for every entropy split}\label{unbal:module-collar_scalar-endpoint-gain-for-every-entropy-split}

The following are sufficient local hypotheses for the estimates in this
subsection and the subsequent mean-interval and parent comparisons.  For
a feasible tuple with ordered means $a\le b$ and positive entropies $e,f$,
put $d=b-a$, $q=|1-a-b|$, and $E=(e+f)/2$, and assume
\begin{equation}\label{unbal:eq:collar-local-domain}
 0<E\le\frac18,\qquad q\le2E,\qquad d\ge8E.
\end{equation}
These hypotheses imply $q<d$, so the means lie on opposite sides of $1/2$.
For the strict cross-half endpoint existence test in
Proposition~\ref{bal:prop:fixed-difference-minimum}, feasibility gives
$d\le1-H^{-1}(e)-H^{-1}(f)$.  In bit units its lower threshold is
\[
 D_{\mathrm{cross}}:=\max(e,f)\left\{\frac12-
 H^{-1}\!\left(\frac{\min(e,f)}{\max(e,f)}\right)\right\}
 \le\frac{\max(e,f)}2\le E<d.
\]
Thus the strict cross-half contact exists.  This $D_{\mathrm{cross}}$ is distinct from
the lower endpoint $|H^{-1}(e)-H^{-1}(f)|$ for general oriented contacts.

Let \(p\) be the endpoint contact mass and \(p_{\mathrm{bar}}\) the equal-entropy contact mass for \((d,E,E)\). The previous endpoint proof gives \(p\ge p_{\mathrm{bar}}\ge d\). Hence
\[
e/p_{\rm bar},f/p_{\rm bar}\le2E/d\le1/4.
\]

Apply \eqref{unbal:eq:collar_scalar:4} at this common mass, using convexity of \(Q(h)+(1/L)\ln h\). With
\[
\tau=(e-f)/(2E),\qquad \mathcal J(\tau)=-\ln(1-\tau^2),
\]
the result is
\begin{equation}\label{unbal:eq:collar_scalar:5}
\boxed{B_{\rm end}\ge F(d,E)+\frac{d}{2L}\mathcal J(\tau).}
\end{equation}

This doubles the earlier logarithmic coefficient and uses the larger entropy interval \(h\le1/4\).

\subsubsection{A mixed derivative bound along the mean interval}\label{unbal:module-collar_scalar-a-mixed-derivative-bound-along-the-mean-interval}

For the \(F\)-contact, write \(\ell=\ln((1-v)/v)\). The previously derived exact derivative estimates imply
\begin{equation}\label{unbal:eq:collar_scalar:6}
sF_{ss}(s,h)\le\frac1L
\frac{(1+1/((1-v)\ell))^2}{(1-v)^2}.
\end{equation}

Whenever \(s/h\ge6\), the contact has \(\ell>3\) and \(v<1/20\): at \(\ell=3\) its ratio \((1-2v)/H(v)\) is less than \(4\), since \(H(v)>1/4\). Therefore
\begin{equation}\label{unbal:eq:collar_scalar:7}
sF_{ss}(s,h)\le\frac1L\left(\frac{1540}{1083}\right)^2<3.
\end{equation}

The final strict inequality follows from \(L>69/100\) and is checked exactly. Homogeneity gives \(\Phi_{sh}=(s/h)F_{ss}\), so
\begin{equation}\label{unbal:eq:collar_scalar:8}
0\le\Phi_{sh}(s,h)\le3/h\qquad(s/h\ge6).
\end{equation}

The child radii are \(s_1=d+q\) and \(s_2=d-q\), and every radius between them obeys \(s/E\ge d/E-q/E\ge6\). Thus
\begin{equation}\label{unbal:eq:collar_scalar:9}
|A_e|:=|\Phi_h(s_1,E)-\Phi_h(s_2,E)|\le6q/E.
\end{equation}

\subsubsection{Radial and parent estimates}\label{unbal:module-collar_scalar-radial-and-parent-estimates}

Put \(f(x)=F(x,1)\). The scalar constant
\begin{equation}\label{unbal:eq:collar_scalar:11}
f'(8)/16<1/2
\end{equation}
is certified with exact rational logarithm-series bounds in the accompanying checker. Concavity of \(z\mapsto F(\sqrt z,E)\), together with \(d/E\ge8\), yields
\begin{equation}\label{unbal:eq:collar_scalar:12}
\frac{F(d+q,E)+F(d-q,E)}2-F(d,E)\le\frac{q^2}{2E}.
\end{equation}

Let \(C(q)=1-H((1-q)/2)\). Recall
\[
\frac{q^2}{2L}\le C(q)\le\frac{q^2}{2L(1-q^2)},\quad
\eta(E)-\eta(E+C)\ge2C+\log_2(1+C/E).
\]

As \(q\le2E\le1/4\), we have \(C(q)/E\le16E/5\). Consequently, for \(q>0\),
\[
\frac{E}{q^2}[\eta(E)-\eta(E+C(q))]
\ge\frac{10}{7}E+\frac{50}{49(1+16E/5)}.
\]

The last function decreases on \([0,1/8]\): its derivative is at most
\[
\frac{10}{7}-\frac{4000}{2401}<0.
\]

Therefore
\begin{equation}\label{unbal:eq:collar_scalar:13}
\boxed{\eta(E)-\eta(E+C(q))
\ge\left(\frac5{28}+\frac{250}{343}\right)\frac{q^2}{E}.}
\end{equation}

At \(q=0\) this holds directly with both sides zero.

\clearpage
\subsection{Stronger endpoint log gain and the three inherited collars}
\label{unbal:app:sharper_collars}

\emph{Argument guide.}
The three rows use one common calculation with different ratio
thresholds. Stronger endpoint curvature absorbs the entropy imbalance,
and mixed-derivative and radial estimates bound the two losses. Each row
then supplies its own parent correction; subtracting the losses leaves
the listed margin under the stated common-cap hypothesis.

\noindent\textit{Source:} \path{CK_NO_SEPARATION_EXTENSION/prior_inputs/SHARPER_COLLARS.md}.\par
The estimates below strengthen \path{ANALYTIC_COLLAR.md}. They are analytic continuum proofs, with exact rational verification of their fixed constants. They use the established results collected in Section~\ref{unbal:sec:inputs}.

Write \(E=(e+f)/2\), \(d=|a-b|\), \(q=|1-a-b|\), \(B=\max(\phi,\psi)\), and \(B_{\mathrm{end}}\) for Proposition~\ref{bal:prop:fixed-difference-minimum}'s endpoint lower bound.

\begin{theorem}\label{unbal:thm:sharper_collars:1}
Assume positive feasible entropies and
\[
E\le\min\{H(a),H(b)\}.
\]

Each row below proves \(\max\{R_\phi,B_{\mathrm{end}}\}\ge R_B\) for every entropy split. In the branch where \(\psi\) is active at the parent, the stronger listed margin holds.

\begin{longtable}[]{@{}
  >{\raggedright\arraybackslash}p{(\columnwidth - 2\tabcolsep) * \real{0.5000}}
  >{\raggedright\arraybackslash}p{(\columnwidth - 2\tabcolsep) * \real{0.5000}}@{}}
\toprule\noalign{}
\begin{minipage}[b]{\linewidth}\raggedright
Mean/entropy region
\end{minipage} & \begin{minipage}[b]{\linewidth}\raggedright
Lower bound on \(B_{\mathrm{end}}-R_B\) in the \(\psi\)-parent branch
\end{minipage} \\
\midrule\noalign{}
\endhead
\bottomrule\noalign{}
\endlastfoot
\(q\le E\) and \(d\ge5E\) & \(q^2/(1000E)\) \\
\(q\le2E\) and \(d\ge6E\) & \(q^2/(10E)\) \\
\(q\le4E\) and \(d\ge8E\) & \(q^2/(50E)\) \\
\end{longtable}

There is no extra small-entropy cutoff and no bound on the entropy ratio. The shared-entropy cap condition is part of the theorem. The inequalities include \(q=0\) with zero guaranteed margin.
\end{theorem}

\subsubsection{Logarithmic curvature up to entropy \texorpdfstring{\(19/50\)}{19/50}}\label{unbal:module-sharper_collars-logarithmic-curvature-up-to-entropy-1950}

We strengthen equation \eqref{unbal:eq:collar_scalar:4} of \path{ANALYTIC_COLLAR.md} to
\begin{equation}\label{unbal:eq:sharper_collars:1}
Q''(h)\ge\frac1{Lh^2}\qquad(0<h\le19/50),
\quad Q(h)=J(H^{-1}(h)),\quad L=\ln2.
\end{equation}

Use the same variables \(v=H^{-1}(h)\), \(\ell=\ln((1-v)/v)\), \(t=1/\ell\). The checker proves
\[
H(3/40)>19/50,\qquad 5/2<\ln(37/3)<13/5.
\]

Thus \(v<3/40\) and \(t<2/5\). As in the earlier proof, the relevant numerator difference is
\[
t[1-t-t^2-4v-2vt-v^2/t].
\]

The function \(v^2\ln((1-v)/v)\) increases on \((0,3/40]\), since its derivative is \(v[2\ell-1/(1-v)]>0\) there. Hence
\[
v^2/t\le\frac9{1600}\ln(37/3)<\frac{117}{8000}.
\]

The bracket is consequently at least
\[
1-\frac25-\frac4{25}-\frac3{10}-\frac3{50}-\frac{117}{8000}
=\frac{523}{8000}>0.
\]

This proves \eqref{unbal:eq:sharper_collars:1}.

For an equal-entropy \(F\)-contact at \(x=d/E\ge5\), its lower coordinate \(u\) is at most \(u_5\), the contact at \(x=5\). The exact inequality \(5H(1/40)<19/20\) gives \(u_5>1/40\), and therefore
\[
H(u)\le H(u_5)=\frac{1-2u_5}{5}<19/100.
\]

Let \(p_{\mathrm{bar}}\) be the equal-entropy contact mass. Then \(E/p_{\mathrm{bar}}=H(u)\), so \(e/p_{\mathrm{bar}},f/p_{\mathrm{bar}}<19/50\) for every split. The earlier contact-mass comparison \(p\ge p_{\mathrm{bar}}\) and \eqref{unbal:eq:sharper_collars:1} prove
\begin{equation}\label{unbal:eq:sharper_collars:2}
\boxed{B_{\rm end}\ge F(d,E)+\frac d{2L}\mathcal J(\tau),
\qquad \mathcal J(\tau)=-\ln(1-\tau^2).}
\end{equation}

The endpoint exists throughout all three rows: \(d>E\), \(q<d\), and feasibility gives the upper endpoint test.

\subsubsection{The improved mixed derivative bound}\label{unbal:module-sharper_collars-the-improved-mixed-derivative-bound}

We prove
\begin{equation}\label{unbal:eq:sharper_collars:3}
\boxed{sF_{ss}(s,h)<13/6\qquad(s/h\ge4).}
\end{equation}

At the contact put \(r=1-2v\), \(\nu=1-r^2\), \(h_n=LH(v)\), and \(K=\ell/2-\ln(1-v)\). The exact formula is
\[
sF_{ss}=\frac{2r h_n^2(2K-r^2)}{L\nu^2K^3}.
\]

For \(\ell\ge3\), the previous estimates give \(v\le1/20\), \(r\ge9/10\), and
\[
K\le\ell/2+1/19\le59\ell/114,
\qquad
\frac{h_n}{\nu K}\le\frac{1+1/((1-v)\ell)}{2(1-v)}.
\]

Retaining the factor previously discarded from the derivative formula yields
\begin{equation}\label{unbal:eq:sharper_collars:4}
sF_{ss}\le\frac1L\left(\frac{20}{19}\right)^2
\left(1+\frac{20}{19\ell}\right)^2
\left(1-\frac{4617}{5900\ell}\right).
\end{equation}

For \(\alpha=20/19\) and \(c=4617/5900\), the cubic $(1+\alpha t)^2(1-ct)$ increases on \(0\le t\le1/3\). Its derivative is decreasing on this interval and remains positive at \(1/3\). These are exact polynomial comparisons in the checker. Using \(L>69/100\), \eqref{unbal:eq:sharper_collars:4} gives
\[
sF_{ss}\le\frac{10342547600}{4774831119}<13/6.
\]

Finally, \(s/h\ge4\) implies \(\ell>3\), because the contact ratio at \(\ell=3\) is less than \(4\). This proves \eqref{unbal:eq:sharper_collars:3}.

In each theorem row, every radial coordinate between \(d-q\) and \(d+q\) satisfies \(s/E\ge4\). Homogeneity and \eqref{unbal:eq:sharper_collars:3} imply
\begin{equation}\label{unbal:eq:sharper_collars:5}
|\Phi_h(d+q,E)-\Phi_h(d-q,E)|\le\frac{13q}{3E}.
\end{equation}

\subsubsection{Entropy split and radial losses}\label{unbal:module-sharper_collars-entropy-split-and-radial-losses}

By Theorem~\ref{bal:aux:thm:entropyconvexity}, including its continuous
extension to the upper entropy cap, the shared-entropy cap assumption
makes the two convex supporting lines at $E$ valid on the entire feasible
split interval.  If $E$ itself is a cap, the derivative is the left
derivative specified in that theorem. Combining those lines with \eqref{unbal:eq:sharper_collars:2} and \eqref{unbal:eq:sharper_collars:5} gives a split loss of at most
\begin{equation}\label{unbal:eq:sharper_collars:6}
\frac{169Lq^2}{72d}\le\frac{1183q^2}{720d}.
\end{equation}

This follows by minimizing \(d\tau^2/(2L)-(13/6)q|\tau|\); no entropy ratio is truncated.

For \(f(x)=F(x,1)\), exact scalar certificates give
\begin{equation}\label{unbal:eq:sharper_collars:7}
f'(5)/10<69/100,\qquad f'(6)/12<3/5,
\qquad f'(8)/16<1/2.
\end{equation}

Radial concavity therefore bounds the central \(F\) difference by the corresponding coefficient times \(q^2/E\) in each row.

\subsubsection{Parent correction in the three charts}\label{unbal:module-sharper_collars-parent-correction-in-the-three-charts}

Let \(C(q)=1-H((1-q)/2)\). The estimates from the previous note and \(\ln(1+x)\ge2x/(2+x)\) give
\begin{equation}\label{unbal:eq:sharper_collars:8}
\frac{E}{q^2}[\eta(E)-\eta(E+C(q))]
\ge\frac{10E}{7}+\frac{50}{49(1+C(q)/(2E))}.
\end{equation}

As usual \(q=0\) follows directly without division. Geometry gives \(q+d\le1\).

\textbf{First row.} Here \(q\le E\) and \(d\ge5E\) imply \(q\le1/6\). Thus \(C(q)/E\le27E/35\). The lower bound
\[
10E/7+\frac{50}{49(1+27E/70)}
\]
increases for \(E\ge0\), and is at least \(50/49\).

\textbf{Second row.} Here \(q\le2E\) and \(d\ge6E\) imply \(q\le1/4\), so \(C(q)/E\le16E/5\). The expression
\[
10E/7+\frac{50}{49(1+8E/5)}
\]
is at least \(1\). After multiplication by its positive denominator, this reduces to
\[
\frac{16}{7}E^2-\frac6{35}E+\frac1{49}>0.
\]

Its minimum is \(1/49-9/2800>0\).

\textbf{Third row.} Here \(q\le4E\) and \(d\ge8E\) imply \(q\le1/3\) and \(E\le1/8\). Thus \(C(q)/E\le27E/2\). The expression
\[
10E/7+\frac{50}{49(1+27E/4)}
\]
decreases on \([0,1/8]\), giving the lower bound \(5/28+1600/2891\).

\subsubsection{Completion and scope}\label{unbal:module-sharper_collars-completion-and-scope}

In the \(\psi\)-parent branch, subtract the radial and split losses from these parent corrections. The resulting lower coefficients of \(q^2/E\) are respectively
\[
\frac{50}{49}-\frac{69}{100}-\frac{1183}{3600}
=\frac{317}{176400}>\frac1{1000},
\]
\[
1-\frac35-\frac{1183}{4320}=\frac{109}{864}>\frac1{10},
\]
\[
\frac5{28}+\frac{1600}{2891}-\frac12-\frac{1183}{5760}
=\frac{443467}{16652160}>\frac1{50}.
\]

If \(\phi\) is active at the parent, Theorem~\ref{bal:thm:global-unrestricted-target} proves the target directly. This completes all three rows.

Thus the balanced mean-sum seam has a positive-entropy collar for
$d/E\ge5$ under the stated shared-entropy cap condition. The adjacent
ratio range is treated in the following transverse-strip argument.
Section~\ref{unbal:app:central_assembly} assembles the central regions,
and Section~\ref{unbal:sec:assembly} supplies the global assembly.

\clearpage
\subsection{Global mixed bound and the ratio strip from four to five}
\label{unbal:app:global_mixed_strip}

\emph{Argument guide.}
The ratio strip requires quantitative child curvature in addition to
the endpoint gain. A global mixed-derivative bound supplies that
curvature, after analytic tails and a compact scalar certificate are
joined. Completing the square controls the entropy split, and comparison
with the parent and radial terms closes the strip.

\noindent\textit{Source:} \path{CK_NO_SEPARATION_EXTENSION/prior_inputs/TRANSVERSE_STRIP.md}.\par
Use the definitions and established estimates in the preceding collar argument, whose source is \path{SHARPER_COLLARS.md}.

\begin{theorem}\label{unbal:thm:global_mixed_strip:1}
For every positive feasible entropy pair, if
\[
\boxed{0<E\le11/200,\qquad 4E\le d\le5E,
\qquad q\le E/100,}
\]
then \(\max\{R_\phi,B_{\mathrm{end}}\}\ge R_B\). In the \(\psi\)-parent branch,
\[
\boxed{B_{\rm end}-R_B\ge q^2/(100E).}
\]

No entropy-ratio restriction is imposed. Here the common-entropy cap condition is automatic: \(d+q\le5511/20000\), and \(H((1-d-q)/2)\ge1-d-q>2E\).
\end{theorem}

\subsubsection{A global mixed-derivative bound}\label{unbal:module-global_mixed_strip-a-global-mixed-derivative-bound}

We use the strengthened global statement
\begin{equation}\label{unbal:eq:global_mixed_strip:1}
\boxed{sF_{ss}(s,h)\le13/6\qquad(s,h>0).}
\end{equation}

The contact coordinate \(v\) ranges over \((0,1/2)\). For \(v\le1/22\), its logit exceeds \(3\), so the analytic proof in Section~\ref{unbal:module-sharper_collars-the-improved-mixed-derivative-bound} applies. For \(v\ge1/3\), the exact formula and \(h_n\le\nu K\) give \(sF_{ss}\le4r/L\le4/(3L)<13/6\).

On the remaining compact interval \(1/22\le v\le1/3\), \path{GLOBAL_MIXED_CHECKER.py} certifies the exact expression
\[
\begin{gathered}\frac{2r h_n^2(2K-r^2)}{L\nu^2K^3},\quad
r=1-2v,\quad\nu=4v(1-v),\\
h_n=-v\ln v-(1-v)\ln(1-v),\quad K=-\tfrac12\ln(v(1-v)),\end{gathered}
\]
below \(13/6\) on every accepted interval. The 70-digit directed-interval output includes an exact rational contiguous partition of the full compact interval. Together with the analytic tails, this proves \eqref{unbal:eq:global_mixed_strip:1}. This is a one-dimensional functional certificate, not sampling in the CK moment domain.

\subsubsection{An endpoint gain for \texorpdfstring{\(d/E\ge4\)}{d/E >= 4}}\label{unbal:module-global_mixed_strip-an-endpoint-gain-for-de-4}

On \(0<h\le37/80\),
\begin{equation}\label{unbal:eq:global_mixed_strip:2}
Q''(h)\ge\frac9{10Lh^2}.
\end{equation}

Indeed \(H(1/10)>37/80\), so \(v=H^{-1}(h)<1/10\) and \(\ell=\ln((1-v)/v)>\ln9>2197/1000\). Put \(t=1/\ell\le1000/2197\). The entropy calculation used in the earlier proofs gives
\[
Lh^2Q''(h)\ge\frac{(1-2v-t)(1+t)^2}{(1-v)^2}
\ge\frac{(4/5-t)(1+t)^2}{81/100}.
\]

The second inequality follows because \((1-2v-t)/(1-v)^2\) decreases in \(v\). The cubic in \(t\) has its minimum on this interval at an endpoint: its derivative has the sign of \(3/5-3t\). Both endpoint values after division by \(81/100\) exceed \(9/10\), as checked exactly.

At the equal-entropy contact for \(x=d/E\ge4\), \(u\le u_4\) and the exact inequality \(4H(3/80)<37/40\) gives \(u_4>3/80\). Thus \(H(u)\le H(u_4)<37/160\). At the common mass \(p_{\mathrm{bar}}\), both entropy arguments are consequently below \(37/80\). Applying \eqref{unbal:eq:global_mixed_strip:2} and the existing \(p\ge p_{\mathrm{bar}}\) comparison yields
\begin{equation}\label{unbal:eq:global_mixed_strip:3}
\boxed{B_{\rm end}\ge F(d,E)+\frac{9d}{20L}\mathcal J(\tau).}
\end{equation}

\subsubsection{Quantitative entropy curvature of the children}\label{unbal:module-global_mixed_strip-quantitative-entropy-curvature-of-the-children}

The previously proved $\eta''(h)\ge1/(2Lh^2)$, homogeneity, and \eqref{unbal:eq:global_mixed_strip:1} imply
\[
\Phi_{hh}(z,h)\ge\frac{1/(2L)-(13/6)z}{h^2}.
\]

For either child radius \(z=d+q\) or \(d-q\), put
\[
\gamma=1/(2L)-(13/6)(d+q)>0.
\]

Convexity of \(\Phi(z,h)+\gamma\ln h\) and supporting lines at \(E\) give
\[
\frac{\Phi(d+q,e_1)+\Phi(d-q,e_2)}2
\ge\frac{\Phi(d+q,E)+\Phi(d-q,E)}2
+\frac{E\tau}{2}A_e+\frac\gamma2\mathcal J(\tau),
\]
where \(|A_e|\le13q/(3E)\) by \eqref{unbal:eq:global_mixed_strip:1}. This remains uniform as either entropy tends to zero.

Combining with \eqref{unbal:eq:global_mixed_strip:3}, the coefficient of $\mathcal J$ is
\[
c=\frac1{4L}+d\left(\frac9{20L}-\frac{13}{12}\right)-\frac{13q}{12}
\ge\frac5{14}-\frac{185E}{84}-\frac{13E}{1200}
\ge c_*:=\frac{131833}{560000}>0.
\]

The split loss is therefore at most \(Kq^2\), where
\[
K=\frac{169}{144c_*}.
\]

\subsubsection{Parent and radial comparison}\label{unbal:module-global_mixed_strip-parent-and-radial-comparison}

Here \(C(q)/E\le E/10000\le11/2000000\). The earlier parent estimate gives
\[
\frac{E}{q^2}[\eta(E)-\eta(E+C(q))]
\ge b_*+10E/7,\qquad
b_*:=\frac{50}{49(1+11/2000000)}.
\]

The exact scalar certificate \(f'(4)/8<81/100\) and radial concavity bound the central \(F\) difference by \(81q^2/(100E)\). Consequently, in the \(\psi\)-parent branch,
\[
B_{\rm end}-R_B
\ge\left[b_*-\frac{81}{100}-(K-10/7)E\right]\frac{q^2}{E}
\ge\frac{q^2}{100E}.
\]

The final exact lower coefficient is
\[
\frac{3305965240861}{223610279849775}>1/100.
\]

The \(\phi\)-parent branch follows from \(R_\phi\). At \(q=0\) the argument gives the non-strict target directly.

Combining this theorem with the first row of \path{SHARPER_COLLARS.md} covers \(q\le E/100\), \(E\le11/200\), \(d\ge4E\) whenever \(E\le\min\{H(a),H(b)\}\). The combined guaranteed margin in the \(\psi\)-parent branch is at least \(q^2/(1000E)\). The other regions enter the global assembly in Section~\ref{unbal:sec:assembly}.

\clearpage
\subsection{Global mean-anchor matrix and inherited quantitative cap rectangle}
\label{unbal:app:cap_matrix}

\emph{Argument guide.}
A supporting plane at the means gives a lower bound affine in the
two entropy deficits. Profile convexity first removes their split using
the smaller plane coefficient; strong convexity gives a sharper quadratic
alternative. Concavity in the total deficit then reduces a whole cap
interval or entropy slab to endpoint inequalities in the means.

\noindent\textit{Source:} \path{CK_CAP_REGION_CONTINUATION/PROOF.md}.\par
\subsubsection{Notation and inputs}\label{unbal:module-cap_matrix-notation-and-inputs}

Write
\[
\begin{gathered}H(x)=-x\log_2x-(1-x)\log_2(1-x),\quad J(x)=\log_2((1-x)/x),\\
j(a,b)=\tfrac12(b-a)(J(a)-J(b)),\quad L=\ln2.\end{gathered}
\]
Let \(\zeta(a,b,e,f)\) be the unrestricted infimum of \(\mathbb E\,j(U,W)\) with the two prescribed means and entropy moments. Set
\[
x=H(a)-e,\quad y=H(b)-f,\quad s=(x+y)/2,\quad m=(a+b)/2,
\]
\[
\Delta=H(m)-\tfrac12(H(a)+H(b)),\quad E=(e+f)/2,\quad \Delta+s=H(m)-E.
\]
The entropy profile is
\[
P(I)=\eta(1-I),\qquad \eta(h)=(1-2H^{-1}(h))J(H^{-1}(h)),
\]
where \(H^{-1}\) is the lower inverse and \(P(0)=0\). Thus
\[
R_\psi=P(\Delta+s)-\tfrac12(P(x)+P(y)).
\]
Section~\ref{unbal:app:profile_logsum} proves
\begin{equation}\label{unbal:eq:cap_matrix:1}
P'(0)=4,\quad P''(0)=8L/3,\quad P''>0,\quad P'''>0,
\qquad j(a,b)\ge P(\Delta).
\end{equation}
For clarity, if \(r=1-2H^{-1}(1-I)\), \(w=\operatorname{atanh}(r)\), and \(\nu=1-r^2\), then
\[
P'=2+\frac{2r}{\nu w},\qquad
P''=\frac{2L((1+r^2)w-r)}{\nu^2w^3}.
\]
The derivative formulas and increasing-curvature series are given in Section~\ref{unbal:module-profile_logsum-entropy-profile-facts-with-analytic-constants}.

The cap inequality in \eqref{unbal:eq:cap_matrix:1} has a useful strictness statement. Put \(d=|a-b|\) and \(C(r)=1-H((1-r)/2)\). Then
\[
\Delta=mC(d/(2m))+(1-m)C(d/(2(1-m))),
\]
\[
j=mP(C(d/(2m)))+(1-m)P(C(d/(2(1-m)))).
\]
Strict convexity of \(P\) shows that \(j>P(\Delta)\) whenever \(a\ne b\) and \(a+b\ne1\). Equality occurs on the equal-mean or complementary-mean lines.

\subsubsection{The supporting plane taken at the means}\label{unbal:module-cap_matrix-the-supporting-plane-taken-at-the-means}

For \(0<a<b<1\) determine \(A,D\) by
\begin{equation}\label{unbal:eq:cap_matrix:2}
\begin{pmatrix}
-\ln a&-\ln b\\
-\ln(1-a)&-\ln(1-b)
\end{pmatrix}
\binom A D
=
\binom{(b-a)^2/(2ab)}{(b-a)^2/[2(1-a)(1-b)]}.
\end{equation}
Whenever \(A,D\ge0\), Theorem~\ref{bal:thm:general-endpoint-region}'s pointwise dual proof gives
\[
j(u,w)-c(w-u)+AH(u)+DH(w)\ge0,
\quad c=\frac{j(a,b)+AH(a)+DH(b)}{b-a}.
\]
Taking expectations yields the global lower bound
\begin{equation}\label{unbal:eq:cap_matrix:3}
\boxed{\zeta(a,b,e,f)\ge B_{\rm cap}:=j(a,b)+Ax+Dy.}
\end{equation}
It is global in the entropy moments for these means. The equality-window conditions on the four moments are not needed for a lower bound. On \(0<a<1/2<b<1\), both coefficients are positive by Proposition~\ref{bal:prop:fixed-difference-minimum}. Half-contact limits follow from the continuous dual plane; the interval checker also verifies coefficient positivity directly. For same-side contacts one must check \(A,D\ge0\), or equivalently the ratio test in Theorem~\ref{bal:thm:general-endpoint-region}.

Let \(k=\min(A,D)\). Convexity of \(P\) gives
\begin{equation}\label{unbal:eq:cap_matrix:4}
\boxed{\zeta-R_\psi\ge G(s):=j+2ks-P(\Delta+s)+P(s).}
\end{equation}
Since $P''$ is increasing,
\[
G''(s)=P''(s)-P''(\Delta+s)\le0.
\]
Consequently, if \(0<s_0<1-\Delta\) and \(G(s_0)\ge c_0\ge0\), then simultaneously for every feasible entropy split with \(0\le s\le s_0\),
\begin{equation}\label{unbal:eq:cap_matrix:5}
\boxed{\zeta-R_\psi\ge (1-s/s_0)[j-P(\Delta)]+(s/s_0)c_0.}
\end{equation}
This reduces a whole cap neighborhood to an endpoint inequality in just the two means. In particular, a grid in the entropy imbalance is unnecessary.

An alternative slope is supplied by the previously proved log-sum bound:
\[
B_\chi=j+\frac{(b-a)^2}{4b(1-a)}[\kappa(a)x+\kappa(b)y],
\]
\[
\kappa(z)=\min\left\{\frac{z}{-(1-z)\ln(1-z)},
\frac{1-z}{-z\ln z}\right\}.
\]
Thus \eqref{unbal:eq:cap_matrix:4}--\eqref{unbal:eq:cap_matrix:5} also hold with \(2k\) replaced by
\[
\beta=\frac{(b-a)^2\min(\kappa(a),\kappa(b))}{2b(1-a)},
\]
or by \(\max(\beta,2k)\) when \eqref{unbal:eq:cap_matrix:3} is available. The \(\beta\) version does not require opposite-side means.

\subsubsection{Strong convexity eliminates the split more sharply}\label{unbal:module-cap_matrix-strong-convexity-eliminates-the-split-more-sharply}

The minimum-slope estimate \eqref{unbal:eq:cap_matrix:4} discards the penalty for unequal deficits. Retaining it gives a useful further owner of the residual region.

Let \(t=(x-y)/2\). Since $P''\ge8L/3$,
\[
\tfrac12[P(s+t)+P(s-t)]\ge P(s)+\frac{4L}{3}t^2.
\]
Combining with \eqref{unbal:eq:cap_matrix:3}, and completing a square, proves
\begin{equation}\label{unbal:eq:cap_matrix:11}
\boxed{\zeta-R_\psi\ge Q(s):=
 j+(A+D)s+P(s)-P(\Delta+s)-\frac{3(A-D)^2}{16L}.}
\end{equation}
This holds for every feasible split, whenever \(A,D\ge0\). A sharper version replaces the final negative term with
\[
\min_{t\in[t_-,t_+]}\left((A-D)t+\frac{4L}{3}t^2\right),
\]
where
\[
t_-=\max(-s,s-H(b)),\qquad t_+=\min(s,H(a)-s).
\]
The minimizer is exactly the projection of \(-3(A-D)/(8L)\) onto this interval. This is useful when \(E\) is larger than one child's entropy cap.

Crucially, $Q''(s)=P''(s)-P''(\Delta+s)\le0$. Therefore, for fixed means and an entire average-entropy interval \(E\in[E_0,E_1]\), it suffices to check \eqref{unbal:eq:cap_matrix:11} at \(E=E_0\) and \(E=E_1\). Both the total entropy and the entropy split have then been eliminated analytically. Only a two-dimensional mean cover remains.

Unlike \(G\) in \eqref{unbal:eq:cap_matrix:4}, \(Q(0)\) need not be nonnegative. A positive-entropy slab must have both of its own endpoints checked; it cannot automatically be continued all the way to the caps.

\subsubsection{Concrete interval certificates}\label{unbal:module-cap_matrix-concrete-interval-certificates}

The certificate results and replay counts are recorded in \path{CERTIFICATES.md} and their JSON files. Their claims are:

\begin{enumerate}
\def\labelenumi{\arabic{enumi}.}
\item
  \textbf{Central cross-half cap rectangle:}
\[
  \frac1{10}\le a\le\frac12,\quad
  \frac12\le b\le\frac9{10},\quad 0\le s\le\frac3{20}
  \quad\Longrightarrow\quad \zeta\ge R_\psi.
\]
The checker reuses the completed quantitative subrectangle below wherever it applies, and elsewhere uses either \eqref{unbal:eq:cap_matrix:5} or the log-sum curvature test. In the latter it bounds
\[
  \Delta\le\frac{(b-a)^2}{8L\nu_*},\quad
  \nu_*\le\min_{z\in[a,b]}z(1-z),
\]
and verifies
\[
  \frac{P''(3/20+(b-a)^2/(8L\nu_*))}{8L\nu_*}
  \le\frac{\min(\kappa(a),\kappa(b))}{2b(1-a)}.
\]
The squared mean difference is cancelled before interval evaluation; this includes the equal-mean point \(a=b=1/2\) without a singular division. On each box all bounds are simultaneous, with \(\nu_*\) a lower enclosure of the variance on that box.
\item
  \textbf{A quantitative subrectangle:} \(a\in[1/10,2/5]\), \(b\in[3/5,9/10]\), and \(0\le s\le3/20\). Checking \(G(3/20)>1/1000\) gives
\[
  \zeta-R_\psi\ge s/150.
\]
\end{enumerate}

All boxes have exact rational endpoints. Acceptance uses directed interval arithmetic, not sampled values. The quantitative subrectangle and slab use 96 monotone interval bisection steps for the entropy inverse. The expanded cover uses exact dyadic brackets whose entropy residual signs are verified with directed arithmetic; binary64 is used only to propose brackets. \path{FAST_INVERSE.py} contains that procedure. Each accepted leaf records its exact binary subdivision path; the partition checker requires both children at every internal node. Separate replays evaluate every accepted leaf at 70 decimal working digits. No part of the root rectangle is silently discarded. The wider cover has a geometric owner for boxes contained in the quantitative subrectangle; its replay depends on the separately completed replay of that subrectangle. For tighter enclosures it also uses \(\Delta_a=(J(m)-J(a))/2\le0\) and \(\Delta_b=(J(m)-J(b))/2\ge0\), evaluating \(\Delta\) at the two appropriate corners of an ordered mean box.

The proofs themselves need only nonnegative acceptance margins. Strict margins in certificates give room for interval overestimation; they are not numerical approximations promoted to inequalities.

\clearpage
\subsection{Complete central cap theorem and positive-series mean cost}
\label{unbal:app:restored_cap}

\emph{Argument guide.}
The cap cover is based on a concave comparison anchored at the
deterministic cap. Sharper Jensen-gap and mean-cost estimates make its
endpoint tests stable near coincident means. The two completed roots
then cover cross-half and same-side pairs, with coefficient positivity
checked wherever a mean-contact plane is used.

\noindent\textit{Source:} \path{CK_CAP_REGION_CONTINUATION/RESTORED_CAP_PROOF.md}.\par
This argument gives the analytic acceptance conditions for the cap covers.
The completed certificates and 70-digit replays establish the regional
conclusion below; its role in the central mean square is described in
Section~\ref{unbal:app:central_assembly}.

Use the notation and the established inputs collected in Section~\ref{unbal:sec:inputs}. In particular, \(s\) is the average entropy deficit, \(\Delta\) is the mean Jensen gap, \(P(I)=\eta(1-I)\), and $P'''(I)>0$. A valid lower bound of the form
\[
\zeta\ge j(a,b)+\lambda(a,b)s
\]
reduces the whole-split comparison to
\[
G(s)=j(a,b)+\lambda s-P(\Delta+s)+P(s).
\]
Here \(G(0)\ge0\), and \(G\) is concave. Checking \(G(s_0)\ge0\) proves every feasible split with \(s\le s_0\).

\subsubsection{A stronger bound for the Jensen gap}\label{unbal:module-restored_cap-a-stronger-bound-for-the-jensen-gap}

Put \(d=|a-b|\) and \(\nu_z=z(1-z)\). For distinct means,
\begin{equation}\label{unbal:eq:restored_cap:1}
\boxed{\frac{\Delta}{d^2}\le
K(a,b):=\frac{\nu_a^{-1}+\nu_b^{-1}}{16\ln2}.}
\end{equation}
To prove this, the exact Peano-kernel identity is
\[
\frac{\Delta}{d^2}
=\frac1{2\ln2}\int_0^1
\frac{\min(t,1-t)}{\nu_{a+t(b-a)}}\,dt.
\]
Since \(1/\nu_z=1/z+1/(1-z)\) is convex, it lies below its affine chord between \(a\) and \(b\). Integrating the chord against this symmetric kernel gives \eqref{unbal:eq:restored_cap:1}. This improves the earlier upper bound \(1/(8\ln2\,\min(\nu_a,\nu_b))\).

For a separated interval box \(a\in[a_0,a_1]\), \(b\in[b_0,b_1]\), \(a_1<b_0\), use the further bound
\begin{equation}\label{unbal:eq:restored_cap:2}
\frac{\Delta}{d^2}\le\frac{\Delta(a_0,b_1)}{(b_0-a_1)^2}.
\end{equation}
Indeed \(\Delta\) decreases in its first mean and increases in its second. The checker takes the smaller of the upper enclosures in \eqref{unbal:eq:restored_cap:1} and \eqref{unbal:eq:restored_cap:2}. Both are simultaneous bounds on the entire box.

\subsubsection{A normalized trapezoidal acceptance condition}\label{unbal:module-restored_cap-a-normalized-trapezoidal-acceptance-condition}

Normalize by full label exchange and simultaneous complement. Put
\[
r=\frac{|a-b|}{\min(a+b,2-a-b)}.
\]
The earlier deterministic mean-cost lemma proves \(j\ge(4+r^2/2)\Delta\). Also \(P'\) is convex, so
\[
P(s_0+\Delta)-P(s_0)
=\int_0^\Delta P'(s_0+u)\,du
\le\frac\Delta2[P'(s_0)+P'(s_0+\Delta)].
\]
Suppose that, uniformly on a box,
\[
\lambda\ge d^2\Lambda_*,\qquad
\Delta/d^2\le K_*,\qquad \Delta\le\Delta_*,\qquad r\ge r_*.
\]
Then the following is a sufficient condition for \(G(s_0)\ge0\) throughout the box:
\begin{equation}\label{unbal:eq:restored_cap:3}
\boxed{\Lambda_*s_0\ge K_*\max\left\{0,
\frac{P'(s_0)+P'(s_0+\Delta_*)}{2}-4-\frac{r_*^2}{2}\right\}.}
\end{equation}
For the proof, subtract the trapezoidal upper bound from the mean-cost lower bound and $\lambda s_0$. If the bracket is nonpositive there is no loss to pay. Otherwise $\Delta$ times that bracket is at most $d^2K_*$ times its stated upper bound. This proves \eqref{unbal:eq:restored_cap:3}. Concavity of \(G\) and \(G(0)\ge0\) then prove the whole cap interval, not just \(s=s_0\).

This criterion avoids subtracting several nearly equal cost values. Every transcendental quantity in it is still evaluated with directed intervals. Floating-point guesses only propose inverse-entropy brackets; exact dyadic endpoints and directed sign checks establish their validity.

For the log-sum lower bound \(B_\chi\) one may take
\[
\Lambda_*=\frac{\min\{\kappa(a),\kappa(b)\}}{2V},\quad
V=\max(a,b)(1-\min(a,b)),
\]
using a uniform lower enclosure of this quantity. It is valid for all interior means. For a valid mean-contact plane, take
\[
\Lambda_*=2\min(A/d^2,D/d^2).
\]
The checker computes \(A/d^2\) and \(D/d^2\) directly from the dual system with its \(d^2\) factor removed.

The alternate curvature acceptance condition is
\begin{equation}\label{unbal:eq:restored_cap:4}
\boxed{K_*P''(s_0+\Delta_*)\le\Lambda_*.}
\end{equation}
It follows by integrating $P''$ over an interval of length \(\Delta\), proving \(G'\ge0\). Cancelling \(d^2\) before evaluation also covers boxes touching the diagonal. On the exact diagonal, \(R_\psi\le0\) by convexity of \(P\) and \(\zeta\ge0\) directly.

\subsubsection{Completed-root claims and their combination}\label{unbal:module-restored_cap-completed-root-claims-and-their-combination}

The restored cross-half root is
\[
a\in[1/10,1/2],\quad b\in[1/2,9/10],\quad s\le3/20.
\]
Its acceptance options are the already replayed quantitative subrectangle, the log-sum curvature test, the mean-plane secant, and \eqref{unbal:eq:restored_cap:3} using that plane.

The same-side root is
\[
a,b\in[1/10,1/2],\qquad s\le3/40.
\]
It uses the log-sum curvature test, the log-sum secant, \eqref{unbal:eq:restored_cap:3} using the log-sum slope, and the positive-series refinement in Section~\ref{unbal:module-restored_cap-a-sharper-positive-series-mean-cost-estimate}. A further \texttt{cap\_plane} option uses \eqref{unbal:eq:restored_cap:3} with the mean-contact plane whenever directed intervals certify \(A/d^2>0\) and \(D/d^2>0\) on the whole box. This is valid for same-side contacts by the global supporting-plane proof of Theorem~\ref{bal:thm:general-endpoint-region}; its proof explicitly allows arbitrary ordered contacts satisfying the nonnegative-coefficient test. The checker solves the coefficient system with \(d^2\) cancelled and checks both signs before using the plane. This additional option resolves the narrow band near \(b=1/2\) where the log-sum slope alone gives very small margins.

There is also a symmetry acceptance option for boxes contained in \(a\ge b\). This is not a circular proof: every strict upper-triangle point belongs to a box checked by a direct inequality, because no box contained in \(a\ge b\) can contain it. Strict lower-triangle points follow by swapping the means and entropy labels. The diagonal is covered directly as explained above.

The completed root certificates and their 70-digit replays prove the uniform theorem
\begin{equation}\label{unbal:eq:restored_cap:5}
\boxed{a,b\in[1/10,9/10],\quad
\frac{H(a)-e+H(b)-f}{2}\le\frac3{40}
\quad\Longrightarrow\quad \zeta\ge R_\psi.}
\end{equation}
For opposite-side means use the cross-half root and label exchange. For both means below \(1/2\) use the same-side root. For both above \(1/2\) complement both variables. Thus every pair of means in the stated compact square is included. All feasible entropy splits are included, including their entropy-cap faces.

By Theorem~\ref{bal:thm:global-unrestricted-target}, the usual parent-branch argument transfers each result to \(R_B\) for \(B=\max(\phi,\psi)\). No positive margin against \(R_B\) on a \(\phi\)-parent branch is asserted.

\subsubsection{A sharper positive-series mean-cost estimate}\label{unbal:module-restored_cap-a-sharper-positive-series-mean-cost-estimate}

The same-side checker also uses a stronger version of the mean-cost term in \eqref{unbal:eq:restored_cap:3}. This avoids fine subdivisions where the coarser coefficient \(r^2/2\) loses too much. It introduces no new numerical assumption.

In the canonical orientation \(a<b\) and \(a+b\le1\), put \(m=(a+b)/2\), \(k=m/(1-m)\), \(r=(b-a)/(a+b)\). Define
\[
c_n=\frac1{2n(2n-1)},\qquad
\alpha_n=\frac{2(n-1)}{n(2n-1)}\quad(n\ge2).
\]
The entropy series and mean-cost identities in the previous proof give the exact ratio
\[
\frac{j}{\Delta}-4
=\frac{\sum_{n\ge2}\alpha_n r^{2n-2}(1+k^{2n-1})}
{\sum_{n\ge1}c_n r^{2n-2}(1+k^{2n-1})}.
\]
All coefficients are nonnegative and \(0\le k\le1\). For any integer \(N\ge2\), define
\[
W_N(r,k)=
\frac{\sum_{n=2}^N\alpha_n r^{2n-2}(1+k^{2n-1})}
{\sum_{n=1}^N c_n r^{2n-2}(1+k^{2n-1})+
2c_{N+1}r^{2N}/(1-r^2)}.
\]
Discarding the numerator tail gives a lower bound. Since \(c_n\) decreases and \(1+k^{2n-1}\le2\), the displayed geometric remainder is an upper bound for the denominator tail. Therefore
\[
\boxed{j\ge(4+W_N(r,k))\Delta.}
\]
The same-side checker uses \(N=12\) and replaces \(r_*^2/2\) in \eqref{unbal:eq:restored_cap:3} by a directed lower enclosure of \(W_{12}\) on the entire mean box. The denominator is positive. At the exact diagonal the claim follows with \(\Delta=j=0\); boxes touching it are handled by the curvature condition. This proves the additional \texttt{taylor\_cost} acceptance condition used in the final ledger.

\clearpage
\subsection{Global shifted log-sum planes and the first required same-side cover}
\label{unbal:app:full_entropy}

\emph{Argument guide.}
The cover combines several global lower bounds: a symmetric endpoint
plane, an imbalance-sensitive log-sum bound, and shifted log-sum planes.
For a fixed plane or anchor, concavity reduces each total-entropy interval
to its two endpoints. A separate parent argument handles the full
low-entropy tail before the finite same-side chart is assembled.

\noindent\textit{Source:} \path{CK_FULL_ENTROPY_CONTINUATION/PROOF.md}.\par
\subsubsection{Definitions and the hybrid branch argument}\label{unbal:module-full_entropy-definitions-and-the-hybrid-branch-argument}

Let \(H\) be binary entropy, \(J(x)=\log_2((1-x)/x)\), \(L=\ln 2\), and
\[
j(a,b)=\tfrac12(b-a)(J(a)-J(b)).
\]
The unrestricted value \(\zeta(a,b,e,f)\) is the infimum of \(\mathbb E\,j(U,W)\) over laws with the two specified means and the two specified entropy moments. Set
\[
m=(a+b)/2,\quad C=[H(a)+H(b)]/2,\quad E=(e+f)/2,
\quad s=C-E,\quad\Delta=H(m)-C.
\]
Write \(x=H(a)-e\), \(y=H(b)-f\), so \(x+y=2s\). Let
\[
\eta(h)=(1-2H^{-1}(h))J(H^{-1}(h)),\qquad P(t)=\eta(1-t).
\]
The previously established profile facts are $P(0)=0$, $P'(0)=4$, $P''\ge8L/3$, and $P'''>0$. In particular $P$, $P'$, and $P''$ are increasing, and $P'$ is convex. Define
\[
\psi(m,e)=P(H(m)-e),\quad B=\max(\phi,\psi),
\quad R_B=B(m,E)-\tfrac12[B(a,e)+B(b,f)].
\]
Convexity of \(P\) gives
\begin{equation}\label{unbal:eq:full_entropy:1}
R_\psi\le D_\Delta(s):=P(\Delta+s)-P(s).
\end{equation}
If the parent has \(\phi\ge\psi\), then \(R_B\le R_\phi\), so Theorem~\ref{bal:thm:global-unrestricted-target} applies. If the parent has \(\psi\ge\phi\), then \(R_B\le R_\psi\). Thus any bound against \eqref{unbal:eq:full_entropy:1}, combined with Theorem~\ref{bal:thm:global-unrestricted-target}, proves the hybrid inequality. When the $\psi$ branch is active at the parent, one may instead retain
$\phi$ at the children, using the general bound
\[
 R_B\le\psi(m,E)-\tfrac12\phi(a,e)-\tfrac12\phi(b,f).
\]
This follows from $B(a,e)\ge\phi(a,e)$ and $B(b,f)\ge\phi(b,f)$.
A box accepted by this comparison must certify its right-hand side against
a lower bound for $\zeta$ throughout the relevant feasible intersection.

The stated entropy comparisons concern positive entropy moments. If exactly one child entropy is zero and the means are interior, \(\phi\) at that child diverges to positive infinity and the hybrid comparison is trivial in its extended-value interpretation. Both zero entropies are the separate deterministic case; no undefined infinity-minus-infinity expression is used here.

\subsubsection{A global endpoint bound that ignores the entropy split}\label{unbal:module-full_entropy-a-global-endpoint-bound-that-ignores-the-entropy-split}

For \(d>0\) and \(E>0\), let \(v\) be the unique root in \((0,1/2)\) of
\begin{equation}\label{unbal:eq:full_entropy:2}
dH(v)=E(1-2v),
\qquad F(d,E)=dJ(v).
\end{equation}
The global supporting-plane inequality \eqref{bal:eq:fixed-difference-plane-global}, converted from natural-log units to bits, implies
\begin{equation}\label{unbal:eq:full_entropy:3}
\boxed{\zeta(a,b,e,f)\ge F(|b-a|,(e+f)/2).}
\end{equation}
Here is the derivation, including why a same-side pair of means is allowed. Take the global plane at the symmetric contact \((v,1-v)\). Its two entropy coefficients are equal and positive. With \(r=1-2v\), its coefficient \(c\) satisfies
\[
cr=j(v,1-v)+2AH(v).
\]
The expected plane gives \(\zeta\ge cd-2AE\). Equation \eqref{unbal:eq:full_entropy:2} makes the entropy terms cancel, leaving \(d\,j(v,1-v)/r=dJ(v)\). The contact need not be the pair of prescribed means. There is no assumption on how \(e\) and \(f\) are split and no assumption that \(E\) is feasible separately at both children.

Combining \eqref{unbal:eq:full_entropy:1} and \eqref{unbal:eq:full_entropy:3} gives a sufficient whole-fiber test
\begin{equation}\label{unbal:eq:full_entropy:4}
\boxed{F(d,E)\ge P(H(m)-E)-P(C-E).}
\end{equation}
The inverse in \eqref{unbal:eq:full_entropy:2} decreases in \(d\) and increases in \(E\). \(F\) increases in \(d\) and decreases in \(E\). These facts follow by implicit differentiation: if \(T=dJ(v)+2E>0\),
\[
v_E=(1-2v)/T>0,\quad v_d=-H(v)/T<0,
\quad F_E=dJ'(v)v_E<0,\quad F_d=J(v)+dJ'(v)v_d>0.
\]
The normalized identity
\begin{equation}\label{unbal:eq:full_entropy:5}
\boxed{\frac{F(d,E)}{d^2}=\frac{H(v)J(v)}{(1-2v)E}}
\end{equation}
is also useful in interval arithmetic because it avoids unnecessary \(d^2\) variation on the two sides of an inequality. No monotonicity of \(F/d^2\) is assumed.

\subsubsection{Retaining the imbalance penalty in the global log-sum bound}\label{unbal:module-full_entropy-retaining-the-imbalance-penalty-in-the-global-log-sum-bound}

For ordered interior means define \(V=b(1-a)\) and
\[
\kappa(z)=\min\left\{\frac{z}{-(1-z)\ln(1-z)},
\frac{1-z}{-z\ln z}\right\},
\quad A_\chi=\frac{d^2\kappa(a)}{4V},\quad D_\chi=\frac{d^2\kappa(b)}{4V}.
\]
The earlier global log-sum theorem is
\[
\zeta\ge j+A_\chi x+D_\chi y.
\]
It yields the new whole-fiber bound
\begin{equation}\label{unbal:eq:full_entropy:6}
\boxed{
\zeta-R_\psi\ge j+(A_\chi+D_\chi)s
-D_\Delta(s)-\frac{3(A_\chi-D_\chi)^2}{16L}.
}
\end{equation}
Unlike a plane taken at the means, these coefficients are valid for every ordered interior pair.

For the proof put \(t=(x-y)/2\). The strong convexity $P''\ge8L/3$ gives
\[
\tfrac12[P(s+t)+P(s-t)]\ge P(s)+\frac{4L}{3}t^2.
\]
Adding the log-sum bound and minimizing the resulting quadratic over all real \(t\) gives \eqref{unbal:eq:full_entropy:6}. Minimizing over the actual feasible interval can only strengthen it. Thus discarding the feasible interval here introduces no invalid extension.

This retains the average of the two entropy coefficients, rather than replacing both by their minimum. The penalty is quadratic in their difference. It is particularly effective in the transition where the endpoint bound and the older minimum-coefficient bound each have small margins.

\paragraph{A global family of affine bounds from different anchors}\label{unbal:module-full_entropy-a-global-family-of-affine-bounds-from-different-anchors}

The underlying pointwise log-sum remainder inequality permits arbitrary interior anchors \(\alpha<\beta\); they need not equal the prescribed means \(a,b\). Put
\[
A=\frac{(\beta-\alpha)^2\kappa(\alpha)}{4\beta(1-\alpha)},\quad
D=\frac{(\beta-\alpha)^2\kappa(\beta)}{4\beta(1-\alpha)}.
\]
Its pointwise form is
\[
j(u,w)\ge j(\alpha,\beta)+j_\alpha(u-\alpha)+j_\beta(w-\beta)
+A D_2(u\Vert\alpha)+D D_2(w\Vert\beta).
\]
To see that the stronger \(\kappa\) coefficients are allowed pointwise, use the proved quadratic remainder bound together with \(D_2(u\Vert\alpha)\le C_\alpha(u-\alpha)^2\) and \(\kappa(\alpha)=1/[L\alpha(1-\alpha)C_\alpha]\), and similarly for \(\beta\).

Use the exact entropy identity
\[
D_2(u\Vert\alpha)=H(\alpha)+J(\alpha)(u-\alpha)-H(u).
\]
Define
\[
p=j_\alpha+A J(\alpha),\qquad q_1=j_\beta+D J(\beta),
\quad k=j(\alpha,\beta)-p\alpha-q_1\beta+A H(\alpha)+D H(\beta).
\]
Taking expectations proves the global affine lower bound
\begin{equation}\label{unbal:eq:full_entropy:6a}
\boxed{\zeta(a,b,e,f)\ge k+pa+q_1b-Ae-Df.}
\end{equation}
No endpoint-contact ratio test is needed for this family. Combining it with the same strong-convexity argument gives
\begin{equation}\label{unbal:eq:full_entropy:6b}
\boxed{\begin{aligned}
\zeta-R_\psi\ge{}&k+pa+q_1b+\tfrac12(D-A)[H(a)-H(b)]-(A+D)E\\
&-P(H(m)-E)+P(C-E)-\frac{3(A-D)^2}{16L}.
\end{aligned}}
\end{equation}
This is concave in \(E\) for fixed means and fixed anchors. Choosing \(\alpha=a\), \(\beta=b\) recovers \eqref{unbal:eq:full_entropy:6}; choosing other anchors can be strictly stronger. \path{SHIFTED_LOGSUM.py} proposes anchors by a one-dimensional rescaling search. The resulting decimal anchors are treated as exact rationals, recorded in the leaf, and the entire bound is recomputed with directed intervals. Replay does not run the search. No optimality or completeness of the anchor search is assumed.

\subsubsection{Normalized box inequalities}\label{unbal:module-full_entropy-normalized-box-inequalities}

For ordered means the earlier mean-cost expansion gives \(j\ge(4+W)\Delta\) with \(W\ge0\). We use either \(W=r^2/2\), where \(r=d/(a+b)\), or the stronger positive-series estimate proved in Section~\ref{unbal:module-restored_cap-a-sharper-positive-series-mean-cost-estimate}. For completeness, for \(k=(a+b)/(2-a-b)\le1\),
\[
W_N=\frac{\sum_{n=2}^N\frac{2(n-1)}{n(2n-1)}r^{2n-2}(1+k^{2n-1})}
{\sum_{n=1}^N\frac{r^{2n-2}(1+k^{2n-1})}{2n(2n-1)}
+\frac{2r^{2N}}{2(N+1)(2N+1)(1-r^2)}}
\]
satisfies \(j\ge(4+W_N)\Delta\). The numerator tail is positive and the displayed denominator tail is a geometric upper bound; the checker takes \(N=12\).

Write \(\nu_z=z(1-z)\). The Jensen-gap estimates are
\[
\frac{\Delta}{d^2}\le\frac{\nu_a^{-1}+\nu_b^{-1}}{16L},
\quad
\frac{\Delta}{d^2}\le\frac{\Delta_*}{d_*^2}
\]
whenever $\Delta\le\Delta_*$ and $d\ge d_*>0$. Let $K_*$ be the smaller available upper bound. Convexity of \(P'\) gives
\[
D_\Delta(s)\le\frac\Delta2[P'(s)+P'(s+\Delta)].
\]
On a box let $s\le s_*$, $s+\Delta\le I_*$ and put $T_*=[P'(s_*)+P'(I_*)]/2$. Then a normalized log-sum sufficient test is
\[
\frac{\min(\kappa(a),\kappa(b))}{2V}s
\ge K_*\max(0,T_*-4-W).
\]
The improved test from \eqref{unbal:eq:full_entropy:6} replaces its left side by a lower enclosure of
\[
\frac{\kappa(a)+\kappa(b)}{4V}s
-\frac{3d^2}{16L}\left(\frac{\kappa(a)-\kappa(b)}{4V}\right)^2.
\]
The normalized endpoint test is a lower enclosure of \eqref{unbal:eq:full_entropy:5} $\ge K_*T_*$.

The direct endpoint and log-sum options use the smaller of three upper bounds on \(D\):
\[
P(I_*)-P(s_{\min}),\qquad\Delta_*T_*,\qquad
P(s_*+\Delta_*)-P(s_*),
\]
where the third is used only if $s_*+\Delta_*<1$. It is valid because \(D\) increases in each of \(\Delta\) and \(s\). A mean-plane option solves the two-by-two coefficient system \eqref{unbal:eq:cap_matrix:2} with \(d^2\) cancelled and checks both coefficients positive before using its minimum slope. All these tests are sufficient conditions on the whole box.

\paragraph{Eliminating an entire interval of total entropies}\label{unbal:module-full_entropy-eliminating-an-entire-interval-of-total-entropies}

The two additional endpoint options avoid separately subtracting quantities evaluated at opposite ends of an entropy interval. Fix any rational symmetric contact \(v\in(0,1/2)\), and let its global plane be \(\zeta\ge cd-2AE\) as in Section~\ref{unbal:module-full_entropy-a-global-endpoint-bound-that-ignores-the-entropy-split}. For fixed means,
\[
G_v(E)=c d-2AE-P(H(m)-E)+P(C-E)
\]
satisfies
\[
G_v''(E)=-P''(H(m)-E)+P''(C-E)\le0.
\]
Thus if this same plane gives a nonnegative gap at both endpoints of an \(E\) interval, it gives a nonnegative gap throughout. The checker proposes a contact from the center of a box, treats that contact as an exact rational, computes its positive coefficients with directed intervals, and checks both endpoint inequalities uniformly in the means.

Exactly the same concavity argument applies to the right side of \eqref{unbal:eq:full_entropy:6}, because its remaining terms are affine or constant in \(E\). Accordingly, \texttt{logsum\_quad\_endpoints} verifies the quadratic log-sum bound at both endpoints of the box's \(t\) interval, and \texttt{endpoint\_plane\_endpoints} verifies one fixed global plane at both endpoints. Since \(E\) is affine in \(t\) for each fixed pair of means, endpoint sufficiency holds in these coordinates as well. Different contacts or different lower-bound families are not mixed across the two endpoints of one such acceptance condition.

These tests eliminate both the individual entropy split and a whole interval of total entropies analytically. Only the uniform mean dependence in the endpoint expressions remains to be enclosed.

\paragraph{Centered enclosures in the two means}\label{unbal:module-full_entropy-centered-enclosures-in-the-two-means}

For the fixed-plane gap \(G_v\) at a fixed \(t\), write \(p_I=P'(I)\), \(p_s=P'(s)\). Its two derivatives are
\[
\begin{gathered}\partial_aG_v=-c+\tfrac12J(a)T-\tfrac12J(m)p_I,
\\\partial_bG_v=c+\tfrac12J(b)T-\tfrac12J(m)p_I,
\\T=-2At+tp_I+(1-t)p_s.\end{gathered}
\]
For the gap \eqref{unbal:eq:full_entropy:6b}, the corresponding derivatives are
\[
\begin{gathered}\partial_aG=p+\tfrac12J(a)(T+D-A)-\tfrac12J(m)p_I,
\\\partial_bG=q_1+\tfrac12J(b)(T+A-D)-\tfrac12J(m)p_I,
\\T=-(A+D)t+tp_I+(1-t)p_s.\end{gathered}
\]
These follow by differentiating \(E=\epsilon+t(C-\epsilon)\). On a mean box with center \((a_0,b_0)\) and half-widths \(w_a,w_b\), directed bounds on these derivatives give
\[
G(a,b,t)\ge G(a_0,b_0,t)-w_a\sup|G_a|-w_b\sup|G_b|.
\]
This is the ordinary mean-value theorem on a convex rectangle. The checker uses it at both \(t\) endpoints, keeping the same rational anchor throughout, then applies entropy concavity. Every coefficient and derivative enclosure is checked again in replay.

\subsubsection{The complete entropy tail}\label{unbal:module-full_entropy-the-complete-entropy-tail}

Put \(q=|1-2m|\) and \(\epsilon=10^{-6}\). We prove
\begin{equation}\label{unbal:eq:full_entropy:7}
\boxed{1/20\le q\le4/5,\quad0<E\le\epsilon
\quad\Longrightarrow\quad\phi(m,E)>\psi(m,E).}
\end{equation}
Let \(v=H^{-1}(E)\). Concavity of \(H\) gives \(H(v)\ge2v\), so \(v\le E/2\) and \(1-2v\ge1-E\). Feasibility implies \(q\le1-2v\). The contact defining \(F(q,E)\) is at least \(v\), hence
\[
\phi(m,E)=\eta(E)-F(q,E)
\ge(1-2v-q)J(v)
\ge(1-E-q)J(E/2).
\]
Both positive factors in the last product decrease with \(E\). On a \(q\)-interval \([q_0,q_1]\) this is at least \((1-\epsilon-q_1)J(\epsilon/2)\). Meanwhile
\[
\psi(m,E)\le\eta(1-H((1-q_0)/2)).
\]
The four intervals \([0.05,0.2]\), \([0.2,0.4]\), \([0.4,0.6]\), and \([0.6,0.8]\) give strictly positive differences. \path{LOW_ENTROPY_CHECK.py} verifies these four fixed comparisons at 70 digits; the smallest lower enclosure exceeds \(0.292\). This proves every \(E\) in the stated interval analytically, with no entropy grid.

\subsubsection{Full-entropy cover: exact domain and gates}\label{unbal:module-full_entropy-full-entropy-cover-exact-domain-and-gates}

\begin{theorem}[Completed larger same-side region]\label{unbal:thm:full_entropy:1}
Using the established inputs of Section~\ref{unbal:sec:inputs}, we obtain
\begin{equation}\label{unbal:eq:full_entropy:8}
\boxed{
a,b\in[1/10,1/2],\quad |a-b|\ge1/20
\quad\Longrightarrow\quad\zeta\ge R_B
}
\end{equation}
for every positive feasible entropy pair. The exact root partition has 33572 final leaves, zero unresolved leaves, and a complete 70-digit directed-interval replay. These are recorded in \path{FULL_ENTROPY_RESULT.json} and \path{FULL_ENTROPY_REPLAY.json}.
\end{theorem}

Exchange full labels to assume \(b-a\ge1/20\). Then \(q=1-a-b\ge b-a\ge1/20\) and \(q\le4/5\). Thus \eqref{unbal:eq:full_entropy:7} covers \(E\le\epsilon\). For \(E\ge\epsilon\) the checker uses the exact coordinates
\[
(a,b,t)\in[1/10,1/2]^2\times[0,1],\qquad
E=\epsilon+t(C-\epsilon),\quad s=(1-t)(C-\epsilon).
\]
Every feasible total entropy in \([\epsilon,C]\) has such a \(t\). Boxes entirely below \(b-a=1/20\) are irrelevant and explicitly marked. A box is otherwise accepted only by one of the proved tests, by \(\phi\)-parent dominance, or by the already completed cap theorem \(s\le3/40\). Infeasible orientations are never used to infer an inequality on a feasible point.

The binary paths reconstruct exact rational boxes. The full binary-tree partition check rules out gaps. Discovery uses directed intervals at 40 digits; replay reconstructs every final box and repeats its recorded acceptance condition at 70 digits. A stopped or budget-limited search is not a certificate for \eqref{unbal:eq:full_entropy:8}. Checkpoint compaction may replace descendants only after the parent independently passes a directed test, and the complete frontier partition must still pass.

Simultaneous complement gives the corresponding upper-half region \([1/2,9/10]^2\), with the same separation bound \(|a-b|\ge1/20\). The remaining central configurations are included in Theorem~\ref{unbal:thm:central}, whose proof assembles the component regions in Section~\ref{unbal:sec:central}.

\subsubsection{A monotone \texorpdfstring{\(\phi\)}{phi}-parent enclosure for the last low-entropy block}\label{unbal:module-full_entropy-a-monotone-phi-parent-enclosure-for-the-last-low-entropy-block}

Write the bit-normalized parent candidate as
\[
\Phi(q,E)=\eta(E)-F(q,E)=\phi((1-q)/2,E).
\]
Theorem~\ref{bal:aux:thm:entropyconvexity} establishes convexity in \(E\) on \(0<E<H((1-q)/2)\). The function \(F\) increases in \(q\), so \(\Phi\) decreases in \(q\). For a feasible pair, put \(v=H^{-1}(E)\) and \(r_E=1-2v\). Equation \eqref{unbal:eq:full_entropy:2} gives \(F(r_E,E)=r_EJ(v)=\eta(E)\), and feasibility gives \(q\le r_E\). Consequently \(\Phi(q,E)\ge0\), with \(\Phi(q,H((1-q)/2))=0\).

These facts also imply that \(\Phi\) is nonincreasing in \(E\) throughout its feasible interval. Indeed, with \(h_q=H((1-q)/2)\) and \(0<x<y<h_q\), convexity and the endpoint value give
\[
0\le\Phi(q,y)\le\frac{h_q-y}{h_q-x}\Phi(q,x)\le\Phi(q,x).
\]
The endpoint value follows by continuity from equation \eqref{unbal:eq:full_entropy:2}.

Thus, on a rectangle \(q\in[q_-,q_+]\) and \(E\in[E_-,E_+]\), provided \(E_+\le H((1-q_+)/2)\), the rigorous bound is
\[
\boxed{\Phi(q,E)\ge\eta(E_+)-F(q_+,E_+).}
\]
Every intermediate pair in the monotonic comparison is feasible, since the entropy cap decreases with \(q\). In the program this rectangle-feasibility test is strict and directed: the upper endpoint \(E_+\) must lie below a certified lower bound for \(H((1-q_+)/2)\). Only then is this enclosure used. The directed upper enclosure of \(P(H(m)-E)\) is subtracted to test \(\Phi\ge\psi\) at the parent.

This evaluates the cancellation between \(\eta\) and \(F\) at one common entropy endpoint. The old independent interval evaluation remains as a fallback. No other mathematical acceptance rule was changed. \path{MONOTONE_MIGRATION.json} records the previous and final source hashes and the unchanged leaf records; the previous core source is retained under \path{history/}. The complete 70-digit replay of the final partition is recorded in \path{FULL_ENTROPY_REPLAY.json}, as stated in Theorem~\ref{unbal:thm:full_entropy:1}. The earlier partial frontier replay retains its original hash and historical scope. This enclosure uses the preceding entropy-convexity and monotonicity results.

\clearpage
\subsection{Central small-ratio theorem and the required one-hundredth same-side cover}
\label{unbal:app:small_ratio}

\emph{Argument guide.}
First remove the entropy split and normalize by squared mean
separation to obtain a stable small-ratio inequality. A compact cover
and an analytic entropy tail prove it. The subsequent arguments extend
the same-side separation range and retain the actual parent radius to
cover the central small-ratio region more generally.

\noindent\textit{Source:} \path{CK_SMALL_RATIO_EXTENSION/PROOF.md}.\par
\subsubsection{Notation and retained inputs}\label{unbal:module-small_ratio-notation-and-retained-inputs}

Entropies and costs are in bits. Write \(L=\ln 2\),
\[
H(u)=-u\log_2u-(1-u)\log_2(1-u),\quad
J(u)=\log_2((1-u)/u),
\]
\[
\eta(h)=(1-2H^{-1}(h))J(H^{-1}(h)),\quad
C(r)=1-H((1-r)/2).
\]
Here \(H^{-1}\) is the lower inverse, and \(C\) is even. Put
\[
F(z,h)=zJ(v),\qquad zH(v)=h(1-2v),\quad F(0,h)=0,
\]
\[
\phi(m,h)=\eta(h)-F(|1-2m|,h),\qquad
\psi(m,h)=\eta(h+C(1-2m)),\qquad B=\max(\phi,\psi).
\]
For feasible means and positive entropy moments, define
\[
m=(a+b)/2,\quad E=(e+f)/2,\quad
q=|1-a-b|,\quad d=|a-b|,
\]
\[
\Delta=H(m)-[H(a)+H(b)]/2,
\quad R_A=A(m,E)-[A(a,e)+A(b,f)]/2.
\]
The conditions on the original moments are \(0<a,b<1\) and \(0<e\le H(a)\), \(0<f\le H(b)\).

Theorem~\ref{bal:thm:global-unrestricted-target}, Theorem~\ref{bal:thm:four-moment-lower-bound}, and the decreasing radial-curvature property in Sections~\ref{bal:aux:sec:fourpoint} and~\ref{bal:imp:e8} provide the lower bounds used here. The latter implies concavity of \(z\mapsto F(\sqrt z,h)\). The same-side cover also retains entropy convexity of \(\phi\) (Section~\ref{bal:aux:sec:entropy}), the already proved global log-sum/profile bounds, and the completed cap and separated same-side certificates. The earlier proof package is included unchanged under \path{dependencies/}. The balanced radial verification is described in Section~\ref{bal:sec:radial-verification}.

\subsubsection{A uniform new region with a quantitative margin}\label{unbal:module-small_ratio-a-uniform-new-region-with-a-quantitative-margin}

\begin{theorem}\label{unbal:thm:small_ratio:1}
For every feasible entropy split,
\begin{equation}\label{unbal:eq:small_ratio:1}
\boxed{
0<E\le11/200,\qquad q\le E,\qquad d\le4E
\quad\Longrightarrow\quad
\zeta-R_\psi\ge\frac{d^2}{100E}.
}
\end{equation}
Consequently \(\zeta\ge R_B\) on this region, using the unrestricted \(R_\phi\) input in the \(\phi\)-parent branch. The condition on the means forces \(q+d\le11/40\), so the entire interval joining the means stays inside \([29/80,51/80]\). No lower bound on \(d\) or on the entropy ratio is imposed.
\end{theorem}

The proof below includes both \(d=0\) and the entire limit \(E\to0\). Its only computational component is a complete two-variable directed-interval certificate with 205 leaves, replayed at 70 decimal digits.

\paragraph{Eliminating the entropy split}\label{unbal:module-small_ratio-eliminating-the-entropy-split}

Convexity of \(\eta\) gives
\begin{equation}\label{unbal:eq:small_ratio:2}
R_\psi\le\eta(y)-\eta(y+\Delta),\qquad y=E+C(q).
\end{equation}
Indeed the average of the two child \(\eta\) arguments is \(y+\Delta\). This applies to every feasible split. Let \(x=d/E\) and, for \(d>0\), define
\[
A(E,x)=\frac1{2L[1-(x+1)^2E^2]},\qquad
h_*(E,x)=E+C((x+1)E).
\]
The entropy second derivative is $-H''(u)=1/[Lu(1-u)]$. Since the largest child radial coordinate is \(q+d<1\), the central second-difference estimate is
\begin{equation}\label{unbal:eq:small_ratio:3}
\Delta\le\frac{d^2}{2L[1-(q+d)^2]}\le A(E,x)d^2.
\end{equation}
Also
\begin{equation}\label{unbal:eq:small_ratio:4}
E\le y\le y+\Delta
=E+\tfrac12[C(q+d)+C(q-d)]\le h_*(E,x)<1.
\end{equation}
Evenness and monotonicity of \(C\) on \([0,1)\) justify the last inequality.

\paragraph{An explicit upper bound on the logarithmic entropy slope}\label{unbal:module-small_ratio-an-explicit-upper-bound-on-the-logarithmic-entropy-slope}

For \(0<h<1\) set \(v=H^{-1}(h)\), \(\ell=\ln((1-v)/v)\), and \(r=1-2v\). Direct differentiation and the entropy identity give
\[
-\eta'(h)=2+\frac{r}{v(1-v)\ell},\qquad
Lh=v\ell-\ln(1-v).
\]
Using \(r/(1-v)\le1\) and \(-\ln(1-v)\le v/(1-v)\), we obtain
\begin{equation}\label{unbal:eq:small_ratio:5}
h[-\eta'(h)-2]
\le\frac1L\left(1+\frac1{(1-v)\ell}\right).
\end{equation}
The function \((1-v)\ln((1-v)/v)\) decreases on \((0,1/2)\), since its derivative is \(-\ell-1/v<0\). Consequently, with $v_*=H^{-1}(h_*)$,
\[
K(E,x)=\frac1L\left(1+\frac1{(1-v_*)\ln((1-v_*)/v_*)}\right)
\]
satisfies \(-\eta'(h)\le2+K(E,x)/h\) for every \(0<h\le h_*\). Integrating on \([y,y+\Delta]\) and applying \eqref{unbal:eq:small_ratio:3}--\eqref{unbal:eq:small_ratio:4} yields
\begin{equation}\label{unbal:eq:small_ratio:6}
R_\psi\le2A d^2+K\ln\left(1+\frac{A d^2}{E}\right).
\end{equation}

\paragraph{The normalized two-variable inequality}\label{unbal:module-small_ratio-the-normalized-two-variable-inequality}

Put \(f(x)=F(x,1)\) and
\[
\ell_0(z)=\begin{cases}\ln(1+z)/z,&z>0,\\1,&z=0.\end{cases}
\]
Homogeneity gives \(F(d,E)=Ef(d/E)\). For \(d>0\), \eqref{unbal:eq:small_ratio:6} implies
\[
\frac E{d^2}[F(d,E)-R_\psi]
\ge\mathcal G(E,x),
\]
\begin{equation}\label{unbal:eq:small_ratio:7}
\boxed{\mathcal G(E,x)=\frac{f(x)}{x^2}
-2A(E,x)E-K(E,x)A(E,x)\ell_0(A(E,x)x^2E).}
\end{equation}
At \(x=0\) the first term has the continuous value \(2/L\). Concavity of \(z\mapsto F(\sqrt z,1)\), with value zero at \(z=0\), implies that \(f(x)/x^2\) is nonincreasing for \(x>0\). Also
\[
\ell_0(z)=\int_0^1\frac{dt}{1+zt}
\]
is positive and nonincreasing on \([0,\infty)\).

On a rational box \(E\in[E_-,E_+]\) and \(x\in[x_-,x_+]\), the checker therefore bounds the first term from below by \(F(x_+,1)/x_+^2\). It encloses $A$ and $h_*$ with directed intervals, obtains an upper enclosure for $K$ using a certified inverse-entropy bracket at the upper endpoint of $h_*$, and bounds \(\ell_0\) from above at a lower enclosure of \(Ax^2E\). If that endpoint is zero, \(\ell_0\le1\) is used directly. Subtracting the resulting upper enclosure for the last two terms is a lower enclosure for \(\mathcal G\) on the entire box.

\path{NORMALIZED_COLLAR_RESULT.json} contains an exact binary-tree partition of
\[
[10^{-6},11/200]\times[0,4]
\]
into \textbf{205 leaves}, with zero unresolved leaves. On every leaf, the lower enclosure of \(\mathcal G\) is at least \(1/100\). \path{NORMALIZED_COLLAR_REPLAY.json} reconstructs all leaves and repeats every inequality at 70 decimal digits. The complete partition check and the replay both passed. Inverse roots are proposed in binary64 but their exact dyadic brackets are checked by directed signs; proposals and samples are never acceptance criteria.

\paragraph{The entire small-entropy tail}\label{unbal:module-small_ratio-the-entire-small-entropy-tail}

Let \(\epsilon=10^{-6}\), and define
\[
A_*=\frac1{2L(1-25\epsilon^2)},\quad
h_*=\epsilon+C(5\epsilon),
\]
with \(K_*\) obtained from this \(h_*\) by \eqref{unbal:eq:small_ratio:5}. For every \(0<E\le\epsilon\) and \(0\le x\le4\), monotonicity gives $A\le A_*$, $K\le K_*$, \(\ell_0\le1\), and hence
\[
\mathcal G(E,x)\ge \frac{F(4,1)}{16}-2A_*\epsilon-K_*A_*.
\]
\path{SMALL_RATIO_TAIL.py} verifies at 70 digits that the last quantity exceeds \(0.0677875699786601\), in particular \(1/100\). This one fixed comparison proves the entire continuum \(0<E\le\epsilon\); it is not a grid in \(E\).

Combining this tail with the normalized two-variable inequality of Section~\ref{unbal:module-small_ratio-the-normalized-two-variable-inequality} proves \eqref{unbal:eq:small_ratio:1} for \(d>0\). When \(d=0\), the means are equal, \(\Delta=0\), and \eqref{unbal:eq:small_ratio:2} gives \(R_\psi\le0\), while \(\zeta\ge F(0,E)=0\). Thus \eqref{unbal:eq:small_ratio:1} also holds there. Finally, if \(B(\mathrm{parent})=\psi(\mathrm{parent})\), then \(R_B\le R_\psi\); if \(B(\mathrm{parent})=\phi(\mathrm{parent})\), then \(R_B\le R_\phi\). This proves the hybrid conclusion.

\subsubsection{Closing the missing ratio range near complementary means}\label{unbal:module-small_ratio-closing-the-missing-ratio-range-near-complementary-means}

\begin{corollary}\label{unbal:thm:small_ratio:2}
For every feasible entropy split,
\begin{equation}\label{unbal:eq:small_ratio:8}
\boxed{
0<E\le11/200,\quad q\le E/100,\quad
E\le\min\{H(a),H(b)\}
\quad\Longrightarrow\quad\zeta\ge R_B.
}
\end{equation}
There is \textbf{no condition on the mean difference \(d\)}.
\end{corollary}

For \(d\le4E\), use Theorem~\ref{unbal:thm:small_ratio:1}. For \(4E\le d\le5E\), use the previously proved transverse strip in \path{prior_inputs/TRANSVERSE_STRIP.md} (Theorem~\ref{unbal:thm:global_mixed_strip:1}), whose guaranteed \(\psi\)-parent margin is \(q^2/(100E)\). For \(d\ge5E\), use the first collar in \path{prior_inputs/SHARPER_COLLARS.md} (\(q\le E\) and \(d\ge5E\); Theorem~\ref{unbal:thm:sharper_collars:1}), whose margin is \(q^2/(1000E)\). Endpoints of these ranges are included, so the union has no missing ratio strip. The old common-entropy-cap condition is retained.

In particular, if \(a,b\) are in \([1/10,9/10]\), then \(H(a),H(b)\ge H(1/10)\ge1/5>11/200\). For those means, the common-cap condition in \eqref{unbal:eq:small_ratio:8} is automatic. Thus the whole complementary collar \(q\le E/100\), \(0<E\le11/200\) is covered throughout this central square, for every mean difference and every feasible entropy split.

\subsubsection{Low-entropy tail for smaller same-side separations}\label{unbal:module-small_ratio-low-entropy-tail-for-smaller-same-side-separations}

For \(a,b\in[1/10,1/2]\), \(b-a\ge1/100\) implies \(q=1-a-b\ge b-a\ge1/100\) and \(q\le4/5\). Put \(\epsilon=10^{-6}\) and \(v=H^{-1}(E)\). Concavity of \(H\) gives \(v\le E/2\). From the contact definition and feasibility,
\[
\phi(m,E)\ge(1-E-q)J(E/2).
\]
Both positive factors decrease with \(E\), so on any \(q\) interval \([q_0,q_1]\) and every \(0<E\le\epsilon\),
\begin{equation}\label{unbal:eq:small_ratio:9}
\phi(m,E)-\psi(m,E)
\ge(1-\epsilon-q_1)J(\epsilon/2)
-\eta(1-H((1-q_0)/2)).
\end{equation}
\path{SAME_SIDE_TAIL.py} checks that all 79 fixed comparisons \eqref{unbal:eq:small_ratio:9}, for \([q_0,q_1]=[n/100,(n+1)/100]\), \(1\le n\le79\), are positive at 70 digits. These cover every \(q\in[1/100,4/5]\) and arbitrarily small positive \(E\). Therefore the unrestricted \(R_\phi\) theorem owns the full same-side tail.

\subsubsection{Extending the same-side cover from separation \texorpdfstring{\(1/20\)}{1/20} to \texorpdfstring{\(1/100\)}{1/100}}\label{unbal:module-small_ratio-extending-the-same-side-cover-from-separation-120-to-1100}

\begin{theorem}[Completed smaller-separation cover]\label{unbal:thm:small_ratio:3}
Using the established inputs of Section~\ref{unbal:sec:inputs}, we obtain
\begin{equation}\label{unbal:eq:small_ratio:10}
\boxed{
a,b\in[1/10,1/2],\quad |a-b|\ge1/100,
\quad0<e\le H(a),\quad0<f\le H(b)
\ \Longrightarrow\ \zeta\ge R_B.
}
\end{equation}
The full partition has 76547 leaves and zero unresolved leaves. Every leaf passed a 70-digit replay, recorded in \path{SMALL_DIFFERENCE_RESULT.json} and \path{SMALL_DIFFERENCE_REPLAY.json}. Thus \eqref{unbal:eq:small_ratio:10} follows from these certificates and the preceding lower bounds.
\end{theorem}

The computational root is \([1/10,1/2]^2\times[0,1]\), with \(E=10^{-6}+t((H(a)+H(b))/2-10^{-6})\). Full label exchange permits \(b-a\ge1/100\). An \texttt{outside} leaf has \(b-a<1/100\) throughout its box. A \texttt{prior\_same\_side} leaf has \(b-a\ge1/20\) throughout and invokes the preceding complete 33,572-leaf cover, without repeating it. Every other leaf is certified by a previously proved acceptance condition, Theorem~\ref{unbal:thm:small_ratio:1}, the central extension \eqref{unbal:eq:small_ratio:11}, or the logarithmic parent rule \eqref{unbal:eq:small_ratio:13}. Theorem~\ref{unbal:thm:small_ratio:1} is used only when \(q_{\mathrm{upper}}\le E_{\mathrm{lower}}\), \(d_{\mathrm{upper}}\le4E_{\mathrm{lower}}\), and \(E_{\mathrm{upper}}\le11/200\). The central extension requires only the latter two conditions on this fixed mean square; the logarithmic parent rule is checked with the rectangle conditions in Section~\ref{unbal:module-small_ratio-a-direct-lower-bound-for-phi-parent-dominance}.

The copied core sets its minimum-difference parameter to exactly \(1/100\); in particular its clipped radial lower endpoint is valid because \(q\ge d\ge1/100\) on every relevant same-side point. Its old cap test is used only on the original compact square where that cap theorem was proved. All entropy-split and endpoint concavity arguments remain unchanged. The exact rational tree paths check the complete root partition, and the replay checks each recorded acceptance rule at 70 decimal digits. Section~\ref{unbal:module-small_ratio-low-entropy-tail-for-smaller-same-side-separations} covers every remaining positive \(E\) below \(10^{-6}\).

Simultaneous complement supplies the upper-half square \([1/2,9/10]^2\). An individual reflection of just one mean is not used as a symmetry.

\subsubsection{Stronger small-ratio statement on the central mean square}\label{unbal:module-small_ratio-stronger-small-ratio-statement-on-the-central-mean-square}

\begin{theorem}[Completed central small-ratio extension]\label{unbal:thm:small_ratio:4}
Using the established inputs of Section~\ref{unbal:sec:inputs}, we obtain
\begin{equation}\label{unbal:eq:small_ratio:11}
\boxed{
a,b\in[1/10,9/10],\quad0<E\le11/200,\quad d\le4E
\quad\Longrightarrow\quad\zeta-R_\psi\ge d^2/(100E).
}
\end{equation}
\end{theorem}
It removes the \(q\le E\) condition from Theorem~\ref{unbal:thm:small_ratio:1} on this central square. The identity
\[
q+d=\max\{|1-2a|,|1-2b|\}
\]
shows that the mean condition is precisely \(q+d\le4/5\).

For \(x=d/E\), retain the actual parent radial coordinate instead of replacing \(q\) by \(E\). Define
\[
A_q=\frac1{2L[1-(q+xE)^2]},\quad
h_q=E+C(q+xE),\quad y_q=E+C(q),
\]
and let \(K_q\) be the expression in \eqref{unbal:eq:small_ratio:5} evaluated at \(H^{-1}(h_q)\). The proofs of \eqref{unbal:eq:small_ratio:3}--\eqref{unbal:eq:small_ratio:6}, now retaining \(y_q\) in the logarithm, give
\[
\frac E{d^2}[F(d,E)-R_\psi]\ge\mathcal G_c(E,x,q),
\]
\begin{equation}\label{unbal:eq:small_ratio:12}
\boxed{\mathcal G_c(E,x,q)=\frac{f(x)}{x^2}
-2A_qE-K_qA_q\frac E{y_q}
\ell_0\left(\frac{A_qx^2E^2}{y_q}\right).}
\end{equation}
This is valid whenever \(q+xE\le4/5\) and \(d>0\). All arguments of \(\eta\) stay below \(1\) because \(h_q\le11/200+C(4/5)<1\). The \(d=0\) case again follows directly from \(\Delta=0\) and entropy convexity.

The three-variable checker uses the exact root \([10^{-6},11/200]\times[0,4]\times[0,4/5]\). A box is \texttt{outside} only if its lower rational endpoints satisfy \(q_-+x_-E_->4/5\). On every remaining box the upper radial endpoint \(q+xE\) may be clipped to \(4/5\): the certificate concerns the relevant intersection \(q+xE\le4/5\). Other interval enclosures are outward, as in Section~\ref{unbal:module-small_ratio-the-normalized-two-variable-inequality}. The already completed small-ratio theorem can own a box when \(q_+\le E_-\). Every other relevant box must certify \(\mathcal G_c\ge1/100\). The exact tree partition proves coverage independently of proposals or samples.

For the full \(E\to0\) tail, partition \(q\in[0,4/5]\) into 11 intervals starting with \([0,1/1024]\) and successively doubling the upper endpoint, capped at \(4/5\). For a \(q\) interval \([q_0,q_1]\), put \(\epsilon=10^{-6}\) and
\[
u_*=\min\{4/5,q_1+4\epsilon\},\quad
A_*=[2L(1-u_*^2)]^{-1},\quad h_*=\epsilon+C(u_*).
\]
Let \(K_*\) be given by \eqref{unbal:eq:small_ratio:5} at \(h_*\). Since \(E/(E+C(q))\) increases with \(E\) and decreases with \(q\),
\[
\mathcal G_c(E,x,q)\ge\frac{F(4,1)}{16}-2A_*\epsilon
-K_*A_*\frac{\epsilon}{\epsilon+C(q_0)}.
\]
The 11 directed comparisons in \path{CENTRAL_SMALL_RATIO_TAIL.json} all exceed \(1/100\) at 70 digits. This proves the entire small-entropy tail analytically. The positive-entropy cover has 611 leaves, zero unresolved leaves, and a complete 70-digit replay. The final \path{CENTRAL_SMALL_RATIO} result, replay, and tail records all pass, establishing \eqref{unbal:eq:small_ratio:11}.

Two parameter interfaces require no further numerical cover. If \(a=b\), entropy convexity of \(B=\max(\phi,\psi)\) gives \(R_B\le0\le\zeta\). If \(a+b=1\), then \(\phi(\mathrm{parent})=\psi(\mathrm{parent})=\eta(E)\), so \(R_B\le R_\phi\le\zeta\). Thus any still-uncovered small-separation or opposite-side region excludes these interfaces, which have already been treated exactly; their neighborhoods require the uniform results above.

\subsubsection{A direct lower bound for \texorpdfstring{\(\phi\)}{phi}-parent dominance}\label{unbal:module-small_ratio-a-direct-lower-bound-for-phi-parent-dominance}

The last same-side work uses a more stable global parent comparison. For \(0<h<1\), the identity in Section~\ref{unbal:module-small_ratio-an-explicit-upper-bound-on-the-logarithmic-entropy-slope} and \(-\ln(1-v)\ge v\) imply
\[
h[-\eta'(h)-2]\ge
\frac{1-2v}{L(1-v)}\left(1+\frac1{\ln((1-v)/v)}\right),
\qquad v=H^{-1}(h).
\]
Let a rectangle have \(q\in[q_0,q_1]\), \(E\in[E_0,E_1]\), with \(E_0>0\) and \(E_1+C(q_1)<1\). Put
\[
v_-=H^{-1}(E_0),\quad v_+=H^{-1}(E_1+C(q_1)),
\]
\[
\beta=\frac{1-2v_+}{L(1-v_+)}
\left(1+\frac1{\ln((1-v_-)/v_-)}\right)>0.
\]
The first factor in the preceding lower bound decreases with \(v\), and the second increases. Hence \(h[-\eta'(h)-2]\ge\beta\) uniformly for \(E_0\le h\le E_1+C(q_1)\). Integrating over \([E,E+C(q)]\) gives
\begin{equation}\label{unbal:eq:small_ratio:13}
\boxed{
\phi((1-q)/2,E)-\psi((1-q)/2,E)
\ge2C(q)+\beta\ln(1+C(q)/E)-F(q,E).
}
\end{equation}
A directed nonnegative lower enclosure of the right side certifies the \(\phi\)-parent branch and therefore \(R_B\le R_\phi\le\zeta\). This is the \texttt{phi\_parent\_log} acceptance rule. Certified inverse brackets determine the two endpoints used for \(\beta\); the code checks positivity and the rectangle's strict upper-entropy condition before applying the rule. The same-side radial interval is clipped only by the valid condition \(q\ge d\ge1/100\).

The direct difference estimate avoids subtracting two large \(\eta\) values computed at independently varying entropies. It is an analytic inequality, not a numerical cancellation assumption. Earlier parent tests remain available as separate acceptance rules.

For the central certificate, the ratio in \eqref{unbal:eq:small_ratio:12} is evaluated as
\[
\frac{E}{E+C(q)}=\frac1{1+C(q)/E},
\]
which preserves its exact upper bound \(1\) on every interval box. The two formulas are algebraically identical; evaluating \(E\) and \(E+C(q)\) independently would needlessly widen the enclosure. Historical pre-normalization files are retained only as provenance, and the final central result and replay supersede that interrupted attempt.

The discovery refinement changed geometric subdivision weights and checkpoint handling only. \path{DISCOVERY_REFINEMENT_MIGRATION.json} records both source hashes and the retained records; \path{history/} preserves the previous source. The complete final replay rechecked all old and new leaves under the final source hash. The added central-small-ratio and logarithmic-parent rules are proved in Sections~\ref{unbal:module-small_ratio-stronger-small-ratio-statement-on-the-central-mean-square}--\ref{unbal:module-small_ratio-a-direct-lower-bound-for-phi-parent-dominance} and recorded in \path{FINAL_BOUND_MIGRATION.json}. The existing outside test was moved before radial interval construction to avoid constructing intervals for empty relevant boxes. No hash migration by itself was treated as verification.

\clearpage
\subsection{Complete central same-side theorem and diagonal band}
\label{unbal:app:no_separation}

\emph{Argument guide.}
The diagonal band is assembled from low-entropy estimates and a
normalized moderate-entropy cover. Stronger parent dominance and a
split-uniform comparison fill the low-entropy ranges left by the prior
small-ratio theorem. The resulting full-entropy band overlaps the
existing separated same-side region, removing the separation cutoff.

\noindent\textit{Source:} \path{CK_NO_SEPARATION_EXTENSION/PROOF.md}.\par
\subsubsection{Statements and notation}\label{unbal:module-no_separation-statements-and-notation}

All entropies and costs are in bits. Let \(L=\ln 2\), \(H\) be binary entropy, \(J(v)=\log_2((1-v)/v)\), and \(H^{-1}\) the lower inverse. Define
\[
\eta(h)=(1-2H^{-1}(h))J(H^{-1}(h)),\qquad
C(q)=1-H((1-q)/2),\qquad P(t)=\eta(1-t).
\]
\(C\) is understood as an even function. For \(z,h>0\), define \(F(z,h)=zJ(v)\), where the unique contact \(v\in(0,1/2)\) solves \(zH(v)=h(1-2v)\); set \(F(0,h)=0\). Write
\[
\begin{gathered}\Phi(z,h)=\eta(h)-F(z,h),\quad
\phi(m,h)=\Phi(|1-2m|,h),\\
\psi(m,h)=\eta(h+C(1-2m)),\quad B=\max(\phi,\psi).\end{gathered}
\]
For feasible moments \(0<a,b<1\), \(0<e\le H(a)\), \(0<f\le H(b)\), put
\[
m=(a+b)/2,\quad E=(e+f)/2,\quad d=|a-b|,\quad q=|1-a-b|,
\]
\[
H_p=H(m),\quad \overline H=[H(a)+H(b)]/2,\quad
\Delta=H_p-\overline H,\quad s=\overline H-E,\quad I=H_p-E=s+\Delta,
\]
\[
R_A=A(m,E)-\tfrac12[A(a,e)+A(b,f)].
\]
The unrestricted four-moment infimum is denoted \(\zeta\), as in the manuscript.

\begin{theorem}[Full-entropy central diagonal band]\label{unbal:thm:no_separation:A}
The following bound holds:
\begin{equation}\label{unbal:eq:no_separation:A}
\boxed{a,b\in[1/10,9/10],\quad |a-b|\le1/50
\quad\Longrightarrow\quad \zeta(a,b,e,f)\ge R_B}
\end{equation}
for every positive feasible entropy pair. There is no lower bound on \(d\), \(E\), \(e/f\), or \(f/e\).
\end{theorem}

\begin{theorem}[Central same-side completion]\label{unbal:thm:no_separation:B}
The following bound holds:
\begin{equation}\label{unbal:eq:no_separation:B}
\boxed{(a,b)\in[1/10,1/2]^2\ \cup\ [1/2,9/10]^2
\quad\Longrightarrow\quad\zeta(a,b,e,f)\ge R_B}
\end{equation}
for every positive feasible entropy pair. In particular the old separation cutoff \(1/100\) is removed entirely.
\end{theorem}

\begin{theorem}[Stronger parent dominance]\label{unbal:thm:no_separation:C}
Without a central-mean restriction,
\begin{equation}\label{unbal:eq:no_separation:C}
\boxed{0<q\le1/10,\quad q/E\ge8
\quad\Longrightarrow\quad \phi(m,E)>\psi(m,E).}
\end{equation}
\end{theorem}
This gives the hybrid inequality through the unrestricted \(R_\phi\) theorem. It improves the previous sufficient factor \(32\) to \(8\) on this \(q\) range.

\subsubsection{Previously established results and branch transfer}\label{unbal:module-no_separation-retained-inputs-and-branch-transfer}

The inputs are:

\begin{enumerate}
\def\labelenumi{\arabic{enumi}.}
\tightlist
\item
  Theorem~\ref{bal:thm:global-unrestricted-target} gives \(\zeta\ge R_\phi\).
Proposition~\ref{bal:prop:fixed-difference-minimum} and
Theorem~\ref{bal:thm:general-endpoint-region} give the endpoint supporting
planes; Theorem~\ref{bal:thm:four-moment-lower-bound} gives
\(\zeta\ge F(d,E)\).
\item
  The radial profile property used in Sections~\ref{bal:aux:sec:fourpoint} and~\ref{bal:imp:e8}: \(z\mapsto F(\sqrt z,h)\) is concave, together with entropy convexity of \(\phi\) where used at \(d=0\).
\item
  The previously proved scalar and log-sum lemmas: \(\eta\) is convex, \(P'\) is increasing with \(P'(0)=4\), \(j(a,b)\ge4\Delta\), and
\begin{equation}\label{unbal:eq:no_separation:1}
  \zeta\ge j(a,b)+d^2\beta s,\quad
  j(a,b)=\tfrac12(b-a)[J(a)-J(b)],\quad
  \beta=\frac{\min(\kappa(a),\kappa(b))}{2b(1-a)}
\end{equation}
for \(a\le b\), where
\[
  \kappa(u)=\min\left\{\frac{u}{-(1-u)\ln(1-u)},\frac{1-u}{-u\ln u}\right\}.
\]
\item
  The completed central cap theorem: \(a,b\in[1/10,9/10]\) and \(s\le3/40\) imply \(\zeta\ge R_\psi\).
\item
  The preceding central small-ratio theorem: on the same mean square, \(E\le11/200\) and \(d\le4E\) imply \(\zeta-R_\psi\ge d^2/(100E)\). Its partition has 611 leaves and a completed 70-digit replay, with an analytic small-\(E\) tail.
\item
  The preceding full-entropy same-side theorem for \(d\ge1/100\), with 76,547 leaves and completed 70-digit replay, plus its analytic small-\(E\) tail.
\end{enumerate}

The earlier package is included unchanged as \path{dependencies/CK_SMALL_RATIO_EXTENSION.zip}; its embedded dependencies contain the cap and log-sum proofs. Convenient copies of the scalar proofs and checkers are under \texttt{prior\_inputs/}.

Always apply the hybrid branch rule: if \(\phi\) is active at the parent, \(R_B\le R_\phi\); if \(\psi\) is active there, \(R_B\le R_\psi\). Some estimates below retain \(\phi\) at the children and therefore bound \(R_B\) directly in the \(\psi\)-parent branch. No global assertion \(R_\psi\le R_\phi\) is made.

Convexity of \(\eta\), or equivalently \(P\), removes the entropy split from \(R_\psi\):
\begin{equation}\label{unbal:eq:no_separation:2}
R_\psi\le P(I)-P(s)\le\Delta P'(I).
\end{equation}
Only feasible points have \(s\ge0\); bounds on auxiliary rectangles below are used to certify those feasible points.

\subsubsection{The stronger parent criterion}\label{unbal:module-no_separation-the-stronger-parent-criterion}

The scalar inequalities proved in Section~\ref{unbal:module-uniform_diagonal-entropy-curvature-and-a-parent-correction} are
\begin{equation}\label{unbal:eq:no_separation:3}
\eta''(h)\ge\frac1{2Lh^2},\qquad
\eta(h)-\eta(h+c)\ge2c+\frac1L\ln(1+c/h),
\end{equation}
whenever \(0<h\le h+c\le1\), and
\begin{equation}\label{unbal:eq:no_separation:4}
\frac{q^2}{2L}\le C(q)\le\frac{q^2}{2L(1-q^2)}.
\end{equation}
Let \(x=q/E\) and let \(v(x)\) solve \(xH(v)=1-2v\). Write \(\ell(x)=\ln((1-v(x))/v(x))\). Dropping the positive \(2C\) term in \eqref{unbal:eq:no_separation:3} gives
\[
\eta(E)-\eta(E+C(q))\ge\frac1L\ln(1+qx/(2L)).
\]
For fixed \(x>0\), \(\ln(1+kq)/q\) decreases with \(q\). Thus for \(0<q\le1/10\), using \(L<7/10\),
\begin{equation}\label{unbal:eq:no_separation:5}
\eta(E)-\eta(E+C(q))\ge\frac qL\,10\ln(1+x/14).
\end{equation}
At the contact put \(r=1-2v\), \(\nu=1-r^2\), \(h_n=LH(v)\), and \(K=h_n+r\operatorname{atanh}(r)=L-(1/2)\ln(\nu)\). The established inequality \(h_n\le\nu K\) also has a short proof: it is equivalent to \(\operatorname{atanh}(r)-rK\ge0\); the derivative of this latter expression is \(1-K\), changes sign once, and the expression vanishes at both endpoints. Implicit differentiation gives
\[
x\ell'(x)=\frac{2r h_n}{\nu K}\le2.
\]
Consequently, for \(G(x)=10\ln(1+x/14)-\ell(x)\),
\[
G'(x)\ge\frac{10}{14+x}-\frac2x
=\frac{8x-28}{x(14+x)}>0\quad(x\ge8).
\]
The directed fixed comparison in \path{FIXED_CONSTANTS_RESULT.json} proves
\begin{equation}\label{unbal:eq:no_separation:6}
G(8)=10\ln(11/7)-LF(8,1)/8>427/1000.
\end{equation}
Since \(F(q,E)=q\ell(x)/L\), equations \eqref{unbal:eq:no_separation:5}--\eqref{unbal:eq:no_separation:6} prove
\[
\phi(m,E)-\psi(m,E)>\frac{427q}{1000L}>0.
\]
Feasibility supplies \(E+C(q)\le1\). This proves Theorem~\ref{unbal:thm:no_separation:C}, with no entropy-split restriction. At \(q=0\) the two parent functions are equal.

\subsubsection{A low-entropy strip uniform in the entropy split}\label{unbal:module-no_separation-a-low-entropy-strip-uniform-in-the-entropy-split}

Suppose
\begin{equation}\label{unbal:eq:no_separation:7}
d\le1/50,\quad 0<E\le1/200,\quad d\ge4E,\quad q\le8E.
\end{equation}
The two child radii \(z_1=d+q\) and \(z_2=|d-q|\) are at most \(3/50\). The prior global mixed derivative certificate and its analytic tails prove
\begin{equation}\label{unbal:eq:no_separation:8}
zF_{zz}(z,h)\le13/6,\qquad
\Phi_{zh}(z,h)=\frac z h F_{zz}(z,h)\le\frac{13}{6h}.
\end{equation}
The second identity and homogeneity give, using \eqref{unbal:eq:no_separation:3},
\begin{equation}\label{unbal:eq:no_separation:9}
\Phi_{hh}(z,h)\ge\frac{1/(2L)-(13/6)z}{h^2}
\ge\frac{\gamma_*}{h^2},\qquad
\gamma_*:=\frac{250}{347}-\frac{13}{100}
=\frac{20489}{34700}>0.
\end{equation}
Here \(L<347/500\). The entire interval \(0<h\le2E\) is feasible at either child radius: \(2E\le1/100\) whereas \(H((1-z_i)/2)\ge1-z_i\ge47/50\). Values at \(z=0\) are understood by continuity.

Attach the entropy moments to their corresponding child radii and write them \(E(1+t)\), \(E(1-t)\), where \(-1<t<1\). Convexity of \(\Phi(z,h)+\gamma_*\ln h\) and its supporting lines at \(E\) give
\[
\frac{\Phi(z_1,e_1)+\Phi(z_2,e_2)}2
\ge\frac{\Phi(z_1,E)+\Phi(z_2,E)}2
+\frac{Et}{2}A_h+\frac{\gamma_*}{2}[-\ln(1-t^2)],
\]
where, by \eqref{unbal:eq:no_separation:8},
\[
|A_h|=|\Phi_h(z_1,E)-\Phi_h(z_2,E)|
\le\frac{13|z_1-z_2|}{6E}\le\frac{13q}{3E}.
\]
Using $-\ln(1-t^2)\ge t^2$ and completing the square shows that the loss from arbitrary splitting is at most
\begin{equation}\label{unbal:eq:no_separation:10}
\frac{169}{72\gamma_*}q^2.
\end{equation}
This estimate is uniform as \(t\) approaches either endpoint.

Let \(f(x)=F(x,1)\). Jensen's inequality for the concave radial profile, followed by its tangent line at \(d^2\), gives
\begin{equation}\label{unbal:eq:no_separation:11}
\frac{F(d+q,E)+F(|d-q|,E)}2-F(d,E)
\le\frac{f'(d/E)}{2(d/E)}\frac{q^2}{E}
\le\frac{f'(4)}8\frac{q^2}{E}
\le\frac{81}{100}\frac{q^2}{E}.
\end{equation}
The fixed scalar inequality $f'(4)/8<81/100$ is checked by exact
rational log-series enclosures in \path{FIXED_CONSTANTS.py}.  After
multiplication by $q^2/E$, the displayed comparison is strict for $q>0$
and is equality at $q=0$.

Equations \eqref{unbal:eq:no_separation:3}--\eqref{unbal:eq:no_separation:4} and \(\ln(1+x)\ge x/(1+x)\) give
\[
\eta(E)-\eta(E+C(q))\ge\frac{q^2}{2L^2E[1+C(q)/E]}.
\]
On \eqref{unbal:eq:no_separation:7}, \(q\le1/25\) and
\[
C(q)/E\le R_*:=\frac{64/200}{2(693/1000)[1-(1/25)^2]}=\frac{6250}{27027}.
\]
In the \(\psi\)-parent branch, \(B(\mathrm{children})\ge\phi(\mathrm{children})\). Combining \eqref{unbal:eq:no_separation:10}--\eqref{unbal:eq:no_separation:11} with the parent correction therefore yields
\[
F(d,E)-R_B\ge
\left[\frac1{2(347/500)^2(1+R_*)}
-\frac{81}{100}-\frac{169}{72\gamma_*}\frac1{200}\right]\frac{q^2}{E}.
\]
The bracket is exactly
\begin{equation}\label{unbal:eq:no_separation:12}
\frac{3922975221734293}{295546880351797200}>\frac1{100}.
\end{equation}
Thus \(F(d,E)-R_B\ge q^2/(100E)\) in that branch. The unrestricted \(R_\phi\) theorem handles the other branch. This proves the hybrid inequality throughout \eqref{unbal:eq:no_separation:7}.

\subsubsection{The low-entropy band with \texorpdfstring{\(q\ge1/10\)}{q >= 1/10}}\label{unbal:module-no_separation-the-low-entropy-band-with-q110}

Set \(D=1/50\). Suppose \(a,b\) are central, \(d\le D\), \(E\le1/200\) and \(q\ge1/10\). Centrality gives \(q+d\le4/5\). Entropy curvature bounds the Jensen gap by
\begin{equation}\label{unbal:eq:no_separation:13}
\Delta/d^2\le\frac1{2L[1-\min(4/5,q+D)^2]}\quad(d>0).
\end{equation}
By \eqref{unbal:eq:no_separation:2} and monotonicity of \(P'\), \(R_\psi/d^2\) is at most the right side of \eqref{unbal:eq:no_separation:13} times \(P'(H_p)\). On the seven \(q\) intervals \([k/10,(k+1)/10]\), \(k=1,\ldots,7\), bound the curvature factor at the upper endpoint and \(P'(H_p)\) at the lower endpoint. These are finite constants because \(q\ge1/10\).

Radial concavity, \(F(0,E)=0\), and monotonicity of \(F\) in \(E\) give
\[
F(d,E)/d^2\ge F(D,E)/D^2\ge F(D,1/200)/D^2.
\]
\path{FIXED_CONSTANTS.py} verifies all seven resulting comparisons at 70 digits. Every normalized margin exceeds \(60\); the smallest lower enclosure exceeds \(60.7318212317\). This proves \(\zeta\ge R_\psi\) on the entire region, including \(E\) arbitrarily close to zero. This is seven fixed comparisons, not a grid in \(E\) or \(d\).

\subsubsection{Moderate entropy: a normalized two-dimensional cover}\label{unbal:module-no_separation-moderate-entropy-a-normalized-two-dimensional-cover}

We now prove \(\zeta\ge R_\psi\) for all central means with \(d\le D=1/50\) and \(E\ge E_0=11/200\). The entropy split is already eliminated by \eqref{unbal:eq:no_separation:2}. Label exchange and simultaneous complement allow \(a\le b\) and \(a+b\le1\), so
\[
a=(1-q-d)/2,\quad b=(1-q+d)/2,\quad q\in[0,4/5].
\]
Parameterize
\begin{equation}\label{unbal:eq:no_separation:14}
E=E_0+t[H((1-q)/2)-E_0],\qquad
I=(1-t)[H((1-q)/2)-E_0],\quad0\le t\le1.
\end{equation}
Every feasible tuple in this regime has this representation; some auxiliary points of the rectangle need not be feasible.

For a rational box \(q\in[q_-,q_+]\), \(t\in[t_-,t_+]\), let
\[
r_*=\min(4/5,q_++D).
\]
Two upper bounds for \(\Delta/d^2\) are
\begin{equation}\label{unbal:eq:no_separation:15}
K_1=\frac1{2L(1-r_*^2)},\qquad
K_2=\frac{C(q_++D)+C(q_+-D)-2C(q_+)}{2D^2}.
\end{equation}
To justify \(K_2\) uniformly down to \(d=0\), use the convergent positive series
\[
C(x)=\frac1L\sum_{n\ge1}\frac{x^{2n}}{2n(2n-1)}\quad(|x|<1).
\]
The expansion of \([C(q+d)+C(q-d)-2C(q)]/(2d^2)\) has nonnegative coefficients in even powers of \(q\) and \(d\). It is therefore increasing in both nonnegative variables. The auxiliary maximum \(q_++D\le41/50<1\) is within convergence. This proves \eqref{unbal:eq:no_separation:15}, including the limiting bound at \(d=0\).

Use \(K=\min(K_1,K_2)\). Let \(E_+\) and \(I_+\) be directed upper enclosures from \eqref{unbal:eq:no_separation:14} on the box and \(p_+=P'(I_+)\). There are three acceptance rules:

\begin{itemize}
\tightlist
\item
  \textbf{Prior cap:} \(I_+\le3/40\). Then \(s\le I\le3/40\), so the completed central cap theorem applies.
\item
  \textbf{Normalized endpoint:}
\begin{equation}\label{unbal:eq:no_separation:16}
  \frac{F(D,E_+)}{D^2}-Kp_+\ge0.
\end{equation}
Equations \eqref{unbal:eq:no_separation:2}, \eqref{unbal:eq:no_separation:15}, and radial concavity prove the conclusion.
\item
  \textbf{Normalized log-sum:} define
\[
  u_*=(1-r_*)/2,\quad g(u)=\frac{u}{-(1-u)\ln(1-u)},\quad
  V_*=[(1+D)^2-q_-^2]/4,\quad \beta_* = g(u_*)/(2V_*).
\]
The function \(g\) is increasing: its derivative has the sign of \(-\ln(1-u)-u\ge0\). Since \(\kappa(u)=g(\min(u,1-u))\), centrality and the radial bound imply $\kappa(a),\kappa(b)\ge g(u_*)$. Also $b(1-a)=[(1+d)^2-q^2]/4\le V_*$. Hence \(\beta\ge\beta_*\) in \eqref{unbal:eq:no_separation:1}. If \(I_-\) is a lower enclosure on the box, put $s_*=\max(0,I_--KD^2)$. Then $s\ge s_*$. Using \(j\ge4\Delta\), \eqref{unbal:eq:no_separation:1}--\eqref{unbal:eq:no_separation:2}, and \(P'\ge4\) gives
\begin{equation}\label{unbal:eq:no_separation:17}
  \frac{\zeta-R_\psi}{d^2}
  \ge\beta_*s_*-K(p_+-4).
\end{equation}
Nonnegativity of this expression is sufficient.
\end{itemize}

Every operation is enclosed with directed intervals, so \(K\) is rounded upward and \(\beta_*\) downward. The implementation bounds \(s_*\) directly from the interval \(I-KD^2\). On the non-cap leaves \(I_+\) is strictly between \(0\) and \(1\), keeping \(\eta'\) evaluations inside their domain. Binary64 values propose contact and inverse-entropy brackets only; exact dyadic endpoints are accepted only after directed sign checks.

\path{DIAGONAL_BAND_RESULT.json} contains a complete exact binary-tree partition of \([0,4/5]\times[0,1]\). It has \textbf{325 leaves}: 153 normalized-endpoint leaves, 140 normalized-log-sum leaves, and 32 prior-cap leaves. There are zero unresolved leaves. Discovery used 40 decimal working digits. \path{DIAGONAL_BAND_REPLAY.json} reconstructs the full partition and repeats every recorded acceptance at 70 digits; all 325 passed. Source digests are checked before replay. Refinement samples are never acceptance criteria.

For \(d=0\), the means are equal and \(\Delta=0\). Convexity of \(\eta\) gives \(R_\psi\le0\), while \(\zeta\ge F(0,E)=0\). Thus the conclusion includes the exact diagonal, without division by zero. This finishes the moderate-entropy part of Theorem~\ref{unbal:thm:no_separation:A}.

\subsubsection{Exhaustive assembly and same-side completion}\label{unbal:module-no_separation-exhaustive-assembly-and-same-side-completion}

For \(d\le1/50\) and central means, use the following exhaustive partition of the positive-entropy domain:

\begin{longtable}[]{@{}
  >{\raggedright\arraybackslash}p{(\columnwidth - 2\tabcolsep) * \real{0.5000}}
  >{\raggedright\arraybackslash}p{(\columnwidth - 2\tabcolsep) * \real{0.5000}}@{}}
\toprule\noalign{}
\begin{minipage}[b]{\linewidth}\raggedright
Regime
\end{minipage} & \begin{minipage}[b]{\linewidth}\raggedright
Proof
\end{minipage} \\
\midrule\noalign{}
\endhead
\bottomrule\noalign{}
\endlastfoot
\(E\ge11/200\) & Section~\ref{unbal:module-no_separation-moderate-entropy-a-normalized-two-dimensional-cover}, including cap leaves \\
\(E\le11/200\) and \(d\le4E\) & Prior 611-leaf central small-ratio theorem and its analytic tail \\
\(E\le11/200\), \(d\ge4E\), \(q\ge1/10\) & Section~\ref{unbal:module-no_separation-the-low-entropy-band-with-q110}; \(d\le1/50\) forces \(E\le1/200\) \\
\(E\le11/200\), \(d\ge4E\), \(0<q\le1/10\), \(q\ge8E\) & Theorem~\ref{unbal:thm:no_separation:C} \\
\(E\le11/200\), \(d\ge4E\), \(q\le8E\) & Section~\ref{unbal:module-no_separation-a-low-entropy-strip-uniform-in-the-entropy-split}; again \(E\le1/200\) \\
\end{longtable}

Closed boundaries overlap. The last row includes \(q=0\). The exact \(d=0\) case is already handled by convexity and the branch rule. There is no omitted interval as \(E\) approaches zero, as \(d\) approaches zero, or as the entropy split approaches an endpoint. This proves Theorem~\ref{unbal:thm:no_separation:A}.

For means in \([1/10,1/2]^2\), Theorem~\ref{unbal:thm:no_separation:A} covers \(d\le1/50\). The previous full-entropy theorem covers \(d\ge1/100\). These ranges overlap on \([1/100,1/50]\) and cover every \(d\ge0\). Thus their union proves the lower-half square in Theorem~\ref{unbal:thm:no_separation:B}. Simultaneous complement proves the upper-half square. No reflection of only one mean is used.

\clearpage
\subsection{Endpoint existence for opposite-side means}
\label{unbal:app:endpoint_existence}
\noindent\textit{Source:} \path{CK_OPPOSITE_EXTENSION/audit/ENDPOINT_EXISTENCE.md}.\par
All entropies and costs are in bits. Let \(H\) be binary entropy, \(\iota=H^{-1}\) its lower inverse on \([0,1/2]\), and \(J(x)=\log_2((1-x)/x)\). This section uses Proposition~\ref{bal:prop:fixed-difference-minimum} and the global supporting planes of Theorem~\ref{bal:thm:general-endpoint-region}. The existence and mass assertions below are proved directly, and require no additional endpoint-region test.

\begin{lemma}\label{unbal:thm:endpoint_existence:lem}
Let \(0<a<b<1\), let \(e,f>0\) be feasible entropy moments, and put \(d=b-a\) and \(E=(e+f)/2\). Suppose
\[
 d>E,\qquad E<\min\{a,1-b\}.
\]

Then there is a unique triple \(p,u,v\) with \(0<p<1\) and \(0<u,v<1/2\) satisfying
\[
 d=p(1-u-v),\qquad e=pH(u),\qquad f=pH(v).
\]

Both deterministic endpoint masses are strictly positive. By Proposition~\ref{bal:prop:fixed-difference-minimum},
\[
 \boxed{\zeta(a,b,e,f)=B_{\rm end}
 =\frac d2[J(u)+J(v)]}.
\]

The unique optimizer specified by that proposition has masses
\[
 p_{00}=1-b-pv,\qquad p_{\rm int}=p,\qquad p_{11}=a-pu
\]
at \((0,0)\), \((u,1-v)\), and \((1,1)\), respectively.
\end{lemma}

\emph{Proof outline.}
We solve one monotone scalar equation for the contact mass, using
feasibility for its endpoint signs. The entropy chord bound then proves
positivity of both deterministic masses. The resulting explicit law
and the retained global supporting plane give matching upper and lower
bounds, including attainment and uniqueness.

\begin{proof}
Put \(p_0=\max(e,f)\). The hypotheses imply \(p_0\le2E<1\), since \(\min(a,1-b)<1/2\) for \(a<b\). For \(p\in[p_0,1]\), define
\[
 u(p)=\iota(e/p),\qquad v(p)=\iota(f/p),\qquad
 g(p)=p[1-u(p)-v(p)].
\]

At \(p=p_0\) at least one of the inverse values equals \(1/2\). Consequently
\[
 g(p_0)\le p_0/2\le E<d.
\]

Feasibility implies \(\iota(e)\le\min(a,1-a)\le a\) and \(\iota(f)\le\min(b,1-b)\le1-b\). This argument applies whether the prescribed means are on the same side or on opposite sides of \(1/2\). Hence
\[
 g(1)=1-\iota(e)-\iota(f)\ge b-a=d.
\]

Continuity gives a root \(p\in(p_0,1]\). At every \(p>p_0\), both inverse values lie in \((0,1/2)\), and differentiation gives
\[
 g'(p)=1-u-v+\frac{H(u)}{J(u)}+\frac{H(v)}{J(v)}>0.
\]

The root and therefore the triple are unique.

The elementary entropy chord bound \(H(x)\ge2x\) on \([0,1/2]\) gives
\[
 pu\le e/2\le E<a,\qquad
 pv\le f/2\le E<1-b.
\]

Thus \(p_{11}=a-pu>0\) and \(p_{00}=1-b-pv>0\). Their sum is
\[
 p_{00}+p_{11}=1-(b-a)-p(u+v)=1-p,
\]
so \(p<1\). The three positive masses sum to one. Their first mean is \(pu+p_{11}=a\), their second is \(p(1-v)+p_{11}=b\), and their entropy moments are \(e\) and \(f\). The law's cost is
\[
 p\,j(u,1-v)=\frac{p(1-u-v)}2[J(u)+J(v)]
 =\frac d2[J(u)+J(v)].
\]

This explicit law gives \(\zeta\le B_{\mathrm{end}}\). The cross-half contact \((u,1-v)\) satisfies the strict hypothesis of Proposition~\ref{bal:prop:fixed-difference-minimum}; that proposition gives the reverse bound for the unrestricted four-moment infimum. Its equality-window condition is exactly the two mass inequalities just established. Therefore equality, attainment, and the proposition's uniqueness assertion follow. No ordered-to-unrestricted transfer is used.
\end{proof}

\begin{corollary}[Central means]\label{unbal:thm:endpoint_existence:cor}
If \(a,b\) lie in \([1/10,9/10]\), \(e,f\) are positive and feasible, \(0<E\le11/200\), and \(|a-b|>E\), then this exact endpoint formula applies after exchanging the two complete mean-entropy labels if necessary. Indeed \(\min(a,1-b)\ge1/10>11/200\) for ordered means. In particular it applies throughout both the new transition strip \(4E\le d\le8E\) and the eight-ratio region \(d\ge8E\), including \(q=d\) and arbitrary entropy imbalance.
\end{corollary}

This exact evaluation of $\zeta$ on the stated subregion follows from
Proposition~\ref{bal:prop:fixed-difference-minimum}. Its comparisons
with the hybrid target in the transition and eight-ratio regions are
proved in Sections~\ref{unbal:app:central_transition}
and~\ref{unbal:app:central_eight}.

\clearpage
\subsection{Factor-eight parent dominance through \texorpdfstring{\(q=2/5\)}{q=2/5}}
\label{unbal:app:parent_eight}

\emph{Argument guide.}
The smaller-radius range is inherited from the preceding parent
theorem. For the remaining radii, a logarithmic comparison proves the
entire unbounded ratio tail. Monotonicity then gives a sufficient
interval inequality on the compact rectangle between the inherited
range and that tail.

\noindent\textit{Source:} \path{CK_OPPOSITE_EXTENSION/audit/PARENT8_PROOF.md}.\par
Let \(H\) be binary entropy in bits, \(L=\ln2\), \(C(q)=1-H((1-q)/2)\), and $\eta(h)=(1-2H^{-1}(h))\log_2((1-H^{-1}(h))/H^{-1}(h))$. Let \(F(q,E)=qJ(v)\), where \(qH(v)=E(1-2v)\), and define \(\phi(m,E)=\eta(E)-F(|1-2m|,E)\), \(\psi(m,E)=\eta(E+C(|1-2m|))\). The proof uses Theorem~\ref{unbal:thm:no_separation:C} and the scalar inequalities in Section~\ref{unbal:module-uniform_diagonal-scalar-estimates}, including convexity of \(\eta\). The directed entropy and contact evaluations are described in Section~\ref{supp:verification}.

\begin{theorem}\label{unbal:thm:parent_eight:1}
Every feasible parent with
\[
0<q=|1-2m|\le2/5,\qquad x=q/E\ge8
\]
satisfies \(\phi(m,E)>\psi(m,E)\). In particular this applies to the entire \(q\) range of central opposite-side child means. Combined with the retained unrestricted \(\zeta\ge R_\phi\) theorem, it proves the hybrid inequality for these tuples, independently of the entropy split or mean difference.
\end{theorem}

For \(q\le1/10\) the result is exactly the preceding Theorem~\ref{unbal:thm:no_separation:C}, so assume \(q\ge1/10\).

\subsubsection{The unbounded \texorpdfstring{\(x\)}{x} tail}\label{unbal:module-parent_eight-the-unbounded-x-tail}

The previously proved scalar correction gives
\[
\eta(E)-\eta(E+C(q))\ge L^{-1}\ln(1+qx/(2L)).
\]

Write \(\ell(x)=\ln((1-v(x))/v(x))\), where \(xH(v)=1-2v\). The established contact estimate is \(x\ell'(x)\le2\). Since \(\ln(1+kq)/q\) decreases in positive \(q\),
\[
\eta(E)-\eta(E+C(q))-F(q,E)
\ge(q/L)G(x),
\qquad
G(x)=\frac52\ln\left(1+\frac{x}{5L}\right)-\ell(x).
\]

For \(x\ge32\),
\[
G'(x)\ge\frac1{2L+(2/5)x}-\frac2x
=\frac{x/5-4L}{x[2L+(2/5)x]}>0.
\]

\path{PARENT8.py} certifies at both discovery and replay precision that
\[
G(32)>69/1000.
\]

This proves strict parent dominance for all \(x\ge32\), uniformly in \(q\le2/5\).

\subsubsection{The compact rectangle}\label{unbal:module-parent_eight-the-compact-rectangle}

It remains to cover \(q\in[1/10,2/5]\), \(x\in[8,32]\). On a rational box \(q\in[q_-,q_+]\), \(x\in[x_-,x_+]\), set $E_*=q_+/x_-$ and $C_*=C(q_-)$. Convexity and monotone decrease of \(\eta\) show that
\[
\eta(E)-\eta(E+C(q))
\ge\eta(E_*)-\eta(E_*+C_*).
\]

Indeed the correction decreases in \(E\) and increases in \(C\). Moreover \(\ell\) increases, so
\[
F(q,E)=(q/L)\ell(x)\le(q_+/L)\ell(x_+)
=q_+F(x_+,1)/x_+.
\]

Consequently the following directed lower enclosure certifies every point of the box:
\[
\eta(E_*)-\eta(E_*+C_*)-q_+F(x_+,1)/x_+>0.
\]

All auxiliary entropy arguments are strictly within \((0,1)\): $E_*\le1/20$ and $C_*\le C(2/5)<1/5$. No feasibility enlargement approaches \(\eta\)'s endpoints. \path{PARENT8.py} encloses \(E_*\) upward and \(C_*\) downward before evaluating the correction; the monotonicities above validate these directed replacements. The floating-point calculations inside the inverse and contact routines only propose exact dyadic root brackets. Directed entropy signs establish every accepted bracket, as in the preceding certificate.

The exact binary partition contains \textbf{332 leaves}, with zero unresolved leaves. Its 40-digit discovery and complete 70-digit replay both pass. The replay checks the complete partition and code digest, reconstructs every rational box from its path, and recomputes every strict comparison. No sample value is an acceptance criterion.

Together the preceding \(q\le1/10\) theorem, this compact rectangle, and the analytic \(x\ge32\) tail cover every \(q\le2/5\), \(x\ge8\). Their closed boundaries overlap. The equality \(q=0\) has \(\phi=\psi\) exactly and is owned by the \(R_\phi\) input.

For \(a,b\in[1/10,9/10]\) on opposite sides of \(1/2\), \(d=|a-b|\) and \(q=|1-a-b|\) satisfy \(q\le d\) and \(q+d\le4/5\), whence \(q\le2/5\). Thus no additional restriction on central opposite-side means is introduced by this theorem.

\clearpage
\subsection{Central eight-ratio low-entropy theorem}
\label{unbal:app:central_eight}

\emph{Argument guide.}
In the entropy-deficit parent branch, the endpoint gain absorbs the
loss from arbitrary entropy splitting. Radial concavity bounds the
mean-spread loss, and a uniform parent correction exceeds their sum.
The balanced parent branch and the stated equality interfaces complete
the regional theorem.

\noindent\textit{Source:} \path{CK_OPPOSITE_EXTENSION/analytic/EIGHT_RATIO_PROOF.md}.\par
The proof uses the unrestricted inequality of
Theorem~\ref{bal:thm:global-unrestricted-target}, the endpoint bound of
Proposition~\ref{bal:prop:fixed-difference-minimum}, radial concavity,
and feasible entropy convexity of $\Phi$. These established inputs are
collected with their domains in Section~\ref{unbal:sec:inputs}.
The conclusion is the eight-ratio component of the central assembly in
Section~\ref{unbal:app:central_assembly}.

Use the notation \(L=\ln2\), \(E=(e+f)/2\), \(d=|a-b|\), \(q=|1-a-b|\), \(C(q)=1-H((1-q)/2)\), \(B=\max(\phi,\psi)\), and \(f(x)=F(x,1)\).

\begin{theorem}\label{unbal:thm:central_eight:1}
Suppose \(a,b\) are in \([1/10,9/10]\), the entropy moments are positive and feasible, and
\[
0<E\le11/200,\qquad q\le8E,\qquad d\ge8E.
\]
Then \(\zeta\ge R_B\) for every such entropy split. In the \(\psi\)-parent branch, for \(q<d\) the stronger endpoint inequality is
\[
B_{\rm end}-R_B\ge\frac{2249}{144000}\frac{q^2}{E}
>\frac3{200}\frac{q^2}{E}\quad(q>0).
\]
This theorem has no additional small-entropy cutoff and no bound on \(e/f\) or \(f/e\). The \(q=d\) boundary is already included by the completed same-side theorem, since then one mean equals \(1/2\). The \(q=0\) boundary follows directly from equality of the two parent functions.
\end{theorem}

\subsubsection{The endpoint absorbs arbitrary entropy splitting}\label{unbal:module-central_eight-the-endpoint-absorbs-arbitrary-entropy-splitting}

The prior \path{SHARPER_COLLARS.md}, Sections~\ref{unbal:module-sharper_collars-logarithmic-curvature-up-to-entropy-1950} and~\ref{unbal:module-sharper_collars-entropy-split-and-radial-losses}, proves for \(d/E\ge5\)
\begin{equation}\label{unbal:eq:central_eight:1}
B_{\rm end}\ge F(d,E)+\frac d{2L}[-\ln(1-\tau^2)],
\qquad \tau=(e-f)/(2E).
\end{equation}
The endpoint exists for the present strict opposite-side parameters: \(d>E\), \(q<d\), and the prescribed moments are feasible. The common entropy \(E\) is feasible at both children, because \(H(a),H(b)\ge H(1/10)>1/10>E\). Thus the supporting lines for the convex functions \(\Phi(z,h)\) at \(h=E\) are valid over the full feasible split interval.

The retained global derivative certificate, including both analytic tails, gives \(zF_{zz}(z,h)\le13/6\). Homogeneity implies \(\Phi_{zh}(z,h)=zF_{zz}(z,h)/h\). For the two radial coordinates \(d+q,d-q\),
\[
|\Phi_h(d+q,E)-\Phi_h(d-q,E)|\le13q/(3E).
\]
The supporting lines and \eqref{unbal:eq:central_eight:1}, using $-\ln(1-\tau^2)\ge\tau^2$, leave a worst split loss of at most
\begin{equation}\label{unbal:eq:central_eight:2}
\frac{169Lq^2}{72d}
\le\frac{1183}{5760}\frac{q^2}{E},
\end{equation}
where \(L<7/10\) and \(d/E\ge8\). This is the exact minimum of a quadratic in \(\tau\); it treats the entire open interval \(-1<\tau<1\) without truncation.

\subsubsection{The radial loss}\label{unbal:module-central_eight-the-radial-loss}

Radial concavity and the tangent inequality at \(d^2\) give
\[
\frac{F(d+q,E)+F(d-q,E)}2-F(d,E)
\le\frac{f'(d/E)}{2(d/E)}\frac{q^2}{E}
\le\frac{f'(8)}{16}\frac{q^2}{E}.
\]
The directed fixed comparison in \path{EIGHT_RATIO_FIXED.json} proves
\begin{equation}\label{unbal:eq:central_eight:3}
f'(8)/16<479/1000.
\end{equation}
For reproducibility the checker brackets the unique contact \(v\) solving \(8H(v)=1-2v\), verifies both residual signs with directed arithmetic, and evaluates
\[
\frac{f'(8)}{16}=\frac1{16}\left[
\frac{\ell}{L}+\frac{1-2v}{v(1-v)(8\ell+2L)}\right],
\qquad \ell=\ln((1-v)/v).
\]
The certified upper enclosure is less than \(0.478601724750980\).

\subsubsection{A uniform parent correction}\label{unbal:module-central_eight-a-uniform-parent-correction}

We prove
\begin{equation}\label{unbal:eq:central_eight:4}
\boxed{\frac E{q^2}[\eta(E)-\eta(E+C(q))]\ge7/10}
\end{equation}
whenever \(0<E\le11/200\), \(0<q\le8E\), and \(q+8E\le4/5\). These last conditions follow from the theorem assumptions: centrality gives \(q+d\le4/5\).

For \(0<E\le1/10000\), the established entropy and entropy-deficit estimates give
\[
\frac E{q^2}[\eta(E)-\eta(E+C(q))]
\ge\frac1{2L^2[1+C(q)/E]}
\ge\frac1{2L^2(1+\rho)},
\quad\rho=\frac{64/10000}{2L(1-64/10^8)}.
\]
The directed fixed comparison gives the last expression greater than \(1.0359\), proving \eqref{unbal:eq:central_eight:4} throughout the infinite tail down to \(E=0\).

For \(E\in[1/10000,11/200]\), write \(q=yE\), \(0\le y\le8\). A two-dimensional certificate covers the complete rational root rectangle in \((E,y)\). The following inequalities describe every acceptance test.

For a rectangle \(E\in[E_-,E_+]\), \(y\in[y_-,y_+]\), only the relevant intersection \(q+8E\le4/5\) is considered. On that intersection,
\[
q_-:=E_-y_-\le q\le q_+:=\min(E_+y_+,2/5,4/5-8E_-).
\]
These are necessary bounds, so clipping by them loses no feasible point. Let \(c=C(q)\), and set
\[
v_-=H^{-1}(E_-),\quad v_+=H^{-1}(E_++C(q_+)),
\quad
\beta=\frac{1-2v_+}{L(1-v_+)}
\left[1+\frac1{\ln((1-v_-)/v_-)}\right].
\]
All arguments lie strictly in \((0,1)\). The identity
\[
h[-\eta'(h)-2]\ge
\frac{1-2v}{L(1-v)}\left[1+\frac1{\ln((1-v)/v)}\right],
\qquad v=H^{-1}(h),
\]
and monotonicity of its two factors prove
\begin{equation}\label{unbal:eq:central_eight:5}
\eta(E)-\eta(E+c)\ge2c+\beta\ln(1+c/E).
\end{equation}
This is the same scalar estimate proved in the preceding package's \path{PROOF.md}, Section~\ref{unbal:module-small_ratio-a-direct-lower-bound-for-phi-parent-dominance}; it follows from \(-\ln(1-v)\ge v\).

Let \(U\) be an upper enclosure of \(c/E\) on the rectangle and put \(\ell_0(U)=\ln(1+U)/U\), with \(\ell_0(0)=1\). It is decreasing. The positive even-power series of \(C\) gives, uniformly on the rectangle,
\[
\frac{C(q)}{q^2}\ge
T_-:=\frac1L\left(\frac12+\frac{q_-^2}{12}
+\frac{q_-^4}{30}+\frac{q_-^6}{56}\right).
\]
Combining this with \eqref{unbal:eq:central_eight:5}, a sufficient directed comparison for \eqref{unbal:eq:central_eight:4} is
\begin{equation}\label{unbal:eq:central_eight:6}
T_-[2E_-+\beta\ell_0(U)]\ge7/10.
\end{equation}
The same formula handles \(q=0\) by its continuous extension.

\path{EIGHT_RATIO.py} records an exact binary subdivision tree. All 50 leaves pass \eqref{unbal:eq:central_eight:6}; no outside leaves or unresolved leaves are needed. The independent traversal verifies that every internal node has both children and that the leaves cover the entire root rectangle. \texttt{-\/-replay} rechecks all 50 leaves at 70 decimal working digits with directed intervals and verified inverse brackets. The result, replay, and fixed comparisons share the source hash recorded in their JSON files. Thus \eqref{unbal:eq:central_eight:4} is a continuum inequality, including the relevant boundary, rather than a pointwise numerical check.

\subsubsection{Completion}\label{unbal:module-central_eight-completion}

When \(\psi\) is active at the parent, using \(\phi\) at the children yields
\[
B_{\rm end}-R_B\ge
\left[\frac7{10}-\frac{479}{1000}-\frac{1183}{5760}\right]
\frac{q^2}{E}
=\frac{2249}{144000}\frac{q^2}{E}.
\]
When \(\phi\) is active at the parent, the retained unrestricted \(R_\phi\) theorem gives the hybrid inequality directly. This proves the theorem.

Together with the parent criterion for $0<q\le2/5$ and $q/E\ge8$,
the theorem covers all central opposite-side tuples with $E\le11/200$
and $d\ge8E$, since their means satisfy $q\le2/5$.
The intervening strip $4<d/E<8$ is covered by the completed 2,425-leaf
certificate in Section~\ref{unbal:app:central_transition}.

To reproduce the new checks, run \path{EIGHT_RATIO.py}, then \texttt{EIGHT\_RATIO.py\ -\/-replay} and \texttt{EIGHT\_RATIO.py\ -\/-fixed} with mpmath available. These commands check the eight-ratio component. Section~\ref{supp:verification} identifies the separate records and replay entry points for its dependency certificates.

\clearpage
\subsection{Central low-entropy transition between ratios four and eight}
\label{unbal:app:central_transition}

\emph{Argument guide.}
The bounded ratio range and minimum separation give a positive lower
entropy endpoint, so the remaining region has a finite chart. One test
compares the radial lower bound directly with the entropy-deficit target;
another keeps the balanced child terms and controls their split loss.
Together with the inherited collars, the recorded partition covers the
entire constrained transition region.

\noindent\textit{Source:} \path{CK_OPPOSITE_EXTENSION/transition/PROOF.md}.\par
Use the definitions and the established results recalled in Section~\ref{unbal:module-central_assembly-retained-inputs}. This component proves the hybrid inequality \(\zeta\ge R_B\) for all positive feasible entropy splits when \(a,b\) are central opposite-side means and
\begin{equation}\label{unbal:eq:central_transition:1}
d\ge1/50,\qquad E\le11/200,\qquad 4\le d/E\le8.
\end{equation}
The exact same-side boundary, where \(q=d\) and one mean is \(1/2\), is already owned by the preceding same-side theorem. The following endpoint argument handles strict opposite-side means; the endpoint-existence note also justifies that boundary directly under the retained supporting-plane theorem.

\subsubsection{A finite parametrization}\label{unbal:module-central_transition-a-finite-parametrization}

After label exchange and simultaneous complement put \(a\le b\) and \(a+b\le1\). With \(x=d/E\) and \(y=q/E\),
\[
a=[1-(x+y)E]/2,\qquad b=[1+(x-y)E]/2.
\]
The computational rectangle is
\begin{equation}\label{unbal:eq:central_transition:2}
E\in[1/400,11/200],\quad x\in[4,8],\quad y\in[0,8].
\end{equation}
Every relevant tuple in \eqref{unbal:eq:central_transition:1} occurs here: \(d\ge1/50\) and \(x\le8\) force \(E\ge1/400\). Its necessary feasibility constraints are
\begin{equation}\label{unbal:eq:central_transition:3}
y\le x,\qquad (x+y)E\le4/5,\qquad xE\ge1/50.
\end{equation}
These are also the mean constraints for this orientation. Strictly irrelevant boxes are labelled \texttt{outside}. For boxes meeting the domain, interval values of \(d\), \(q\), \(d+q\), \(d-q\) are clipped only by necessary inequalities: \(d\ge1/50\), \(q\le2/5\), \(d+q\le4/5\), \(d-q\ge0\).

\subsubsection{Endpoint comparison against \texorpdfstring{\(R_\psi\)}{R_psi}}\label{unbal:module-central_transition-endpoint-comparison-against-r_psi}

Convexity of \(\eta\) gives, for every feasible entropy split,
\begin{equation}\label{unbal:eq:central_transition:4}
R_\psi\le\eta(E+C(q))
-\eta\left(E+\frac{C(d+q)+C(d-q)}2\right).
\end{equation}
The global supporting-plane bound is \(\zeta\ge F(d,E)=EF(x,1)\). The \texttt{endpoint} owner verifies directly that this bound is at least the right side of \eqref{unbal:eq:central_transition:4}, using directed intervals. Evenness of \(C\) justifies the oriented \(d-q\) coordinate.

\subsubsection{A normalized comparison retaining \texorpdfstring{\(\phi\)}{phi} at both children}\label{unbal:module-central_transition-a-normalized-comparison-retaining-phi-at-both-children}

This second inequality is used only when \(\psi\) is active at the parent; the unrestricted \(R_\phi\) theorem handles the other branch. Write \(\Phi(z,h)=\eta(h)-F(z,h)\). The prior \path{TRANSVERSE_STRIP.md}, Section~\ref{unbal:module-global_mixed_strip-an-endpoint-gain-for-de-4}, and its exact \path{STRIP_CONSTANT_CHECKER.py} prove
\begin{equation}\label{unbal:eq:central_transition:5}
B_{\rm end}\ge F(d,E)+\frac{9d}{20L}[-\ln(1-t^2)],
\qquad t=(e-f)/(2E),\quad L=\ln2,
\end{equation}
for \(x=d/E\ge4\) in the present endpoint region. Its scalar input is $Q''(h)\ge9/(10Lh^2)$ for \(Q(h)=J(H^{-1}(h))\) and \(h\le37/80\). The equal-entropy \(x=4\) contact has \(H(v)<37/160\), so doubling for either entropy split stays in that range. This is a retained proved lemma, not a numerical assumption at the edge \(x=4\).

Entropy convexity of \(\Phi\) on its feasible interval, together with the global estimates
\[
\eta''(h)\ge\frac1{2Lh^2},\qquad zF_{zz}(z,h)\le13/6,
\]
implies
\begin{equation}\label{unbal:eq:central_transition:6}
\Phi_{hh}(z,h)\ge\frac{\gamma}{h^2},\qquad
\gamma=\max\{0,1/(2L)-(13/6)(d+q)\}.
\end{equation}
Every \(h\in(0,2E]\) is feasible at both central child radii: \(2E\le11/100<H(1/10)\). Thus both \eqref{unbal:eq:central_transition:6} and the supporting lines at \(E\) apply to the entire entropy split.

Let \(A_h=\Phi_h(d+q,E)-\Phi_h(d-q,E)\). Homogeneity and the mixed derivative bound give
\[
|A_h|\le13q/(3E).
\]
The two supporting lines for \(\Phi(z,h)+\gamma\ln h\), added to \eqref{unbal:eq:central_transition:5}, leave a linear term \(EtA_h/2\) and a coefficient
\[
c=\frac{9d}{20L}+\frac\gamma2>0
\]
in front of \(-\ln(1-t^2)\). Since $-\ln(1-t^2)\ge t^2$, minimizing the resulting quadratic over all real \(t\) loses at most
\begin{equation}\label{unbal:eq:central_transition:7}
\frac{169q^2}{144c}.
\end{equation}
This is uniform as either positive entropy tends to zero relative to the other.

Radial concavity and its tangent inequality at \(d^2\) give
\begin{equation}\label{unbal:eq:central_transition:8}
\frac{F(d+q,E)+F(d-q,E)}2-F(d,E)
\le W(x)\frac{q^2}{E},\qquad W(x)=\frac{f'(x)}{2x},\quad f(x)=F(x,1).
\end{equation}
\(W\) is nonincreasing. At the contact \(xH(v)=1-2v\),
\[
f'(x)=J(v)+\frac{1-2v}{Lv(1-v)[xJ(v)+2]},
\]
which is evaluated with directed root brackets.

For a box let \(E_-,E_+\) be its entropy endpoints and let \(q_+\) bound \(q\) from above. Put \(C_+=C(q_+)\), \(v_-=H^{-1}(E_-)\), \(v_+=H^{-1}(E_++C_+)\), and
\[
\beta=\frac{1-2v_+}{L(1-v_+)}
\left[1+\frac1{\ln((1-v_-)/v_-)}\right].
\]
The retained logarithmic slope inequality yields
\[
\eta(E)-\eta(E+C(q))\ge2C(q)+\beta\ln(1+C(q)/E).
\]
Using \(C(q)/q^2\ge1/(2L)\) and the decreasing function \(\ell_0(u)=\ln(1+u)/u\), \(\ell_0(0)=1\), therefore proves the normalized lower bound
\begin{equation}\label{unbal:eq:central_transition:9}
\frac E{q^2}[\eta(E)-\eta(E+C(q))]
\ge\frac{E_-}{L}+\frac\beta{2L}\ell_0(C_+/E_-).
\end{equation}
This extends continuously to \(q=0\). All auxiliary entropy arguments remain inside \((0,1)\).

For \(c_*>0\) a directed lower bound on \(c\), equations \eqref{unbal:eq:central_transition:7}--\eqref{unbal:eq:central_transition:9} show that the following is a sufficient box inequality:
\begin{equation}\label{unbal:eq:central_transition:10}
\frac{E_-}{L}+\frac\beta{2L}\ell_0(C_+/E_-)
-W(x_-)-\frac{169E_+}{144c_*}\ge0.
\end{equation}
The \texttt{normalized\_phi\_children} owner verifies \eqref{unbal:eq:central_transition:10}. The implementation takes
\[
c_*\le\frac{9d_-}{20L}
+\frac12\max\{0,1/(2L)-(13/6)\min(4/5,(x_++y_+)E_+)\}.
\]
Every replacement has the conservative sign. At \(q=0\), parent equality \(\phi=\psi\) gives the hybrid conclusion directly; no numerical division by \(q\) is performed.

\subsubsection{Prior collars and complete certificate}\label{unbal:module-central_transition-prior-collars-and-complete-certificate}

The \texttt{prior\_collar} owner reuses the exact, previously proved domains
\[
y\le1,\ x\ge5;\qquad y\le2,\ x\ge6;\qquad y\le4,\ x\ge8.
\]
Their shared-entropy cap condition is automatic here because \(E\le11/200<H(1/10)\). The corresponding proofs and exact scalar checks are in the dependency archive.

The complete rational binary partition has \textbf{2,425 leaves}:

\begin{longtable}[]{@{}lr@{}}
\toprule\noalign{}
Owner & Leaves \\
\midrule\noalign{}
\endhead
\bottomrule\noalign{}
\endlastfoot
Endpoint comparison \eqref{unbal:eq:central_transition:4} & 1,687 \\
Normalized \(\phi\)-child comparison \eqref{unbal:eq:central_transition:10} & 508 \\
Strictly outside the relevant domain & 226 \\
Prior collar & 4 \\
\end{longtable}

There are zero unresolved leaves. Discovery used 40 decimal working digits; \path{TRANSITION_REPLAY.json} records the successful full 70-digit replay. The replay checks source digest, root, complete tree partition and each owner. Binary64 calculations only propose exact dyadic root brackets; directed signs and interval inequalities are the acceptance gates.

Together with the prior small-ratio theorem for \(d\le4E\) and the new eight-ratio theorem/parent criterion for \(d\ge8E\), this closes the whole low-entropy central opposite-side domain once the preceding \(d\le1/50\) band is included.

\clearpage
\subsection{Central opposite-side pairs at moderate entropy}
\label{unbal:app:central_opposite_moderate}

\emph{Argument guide.}
Convexity removes the individual entropy split, and the total-entropy
coordinate includes its full feasible interval up to the cap. The cover
uses global lower bounds and fixed-plane endpoint comparisons on mean
rectangles, while the prior diagonal and cap theorems handle their
assigned regions. The final partition records how these sufficient
conditions exhaust the stated domain.

\noindent\textit{Source:} \path{CK_OPPOSITE_EXTENSION/opposite_cover/PROOF.md}.\par
This component proves the following bound:
\[
\begin{gathered}a\in[1/10,1/2],\quad b\in[1/2,9/10],\\
b-a\ge1/50,\quad E=(e+f)/2\ge11/200
\\\Longrightarrow\quad \zeta(a,b,e,f)\ge R_\psi(a,b,e,f)\end{gathered}
\]
for every positive feasible pair \(e\le H(a)\), \(f\le H(b)\). Combining this with Theorem~\ref{bal:thm:global-unrestricted-target} gives the hybrid inequality \(\zeta\ge R_{\max(\phi,\psi)}\) there. The existing diagonal-band theorem owns \(b-a\le1/50\). Combined with the prior completed same-side half-squares, the result therefore covers \textbf{every central mean pair \(a,b\in[1/10,9/10]\) whenever \(E\ge11/200\)}.

The bound is independent of the entropy split. Write
\[
m=(a+b)/2,\quad d=b-a,\quad C=(H(a)+H(b))/2,\quad
s=C-E,\quad \Delta=H(m)-C,\quad I=s+\Delta.
\]

For \(P(t)=\eta(1-t)\) the retained scalar lemma gives increasing convex \(P'\), increasing \(P''\), and \(P''\ge8\ln(2)/3\). Convexity eliminates the entropy split:
\[
R_\psi\le D_\Delta(s):=P(s+\Delta)-P(s).
\]

The parameter box is exactly
\[
(a,b,t)\in[1/10,1/2]\times[1/2,9/10]\times[0,1],\qquad
E=11/200+t(C-11/200),\quad s=(1-t)(C-11/200).
\]

Every feasible tuple with \(E\ge11/200\) occurs in this parametrization. Since centrality gives \(C\ge H(1/10)>11/200\), no denominator vanishes. Boxes whose maximum \(b-a\) is strictly below \(1/50\) are irrelevant. All other boxes enclose their relevant subset using the exact lower bound \(d\ge1/50\).

\subsubsection{Sufficient inequalities}\label{unbal:module-central_opposite_moderate-sufficient-inequalities}

The checker uses the globally valid inequalities proved in Sections~\ref{unbal:module-full_entropy-a-global-endpoint-bound-that-ignores-the-entropy-split}--\ref{unbal:module-full_entropy-normalized-box-inequalities}:

\begin{enumerate}
\def\labelenumi{\arabic{enumi}.}
\tightlist
\item
  The supporting-plane argument in Section~\ref{unbal:module-full_entropy-a-global-endpoint-bound-that-ignores-the-entropy-split} gives \(\zeta\ge F(d,E)\), where \(F(d,E)=dJ(v)\) and \(dH(v)=E(1-2v)\).
\item
  The global log-sum theorem implies
\[
  \zeta\ge j(a,b)+d^2\frac{\min(\kappa(a),\kappa(b))}{2b(1-a)}s,
  \quad j(a,b)=\tfrac12d[J(a)-J(b)]\ge4\Delta,
\]
where \(\kappa(u)=\min\{u/[-(1-u)\ln(1-u)],(1-u)/[-u\ln u]\}\).
\item\label{unbal:item:central_opposite_moderate:3}
  Retaining the unequal entropy coefficients strengthens this to
\[
  \zeta-R_\psi\ge j+(A+D)s-D_\Delta(s)
      -\frac{3(A-D)^2}{16\ln2},\quad
  A=\frac{d^2\kappa(a)}{4b(1-a)},\quad
  D=\frac{d^2\kappa(b)}{4b(1-a)}.
\]
\item
  A contact plane at the prescribed means is used only after directed arithmetic proves both entropy coefficients positive. Arbitrary shifted log-sum anchors remain valid globally and are recorded as exact rationals. Symmetric endpoint-contact planes have equal entropy coefficients and are valid globally.
\item
  The previously completed central cap theorem owns \(s\le3/40\).
\end{enumerate}

All these inequalities apply to opposite-side means. The former same-side positive-series enhancement \(j\ge(4+W)\Delta\) is \textbf{disabled} here by setting \(W=0\). No condition involving the sign of \(1-a-b\) is used, and the former parent-\(\phi\) test is removed. Thus the entire cross-half rectangle, including \(a+b=1\), is covered directly.

For a box the entropy Jensen gap obeys
\[
\frac{\Delta}{d^2}\le K_*:=\min\left\{
\frac{[a(1-a)]^{-1}+[b(1-b)]^{-1}}{16\ln2},
\frac{\Delta_*}{d_*^2}\right\}.
\]

The first right hand side is enclosed above on the box. The second uses directed bounds \(\Delta\le\Delta_*\) and \(d\ge d_*\). The gap increases as the smaller mean decreases or the larger mean increases, giving its directed corner bounds. Convexity of \(P'\) yields
\[
D_\Delta(s)\le\Delta T_*,\qquad
T_*=[P'(s_*)+P'(I_*)]/2.
\]

Consequently the normalized endpoint test is \(F(d,E)/d^2\ge K_*T_*\), and the normalized log-sum test is \(\beta_*s_{\min}\ge K_*\max(0,T_*-4)\). The quadratic-penalty version is the normalized form of item~\ref{unbal:item:central_opposite_moderate:3}. Other owners compare the same lower bounds directly against the smallest of
\[
P(I_*)-P(s_{\min}),\quad \Delta_* T_*,\quad
P(s_*+\Delta_*)-P(s_*),
\]
where the last expression is used only when \(s_*+\Delta_*<1\).

For a fixed symmetric contact plane or fixed shifted log-sum anchor, its lower bound minus \(D_\Delta(C-E)\) is concave in \(E\), since its second derivative is
\[
-P''(H(m)-E)+P''(C-E)\le0.
\]

Therefore checking the two entropy endpoints proves the full entropy interval. When mean-value bounds are used at these endpoints, the first derivatives are enclosed over the entire mean rectangle. The exact formulas and derivations are given in Section~\ref{unbal:module-full_entropy-normalized-box-inequalities}; only the root domain and constants are changed.

\subsubsection{Arithmetic and completeness}\label{unbal:module-central_opposite_moderate-arithmetic-and-completeness}

\path{FULL_ENTROPY_COVER.py} writes a complete binary subdivision partition, including exact path digits and the sufficient inequality used at each leaf. Each digit records the axis and side. \texttt{check\_partition} independently requires exactly two complementary children at each internal node and exactly one owner per terminal node. The leaf union is the entire declared root box.

Discovery uses 40 decimal digits with directed \texttt{mpmath.iv} arithmetic. Every inverse-entropy and endpoint-contact root proposal is accepted only after directed interval sign checks bracket the unique root. Floating-point calculations are allowed only for proposals; they never accept an inequality or exclude a cell.

\texttt{-\/-replay} rebuilds every exact rational box from its path, checks the code hash, status, root and full partition, and recomputes its recorded owner at 70 decimal digits. A completed result and a passed replay, both with zero unresolved boxes, are the required gates. A stopped or budget-exhausted run is expressly incomplete.

This certificate establishes the $R_\psi$ bound for $1/10\le a\le1/2\le b\le9/10$, $b-a\ge1/50$, and $E\ge11/200$, with positive feasible entropy coordinates. Section~\ref{unbal:sec:central} combines this result with the other central-region estimates; Section~\ref{unbal:sec:assembly} assembles the full domain to prove Theorem~\ref{unbal:thm:global}.

\subsubsection{Completed result}\label{unbal:module-central_opposite_moderate-completed-result}

Discovery completed with 9,459 nodes and 4,730 accepted leaves in 523.24 seconds. There are zero unresolved leaves. The full 70-digit replay passed all 4,730 leaves in 145.58 seconds. The source hash checked by replay is \path{11988425c484cbbeee5cc3161a46b498d5473fe0298327acc2a70594c84f7b25}.

The recorded owner counts are:

\begin{longtable}[]{@{}lr@{}}
\toprule\noalign{}
Owner & Leaves \\
\midrule\noalign{}
\endhead
\bottomrule\noalign{}
\endlastfoot
\texttt{endpoint\_plane\_taylor} & 3662 \\
\texttt{endpoint\_plane\_endpoints} & 49 \\
\texttt{shifted\_logsum} & 247 \\
\texttt{cap} & 530 \\
\texttt{endpoint\_sum} & 2 \\
\texttt{sum\_normalized} & 35 \\
\texttt{outside} & 6 \\
\texttt{logsum\_quad\_endpoints} & 199 \\
\end{longtable}

The six \texttt{outside} leaves have maximum mean difference strictly below \(1/50\); the old diagonal band owns their feasible tuples. The 530 \texttt{cap} leaves invoke the retained central cap theorem. Every remaining leaf passes one of the directed sufficient inequalities above.

\clearpage
\subsection{Exhaustion of the entire central mean square}
\label{unbal:app:central_assembly}

\emph{Argument guide.}
This is a domain-exhaustion argument using the preceding component
theorems. Same-side means are already covered; opposite-side means are
split by separation, total entropy, and then separation-to-entropy ratio.
The final table joins the diagonal band, the moderate-entropy cover,
the low-entropy transition, and the large-ratio estimates, including
their equality interfaces.

\noindent\textit{Source:} \path{CK_OPPOSITE_EXTENSION/PROOF.md}.\par
\subsubsection{Main theorem}\label{unbal:module-central_assembly-main-theorem}

Let \(H\) be binary entropy in bits, \(J(u)=\log_2((1-u)/u)\), \(L=\ln2\), and \(H^{-1}\) the lower inverse on \([0,1/2]\). Define
\[
\eta(h)=(1-2H^{-1}(h))J(H^{-1}(h)),\qquad C(q)=1-H((1-q)/2),
\]
\[
F(z,h)=zJ(v),\qquad zH(v)=h(1-2v),\qquad F(0,h)=0,
\]
\[
\phi(m,h)=\eta(h)-F(|1-2m|,h),\quad
\psi(m,h)=\eta(h+C(1-2m)),\quad B=\max(\phi,\psi).
\]
For four moments \(M=(a,b,e,f)\), put
\[
m=(a+b)/2,\quad E=(e+f)/2,\quad d=|a-b|,\quad q=|1-a-b|,
\]
\[
R_A(M)=A(m,E)-\tfrac12[A(a,e)+A(b,f)].
\]
The unrestricted four-moment infimum \(\zeta(M)\) is the infimum of the expected cost
\[
j(U,W)=\tfrac12(W-U)[J(U)-J(W)]
\]
over joint laws with means \(a,b\) and expected binary entropies \(e,f\), with the usual diagonal endpoint interpretation.

\begin{theorem}[Central square, all positive feasible entropies]\label{unbal:thm:central_assembly:1}
The following bound holds:
\begin{equation}\label{unbal:eq:central_assembly:1}
\boxed{
 a,b\in[1/10,9/10],\quad
 0<e\le H(a),\quad0<f\le H(b)
 \quad\Longrightarrow\quad
 \zeta(a,b,e,f)\ge R_B(a,b,e,f).
}
\end{equation}
There is no condition on the mean difference, the average entropy, the entropy deficit, or the ratio of the two entropies beyond feasibility and positivity. Both same-side and opposite-side means are included, along with the individual entropy-cap faces.
\end{theorem}

This establishes the hybrid Bellman inequality on the entire stated mean
square. Section~\ref{unbal:sec:assembly} combines this central result
with the same-side and outer opposite-side estimates to cover all feasible
means.

\subsubsection{Previously established results}\label{unbal:module-central_assembly-retained-inputs}

Theorem~\ref{bal:thm:global-unrestricted-target} gives \(\zeta\ge R_\phi\).
Proposition~\ref{bal:prop:fixed-difference-minimum} and
Theorem~\ref{bal:thm:general-endpoint-region} provide the endpoint minimum
and its global supporting planes. Theorem~\ref{bal:thm:four-moment-lower-bound}
gives \(\zeta\ge L_4\ge F(d,E)\). The radial profile estimates in
Sections~\ref{bal:aux:sec:fourpoint} and~\ref{bal:imp:e8} give concavity of
\(z\mapsto F(\sqrt z,h)\), and
Theorem~\ref{bal:aux:thm:entropyconvexity} gives convexity of
\(\Phi(z,h)=\eta(h)-F(z,h)\) in its feasible entropy coordinate.
Lemma~\ref{unbal:lem:global-mixed} supplies \(zF_{zz}\le13/6\).

The earlier proof package, included unchanged as \path{dependencies/CK_NO_SEPARATION_EXTENSION.zip}, supplies:

\begin{itemize}
\tightlist
\item
  The complete same-side half-squares \([1/10,1/2]^2\) and \([1/2,9/10]^2\), all positive feasible entropies.
\item
  The complete central diagonal band \(d\le1/50\), all positive feasible entropies.
\item
  The central small-ratio theorem \(E\le11/200\), \(d\le4E\), including its analytic \(E\to0\) tail and 611-leaf 70-digit replay.
\item
  The prior central cap, log-sum, scalar profile, and narrow-collar lemmas and certificates, through its embedded dependencies.
\end{itemize}

The branch transfer is used throughout: if \(\phi\) is active at the parent, \(R_B\le R_\phi\); if \(\psi\) is active there, \(R_B\le R_\psi\). When a component retains \(\phi\) at the children, it compares directly with \(R_B\) in the \(\psi\)-parent branch. It never assumes that the stronger global inequality \(\zeta\ge R_\psi\) must hold.

\subsubsection{Four new components}\label{unbal:module-central_assembly-four-new-components}

\paragraph{Parent dominance over the full central opposite-side range}\label{unbal:module-central_assembly-parent-dominance-over-the-full-central-opposite-side-range}

Theorem~\ref{unbal:thm:parent_eight:1} gives
\begin{equation}\label{unbal:eq:central_assembly:2}
\boxed{0<q\le2/5,\quad q/E\ge8
\quad\Longrightarrow\quad \phi(m,E)>\psi(m,E).}
\end{equation}
This theorem does not require central child means. It strengthens the preceding factor-eight result, which required \(q\le1/10\).

The range \(q\le1/10\) is retained from the previous theorem. A complete directed cover handles \(1/10\le q\le2/5\) and \(8\le q/E\le32\). The entire unbounded \(q/E\) tail is analytic: with \(x=q/E\) and \(\ell(x)\) the natural contact logit, \(x\ell'(x)\le2\), and
\[
G(x)=\frac52\ln(1+x/(5L))-\ell(x)
\]
is strictly increasing for \(x\ge32\). A fixed directed comparison proves \(G(32)>69/1000\), implying strict dominance there. Thus no positive lower bound on \(E\) is introduced by \eqref{unbal:eq:central_assembly:2}.

The finite part has 332 leaves and a complete 70-digit replay. At \(q=0\), \(\phi\) and \(\psi\) are equal at the parent, so \(R_\phi\) owns that face.

\paragraph{Large \texorpdfstring{\(d/E\)}{d/E} with arbitrary entropy splitting}\label{unbal:module-central_assembly-large-de-with-arbitrary-entropy-splitting}

Theorem~\ref{unbal:thm:central_eight:1} gives
\begin{equation}\label{unbal:eq:central_assembly:3}
\boxed{
 a,b\in[1/10,9/10],\quad E\le11/200,\quad
 q\le8E,\quad d\ge8E
 \quad\Longrightarrow\quad\zeta\ge R_B.
}
\end{equation}
In the strict opposite-side \(\psi\)-parent branch the stronger bound is
\[
B_{\rm end}-R_B\ge\frac{2249}{144000}\frac{q^2}{E}.
\]
The endpoint's entropy-split curvature absorbs the loss from unequal \(e\) and \(f\). A 50-leaf two-variable cover proves a normalized parent correction of at least \(7/10\). The radial and split losses sum to at most
\[
\frac{479}{1000}+\frac{1183}{5760}<\frac7{10}.
\]
An analytic tail treats every \(0<E\le10^{-4}\); the finite cover treats the remaining \(E\) values. Thus the full \(E\to0\) limit and arbitrary entropy imbalance are included.

For opposite-side central means, \(q\le d\) and \(q+d\le4/5\) imply \(q\le2/5\). Combining \eqref{unbal:eq:central_assembly:2} and \eqref{unbal:eq:central_assembly:3} therefore covers every such tuple with \(E\le11/200\) and \(d\ge8E\).

\paragraph{The remaining low-entropy transition}\label{unbal:module-central_assembly-the-remaining-low-entropy-transition}

Section~\ref{unbal:app:central_transition} proves
\begin{equation}\label{unbal:eq:central_assembly:4}
\boxed{
 a,b\text{ central and opposite-side},\quad d\ge1/50,
 \quad E\le11/200,\quad4\le d/E\le8
 \quad\Longrightarrow\quad\zeta\ge R_B.
}
\end{equation}
The finite parameters are \(E\in[1/400,11/200]\), \(x=d/E\in[4,8]\), \(y=q/E\in[0,8]\), with the exact mean constraints \(y\le x\), \((x+y)E\le4/5\), \(xE\ge1/50\). The lower bound \(E\ge1/400\) follows from \(d\ge1/50\) and \(x\le8\); it is not a discarded tail.

The decisive new acceptance inequality retains the positive entropy curvature of both \(\phi\) children in addition to the endpoint's split correction. If
\[
\gamma=\max\{0,1/(2L)-(13/6)(d+q)\},\qquad
c=9d/(20L)+\gamma/2,
\]
then the arbitrary-split loss is at most \(169q^2/(144c)\). After division by \(q^2/E\), the comparison is the normalized parent correction against
\[
\frac{f'(d/E)}{2(d/E)}+\frac{169E}{144c},\qquad f(x)=F(x,1).
\]
This bound complements the direct \(F(d,E)\) comparison against \(R_\psi\) and the previous narrow collars. No grid in the entropy split is used.

The complete partition has 2,425 leaves, all replayed at 70 digits. Of these, 226 are proved irrelevant to the constrained mean domain; none are unresolved.

\paragraph{Moderate entropy for every opposite-side central pair}\label{unbal:module-central_assembly-moderate-entropy-for-every-opposite-side-central-pair}

Section~\ref{unbal:app:central_opposite_moderate} proves
\begin{equation}\label{unbal:eq:central_assembly:5}
\boxed{
 a\in[1/10,1/2],\quad b\in[1/2,9/10],\quad
 d\ge1/50,\quad E\ge11/200
 \quad\Longrightarrow\quad\zeta\ge R_\psi.
}
\end{equation}
The parameter \(E=11/200+t(([H(a)+H(b)]/2)-11/200)\), \(t\in[0,1]\), covers its entire feasible range. Convexity removes the entropy split. Global symmetric contact planes, shifted log-sum planes and their entropy-endpoint concavity give sufficient bounds over whole intervals of \(E\). Directed derivative bounds extend these comparisons over mean rectangles. The prior cap theorem owns its previously completed portion.

The former same-side positive-series enhancement is disabled; every retained inequality is valid on the whole cross-half rectangle. The new complete partition has 4,730 leaves, with zero unresolved leaves. Section~\ref{unbal:module-central_opposite_moderate-completed-result} records the completed calculation and the full 70-digit replay of all 4,730 leaves.

\subsubsection{Exhaustive assembly of the main theorem}\label{unbal:module-central_assembly-exhaustive-assembly-of-the-main-theorem}

If the means are on the same side of \(1/2\), the preceding complete half-square theorem applies. Otherwise exchange the complete mean-entropy labels if necessary so that \(a\le1/2\le b\). The following table exhausts the remaining positive feasible entropy tuples:

\begin{longtable}[]{@{}
  >{\raggedright\arraybackslash}p{(\columnwidth - 2\tabcolsep) * \real{0.5000}}
  >{\raggedright\arraybackslash}p{(\columnwidth - 2\tabcolsep) * \real{0.5000}}@{}}
\toprule\noalign{}
\begin{minipage}[b]{\linewidth}\raggedright
Domain
\end{minipage} & \begin{minipage}[b]{\linewidth}\raggedright
Owner
\end{minipage} \\
\midrule\noalign{}
\endhead
\bottomrule\noalign{}
\endlastfoot
\(d\le1/50\) & Previous full-entropy diagonal band \\
\(d\ge1/50\) and \(E\ge11/200\) & Moderate-entropy theorem \eqref{unbal:eq:central_assembly:5} \\
\(d\ge1/50\), \(E\le11/200\), \(d\le4E\) & Previous central small-ratio theorem and tail \\
\(d\ge1/50\), \(E\le11/200\), \(4E\le d\le8E\) & Transition theorem \eqref{unbal:eq:central_assembly:4} \\
\(d\ge1/50\), \(E\le11/200\), \(d\ge8E\), \(q\ge8E\) & Parent dominance \eqref{unbal:eq:central_assembly:2}, since \(q\le2/5\) \\
\(d\ge1/50\), \(E\le11/200\), \(d\ge8E\), \(q\le8E\) & Eight-ratio theorem \eqref{unbal:eq:central_assembly:3} \\
\end{longtable}

All interfaces overlap at equality. The exact diagonal is included in the previous band. The \(q=d\) boundary lies in the already completed same-side half-squares because one mean is \(1/2\); the endpoint-existence lemma in Section~\ref{unbal:app:endpoint_existence} also handles it directly. The \(q=0\) face uses exact parent equality. No limiting entropy ratio or average-entropy tail is left between these cases. This proves \eqref{unbal:eq:central_assembly:1}.

\section{Source and artifact inventory}
\label{unbal:app:inventory}
This inventory identifies the component sources, certificate archives,
and integrity records for the proof. The machine-readable archives are
separate from this \LaTeX{} source. Section~\ref{bal:sec:verification}
specifies the verification scope, and Section~\ref{supp:verification}
gives the unbalanced certificate counts and reproduction commands.

The publication source archive contains the editable integrated LaTeX,
the source documents used for the preceding technical modules, their excerpt
manifest, the final completion records, and the source PDF of the balanced
manuscript. The accompanying artifact archive contains the complete sealed
extension package, including its unchanged prior dependency archives.

The following table maps every retained technical module to its role.
\begin{longtable}{p{0.06\textwidth}p{0.35\textwidth}p{0.49\textwidth}}
\toprule No. & Module & Role\\\midrule\endhead
1 & Global log-sum inequality and entropy-profile calculus & Global Bregman/log-sum lower bound; strengthened variance term; analytic \(P\) curvature, \(P\) third-derivative monotonicity, and deterministic cap \(j\ge P(\Delta)\). \\
2 & Uniform entropy curvature and normalized small-difference theorem & Scalar \(\eta\) curvature and derivative estimates, radial perspective facts, exact \(f(3)>12\) and \(f'(3)/6<1\), and the uniform small-difference theorem required by the practical low-entropy cutoff. \\
3 & Opposite deterministic corner and endpoint mass comparison & Endpoint existence/contact-mass comparison, log-convexity entropy-imbalance gain, and the complete mean-only opposite deterministic corner. Sections~\ref{unbal:module-opposite_corner-definitions-and-theorem}--\ref{unbal:module-opposite_corner-proof-of-the-complete-mean-only-corner} are needed; the older global-cutoff corollary in section 7 of the source document is superseded. \\
4 & Strong mean cost and practical uniform low-entropy theorem & Positive-series proof \(j\ge(4+r^2/2)\Delta\), global parent-deficit \(I\le1/100\) theorem, and the practical \(E\le10^{-6}\), \(q\le32E\) theorem. The new global-low-entropy theorem invokes the latter. \\
5 & Scalar formulas underlying the stronger endpoint and mixed-derivative estimates & Retain the explicit \(Q\) double-derivative formula, logarithmic normalization, common-contact endpoint log-gain argument, radial mixed-derivative inequality, and parent-correction derivation used by the subsequent stronger collar proofs. The original weaker regional theorem is excluded. \\
6 & Stronger endpoint log gain and the three inherited collars & \(Q\) curvature up to entropy \(19/50\); equal-contact control for \(d/E\ge5\); endpoint gain \(d/(2\ln2)\) times \(-\ln(1-\tau^2)\); and the three explicit collars reused as geometric owners in the transition cover. \\
7 & Global mixed bound and the ratio strip from four to five & \path{GLOBAL_MIXED} supplies the 105-leaf compact interval calculation, while Section~\ref{unbal:module-global_mixed_strip-a-global-mixed-derivative-bound} supplies the indispensable analytic tails. Together they prove the global \(13/6\) bound. The same document proves the \(d/E\ge4\) log gain and transverse strip. \\
8 & Global mean-anchor matrix and inherited quantitative cap rectangle & The two-by-two capmatrix supporting plane, requiring nonnegative solved coefficients \(A,D\); the slope/secant argument; strong-convexity split elimination; and the quantitative subrectangle required by the restored expanded cap cover. \\
9 & Complete central cap theorem and positive-series mean cost & Cross-half cap \(s\le3/20\), same-side cap \(s\le3/40\), hence the full central square at \(s\le3/40\). Includes the normalized Peano/trapezoid estimates and \(W_N\) positive-series strengthening needed by later covers. \\
10 & Global shifted log-sum planes and the first required same-side cover & Global symmetric \(F(d,E)\) bound; arbitrary rational shifted anchors and affine entropy costs; analytic split and entropy-interval elimination; centered mean derivatives; complete \(E\to0\) tail; and the required 33,572-leaf central same-side cover at separation \(\ge1/20\). Section~\ref{unbal:module-full_entropy-a-monotone-phi-parent-enclosure-for-the-last-low-entropy-block} supplies the monotone \(\phi\)-parent enclosure. \\
11 & Central small-ratio theorem and the required one-hundredth same-side cover & Normalized \(q\le E\) theorem (205 leaves) and its analytic entropy tail; central \(d\le4E\) theorem (611 leaves) and the 11-interval \(E\to0\) estimate; 79-interval same-side tail; the required 76,547-leaf separation \(\ge1/100\) cover; and the stable logarithmic \(\phi\)-parent lower bound. \\
12 & Complete central same-side theorem and diagonal band & \(q\le1/10\) factor-eight parent criterion; low-entropy diagonal strip; and the 325-leaf moderate-entropy diagonal band. These remove the central same-side separation cutoff and also cover all opposite-side central pairs with \(d\le1/50\). \\
13 & Endpoint existence for opposite-side means & Derives automatic endpoint existence when \(E<d\). The included three-atom exactness observation is optional for the global proof but belongs to this short self-contained document. \\
14 & Factor-eight parent dominance through \(q=2/5\) & Extends the \(q\le1/10\) parent theorem to \(q\le2/5\) using a 332-leaf compact rectangle and a proved unbounded \(x=q/E\) tail. \\
15 & Central eight-ratio low-entropy theorem & The central shared-cap eight-ratio theorem: endpoint gain absorbs every entropy split, radial and parent estimates give the required inequality. Its 50-leaf normalized-parent certificate is also a retained input to the newer cap-free proof. \\
16 & Central low-entropy transition between ratios four and eight & Exact three-coordinate cover of the residual central low-entropy transition, 2,425 leaves. Includes the normalized \(\phi\)-child inequality and the inherited collar owners; those four collar-owned leaves require \path{SHARPER_COLLARS}. \\
17 & Central opposite-side pairs at moderate entropy & 4,730-leaf complete central opposite-side cover for \(d\ge1/50\) and \(E\ge11/200\), every feasible entropy split. Reuses the global shifted and unshifted supporting planes, with the same-side-only enhancement disabled. \\
18 & Exhaustion of the entire central mean square & Exhaustive assembly of \(\zeta\ge R_B\) for \(a,b\in[1/10,9/10]\) and all positive feasible entropy pairs. The outer regions enter the global assembly in Section~\ref{unbal:sec:assembly}. \\
\bottomrule\end{longtable}

\subsection{Balanced certificate archive set}
The balanced mathematical content is developed in Part~\ref{part:balanced}
and its supplementary proofs. The archive integrity record reports that
the following three ZIPs match the publication checksum manifest and pass
all member CRC32 checks. Their paths in the artifact bundle are under
\path{upstream/}:
\begin{itemize}
\item \path{BALANCED_CK_PUBLICATION_CORE_2026-09-07.zip}:
253,422,817 bytes, 7,534 entries.
\item \path{ck_radial_fresh_part1_g000-g001.zip}:
416,891,275 bytes, 1,753 entries.
\item \path{ck_radial_fresh_part2_g002-g008.zip}:
356,717,868 bytes, 5,259 entries.
\end{itemize}
The original checksum manifest and \path{RECOVERED_UPSTREAM_INTEGRITY.json}
record the exact hashes. \path{UPSTREAM_REPRODUCTION.md} explains the original
archive placement and verification procedure.

The archive identities used here are those in
\path{RECOVERED_UPSTREAM_INTEGRITY.json}. The separate historical integrity
report, \path{UPSTREAM_ARCHIVE_INTEGRITY.md}, is not the availability record
for this archive set.

The verification scope of each balanced component is recorded in
Section~\ref{bal:sec:verification}; archive-integrity checks authenticate
the files and are distinct from arithmetic replay. The extension uses
the proved mathematical results listed in Section~\ref{unbal:sec:inputs}.


\section{Certificates, arithmetic, and reproducibility}
\label{unbal:sec:verification}\label{supp:verification}

\subsection{Completed extension certificates}

The following records refer to complete continuum partitions. Counts include
leaves assigned to previously proved owners and, where applicable, leaves
proved disjoint from the constrained feasible domain. They are not counts
of sampled tuples.

\begin{center}\small
\begin{tabular}{p{0.56\textwidth}r l}
\toprule
Certificate & Leaves & Final check \\
\midrule
Entire remaining same-side chart & 160,789 & 70-digit replay \\
Entire remaining opposite-side chart & 29,495 & 70-digit replay \\
Small-mean theorem $a+b\le1/16$ & 20 & 70-digit replay \\
Global low entropy $E\le10^{-6}$ & 5 & 70-digit replay \\
Extreme opposite strip $a\le2^{-28}$ & 17 & 70-digit replay \\
Factor-sixteen parent theorem & 69 & 70-digit replay \\
Eight-ratio parent correction & 77 & 70-digit replay \\
Eight-ratio split loss & 2 & 70-digit replay \\
Additional high-$q$ parent comparison & 59 & 70-digit replay \\
\midrule
Total final partition leaves & 190,533 & All passed \\
\bottomrule
\end{tabular}
\end{center}

There are no pending or unresolved leaves in either main partition.
Separate fixed-constant checks prove the same-side ratio tail and the
eight-ratio small-entropy tail and radial derivative constant. Exact rational
checkers also reproduce the retained opposite-corner, normalized-low-entropy,
and uniform-diagonal constants.

The two main source digests are
\begin{quote}\small
Same-side: \path{133194ae93b0e15793e6e93212112844475b3dcdf35b98dff07ccaf6db4f6ad0}.

Opposite-side: \path{a8ac3a0f2966c4cd52bac62c7abaa46b98e089d9859f5d2f0d0439721ef503a1}.
\end{quote}
The sealed extension archive is \path{CK_GENERAL_COMPLETION.zip},
19,210,912 bytes, SHA256
\begin{quote}\small
\path{afeb9e82531f20ade5c688056b4d35dd08e36dd25731f481fdd6e92085528160}.
\end{quote}
It contains the complete new partitions and the unchanged preceding
extension archives. Its manifest has 88 file entries, all of which were
checked before the archive was delivered. The final status is reproduced
as \path{GENERAL_CK_STATUS.json} in the publication bundles.

\subsection{Why a finite ledger proves the full box}

A bisection path character encodes an axis and a child side. The root
endpoints are exact rational numbers. Reconstruction applies the indicated
midpoint bisections exactly. The partition checker builds a prefix tree,
rejects an accepted leaf that has descendants, rejects duplicate leaves,
and requires each internal node to have exactly the two opposite sides
of a split along one common axis. Induction on this tree proves that the
accepted closed boxes cover the entire root.

For the opposite chart, powers such as $2^{-u}$ are evaluated as outward
intervals. Their binary endpoint representations are converted exactly
to rationals before passing to the inherited mean-rectangle checker.
Canonical clipping retains necessary enclosures of the feasible
intersection. Including extra points can make acceptance harder but
cannot remove an actual feasible tuple. An outside leaf requires a
proved empty intersection.

For each nongeometric leaf, the recorded owner specifies one of the
analytic sufficient conditions proved in the text. Replay reconstructs
the box and re-evaluates that owner's inequalities. The same-side replay
records the actual imported primitive hashes, checks them again at the
end, and binds its replay-script hash. The opposite replay checks its
combined source and primitive digest at both the beginning and end.
The final gates check the root, thresholds, complete partition, source
bindings, replay precision, and exact leaf and owner counts.

Some discovery stages used strengthened source versions while retaining
already accepted leaves. The final proof does not depend on informal
reasoning that those migrations preserve validity: every retained leaf
was checked anew against the final frozen source. Historical owner names
in the opposite ledger use the final low-entropy union semantics described
in the opposite-side chapter. They need not assert a bare $R_\psi$ bound
where the final proved assertion is the hybrid $R_B$ bound.

\subsection{Directed arithmetic}

The new checkers use Python's exact \texttt{fractions.Fraction} for rational
geometry and \texttt{mpmath.iv} for outward arithmetic. The recorded runtime
uses Python 3.12 and \texttt{mpmath==1.4.1}. Entropy inverses and scalar
contacts are enclosed with certified sign brackets. A binary64 computation
may suggest a bracket or a subdivision axis, but it cannot accept a
mathematical inequality. Positive-series evaluations include explicit
remainder bounds. Increasing the working precision in replay does not
replace the need for outward endpoints or validated brackets.

The mathematical inference from accepted inequalities relies on the
correct implementation of this interval arithmetic and the stated
certifiers. Section~\ref{unbal:sec:inputs} identifies the analytic
results from Part~\ref{part:balanced} and its supplementary proofs
used in the extension.

\subsection{Running the checks}

The artifact bundle preserves \path{CK_GENERAL_COMPLETION.zip} unchanged.
After extracting it, install its dependency from \path{requirements.txt}
and run
\begin{Verbatim}[fontsize=\small]
python VERIFY_ALL.py
\end{Verbatim}
from the extracted \path{CK_GENERAL_COMPLETION} directory. This verifies
the delivered file hashes and runs all new replays in a temporary copy,
including both full covers, the seven supporting covers, the fixed
eight-ratio constants, the same-side tail, and the three restored exact
constant checkers. It does not alter the sealed records. The same-side
and opposite replays use four and three workers respectively.

The retained extension certificates have completed execution and replay
records in their individual packages. Their exact scripts, ledgers, and
verification records are preserved in the nested archives; the recorded
replays refer to those component executions. The companion
\path{PRIOR_REPLAY_COMMANDS.json} and \path{PRIOR_REPLAY_COMMANDS.md}
identify the actual entry points, arguments, working directories,
and archive chains for replaying those retained certificates.

The publication source archive contains the integrated \LaTeX{},
the unchanged source documents from which the technical appendices were
drawn, an exact excerpt manifest, and a PDF build script. This allows
the mathematical write-up and the interval computations to be reviewed
and reproduced independently.

\section{Further structural questions for the four-moment problem}
\label{supp:structural}
This section is not used in the proof of Theorem~\ref{unbal:thm:ck}.
The additional support hypothesis in Proposition~\ref{bal:prop:conditional-two-atom-integration}
is a separate assumption; the unconditional endpoint theorem does not require it.
All entropies and costs in this section use natural logarithms.

\begin{proposition}[Integration with a two-atom support theorem: conditional part]
\label{bal:prop:conditional-two-atom-integration}
On the moment domain of Theorem~\ref{bal:thm:general-endpoint-region}, let
$\mathcal E_3$ be the set where its contact triple exists, its ratio test
holds, and both endpoint masses are strictly positive.  Let $T_2(M)$ be
the infimum of $\mathbb E j_e(U,W)$ over laws with the prescribed moments
supported on at most two points, with value $+\infty$ if there are none.
Assume the following \emph{additional support theorem}: for every feasible
$M$ of finite Bellman value in this domain, some unrestricted optimizer
has either at most two support points or exactly the three support points
$(0,0),z,(1,1)$ with all three masses positive.  Then
\[
 \zeta_e(M)=
 \begin{cases}
 \displaystyle\frac D2\{J_e(x)-J_e(y)\},&M\in\mathcal E_3,\\[1mm]
 T_2(M),&M\notin\mathcal E_3.
 \end{cases}
\]
On $\mathcal E_3$ the unique optimizer is the displayed three-atom law,
and $T_2(M)>\zeta_e(M)$.  The explicit value and this strict inequality
on $\mathcal E_3$ hold \emph{without} the additional support theorem;
only the assertion on its complement uses that hypothesis.
\end{proposition}
\begin{proof}
The general endpoint theorem proves both the value and uniqueness on
$\mathcal E_3$.  The class of probability laws on the compact square
with at most two atoms is weakly compact: it is the continuous image of
$[0,1]\times([0,1]^2)^2$ under the mixture map.  Its intersection with
the four moment constraints is closed.  Lower semicontinuity of the cost
therefore makes any finite $T_2$ attained.  Equality with $\zeta_e$ would
contradict uniqueness of its genuine three-atom optimizer.
Outside $\mathcal E_3$, the endpoint theorem excludes every genuine
three-atom endpoint optimizer.  Under the stated support hypothesis,
an optimizer must consequently have at most two atoms, giving $\zeta_e=T_2$.
\end{proof}

The remaining two-atom calculation has the concrete form
\[
 \inf\bigl\{\lambda j_e(x,y)+(1-\lambda)j_e(X,Y)\bigr\},
 \qquad
 X=\frac{\mu_u-\lambda x}{1-\lambda},\quad
 Y=\frac{\mu_w-\lambda y}{1-\lambda},
\]
subject to $0<\lambda<1$, $(x,y),(X,Y)\in[0,1]^2$, and
\[
 \lambda h(x)+(1-\lambda)h(X)=e_u,\qquad
 \lambda h(y)+(1-\lambda)h(Y)=e_w.
\]
One-atom laws are included by allowing coincident atoms; the equivalent
compact parametrization also allows $\lambda=0,1$ before eliminating
$(X,Y)$.  Thus diagonal atoms, endpoint atoms, and limiting branches
must be retained.  This is a reduction to three scalar variables with
two entropy equations, not an evaluation of their constrained infimum.

\begin{remark}[Why the ordered reduction does not yet prove the unrestricted split]
\label{bal:rem:ordered-unrestricted-transfer}
Write $\widehat\zeta_e$ for the ordered Bellman infimum,
which requires $U\le W$ almost surely in addition to the same four
moments.  Then $\zeta_e\le\widehat\zeta_e$.  On the endpoint equality
region of Theorem~\ref{bal:thm:general-endpoint-region}, its explicit law is
ordered, so the same law and lower bound give
$\widehat\zeta_e=\zeta_e$.  This establishes the transfer on that region.

The transfer fails in general, even at the level of feasibility.  Take
$U\equiv1/2$ and let $W=1/4$ with probability $3/8$ and $W=3/4$ with
probability $5/8$.  These give
\[
 M=\left(\frac12,\frac9{16},\ln2,h(1/4)\right),\qquad
 \zeta_e(M)\le\frac18\ln3<+\infty.
\]
In an ordered law with the same moments, the entropy cap forces
$U\equiv1/2$, so $W\ge1/2$.  Concavity of $h$ gives the chord inequality
$h(W)\ge2(1-W)\ln2$ on this interval.  Consequently
\[
 e_w\ge\frac78\ln2,
 \qquad
 h(1/4)=2\ln2-\frac34\ln3<\frac78\ln2,
\]
where the last strict inequality is equivalent to $8<9$.
Thus $\widehat\zeta_e(M)=+\infty$.  This example rules out a blanket
identity between the two Bellman functions; it does not refute a
possible unrestricted two-support-type theorem.

The additional support hypothesis in
Proposition~\ref{bal:prop:conditional-two-atom-integration} is therefore
needed for the proposed classification outside $\mathcal E_3$; it does not
follow from the endpoint theorem or from an ordered reduction.
The proposition isolates this optional structural question from the
explicit endpoint theorem and the proof of
Theorem~\ref{bal:hyp:SCplus}.
\end{remark}

\section{Repository release notes and source locators}\label{supp:release-notes}
This section collects source locators and reproduction commands for the
computations used in the proof. The corresponding execution, audit, and
replay scopes are recorded in Sections~\ref{bal:sec:verification}
and~\ref{supp:verification}. Archive filenames refer to the separate
machine-readable certificate packages listed in
Sections~\ref{bal:sec:records} and~\ref{unbal:app:inventory}.

\subsection{Verification scope of the extension}

The extension uses the unrestricted inequality $\zeta\ge R_\phi$ of
Theorem~\ref{bal:thm:global-unrestricted-target}, which is stronger than
the balanced-information conclusion of Part~\ref{part:balanced}.
Together with the extension's analytic arguments and computational lemmas,
this proves the hybrid comparison. The verification scope of the balanced
inputs, including the radial ledger, is stated in
Section~\ref{bal:sec:verification}; the scope of the extension certificates
is stated in Section~\ref{supp:verification}.

\subsection{Finite-cover verification}

Each finite cover partitions an explicitly stated compact parameter box
with exact rational bisections. A leaf is accepted only by a proved
inequality evaluated with outward interval arithmetic, or by a proved
region that owns its entire feasible intersection. The verifier reconstructs
the full binary tree, checks every branch, and re-evaluates every leaf at
70 decimal working digits. An entropy inverse approximation merely proposes
a bracket; directed signs prove the bracket. No sampled point is used as a
substitute for a leaf inequality.

Analytic tails treat the limits that a finite compact box omits. Closed
overlapping owners treat interfaces and equality faces. This distinction
between continuum estimates and finite certificates is part of the proof,
and is recorded explicitly in the final assembly.

\subsection{Internal proofs of the structural inputs}
\label{supp:four-moment-provenance}
The following results and their analytic arguments are included in this
manuscript. The references below locate their statements, proofs, and
computational lemmas.

\paragraph{Four-moment lower bound and reflection.}
Theorem~\ref{bal:thm:four-moment-lower-bound} proves $\zeta\ge L_4$
by the pointwise reflection comparison and Jensen's inequality for the
closed convex extensions of $F$ and $g$. The reflection proof is in
Supplementary Section~\ref{bal:imp:reflection}; the joint-convexity proof
and its endpoint closure are in
Supplementary Section~\ref{bal:imp:jointconvexity}. Individual mean
reflection for the pure gap is proved separately in
Lemma~\ref{bal:lem:pure-gap-reflection}, using the two lifts defined in
Section~\ref{bal:sec:orientation}. It is not used as a symmetry of
the unrestricted Bellman value $\zeta$.

\paragraph{Zero-entropy and maximal-entropy boundaries.}
Theorem~\ref{bal:thm:zero-target} proves the zero-entropy Bellman
inequality in addition form by characterizing all finite-cost laws on
that boundary. The zero-entropy limits of the distinct pure-gap
expression on the fixed-sum seam are proved in
Lemma~\ref{bal:lem:seam-zero}.
Lemma~\ref{bal:lem:maximal-continuity} gives the maximal-entropy
continuation, using the finite endpoint limits established in
Supplementary Section~\ref{bal:imp:jointconvexity}.

\paragraph{Mean-interior reduction.}
The proof of Theorem~\ref{bal:thm:e8-main} is completed in
Supplementary Section~\ref{bal:imp:e8}. The stationarity equations
\eqref{bal:imp:e8:eq:stationarity} and ratio identity
\eqref{bal:imp:e8:eq:ratio-identity} contradict the strict
$Q$-inequality of Theorem~\ref{bal:imp:e8:thm:eq8}.
The proof treats the origin, axis strips, compact regions, and infinite
tails, then joins them in the global case split. Its hypotheses fix
positive strictly submaximal entropies; zero entropy, maximal entropy,
and the retained interface have their separate boundary arguments.

\paragraph{Half-mean exclusion and the deterministic-cap intersection.}
Supplementary Theorem~\ref{bal:aux:thm:D0} proves the scalar inequality
$D_0(s)>0$ by origin, compact, and tail estimates.
Lemma~\ref{bal:aux:lem:halfmean} uses it to exclude a smooth half-mean
local minimum under strict marginal entropy constraints. At a
deterministic-cap intersection, the exact correction identity is
\eqref{bal:eq:cap-3}; the proof of
Theorem~\ref{bal:thm:m1m2} establishes its positive sign by the derivative
estimates and overlapping value regions. Together with clause (iv),
this proves clause (v) of Theorem~\ref{bal:hyp:SCplus}.

\paragraph{Small-seam estimate.}
Supplementary Section~\ref{supp:small-boundary-proof} proves the
same-side bound $R_\phi\le4\Delta_H(a,c)$ used in
Theorem~\ref{bal:thm:small-boundary}, where
$\Delta_H(a,c)=H_2((a+c)/2)-[H_2(a)+H_2(c)]/2$.
It applies on $0\le a\le c\le1/100$; the fixed-sum seam lies in this
domain because $c\le S<1/100$.

\subsection{Opposite-side partition provenance}
Discovery completed with 58,989 nodes and 29,495 leaves, with no
pending or unresolved leaves. The final replay reconstructed the
entire rational partition and evaluated every leaf at 70 decimal
working digits. Its source and primitive digest was checked before
and after the replay. The frozen acceptance digest is
\begin{center}\small
\path{a8ac3a0f2966c4cd52bac62c7abaa46b98e089d9859f5d2f0d0439721ef503a1}.
\end{center}
The order in which discovery accumulated or strengthened acceptance
rules is immaterial: every final leaf was checked against this one
frozen acceptance source.

\subsection{Opposite-side certificate file locations}
\sourcefile{opposite/compact/OUTER_OPPOSITE.py}
\sourcefile{opposite/compact/OUTER_OPPOSITE_RESULT.json}
\sourcefile{opposite/compact/OUTER_OPPOSITE_REPLAY.json}

\subsection{Opposite-side partition sizes}
\begin{center}
\begin{tabular}{lrr}
\toprule
Component & Leaves & Replay digits\\
\midrule
Extreme opposite-side strip & 17 & 70\\
Remaining opposite-side compact domain & 29,495 & 70\\
\bottomrule
\end{tabular}
\end{center}

\subsection{Dependencies of the domain-exhaustion argument}
The domain-exhaustion argument uses the established balanced results and
the central theorem. Their verification scopes are recorded in
Sections~\ref{bal:sec:verification} and~\ref{supp:verification}.

\subsection{Small-mean verification records}
The complete directed certificate for \eqref{unbal:ar:small_W} has 20 leaves.
On a rational interval \([r_-,r_+]\), the checker uses the proved
monotonicity of \(\gamma\), directed arithmetic, and the continuous
value \(C_n(1)=L\) to enclose \(W\). Every strict comparison passes
its full 70-digit replay, as do the fixed comparisons
\eqref{unbal:ar:small_fixed}. Exact subdivision paths reconstruct a complete
partition of \([1/6,1]\). The certificate is
\texttt{reduction/SMALL\_MEAN\_CORNER\_RESULT.json}; its checker and
replay are in the same directory.

\subsection{Factor-sixteen certificate implementation}
The root is $(q,x)\in[2/5,1/2]\times[16,128]$ in the proof of
Theorem~\ref{unbal:ar:parent16}. The directed checker
\texttt{parent/PARENT16.py} proves positivity of
\eqref{unbal:ar:parent16_box} on a complete 69-leaf
partition and replays all leaves and fixed comparisons at 70 digits.
Inverse and contact roots are enclosed by verified signs.

\subsection{Factor-eight certificate implementation}
The checker \texttt{reduction/eight\_global/EIGHT\_RATIO.py} verifies
that at least one of \eqref{unbal:ar:parent_log_box} and
\eqref{unbal:ar:parent_direct_box} is at least \(7/10\) on every leaf of a
complete 77-leaf partition. All leaves pass their full 70-digit replay.
Its fixed-check record also verifies the analytic tail and the radial
constant. No restriction such as \(q\le2/5\) is imposed in this check.

\subsection{Additional parent-comparison certificate}
The directed checker \texttt{HIGH\_Q\_PARENT.py}, in
\texttt{reduction/eight\_global}, proves strict positivity of
\eqref{unbal:ar:highq_box} on
$(q,E)\in[2/5,1/2]\times[1/40,1/25]$ using a complete 59-leaf
partition, with a full 70-digit replay.

\subsection{Low-entropy certificate implementation}
The directed checker \texttt{residual/GLOBAL\_LOW\_ENTROPY.py}
proves \eqref{unbal:ar:global_lowentropy_box} on a complete partition
of five rational intervals.
All five leaves pass their full 70-digit replay; the smallest lower
enclosure exceeds \(0.1083448150583482\).

\subsection{Entropy-split certificate and endpoint convention}
The directed certificate \texttt{GLOBAL\_SPLIT.py} proves
\eqref{unbal:ar:split_loss}
on \(0<E\le11/200\), \(0\le q\le11/25\), using two leaves
and a full 70-digit replay. All expressions at \(q=0\) use the
continuous extension specified in the proof of Theorem~\ref{unbal:ar:eight}.

\subsection{Auxiliary scalar partitions}
The finite scalar comparisons supporting the outer cover comprise
\(20+69+77+2+59+5+17=249\) supporting leaves, all covered by complete
70-digit directed replays, in addition to the retained proofs they cite.

\subsection{Central-square proof modules}
The central-square theorem, Theorem~\ref{unbal:thm:central},
is proved by the component estimates assembled in Section~\ref{unbal:sec:central}.
The complete constituent proof modules are reproduced in
Section~\ref{unbal:app:modules}; their original checkers, partitions and replay
records are contained in the retained artifact archive.

\subsection{Central-square partition records}
The retained four final component partitions have 332, 50, 2,425, and
4,730 leaves, with completed 70-digit replays; their earlier components and
fixed constants are also included in the archive. These records accompany
the region decomposition in Section~\ref{unbal:sec:central} of the main text.

\subsection{Interpretation of the opposite-side certificate labels}
A historical acceptance label may denote the low-entropy hybrid region
\eqref{unbal:eq:opp-lowentropy-owner} rather
than its formerly named bare $R_\psi$ comparison. During a verification run, a box
wholly in \eqref{unbal:eq:opp-lowentropy-owner} may retain its historical label
and be accepted with zero margin. This establishes $\zeta\ge R_B$;
it does not assert that the old stronger $R_\psi$ comparison was
reevaluated there. The final theorem requires only the hybrid target.

\subsection{Extreme-mean strip implementation}
Use the constants defined in the proof of Theorem~\ref{unbal:ar:extreme_strip}.
The checker \texttt{opposite/BOUNDARY\_STRIP.py} starts with the
exactly adjacent intervals
\[
 [K_*-A_*,2K_*],\quad
 [2^{-k},2^{-k+1}]\ (k=12,11,\ldots,3),\quad
 [1/4,2/5],\quad[2/5,1/2].
\]
The first twelve use the sharper target upper bound, and the last uses
\eqref{unbal:ar:strip_parent}. Bisecting only failed comparisons produces a
complete certificate with 17 partition cells. Every partition cell passes its full 70-digit verification run,
which separately checks each binary partition and adjacency of the
initial roots. 
\subsection{Internal certificate identifiers}
\begin{longtable}{lp{0.7\textwidth}}
\toprule Identifier & Mathematical result\\
\midrule\endhead
LB-1 & global four-moment bound (\ref{bal:thm:four-moment-lower-bound})\\
SB-1 & small-boundary theorem (\ref{bal:thm:small-boundary})\\
ZT-1 & zero-entropy theorem (\ref{bal:thm:zero-target})\\
E8-1 & mean-interior reduction (\ref{bal:thm:e8-main})\\
EE-1 & equal-entropy lemma (\ref{bal:lem:global-equal-entropy})\\
EM-1 & equal-mean lemma (\ref{bal:lem:global-equal-mean})\\
DB-1 & double-cap theorem (\ref{bal:thm:double-cap})\\
HF-1 & cap--half-mean theorem (\ref{bal:thm:cap-half-mean})\\
DC-1 & double-cap entropy bound (\ref{bal:prop:double-cap-entropy})\\
FM-1 & maximal-entropy continuity lemma (\ref{bal:lem:maximal-continuity})\\
OM-1 & domain-exhaustion proposition (\ref{bal:prop:owner-exhaustion})\\
SK-1 & seam-closure theorem (\ref{bal:thm:seam-closure})\\
PG-1 & pure-gap theorem (\ref{bal:thm:pure-gap})\\
BT1\texttt{+} & unrestricted Bellman theorem (\ref{bal:thm:global-unrestricted-target})\\
FW-1 & Bellman-to-information theorem (\ref{bal:thm:unrestricted-framework})\\
SC\texttt{+} & seam and endpoint theorem (\ref{bal:hyp:SCplus})\\
ZE-seam & zero-entropy seam lemma (\ref{bal:lem:seam-zero})\\
HM-1 & half-mean estimate (Theorem~\ref{bal:thm:boundary-estimates}(g))\\
CE-stat & stationary cap reduction (Theorem~\ref{bal:thm:boundary-estimates}(h))\\
RA-stat & radial stationary-point estimate (Theorem~\ref{bal:thm:boundary-estimates}(i))\\
MI-1 & middle-cap inequality (Theorem~\ref{bal:thm:boundary-estimates}(j))\\
\bottomrule
\end{longtable}

\clearpage


\addcontentsline{toc}{part}{References}
\begin{thebibliography}{99}

\bibitem{KumarCourtade2013}
G.~R.~Kumar and T.~A.~Courtade,
\emph{Which Boolean functions are most informative?},
in Proc. IEEE Int. Symp. Inf. Theory (ISIT), 2013, pp.~226--230.

\bibitem{chandar2014}
V. Chandar and A. Tchamkerten, ``Most informative quantization functions,'' unpublished note, presented at the Inf. Theory Appl. Workshop, San Diego, CA, USA, Feb. 2014. [Online]. Available: \url{https://perso.telecom-paristech.fr/tchamker/CTAT.pdf}

\bibitem{CourtadeKumar2014}
T.~A.~Courtade and G.~R.~Kumar,
\emph{Which Boolean functions maximize mutual information on noisy inputs?},
IEEE Trans. Inf. Theory, vol.~60, no.~8, pp.~4515--4525, 2014.


\bibitem{ordentlich2016improved}
O.~Ordentlich, O.~Shayevitz, and O.~Weinstein,
\emph{An improved upper bound for the most informative Boolean function
conjecture}, in Proc. IEEE Int. Symp. Inf. Theory (ISIT), pp.~500--504, 2016.

\bibitem{samorodnitsky2016entropy}
A.~Samorodnitsky,
\emph{On the entropy of a noisy function},
IEEE Trans. Inf. Theory, vol.~62, no.~10, pp.~5446--5464, 2016.

\bibitem{anantharam2017hellinger}
V.~Anantharam, A.~Bogdanov, A.~Chakrabarti, T.~S.~Jayram, and C.~Nair,
\emph{A conjecture regarding optimality of the dictator function under Hellinger distance},
in Proc.\ Information Theory and Applications Workshop, 2017.

\bibitem{li2021boolean}
J.~Li and M.~M\'edard,
\emph{Boolean functions: noise stability, non-interactive correlation
distillation, and mutual information}, IEEE Trans. Inf. Theory, vol.~67,
no.~2, pp.~778--789, 2021.

\bibitem{gon21}
A.~Gohari and C.~Nair,
\emph{Outer bounds for multiuser settings: The auxiliary receiver approach},
IEEE Transactions on Information Theory, vol.~68, no.~2, pp.~701--736, 2022.

\bibitem{yu2023phi}
L.~Yu,
\emph{On the $\Phi$-stability and related conjectures},
Probability Theory and Related Fields, vol.~186, nos.~3--4, pp.~1045--1080, 2023, Springer.

\bibitem{barnes2020courtade}
L.~P.~Barnes and A.~\"Ozg\"ur,
\emph{The Courtade--Kumar most informative Boolean function conjecture and a symmetrized Li--M\'edard conjecture are equivalent},
in Proc.\ 2020 IEEE International Symposium on Information Theory (ISIT), pp.~2205--2209, 2020, IEEE.

\bibitem{chen2024optimality}
Z.~Chen and C.~Nair,
\emph{On the optimality of dictator functions and isoperimetric inequalities on Boolean hypercubes},
in Proc.\ 2024 IEEE International Symposium on Information Theory (ISIT), pp.~3380--3385, 2024, IEEE.



\bibitem{yu2024localoptimalitydictatorfunctions}
L.~Yu,
\emph{Local Optimality of Dictator Functions with Applications to Courtade--Kumar and Li--M\'edard Conjectures},
arXiv preprint arXiv:2410.10147, 2024; extended version v5, April 7, 2026,
\url{https://arxiv.org/abs/2410.10147v5}.

\bibitem{javanmard2026progresscourtadekumarconjectureoptimal}
A.~Javanmard and D.~P.~Woodruff,
\emph{Progress on the Courtade--Kumar conjecture: Optimal high-noise
entropy bounds and generalized coordinate-wise mutual information},
arXiv preprint arXiv:2601.09679, 2026,
\url{https://arxiv.org/abs/2601.09679}.

\bibitem{KyTran2026}
V.~K. Ky and T.~Tran,
\newblock \emph{Dictators are most informative},
\newblock manuscript in preparation, 2026.

\bibitem{pichler2018dictator}
G.~Pichler, P.~Piantanida, and G.~Matz,
\emph{Dictator functions maximize mutual information},
The Annals of Applied Probability, vol.~28, no.~5, pp.~3094--3101, 2018.

\bibitem{Johansson2017}
F.~Johansson,
\newblock Arb: Efficient Arbitrary-Precision Midpoint-Radius Interval Arithmetic.
\newblock \emph{IEEE Transactions on Computers}, 66(8):1281--1292, 2017.

\bibitem{ChenGohariNair2025}
Z.~Chen, A.~Gohari, and C.~Nair,
\emph{A differential equation approach to the most-informative Boolean
function conjecture},
in Proc.\ 2025 IEEE Int.\ Symp.\ Inf.\ Theory (ISIT),
2025, pp.~1--6,
doi:~10.1109/ISIT63088.2025.11195377.
Full version available as
arXiv preprint arXiv:2502.10019 [cs.IT],
\url{https://arxiv.org/abs/2502.10019}.

\bibitem{AnantharamGohariKamathNair2013}
V.~Anantharam, A.~Gohari, S.~Kamath, and C.~Nair,
\emph{On hypercontractivity and the mutual information between Boolean
functions},
in Proc.\ 51st Annual Allerton Conference on Communication, Control,
and Computing (Allerton), 2013, pp.~13--19,
doi:~10.1109/ALLERTON.2013.6736499.

\bibitem{stellar-colloseum-paper}
H.~Lin, D.~P.~Woodruff, Y.~Deng, J.~Mao, S.~Zuo, and V.~Mirrokni,
\emph{Stellar Colosseum: A many-agent harness for long-horizon research
in mathematics and theoretical computer science},
arXiv preprint arXiv:2609.15983, 2026.
\url{https://arxiv.org/abs/2609.15983}

\bibitem{acceleration-paper}
D.~P.~Woodruff, V.~Cohen-Addad, L.~Jain, J.~Mao, S.~Zuo,
M.~Bateni, S.~Branzei, M.~P.~Brenner, L.~Chen, Y.~Feng,
L.~Fortnow, G.~Fu, Z.~Guan, Z.~Hadizadeh, M.~T.~Hajiaghayi,
M.~JafariRaviz, A.~Javanmard, Karthik~C.~S.,
K.-i.~Kawarabayashi, R.~Kumar, S.~Lattanzi, E.~Lee, Y.~Li,
I.~Panageas, D.~Paparas, B.~Przybocki, B.~Subercaseaux,
O.~Svensson, S.~Taherijam, X.~Wu, E.~Yogev,
M.~Zadimoghaddam, S.~Zhou, Y.~Matias, J.~Manyika, and
V.~Mirrokni,
\emph{Accelerating scientific research with Gemini: Case studies and
common techniques},
arXiv preprint arXiv:2602.03837, 2026.
\url{https://arxiv.org/abs/2602.03837}

\bibitem{AGY26teamwork}
Google Antigravity Team,
\emph{Teamwork: When AI becomes a research partner},
Google Antigravity Blog, August 27, 2026.
\url{https://antigravity.google/blog/teamwork-when-ai-becomes-a-research-partner}

\bibitem{mahdavifar2026}
H.~Mahdavifar and A.~Beirami,
``The Most Informative Bit and Beyond: A Proof of the Courtade--Kumar
Conjecture and Multibit Extensions,''
\emph{arXiv preprint} arXiv:2609.26444 [cs.IT], 2026.
\url{https://arxiv.org/abs/2609.26444}
\end{thebibliography}
\end{document}